\documentclass[twocolumn,iop]{emulateapj}
\usepackage[applemac]{inputenc} 
\usepackage{hyperref}
\usepackage{appendix}
\usepackage{longtable}
\usepackage{subfloat} 

\usepackage{amsmath, amssymb, rotating, graphicx, natbib}
\hypersetup{colorlinks, citecolor=blue}

\slugcomment{Accepted for publication on ApJ}

\shorttitle{Recoiling Supermassive Black Holes: a search in the Nearby Universe}
\shortauthors{Lena et al.}

\begin{document}
\bibliographystyle{plainnat}

\title{Recoiling Supermassive Black Holes: a search in the Nearby Universe}

\author{
D. Lena\altaffilmark{1}, 
A. Robinson\altaffilmark{1}, 
A. Marconi\altaffilmark{2},  
D. J. Axon\altaffilmark{1,3,$\dag$},
A. Capetti\altaffilmark{4},\\
D. Merritt\altaffilmark{1,5} \&
D. Batcheldor\altaffilmark{6}}
\email{dxl1840@g.rit.edu}

\altaffiltext{1}{School of Physics and Astronomy, Rochester Institute of Technology, 84 Lomb Memorial Drive, Rochester, NY 14623-5603, USA}
\altaffiltext{2}{Dipartimento di Fisica e Astronomia, Università degli Studi di Firenze, Largo E. Fermi 2, 50125, Firenze, Italy}
\altaffiltext{3}{School of Mathematical and Physical Sciences, University of Sussex, Sussex House, Brighton, BN1 9RH, UK}
\altaffiltext{4}{INAF - Osservatorio Astronomico di Torino, Strada Osservatorio 20, 10025 Pino Torinese, Italy}
\altaffiltext{5}{Center for Computational Relativity and Gravitation, Rochester Institute of Technology, Rochester, New York 14623, USA}
\altaffiltext{6}{Department of Physics and Space Sciences, Florida Institute of Technology, 150 W. University Blvd, Melbourne, FL 32901, USA}
\altaffiltext{$\dag$}{1951 - 2012}

\begin{abstract}

The coalescence of a binary black hole can be accompanied by a large
gravitational recoil due to anisotropic emission of gravitational waves.
A recoiling supermassive black
hole (SBH) can subsequently undergo long-lived oscillations in the
potential well of its host galaxy, suggesting that offset SBHs may be
common in the cores of massive ellipticals. We have analyzed HST
archival images of 14 nearby core ellipticals, finding evidence for
small ($\lesssim 10$\,pc) displacements between the AGN (locating the
SBH) and the center of the galaxy (the mean photocenter) in 10 of them.
Excluding objects that may be affected by large-scale isophotal
asymmetries, we consider six galaxies to have detected displacements, including
M87, where a displacement was previously reported by \citet{batch}. In
individual objects, these displacements can be attributed to
residual gravitational recoil oscillations following a major or minor
merger within the last few Gyr. For plausible merger rates, however,
there is a high probability of larger displacements than those observed,
if SBH coalescence took place in these galaxies. Remarkably, the AGN-photocenter
displacements are approximately aligned with the radio source axis in four
of the six galaxies with displacements, including three of the four
having relatively powerful kpc-scale jets.
This suggests intrinsic asymmetries in radio jet power as a possible
displacement mechanism, although approximate alignments are also expected for
gravitational recoil.  Orbital motion in SBH binaries and
interactions with massive perturbers can produce the observed
displacement amplitudes but do not offer a ready explanation for the
alignments.

\end{abstract}


\section{Introduction}
\setcounter{footnote}{0}
\label{sec: intro}
Supermassive black holes (SBHs) have long been identified as the engines of active
galactic nuclei (AGNs). More recently, compelling evidence has been amassed that they are
present in the centers of almost all galaxies \citep{krich,FerrareseFordRev05} and furthermore 
that the growth of the SBH is intimately linked to galaxy evolution 
(\citealp{FerrareseM00, GBBetAl00}; see \citealp{Alexander2012} for a recent review).

As a consequence of energy losses due to dynamical friction, SBHs are generally expected
to reside at the large-scale potential minimum of the host galaxy. Nevertheless, there are
several mechanisms that seem capable of displacing the SBH from its equilibrium position
\citep[hereafter B10]{batch}. Most recently, interest has focused on 
gravitational recoil resulting from the coalescence of an SBH-SBH binary
\citep[e.g.,][]{Fav04, Mer04, camp06, Bak06}. Other possibilities include orbital motion
of SBH-binaries \citep{binary}, asymmetric radio jets \citep{jet} and interactions with
massive perturbers such as globular clusters, or massive molecular clouds.

The first two of these mechanisms, in particular, might be expected  to occur as an
inevitable consequence of galaxy evolution. Cold dark matter hierarchical cosmologies
predict that galaxies experience multiple mergers during their lifetime
\citep[e.g.,][]{ColesLucchin, Springel05}, with minor mergers being more common than major mergers \citep[e.g.][]{HilzNaabOst2012}. 
When two galaxies merge, their central SBHs are expected to form a binary system
\citep{BBR80} at the center of the merged system due to dynamical friction.
Further tightening of the binary ensues through three-body interactions with stars,
or gravitational drag from gas \citep{mm, mayer}, until finally, energy loss due to 
gravitational wave emission irreversibly drives the two SBHs to coalescence\footnote{\citet{BBR80} 
suggested that the evolution of a binary SBH might stall at separations of $\sim 1$ \,pc once the supply of ``loss-cone stars" (stars
intersecting the orbit of the binary) was exhausted. However, gas dynamical torques
 can drive continued shrinking of the binary orbit \citep[e.g.,][]{Esc05, Cua09, mayer}, and recent
\textit{N}-body simulations suggest that hardening of the binary due to interactions with stars can 
be efficient in non spherical potentials \citep{PretoBBS11, GualandrisM11, VAM14}.}.

In the final stage of SBH-binary coalescence, anisotropic emission of gravitational waves
will, in general, impart a recoil velocity to the coalesced object \citep{Bek73}. 
Following a recent breakthrough allowing the orbits of spinning black holes to be computed to
coalescence \citep{Pre05, camp06, Bak06}, numerical relativity simulations have produced
recoil velocities $v \sim 10^3$ km s$^{-1}$, even reaching  $\sim 5\times 10^3$ km s$^{-1}$ for
certain spin configurations \citep{camp, Gon07, TM07, LZ11, LoustoHangUp}. Such velocities
would cause large displacements of the coalesced SBH from the center of the galaxy or, in
the most extreme cases, eject it entirely from the host \citep{Mer04, campA, Vol10}.

Recoils exceeding the escape velocity of the host galaxy are expected to be relatively
rare \citep[e.g.,][hereafter L12]{LoustoHangUp} and, more frequently, the SBH will undergo damped
oscillations in the galaxy potential. $N$-body simulations \citep[hereafter GM08]{gualam} have shown that
moderately large kicks ($\sim$ few $10^{2}$ km s$^{-1}$, sufficient to eject the SBH from
the core\footnote{The core (or break) radius, $r_{c}$, is defined as the radius where the
projected surface brightness distribution of the galaxy becomes shallower than the inward extrapolation
of the outer surface brightness profile.} but not from the galaxy) result in long-lived
oscillations which damp on a timescale $\Delta t \sim 1$ Gyr. GM08
identified three dynamical phases for the particular case where the SBH is initially
displaced beyond the core radius ($r_{c} \sim 10^{2}$ pc). In ``phase I" the SBH
undergoes damped harmonic motion at amplitudes greater than the core radius. These
oscillations are damped relatively quickly ($\sim 10^7$\,yr) by dynamical friction.
``Phase II" is characterized by oscillations of amplitude $\sim 10 - 100$\,pc 
between the SBH and the stellar core about the center of mass, which persist for up to
$\sim 1$ Gyr. Finally, in ``phase III", the SBH has reached thermal equilibrium
with the surrounding stars and experiences Brownian motion, signifying the end of the
event.

Large gravitational recoil kicks would have a variety of observable consequences. 
In the case of SBHs associated with AGNs, most of the accretion disk and broad 
emission line gas will remain bound for recoils with velocities $v_\mathrm{kick} \lesssim 10^3$ \ km\,s$^{-1}$
and such systems might be observed as a displaced, or velocity-shifted AGNs \citep[e.g.,][]{Mer04,
Mad04, Loeb07, Bon07, ZanottiRDZP10}. Stars moving with orbital velocities $v < v_\mathrm{kick}$ 
would also remain bound to the SBH after the kick, so that a recoiling SBH would be associated with
 a ``hypercompact stellar system” of high internal velocity dispersion \citep{MSK2009}.

Possible spectroscopic signatures of
recoiling SBHs have been identified in a few individual active galaxies \citep{KZL08,
SRS09, Andy10, Stein12}, which would indicate large ($\gtrsim 10^3$ \ km\,s$^{-1}$) kick
velocities. In addition, systematic searches of large SDSS AGN samples have revealed $\sim
100$ objects that exhibit velocity shifts $\sim 10^3$ \ km\,s$^{-1}$ between the broad and
narrow lines \citep{Bon07, Erac11, Tsal11}. Some of these may be recoiling SBHs, or SBH
binaries, but alternative explanations for the shifts involving extreme broad line region kinematics
cannot be excluded. Perhaps the most interesting case is that
of CIS-42, a galaxy which hosts two compact optical sources separated by $\approx
2.5$ kpc and also exhibits a large velocity shift ($\sim 1300$\,km\,s$^{-1}$) between
broad and narrow  H$\beta$ emission lines \citep{Civano10, Civano12, Blecha12}. 

The gravitational recoil candidates discovered to date have been identified on the basis of 
anomalously large  velocity shifts  ($\gtrsim 10^3$ \ km\,s$^{-1}$) or a large spatial offset
($\gtrsim 1$\,kpc). If they are indeed recoiling SBHs, we are observing them during the
early stages in their dynamical evolution following a large kick; that is, during
the large amplitude phase I oscillations of GM08, if the SBH remains bound to the galaxy.
However, special configurations of the progenitor binary are required to produce
large kicks, which are therefore expected to be relatively rare events (L12)
and furthermore, the subsequent large amplitude phase I oscillations 
are relatively short-lived. 

On the other hand, the $N$-body simulations of GM08 predict that, following a
recoil kick sufficiently large to remove the SBH from the galaxy core, the decaying
oscillations persist in phase II for $\sim$ 1 Gyr, comparable with the time between
galaxy mergers \citep[hereafter H10]{Hopkins2010}. This suggests that low amplitude SBH displacements resulting from
post-merger SBH binary coalescence should be relatively common in the cores of bright
elliptical galaxies. Thus, rather than searching for short-lived phase I oscillations
following rare large kicks, an alternative approach to studying the incidence of SBH
binary coalescence is to look for phase II oscillations in nearby ellipticals.  

Spectroscopic identification of phase II oscillations would be extremely difficult, if
not impossible, as the associated velocity shifts would be comparable with gas motions
due to rotation, or flows or turbulence driven by the AGN, or starburst activity. However,
it is feasible to directly detect small-amplitude displacements between the AGN and the
galaxy photocenter in high spatial resolution images of nearby galaxies. We recently 
analyzed archival Hubble Space Telescope Advanced
Camera for Surveys images of the active giant elliptical M87 and found a projected offset
of $6.8\pm0.8$\,pc ($\approx 0\farcs1$) between the stellar photocenter and the AGN
(B10; we note, however, that \citealt{Geb11} did not find a significant offset in their
near-infrared integral field spectroscopy data).

Here we describe a photometric analysis of archival HST images of 14 nearby (d\textless100
Mpc), bright ellipticals hosting low-luminosity AGNs, with the aim of directly measuring
spatial offsets comparable with the core radius of the galaxy (typically $< 500$\,pc).
Our method consists in measuring the relative positions of the AGN point source and the
photocenter of the inner 2-3 kpc of the galaxy, assuming that the former marks the SBH
position and the latter the minimum of the galactic potential.  As long as the AGN is
bright enough to be easily detected and faint enough that it does not overwhelm the host
galaxy, it is possible to perform standard photometric analysis to locate the photocenter.
The ``displacement" is then computed as the relative distance between the AGN and
photocenter.

 The plan of this paper is as follows: in \textsection \ref{sec: sample} we describe the
 sample; \textsection \ref{sec: analysis} describes the analysis methods used to determine
 the photocenter, SBH position and offset, and their uncertainties; in \textsection
 \ref{sec: results} the measured displacements are presented for each galaxy. In
 \textsection \ref{sec: disc}, we discuss the results in the context of gravitational
 recoil and  other possible displacement mechanisms. Finally, our conclusions are
 summarized in  \textsection \ref{sec: conclusion}. The results for selected galaxies are
 discussed in more detail in an Appendix.

\section{The sample} \label{sec: sample}
 
The sample selected for this study consists of 14 nearby, regular, core elliptical
galaxies which host AGNs and for which HST images obtained with ACS, NICMOS2, WFPC2 or WFC3 are
available in the Hubble Legacy Archive (HLA).

HST observations have revealed that the central light distribution in
nearly all nearby early type galaxies 
can be described by a singular surface brightness profile:
$\Sigma(r) \sim r^{-\gamma}$ as $r \rightarrow 0\farcs1$, where $\gamma \geq 0$. The slope of
the innermost surface brightness distribution
is bimodal in the sense that there is a 
paucity of galaxies which have $0.3 < \gamma < 0.5$. Hence,
early type galaxies have been classified in two families: 
``core galaxies", which have a shallower ($\gamma \leq 0.3$) inner cusp within
a ``break'' (or ``core'') radius r$_{c}\sim 100$\,pc,
and ``power law galaxies", which have a steep cusp ($\gamma \geq 0.5$) continuing 
into the HST resolution limit \citep{lau95}. These classifications 
correlate with several other galaxy properties \citep[e.g.,][]{Faber1997}, 
in particular, core galaxies are on average more luminous than power-law galaxies, 
with essentially all bright ($M_V <-22$) ellipticals having cores. 

Core galaxies are promising systems in which to search for SBH binaries or SBHs
displaced by gravitational wave-induced recoils. They are bright ellipticals
which are often the dominant components of clusters or groups and are thus likely to have
experienced a recent major merger leading to the formation of an SBH binary.
The shallow inner surface brightness profile indicates a mass deficit relative to that implied
by inward extrapolation of the steeper brightness profile prevailing at larger radii.
This is predicted as a natural consequence of depopulation of the
inner region of the galaxy by
3-body interactions between an SBH binary and stars
crossing the orbit of the binary \citep[``loss-cone stars"][]{mm}. Therefore, ``flat'' 
inner surface brightness profiles have been proposed as a clear footprint of the formation 
of a tightly bound SBH binary \citep[e.g.,][]{Faber1997, Merritt06radii}.

In addition, core galaxies tend to have regular photometric
structures, and hence well-defined photocenters. 
In order to accurately locate the
position of the SBH, we further require that each galaxy hosts a low luminosity AGN
visible as a point source in the HST images.

In order to resolve the low amplitude offsets between photocenter and AGN expected in the phase II oscillations predicted by GM08, we select only galaxies within 100 Mpc.

The sample analyzed in this work was extracted from a sample of 29 core elliptical galaxies previously studied
by Capetti and Balmaverde, in a series of three articles \citep[hereafter CB05, BC06 and CB06 respectively]{CapBal05,
Bal06, CapettiB06}. CB05 compiled a radio-flux limited sample of AGNs in early type hosts by
selecting radio-detected sources from the VLA surveys of \citet{1991WH} and
\citet{Sadler1989}. The former is a northern sample of 216 early-type galaxies extracted
from the CfA redshift survey \citep{quattro}, with the following criteria:  declination
$\delta_{1950} \geq 0 $; photometric magnitude $B \leq 14$; heliocentric velocity $\leq
3000$ km s$^{-1}$; morphological Hubble type $T \leq -1$. The latter is a similar southern
sample of 116 E and S0 galaxies with declination $-45 \leq \delta \leq -32 $. Both were observed at 5 GHz with a flux limit of $\sim 1$ mJy.  

The 65 objects with available archival HST images were classified as core or power-law
galaxies on the basis of the slopes of their nuclear brightness profile as obtained by
fitting a broken power law (the so-called ``Nuker law''; \citealt{lau95}; CB05; CB06). 
For the purpose of this study, we focus on the core galaxies listed in Table 1 of BC06
and impose two additional selection criteria, based on visual inspection of HST images:
(i) the presence of an optically bright central point-like source (a low luminosity AGN);
(ii) the absence of heavy nuclear obscuration and other photometric
irregularities. 

The 14 galaxies selected for this study are listed
in Table \ref{tab: distanze} together with their distances, core radii and details of the
archival data sets that were retrieved for photometric analysis. 
The sample includes two galaxies (NGC 4696 and 5419)
that exhibit double nuclei, separated by $\sim 0\farcs25$. In these cases, it is not
known which of the two point sources is the AGN and therefore we assume that the brightest
is the AGN (although we have measured displacements relative to both).

The radio source
properties, including the total power at 5 GHz and the position angle of any extended
structure, are summarized in Table \ref{tab: radio}.

\begin{deluxetable*}{ ll | cc cc c lc c l}
\tablecaption{THE SAMPLE}
\tablewidth{0pt}
\tablehead{
\multicolumn{2}{c}{Galaxy}		& \multicolumn{2}{c}{Distance} 			& \colhead{r$_{c}$} 	& \colhead{r$_{h}$}	& \colhead{Instrument} 	& \colhead{Pixel Scale} 	 & \colhead{ID} &\\
\colhead{(1)}	& \colhead{(2)}		& \colhead{(3)}		&	\colhead{(4)} 	& \colhead{(5)}		& \colhead{(6)}		& \colhead{(7)}			& \colhead{(8)} 			 & \colhead{(9)}	&\\
}

\startdata
                     	&				&  				&				&  				&				& 					&					&\\
  NGC 1399	&				& $18\pm1$		&	sbf			& 2.19 (189)		& 	0.99 (86)		& WFPC2/PC/F606W	& 	0.05 				& 8214 \\
			&				& 				& 				&				&				& WFPC2/PC/F814W	&	0.05				& 5990 \\
			&				&				&				&				&				& WFC3/IR/F110W		&	0.09				& 11712\\
			&				&				&				&				&				& WFC3/IR/F160W		&	0.09				& 11712\\
 NGC 4168	& UGC 7203		& $31\pm6$		& 	sbf			& 2.02 (303)		&	0.09 (14)		& WFPC2/PC/F702W	&	0.05				& 6357\\
 NGC 4261	& UGC 7360		& $30\pm2$		& 	sbf			& 1.62 (237)		&	0.428 (62)		& NICMOS2/F160W		& 	0.05				& 7868\\
 NGC 4278  	& UGC 7386		& $18\pm1$		& 	sbf			& 0.97 (83)		&	0.334 (29)		& ACS/WFC/F850LP  	& 	0.05				& 10835\\ 
 		 	&	 			& 				& 				&				&				& WFPC2/PC/F814W	&	0.05				& 5454\\ 	
 		 	&	 			& 				& 				&				&				& NICMOS2/F160W		& 	0.05				& 7886 \\
 NGC 4373 	&				& $40\pm2$		& 	sbf			& 1.19 (269)		&	0.167 (32) 	& WFPC2/PC/F814W	&	0.05				& 5214\\
 NGC 4486	& UGC 7654 (M87)	& $16\pm1$		& 	sbf			& 9.41 (733)		&	1 (78) 		& ACS/HRC/F606W		& 	0.025			& B10 		& $^{\star}$\\
 			& 				& 				& 				& 				&				& ACS/HRC/F814W		& 	0.025			& B10 		& $^{\star}$\\
  			&				&				&				& 				&				& NICMOS2/F110W		&	0.05				& 7171\\
			&				& 				&				&				&				& NICMOS2/F160W		& 	0.05				& 7171\\
  			&				&				&				&				&				& NICMOS2/F222M		&	0.05				& 7171\\
			&				&				&				& 				&				& WFPC2/PC/F814W	&	0.05				& 8592 \\
 NGC 4552	& UGC 7760		& $15\pm1$		& 	sbf			& 0.49 (36)		&	0.48 (35) 		& WFPC2/PC/F814W	&	0.05				& 6099\\
 NGC 4636 	& UGC 7878		& $13\pm1$		& 	sbf			& 3.44 (219)		&	0.297 (19)		& WFPC2/PC/F814W	&	0.05				& 8686\\
 NGC 4696 	&				& $37\pm2$		& 	sbf			& 1.4 (251)		&	0.2 (36)		& ACS/WFC/F814W	 	&	0.05				& 9427		& $^{\dagger}$\\
  NGC 5419	&				& $49\pm9$		& 	D$_n - \sigma$	& 2.11 (499)		&	0.38 (90)		& WFPC2/PC/F555W	&	0.05				& 6587\\
 NGC 5846	& UGC 9706		& $25 \pm 2$		& 	sbf			& 1.52 (183)		&	0.25 (30)		& WFPC2/PC/F814W	&	0.05				& 5920\\
 IC 1459		&				& $29\pm4$		& 	sbf			& 1.82 (258)		&	0.43 (60)		& WFPC2/PC/F814W	&	0.05				& 5454\\
 IC 4296		& 				& $50 \pm 3$		& 	sbf			& 1.44 (347)		&	0.32 (77)		& ACS/HRC/F625W		& 	0.025			& 9838\\
			&				& 				& 				&				&				& WFPC2/PC/F814W	&	0.05				& 5910\\
 		 	&	 			& 				& 				& 				&				& NICMOS2/F160W		& 	0.05				& 7453\\
 IC 4931		&				& $\sim 84$		& 	vvir			& 0.99 (403)		&	0.13 (51)		& WFPC2/PC/F814W	&	0.05				& 8683\\
  \enddata
  \tablecomments{(1) Source name; (2) name as it appears in BC06 if different than the name given in the first column; (3) adopted distance in Mpc (from the \textit{HyperLeda} extragalactic database); (4) method used to determine the distance: ``sbf", surface brightness fluctuation; ``vvir", from radial velocity, corrected for Local Group infall into Virgo; ``D$_n - \sigma$", size-sigma relation \citep{DresslerEtAl87}; (5) core radius as derived by CB05, in arcsec (pc); (6) SBH influence radius ($GM_{\bullet}$/$\sigma^{2}$) in arcsec (pc). Central velocity dispersions are given in Table \ref{tab: param}. Black hole masses have been derived from the $M_{\bullet}-\sigma$ relation given in \cite{FerrareseFordRev05}; (7) HST instrument/camera/filter used; (8) pixel scale in arcseconds/pixels as determined from direct measurement on the frames; (9) proposal identification number. B10$^{\star}$: this is a combined image, see Table 1 in B10 for further details. $^{\dagger}$ A WFC3/IR image is available for NGC 4696, however the complex nuclear features and the pixel size of the camera do not allow us to infer the SBH position with the precision necessary for this work.}
\label{tab: distanze}
\end{deluxetable*}

\begin{table*}[tb]
\caption{NOTES ON THE RADIO SOURCES OF INDIVIDUAL GALAXIES} 
\label{tab: radio} \centering
\scalebox{.86}{
\begin{tabular}{ l    cccl l l}
\hline \hline
& & &\\
\multicolumn{1}{c}{\textbf{Galaxy}}	& 			\multicolumn{4}{c}{\underbar{\hbox to 330pt{\hfill Radio Source\hfill}}} 										& \multicolumn{1}{c}{Nature of the galaxy} \\
							& Scale		& PA						& $P_\mathrm{5GHz}$				&	Comments							&  		  &\\ \hline
& & & & \\
NGC 1399			 		& 10 kpc$^{e}$	& $-10^{\circ}$				& $137^{a}$				&	Linear, almost symmetric.					& 	Dominant component of Fornax cluster.\\
NGC 4168					& - 			& -						& $7 \pm 0.3^{b}$			&  	Unresolved radio emission				&   	E2 in Virgo cluster.\\
NGC 4261					& 30 kpc$^{f}$	 & $87 \pm 8 ^{\circ}$$^{p}$	& $12657 \pm 610^{c}$		& 	Linear, symmetric lobes. W jet brighter than E. 	& 	Massive E in the outskirts of Virgo cluster. \\
NGC 4278					& 1.4 pc$^{g}$	& $100^{\circ} - 155^{\circ}$	& $69 \pm 2^{c}$			&	S-shaped, inner jet brightest to the N. 		&	Large E near the center of Coma I cloud.\\
NGC 4373					& -			& -						& $40^{d}$				& 	Compact radio source.					& 	Outskirts of Hydra-Centaurus supercluster.\\	
NGC 4486					& 1.5 kpc$^{h}$& $260^{\circ}$				& $43552 \pm 242^{c}$		&	Linear on kpc scale.						&	Virgo Cluster BCG.\\
NGC 4552					& 0.5 pc$^{i}$	& $\sim 50^{\circ}$			& $12 \pm 0.4^{c}$			& 	Twin pc-scale extension. 					&	Massive E in Virgo cluster.\\
NGC 4636					& $\lesssim$1 kpc$^{l}$& $\sim 35^{\circ}$	& $16 \pm 3^{c}$			&	Z shaped. No radio lobes.					& 	Massive E in the outskirts of Virgo cluster. \\
NGC 4696					& 1.6 kpc$^{m}$	& $-150^{\circ}$ $^{\dagger}$ 	& $3345^{a}$			&	Broad, one sided on pc-scale.				&	Centaurus Cluster BCG.\\
NGC 5419					& -			& -						& $1719^{a}$				&	Low surface brightness radio relic. 			& 	Dominant component of group S753. \\			
NGC 5846					& -			& -						& $4 \pm 0.2^{b}$			&	Complex morphology.					& 	Dominant component of group [HG82]50.\\			
IC 1459						& -			& - 						& $541^{a}$ 				& 	Compact source.						&	Giant E in group [HG82]15.\\
IC 4296						& 240 kpc$^{n}$& $130^{\circ}$$^{n}$		& $5912^{a}$				&	Linear, NW component slightly brighter.		&	Giant E, BCG of group A3565.\\
IC 4931						& -			& -						& $7^{d}$					&	-									&	BCG of A3656.\\
& & & & \\
\hline
\end{tabular}}
\tablecomments{(1) Optical identification; (2) approximate length of the radio jet/extension; (3) position angle (degrees E of N);  (4) total power at 5\,GHz [$10^{20}$ W]; (5) comments on the radio source morphology; (6) comments on the nature of the galaxy. $^{\dagger}$ This value is computed for the inner 25 pc. \textbf{References}: $^{a}$\citet{Slee94}, $^{b}$\citet{1991WH}, $^{c}$\citet{wrobel91}, $^{d}$\citet{Sadler1989}, $^{e}$\citet{Shurkin08}, $^{f}$\citet{Cavagnolo10}, $^{g}$\citet{Giro2005}, $^{h}$\citet{BaadeM1954}, $^{i}$\citet{Nagar2002}, $^{l}$\citet{StangerW86}, $^{m}$\citet{Taylor06}, $^{n}$\citet{KBC86a}, $^{p}$\citet{JonesWMP00}. [HG82]: \citet{HG82}.}
\end{table*}

\section{Data Analysis} \label{sec: analysis}

Broad-band images of the sample galaxies acquired with ACS, NICMOS2, WFPC2 and WFC3 were
retrieved from the \href{http://hla.stsci.edu}{Hubble Legacy Archive} (HLA), giving
preference to images obtained with the highest spatial resolutions and red filters (i.e.,
F606W and longer wavelengths).

Whenever available, several images taken with different instruments and filters, covering
both optical and NIR bands, were obtained for each galaxy. A list of the images analyzed
for each galaxy is given in Table~\ref{tab: distanze}. No additional processing steps were
applied beyond the standard HLA reduction pipeline. As the goal of our analysis is simply
to determine the relative positions of the isophotal center of the inner few kiloparsecs
of the galaxy and the AGN point source within each individual image, this is sufficient
for our purposes.

Each image was analyzed according to the following main steps: (1) a mask was constructed
to block image defects and distorting features such as dust lanes, jets or bright stars
and globular clusters; (2) elliptical isophotes were fitted to the galaxy surface
brightness distribution and the photocenter computed as the flux-weighted average of
the isophote centers; (3) the position of the point source (assumed to locate the SBH) was
determined by fitting a gaussian profile. See \textsection \ref{subsec: photo} and
\textsection \ref{subsec: AGN} for a more detailed description of these procedures.  Sky
subtraction was not performed as a uniform background will not affect the photometric
analysis and in many cases it is not possible to measure the sky background as the galaxy
covers the frame.

\subsection{The inner photocenter}	
\label{subsec: photo}

Each image was analyzed with the  IRAF task \textit{ellipse} \citep{cinque}, which was
used to fit elliptical isophotes to the surface brightness distribution. The first step
was to construct an image mask in order to minimize distortions due to image defects and
intrinsic photometric irregularities such as dust features, optical or NIR knots
associated with jets, globular clusters, foreground stars, etc. If an exposure time map
(the ``weight image") was provided in the retrieved data set, this was used as an initial
mask to eliminate cosmic rays, bad pixels, null areas and other image frame blemishes.
Applying the initial mask, a first run of \textit{ellipse} was performed to create a
zeroth-order photometric model. To account for intrinsic irregularities in the surface
brightness distribution, we then created a second mask by subtracting the zeroth-order
photometric model from the original image and masking residuals exceeding $\pm n \sigma$,
where $\sigma$ is the standard deviation of the residual image and \textit{n} is a
clipping parameter. 

To allow \textit{ellipse} to find the photometric peak, the central region was unmasked
and a new run performed. This process was iterated, adjusting \textit{n} in order to map
distorting features in the maximum possible detail. The iteration was terminated when the
mask converged and further runs did not add any significant detail. 

Once the final mask was constructed, \textit{ellipse} was re-run with the mask
applied to determine the photocenter of the galaxy. Elliptical isophotes were fitted
between minimum and maximum values of the semi-major axis (SMA) which were usually
determined by the core radius (r$_c$) and the image size, respectively. In almost all
cases, the minimum SMA was set equal to or greater than the value of r$_c$ as determined
by CB05; Table~\ref{tab: distanze}. This was typically $\sim 2$\arcsec, or
$\sim 300$\,pc. In some cases, where a dusty disk is present in the center of the galaxy,
the initial SMA was chosen such that the inner ellipse is larger than the disk. The only
exception was NGC\,4486 (M\,87), which has a very large core. In this case, we adopted the
SMA range 1--3\arcsec\ used by B10, who found that isophotes within 1\arcsec\ are
influenced by the AGN point source. The maximum SMA was that of the largest ellipse that
completely fits within the image area -- that is, only complete isophotes were fitted.
This limits our isophotal analysis to the inner few kiloparsecs of each galaxy; the SMA of
the outermost ellipse fit varies in the range $1 \lesssim r \lesssim 4$ kpc. 

Between these limits, a series of ellipses was fitted to the 2-D surface brightness
distribution, with the SMA being increased by two pixels at each step. Each fit returned
values of the position angle, eccentricity and the pixel $x-y$ coordinates of the center
of the ellipse. Finally, the photocenter was obtained as the flux-weighted average of the
centers of the elliptical annuli generated by \textit{ellipse}:
	
	\begin{align} \label{eq: cm}
	(x_\mathrm{pc}, y_\mathrm{pc}) 	&= \left(  \frac{\sum_{i} x_{i} j_{i}} {\sum_{i} j_{i}}  ,  \frac{\sum_{i} y_{i} j_{i}} {\sum_{i} j_{i}} \right)
	\end{align}

\noindent where $x_{i}$ and $y_{i}$ are the photocenter coordinates of the i-th isophote and $j_{i}$
is a weight, given by the product of the area of the annulus defined by neighbouring
isophotes, with the mean intensity within the annulus. 

\begin{figure}[t]
\centering
\includegraphics[draft=false,  scale=0.75, trim = 2cm 0cm 0cm 0cm]{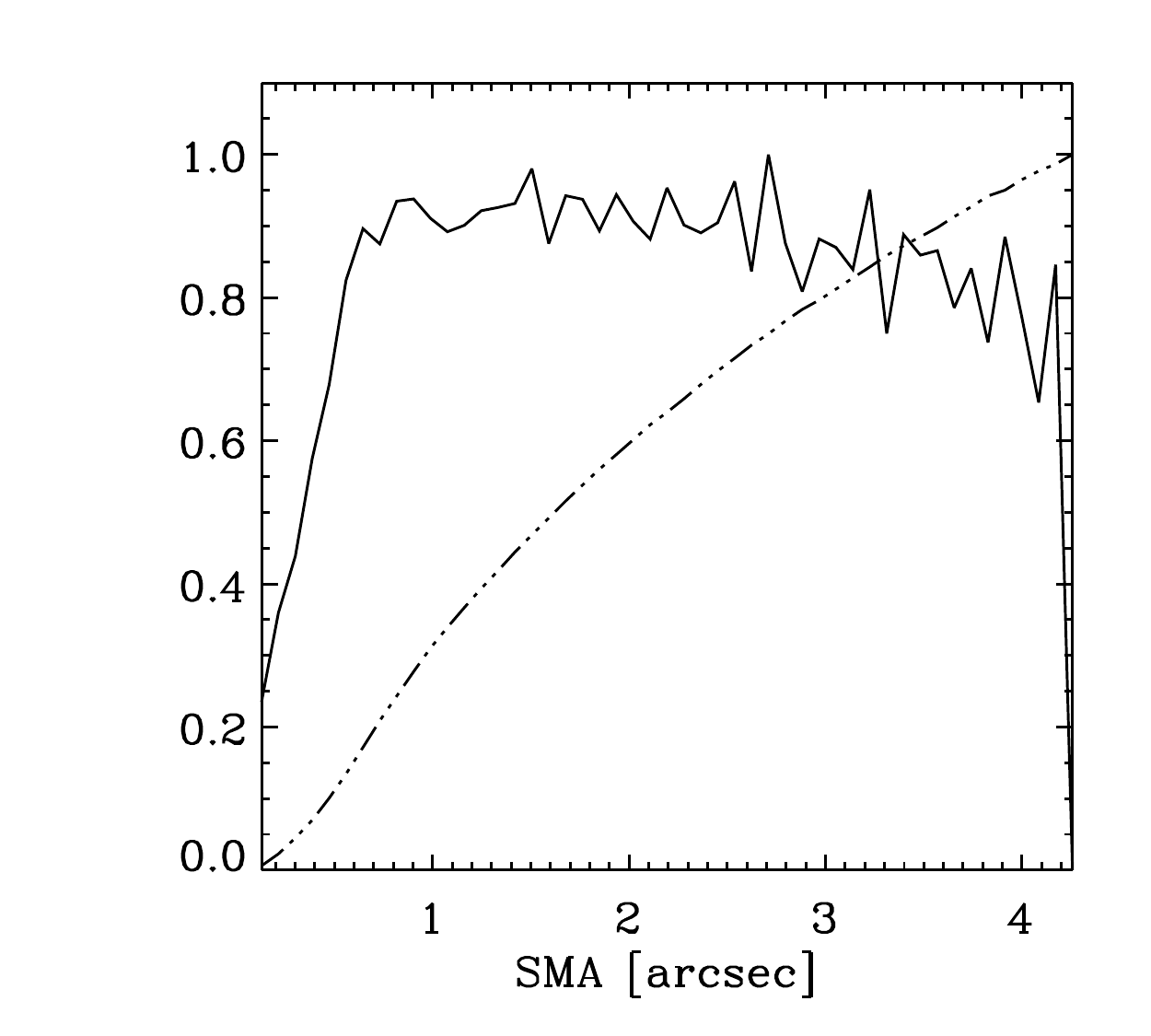}
\caption[Weight functions]{Alternative weighting functions for the mean photocenter, derived from the NICMOS2/F160W image of NGC 4261. Both functions are normalized to their maximum values. \textit{Solid line}: each isophote center is weighted by the light content of the corresponding elliptical annulus. \textit{Dashed dotted}: each isophote center is weighted by the light content enclosed by the ellipse of the corresponding SMA. }
\label{fig: weight}
\end{figure}

To assess the sensitivity of the mean photocenter to the weighting method, we computed an
alternative weight function using the flux enclosed {\em within} each successive
elliptical isophote. For comparison, the two weight functions are plotted as a function of
SMA in Fig.\ref{fig: weight}, for the NICMOS image of NGC\,4261. The annular flux method
produces an approximately uniform weight distribution beyond the innermost region (SMA $<
0\farcs5$), but typically decreases slowly at large SMA values. Not surprisingly, the
enclosed flux method produces weights that increase monotonically with SMA, thus assigning
much greater weight to the outer, fainter isophotes. This may be an undesirable
characteristic in that isophotal distortions which may be present at fainter surface
brightness levels (due to interactions with nearby companions, for example) might unduly
influence the mean photocenter. Nevertheless, for the sources in our sample, the
photocenters computed using the two methods are consistent within the errors in all cases.
This result reflects the rather regular nature of the sample galaxies at the wavelengths
under consideration.

We explored alternative methods for the measurement of the photocenter position, 
namely a moment based technique and a 2-D decomposition of the surface brightness distribution.
In the first case, the photocenter position is obtained by weighting the coordinate of each pixel by its own intensity.
This method seems to be particularly sensitive to the masking details and to the presence of asymmetries in the light distribution.
The second approach, performed with the GALFIT
surface brightness fitting program \citep{GALFIT2002}, requires, at least in some cases,
multiple components for a successful fit (e.g. two or more S\'{e}rsic profiles) casting doubt on the physical meaning of the
recovered photocenter and the errors associated with it.

The IRAF task \textit{ellipse} was specifically built to fit the light profile of elliptical galaxies with minimal assumptions;
the results obtained seem to be weakly dependent on the masking details and errors associated with the fitted parameters have 
been thoroughly studied \citep{Busko96}. Therefore, we prefer to adopt the photocenter position determined with this approach.

\subsection{The SBH position} \label{subsec: AGN}

We make the key assumption that the point-like source near the center of each galaxy is an
AGN, which therefore locates the position of the SBH (see \textsection \ref{sec: disc}
for a detailed discussion of this assumption).

The AGN point sources are bright enough to be easily detected, but in our sample they are not so bright that the
point spread function dominates the host galaxy. In order to determine the position of the
SBH, we proceeded as follows. First, a median filtered version of the image
was subtracted from the original producing a residual map in which the AGN is a
prominent feature\footnote{We prefer to subtract a median filtered image instead of the
photometric model because in the former method we can control the size of the structures
that are removed. That is we can change the dimensions of the box over which the median is
computed. This determines the degree of smoothing of the filter and hence the size of the
structures removed after the subtraction.}. In the second step, we used the IRAF task
\textit{center} to fit gaussians to the marginal distributions of the residual point
source along the \textit{x} and \textit{y} axes.

\subsection{The photocenter--AGN displacement} \label{sec: off}

For each image of each galaxy, the \textit{x} and \textit{y} components of the
displacement were measured as the difference in pixels between the \textit{x} and
\textit{y} co-ordinates of the photocenter and the AGN point source. The significance
of the displacement is classified in two steps, considering the \textit{x} and
\textit{y} components separately. Displacements smaller than 3$\sigma$, where $\sigma$
is the error on the displacement (Section \ref{subsubsec: err}), are considered
non-significant. If the displacement exceeds $3\sigma$ we make a further
classification based on the distribution of the isophote center coordinates produced
by \textit{ellipse}. For each coordinate, \textit{x} and \textit{y}, the
inter-quartile range (IQR, the difference between the upper and lower quartiles) of
the isophote centers was computed. The corresponding displacement is then assigned
significance levels of ``null", ``low", ``intermediate" or ``high"  depending on its
magnitude relative to the IQR:  
	\begin{itemize}
	\item null, $\Delta x,  \Delta y < 0.8$ IQR
	\item low, 0.8 IQR $\leq \Delta x,  \Delta y < 1.6$ IQR
	\item intermediate, 1.6 IQR $\leq \Delta x,  \Delta y < 2.4$ IQR
	\item high, $\Delta x,  \Delta y \geq 2.4$ IQR.	
	\end{itemize}

The threshold values are based on the fact that, for a normal distribution, 0.8 IQR is
equivalent to 1$\sigma$. For an ideal, symmetric galaxy the isophotes would be concentric,
giving IQR = 0. Any irregularities in the surface brightness distribution will cause a
dispersion in the centers of the fitted elliptical isophotes, resulting in a non-zero IQR.
Hence, by normalizing to the IQR, we allow for uncertainties in the recovered offsets due
to asymmetric surface brightness distributions arising from, for example, dust lanes or
tidal distortions.

\subsection{Errors and biases}

\subsubsection{Position uncertainties}
\label{subsubsec: err}

The final error on the measured projected displacement is the combination of the
uncertainties in the positions of the photocenter and the AGN point source. As the
photocenter is determined by fitting elliptical isophotes, it is important to understand
if the errors provided by \textit{ellipse} are reasonable and robust. To investigate this,
we used \textit{ellipse} to analyze synthetic galaxy images constructed with GALFIT. 
We also used these simulated galaxies to estimate the uncertainties on both the AGN 
and photocenter positions.	

The model images were constructed by fitting a S{\'e}rsic profile and a nuclear point source (a PSF generated with Tiny Tim, \citealt{TinyTim93}) to selected  
galaxies in our sample. We used NGC 4261 and 4278 to build models of NICMOS images, IC 1459 for WFPC2 images, NGC 1399 for WFC3 and IC 4296 for ACS images.
The resulting model surface brightness distributions were populated with gaussian random noise 
so as to match the signal-to-noise ratio (SN) of the original galaxy image\footnote{For this purpose,
the signal-to-noise ratio is defined as the ratio between the flux of the nuclear point source to the 
standard deviation of the residual measured in the outer regions of the galaxy after subtracting the
photometric model from the image.}. Four realizations of each synthetic galaxy image were generated
and each realization was analyzed using the methods described in 
Sections~\ref{subsec: photo} and~\ref{subsec: AGN}, superposing the masks obtained from the
corresponding real galaxy images.

\textbf{Ellipse errors:} for each isophote fit, the ``true error'' on the
isophote center is computed as the difference between the position returned by \textit{ellipse} and
the known photocenter of the synthetic galaxy (i.e., the center position of the S{\'e}rsic component fitted by {\sc Galfit}).
These ``true errors'' are compared to the errors on the positions of the isophote centers 
returned by \textit{ellipse} in Fig.\ref{fig:
ellipse_err}. The ``true'' error distribution is slightly broader than that of the
\textit{ellipse} errors, but both distributions are highly peaked, with $\sim 50\%$ of fits
producing errors $\le 0.1$ pixels. A detailed study by \citet{Busko96} of the errors returned by
\textit{ellipse} shows that errors on ellipticity, position angle and center position are
unbiased and accurate when the radial gradient relative error\footnote{The radial
gradient relative error ($\sigma^{'}/I^{'}$) is the relative error on the radial gradient
of the intensity. This is a default output of \textit{ellipse}.} is less than $50\%$. 
This condition is satisfied for our sample galaxies. 

\begin{figure*}[t]
\centering
\includegraphics[draft=false, trim = 0cm 0cm 0cm 1cm, clip=true, scale=0.7]{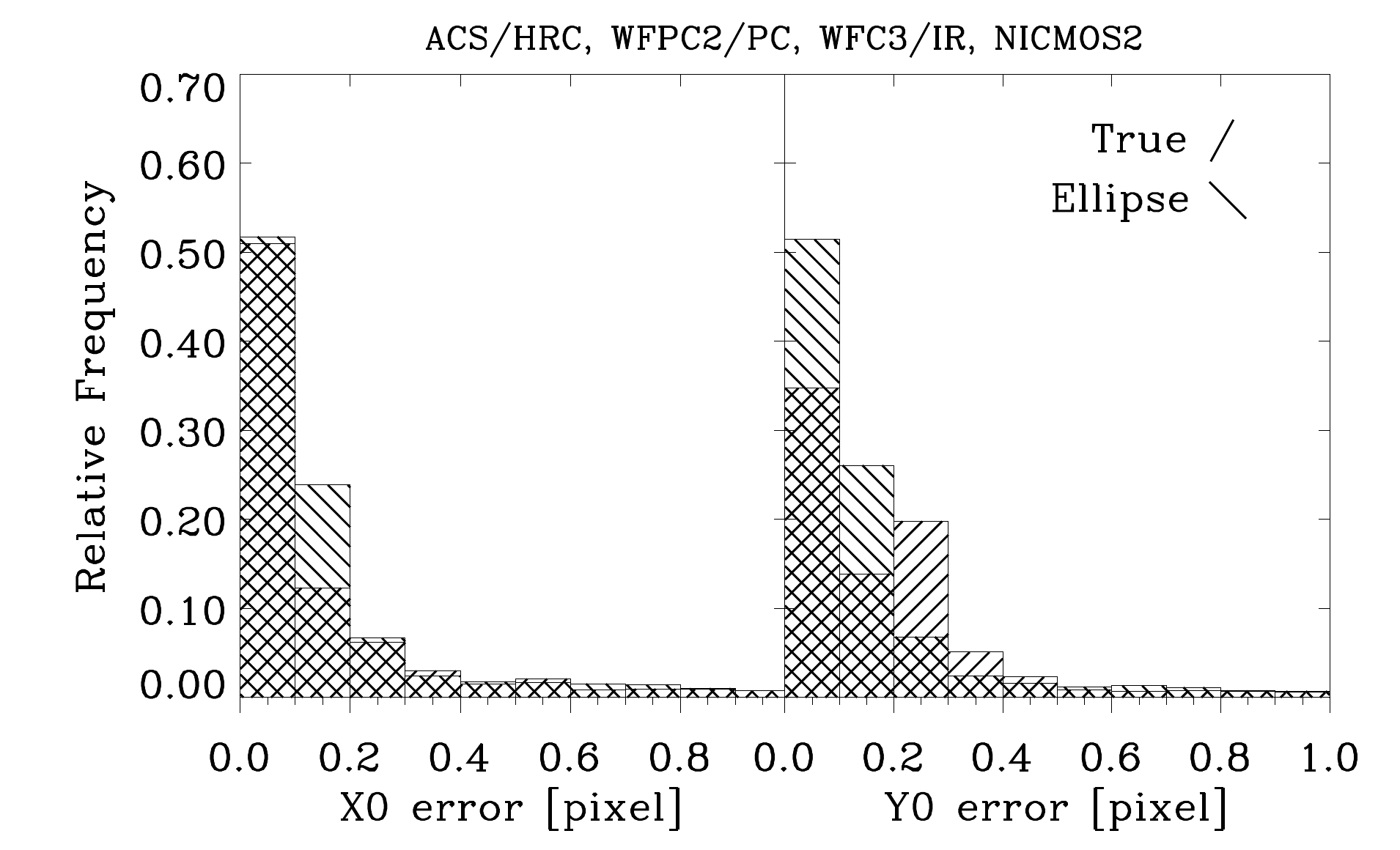}
\caption[ellipse_err]{Error distributions for the $(x,y)$ co-ordinates of isophote center positions derived from \textit{ellipse} fits to sixteen simulated galaxy images. For each galaxy, a series of ellipses was fitted to the surface brightness distribution, incrementing the SMA by two pixels at each step, as described in Sec.~\ref{subsec: photo}. The errors returned by the IRAF \textit{ellipse} task are compared with the ``true errors", defined as the difference between the known position of the center of the model galaxy and the center position returned by the \textit{ellipse} fit. 
The errors on each isophote center returned by \textit{ellipse} are broadly consistent with the true errors. 	
The error distributions combine results from all sixteen simulated galaxy images. The model images were derived from Galfit fits to the following data: NGC 4261/NICMOS2, IC 1459/WFPC2-PC, NGC 1399/WFC3-IR, IC 4296/ACS-HRC.}
\label{fig: ellipse_err}
\end{figure*}

\textbf{Position errors:} the ``true errors'' on the ($x$,$y$) coordinates of the AGN and galaxy
photocenter positions were also determined for each realization of the synthetic galaxy images, 
where the true error in each case is defined as the difference 
between the known positions of the AGN and photocenter (i.e., the positions returned by Galfit) 
and the values recovered using our analysis  methods. 

For signal-to-noise ratios characteristic of our data ($S/N \sim 100$), the distributions of
the ``true errors'' on the AGN and photocenter positions derived from both the ACS and WFPC2 
synthetic galaxy images are narrow and confined within $0.1$ pixel. We therefore adopt $0.1$ pixel as a conservative 
estimate of the uncertainty on both the AGN and photocenter positions derived from ACS and WFPC2 data.
The ``true error'' obtained from the WFC3 and NICMOS2 images exhibit broader distributions, 
with tails extending to $0.4$ pixels. We therefore adopt the median values of
the ``true errors'' as the uncertainties for WFC3 and NICMOS2 data, yielding 0.2 pixels for the photocenter
position in both cameras and respectively 0.1 and 0.2 pixels for the AGN.

These uncertainties represent the precision with which we can determine the AGN and photocenter positions
for an ideal galaxy, in an image with noise characteristics representative of our data.
The adopted uncertainties are summarized in Table
\ref{tab: precision}, together with the equivalent angular distances and linear distances 
corresponding to the closest and furthest of
our sample galaxies.  

\begin{table}[tb]
\caption{PRECISION ON ISOPHOTAL CENTER AND AGN POSITION} 
\label{tab: precision} \centering
\scalebox{.8}{
\begin{tabular}{ l    cc cc c c l}
\hline \hline
& & &\\
\multicolumn{1}{c}{\textbf{Precision}}	& \multicolumn{2}{c}{\underbar{\hbox to 65pt{\hfill ACS \hfill}}} 	& \multicolumn{2}{c}{\underbar{\hbox to 65pt{\hfill WFPC2 \hfill}}} 	& NICMOS2 	&	WFC3	&	\\
							& HRC			& WFC									& PC 				& WF						& \\ \hline
& & & & \\
		0.1 pxl			 	& 2.5 			& 5										& 5					& 10	 						&  			&	9		& mas\\
						 	& 0.2 			& 0.3										& 0.3					& 0.6 						&  			&	0.6		& pc\\
							& 1	 			& 2										& 2					& 4.1 						&  			&	3.7		& pc \\
		0.2 pxl				& 				&										&					&							& 10			&	18		& mas\\
							& 				&										&					&							& 0.6			&	1.1		& pc\\
							& 				&										&					&							& 4			&	7.3		& pc\\					
& & & & \\
\hline
\end{tabular}}
\tablecomments{Values are given in mas (first row), in pc at the nearest galaxy (13 Mpc, second row) and at the farthest (84 Mpc, third row). Similarly for the other rows. Values for NICMOS2 are computed for a pixel scale of 0\farcs05/pixel.}
\end{table}

\textbf{Minimum detectable displacement:} the uncertainties on the AGN and photocenter
coordinates determine the minimum
detectable displacement. We require a 3$\sigma$ detection for a displacement to be
considered significant.  Combining the uncertainties on the photocenter and AGN in
quadrature, the error on each component of the displacement is $\sigma = 0.14$ pxl, for ACS and
WFPC2 data, $\sigma = 0.22$ pxl for WFC3, and $\sigma = 0.28$ pxl for NICMOS2 data. Therefore, the minimum displacements considered
significant are $\Delta x = \Delta y \approx 0.4$ pxl, for ACS and WFPC2, $\Delta x = \Delta y \approx 0.7$ pxl for WFPC3, and $\Delta x =
\Delta y \approx 0.9$ pxl for NICMOS2.

\subsubsection{Possible biases}
\label{subsubsec: bias}

\textbf{Asymmetric surface brightness distributions:} the galaxies in our sample were
selected to be symmetric and regular, based on visual inspection of the images. However, 
in some cases, the galaxy extends beyond the edges of the frame and/or is not centered in
the image frame. To eliminate
the possibility of spurious offsets due to truncation of isophotes by the frame edge,
the upper limit on the SMA range used in the surface brightness fits is taken to be that
of the largest ellipse that fits completely within the frame (Sec.~\ref{subsec: photo}).

In addition, subtle intrinsic asymmetries might be present that were not revealed by
visual inspection. As discussed by \citet{cinque}, asymmetric surface brightness distributions
(such as large-scale lopsidedness in the isophotes) will cause shifts in the  isophote
centers, which in turn will increase the IQR. Therefore, intrinsic
surface brightness asymmetries (or other irregularities) will tend to reduce the level
of significance assigned to any measured displacement (Sec.~\ref{sec: off}).

A similar possibility is that the outer isophotes are distorted by tidal interactions with nearby galaxies.
As our analysis is confined to the inner
few kiloparsecs, such interactions are unlikely to significantly affect our results. 
Nevertheless, we have verified that the photocenter
position does not change significantly when recomputed excluding isophote centers
corresponding to ellipse SMAs greater than  $\sim$ 85, 90 and 95\% of that 
of the largest complete ellipse that fits within the image frame. \vskip 10pt

\textbf{Isophote twists and PSF effects:} 
core galaxies are often characterized by strong isophote twisting at radii interior to
the core radius, $r_{c}$ \citep{lauer}. In addition, the inner isophotes are distorted by
the point spread function; TINY TIM simulations show that for the HST instruments used,
the PSF can affect the region within a radius r$_\mathrm{PSF}\sim15$ pixels. 
As typically r$_\mathrm{PSF} \lesssim r_{c}$, we mitigate
both effects by setting r$_c$ as the minimum SMA for the \textit{ellipse} fits. 
\vskip 10pt

\textbf{Lopsided stellar nuclei (LSN)}: observations of the nuclear regions of nearby
galaxies have revealed the presence of double nuclei with separations in the range 1--10 pc
\citep{Lauer93, LauerEtAl96, lauer, ThatteEtAl2000, Debattista2006}. 
One explanation for such configurations is that lopsided stellar orbit distributions
tend to persist within the SBH sphere of gravitational influence 
since orbits do not precess in Keplerian potentials and are not, therefore, axisymmetrized by phase mixing \citep{PeirisTrem03}. 
The best studied example is M31, in which the components of the double nucleus are separated by $\approx 0\farcs5$ ($\approx 2$\,pc).
\cite{PeirisTrem03} proposed a model for M31 in which the nucleus consists of an eccentric
disk of stars orbiting the SBH, the latter being coincident with the fainter component.

The WFPC2 Planetary Camera (with which most of the images studied here were
obtained) resolves spatial separations of $\approx 5$ pc only for galaxies within 20 Mpc.
It is possible, therefore, that some of the galaxies in our sample harbor 
unresolved double nuclei. This will not affect the determination of the galaxy photocenter,
since the core is excluded from the isophote fitting. However, if the unresolved nuclei are of
unequal brightness as in M31, this could result in a systemic error in the AGN position
which may approach the PSF FWHM ($\sim 1 - 2$ pixels). This in turn would result in a spurious displacement.

\section{Results} \label{sec: results}

The results of the photometric analysis are presented in Table~\ref{tab: snsummary}, where we
list the (\textit{x, y}) pixel co-ordinates of the mean photocenter and the point source, the magnitudes of the corresponding 
components of the displacement, 
the direction of each displacement component on the sky and the significance level of the displacement, 
determined as outlined in Sec.~\ref{sec: off}. 

\begin{figure*}[hp]
\begin{center}$
\begin{array}{ccc}
\includegraphics[trim=3.75cm 1cm 3cm 0cm, clip=true, scale=0.48]{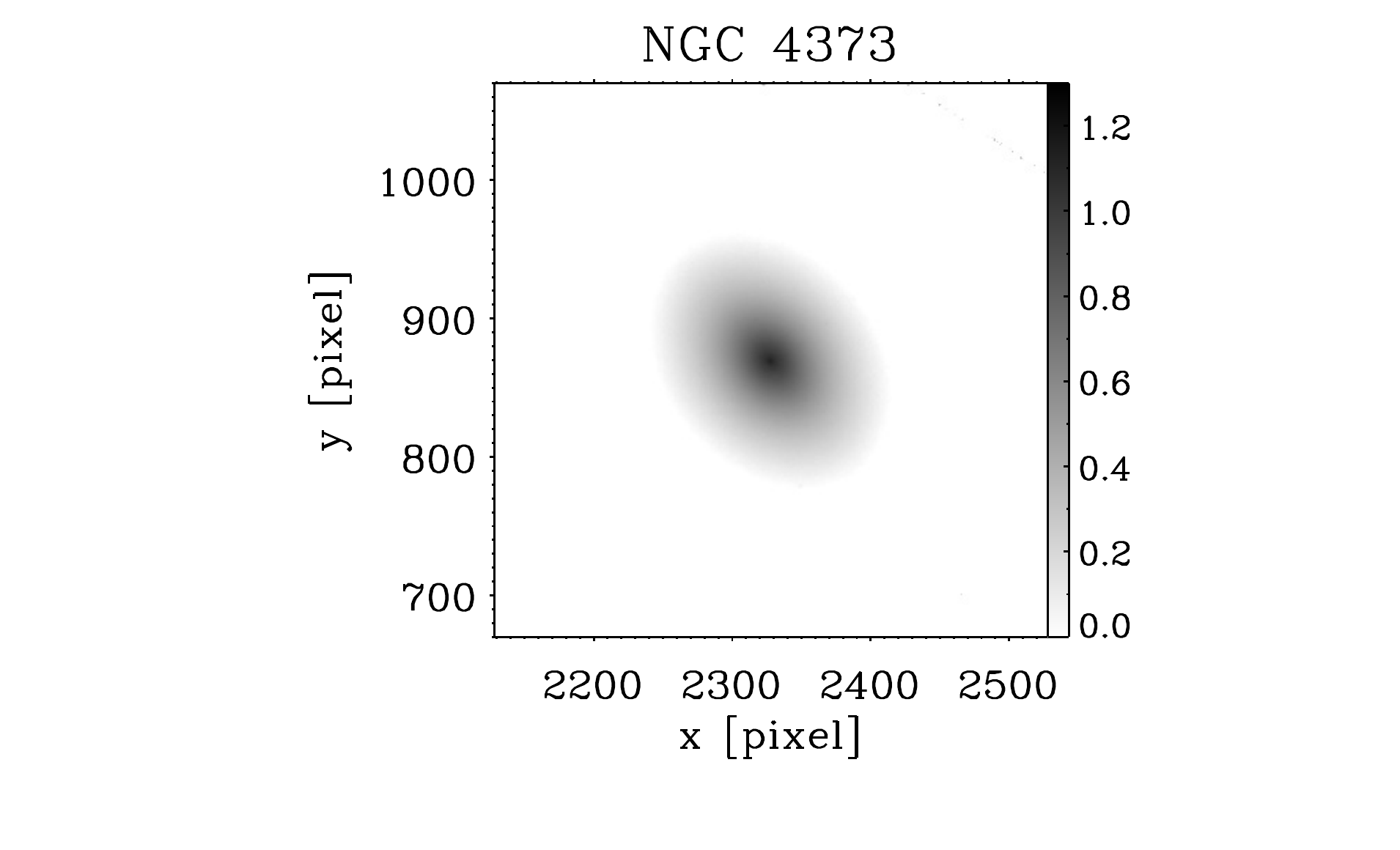} & \includegraphics[trim= 4cm 1cm 3cm 0cm, clip=true, scale=0.48]{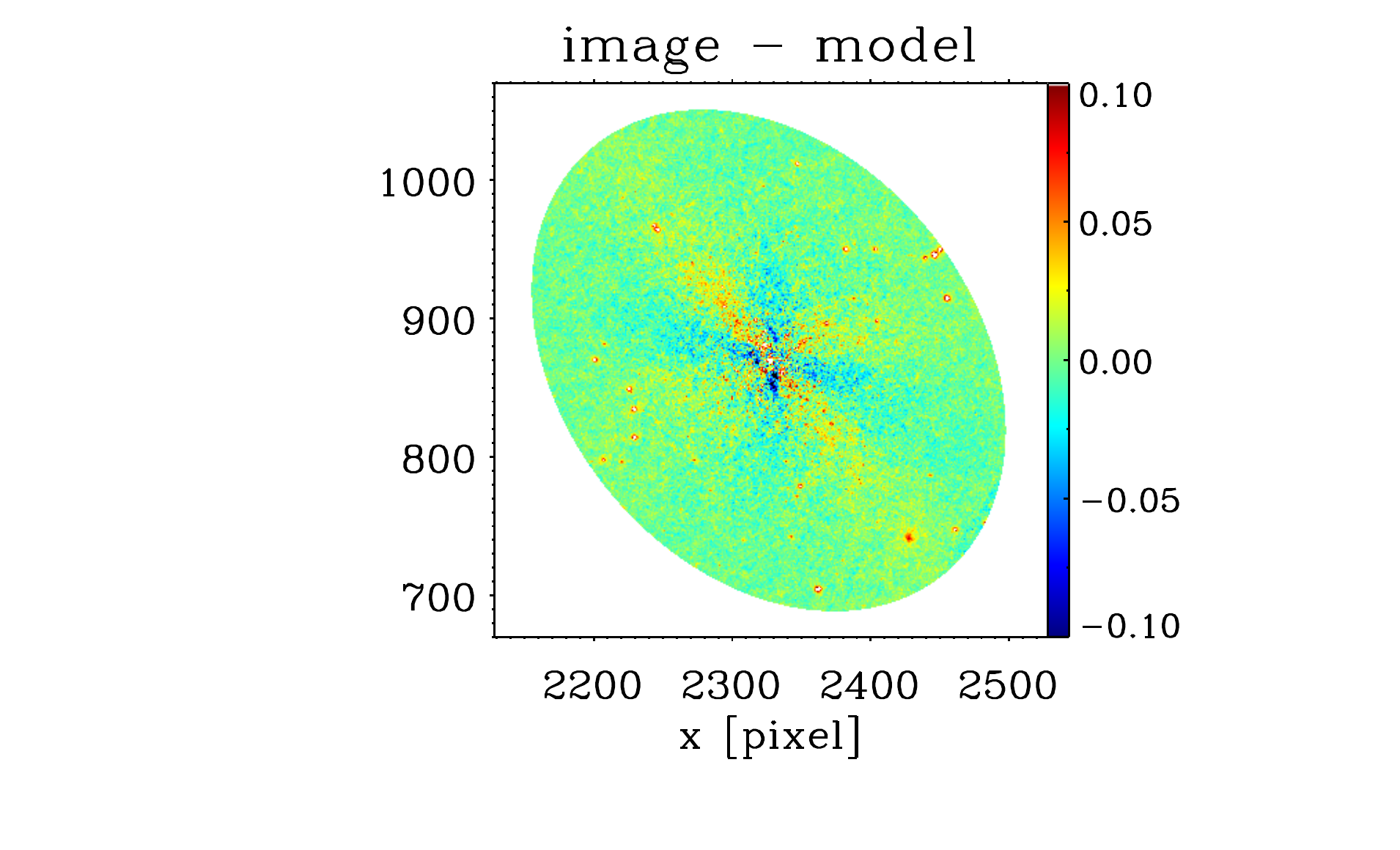}	& \includegraphics[trim= 4cm 1cm 3cm 0cm, clip=true, scale=0.48]{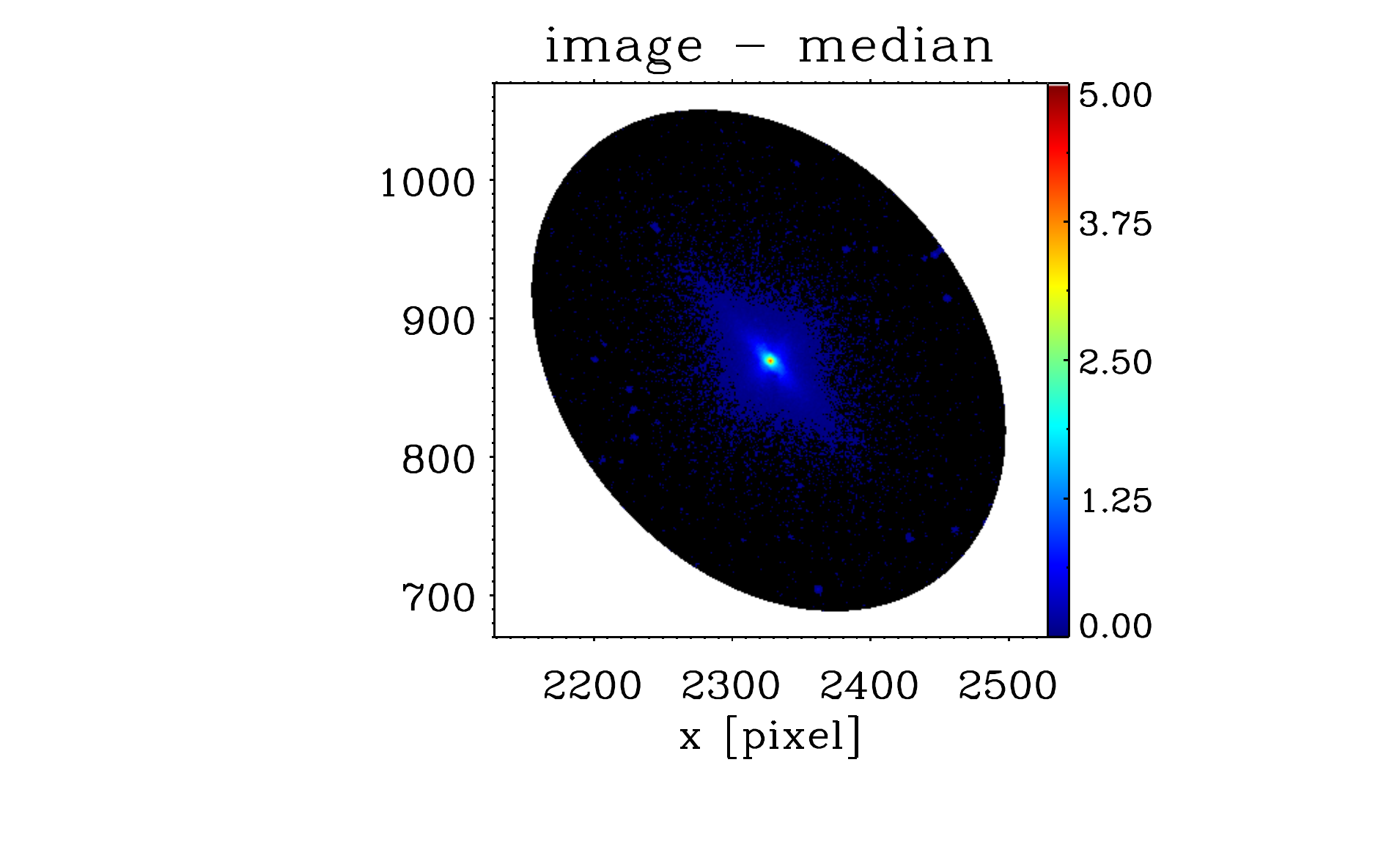} \\
\includegraphics[trim=0.7cm 0cm 0cm 0cm, clip=true, scale=0.46]{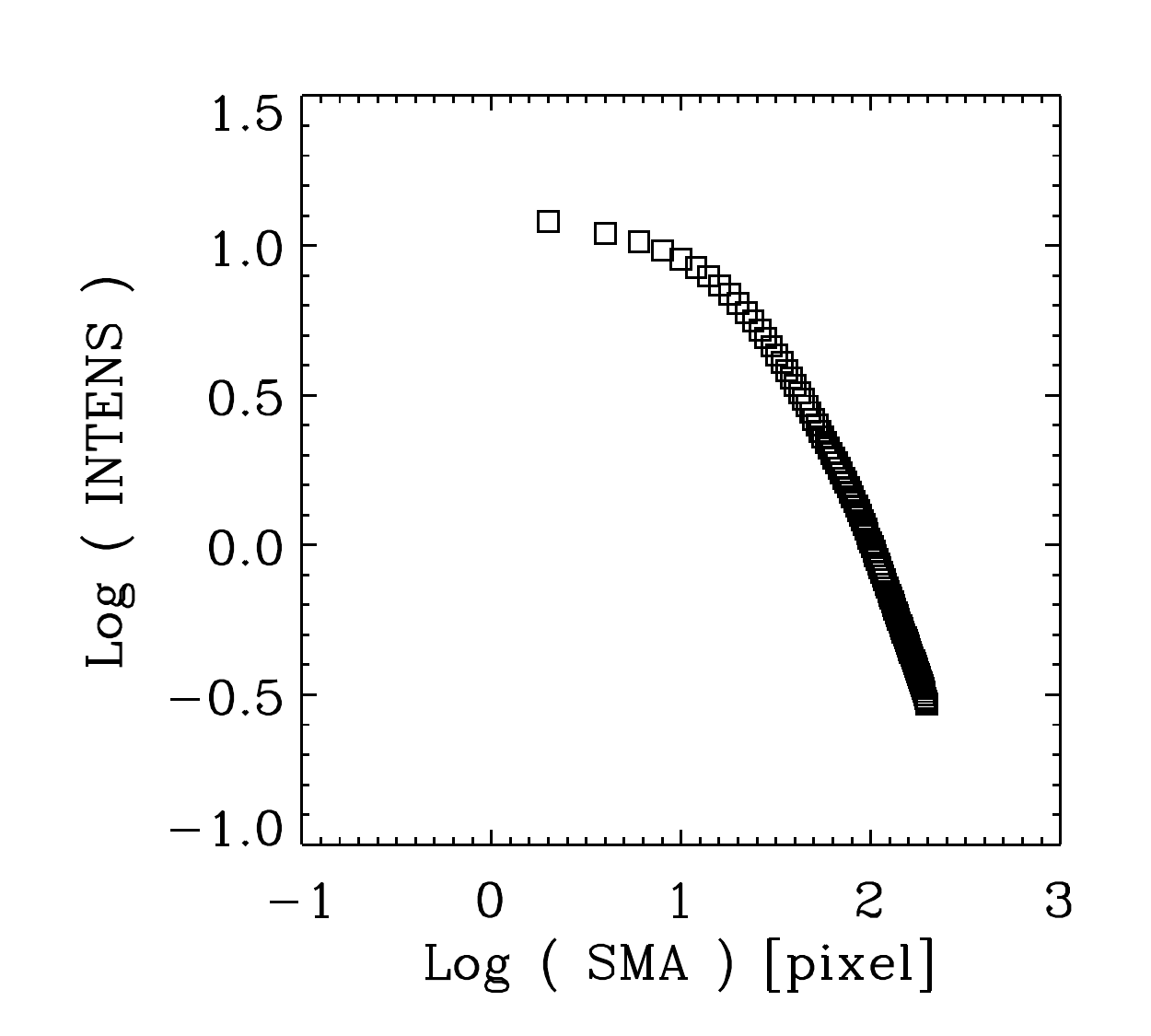}	    &  \includegraphics[trim=0.cm 0cm 0cm 0cm, clip=true, scale=0.43]{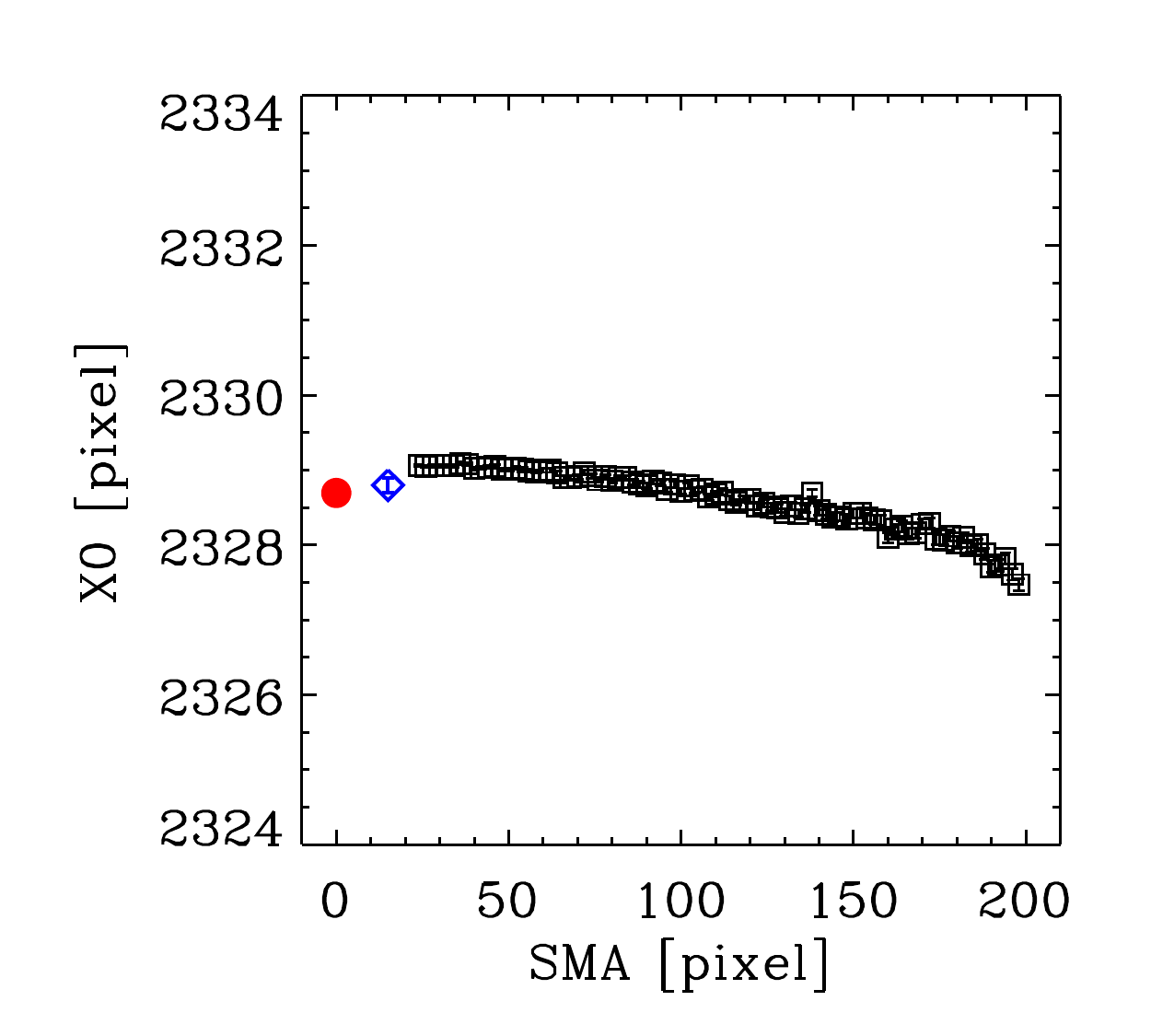}  &  \includegraphics[trim=0cm 0cm 0cm 0cm, clip=true, scale=0.43]{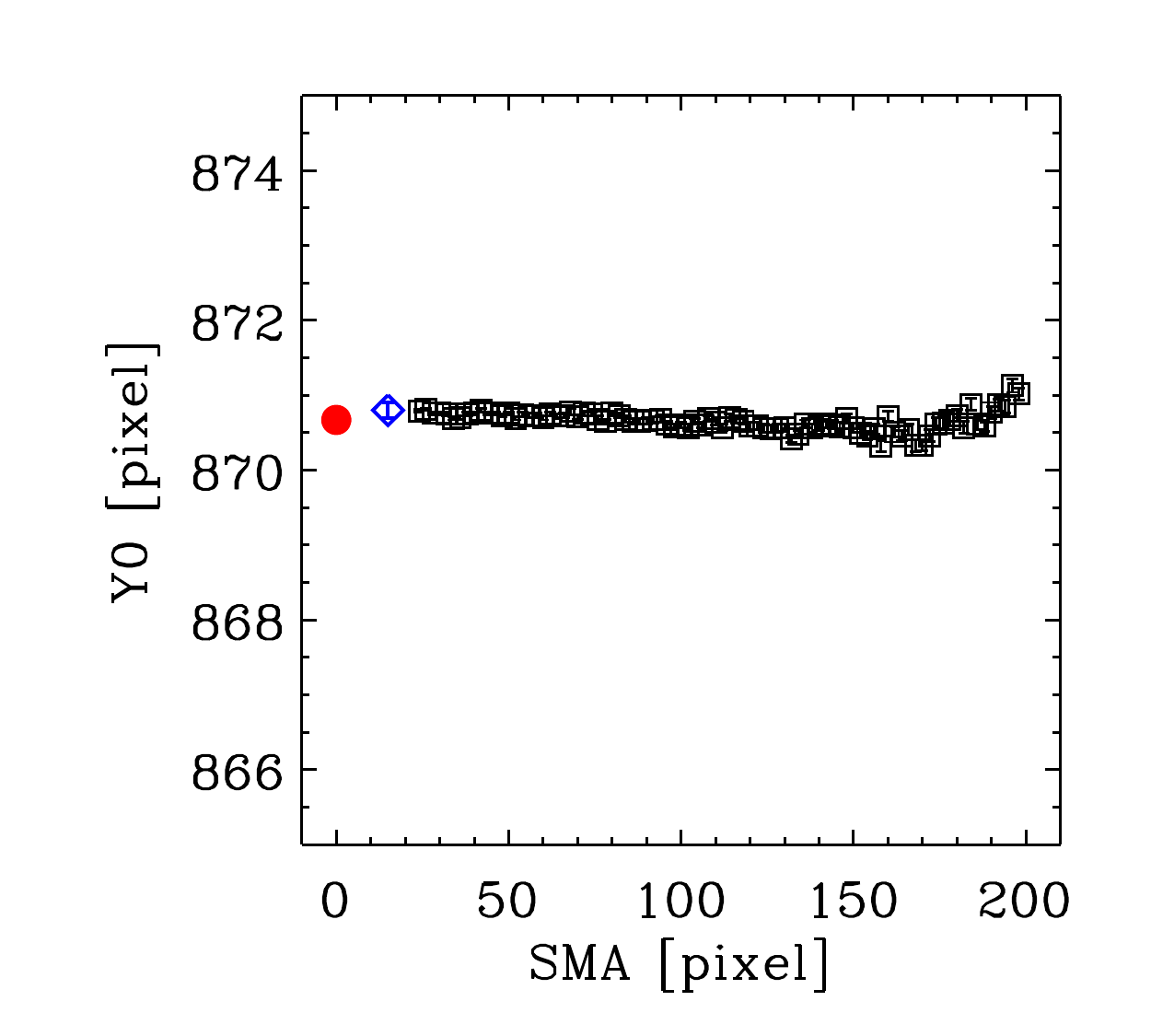} \\	
 \includegraphics[trim=0.65cm 0cm 0cm 0cm, clip=true, scale=0.46]{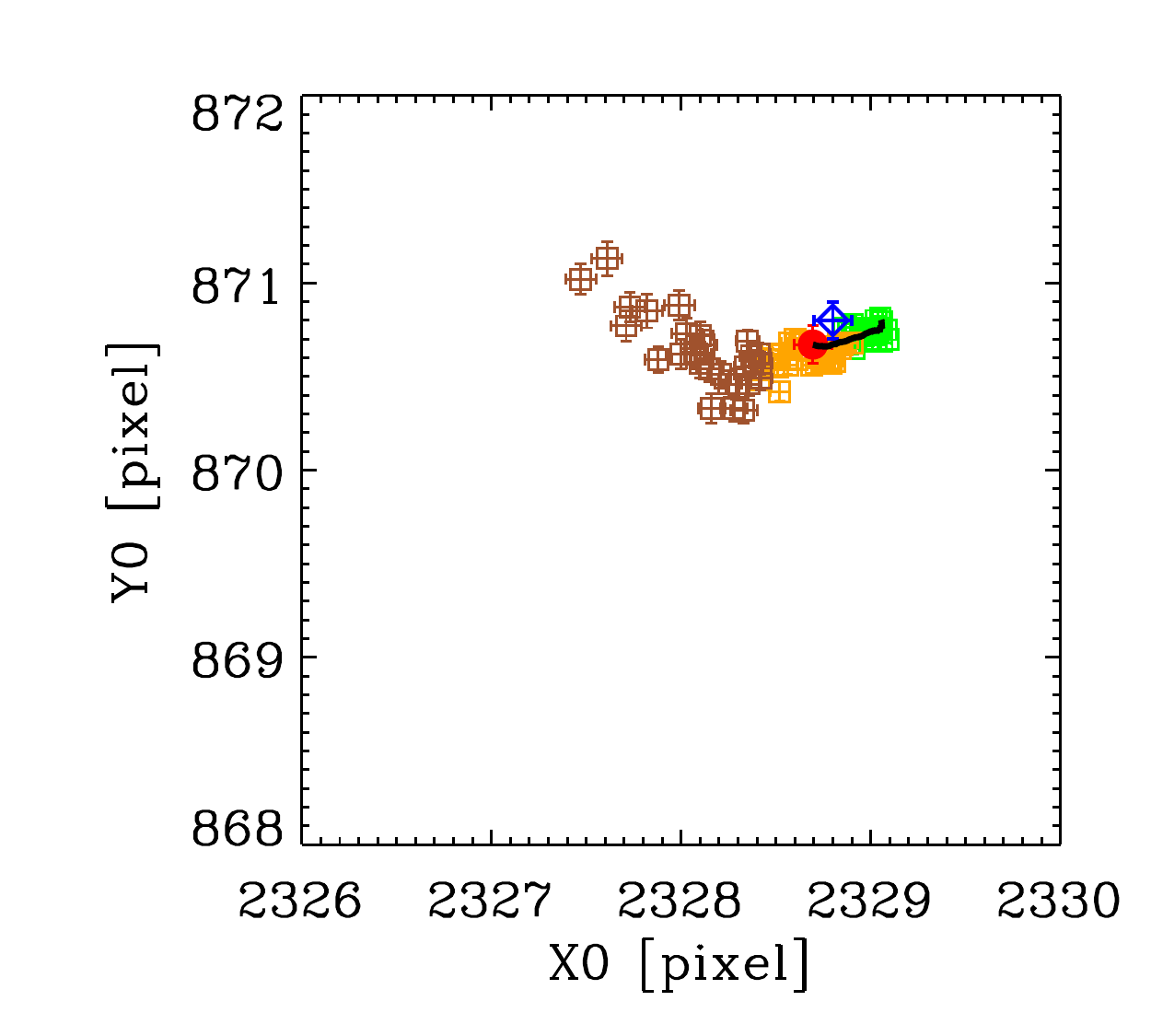}	&  \includegraphics[trim=0.6cm 0cm 0cm 0cm, clip=true, scale=0.46]{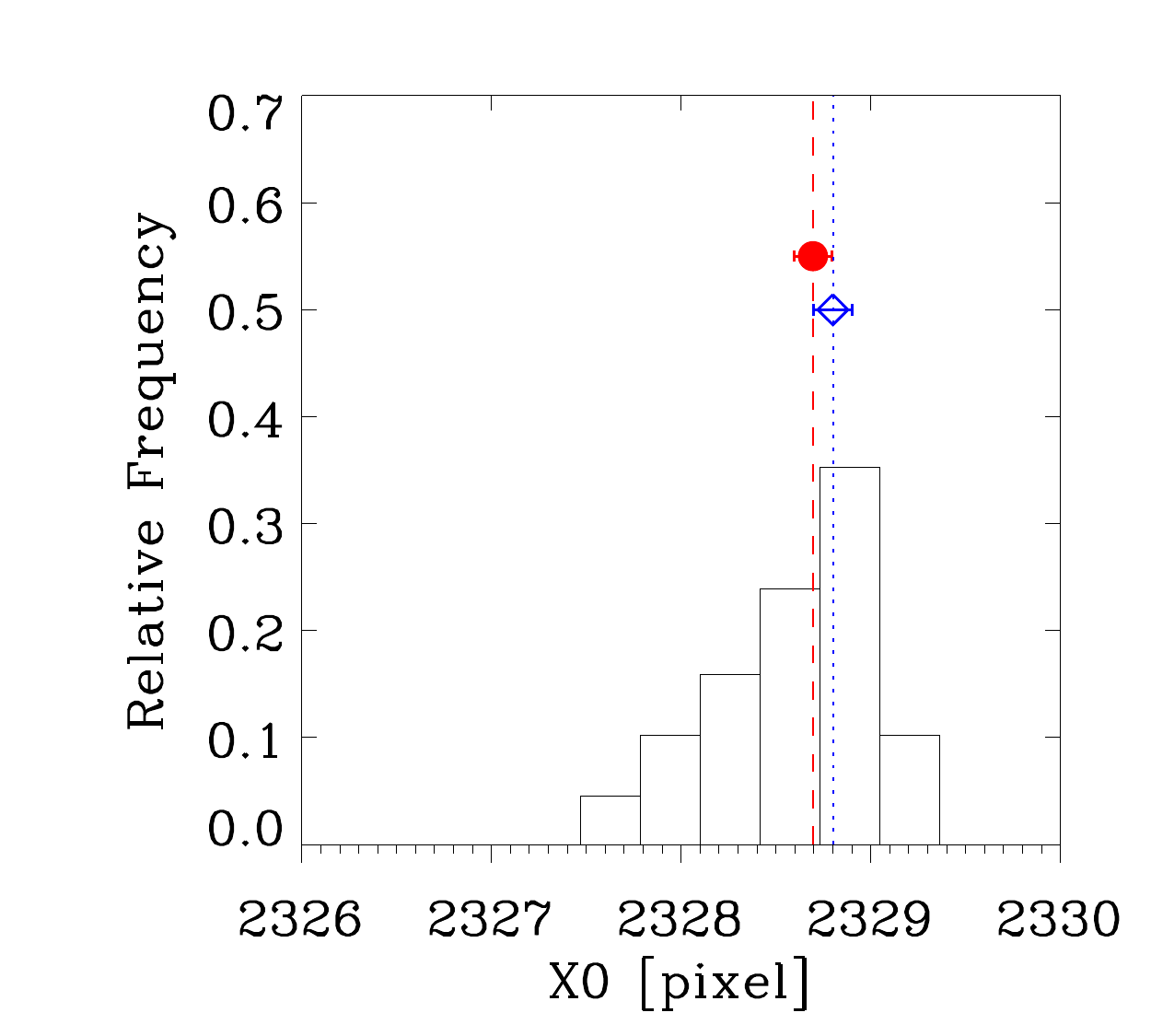}	& \includegraphics[trim=0.6cm 0cm 0cm 0cm, clip=true, scale=0.46]{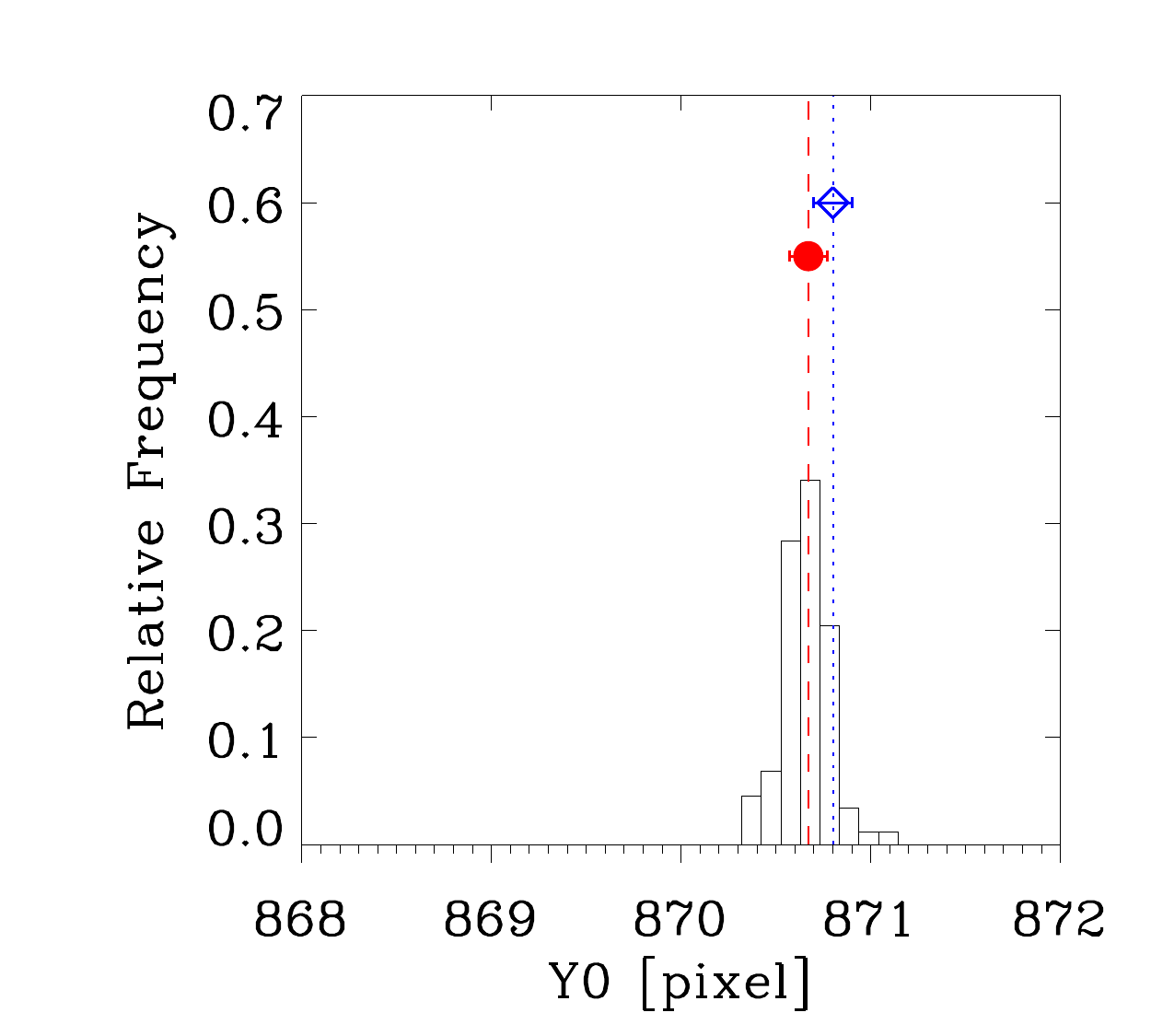}\\
\end{array}$
\end{center}
\caption[NGC 4373]{Example of a galaxy where the displacement is not significant. NGC 4373, WFPC2/PC - F814W, scale=$0\farcs05$/pxl. \textbf{First row} - left to right: the galaxy Log(counts s$^{-1}$), note that the color scale is adjusted to highlight the presence of the nuclear point source, emission is present over the whole field of view; galaxy after subtraction of the isophotal model; galaxy after subtraction of the median filter. \textbf{Second row} - surface brightness profile in counts s$^{-1}$; second and third panels: $x$ and $y$ coordinates of the isophotal centers versus semi-major axis length. Blue diamond: AGN; red bullet: photocenter. The blue diamond and the red bullet are not positioned both at SMA=0 to make the figure clear and avoid superpositions. \textbf{Third row} - scatter plot of the isophote centers. The solid line (near the photocenter) is the cumulative photocenter computed including progressively all data from the core radius outward. Different colors for the data points are used to represent the centers of isophotes with SMA length in a given range. Defining $w = 33\% (SMA_\mathrm{max} - r_\mathrm{c})$: green: $r_\mathrm{c} < SMA \leq r_\mathrm{c} + w$, orange: $r_\mathrm{c} + w < SMA \leq r_\mathrm{c} + 2w$, brown: $r_\mathrm{c} + 2w < SMA < SMA_\mathrm{max}$. Second and third panels: histograms of the distributions of the $x$ and $y$ coordinates of the isophotal centers.}
\label{fig: NGC4373_W2}
\end{figure*}

\begin{figure*}[h]
\begin{center}$
\begin{array}{ccc}
\includegraphics[trim=3.75cm 1cm 3cm 0cm, clip=true, scale=0.48]{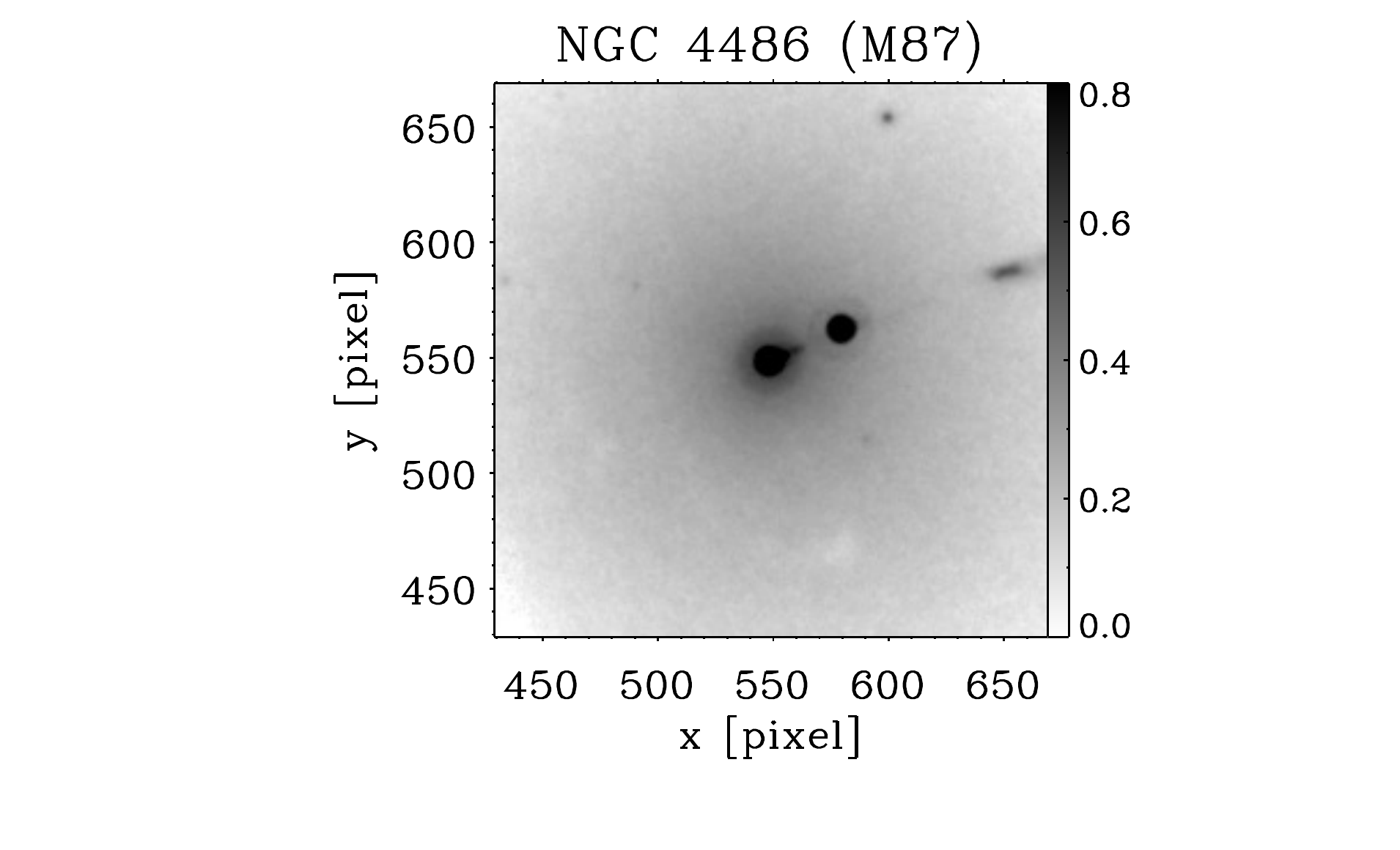} & \includegraphics[trim= 4.cm 1cm 3cm 0cm, clip=true, scale=0.48]{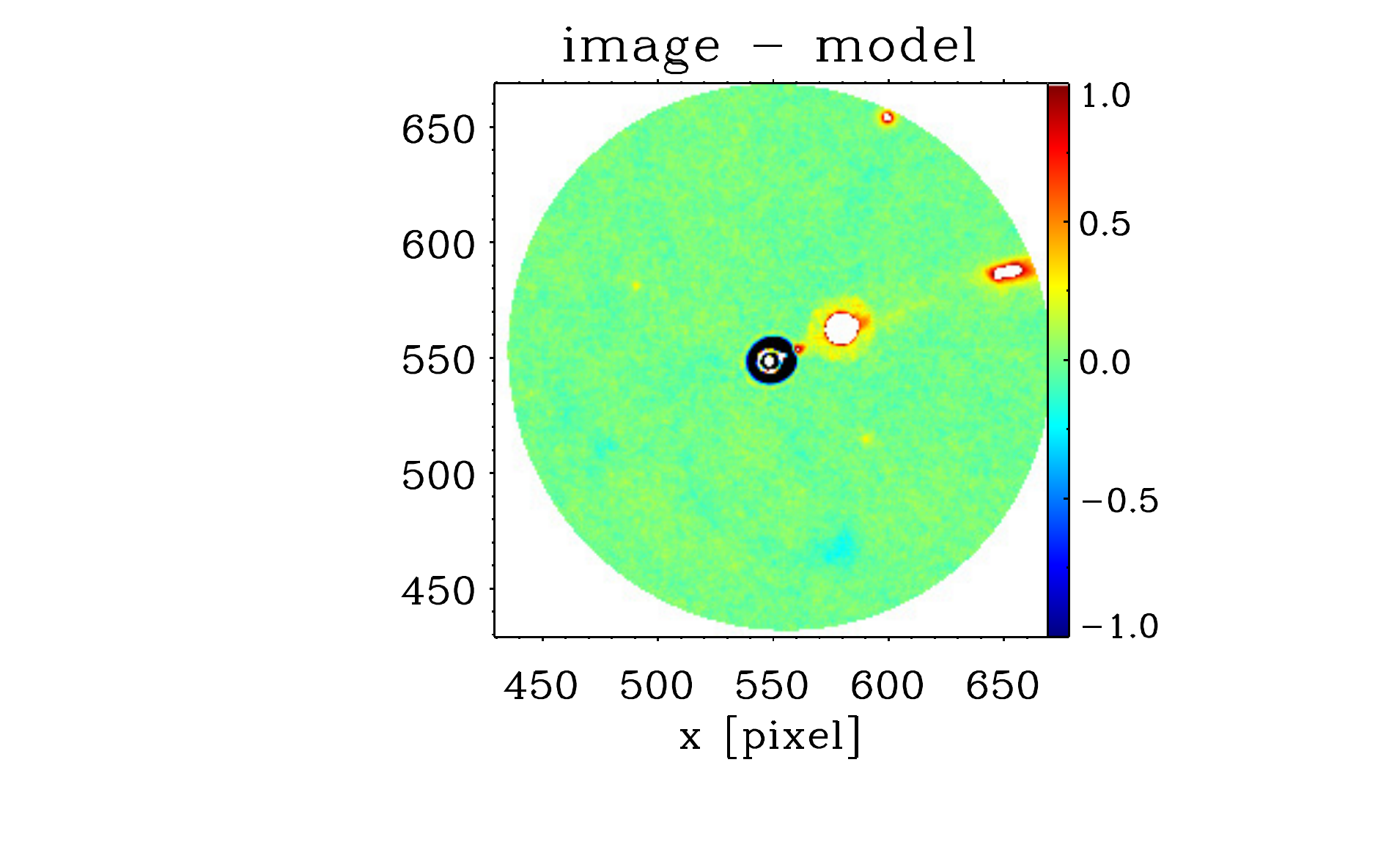}	& \includegraphics[trim= 4.cm 1cm 3cm 0cm, clip=true, scale=0.48]{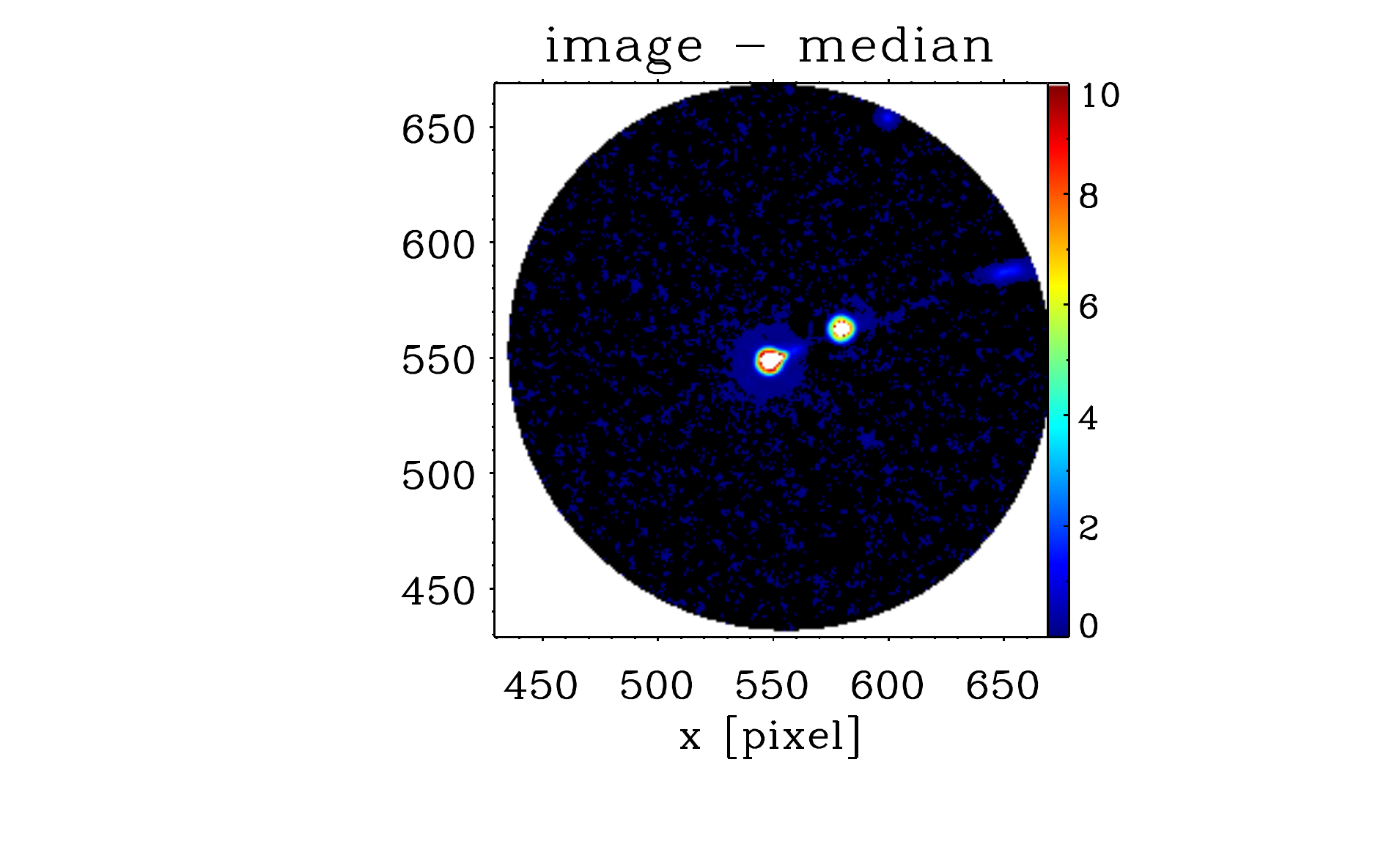} \\
\includegraphics[trim=0.7cm 0cm 0cm 0cm, clip=true, scale=0.46]{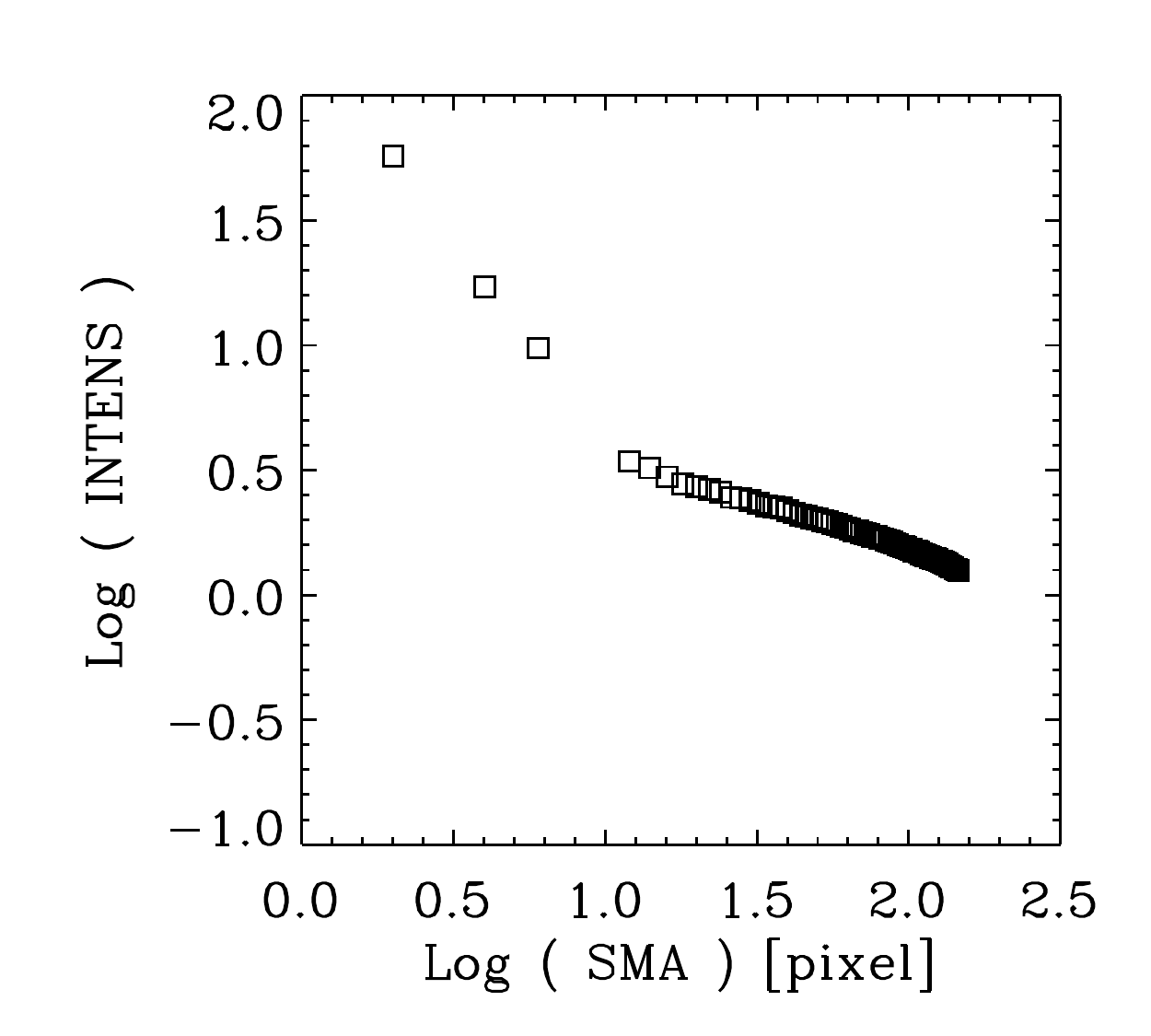}	    &  \includegraphics[trim=0.6cm 0cm 0cm 0cm, clip=true, scale=0.46]{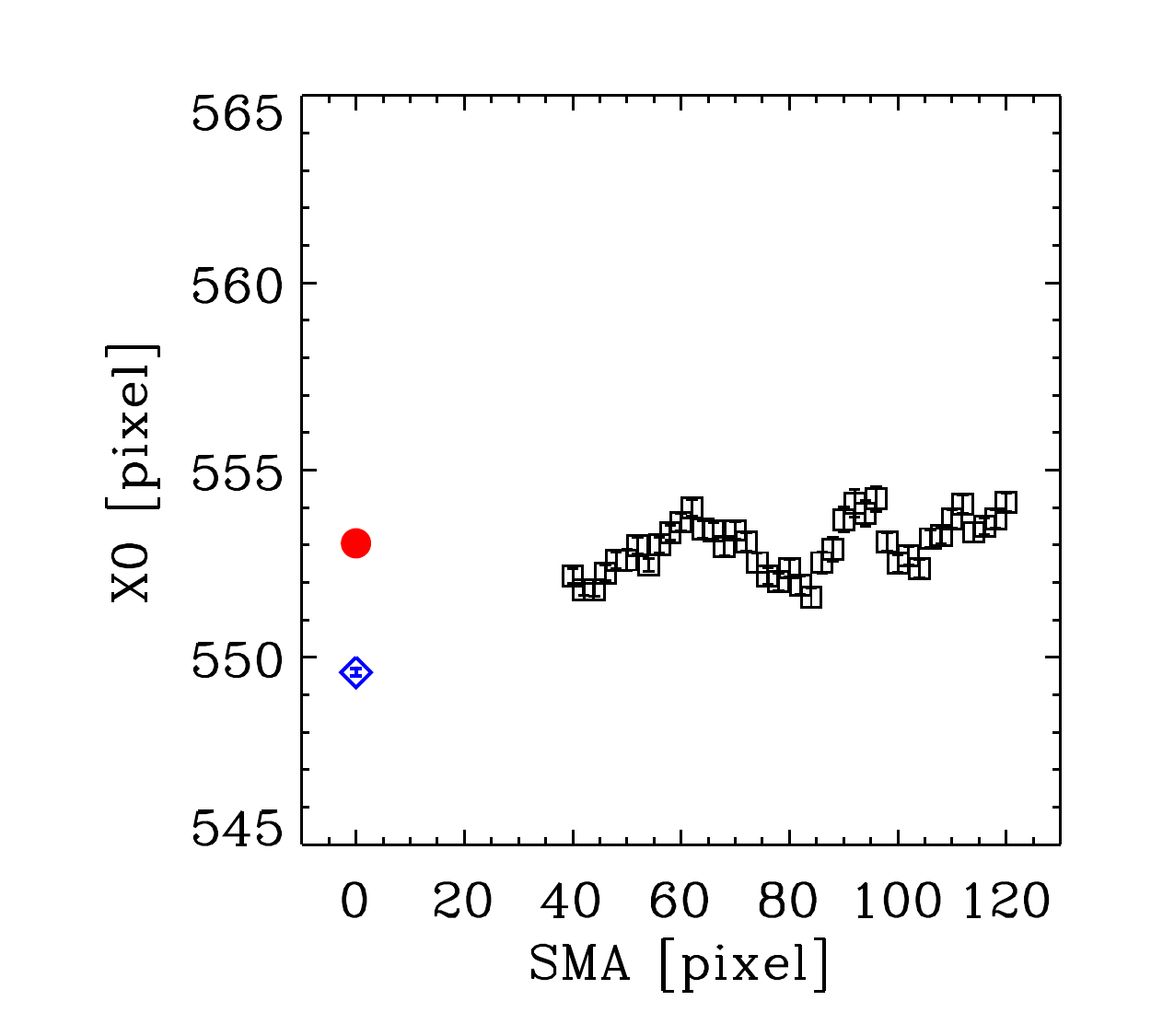}  &  \includegraphics[trim=0.6cm 0cm 0cm 0cm, clip=true, scale=0.46]{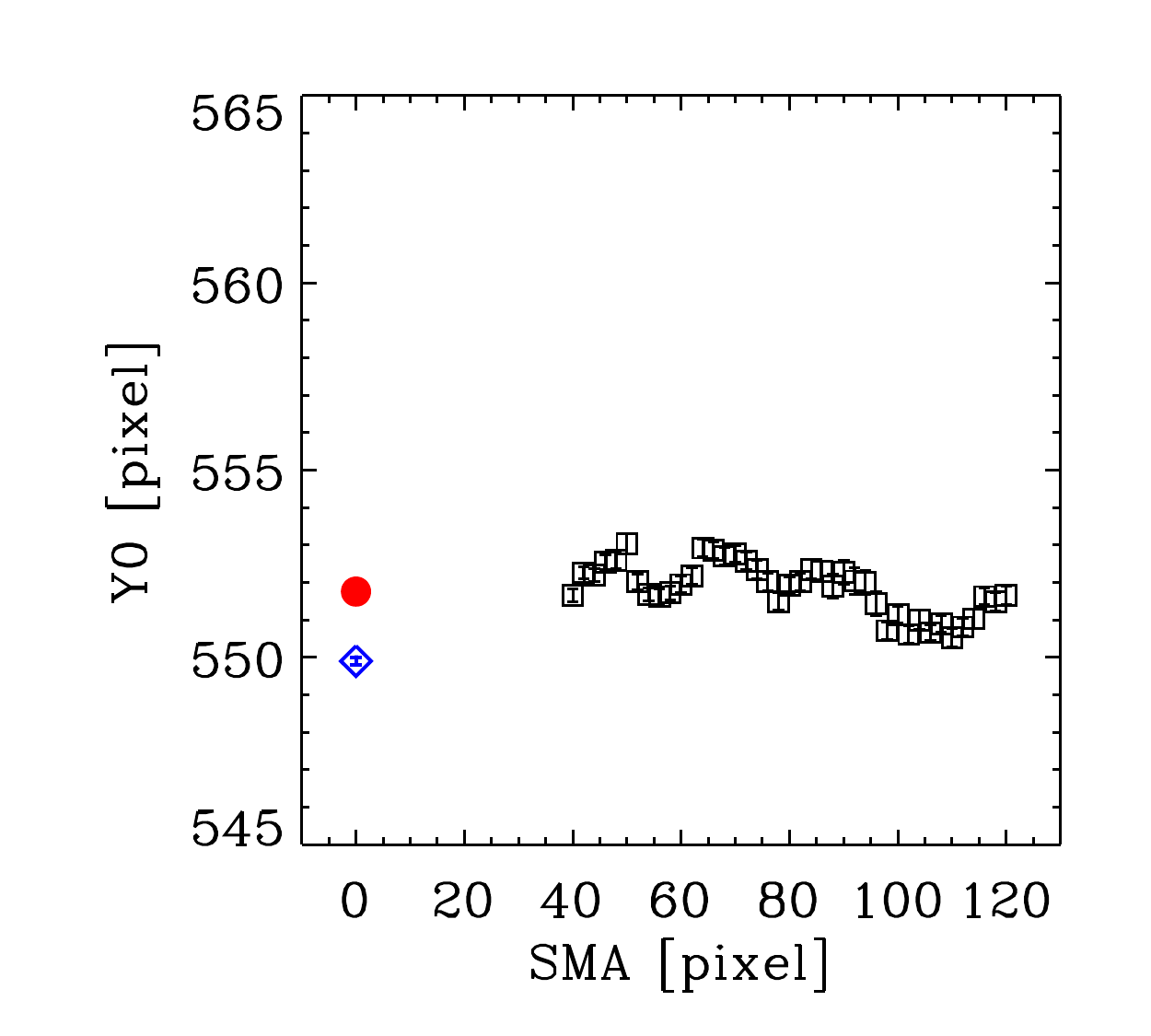} \\	
 \includegraphics[trim=0.65cm 0cm 0cm 0cm, clip=true, scale=0.46]{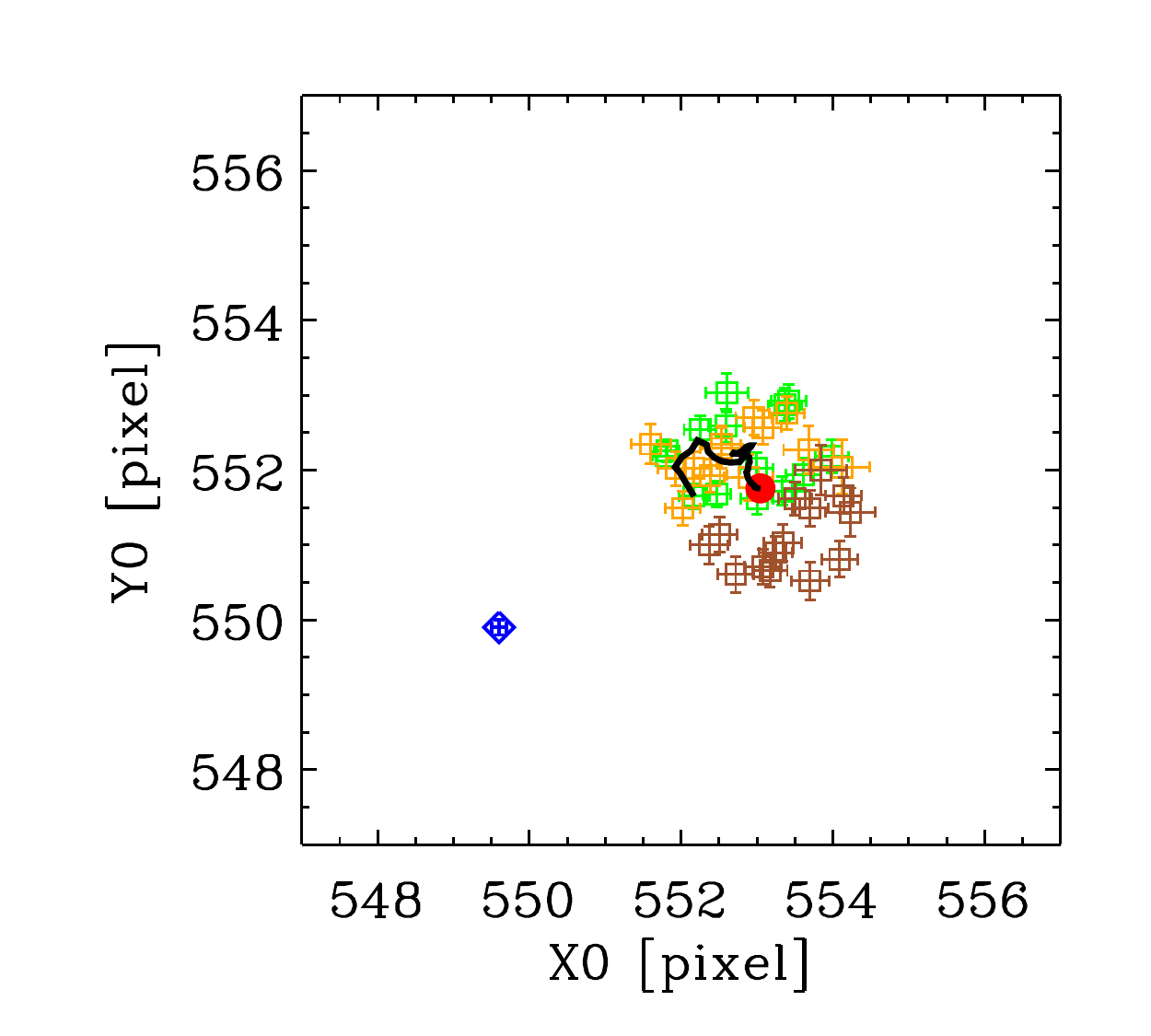}	&  \includegraphics[trim=0.6cm 0cm 0cm 0cm, clip=true, scale=0.46]{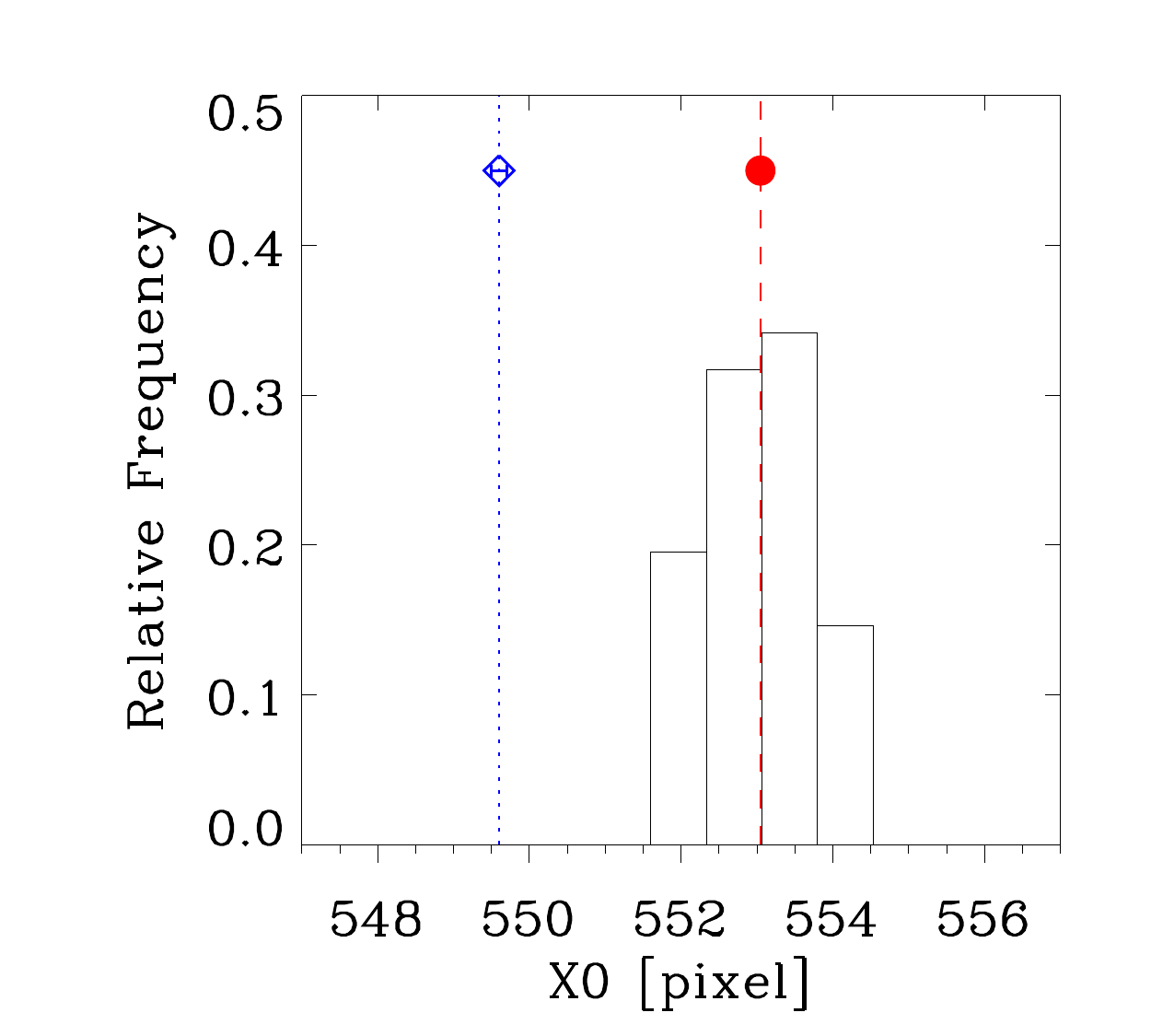}	& \includegraphics[trim=0.6cm 0cm 0cm 0cm, clip=true, scale=0.46]{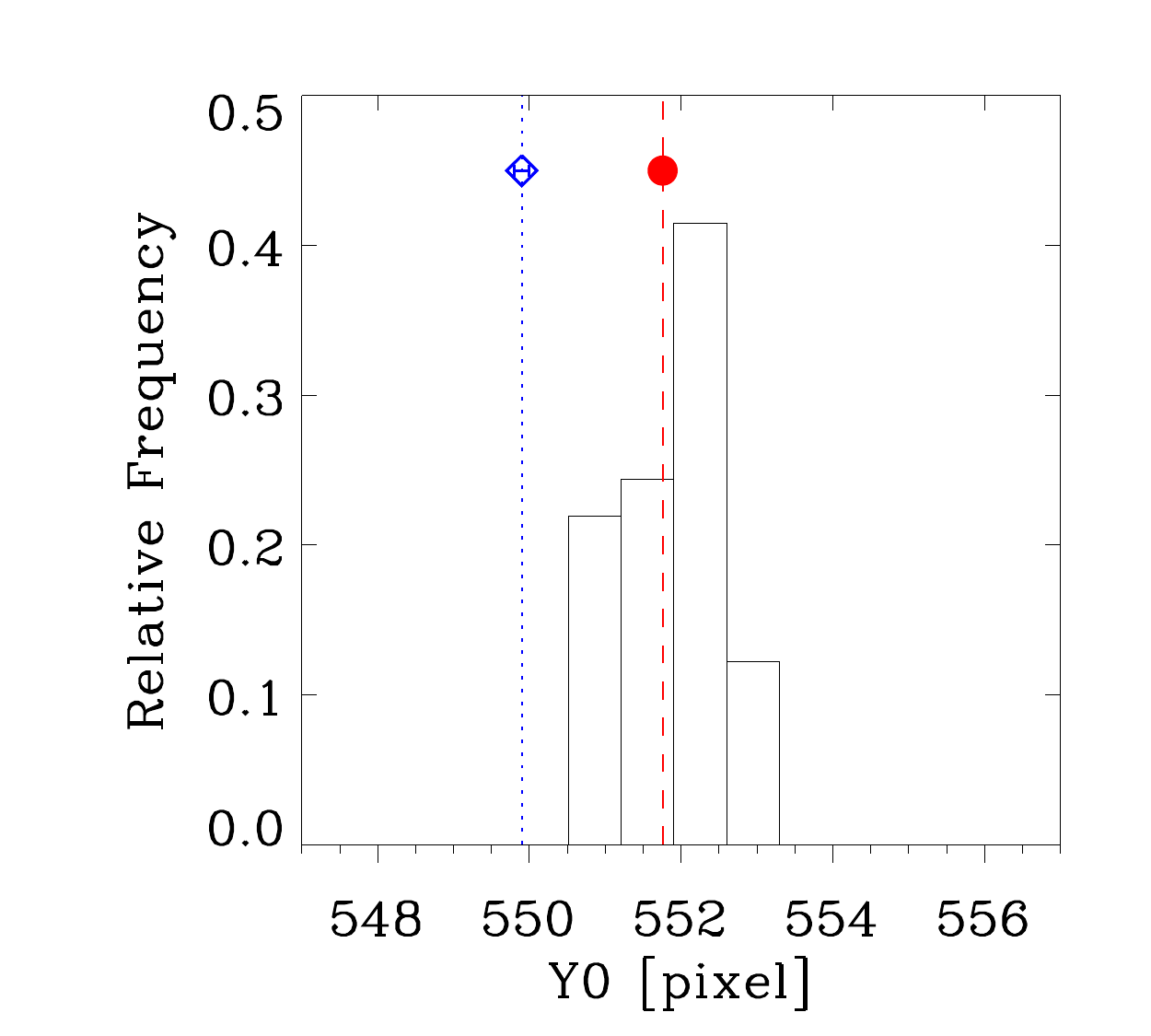}\\
\end{array}$
\end{center}
\caption[M87 (ACS)]{Example of a galaxy where the displacement is significant. NGC 4486 (M 87), ACS/HRC/F814W, scale=$0\farcs025$/pxl. Caption as in Fig.\ref{fig: NGC4373_W2}.} 
\label{fig: M87_ACS_F814W}
\end{figure*}

Figures~\ref{fig: NGC4373_W2} and \ref{fig: M87_ACS_F814W} illustrate examples of the
results for galaxies in which, respectively, no significant displacement was found
(NGC4373, Fig. \ref{fig: NGC4373_W2}) and in which the measured displacement is considered significant 
(NGC 4486, Fig. \ref{fig: M87_ACS_F814W}). In each figure, the top row of panels shows the
original image and the residual images after subtraction of, respectively, the isophotal
model generated by \textit{ellipse} fits and a median filtered image. In the middle row
are plotted the surface brightness profile and the $x$ and $y$ pixel coordinates of
the isophotal centers, all as functions of the ellipse SMA. The positions of the 
photocenter and the AGN point source are also plotted. The leftmost panel of the bottom
row is a scatter plot showing the distribution of the isophote centers, colour coded
according to SMA. The locus of the cumulative mean photocenter position, computed outwards
from the inner SMA limit, is also shown as a solid black line. The last two panels show
the distributions of the ($x, y$) isophote center co-ordinates. The co-ordinates of the
mean photocenter and the AGN point source are also plotted in all three panels. Similar
figures for the remaining galaxies are presented in the Appendix (Sec. \ref{app: images}),
where the results for individual galaxies are also described in more detail (Sec.
\ref{app: galaxies}).

In the case of NGC 4373, the $x$ and $y$ components of the displacement are $< 1\sigma$ and thus consistent with zero.
In the case of the ACS/HRC F814W image of NGC 4486, both the $x$ and $y$ components of the displacement
are significant at the $>3\sigma$ level {\em and} their magnitudes are  $\ge 2.4\times $IQR and accordingly classified as 
having a ``high'' significance level (Table~\ref{tab: snsummary}).

For four galaxies, two or more images obtained with different instruments and/or filters were
analyzed, providing various combinations of wavelength and spatial resolution. The
photocenter displacements in milliarcseconds, relative to the AGN point source, are plotted
for these cases in Fig.~\ref{fig: multif}. There are two galaxies, NGC 4696 and 5419, in
which a second point-like brightness peak is present. 
In these cases, the following discussion refers to the displacement relative to the
primary point source (the brightest, which we take to be the AGN), unless otherwise noted.

AGN--photocenter displacements exceeding the minimum detectable value (i.e., significant
at 3$\sigma$ level; Sec.~\ref{subsubsec: err}) were measured in at least one direction
($x$ or $y$), in at least one image, for ten out of the 14 galaxies in the sample: NGC
1399, 4168, 4278, 4486 (M87), 4636, 4696, 5419, 5846, IC 4296, and IC 4931. In the remaining galaxies,
 NGC 4261, 4373, 4552, and IC 1459, the AGN and photocenter are coincident within the position
uncertainties ($< 3\sigma$). 

The distribution of the isophote center co-ordinates provides an indication of the
systematic uncertainty arising from photometric distortions due to effects such as those
outlined in Sec.~\ref{subsubsec: bias}. As discussed in Sec.~\ref{sec: off}, we assign a
significance level for each measured displacement based on its magnitude relative to the
IQR, which characterizes the width of the isophote center distribution. Three galaxies
exhibit displacements (in at least one direction, in one image)  $\Delta x, \Delta y \ge
1.6\times$IQR, that we classify as having intermediate or high significance. These are NGC
4278, 4486 (M87) and 5846. Three more have displacements classified as having low significance:
NGC 1399, 5419 and IC 4296 ($\Delta x, \Delta y \ge 0.8\times$IQR).

 In the two double nucleus galaxies, the photocenter position is much closer to the
 primary point source than the secondary. The offsets (relative to the primary) are
 $\approx 0.5$ pxl, but given the large IQR in NGC 4696 the displacement is classified as
 not significant in this case. In neither galaxy does the photocenter lie between the two
 ``nuclei'' (Fig. \ref{fig: multif_DN}, \ref{fig: NGC4696_ACS} and \ref{fig: NGC5419_WF}).

For the galaxies where multiple images were analyzed, the displacements in mas derived
from the different images are consistent to within $2\sigma$ for two of the four galaxies (NGC
1399 and 4278). 

In the remaining two galaxies (NGC 4486 and IC 4296), the results from one or more images differ significantly from the others
(Fig.~\ref{fig: multif}). 
In IC 4296, the offset measured from the red (F814W) WFPC2 image is significantly
different from those measured from the ACS-F625W and NICMOS2 F160W images, which are
consistent to within $2\sigma$. This galaxy has a warped dusty disk oriented approximately
E-W, extending $\approx 1^{\prime\prime}$ either side of the nucleus but it seems unlikely that this
structure is the cause of the discrepancy, the origin of which remains unclear. 

Several ACS, NICMOS2 and WFPC2 images were analyzed for NGC 4486 (M87). For the ACS red
optical image (F814W) we recover the B10 result, which indicates that the
photocenter is displaced by $\approx 100$\,mas to the north-west of the nucleus. A smaller
but less significant offset in approximately the same direction is measured from the WFPC2
F814W image. However, the photocenters derived from the NICMOS2 F110W, F160M and F222W
image are consistent with the AGN position. As discussed in detail in Appendix~\ref{app:
galaxies}, the origin of these differences is unclear. Neither dust, nor the prominent
optical jet (also visible in the NIR) appear to provide satisfactory explanations. The
central region of the galaxy where the isophote fitting was performed ($1^{\prime\prime} \le r \le
3^{\prime\prime}$) appears to be free of large scale dust features. The masking procedure should
prevent the jet from distorting the isophote fits but in any case, even when the fits were
repeated with different levels of masking and even with no mask at all, no significant differences in the mean photocenter
position were found. As there is no compelling reason to favor or discard the results from any given
image, we compute a weighted average displacement.

Whether classified as significant, or not, the measured angular photocenter
displacements are always small, typically a few$\times 10$\,mas and almost without exception
$\lesssim 100$ mas. The only cases that exceed 100 mas  are the displacements relative to
the secondary point sources in the two double nucleus galaxies. Disregarding these, the
projected linear displacements are all $\lesssim 10$\,pc. Therefore, for every galaxy analyzed, the measured displacement is a small
fraction of the galaxy core radius; as a fraction of $r_{c}$, the range in displacement
magnitude is approximately 1--10\%. 

In our previous analysis of NGC 4486 (B10), we found that the galaxy photocenter
(in ACS and WFPC images) is displaced in approximately the same direction as the jet, implying that the SBH is
displaced in the counter-jet direction. It is therefore of interest to compare the
direction of the measured photocenter displacements with the radio source axis for this
sample. In Table \ref{tab: PA} we give the position angle (PA) of the radio source, for those
galaxies in which jets or jet-like extended structures have been observed, along with the PA of
the measured displacement (as derived from the \textit{x, y} components). Five galaxies
are associated with relatively powerful, extended ($\sim 10$ kpc) radio sources: NGC 1399,
4261, 4486, 4696 and IC 4296\footnote{We omit NGC 5419 from this list, since the origin of
its unusual radio source is unclear, see Section~\ref{app: galaxies}.}.  Of these, NGC 4486
(M 87; 3C 274), NGC 4261 (3C 270) and IC 4296 (PKS 1333-33) are Fanaroff-Riley Type I
(FRI) radio sources, characterized by prominent twin jets ending in kiloparsec-scale lobes
\citep{FR74}. All three also have parsec-scale jets approximately aligned with the
kpc-scale structure. 

In NGC\,4486, the displacement PA derived from our reanalysis of the ACS F814W image
agrees with the B10 result. The direction of the weighted mean displacement,
derived from displacements obtained from all the images, is also closely aligned with the
jet direction. 
We find a similar result for IC 4296. As already noted, the results  from
the ACS and NICMOS2 images are consistent and indicate a displacement to the NW. If the
discrepant WFPC2 result is disregarded, the weighted mean of the ACS and NICMOS2
displacements yields a PA only $\sim 4^{\circ}$ different from that of the large-scale
radio axis, with the photocenter displaced on the side of the brighter north-western jet. 

The displacement measured in NGC 4261 is not considered to be significant ($< 3\sigma$). Nevertheless,
the derived PA is consistent (albeit within the large uncertainty) with that of the radio axis, and
again, the offset is in the direction of the brighter western jet. 

NGC\,1399 also has an FRI-like radio source morphology with a well-defined axis in
PA$\approx -10^{\circ}$. The measured displacement (classified as low significance) has a consistent 
PA with a value of $-17\pm16^{\circ}$. NGC\,4696, is one of the galaxies with two optical
point sources. Its large scale radio source does not exhibit well-defined jets, but is
elongated approximately E--W over $\sim 10$\,kpc, with the ends of both ``arms'' bending
south. At parsec scales there is a compact core with a one-sided jet emerging to the SW in
PA$\sim -150^{\circ}$. However, it is not clear which of the two optical nuclei hosts the core
radio source \citep{Taylor06}, complicating the comparison with the photocenter
displacement. If the secondary point source is identified as the AGN, then the
photocenter, which is close to the primary point source, is displaced
approximately in the counter-jet direction. On the other hand, if the primary point source is
indeed the AGN, the displacement (which, in any case, is classified as having ``null'' significance) would be
almost perpendicular to the jet. 

Three more galaxies have relatively weak, small radio  sources that have jet-like features
or elongations on sub-kiloparsec scales. The measured photocenter displacements are not
aligned with  the radio source axis in NGC 4552 or NGC 4636. NGC 4278 has a compact ($\sim
3$\,pc) source consisting of a core with jet-like features emerging along a SE--NW axis on either side. These features gradually bend to 
the east and west, respectively, becoming fainter and more diffuse. The weighted mean photocenter position is displaced
approximately in the initial direction of the SE jet. According to \citet{Giro2005}'s analysis of this source, the jet axis is closely aligned with our line of sight,
with the SE component being the oppositely directed counter-jet.

\begin{table*}[p]
\centering
\caption{MEASURED PROJECTED OFFSETS} 
\label{tab: snsummary} 
\centering
\scalebox{.85}{
\begin{tabular}{ cl  |  cccc  | cccc  |  cll}
\hline \hline
&&&&&&&&&&&&\\ 
 \textbf{Galaxy} &\multicolumn{1}{c |}{ \textbf{Instrument}} & \multicolumn{1}{c}{ \textbf{region}} & \multicolumn{3}{c|}{$\dbinom{X}{Y}$} & \multicolumn{4}{c|}{Offset} 									& 	Dir 		& Type\\
&&&&&&&&&&&&\\ 
                   	&HST					& 	 		& IQR 	&       PC 			&	NPS			& [pxl] 				& [mas] 			& 		[IQR]		&	[pc]				& 			& \\ 
	(1)		&	(2)					& 	(3)		& 	(4)	&	(5)			&	(6)			& (7)					&	(8)			&		(9)		&	(10)				&	(11)		& (12) \\ \hline	
&&&&&&&&&&&&\\ 
 NGC 1399 	& WFPC2/PC - F606W		& [2.5:15]    	& 0.19 	& 520 $\pm$ 0.1	& 519.9 $\pm$ 0.1	& 0.1 $\pm$ 0.14 		& 5 $\pm$ 7 	 	& 0.5 $\pm$ 0.7	&	0.4 $\pm$ 0.6		& W			& non sig\\ 
  		  	& 						&  	 		& 0.28	& 465.4 $\pm$ 0.1 	& 465 $\pm$ 0.1	& \textbf{0.4 $\pm$ 0.14}	& 20 $\pm$ 7		& 1.4 $\pm$ 0.5 	&	1.7 $\pm$ 0.6		& N			& \textbf{low}\\ 
&&&&&&&&&&&&\\ 
			&WFPC2/PC - F814W		& 		 	& 0.29	& 502.9 $\pm$ 0.1	& 502.9 $\pm$ 0.1 	& 0 $\pm$ 0.14 		& 0 $\pm$ 7 		& 0 $\pm$ 0.5 		&	0 $\pm$ 0.6		& - 			& non sig\\ 
			&						&  			& 0.27	& 438.6 $\pm$ 0.1	& 438.3 $\pm$ 0.1 	& 0.3 $\pm$ 0.14 		& 15 $\pm$ 7 		& 1.1 $\pm$ 0.5 	&	1.3 $\pm$ 0.6		& N 			& non sig\\ 
&&&&&&&&&&&&\\ 
			&WFC3/IR - F110W			& 		 	& 0.13	& 759.9 $\pm$ 0.2	& 759.7 $\pm$ 0.1	& 0.2 $\pm$ 0.22 		& 18 $\pm$ 20		& 1.5 $\pm$ 1.7	& 	1.6 $\pm$ 1.8		& W 			& non sig\\
			&						&  			& 0.14	& 725.4 $\pm$ 0.2	& 725.3 $\pm$ 0.1	& 0.1 $\pm$ 0.22		& 9 $\pm$	20		& 0.7 $\pm$ 1.6	& 	0.8 $\pm$	1.8		& N			& non sig\\
&&&&&&&&&&&&\\ 
			&WFC3/IR - F160W			& 		 	& 0.18	& 759.9 $\pm$	0.2	& 759.7 $\pm$ 0.1	& 0.2 $\pm$ 0.22		& 18 $\pm$ 20		& 1.1 $\pm$ 1.2 	&	1.6 $\pm$ 1.8		& W			& non sig\\
			&						&  			& 0.08	& 725.4 $\pm$	0.2	& 725.4 $\pm$	0.1	& 0 $\pm$	 0.22			& 0 $\pm$ 20		& 0 $\pm$ 2.8		& 	0 $\pm$ 1.8 		& - 			& non sig\\
&&&&&&&&&&&&\\ 
NGC 4168 	& WFPC2/PC - F702W 		& [2.1:8.5] 	& 1.76 	& 409.6 $\pm$ 0.1 	& 408.9 $\pm$ 0.1 	& \textbf{0.7 $\pm$ 0.14} 	& 35 $\pm$ 7  		& 0.4 $\pm$ 0.1	&	5 $\pm$ 1			& W			& null \\ 
			&						& 		 	& 1.47 	& 425.1 $\pm$ 0.1 	& 424.6 $\pm$ 0.1 	& \textbf{0.5 $\pm$ 0.14} 	& 25 $\pm$ 7 		& 0.3 $\pm$ 0.1	&	4 $\pm$ 1			& N			& null \\ 
&&&&&&&&&&&&\\ 
 NGC 4261  	&     NICMOS2 - F160W		& [1.7:7.5] 	& 0.08	& 200 $\pm$ 0.2	& 199.6 $\pm$ 0.2	& 0.4 $\pm$ 0.3 		& 20 $\pm$ 15 		& 5 $\pm$ 3.7		&	2.9 $\pm$ 2.2		& W			& non sig\\ 
   			& 						& 		      	& 0.26	& 182.3  $\pm$ 0.2	& 182.1 $\pm$ 0.2	& 0.2 $\pm$ 0.3 		& 10 $\pm$ 15		& 0.8 $\pm$ 1.1  	&	1.5 $\pm$ 2.2		& N			& non sig\\ 
&&&&&&&&&&&&\\ 
NGC 4278 	& ACS/WFC - F850LP 		& [1.2:9.1]		& 0.31	& 1165.6 $\pm$ 0.1 	& 1166.4 $\pm$ 0.1	& \textbf{0.8 $\pm$ 0.14} 	& 40 $\pm$ 7 		& 2.6 $\pm$ 0.5	&	3.4 $\pm$ 0.6		& E			& \textbf{high}\\
			& 						& 		 	& 0.83	& 3160 $\pm$ 0.1	& 3161.1 $\pm$ 0.1	& \textbf{1.1 $\pm$ 0.14} 	& 55 $\pm$ 7 	 	& 1.3 $\pm$ 0.2	&	4.7 $\pm$ 0.6		& S			& \textbf{low}\\
&&&&&&&&&&&&\\ 
			& WFPC2/PC - F814W		&			& 0.44	& 370.2 $\pm$ 0.1	& 370.6 $\pm$ 0.1	& \textbf{0.4 $\pm$ 0.14}	& 22 $\pm$ 7		& 1 $\pm$ 0.3		& 	1.9 $\pm$ 0.6		& E			& \textbf{low}\\
			&						&			& 0.87	& 357.3 $\pm$ 0.1	& 359     $\pm$ 0.1	& \textbf{1.7 $\pm$ 0.14}	& 84 $\pm$ 7		& 1.9 $\pm$ 0.2	&	7.2 $\pm$ 0.6		& S			& \textbf{int} \\
&&&&&&&&&&&&\\ 
			& NICMOS2 - F160W		& 		  	& 0.26	& 245.8 $\pm$ 0.2	& 246.4 $\pm$ 0.2 	& 0.6 $\pm$ 0.3	 	& 30 $\pm$ 15	 	&  2.3 $\pm$ 1.2 	&	2.6 $\pm$ 1.3		& E			& non sig\\ 
    			&  						&  			& 0.62	& 224.8 $\pm$ 0.2	& 225.8 $\pm$ 0.2 	& \textbf{1 $\pm$ 0.3}	& 50 $\pm$ 15 		&  1.6 $\pm$ 0.5	&	4.3 $\pm$ 1.3		& S 			& \textbf{int}\\ 
&&&&&&&&&&&&\\ 						
 NGC 4373  	& WFPC2/PC - F814W 		& [1.2:10]	 	& 0.58	& 2328.7 $\pm$ 0.1 	& 2328.8 $\pm$ 0.1	& 0.1 $\pm$ 0.14		& 5 $\pm$ 7		& 0.2 $\pm$ 0.2	&	2 $\pm$ 2.7		& E			& non sig\\ 
   			& 						&  			& 0.16	& 870.7 $\pm$ 0.1 	& 870.8 $\pm$ 0.1	& 0.1 $\pm$ 0.14		& 5 $\pm$ 7		& 0.6 $\pm$ 0.9	&	2 $\pm$ 2.7		& S			& non sig\\ 
&&&&&&&&&&&&\\ 
  NGC 4486 	& ACS/HRC - F606W $^{\star}$&   [1:3] 	 	& 2.78	& 419.1 $\pm$ 0.1  	& 682.1  $\pm$ 0.1	& \textbf{1.4 $\pm$ 0.14}	& 35 $\pm$ 4		& 0.5 $\pm$ 0.1 	&	2.7 $\pm$ 0.3		& W			& null \\
 		   	&  						&      			& 0.98	& 417.7 $\pm$ 0.1 	& 679.5  $\pm$ 0.1	& \textbf{2.6 $\pm$ 0.14}	& 65 $\pm$ 4		& 2.7 $\pm$ 0.1	&	5.1 $\pm$ 0.3		& N			& \textbf{high}\\
&&&&&&&&&&&&\\  
		 	& ACS/HRC - F814W $^{\star}$&   	 	 	& 1.03	& 553.1 $\pm$ 0.1  	& 549.6 $\pm$ 0.1	& \textbf{3.5 $\pm$ 0.14}	& 87 $\pm$ 4		& 3.4 $\pm$ 0.1 	&	6.8 $\pm$ 0.3		& W			& \textbf{high} \\
 		   	&  						&      			& 0.8		& 551.8 $\pm$ 0.1 	& 549.9 $\pm$ 0.1	& \textbf{1.9 $\pm$ 0.14}	& 47 $\pm$ 4		& 2.4 $\pm$ 0.1	&	3.6 $\pm$ 0.3		& N			& \textbf{high}\\
&&&&&&&&&&&&\\  
			& NICMOS2 - F110W 		&    			& 0.42	& 207.2 $\pm$ 0.2	& 206.8 $\pm$ 0.2	& 0.4 $\pm$ 0.3		& 20 $\pm$ 15  	&  1 $\pm$ 0.7		&	1.6 $\pm$ 1.2		& W 			& non sig \\
 			&  						&    			& 0.96	& 277.9 $\pm$ 0.2	& 277.7 $\pm$ 0.2	& 0.2 $\pm$ 0.3 		& 10 $\pm$ 15  	&  0.2 $\pm$ 0.3  	&	0.8 $\pm$ 1.2		& N 			& non sig \\
&&&&&&&&&&&&\\ 
  			& NICMOS2 - F160W 		&    			&  0.65	& 207.5 $\pm$ 0.2	& 207.4 $\pm$ 0.2	& 0.1 $\pm$ 0.3		& 5 $\pm$ 15 		&  0.2 $\pm$ 0.5 	&	0.4 $\pm$ 1.2		& W			& non sig\\
 			&  						&    			&  0.76	& 277.8 $\pm$ 0.2	& 277.7 $\pm$ 0.2	& 0.1 $\pm$ 0.3		& 5 $\pm$ 15 		&  0.1 $\pm$ 0.4  	&	0.4 $\pm$ 1.2		& N			& non sig\\
&&&&&&&&&&&&\\  
			& NICMOS2 - F222M 		& 		 	& 0.51 	& 207.6 $\pm$ 0.2 	& 207.2 $\pm$ 0.2 	& 0.4 $\pm$ 0.3 		& 20 $\pm$ 15 		& 0.8 $\pm$ 0.6 	&	1.6 $\pm$ 1.2		& W 			& non sig\\ 
			&						&  			& 0.59 	& 277.3 $\pm$ 0.2 	& 277.9 $\pm$ 0.2 	& 0.6 $\pm$ 0.3 		& 30 $\pm$ 15 		& 1 $\pm$ 0.5 		&	2.3 $\pm$ 1.2		& S 			& non sig\\ 
&&&&&&&&&&&&\\ 
  			& WFPC2/PC - F814W		&   	    		& 0.42	& 448.3 $\pm$ 0.1	& 447.6 $\pm$ 0.1 	& \textbf{0.7 $\pm$ 0.14}	& 35 $\pm$ 7 		& 1.7 $\pm$ 0.3	&	2.7 $\pm$ 0.6		& W 			& \textbf{low} \\
 			&  						&      			& 0.58	& 517.2 $\pm$ 0.1 	& 517 $\pm$ 0.1	& 0.2 $\pm$ 0.14		& 10 $\pm$ 7		& 0.3 $\pm$ 0.2	&	0.8 $\pm$ 0.6		&N			& non sig\\
&&&&&&&&&&&&\\ 

\hline
\end{tabular}} 
  \tablecomments{(1) Optical name; (2) instrument/camera/filter on HST; (3) lower and upper limit, in arcseconds, of the region analyzed; (4) inter-quartile range in pixel, this is the difference between 75th and 25th percentile of the isophotal center dataset; (5) photocenter position in pixel on the frame; (6) nuclear point source position in pixels; (7 - 8) offset of the isophotal center with respect to the nuclear point source in pixels and milliarcseconds; (9) offset in pixel normalized by the corresponding IQR; (10) offset in pc; (11) direction of the offset; (12) type of displacement as defined in \textsection \ref{sec: off}. $^{\star}$: for further details see Table 1 in B10. }
\end{table*}

\begin{table*}[!h]
\addtocounter{table}{-1}
\centering
\caption{MEASURED PROJECTED OFFSETS (continued)} 
\centering
\scalebox{.78}{
\begin{tabular}{ cc  |  ccccc  | cccc  |  ll}
\hline \hline
&&&&&&&&&&&&\\ 
 \textbf{Galaxy} &\multicolumn{1}{c |}{ \textbf{Instrument}} & \multicolumn{1}{c}{ \textbf{region}} & \multicolumn{4}{c|}{$\dbinom{X}{Y}$} & \multicolumn{4}{c|}{Offset} 															& Dir 		& Type\\
&&&&&&&&&&&&\\ 
                   	&HST					&  	 			& IQR 	&       PC 			&	NPS1		& NPS2			& [pxl] 				& [mas] 			& 		[IQR]			&	[pc]				& 			& \\ 
	(1)		&	(2)					& 	(3)			& 	(4)	&	(5)			&	(6)			& 	(7)			&	(8)				&	(9)			&		(10)			&	(11)				&	(12)		& (13) \\ \hline	
&&&&&&&&&&&&\\ 
 NGC 4552 	& WFPC2/PC - F814W		& [1:15] 			& 0.42 	& 538.9 $\pm$ 0.1 	& 538.7 $\pm$ 0.1 	&				& 0.2 $\pm$ 0.14		& 10 $\pm$ 7 		& 0.5 $\pm$ 0.3		&	0.7 $\pm$ 0.5		& W 			& non sig\\ 
			& 						&  				& 0.89	& 500.8 $\pm$ 0.1	& 500.4 $\pm$ 0.2 	&				& 0.4 $\pm$ 0.23	 	& 20 $\pm$ 12 		& 0.4 $\pm$ 0.3 		&	1.5 $\pm$ 0.9		& N 			& non sig\\  
&&&&&&&&&&&\\  
NGC 4636 	& WFPC2/PC - F814W 		& [4:14] 			& 0.94 	& 698 $\pm$ 0.1 	& 697.4 $\pm$ 0.1 	&				& \textbf{0.6 $\pm$ 0.14} 	& 30 $\pm$ 7 		& 0.6 $\pm$ 0.2		&	1.9 $\pm$ 0.5		& W			& null\\  
			&						&  				& 0.93 	& 569.9 $\pm$ 0.1 	& 570 $\pm$ 0.1 	&				& 0.1 $\pm$ 0.14 		& 5 $\pm$ 7		& 0.1 $\pm$ 0.2		&	0.3 $\pm$ 0.5		& S 			& non sig\\   
&&&&&&&&&&&&\\ 
 NGC 4696	& ACS/WFC - F814W 		& [1.5:16]	   		& 2.15	& 1582.9 $\pm$ 0.1	& 1583.4 $\pm$ 0.1	&  				&  \textbf{0.5 $\pm$ 0.14}	&  25 $\pm$ 7 		& 0.2 $\pm$ 0.07		&	4.5 $\pm$ 1.3		& E			& null \\ 
 			& 						&  				& 3.78	& 2977.6 $\pm$ 0.1	& 2977.8 $\pm$ 0.1	&  				&  0.2 $\pm$ 0.14		 & 10 $\pm$ 7 		& 0.05 $\pm$ 0.04		&	1.8 $\pm$ 1.3		& S			& non sig\\ 
&&&&&&&&&&&&\\ 
 		    	&  		 				& 				& 		& 				& 				& 1586.1 $\pm$ 0.1 	& \textbf{3.2 $\pm$ 0.14}	&  160 $\pm$ 5		& 1.5 $\pm$ 0.05		&	29 $\pm$ 1.3		& E			& \textbf{low}\\ 
 		  	& 		 				&  				& 		& 				& 				& 2973.2 $\pm$ 0.1 	&  \textbf{4.4 $\pm$ 0.14}	&  220 $\pm$ 5 	& 1.2 $\pm$ 0.03		&	39 $\pm$ 1.3		& N			& \textbf{low}\\ 
&&&&&&&&&&&\\  
  NGC 5419	& WFPC2/PC - F555W 		& [2:13.5]			& 0.5 	& 398.4 $\pm$ 0.1 	&  397.8 $\pm$ 0.1	& 				&  \textbf{0.6 $\pm$ 0.14}	& 30 $\pm$ 7		& 1.2 $\pm$ 0.3		&	7 $\pm$ 2			& W			& \textbf{low} \\
    			& 						&     				& 0.22	& 447.6 $\pm$ 0.1 	& 447.8 $\pm$ 0.1	& 				&  0.2 $\pm$ 0.14		& 10 $\pm$ 7		& 0.9 $\pm$ 0.6		&	2 $\pm$ 2			& S			& non sig\\
&&&&&&&&&&&&\\ 
  	 	    	&    						& 				& 		&  				& 				& 397.1 $\pm$ 0.1	& \textbf{1.3 $\pm$ 0.14}	& 65 $\pm$ 7	& \textbf{2.6 $\pm$ 0.2}		&	15 $\pm$ 2		& W			& \textbf{high} \\
    		     	&   						&      				& 		&  				& 				& 442.5 $\pm$ 0.1 	& \textbf{5.1 $\pm$ 0.14} 	& 255 $\pm$ 7	& \textbf{23.2 $\pm$ 0.5}		&	60 $\pm$ 2		& N			& \textbf{high} \\
&&&&&&&&&&&&\\ 
NGC 5846 	& WFPC2/PC - F814W		& [1.6:15.8] 		& 0.84	& 519.5 $\pm$ 0.1	& 518.2 $\pm$ 0.4 	&				& \textbf{1.3 $\pm$ 0.4}     & 65 $\pm$ 21		& 1.6 $\pm$ 0.5		&	7.8 $\pm$ 2.5		& W 			& \textbf{int}\\  
			& 						&  				& 1.05	& 466 $\pm$ 0.1	& 466.4 $\pm$ 0.1  	&				& \textbf{0.4 $\pm$ 0.14} 	& 20 $\pm$ 7		& 0.4 $\pm$ 0.1 		&	2.4 $\pm$ 0.9		& S 			& null\\
&&&&&&&&&&&&\\ 
IC 1459		& WFPC2/PC - F814W		& [2.5:15] 			& 0.42 	& 536.4 $\pm$ 0.1 	& 536.7 $\pm$ 0.1 	&				& 0.3 $\pm$ 0.14	 	& 15 $\pm$ 7 		& 0.7 $\pm$ 0.3		&	2.1 $\pm$ 1		& E			& non sig\\   
			&						& 				& 0.50	& 510.5 $\pm$ 0.1	& 510.7 $\pm$ 0.1 	&				& 0.2 $\pm$ 0.14 		& 10 $\pm$ 7 		& 0.4 $\pm$ 0.3	 	&	1.4 $\pm$ 1		& S			& non sig \\ 
&&&&&&&&&&&&\\ 
IC 4296 		& ACS/HRC - F625W  		&  [1.5:6.5]  		& 0.49 	& 833.3 $\pm$ 0.1  	& 832.6 $\pm$ 0.2	& 				& \textbf{0.7 $\pm$ 0.22}	&  18 $\pm$ 6		&  1.4 $\pm$ 0.4 		&	4.2 $\pm$ 1.2		& W			& \textbf{low}\\
			& 						&  				& 0.38	& 785.3 $\pm$ 0.1	& 784.8 $\pm$ 0.1	& 				& \textbf{0.5 $\pm$ 0.14}  	&  13 $\pm$ 4		&  1.3 $\pm$ 0.3		&	3 $\pm$ 0.6		& N			& \textbf{low} \\
&&&&&&&&&&&&\\ 
			& WFPC2/PC - F814W		&  				& 0.39 	& 551.5 $\pm$ 0.1	& 552 $\pm$ 0.18 	&				& 0.5 $\pm$ 0.2 		& 25 $\pm$ 10 		& 1.3 $\pm$ 0.5 		&	6 $\pm$ 2			& E 			& non sig\\ 
			&						&  				& 0.13 	& 515 $\pm$ 0.1 	& 514.9 $\pm$ 0.18 	&				& 0.1 $\pm$ 0.2 		& 5 $\pm$ 10 		& 0.8 $\pm$ 1.6 		&	1 $\pm$ 2			& N 			& non sig\\ 
&&&&&&&&&&&&\\ 
			& NICMOS2 - F160W		&  		   		& 0.13   	& 420.8 $\pm$ 0.2 	& 420.6 $\pm$ 0.2	& 				& 0.2 $\pm$ 0.3		& 13 $\pm$ 14 		& 1.9 $\pm$ 2.2 		&	2.8 $\pm$ 3.2 		& W 			& non sig\\
			& 						&   				& 1.07   	& 253.4 $\pm$ 0.2   	& 252.7 $\pm$ 0.1 & 				& \textbf{0.7 $\pm$ 0.2}	&  37  $\pm$ 11	& 0.7 $\pm$ 0.2		&	8.3 $\pm$ 2.5		& N 			& null\\
&&&&&&&&&&&&\\ 
IC 4931		& WFPC2/PC - F814W		& [1:15] 			& 1.29 	& 403.6 $\pm$ 0.1 	& 404.1$\pm$ 0.1 	&				& \textbf{0.5 $\pm$ 0.14} 	& 25 $\pm$ 7 		& 0.4 $\pm$ 0.1 		&	10 $\pm$ 3		& E 			& null\\ 
			&						&  				& 0.66 	& 505.9 $\pm$ 0.1 	& 505.8 $\pm$ 0.1 	&				& 0.1 $\pm$ 0.14 		& 5 $\pm$ 7 		& 0.2 $\pm$ 0.2 		&	2 $\pm$ 3			& N 			& non sig\\ 
&&&&&&&&&&&&\\ 
\hline
\end{tabular}} 
\tablecomments{(1) Optical name; (2) instrument/camera/filter on HST; (3) lower and upper limit, in arcseconds, of the region analyzed; (4) inter-quartile range in pixel, this is the difference between 75th and 25th percentile of the isophotal center dataset; (5) photocenter position in pixel on the frame; (6) and (7) nuclear point source position in pixels; (8 - 9) offset of the isophotal center with respect to the nuclear point source in pixels and milliarcseconds; (10) offset in pixel normalized by the corresponding IQR; (11) offset in pc; (12) direction of the offset; (13) type of displacement as defined in \textsection \ref{sec: off}.}
\end{table*}

\begin{table*}[tb]
\caption{RADIO JET AND PHOTOCENTER POSITION ANGLES}
\label{tab: PA} \centering
\scalebox{1}{
\begin{tabular}{ l    llll}
\hline \hline
& & &\\
\multicolumn{1}{c}{\textbf{Galaxy}}		& 			\multicolumn{3}{c}{\underbar{\hbox to 250pt{\hfill  \textbf{PA (deg)} \hfill}}} 	 & \\
								& jet 							& \multicolumn{3}{c}{\underbar{\hbox to 280pt{\hfill displacement\hfill}}}  		  \\ 
								&							&	NPS1							&	NPS2			& \\ \hline
& & & \\
kpc-scale jets:& & & &\\
NGC 1399			 			& $-10$						& $-17 \pm 16^{a}$						&					&	low	\\
NGC 4261						& $87 \pm 8$					& $297 \pm 38$ ($117 \pm 38$)			& 					& 	non significant \\
NGC 4486						& $290\pm 3$					& $307 \pm 1$ ($127 \pm 1$) 				&					&	$^{\ddagger}$	\\
NGC 4696						& $210$ ($+30$)		 		& $112 \pm 15$						& 	$36 \pm 1$		&	$^{\ddagger}$ \\
IC 4296							& $130$						& $338 \pm 7$ (158 $\pm$ 8)$^{b}$			&					&	low	\\
& & & \\
pc-scale jets: & & & &\\
NGC 4278						& $100-155$					& $152 \pm 3$ 							&					& 	$^{\ddagger}$	\\
NGC 4552						& $50$						& $333 \pm 21$							&					& 	low\\
NGC 4636						& $35$						& $261\pm 9\ (81\pm 9)$					&					& 	null\\
& & & \\
\hline

\end{tabular}}
\tablecomments{Photocenter position angles are computed with respect to the nuclear point source. Angles increase east of north. When multiple values have been derived from different images an error weighted mean is computed. Values in parenthesis indicate the supplementary angles (i.e. $\theta + 180^{\circ}$). $^{\ddagger}$ The significance changes for different instruments (see Table \ref{tab: snsummary}). $^{a}$This value changes to $-8 \pm 16^{\circ}$ when only WFPC2 results are considered. $^{b}$This value changes to $312 \pm 8^{\circ}\ (132 \pm 8^{\circ})$ when only ACS and NICMOS2 results are considered. See Table \ref{tab: radio} for references on the radio jet PA.}
\end{table*}

\begin{subfigures}
\begin{figure*}[!h]
\begin{center}$
\begin{array}{cc}
\includegraphics[width=2.6 in, trim = 0cm 1cm 0cm 0cm, clip,angle=-90]{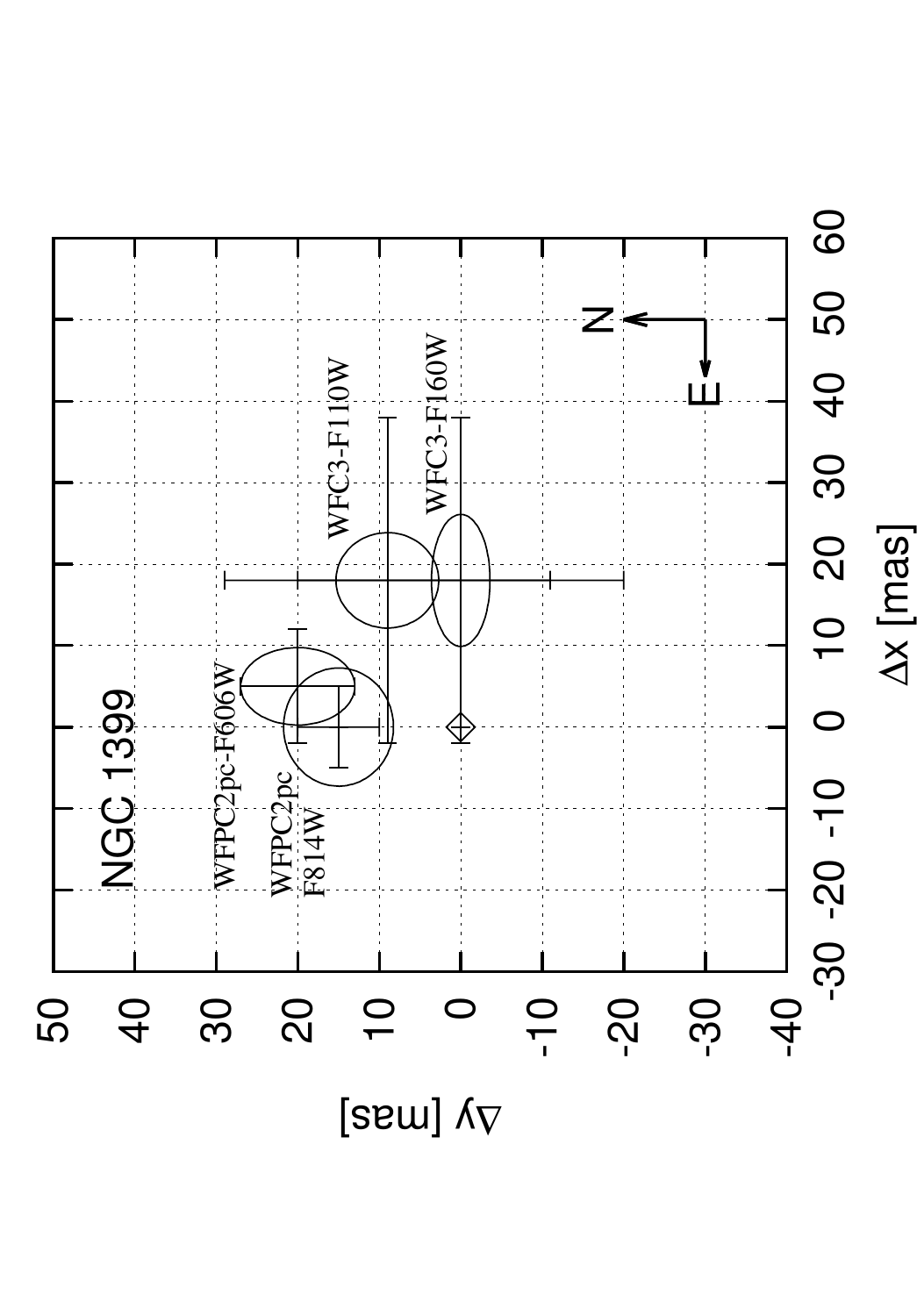} 	& 	\includegraphics[width=2.6 in, trim = 0cm 1cm 0cm 0cm, clip,angle=-90]{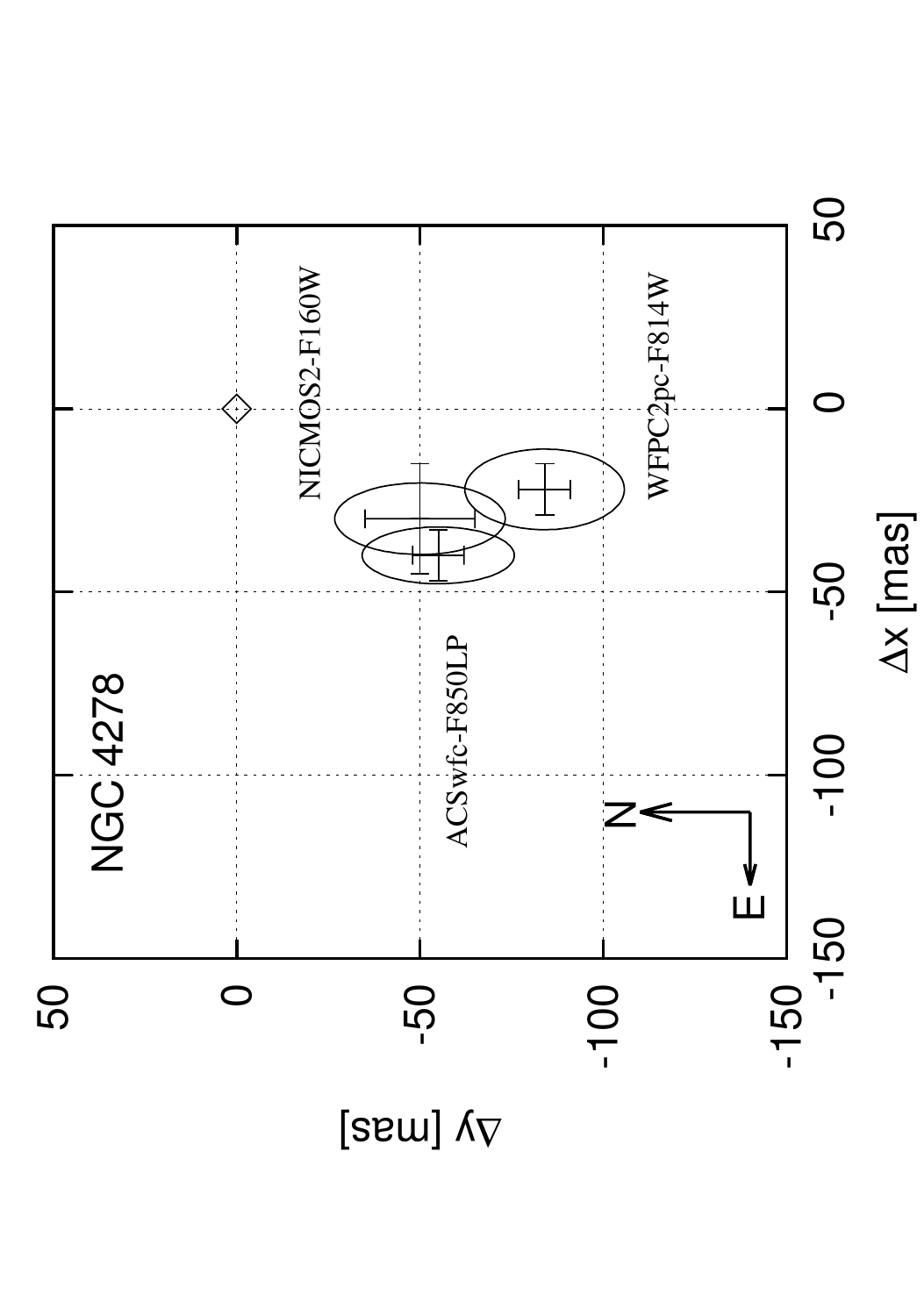} \\
\includegraphics[width=2.6 in, trim = 0cm 1cm 0cm 0cm, clip,angle=-90]{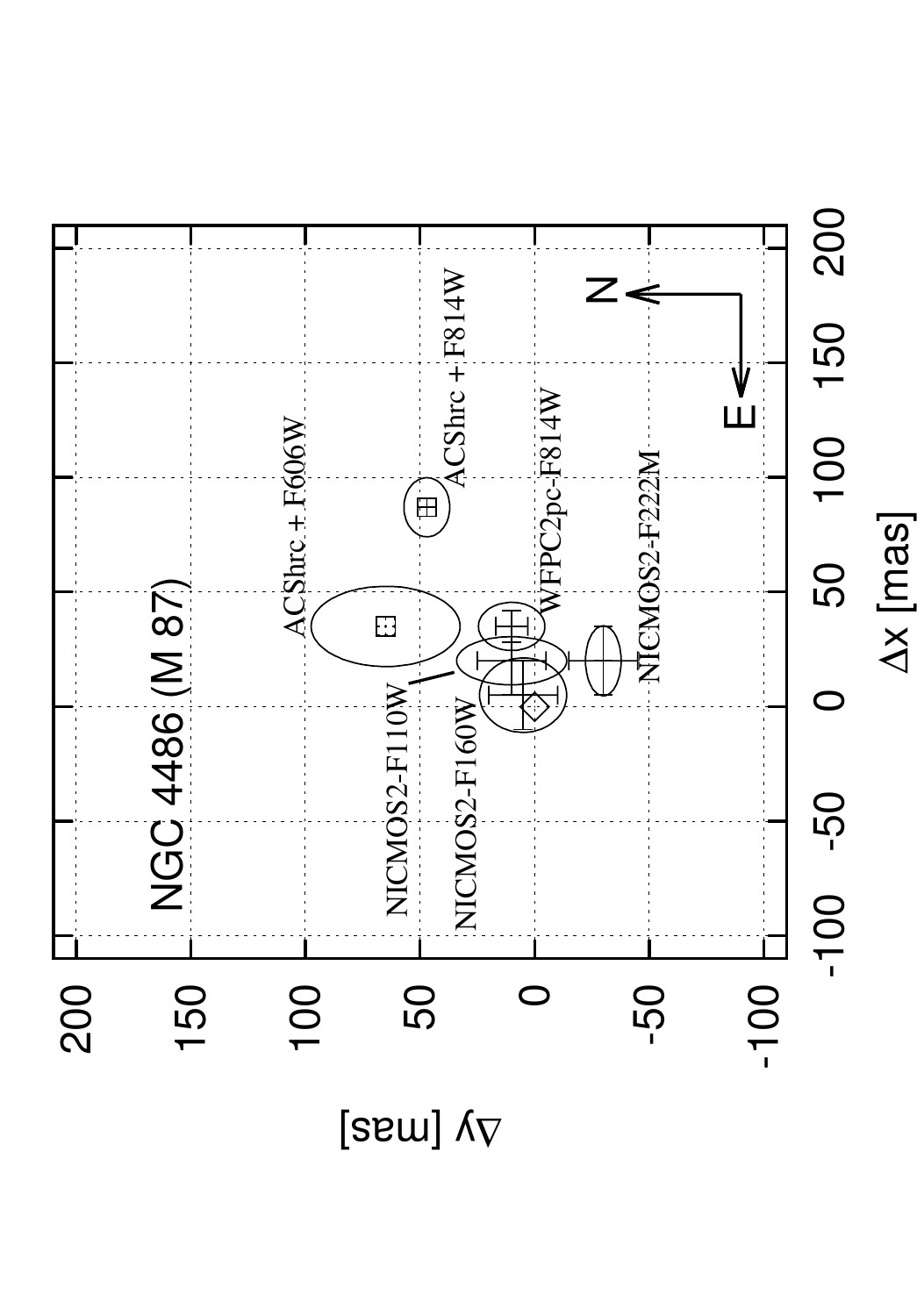} 			&	\includegraphics[width=2.6 in, trim = 0cm 1cm 0cm 0cm, clip,angle=-90]{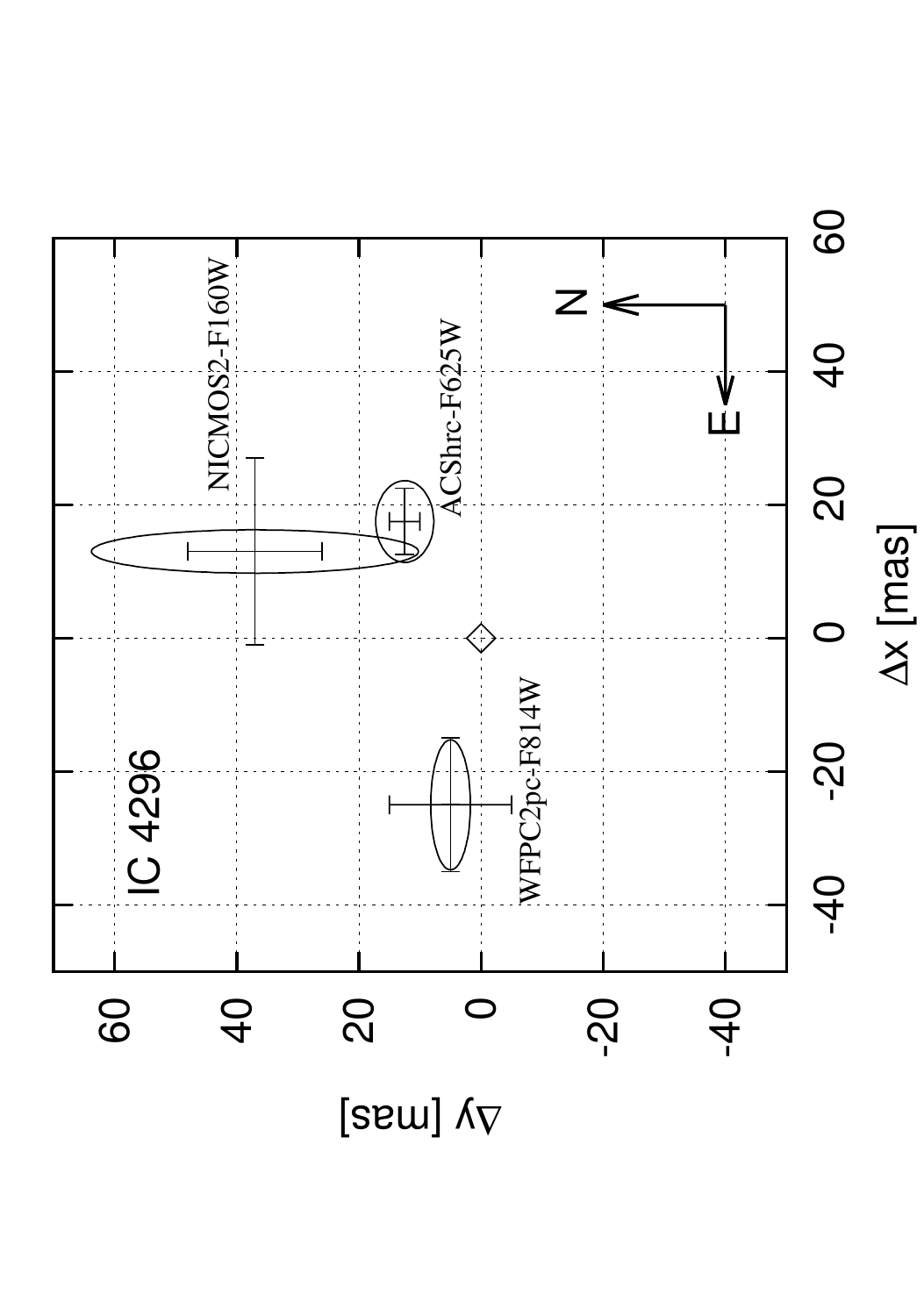} \\ 
\end{array}$
\end{center}
\caption{Offset of the isophotal center with respect to the SBH as measured on different instruments, at different frequencies. Each photocenter is centered on an ellipse whose axes represent the IQR of the isophote centers dataset used to derive the photocenter. Error bars represent the error on the offset. The diamond marks the origin of the reference frame (SBH position).}
 \label{fig: multif}
\end{figure*} 
 \begin{figure*}[!h]
\begin{center}$
\begin{array}{cc}
\includegraphics[width=2.6 in, trim = 0cm 1cm 0cm 0cm, clip,angle=-90]{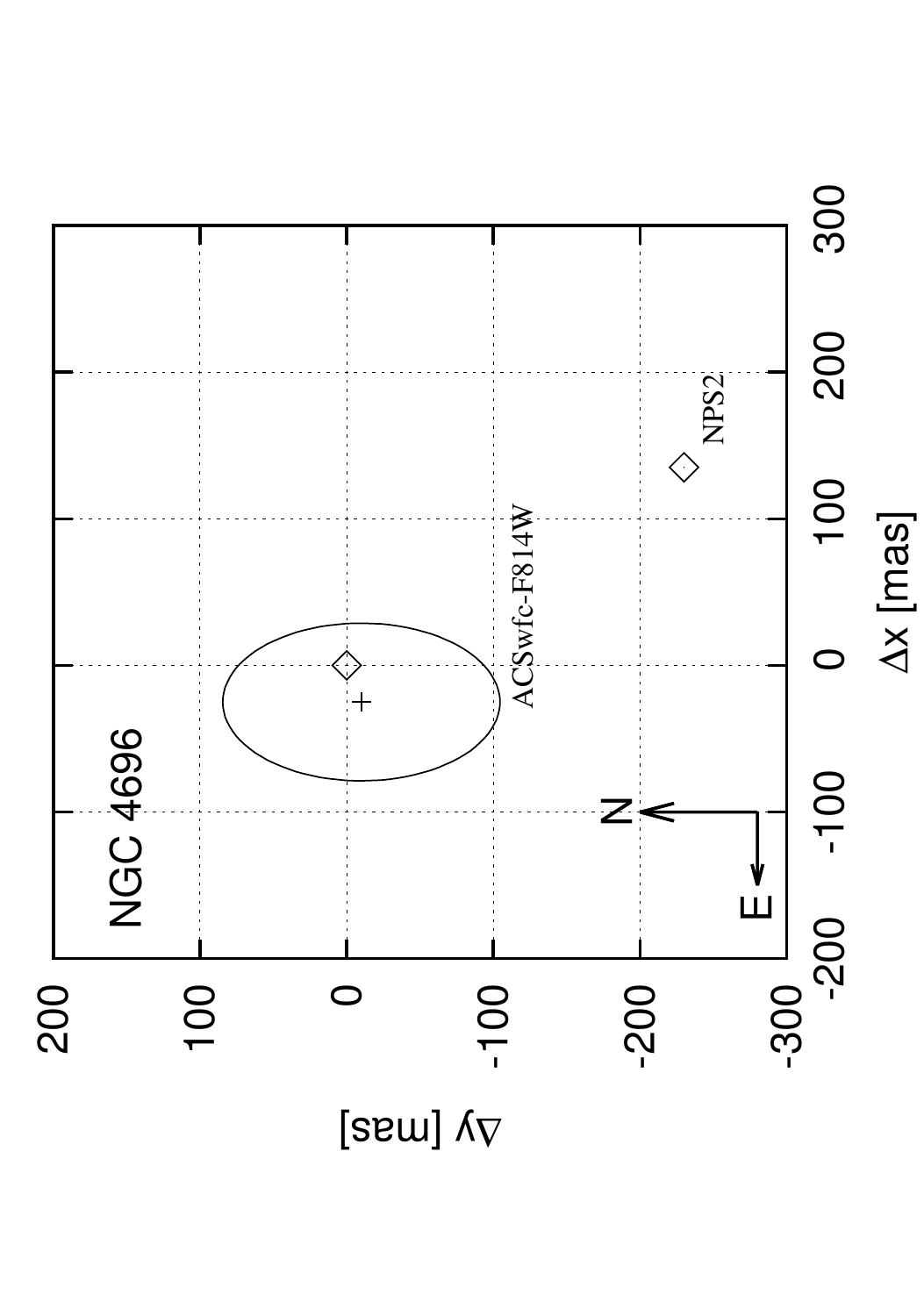} & \includegraphics[width=2.6 in, trim = 0cm 1cm 0cm 0cm, clip,angle=-90]{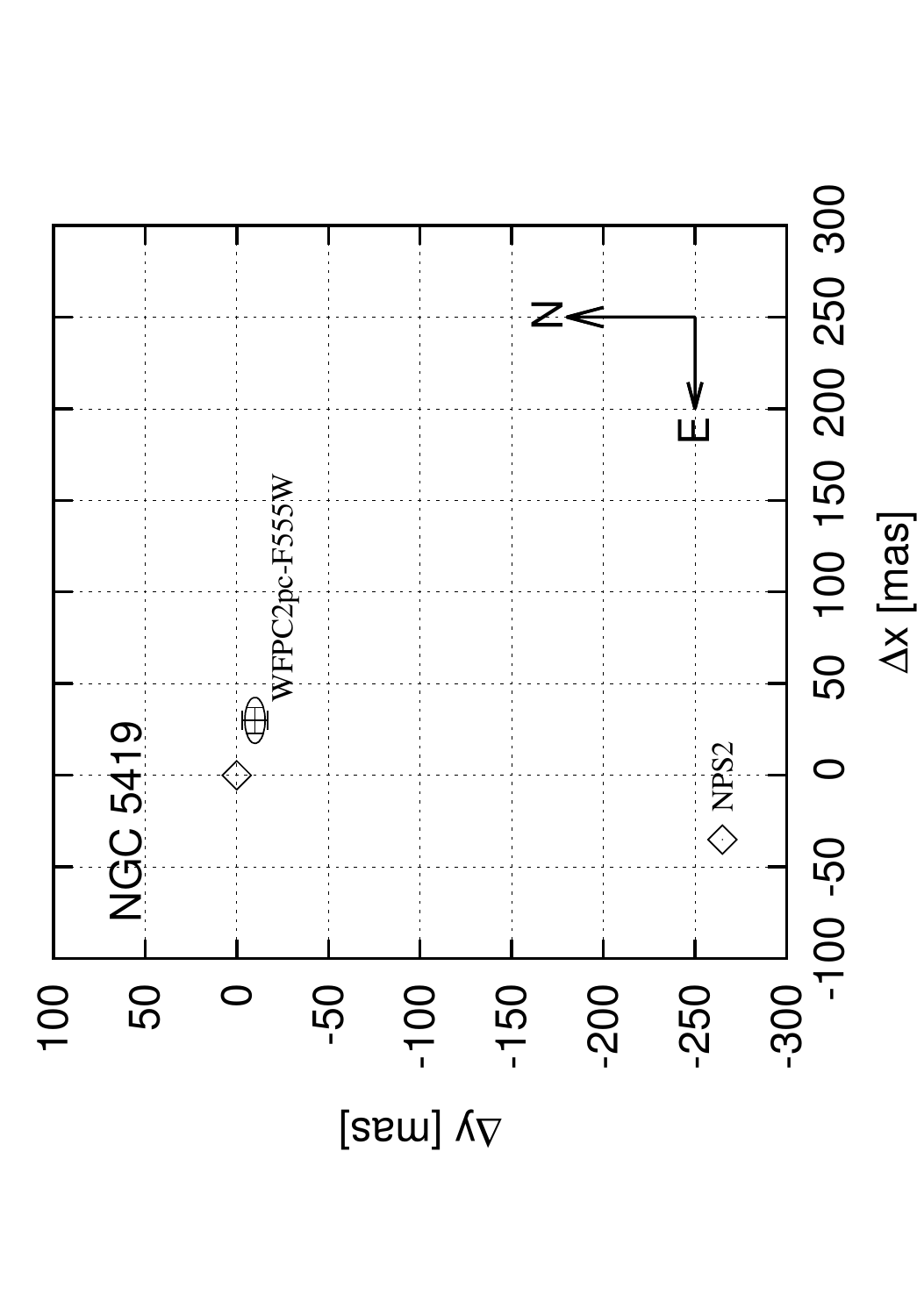} \\
\end{array}$
\end{center}
\caption{As in Fig.\ref{fig: multif} for the galaxies showing a double nuclear point source. The offset of the fainter nuclear point source with respect to the SBH is marked with a diamond and the label NPS2.}
 \label{fig: multif_DN}
\end{figure*} 
\end{subfigures}

\section{Discussion} \label{sec: disc}
In this section we will examine our key assumption that the SBH position
is marked by the nuclear point source. We will then discuss the possible 
origins of the observed offsets.  Most of the displacement mechanisms considered below 
have been previously discussed in B10. 
Nevertheless, we summarize them here in order to provide context for our new results.

\subsection{The AGN nature of the nuclear point sources}
Our sample of 14 core elliptical galaxies was selected from the 26 (out of 29) studied by
BC06 that were identified as containing AGN based 
on the presence of i) an unresolved optical or X-ray source; ii) an
``AGN-like'' optical spectrum, or iii) radio jets. Most of our sample of 14 exhibit two or more
AGN signatures in addition to the presence of a radio source and optical point source
nucleus. Optical or UV variability has been detected in HST observations of NGC 4486
\citep{Perlman2003}, NGC1399 \citep{OConnell2005}, NGC4552 \citep{Cappellari99, Maoz2005}
and NGC 4278 \citep{Cardullo2009}. Hard X-ray point sources have been detected in NGC
4261, 4278, 4486, 4552, 5419 and IC 1459 {\citep{GonzalezMMG09}. Weak broad H$\alpha$
lines (NGC 4278, 4636 and NGC 4168), or emission line ratios indicative of AGN
photoionization (NGC 4261, 4486, 4552, 5846) have also been detected in several galaxies
\citep{Ho97}. In addition, the radio sources in most objects feature compact ($\lesssim 1$
pc) cores (NGC 1399, 4168, 4373, 4552, 5419, IC 1459) and/or parsec or kpc-scale jets (NGC
4261, 4278, 4486, 4696, IC 4296; see Appendix~\ref{app: galaxies} for references). 
Only IC 4931, which by comparison has been relatively little
studied, lacks supporting evidence indicating
the presence of an AGN in the form of X-ray, radio or line emission.

There is, therefore, plenty of evidence that these galaxies (with the possible exception of IC 4931) host AGNs. 
However, it does not necessarily follow (except in cases where
 optical/UV variability has been observed) that the optical or NIR point sources are themselves manifestations of the AGN, rather than,
 for example, nuclear star clusters. 
Low luminosity radio loud AGNs are thought to be powered by radiatively inefficient accretion flows 
 \citep[see, for example,][and references therein]{Ho2008, Balmaverde2008}, with 
 most of the accretion power channeled into the kinetic energy of the radio jets. Detailed studies
 of NGC4486 \citep{DiMatteo2003}, IC 1459 \citep{Fabbiano2003} and  IC 4296 \citep{Pellegrini2003} show
 that this is the case for at least three galaxies of our sample.
 
In low luminosity radio galaxies, 
the optical, near infrared and X-ray luminosities of the
nuclei correlate tightly with the core radio luminosity, implying a common origin in
non-thermal emission from the jet \citep{Chiaberge99, CapBal05, Baldi2010}. This is
supported by polarization measurements: \citet{CapettiEtAl07} found that the optical
nuclei of the nine nearest FR I radio galaxies in the 3C catalogue, including two galaxies
from our sample, NGC 4486 (M87, 3C\,274)  and NGC\,4261 (3C\,270), have high polarizations
($2-11$\%), which they attribute to synchrotron emission. The optical or UV variability of
the point sources in NGC 4486 and three other galaxies, noted above, is also consistent with
jet synchrotron emission. More generally, BC06
found that their sample of core elliptical galaxies, from which this sample is drawn, also exhibit optical-radio and X-ray-radio correlations
and indeed form a continuous distribution with the radio galaxies, extending these
correlations to lower luminosities. Therefore, there is direct evidence, in the form of optical polarization and/or optical/UV variability
that the nuclear sources in five of our sample galaxies are produced by synchrotron
emission associated with the radio core source, presumably the base of the jet. It is reasonable to assume, given the correlations found by BC06, 
that the same is true for the rest of the sample. 

The base of the jet is, in turn, very close to the
SBH. Arguably the best studied jet is that of M87. Using multifrequency observations made with the Very Long Baseline Array,
\citet{HadaDK11} were able to show that the radio core at 43 GHz is located within 14-23
Schwarzchild radii of the SBH and accretion flow. In the optical, the lower optical depth should
move the peak of the emission even closer to the SBH. 

In summary, we contend that there are good reasons to believe that the optical or
near-infrared point sources in our galaxies reveal the position of the SBH.

\subsection{Recoiling SBHs}
\label{sec: recoiling_SBH}

Here we consider if the measured SBH displacements are consistent with residual
oscillations due to gravitational recoils generated by coalescence of an SBH binary.

GM08 used \textit{N}-body simulations to study the post recoil
dynamical evolution of an SBH in representative core elliptical galaxy potentials. They
found that when the recoil velocity is sufficient to eject the SBH from the core
($v_\mathrm{kick}$ in the range $\sim 40 - 90 \%$ of the escape velocity, $v_{esc}$), the subsequent motion is
characterized by three phases. The initial large-amplitide ($>$r$_c$) oscillations are
damped relatively quickly ($\sim 10^7$\,yr) by dynamical friction (phase I). When
the amplitude becomes comparable with the core radius, $r_c$, the SBH and the stellar
core undergo long-lived oscillations about their center of mass, which persist for $\sim
1$ Gyr (phase II). Finally, the SBH reaches thermal equilibrium with the stars,
experiencing low-amplitude Brownian motion (phase III).

As our measured displacements are $\lesssim$ 0.1 $r_c$, it is highly unlikely that they represent phase I oscillations. 
This would require either fortuitous timing (the SBH would have to be caught whilst passing through or close to the equilibrium position) and/or orientation (the oscillation direction would have to be closely aligned with the line of sight) for {\em each} galaxy. We also rule out phase III Brownian motion due to interactions with {\em individual} stars, which will produce negligibly small amplitudes in real galaxies (GM08). The possibility of Brownian oscillation due to interactions with massive perturbers is discussed in Section~\ref{subsec: disp_mech}. 
Here, we focus on phase II which, as already noted, is characterized by long-lived damped oscillations at amplitudes $< r_{c}$. 

In this phase, the characteristic damping time is given by GM08 as:

\begin{align}\label{eq: tdamp}
\tau \approx 15 \frac{\sigma^{3}}{G^{2}\rho M_{\bullet}} \approx 2 \times 10^{9} \ \sigma_{300}^{-3.86} r_{500}^{2}\ \mathrm{yr}
\end{align} 

\noindent where the approximation on the right-hand side makes use of the M$_{\bullet}$-$\sigma$ relation \citep{FerrareseFordRev05}, $\sigma$ being the 1D stellar velocity dispersion, with $r_{500} \equiv r_{c}/(500$ pc) and $\sigma_{300} \equiv \sigma / 300$ km s$^{-1}$.
For our galaxies, eq.~\ref{eq: tdamp} yields values of $\tau$ in the range $0.02-4.85$\,Gyr, with an average of 1.2\,Gyr (Table \ref{tab: param}).
The rms amplitude of the SBH motion with respect to the galaxy center is expected to evolve as:

\begin{align}\label{eq: rms}
r_\mathrm{rms}(t) \approx r_{c} e^{-(t-t_\mathrm{c})/2\tau} , \ t > t_\mathrm{c} 
\end{align} 

\noindent where $t_{c}$ is the time at which the oscillation amplitude has decayed to a scale comparable to the core radius $r_\mathrm{c}$ (B10). 

Simulations indicate that the galaxy merger
rate is a strong function of redshift and galaxy mass \citep[e.g.,][H10]{Fakhouri2010}. The merger rate for brightest cluster galaxies has recently been determined observationally 
by \cite{Lidman2013}, who find $\approx 0.4$ mergers per Gyr at $z\sim 1$, implying a mean time between mergers of $\approx 2.5$\,Gyr\footnote{Note that this quantity was derived by making use of merger time-scales calibrated against virtual galaxy catalogs from the Millennium Simulation \citep{Springel05} and it depends on the assumed fraction of mass accreted by a BCG during a merger. Therefore, this merger rate relies on model-dependent assumptions about the dark-matter haloes surrounding the galaxies.}. If we take this as 
representative of our sample, equations~\ref{eq: tdamp} and~\ref{eq: rms} together suggest that a ``typical'' rms displacement (ignoring projection effects) is $\sim0.2\ r_c$.
This suggests that larger displacements than were actually observed might be expected if due to post gravitational recoil oscillations. 

However, equation~\ref{eq: rms} describes the oscillation amplitude rather than the instantaneous displacement and moreover the above estimate does not account for projection effects, or
the range in damping timescales characterizing our sample galaxies, or for variations in the merger rate.
In order to investigate in more detail the likelihood of obtaining the observed displacements in our sample, if they result from post-recoil oscillations, we have constructed a simple 
Monte-Carlo simulation based on the GM08 \textit{N}-body simulations. Details of the method are given in Appendix~\ref{distribution}. However, we start from the assumption
that, after any kick large enough to move the SBH beyond the core radius, the distance, R, of the SBH from the center of the galaxy is given by:

\begin{align}\label{eq: osc}
R(t) = R_{0}e^{-\Delta t/ \tau} \sin(\omega_{c}\Delta t)
\end{align}

\noindent where $\Delta t$ is the elapsed time since the kick, i.e., the time since the last merger, $\tau$ is the damping time given by eq.\ref{eq: tdamp} and $\omega_{c}$ is the SBH oscillation frequency calculated from eq.~\ref{eq: omega}. We set $R_{0} = r_{c}$, since phase II begins when the oscillation amplitude is roughly equal to the core radius. We then suppose that each galaxy is observed at a random time, $\Delta t$, since its last merger, where the probability that the merger occurred at time $\Delta t$ follows an exponential distribution characterized by a mean time-between-mergers, $t_{m}$. This distribution is sampled to generate corresponding values of $R_i \equiv R(t_i)$, which are then projected onto the sky plane assuming that the recoil kicks have random directions. 
The distribution of projected displacement for each galaxy is then used to compute the 
probability, $p_i$, of observing a displacement larger than that actually measured, given a kick of sufficient magnitude. 

These probabilities were computed for two values of the mean time between mergers: $t_m$ = 5.0 Gyr and $t_m$ = 0.4 Gyr, which were derived from the galaxy merger rate models of H10 for redshifts $z<1$,  mass ratios $q>0.1$ and galaxy masses $\log (M_\mathrm{gal}/M_{\odot})  > 11$ and $\log ( M_\mathrm{gal}/M_{\odot}) > 12$, respectively (see Appendix~\ref{distribution} for details). 
The two values of $t_m$ are representative of the mass range covered by our sample,
and illustrate the effect on the displacement probabilities of the strong galaxy mass dependence of the merger rate. 

The probabilities, $p_i$, are listed for each galaxy in Table \ref{tab: probabilities} for both values of $t_m$. The probabilities $P(d>x)$ of observing a projected displacement $d$ exceeding distance $x$ are plotted as functions of $x$ in Figures~\ref{fig: projD} and \ref{fig: projDb}, for each galaxy, along with the observed displacement. 

As would be expected, the smaller value of $t_m$ results in larger values of $p_i$; clearly if $t_m \lesssim \tau$, there is a greater chance 
of observing a large displacement than when $t_m >> \tau$. For  $t_m = 0.4$\,Gyr, $p_i > 0.5$ for all but two galaxies (NGC\,4278 and NGC\,4552) and $p_i \ge 0.95$ for all
6 galaxies that have masses $\ge 10^{12}$\,M$_{\odot}$ (for which this value of $t_m$ is presumably most relevant). On the other hand, for $t_m = 5.0$\,Gyr, all but four galaxies 
have $p_i < 0.5$, including five of the six having displacements considered significant at the level of $>0.8$ IQR (the exception is M87, for which $p_i=0.75$). 

Considering the sample as a whole, the probability of \textit{not} observing a displacement larger than those actually measured in any of the galaxies in the sample is 
simply $\mathbb{P} = \prod_i (1-p_i)$. For $t_m = 0.4$\,Gyr, this is negligibly small, $\mathbb{P} \sim 10^{-17}$. Even for $t_m = 5.0$\,Gyr, $\mathbb{P} \approx3\times 10^{-4}$,
indicating that it is statistically unlikely that larger displacements were not observed, given the occurrence of a recoil kick sufficient to trigger phase II oscillations.
 
However, it is unlikely that all the galaxies in the sample experienced a recoil kick big enough to move the SBH beyond the core radius. Assuming that the potential experienced by the SBH can be approximated as a harmonic potential ($\phi = 1/2\ \omega^{2}x^{2}$), we estimated the kick velocity required to displace the SBH out to the core radius for each galaxy, obtaining values in the range $159 \lesssim  v_\mathrm{kick} \lesssim 325$\,km s$^{-1}$, with an average of $v_\mathrm{kick} \approx 240$ km s$^{-1}$ (the values for each galaxy are listed in Table \ref{tab: param}). 

L12 have studied kick velocity distributions for ``hang-up'' kicks, which arise in binary configurations where the SBH spins are partially aligned with the orbital angular momentum by pre-merger accretion. For spin magnitude and orientation distributions derived from accretion simulations and a mass ratio distribution based on galaxy merger studies (including both minor and major mergers), they generate probability distributions for the magnitude of the recoil velocity for two extreme cases of ``hot'' and ``cold'' gas disks (i.e., corresponding to adiabatic indices $\gamma = 5/3$ and $7/5$, respectively). The former regime is likely to be
most applicable to the ellipticals in our sample and in this case, probabilities range from $p_k\approx 66$\% for kicks $>100$ km s$^{-1}$ to $\approx 33$\% for kicks $>300$ km s$^{-1}$. 
Our Monte-Carlo simulation assumes that the coalescence event produces a recoil large enough to produce phase II oscillations (i.e., the values given in Table \ref{tab: param}).  As the probabilities
$p_i$ are conditional on the occurrence of a kick of sufficient magnitude, the probability that both events will occur is $p_k\times p_i$. There is thus a smaller probability of observing a displacement
larger than actually measured for each galaxy. Using the kick probabilities given by L12, we find that the probability of \textit{not} observing a displacement larger than those actually measured in the entire sample is $\mathbb{P'} \approx 8\times 10^{-4}$ for $t_m = 0.4$\,Gyr and  $\mathbb{P'} \approx 5\times 10^{-2}$ for $t_m = 5.0$\,Gyr.

As the merger rate increases with redshift, we may have underestimated the time-between-mergers by integrating over the redshift range $z=1 \rightarrow 0$. However, the higher value of $t_m$
used in the simulation implies a merger rate ($\sim 0.2$ mergers Gyr$^{-1}$) consistent with that given by the models of H10 for $\log ( M_\mathrm{gal}/M_{\odot})  > 11$ galaxies at $z=0$ (see their Fig. 3). The rate for major mergers is a factor 2--3 lower than the rate for major and minor mergers combined, which we used to estimate $t_m$. However, although the magnitude of the recoil velocity 
is a function of SBH mass ratio (with the largest recoils occurring for mass ratios characteristic of major mergers), even small ($q>0.1$) mass ratios are capable of producing kicks $\sim 200$ km s$^{-1}$,
large enough to trigger phase II oscillations in our galaxies \citep{Lousto2010}. Minor mergers were, in any case, included in computing the kick velocity distribution presented by L12. 

\textit{Thus, even allowing for the distribution of kick velocities, the Monte-Carlo simulation suggests that it is highly likely that displacements larger than those actually measured would have been observed in this sample if each galaxy has experienced at least one merger leading to an SBH-binary coalescence and gravitational recoil within the last few Gyr. In fact, our simulations suggest that $t_m \gtrsim 30 - 40$\,Gyr, corresponding to a merger rate $\sim 0.03$ mergers Gyr$^{-1}$, is required for a $>50$\% chance of \underline{not} observing larger displacements in the sample}.

If, indeed, the formation of a binary SBH is an inevitable outcome of galaxy mergers, explanations must be sought in the evolution of the binary or that of the recoiling SBH. Possibilities include (1) the binary stalls before reaching the radius at which gravitational wave emission drives rapid coalescence; (2) if coalescence occurs, the mass ratio or spin-orbit properties of the binary are such as to typically preclude recoil velocities large enough to displace the SBH to the core radius ($\gtrsim 240$ km s$^{-1}$); or (3) the damping time for the recoil oscillations may be shorter than predicted by the pure \textit{N}-body simulations.

Alternatively, it is possible that galaxy mergers are much more infrequent than inferred from studies based on the current $\Lambda$CDM framework \citep[see][and references therein for a detailed discussion]{kroupa14}. 

Although our results imply that SBH displacements due to gravitational recoil are much less common than might be inferred from theoretical considerations, this does not preclude the possibility that observed displacements in individual galaxies are due to recoil oscillations. Indeed, the $p_i$ values listed in Table \ref{tab: probabilities} indicate that for $t_m = 5.0$\,Gyr, the chance of {\em not} observing a larger displacement, given a sufficiently large kick, is $\gtrsim 30$\% for all objects having a displacement considered significant. Even for $t_m = 0.4$\,Gyr, there are still three of these galaxies for which $1-p_i \ge 0.25$. The likelihood that the recovered offset is due to SBH coalescence is discussed for individual galaxies in Appendix \ref{app: galaxies}.  

Assuming that the measured displacements are indeed due to phase II recoil oscillations, we have used eq. \ref{eq: rms} to estimate the merger epoch for those galaxies with displacements rated at least as ``low significance'' in terms of the isophote centroid IQR. Allowing for the time taken for the SBH binary to form and subsequently coalesce in the center of the merged galaxy, and for phase I oscillations, the elapsed times since the merger are typically $\sim$ several Gyr, comparable with the mean time between galaxy mergers estimated here for galaxy masses $\log ( M_\mathrm{gal}/M_{\odot})  > 11$,
and consistent with the observationally determined merger rate for brightest cluster galaxies \citep{Lidman2013}.

Eq.\ref{eq: tdamp}, used to estimate the SBH damping time, relies on the $M_{\bullet}$ - $\sigma$ relation \citep{FerrareseFordRev05}.
To test the dependence of our result on the value adopted for $M_{\bullet}$ we use eq.6 in \citet{lauer2007} to estimate SBH masses using the $M_{\bullet}$ - $M_{\mathrm{V}}$ relation (see Table \ref{tab: mass} and Appendix \ref{distribution} for more details). Using these alternative masses, and taking into account the kick velocity distribution computed by L12,  we derive new values for $\mathbb{P'}$ obtaining $1\times 10^{-3}$, instead of  8$\times 10^{-4}$, for $t_m = 0.4$\,Gyr and $\mathbb{P'}\approx$ 0.14 instead of 0.05 for $t_m = 5$\,Gyr. The likelihood of our result is somewhat increased for t$_{m}$ = 5 Gyr, however there is still a large probability of finding displacements larger than those measured, leaving our basic conclusion unchanged.

No useful constraints can be obtained on the merger epoch from consideration of the time necessary to replenish the stellar population of the core. This is roughly the relaxation time, $T_{rel}$, at the influence radius of the central SBH, which can be estimated as:

\begin{align}
T_{r} (r_{i}) \approx 1.2 \times 10^{11} \ \sigma_{100}^{7.47}\ \mathrm{yr}
\end{align}

\noindent \citet[][eq. 3.5]{Merritt13}. Given that the measured velocity dispersion is always $>100$\,km s$^{-1}$ in the centers of our galaxies (see Table \ref{tab: param}), it is clear that the replenishment time is much longer than the timescales involved in the SBH merger-recoil event ($\sim 10^{9}$ yr). 

In general, coalescence of an SBH-binary results in a  re-orientation of the SBH spin axis, leading to a sudden flip in the direction of the associated radio jet \citep{ME2002}. Therefore, the radio source morphology may act as a ``signpost'' of SBH coalescence. 
However, powerful extended radio sources have lifetimes $\sim10^{6} - 10^{8}$ yr \citep[e.g.,][]{Parma1999, ODea2009, Antognini2012}, much shorter than the time-between-mergers, even for massive galaxies. 
Thus even if a spin flip took place, morphological traces of this event, such as a change in jet direction, would not necessarily be evident in the current (post-coalescence) radio source. The radio source properties of individual galaxies are discussed in Appendix~\ref{app: galaxies}.

It is notable that in four of the six galaxies in which significant displacements have been found, the displacements are also  approximately aligned with the radio jet axis
(Section \ref{sec: results} and Table \ref{tab: PA}).
This number includes three of the four that have powerful kpc-scale jets (Section \ref{sec: results} and Table \ref{tab: PA}),
suggesting jet power asymmetries as a possible displacement mechanism (as discussed below). However, such alignments 
do not necessarily argue against gravitational recoil. In their statistical study of the coalescence of
spinning black hole binaries \citet{Lousto2010} find that the recoil velocity is
preferentially aligned (or counter-aligned) with the orbital angular momentum of the progenitor
binary and moreover, that the spin direction distribution of the recoiling SBH peaks at an angle of $\approx 25^{\circ}$
to the orbital angular momentum (for an equal mass progenitor binary; the probability
distribution is broader and peaks at larger angles for smaller mass ratios). This suggests, assuming that the jet traces the spin axis of the recoiling SBH,
a tendency for the displacement to be somewhat, but not greatly, misaligned with the jet axis, consistent with what is observed in these objects.

\begin{table}[t]
\caption[]{PROBABILITIES} 
\label{tab: probabilities} \centering
\scalebox{1}{
\begin{tabular}{ lccc }
\hline \hline
	Galaxy		&	d$_{i}$ [$r_{c}$]	&	p$_{i}$(t$_{m}$ = 0.4 Gyr) 	& p$_{i}$(t$_{m}$ = 5 Gyr) \\ \hline
 &\\
NGC 1399		&	0.01				&	0.75						& 0.11\\
NGC 4168		&	$^{*}$0.02		&	0.98						& 0.89\\	
NGC 4261		&	$<$0.01			&	0.94						& 0.23\\
NGC 4278		&	0.1				&	0.36						& 0.04\\
NGC 4373		&	$<$0.01			&	0.99						& 0.58\\
NGC 4486		&	0.01				&	0.99						& 0.85\\
NGC 4552		&	$<$0.05			&	0.10						& 0.01\\
NGC 4636		& 	$^{*}$0.01		&	0.99						& 0.69\\
NGC 4696		& 	$^{*}$0.02		&	0.97						& 0.42\\
NGC 5419		&	0.02				&	0.97						& 0.46\\
NGC 5846		&	0.04				&	0.90						& 0.25\\
IC 1459			&	$<$0.01			&	0.92						& 0.21\\
IC 4296			&	0.01				&	0.97						& 0.37\\
IC 4931			&	$^{*}$0.03		&	0.95						& 0.46\\
 &    \\
\hline
\end{tabular}}
\tablecomments{Probability to observe a projected displacement larger than the value actually observed in this work (in units of core radii) at a random time after the last kick. Probabilities are computed for a mean time between galactic mergers t$_{m}$ = 0.4 Gyr and 5 Gyr. The symbol ``$^{*}$" indicates 3$\sigma$ offsets that were classified as ``null" after the normalization for the co-ordinates IQR. The symbol ``$<$" indicates offsets that do not reach the 3$\sigma$ level.}
\end{table}

\subsection{Other displacement mechanisms}
\label{subsec: disp_mech}

We now consider several other mechanisms that may plausibly produce SBH displacements. 

\textbf{1. Asymmetric jets}: if the AGN jets are intrinsically asymmetric in power output, the resulting net thrust  can push the SBH away from the original equilibrium position \citep{jet, SW88}. \citet{Korn08} determine the SBH acceleration for this scenario:

\begin{align}
\label{eq: acc_jet}
a_{\bullet} \approx 2.1 \times 10^{-6} f_\mathrm{jet} \dot m \ \mathrm{cm}\ \mathrm{s^{-2}}
\end{align}
 
\noindent where $\dot m \equiv \dot M_{a}/\dot M_\mathrm{edd}$ and $f_\mathrm{jet}=L_\mathrm{jet}/L_{a}$\footnote{$f_\mathrm{jet}\leq$ 1, unless energy is extracted from the SBH rotation \citep{BlandfordZnajek77}.}, $\dot M_{a}$ is the mass accretion rate, $\dot M_\mathrm{edd}\approx 2.2\ \left(M_{\bullet}/10^{8}M_{\odot}\right)$ M$_{\odot} \mathrm{yr}^{-1}$ the Eddington accretion rate, $L_{a}$ the accretion luminosity and $L_\mathrm{jet}$ is the jet luminosity. For asymmetrical, oppositely directed jets $L_\mathrm{jet}$ is the difference in the luminosities of the two jets.

Under the assumption that the restoring force from the galaxy is negligible (a reasonable approximation for the low-density cores of ``core galaxies''), Eq.~\ref{eq: acc_jet} can be integrated to obtain an expression for the displacement:

 \begin{align} \label{eq: dr_norest}
 \Delta r \approx 340 \ f_\mathrm{jet} \dot m \ t_{6}^{2}\ \mathrm{pc}
 \end{align} 

\noindent where $t_{6}$ is the time over which the SBH is accelerated to produce an offset $\Delta$r.
\vskip10pt

The best candidates for jet thrust displacements are the four galaxies (NGC 1399, 4261, 4486 and IC 4296) that have relatively powerful kpc-scale jets (Section \ref{sec: results} and Table \ref{tab: radio}). Of these, we did not detect a significant displacement in NGC 4261, which has an FR\,I radio source. NGC 4486 and IC 4296 also host FR I radio sources and NGC 1399 has an FR\,I-like morphology, despite its relatively low power. Interestingly, all three of these galaxies exhibit photocenter displacements that are at least approximately (within 20$^{\circ}$) in the jet direction (Table \ref{tab: PA}). This implies that the SBH is displaced relative to its equilibrium position in the counter-jet direction, as might be expected if the displacements are related to intrinsic asymmetries in jet power.  NGC\,4486 (M87) has already been discussed in detail by B10, who concluded that a jet power
asymmetry amounting to $\sim 3$\% of the accretion luminosity can explain the observed displacement, for a radio source lifetime $\sim 10^8$ yr.
We find similar results in the cases of NGC 1399 and IC 4296, where the observed displacement can be produced for a jet asymmetry $<1$\% of the Bondi accretion luminosity, again for a radio source lifetime $\sim 10^8$\,yr (See Appendix \ref{app: galaxies} for details). 

A close alignment between the displacement and the (initial) jet direction is also found for the low power pc-scale radio source in NGC 4278. In this case, 
\citet{Giro2005}'s interpretation of the radio data implies that the photocenter is displaced in the counter-jet direction, with the jet axis being closely aligned with the line of sight.
Assuming that the SBH is displaced along the jet axis, the de-projected magnitude of the displacement would be $\sim 100-200$ pc. Nevertheless, it seems possible that
this could result from jet thrust, if sustained for $\sim 10^8$ yr (Appendix \ref{app: galaxies}).  

The double nucleus galaxy NGC 4696 does not show well-defined jets on kpc scales, but has a one-sided pc-scale jet in PA$\approx -150^{\circ}$. The displacement relative to the brighter nucleus is considered non-significant because of the large IQR. However, as already noted, it is not known which of the two optical nuclei hosts the AGN producing the radio jet. If it is the fainter nucleus, the photocenter displacement is approximately in the counter-jet direction, which is not consistent with jet acceleration of the SBH (which would cause the SBH to be displaced in the counter-jet direction). 

The remaining two objects which exhibit displacements considered to be significant (NGC 5419 and 5846) do not have well-defined jets on either parsec or kiloparsec scales.

\textbf{2. Stalled SBH binaries}: in the aftermath of a galaxy-galaxy merger, the SBH binary orbit shrinks at first due to dynamical friction and subsequently through 
slingshot ejection of stars intersecting the orbit. Investigations of quasi-steady spherical models suggested that the evolution of the binary stalls at separations $\sim 1$\,pc, due to a paucity of interacting stars, rather than hardening to the point at which gravitational wave emission drives the final inspiral to coalescence \citep[the so-called ``final parsec problem''; e.g.,][]{mm}. 
Based on $N$-body simulations, \citet{Merritt06radii} estimated ``typical'' semi-major axes for stalled binaries, finding $a_\mathrm{stall} \lesssim 10$ pc for an SBH mass ratio $q=0.5$ and $a_\mathrm{stall} \lesssim 3$ pc for $q=0.1$, respectively. If the binary center of mass is located at the photocenter, the displacements of the primary and secondary components would be  $\Delta r_1 \sim a_\mathrm{stall}\ q/(1+q)$
and $\Delta r_2 \sim a_\mathrm{stall}\ /(1+q)$, respectively, giving  $\Delta r_1 \sim 3\ (0.3)$\, pc and $\Delta r_2 \sim 7\ (3)$\,pc for $q=0.5$ (0.1). 
Thus stalled binaries could produce displacements comparable with our results, particularly if the secondary SBH is accreting and the mass ratio is near unity.

It has been argued that stalling can be avoided in galaxies containing significant amounts of nuclear gas \citep[e.g.,][]{Esc05, Cua09, mayer}. Even in purely stellar nuclei, \textit{N}-body simulations sometimes find that evolution of the binary can continue efficiently due to the presence of centrophilic orbits \citep{KJM11, PretoBBS11, GualandrisM11}. However the existing gas-dynamical simulations probably do not yet have enough spatial resolution to follow the binary’s evolution to sub-parsec scales, and the \textit{N}-body simulations appear to not yet have large enough \textit{N} that their results can be robustly extrapolated to the much larger-\textit{N} regime of real galaxies \citep{VAM14}.

Another possibility, discussed by \citet{FabioAndDavid11}, is accretion of a less massive galaxy by a giant elliptical, such as NGC 4486, which has a pre-existing depleted core (presumably the result of the evolution of SBH binaries formed in previous mergers). In such situations, dynamical friction is very inefficient in the core due to the lack of stars moving slower than the sinking object.  
The orbital eccentricity can increase rapidly while the apoapsis hardly changes, resulting in a slowly evolving SBH binary in a highly eccentric orbit. The simulations presented by 
\citet{FabioAndDavid11} indicate that in an M87-like core, a low mass ratio binary ($q\approx 10^{-3}$) can persist over a Hubble time in an increasingly eccentric orbit with a semi major axis $\sim 10$\,pc.
If the secondary SBH is accreting, it will be visible as an off-center AGN; if both components of the SBH binary are accreting, a double nucleus might be observed (perhaps as in NGC 4696 and NGC 5419).

Helical distortion or ``wiggling'' of parsec-scale radio jets has been linked to putative SBH binaries in several AGNs, with the jet wiggles being variously attributed
 to orbital motion, precession of the accretion disk around the jet-emitting black hole or to geodetic precession \citep[e.g.,][]{BBR80, Roos1993, Katz1997, Romero2000, Lobanov2005}. 
However, the periods ($\sim 3$ years) and separations ($\sim 10^{-2}$\,pc) typically inferred from analyses of jet wiggles are much smaller than would be the case for 
a stalling binary. If the $\sim$ pc-scale displacements measured in this work are interpreted as SBH binary orbits, periods $\sim 10^{4-5}$ years are implied for total masses $\sim 10^{8-9}$\,M$_\odot$. 
Geodetic precession is insignificant at these separations. The wavelength of jet wiggles caused by orbital motion is given by $\lambda_\mathrm{jet} \sim v_\mathrm{app}t_\mathrm{orbit}$, where $v_\mathrm{app} = \beta c \sin{i}/(1-\beta\cos{i})$ is the apparent jet velocity (for a jet speed $v_\mathrm{jet} = \beta c$ and inclination to the line of sight, $i$) and $t_\mathrm{orbit}$ is the orbital period. Therefore, assuming $\beta\sim 1$, a pc-scale binary will produce very long wavelength wiggles in the jet ($\sim 10$\,kpc). In general, due to the combination of orbital and jet velocities, the jet will precess on the surface of a cone which, for a pc-scale binary of mass $\sim 10^9$ M$_{\odot}$ will have a half-opening angle $\lesssim 1^{\circ}$ \citep[see Equation~7 in][]{Roos1993}. Thus, orbital motion would cause only small curvatures in the jet over $\sim $ kpc scales, which would be difficult to discern as the jet loses collimation, or to distinguish from jet bending due to environmental effects, such as ram pressure. This will also be the case for disk precession, since the precession period 
exceeds the orbital period. Therefore, although the jet morphology has been mapped in detail from pc to kpc scales for a number of our sample galaxies (see Table \ref{tab: radio} for references), these observations are unlikely to provide unambiguous clues as to the presence, or not, of a stalled pc-scale binary. 

\textbf{3. Massive perturbers:} galaxy centers host a variety of potential perturbers 
with masses ranging  from $\sim 1 M_{\odot}$ (e.g., stellar mass black holes and neutron stars)
to $\sim 10^{4}-10^{7} M_{\odot}$,  such as giant molecular clouds and stellar clusters.

Gravitational interactions with these objects will cause the SBH to undergo a type of Brownian motion, with the amplitude of the root-mean-square displacement given by:

\begin{align}
\Delta r_\mathrm{rms} \approx \left(\frac{\tilde m}{M_{\bullet}}\right)^{0.5} r_\mathrm{c}
\end{align}

\noindent where $\tilde m$ is proportional to the second moment of the mass distribution of the massive perturbers:

\begin{align*}
\tilde m \equiv \frac{\int n(m)m^{2} dm}{\int n(m)m dm}
\end{align*}

\noindent with $n(m)\ dm$ being the number of perturbers with masses in the range $m, m+dm$ \citep[][eq. 7.63]{Merritt13}.

The offsets measured in this work range from 0.01 to 0.1 $r_{c} $, but are typically $\sim 0.01\ r_{c} $ (Table \ref{tab: probabilities}). For a typical SBH mass of order $10^{9}$ M$_{\odot}$ (Table \ref{tab: mass}), rms displacements 
$\sim 0.01 r_{c} $ can be generated by Brownian motion for $\tilde m \sim 10^{5}$ M$_{\odot}$. 

The mass functions of globular clusters, gas clumps and giant molecular clouds in the inner 100 pc of the Galaxy have been estimated by \citet{PeretsHA07}. 
For comparison, using the mass functions presented in their paper, we find that $\tilde m$ can be as high as $10^{5}$ M$_{\odot}$ for a population of giant molecular clouds. The Milky Way, of course, is very different from the galaxies studied here. However, as recently discussed by \citet{FabioAndDavid11}, 
the low efficiency of dynamical friction in low density cores favors the formation of a population of stalling massive objects in the cores of giant ellipticals.

Therefore, it seems plausible that displacements of order those observed could result from Brownian motion due to a population of massive perturbers.
In several galaxies, the observed displacements are approximately aligned with the kpc-scale radio jets. This suggests that 
the displacement is not random, at least in these cases, favoring other mechanisms (i.e,  gravitational recoil or jet thrust) that are expected to offset the SBH in the jet direction.  Nevertheless, we conclude that Brownian motion cannot be excluded as the cause of the displacement in any individual galaxy, particularly those
where the offsets are not aligned with kpc-scale jets, i.e. NGC 4278, 5846 and 5419.

\begin{table*}[tb]
\caption[]{SUMMARY} 
\label{summary} \centering
\scalebox{1}{
\begin{tabular}{  l    ll ll ccccc c}
\hline \hline
& & &\\
\multicolumn{1}{l}{\textbf{Galaxy}}	& \multicolumn{2}{c}{\underbar{\hbox to 70pt{\hfill \textbf{Offset (pc)}\hfill}}} 				& \multicolumn{2}{c}{\underbar{\hbox to 70pt{\hfill \textbf{PA (deg)}\hfill}}}		& \multicolumn{4}{c}{\underbar{\hbox to 200pt{\hfill \textbf{Suspected Origin} \hfill}}} 	&\\
							& 	NPS1				&	NPS2			&		NPS1		&	NPS2		& Gravitational		& 	Jet 			& Stalling			& Massive			\\
							& 			 			&					&					&				& Recoil	 		& 	 Thrust		& SBH-SBH	& Perturbers		\\ \hline
& & &\\
NGC 1399					& 1.5 $\pm$ 0.4			&					&		-17 $\pm$ 16	&				& $\checkmark$	&	$\checkmark$	& 				& 				\\
NGC 4278  					& 7.6 $\pm$ 0.4			&					&		152 $\pm$ 3	&				& $\checkmark$ 	& 				& $\checkmark$	& $\checkmark$ 	\\ 
NGC 4486					& 4.3 $\pm$ 0.2			&					&		307 $\pm$ 1	&				& $\checkmark$ 	& 	$\checkmark$ 	&			  	&				\\
NGC 5846					& 8.2 $\pm$ 2.5			&					&		253 $\pm$ 8	&				& $\checkmark$	&				& $\checkmark$	& $\checkmark$	\\
IC 4296						& 3.8 $\pm$ 0.7			&					&		338 $\pm$ 7	&				& $\checkmark$ 	& 	$\checkmark$	& 			  	&				\\
& & & & & & \\
\multicolumn{10}{c}{Galaxies with two nuclear sources}\\
\hline
& & & & & & \\
NGC 5419					& $7.5 \pm 1.7$			& $62 \pm 2$			&		252 $\pm$ 13	&	346 $\pm$ 2	& $\checkmark$	&				& $\checkmark$	& $\checkmark$ 	\\
& & & & & & \\
\hline
\end{tabular}}
  \tablecomments{Projected offsets of the photocenter with respect to the AGN and their possible origin. When multiple images have been analyzed, values presented here are the error weighted average. Position angles (PA) are given in degrees East from North. NPS = nuclear point source. }
\end{table*}

\section{Summary and Conclusions} \label{sec: conclusion}

We have analyzed HST archival images of 14 nearby core elliptical galaxies, each of which hosts a central point-like source associated with a low-luminosity AGN, in order to search 
for offsets between the AGN and the galaxy photocenter. Such AGN--photocenter displacements are possible signposts of gravitational recoils 
resulting from the coalescence of an SBH binary.
 
We find significant ($>3\sigma$) differences between the positions of the nuclear optical (or NIR) point source and the
mean photocenter of the galaxy, as determined from isophote fits, in ten of the 14 galaxies in the sample.
Assuming that the mean photocenter locates the minimum of the galactic potential
well and that the point source locates the position of the AGN and hence the SBH, these
results imply that the SBH is displaced from its equilibrium position by angular distances ranging between 20 and 90 mas, 
or projected linear distances in the range $1-10$\,pc. 
As spurious offsets may occur as a result of large-scale isophotal asymmetries, 
only displacements of magnitude $> 0.8$ $\times$ the inter-quartile range of the distribution of isophote center 
coordinates (equivalent to $1\sigma$ for a gaussian distribution) are considered ``real''.
There are six galaxies that exhibit displacements $> 0.8$ IQR with three of these (NGC 4278, NGC 4486 and NGC 5846) 
having displacements $> 1.6$ IQR (equivalent to $2\sigma$).
In every case, the measured displacement of the SBH relative to the galaxy photocenter is a
small fraction (1--10\%) of the galaxy core radius, which is typically $r_c\sim 300$\,pc, for these galaxies. 

Approximate alignments between the SBH--photocenter displacements and the radio source axis were found in four of the six galaxies considered to have significant displacements,
including three of the four that have FR\,I, or FR\,I-like radio sources with relatively powerful and well-defined kpc-scale jets. 
Indeed, in every case in which there is both a significant displacement {\em and} an unambiguous jet, the two are approximately aligned.

Lacking detailed knowledge of the merger history of the galaxies, or of the SBH binary parameters (such as mass ratio and spin configuration) that determine the recoil velocity,
it is not possible to directly test the hypothesis that the displacements are caused by residual gravitational recoil oscillations. Instead,
we used a simple Monte-Carlo model to investigate if the measured displacements are consistent with gravitational recoil. We find that the displacements
in individual objects can plausibly be attributed to residual gravitational recoil oscillations following a major or minor merger within the last few Gyr. However, for
plausible merger rates there is a high probability of larger displacements than actually observed, if SBH coalescence events took place in these galaxies.

That larger displacements were not observed suggests that the frequency of gravitational recoil kicks large enough to trigger long-lived oscillations is lower than predicted,
perhaps because the evolution of the SBH binary typically results in a configuration that suppresses recoil kicks with velocities $\gtrsim 200$ km s$^{-1}$. Alternatively,
the post-recoil oscillations may be damped more quickly than predicted by pure \textit{N}-body simulations. In either case, gas may play an important role \citep[e.g.,][]{Dotti2010, Sijacki2011}. Otherwise, it is possible that galaxy mergers are much more infrequent than implied by the current $\Lambda$CDM paradigm \citep{kroupa14}.

Several other mechanisms are capable of producing the observed displacements, with the observed alignments between the SBH--photocenter displacements and the radio source axis 
favoring jet acceleration in some objects. An approximate displacement--radio  axis  alignment is also expected for gravitational recoil, but not for 
orbital motion in pre-coalescence SBH binaries or interactions with massive perturbers. However, both of the latter mechanisms are capable of producing displacement amplitudes comparable to those observed and cannot be ruled out in individual objects.

In general, it is not possible to unambiguously distinguish between different mechanisms (including recoil) on the basis of the displacement measurements alone for individual galaxies. However, with a larger sample it may be possible to distinguish mechanisms using statistical arguments. Thus, for jet acceleration the displacement direction should be strongly correlated with the radio jet, with the amplitude correlating with jet power. In the case of gravitational recoil, a weaker correlation with jet direction might be expected. However, no such correlations are expected for binary SBH, or massive perturbers.

\section*{Acknowledgments}
We wish to honor the memory of our great friend, colleague and mentor
David Axon. 
We thank the anonymous referee for the valuable and timely comments that improved the paper. 
DL thanks E. Vasiliev, R. Mittal, F. Antonini and M. Richmond for helpful discussions.
AM acknowledges support from the Italian National Institute for Astrophysics (INAF) through 
PRIN-INAF 2011 ``Black hole growth and AGN feedback through the cosmic time" and from the 
Italian ministry for school, university and research (MIUR) through PRIN-MIUR 2010-2011 
``The dark Universe and the cosmic evolution of baryons: from current surveys to Euclid".
DM was supported by the National Science Foundation under grant no. AST 1211602
and by the National Aeronautics and Space Administration under grant no. NNX13AG92G.
Support for program AR-11771 was provided by NASA 
through a grant from the Space Telescope Science Institute, which is operated by the Association of Universities
for Research in Astronomy, Inc., under NASA contract NAS 5-26555. We acknowledge the usage of the HyperLeda
database (http://leda.univ-lyon1.fr) and the NASA/IPAC Extragalactic Database (NED) which is operated by the
Jet Propulsion Laboratory, California Institute of Technology, under contract with the
National Aeronautics and Space Administration. Based on observations made with the NASA/ESA Hubble Space
Telescope, and obtained from the Hubble Legacy Archive, which is a collaboration between the Space
Telescope Science Institute (STScI/NASA), the Space Telescope European Coordinating Facility
(ST-ECF/ESA) and the Canadian Astronomy Data Center (CADC/NRC/CSA).

\begin{figure*}[!hbp]
\begin{center}$
\begin{array}{ccccc}
\includegraphics[width=1.4 in, trim = 6.8cm 2.6cm 4.85cm 1cm, clip]{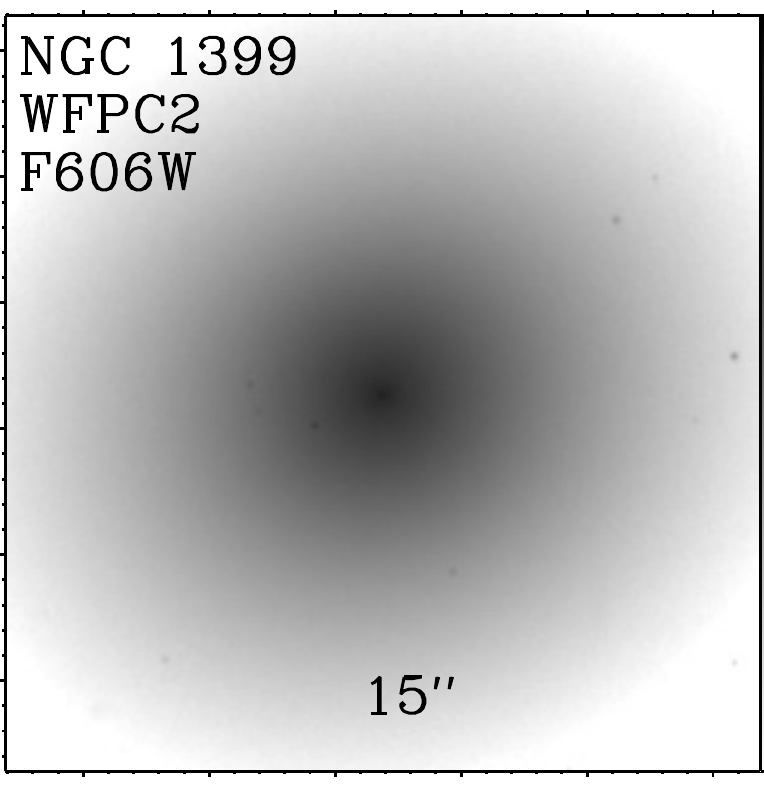} &
 \includegraphics[width=1.4 in, trim = 6.8cm 2.6cm 4.85cm 1cm, clip]{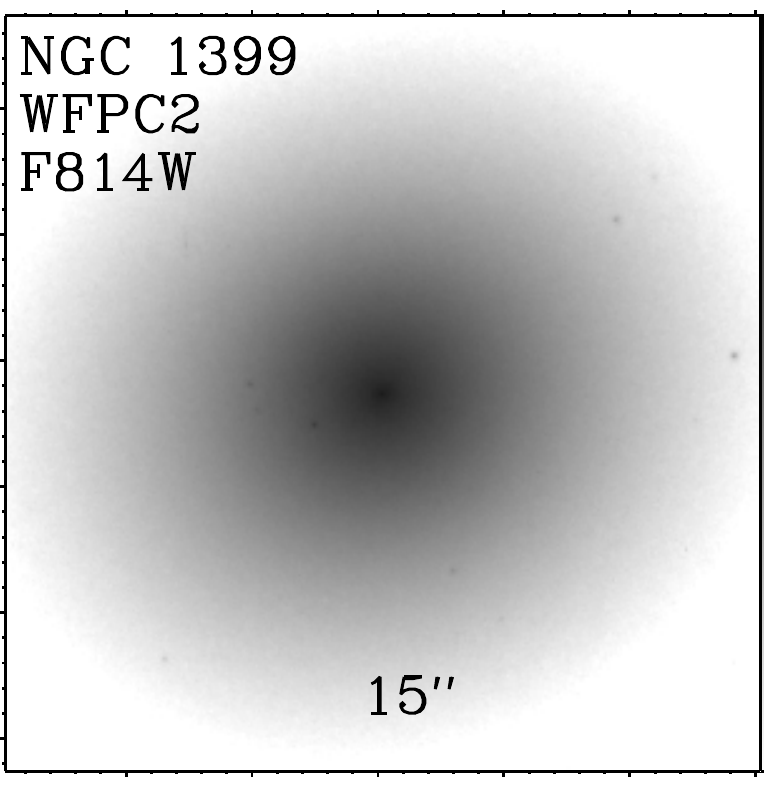}	&
  \includegraphics[width=1.4 in, trim = 6.8cm 2.6cm 4.85cm 1cm, clip]{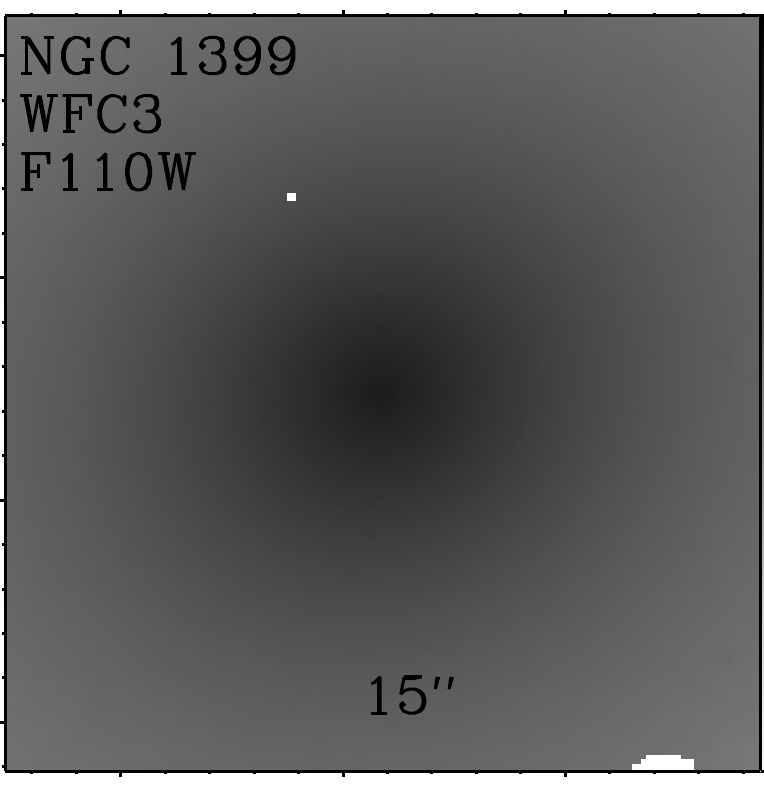}	& 
    \includegraphics[width=1.4 in, trim = 6.8cm 2.6cm 4.85cm 1cm, clip]{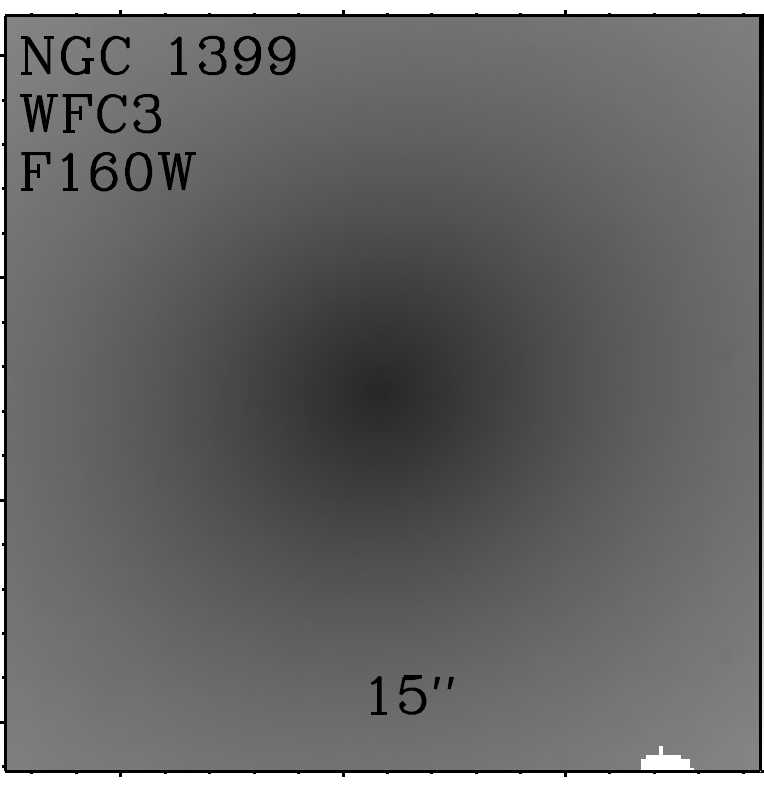}& 
 \includegraphics[width=1.4 in, trim = 6.8cm 2.6cm 4.85cm 1cm, clip]{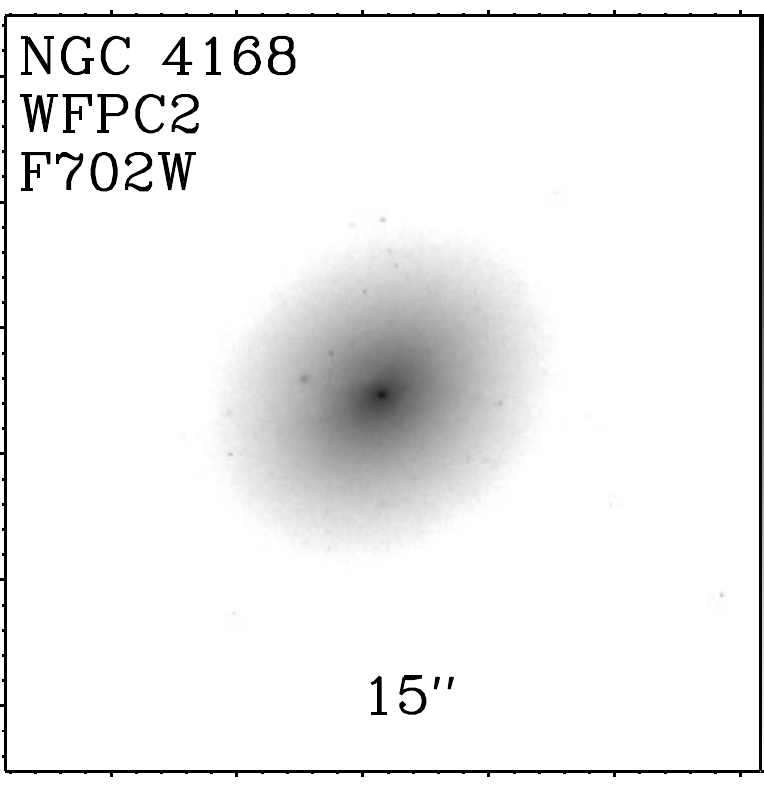} \\ 

  \includegraphics[width=1.4 in, trim = 6.8cm 2.6cm 4.85cm 1cm, clip]{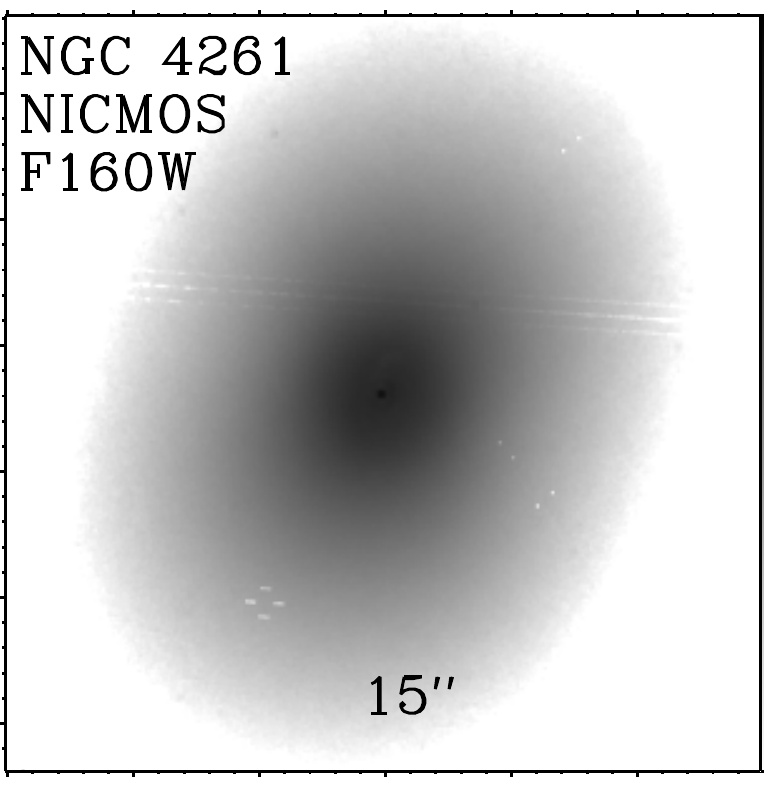} &  
\includegraphics[width=1.4 in, trim = 6.8cm 2.6cm 4.85cm 1cm, clip]{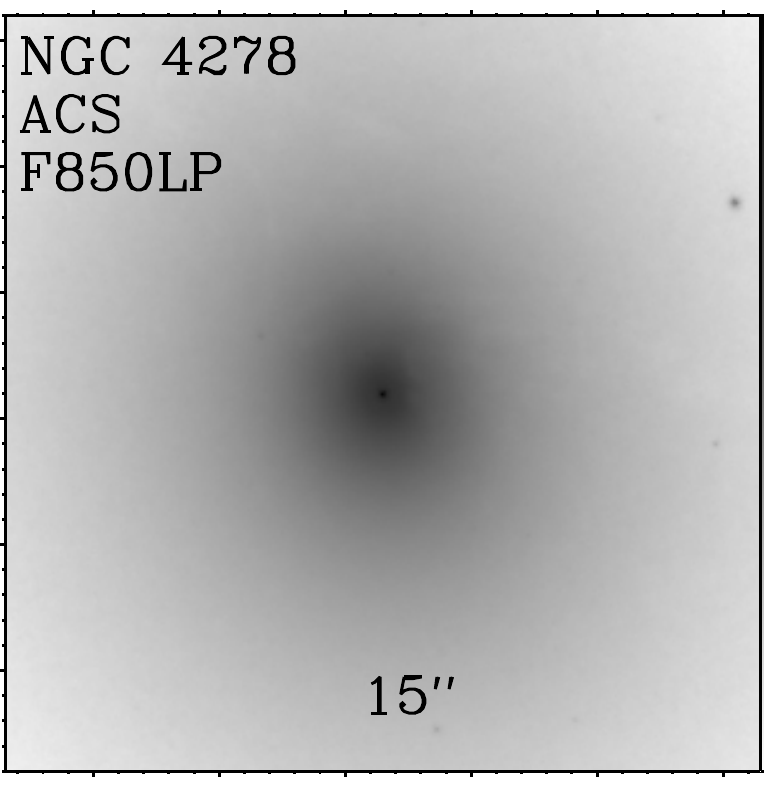}	&
\includegraphics[width=1.4 in, trim = 6.8cm 2.6cm 4.85cm 1cm, clip]{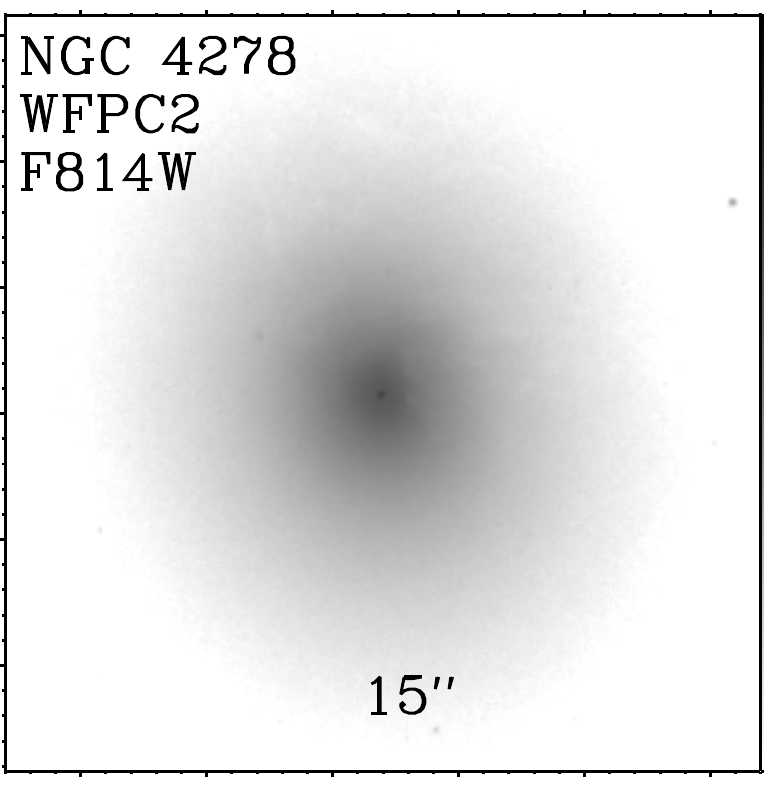} &
\includegraphics[width=1.4 in, trim = 6.8cm 2.6cm 4.85cm 1cm, clip]{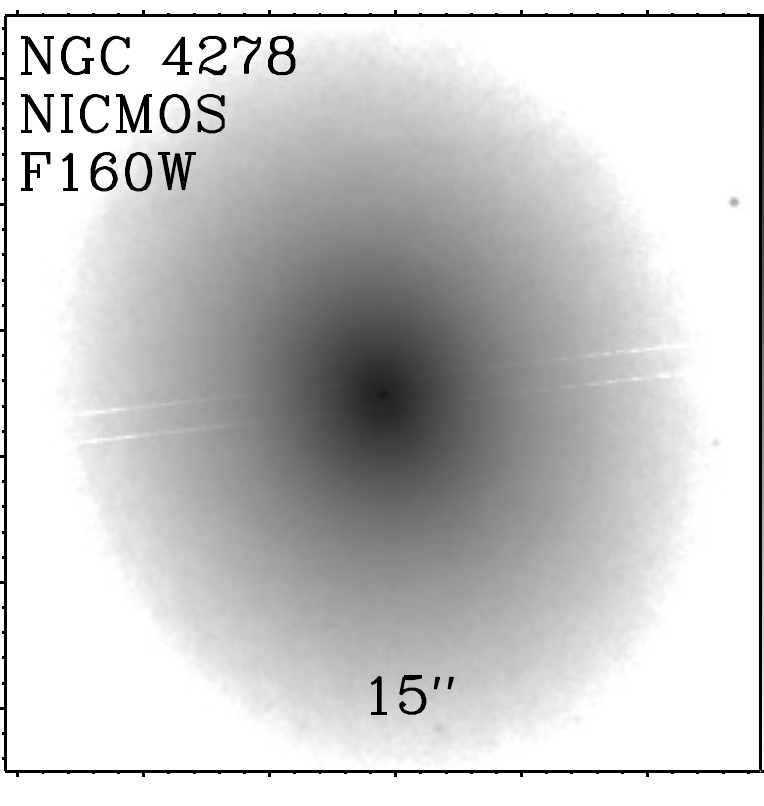} &
\includegraphics[width=1.4 in, trim = 6.8cm 2.6cm 4.85cm 1cm, clip]{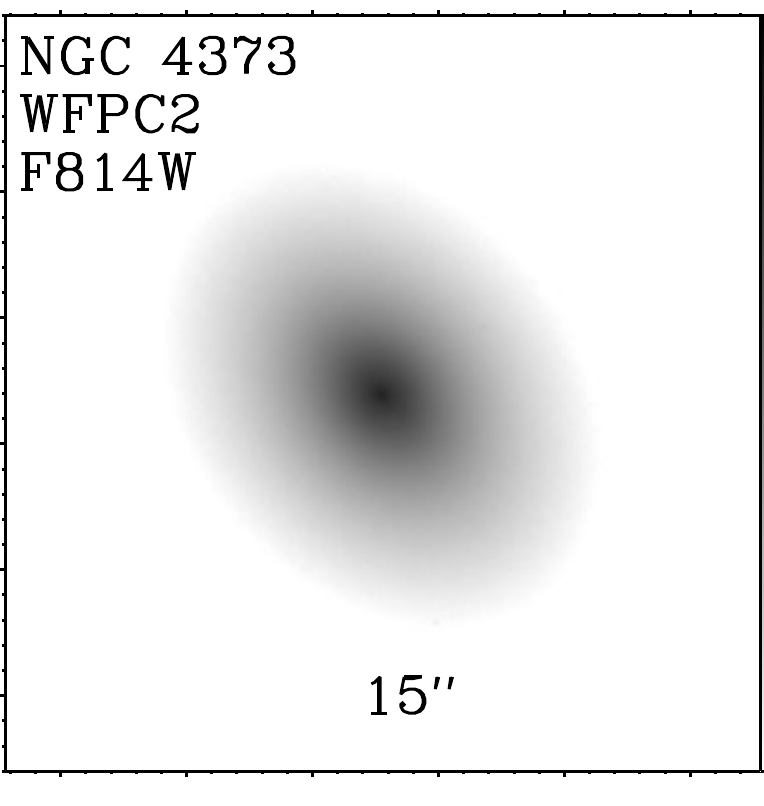}  \\

\includegraphics[width=1.4 in, trim = 6.8cm 2.6cm 4.85cm 1cm, clip]{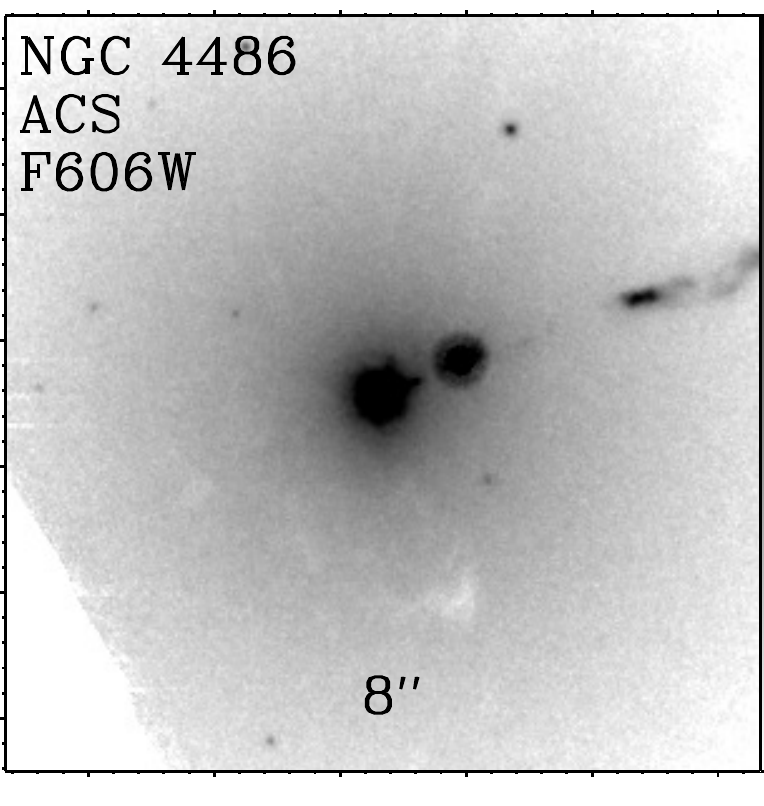} &
\includegraphics[width=1.4 in, trim = 6.8cm 2.6cm 4.85cm 1cm, clip]{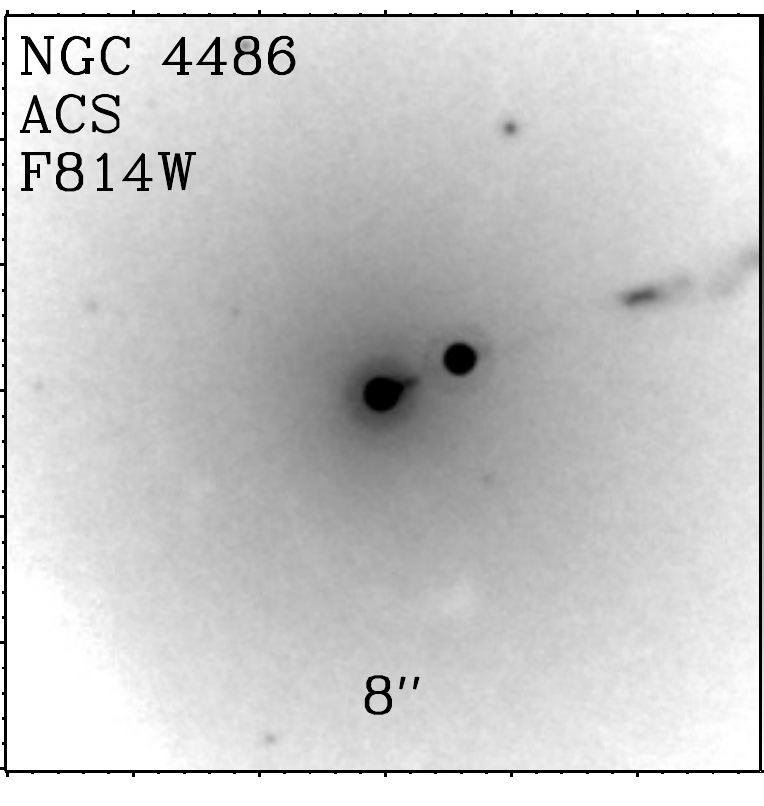} &
\includegraphics[width=1.4 in, trim = 6.8cm 2.6cm 4.85cm 1cm, clip]{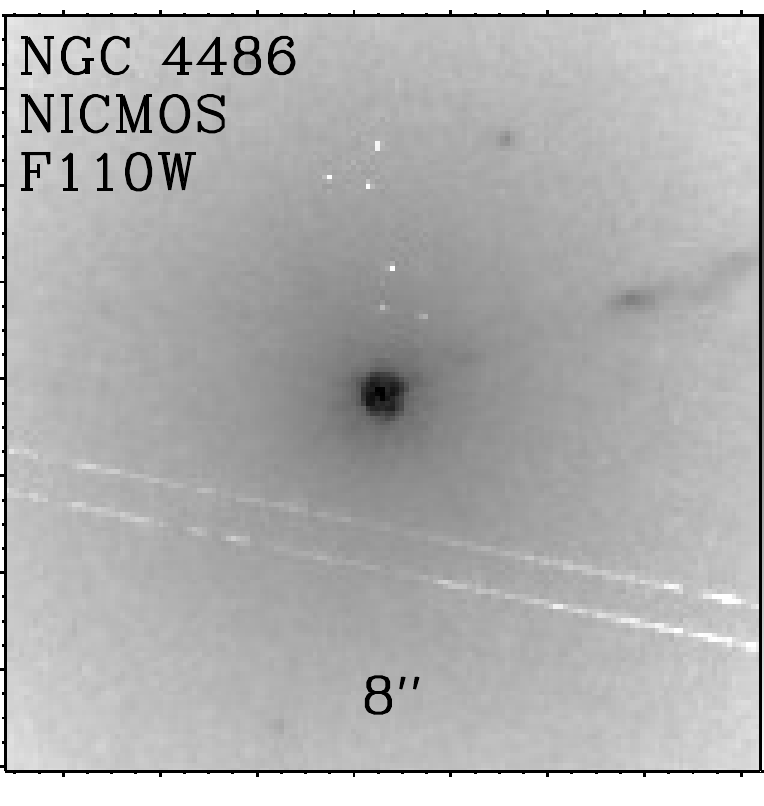} &
\includegraphics[width=1.4 in, trim = 6.8cm 2.6cm 4.85cm 1cm, clip]{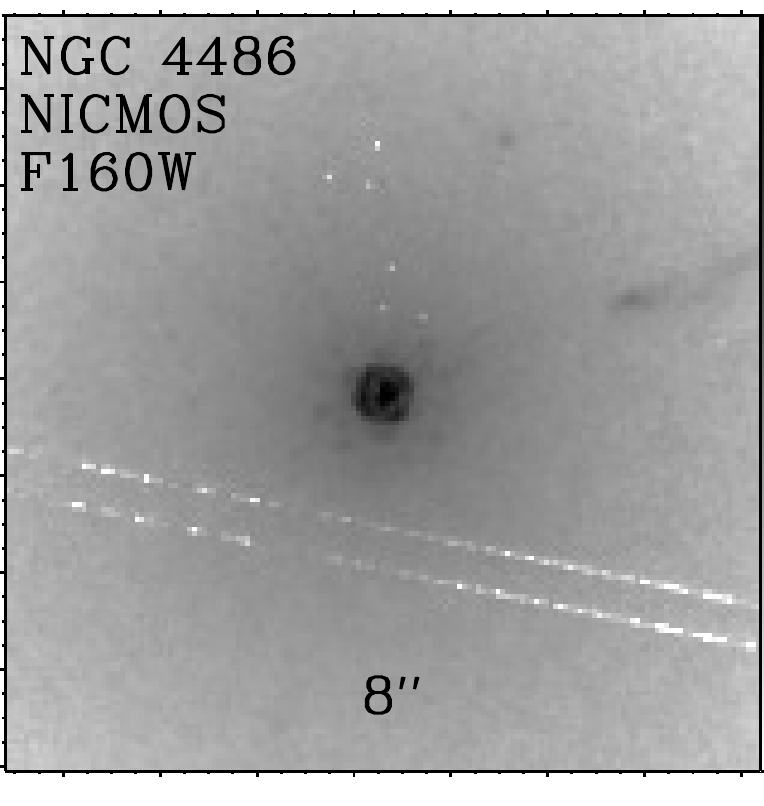} &
\includegraphics[width=1.4 in, trim = 6.8cm 2.6cm 4.85cm 1cm, clip]{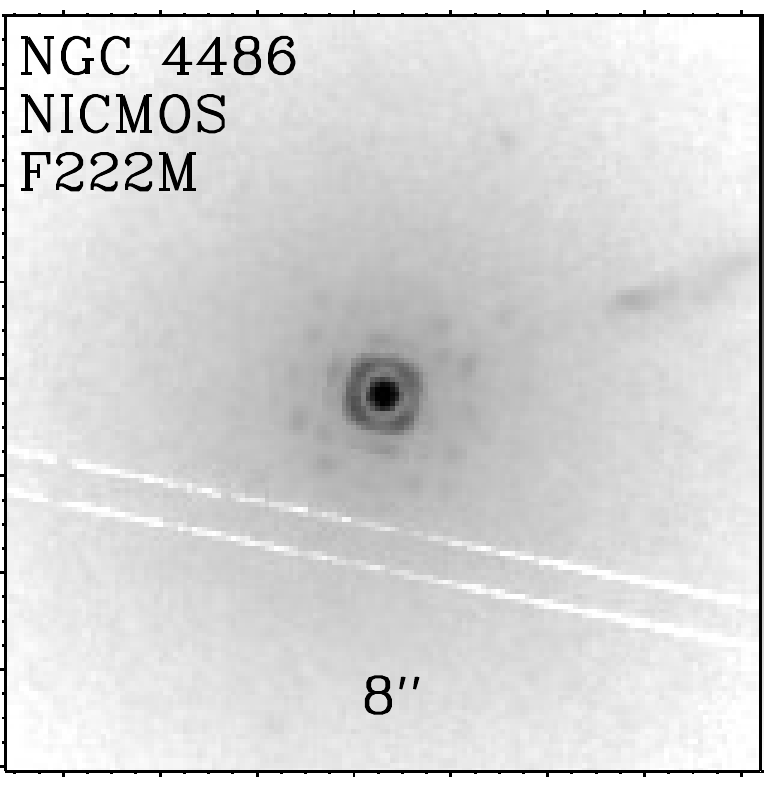} \\

\includegraphics[width=1.4 in, trim = 6.8cm 2.6cm 4.85cm 1cm, clip]{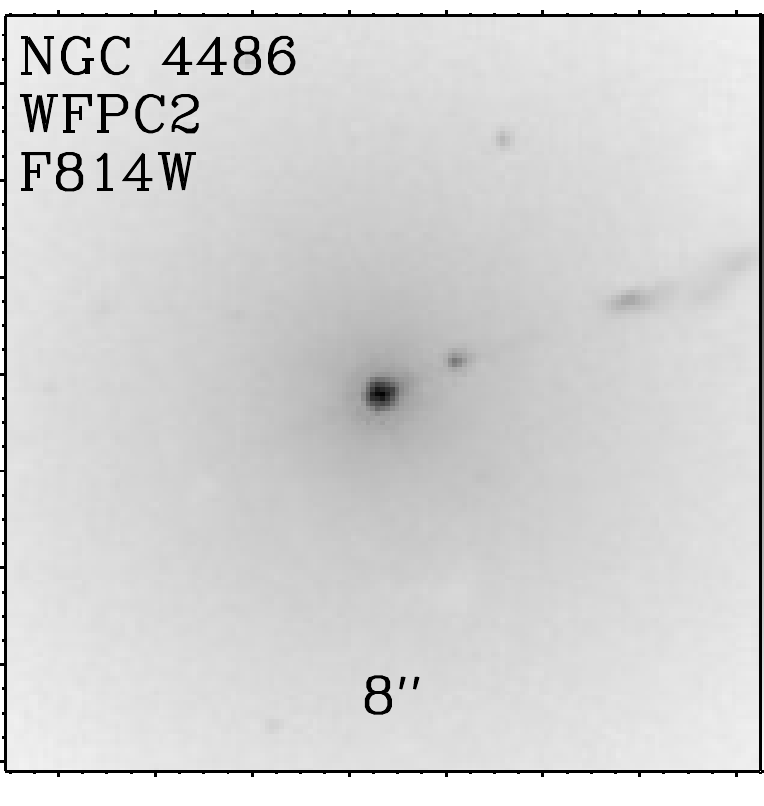} &
\includegraphics[width=1.4 in, trim = 6.8cm 2.6cm 4.85cm 1cm, clip]{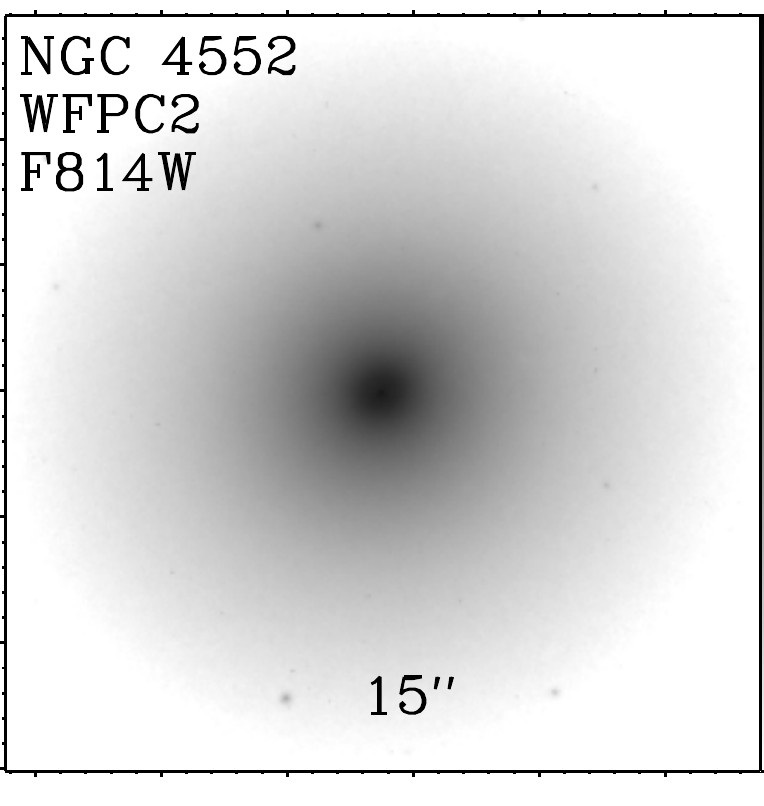} &
\includegraphics[width=1.4 in, trim = 6.8cm 2.6cm 4.85cm 1cm, clip]{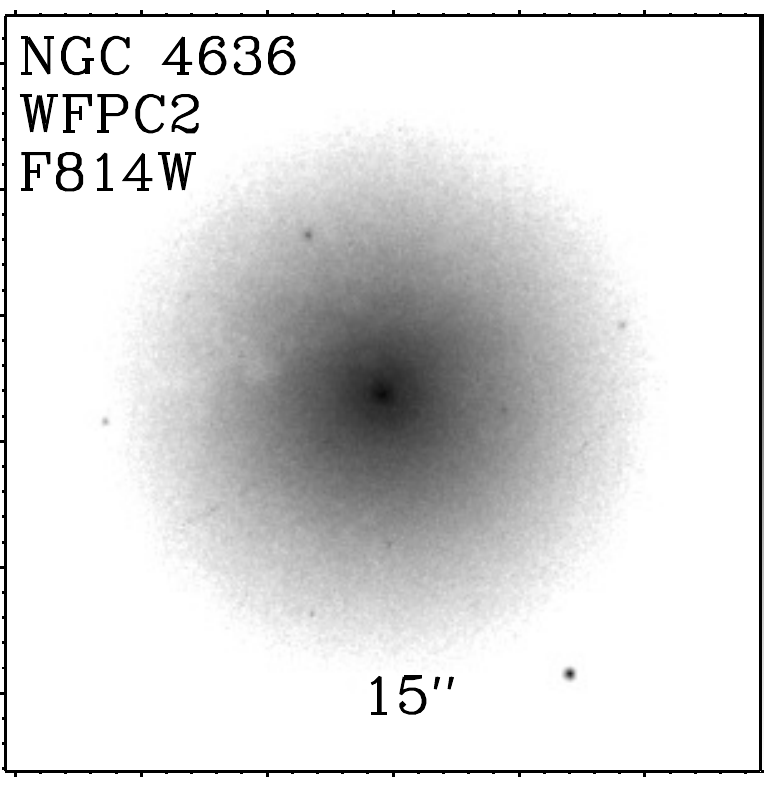} &
\includegraphics[width=1.4 in, trim = 6.8cm 2.6cm 4.85cm 1cm, clip]{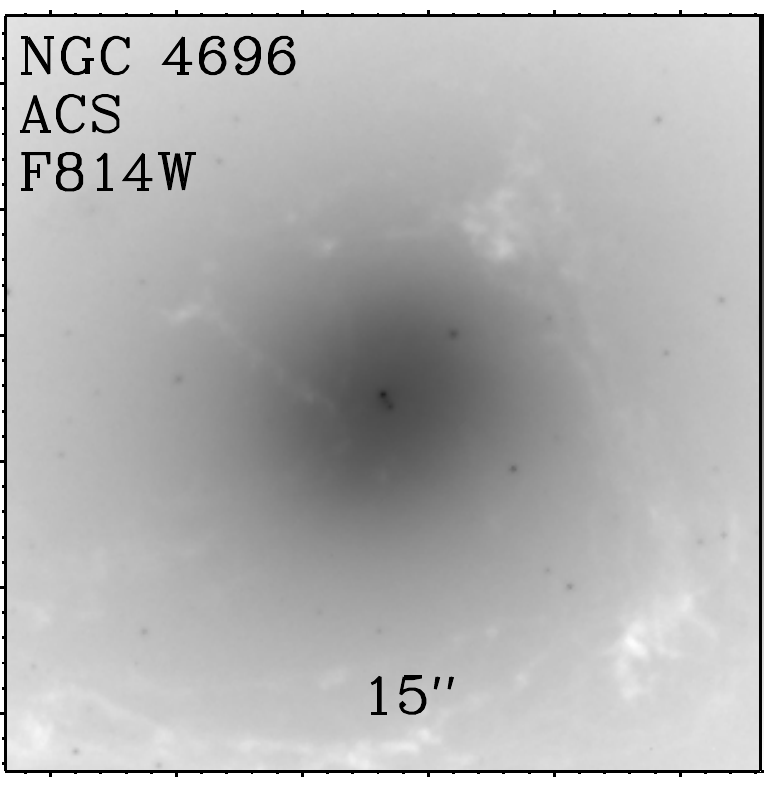} &
\includegraphics[width=1.4 in, trim = 6.8cm 2.6cm 4.85cm 1cm, clip]{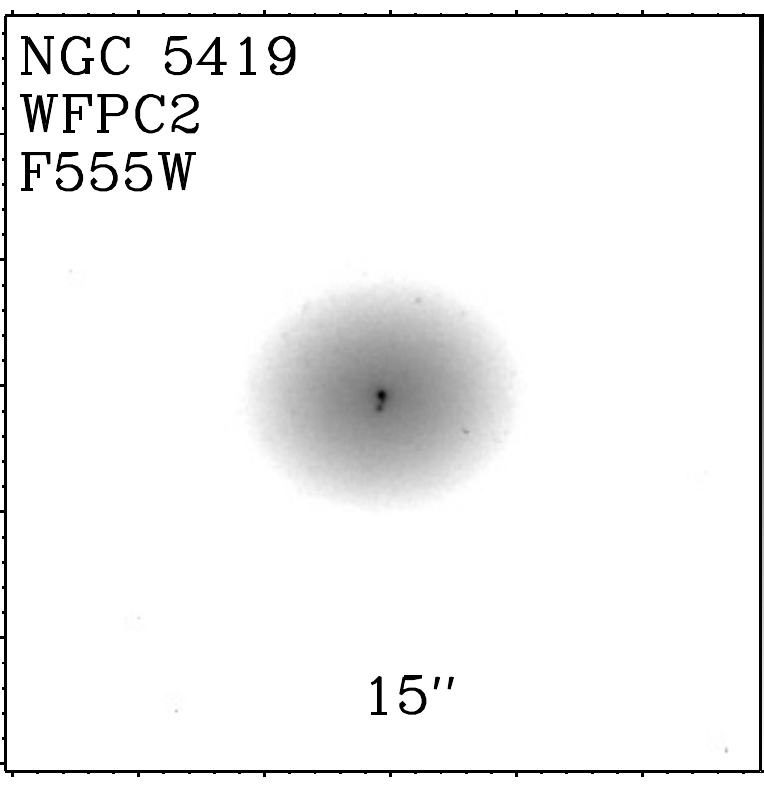} \\

\includegraphics[width=1.4 in, trim = 6.8cm 2.6cm 4.85cm 1cm, clip]{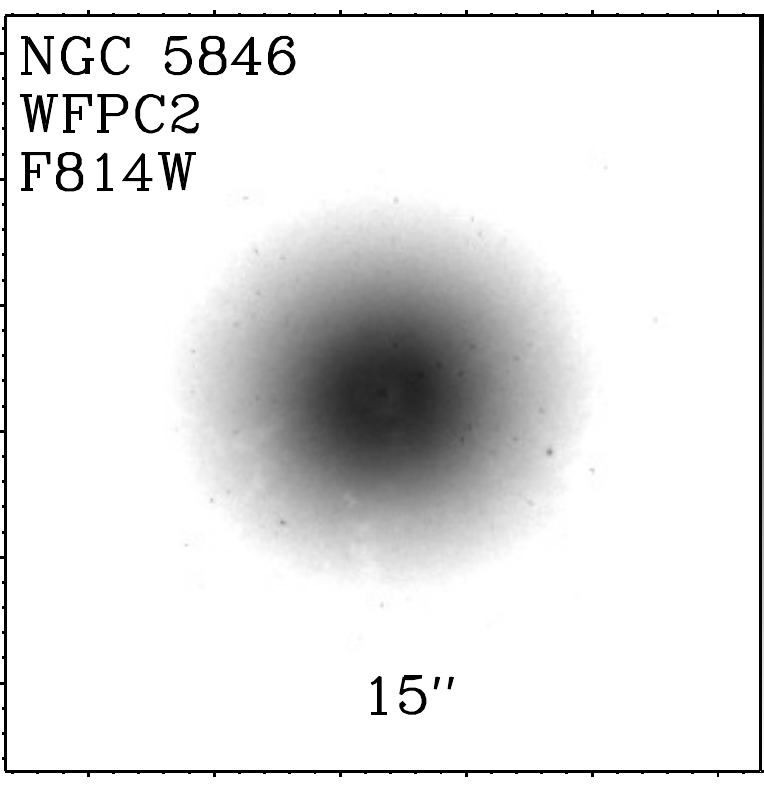} &
\includegraphics[width=1.4 in, trim = 6.8cm 2.6cm 4.85cm 1cm, clip]{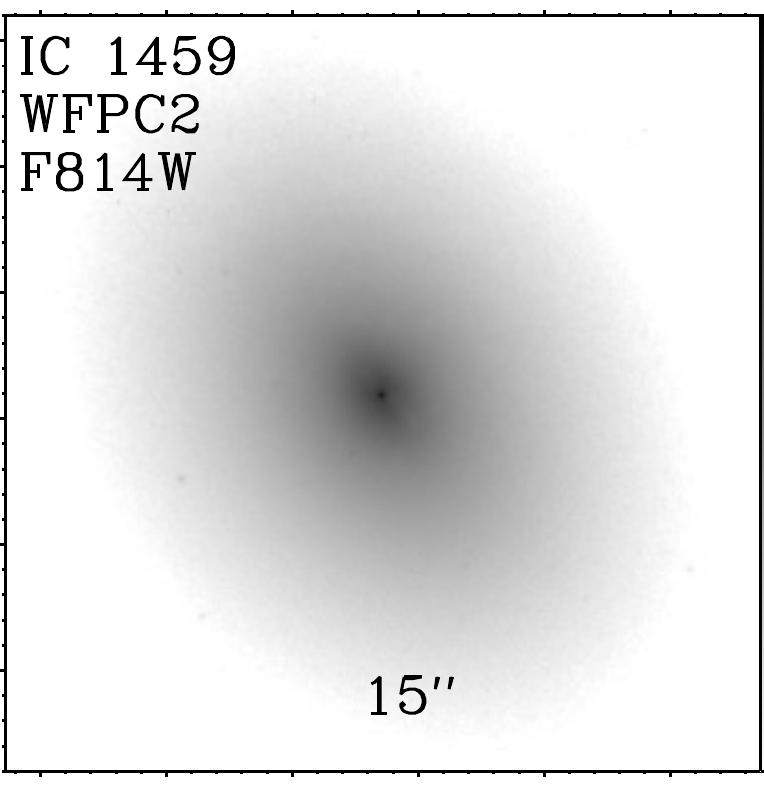}&
\includegraphics[width=1.4 in, trim = 6.8cm 2.6cm 4.85cm 1cm, clip]{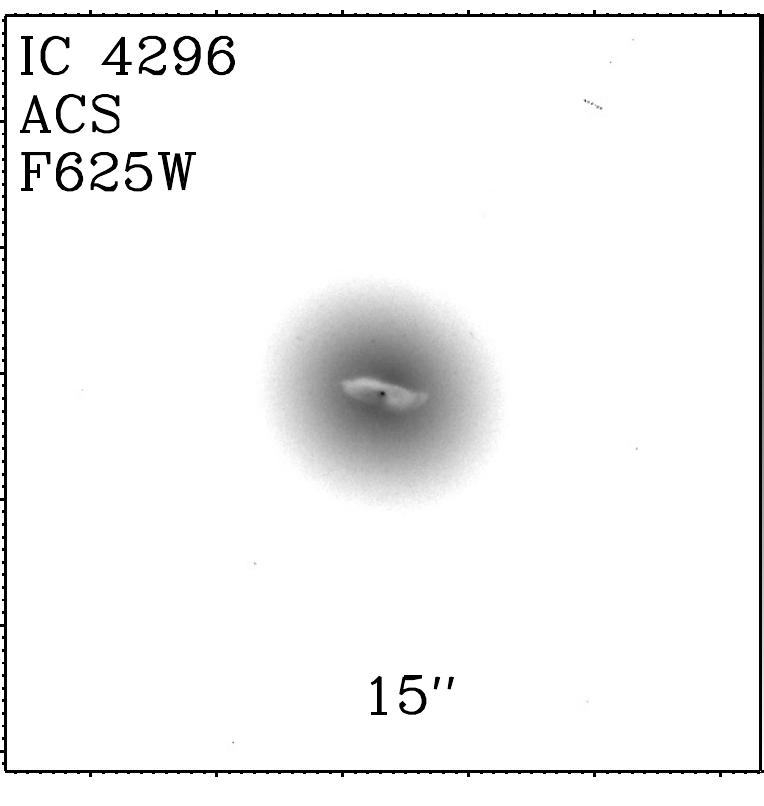} &
\includegraphics[width=1.4 in, trim = 6.8cm 2.6cm 4.85cm 1cm, clip]{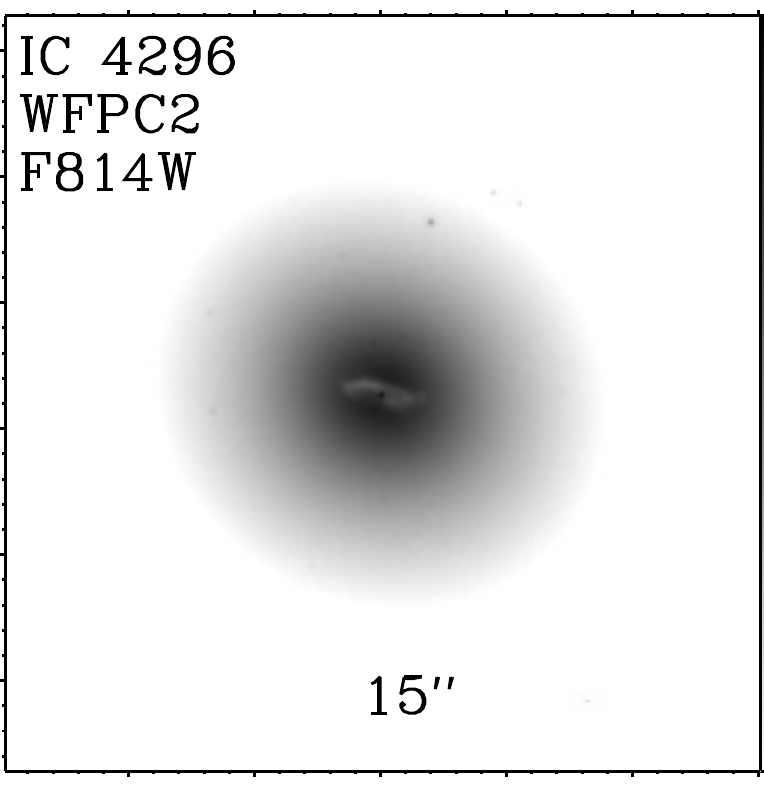} \\

\includegraphics[width=1.4 in, trim = 6.8cm 2.6cm 4.85cm 1cm, clip]{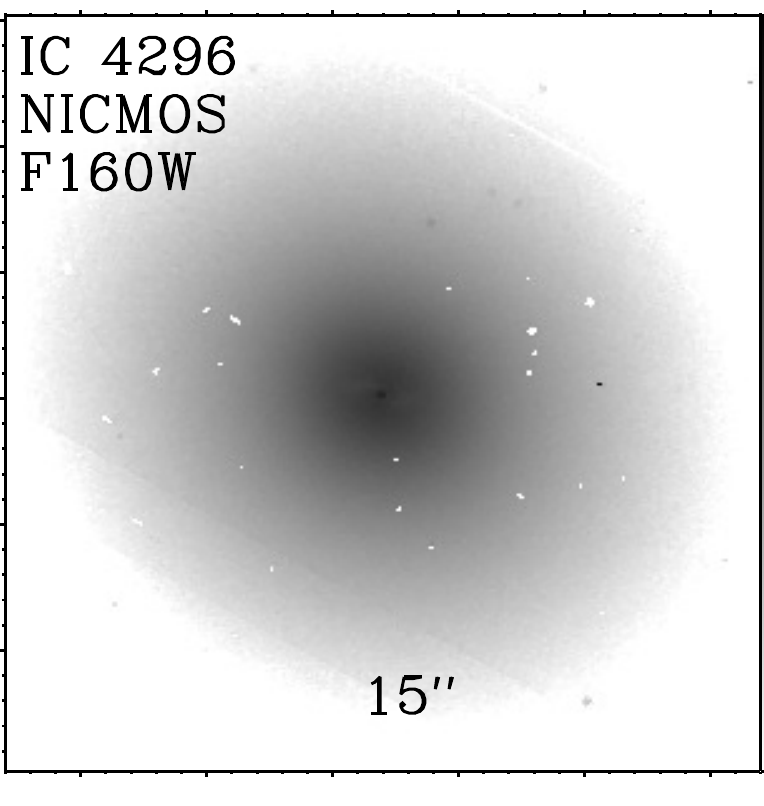} &
\includegraphics[width=1.4 in, trim = 6.8cm 2.6cm 4.85cm 1cm, clip]{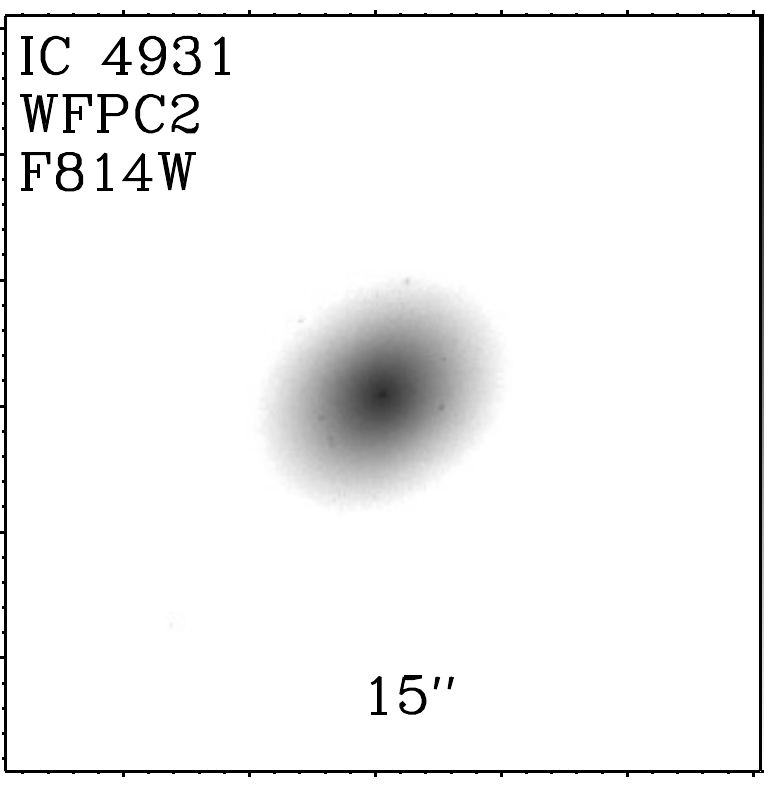}\\

\end{array}$
\end{center}
\caption{HST images of the galaxies in the sample. The number at the bottom represents the length of the whole axes.}
\end{figure*} \label{fig: sample}


\begin{appendix}
\section{A: Distribution of recoil displacements}
\label{distribution}

Here we outline the method used to estimate the probability distributions for displacements resulting from post-recoil oscillations of a kicked SBH. 
For our purposes, of the three phases predicted by GM08, only phase II (when the oscillation amplitude becomes comparable with the galaxy core radius) is of interest. 
The initial, large amplitude phase I oscillations are short-lived, while in phase III the residual oscillations are indistinguishable from Brownian motion (which is considered separately in Section~\ref{subsec: disp_mech}).  

In the GM08 \textit{N}-body simulations, phase II oscillations occur for kick
velocities in the range $v_\mathrm{kick} \sim 0.4 - 0.9\ v_\mathrm{esc}$. The simulations also indicate that the subsequent evolution does not depend strongly on the magnitude of the initial kick. 
Therefore, we assume that, given any initial kick large enough to displace the SBH beyond the core radius, 
the distance, $R$, of the SBH from the center of the galaxy evolves during phase II as a damped harmonic oscillator:

\begin{align}\label{eq: osc}
R(t) = R_{0}e^{-\Delta t/ \tau} \sin(\omega_{c}\Delta t).
\end{align}

The initial amplitude, $R_{0}$, is set by the condition that phase II begins when the oscillation amplitude is comparable to the core radius; 
$\Delta t$ is the time since the kick, i.e., the time since the last merger, and $\omega_{c}$ is the SBH oscillation frequency:

\begin{align}
\label{eq: omega}
\omega_{c} = \sqrt{\frac{3}{4} \frac{\sigma_{c}^{2}}{r_{c}^{2}}}.
\end{align}

Equation~\ref{eq: omega} is obtained by combining the radial oscillation frequency in a spherical core of uniform density, $\omega_{c} = [(4\pi/3)G\rho_{c}]^{1/2}$, with equation 13 in GM08. The damping time, $\tau$, is given by equation~\ref{eq: tdamp}. 
The adopted parameter values are listed in Table \ref{tab: param}.

Suppose that we observe a galaxy at a random time since its last merger. For simplicity, we assume that galaxy mergers occur stochastically and can be represented as a Poisson process. 
The probability that the last merger occurred at time $\Delta t$ ago is then:

\begin{align}
\label{eq: pmerger}
P(\Delta t) = t_{m}^{-1} e^{-\Delta t/t_{m}}, \Delta t > 0
\end{align}

\noindent which is associated with the cumulative distribution:

\begin{align}\label{eq: pcm}
P(<\Delta t) = 1 - e^{-\Delta t/t_{m}}
\end{align}

\noindent where  $t_{m}$ is the mean time between galaxy mergers. 

The galaxy merger rate is uncertain but simulations indicate that it is a function of redshift, galaxy mass and mass ratio \citep[e.g.,][]{Bell2006, Conselice2009, Hopkins2010, Burke2013}.
Here, since the phase II damping times are typically $\sim 1$\,Gyr (Table \ref{tab: param}), we consider redshifts $z\leq 1$ and estimate the mean time between galaxy mergers as the ratio between the lookback time corresponding to $z=1$ and the number of mergers, $N$, during that time. 
The latter was determined using the analytical fits to semi-analytical models for galaxy merger rates given by H10. 
Integrating the merger rate as given by equation~5 in H10 over redshift, we find  for the number of mergers,

\begin{align}
N = t_\mathrm{H}A(M_\mathrm{min})\int_{0}^{1} (1+z)^{\beta(M_\mathrm{min}) - 5/2} dz
\end{align}

\noindent where t$_\mathrm{H}$ = 13.7 Gyr is the Hubble time, the normalization constant $A(M_\mathrm{min})$ and the slope $\beta(M_\mathrm{min})$ are given as functions of the galaxy mass threshold $M_\mathrm{min}$, for mergers
with mass ratio $q>0.1$, by equations 9 \& 10 in H10. For $\log M_\mathrm{gal}\ge \log M_\mathrm{min} = 11$, we find $N\approx 1.5$ for 
the number of mergers per galaxy between $z = 0$ and $z=1$. However, we note that $N$ could vary between 0.75 and 3, allowing for systematic uncertainties of a factor two in $A(M_\mathrm{min})$ and 0.2 in $\beta(M_\mathrm{min})$.
The look back time at $z=1$ is $t_{L} = 7.731$\,Gyr,  for H$_{0}$=71 km s$^{-1}$ Mpc$^{-1}$, $\Omega_{M}=0.27$ and $\Omega_\mathrm{vac}=0.73$,
yielding a mean time between galactic mergers $t_{m}\approx  5$ Gyr.  

The simple analytical functions provided by H10 do not accurately reproduce the numerical results for the most massive galaxies. 
Therefore, we use an average merger rate for $z\leq 1$ 
estimated from their Figure~3, for log M$_\mathrm{gal} > 12$. The rate for $q>0.1$ mergers is roughly constant for $z\leq 1$ and it seems reasonable to adopt an average value of 2.5 mergers/galaxy/Gyr. This corresponds to $t_{m} \approx 0.4$ Gyr. 

The stellar mass, $M_\mathrm{gal}$, is derived for each galaxy from the relation:

\begin{align} \label{eq: mgal}
\Upsilon = 10^{0.4(M_{K} - M_{\sun K})} \frac{M_{\star}}{M_{\sun}}
\end{align}

\noindent where $\Upsilon$ is the mass-to-light ratio and $M_{K}$ is the K-band magnitude given by BC06. 
For the mass-to-light ratio, we use a representative value, $\Upsilon = 3.1$, computed as the average value of the ratio 
between the virial mass and the K-band luminosity for the sample of nearby core ellipticals studied by \cite{SaniMHR2011}. 

As the derived values of $M_\mathrm{gal}$ span the range $11 \lesssim \log M_\mathrm{gal} \lesssim 12$ we compute displacement probability distributions for the limiting cases $t_{m} = 0.4$ Gyr and 
 $t_{m} = 5.0$ Gyr. For comparison, \citet{Lidman2013} find an observationally-derived major merger rate of $\approx 0.38$ mergers\,Gyr$^{-1}$ for brightest cluster galaxies at $z\sim 1$, implying $t_m\approx 2.6$\,Gyr.   

From the cumulative distributions given by eq.~\ref{eq: pcm}, we extract random times in order to generate a set of $R_{i} \equiv R(t_{i})$ from equation~\ref{eq: osc} for each galaxy,
for both values of $t_m$. Projected displacements are then calculated:

\begin{align}\label{eq: projection}
R_{i,p} = R_{i} \sin(\theta_{i})
\end{align}

\noindent where $\theta_{i}$ is the angle between the displacement direction, assumed to be randomly oriented, and the line of sight. Finally, the distribution of the projected offsets is computed. In Figs.~\ref{fig: projD} and \ref{fig: projDb} we plot the probability, $P(d>x)$, of observing a displacement, $d$, larger than a given value, $x$,  as a function of $x$. Values of the measured displacements are also plotted. 
The probability of generating a projected displacement larger than the value actually measured is listed for each galaxy in Table \ref{tab: probabilities}, for both values of $t_m$.  
\vskip10pt

The damping time computed from eq.\ref{eq: tdamp} depends on the mass of the SBH. Measurement uncertainties for SBH masses in galaxies beyond the Local Group are typically large and are probably dominated by poorly-understood systematics, given that none of these galaxies exhibits a clear central rise in the stellar velocities \citep[][section 2.2]{Merritt13}. In Table \ref{tab: mass} we list SBH masses estimated from the $M_{\bullet}$ - $\sigma$ relation \citep{FerrareseFordRev05}, from the $M_{\bullet}$ - $M_{\mathrm{V}}$ relation for bright galaxies \citep[M$_{\mathrm{V}} < -19$,][]{lauer2007} and from direct measurements (stellar or gas dynamics modeling). For most of our galaxies the discrepancy in the masses recovered with different methods is small. We compute damping times using SBH masses from the $M_{\bullet}$ - $M_{\mathrm{V}}$ relation finding a smaller average damping time $\bar \tau$  = 0.66 Gyr, although the new damping times for individual galaxies are not systematically smaller than those computed using the $M_{\bullet}$ - $\sigma$ SBH masses. We then compute new values for the probabilities presented in Table \ref{tab: probabilities} and the corresponding cumulative probabilities of not observing displacements larger than those actually measured in this work ($\mathbb{P}$). We obtain $\mathbb{P}$ = 10$^{-16}$ (10$^{-3}$), instead of  10$^{-17}$ (10$^{-4}$), for an average time between galactic mergers t$_{m}$ = 5 (0.4) Gyr. After taking into account the kick probabilities given by L12, as specified in \textsection \ref{sec: recoiling_SBH}, the probability of \textit{not} observing a displacement larger than those actually measured in the entire sample is $\mathbb{P'}\approx 1\times 10^{-3}$, instead of  8$\times 10^{-4}$, for $t_m = 0.4$\,Gyr and $\mathbb{P'}\approx$ 0.14 instead of 0.05 for $t_m = 5.0$\,Gyr. The different masses adopted for the SBH caused, therefore, a difference in $\mathbb{P}$ of one order of magnitude. Nevertheless, even for $t_{m} = 5$ Gyr, there remains a large probability of finding displacements larger than those actually observed, if these galaxies experienced gravitational recoil events within the last few Gyr.

\begin{subfigures}
\begin{figure}[p] 
\begin{center}$
\begin{array}{cc}
\includegraphics[scale=0.5]{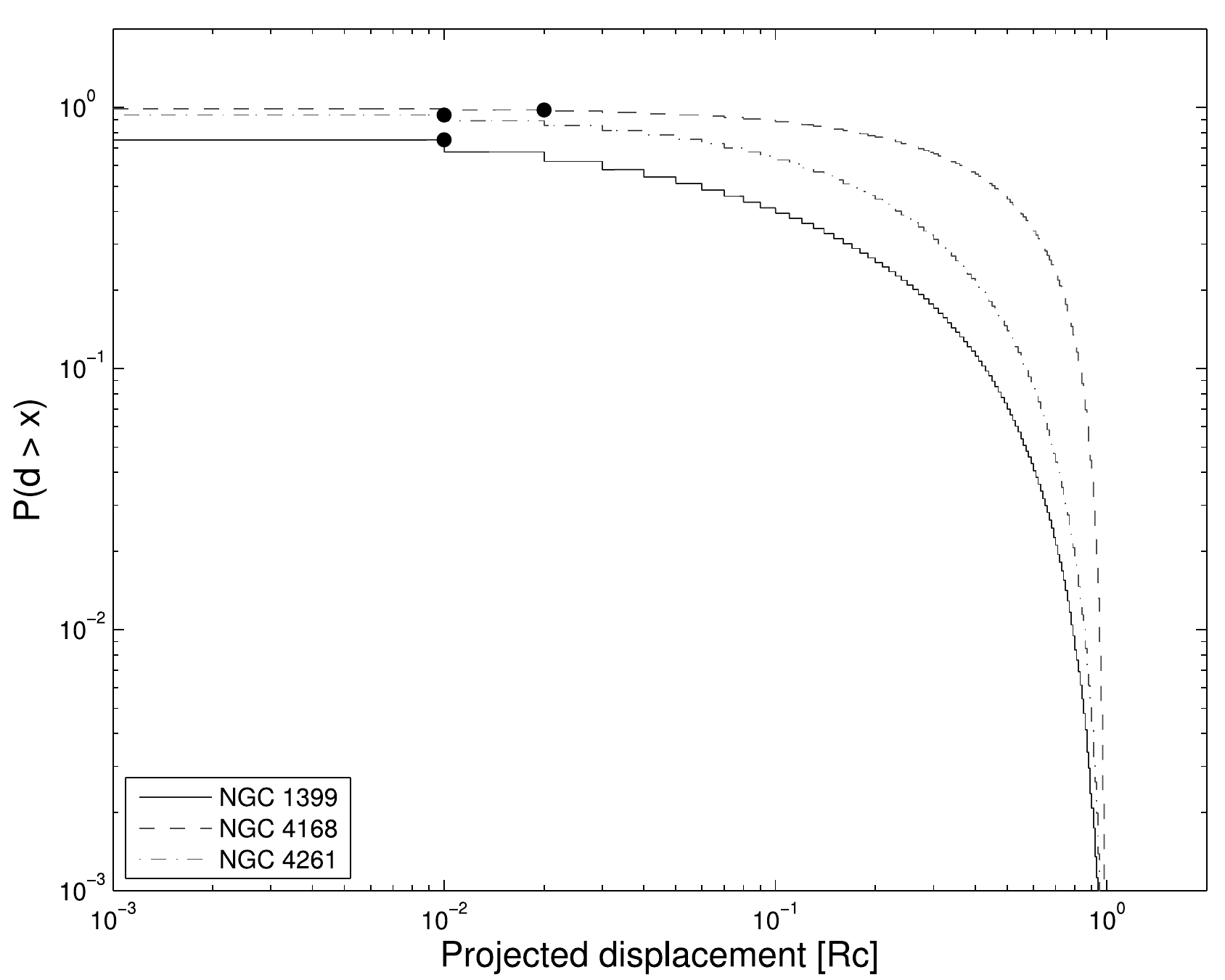} 		& \includegraphics[scale=0.5]{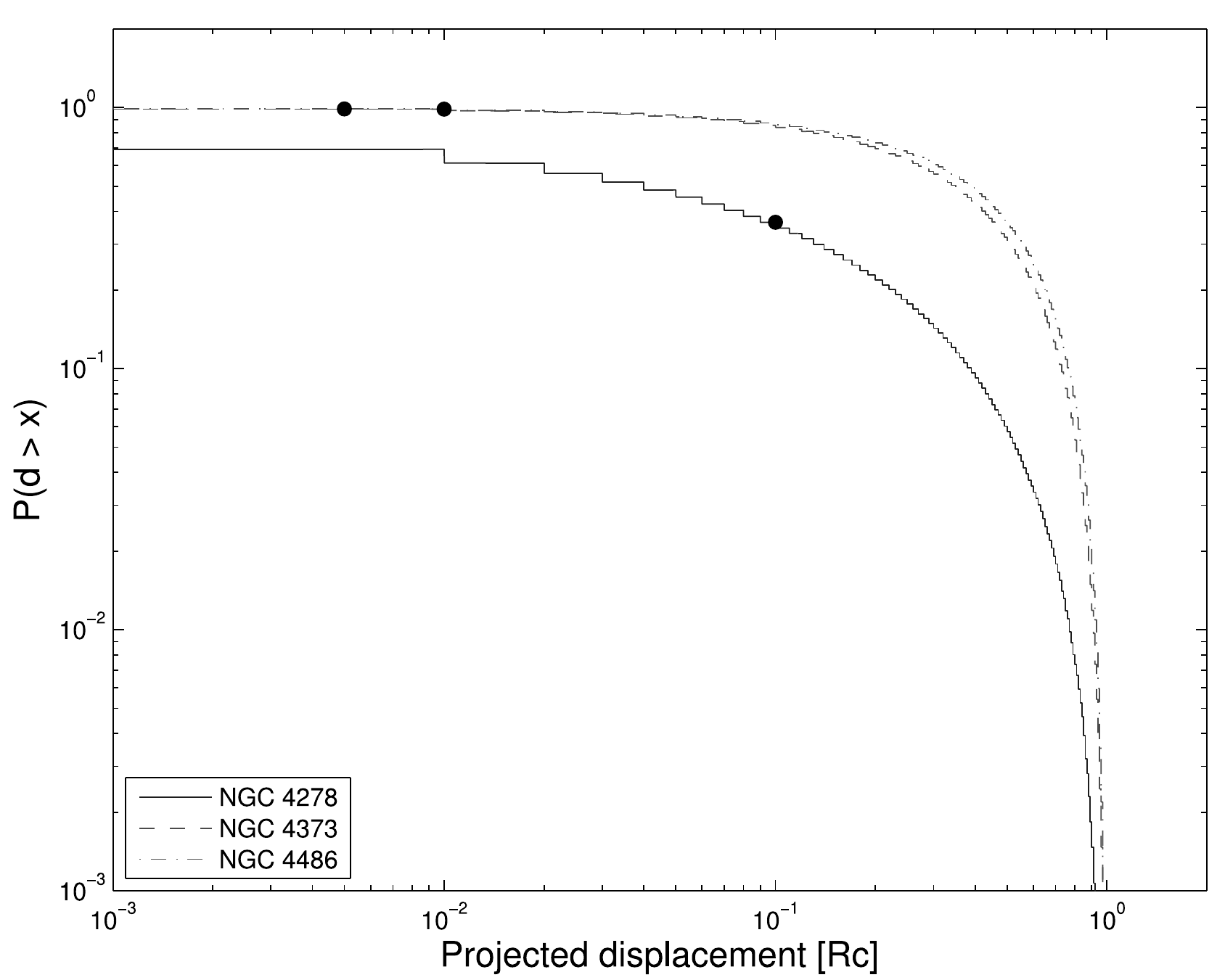} \\
&\\
\includegraphics[scale=0.5]{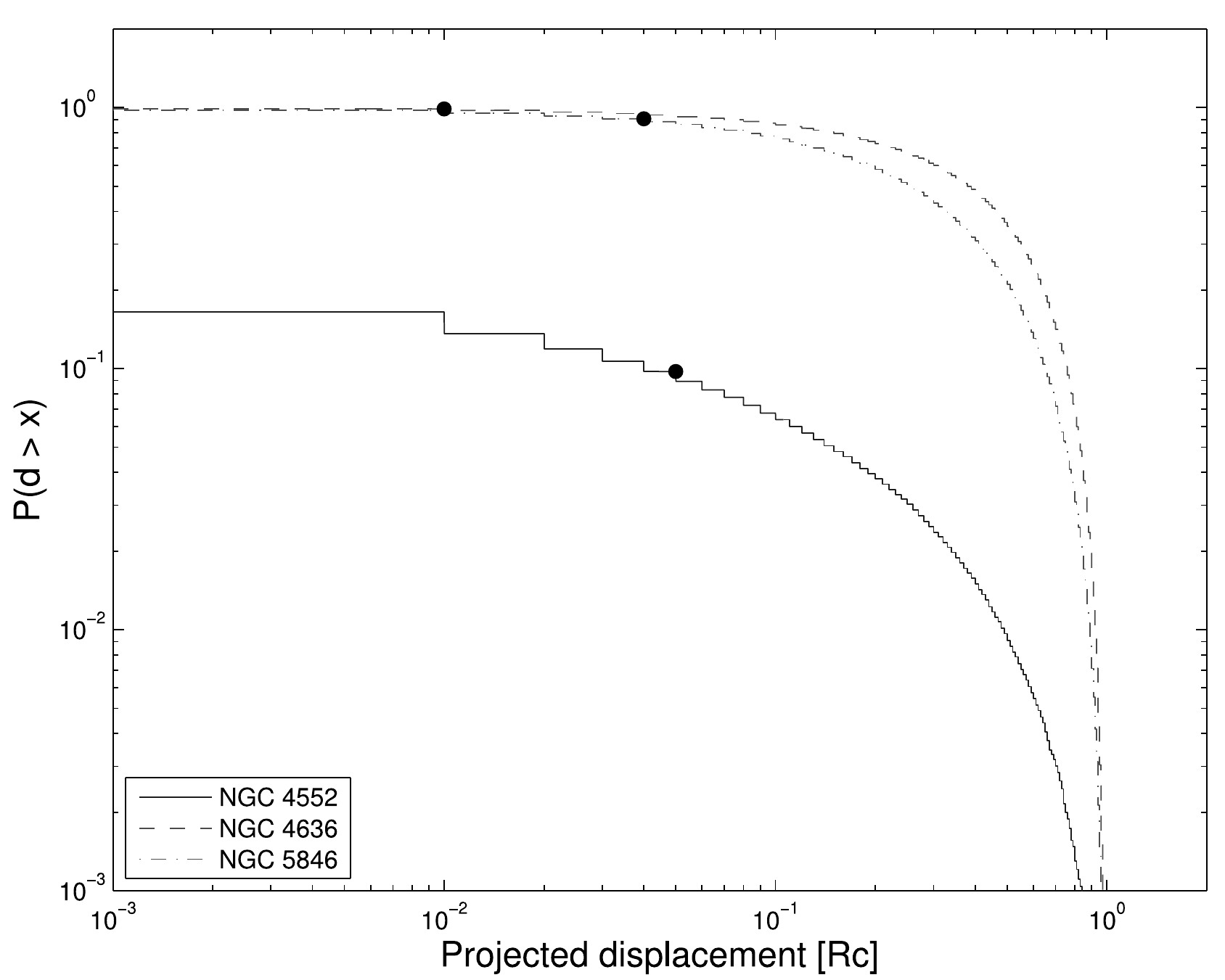} 	& \includegraphics[scale=0.5]{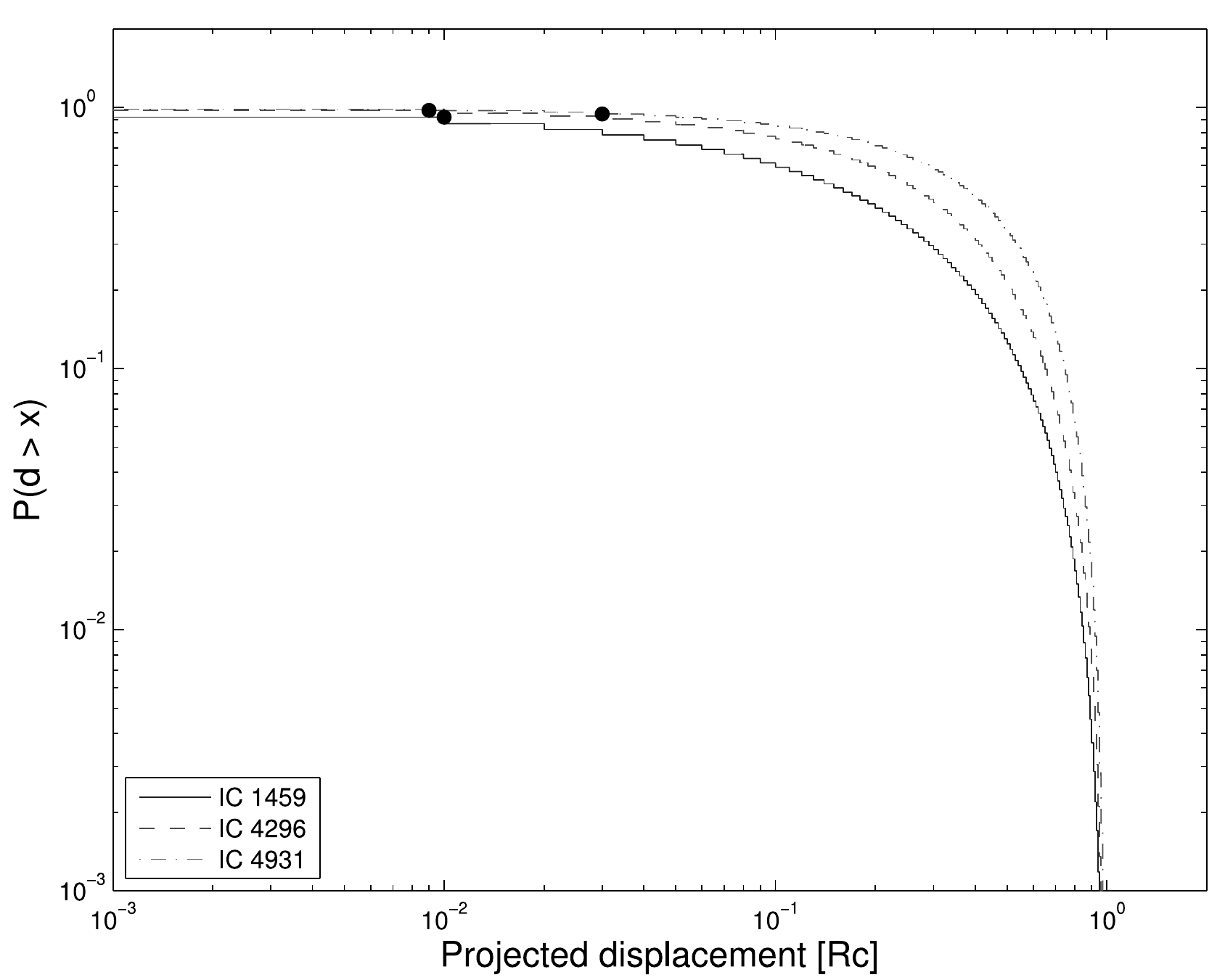} \\
&\\
\includegraphics[scale=0.5]{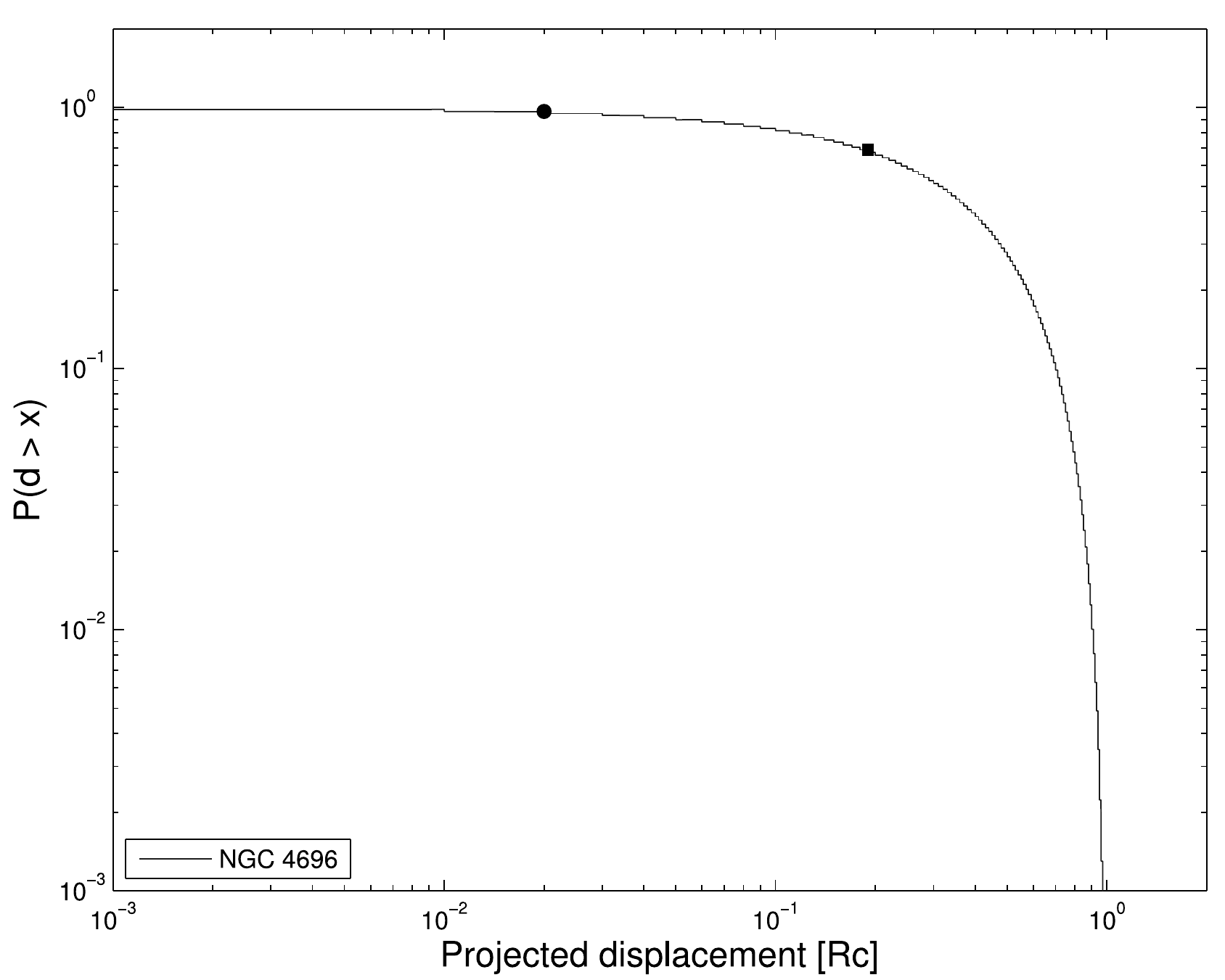} 		& \includegraphics[scale=0.5]{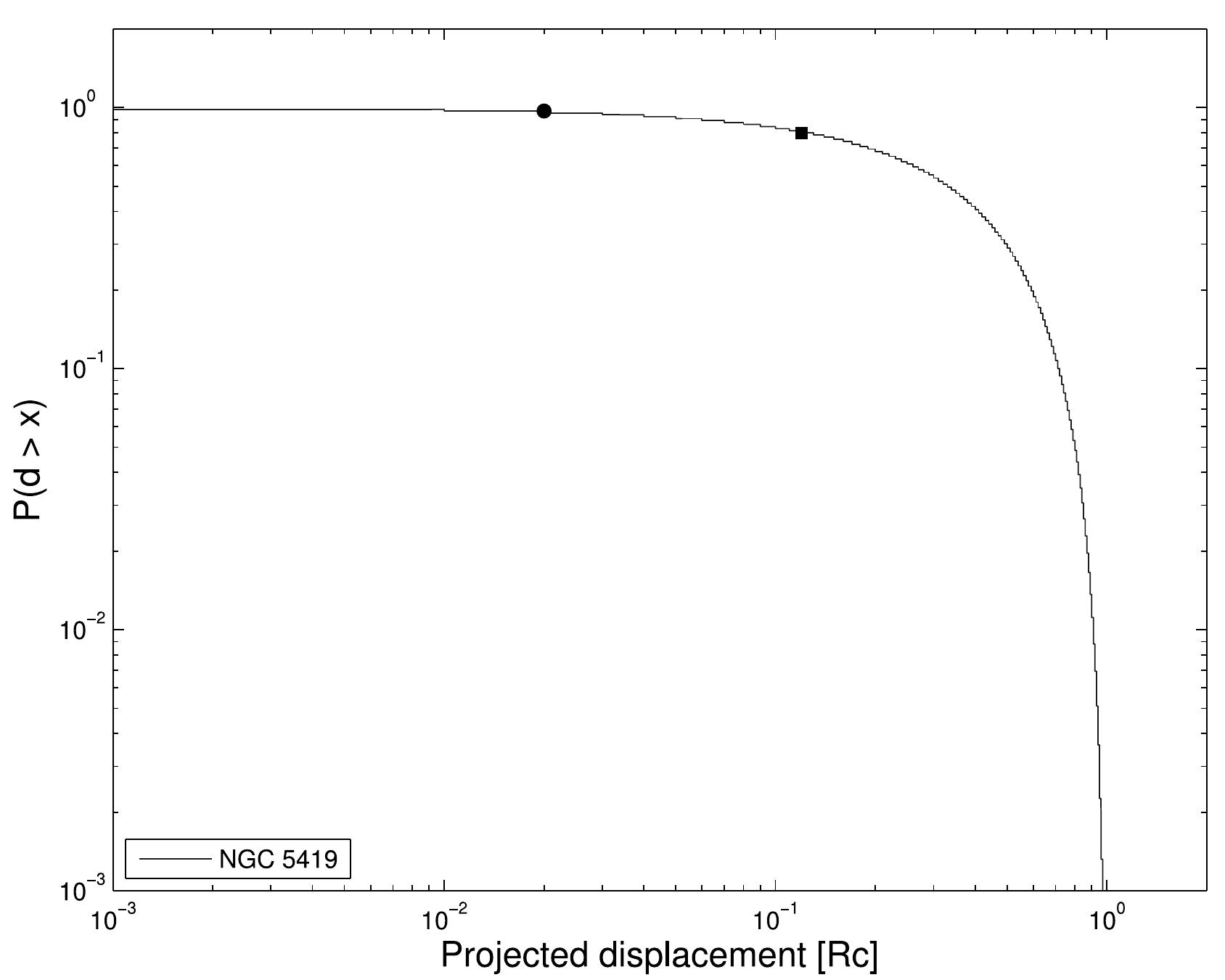} \\
\end{array}$
\end{center}
\caption{Probability to observe a displacement larger than the value specified on the x-axis in units of the core radius for t$_{m}$ = 0.4 Gyr. The observed displacement is marked as a filled circle. The filled square (in NGC 4694 and 5419) marks the offset of the secondary point source.} 
\label{fig: projD}
\end{figure} 
\begin{figure}[p] 
\begin{center}$
\begin{array}{cc}
\includegraphics[scale=0.5]{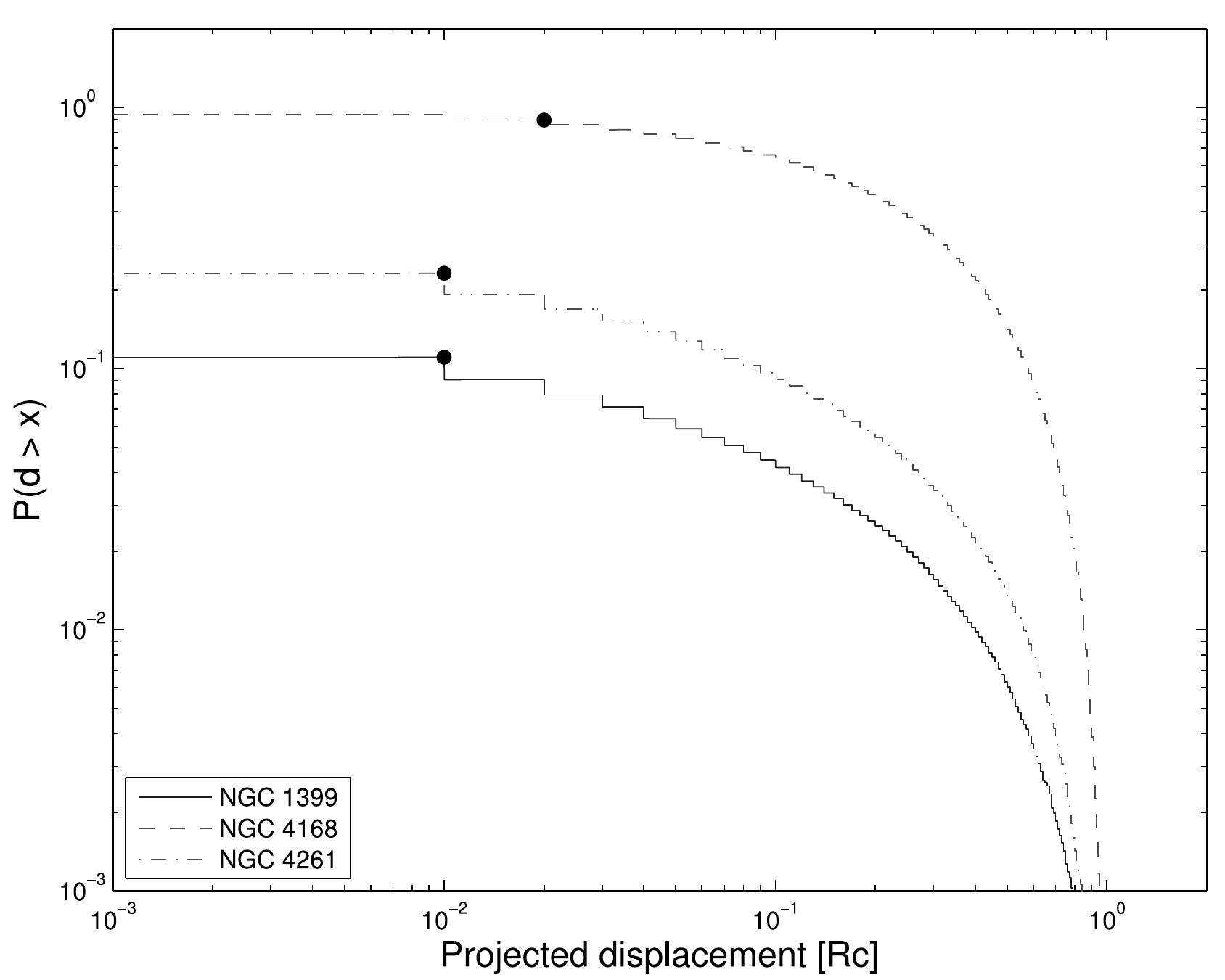} 	& \includegraphics[scale=0.5]{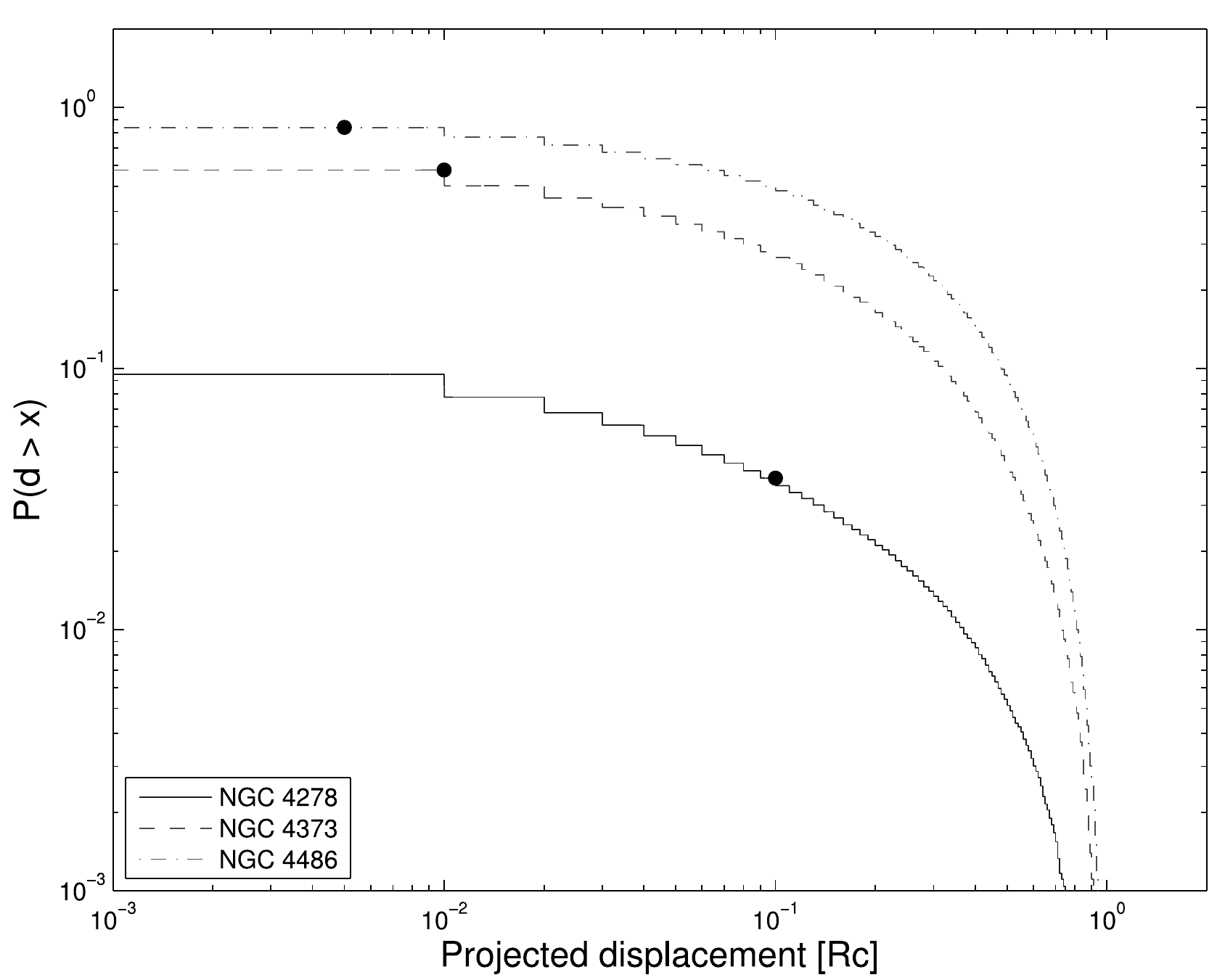} \\
&\\
\includegraphics[scale=0.5]{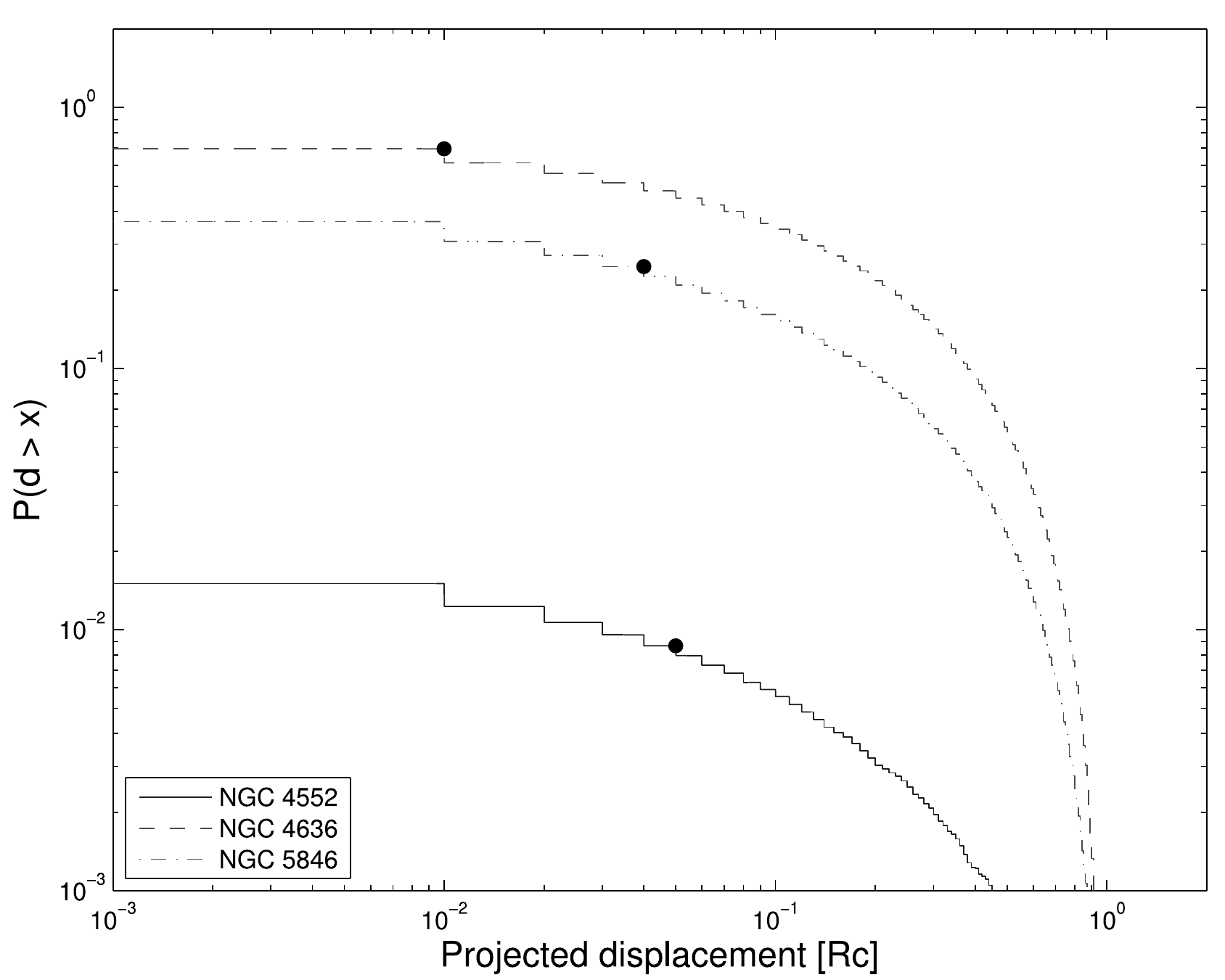} 	& \includegraphics[scale=0.5]{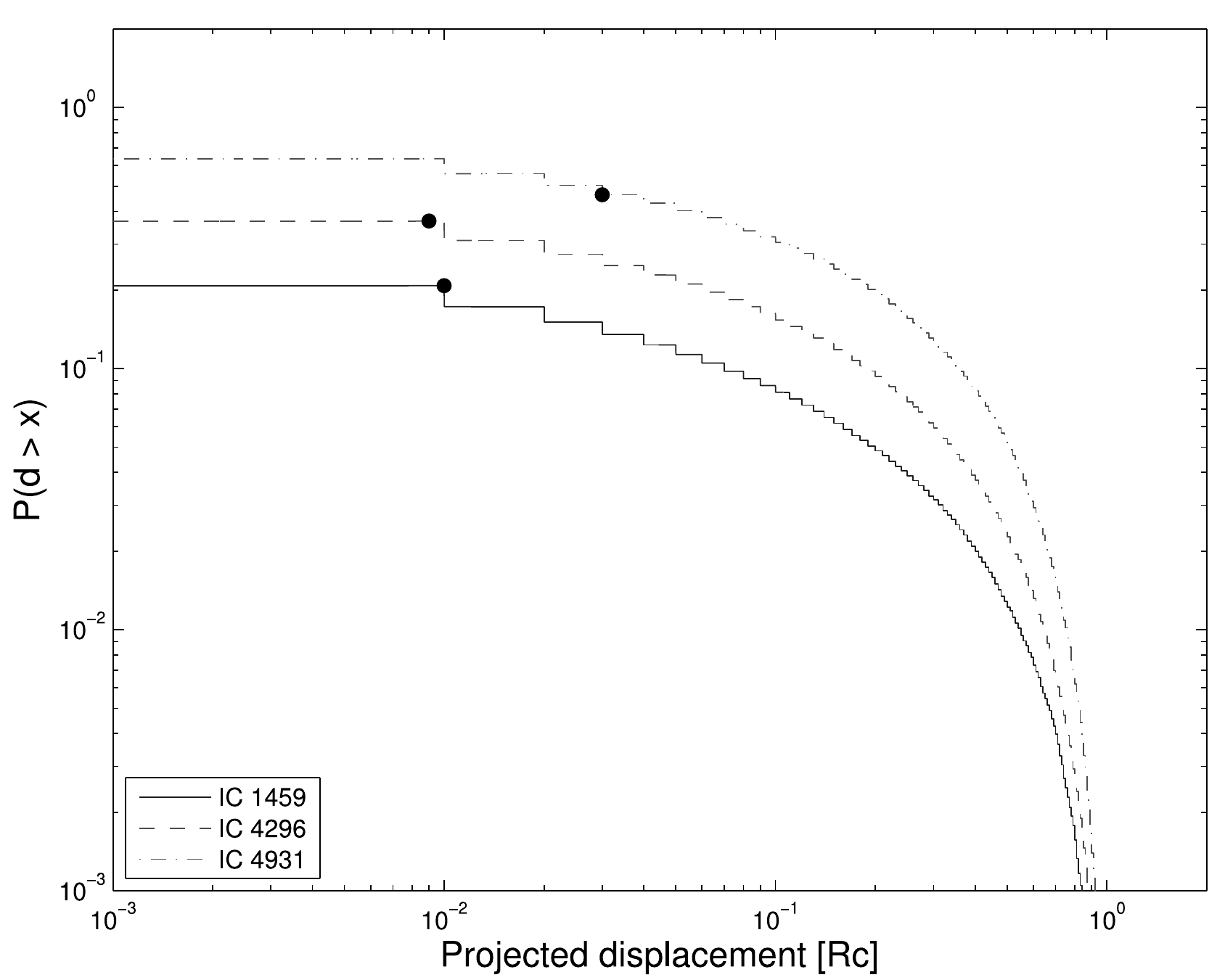} \\
&\\
\includegraphics[scale=0.5]{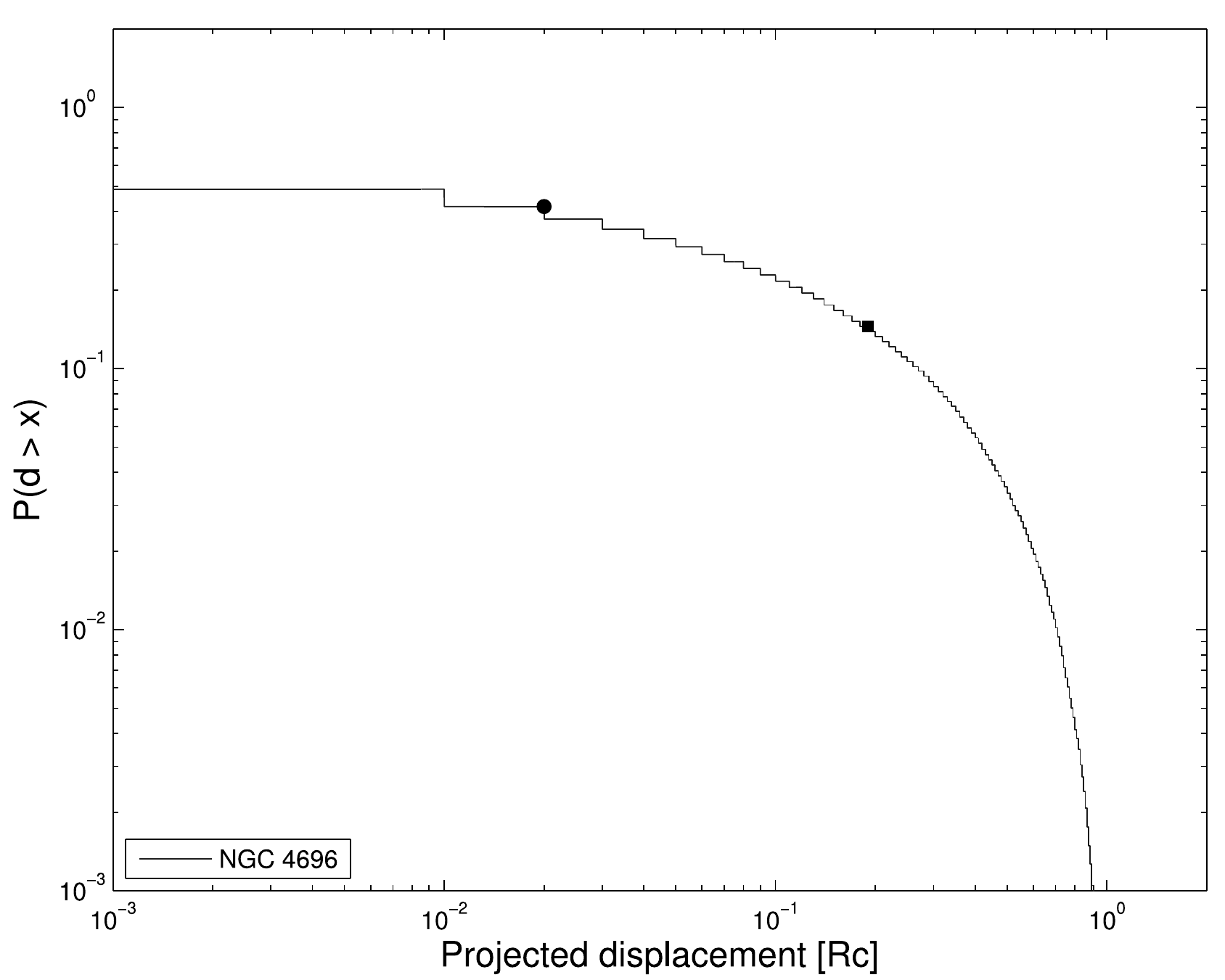} 		& \includegraphics[scale=0.5]{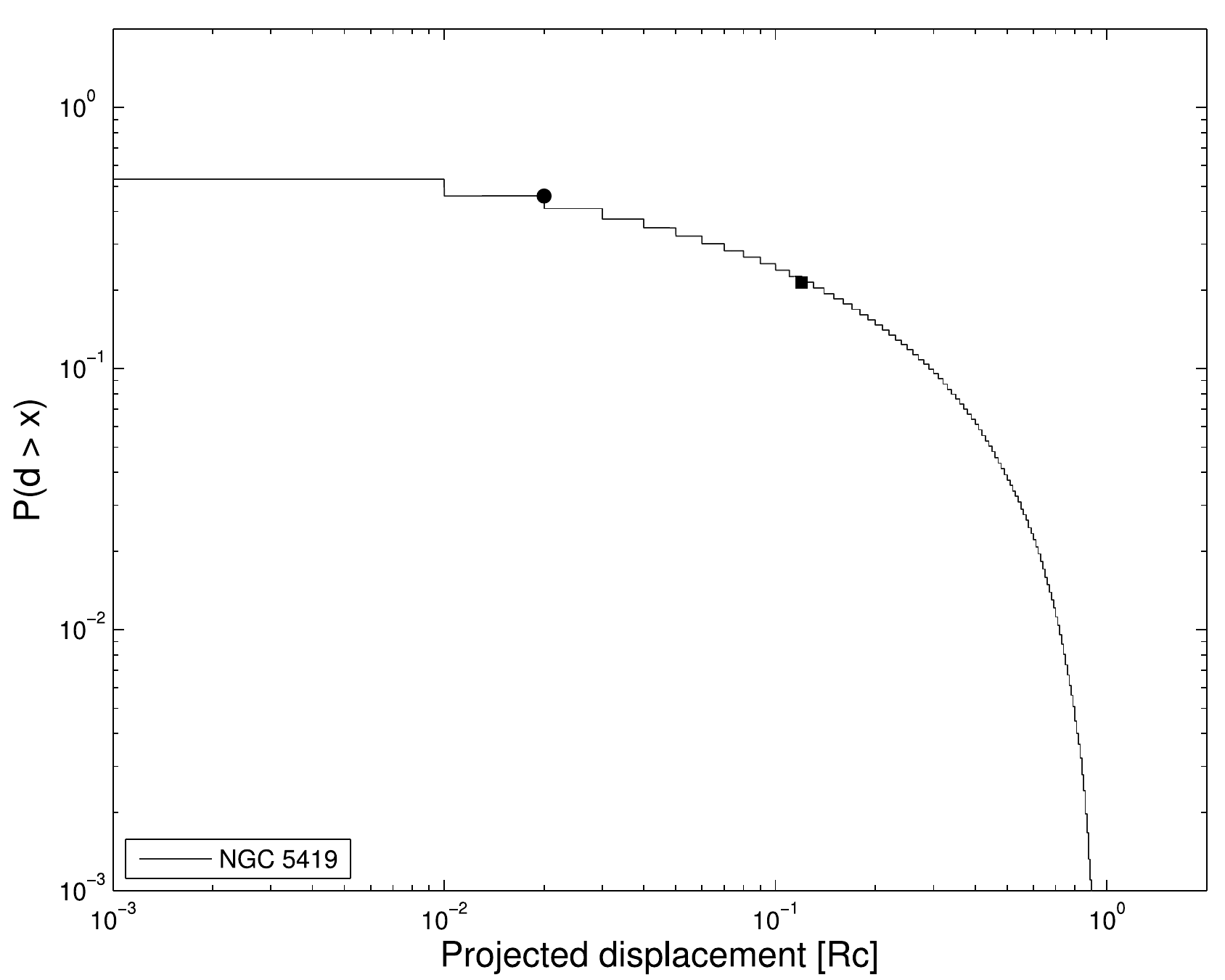} \\
\end{array}$
\end{center}
\caption{Probability to observe a displacement larger than the value specified on the x-axis in units of the core radius for t$_{m}$ = 5 Gyr. The observed displacement is marked as a filled circle. The filled square (in NGC 4694 and 5419) marks the offset of the secondary point source.} 
\label{fig: projDb}
\end{figure}
\end{subfigures}

\begin{table*}[t]
\caption[]{PARAMETERS OF RECOIL OSCILLATION MODEL} 
\label{tab: param} \centering
\scalebox{1}{
\begin{tabular}{ l llllll}
\hline \hline
& &\\

 	Galaxy			&	$r_{c}$ 		& 	$\sigma$ 		& 	log($M_\mathrm{gal}$)	& $\tau$ 			& $2 \pi / \omega_{c}$ 	&	$v_\mathrm{kick}$	\\
					&	 [pc]			& 	 [km s$^{-1}$]	& 	[$M_{\odot}$]			& [Gyr] 			&  [Gyr]				&	[km s$^{-1}$]		\\
	(1)				&	 (2)			&	(3)			& (4)						& (5)				&	(6)				& 	(7)				\\
	& &\\ \hline

	&& &\\
	NGC 1399		&	189			&	346			&	11.70				& 	0.16			& 	0.004			&	300		\\
	NGC 4168		&	303			& 	184	 		&	11.44 				&   	4.85			&	0.012			&	159		\\
	NGC 4261		&	237			& 	309			&	11.85 				&  	0.40			&	0.005			&	273		\\
	NGC 4278		&	83			&	237			&	10.99 				& 	0.14			&	0.002			&	205		\\
	NGC 4373		&	269			&	246			&	12.01 				& 	1.25			&	0.008			&	213		\\
	NGC 4486		&	733			&	334			&	12.00 				&	2.84			& 	0.014			&	325		\\
	NGC 4552		&	36			&	252			&	10.55 				&	0.02			&	0.001			&	218		\\
	NGC 4636		& 	219			&	203			&	11.58 				&	1.73			&	0.008			&	176		\\
	NGC 4696		& 	251			& 	254			&	12.08 				&	0.96			&	0.007			&	220		\\
	NGC 5419		&	499			& 	351			&	12.26 				&	1.09			&	0.01				&	304		\\
	NGC 5846		&	183			& 	239			&	11.83 				& 	0.64			&	0.005			&	207		\\
	IC 1459			&	258			& 	306			&	11.68				&	0.49			&	0.005			&	294		\\
	IC 4296			&	347			& 	333			&	12.17 				&	0.64			& 	0.011			&	196		\\
	IC 4931			&	397			& 	288			&	12.12 				&	1.52			&	0.01				&	249		\\
	&&&&&\\
	average			& 	286			&	277			& 11.73					&	1.20			&	0.007			&	239		\\
	&&&&&\\
\hline
\end{tabular}}
\tablecomments{(1) Galaxy name; (2) core radius from CB05; (3) stellar velocity dispersion from Hyperleda; (4) stellar mass of the galaxy; (5) dynamical time; (6) damping time; (7) kick velocity required to displace the SBH up to the core radius. The dominant source of errors in the quantities presented above is the stellar velocity dispersion.}
\end{table*}

\begin{table}[h]
\caption[]{SBH MASSES} 
\label{tab: mass} \centering
\scalebox{1}{
\begin{tabular}{ l ccccc }
\hline \hline
&\\
Galaxy			&	\multicolumn{4}{c}{\underbar{\hbox to 220pt{\hfill log(M$_{\bullet}$/M$_{\odot}$)\hfill}}} \\ 
				&	M$_{\bullet}$-$\sigma$			&	M$_{\bullet}$-M$_{V}$	& stars				& gas 		\\
(1)				& 	(2)							& 	(3)					& (4)					& (5)\\				
&\\\hline
 &\\
NGC 1399		&	9.38							&	8.71					& 8.64$^{a}$, 9.01$^{b}$	& 			\\
NGC 4168		&	8.04							&	8.56					& 					& 			\\	
NGC 4261		&	9.14							&	8.81					& 					& 8.69$^{c}$	\\
NGC 4278		&	8.58							&	8.17					& 					&			\\
NGC 4373		&	8.66							&	8.72					& 					& 			\\
NGC 4486		&	9.30							&	9.04					& 9.74$^{d}$			& 			\\
NGC 4552		&	8.71							&	8.49					& 					& 			\\
NGC 4636		& 	8.25							&	8.60					& 					& 			\\
NGC 4696		& 	8.72							&	9.90					& 					& 			\\
NGC 5419		&	9.41							&	9.39					& 					& 			\\
NGC 5846		&	8.60							&	8.64					& 					& 			\\
IC 1459			&	9.12							&	8.94					& 9.40$^{e}$			& 8.53 - 8.98$^{e,\dagger}$, 8.15 - 8.75$^{f}$			\\
IC 4296			&	9.30							&	8.89					& 					& 			 \\
IC 4931			&	8.99							&	9.45					& 					& 			\\				
 				&    \\
average			& 	8.87							&	8.88\\
&\\				
\hline
\end{tabular}}
\vskip5pt
\tablecomments{(1) Galaxy name, (2) masses from the $M_{\bullet}-\sigma$ \citep{FerrareseFordRev05} using the velocity dispersion specified in Table \ref{tab: param}, (3) masses from the $M_{\bullet}$-$M_{V}$ relation from eq.6 in \cite{lauer2007}; values for M$_{V}$ have been taken from \cite{lauer2007}, when not available (NGC 4373, 5846 and IC 4296) they have been taken from NED \citep[$V_{\mathrm{T}}$ values from the RC3 catalog,][]{deVac91}, (4) masses from stellar orbit modeling, (5) masses derived from gas kinematics. Masses have been scaled to the distances assumed in this work. $^{\dagger}$The authors warn that these values might be affected by non-gravitational motions in the gas (i.e. inflows or outflows). \textbf{References}: $^{a}$\cite{gebhardt2007}, $^{b}$\cite{hough2006}, $^{c}$\cite{ferrarese1996}, $^{d}$\cite{Geb11}, $^{e}$\cite{cappellari2002}, $^{f}$\cite{verdoes2000}.}
\end{table}

\section{B: Notes on individual sources}
\label{app: galaxies}
 
In this section, the galaxies are ordered according to the significance assigned to the displacement.

\subsubsection{NGC 4278  (high/intermediate)}

NGC 4278 is a large elliptical near the center of the dense Coma I cloud which is characterized by the presence of an extended dust distribution extending over an arc ranging from N to NW relative to the photocenter \citep{Carollo97}. Radio interferometry (VLBA) observations reveal a low-power, parsec-scale source  ($\sim 45$\,mas) with twin mildly-relativistic S-shaped jets emerging from a compact, flat spectrum core \citep{Giro2005}. Within the inner 7 mas the jets are oriented SE--NW but then bend to the east and west, respectively, forming an ``S'' shape. 
\citeauthor{Giro2005} argue that the jets are closely aligned with the line of sight (with a viewing angle $\sim 2 - 4^{\circ}$) with the NW jet approaching and the bends 
being explained by amplification of intrinsically small deviations due to projection effects.
These characteristics and the presence of atomic hydrogen suggest that the galaxy may have experienced a minor merger with another member of the cloud \citep{GerinCasoli94}.   

\textbf{Photometric analysis.} Three images, obtained with ACS/WFC-F850LP, WFPC2/PC-F814W and NICMOS2/F160W, respectively, were analyzed for this galaxy (Figs~\ref{fig: NGC4278_9p_850}, \ref{fig: NGC4278_9p_814} and~\ref{fig: NGC4278_NIC2}). Photocenter--AGN displacements of intermediate or high significance are detected in all three. The uncertainty-weighted average displacement has an amplitude $\Delta$r $\geq 7.6 \pm 0.4$ pc and direction SE (PA = 152 $\pm$ 3$^{\circ}$). The photocenter positions derived from each image are plotted relative to the AGN position in Fig.\ref{fig: multif}. The photometric analysis of the optical ACS and WFPC2 images shows that the $y$ co-ordinate of the isophote centers migrates southward at intermediate SMA, then northward at larger SMA. These irregularities are not present in the NICMOS2 image and are likely due to the dust N-NW of the center.

\textbf{Offset origin.} The extended dust distribution produces a large-scale asymmetry in surface brightness, which is particularly evident in optical and UV wavelengths. This may
bias the photocenter, shifting it in the opposite direction (S-SE) to the dust. As the AGN is a point source, its position should not be significantly affected by the dust distribution and thus it would appear to be shifted N-NW relative to the photocenter. Therefore, the offsets recovered from ACS and WFPC2 data may be partly due to asymmetric extinction. However, the effects of dust are not apparent in the NIR (NICMOS2) image, from which an offset of similar magnitude and direction was recovered. Therefore, we believe that the measured displacement is not simply an artifact of extinction, although this may contribute.

Relative to the core radius, NGC 4278 has the largest displacement ($\approx 0.1r_c$; Table~\ref{tab: probabilities}) in our sample. 
If it is due to phase II oscillations of the SBH following gravitational recoil
(see \textsection \ref{sec: recoiling_SBH}), we estimate that the probability of observing a projected displacement larger than that actually measured is 4 (36)\% for a mean time-between-mergers of $t_m = 5\ (0.4)$ Gyr (Table~\ref{tab: probabilities}) and given a kick large enough to displace the SBH at least as far as the core radius. The ``hot'' disk recoil velocity distribution of L12 indicates a 45\% probability of a sufficiently large kick ($\gtrsim 200$\,km\,s$^{-1}$, Table~\ref{tab: param}) occurring in an SBH coalescence event. Therefore, we consider residual oscillations following gravitational recoil to be a plausible explanation for the displacement.

In this case, we can estimate the merger epoch. Supposing the SBH to be undergoing phase II oscillations, we combine eq.\ref{eq: tdamp} and \ref{eq: rms}. Using the measured (projected) offset
as a lower limit for $r_\mathrm{rms}$, $r_\mathrm{rms}\gtrsim 7.6$ pc, and taking the values of $r_c$ (core radius) and $\sigma$ (stellar velocity dispersion) from Table~\ref{tab: param}, we find that $(t - t_{c}) \lesssim 5 \tau \approx 0.7$ Gyr  has elapsed since the amplitude of the oscillation damped down to $r \sim r_{c}$. This implies that the onset of phase II occurred no later than $\sim 0.7$ Gyr ago. As phase I is shorter than phase II,  this scenario is consistent with an SBH-SBH coalescence event that occurred within $2 (t - t_{c})\sim$ 1.4 Gyr. 
The total evolution time from the beginning of the galactic merger to SBH binary coalescence is suggested to be of order $10^8$ yr \citep{GualandrisM11}.  Adding this time to the estimated kick epoch, we infer that a galaxy merger that took place up to $\sim$ 1.5 Gyr in the past could be responsible for a recoil kick consistent with the observed displacement. 

Although the displacement is approximately in the direction of elongation of the mas-scale radio core (155$^{\circ}$ over the inner $\sim 7$ mas), the low power (P$_\mathrm{5GHz}\approx70\times 10^{20}$ W, \citealt{wrobel91}) and compact nature ($\sim 3$ pc) of the radio source argue against jet thrust as the cause of the offset.

We find no compelling reasons to exclude either a stalled binary on an eccentric orbit, or massive perturbers (Section~\ref{subsec: disp_mech}) as possible causes of the displacement.


\subsubsection{NGC 5846 (intermediate)} 

NGC 5846 is the dominant component of a small, compact group of $\approx 50$ galaxies \citep{HG82}. A companion galaxy, NGC5846A, is located 0.7 arcmin to the S, but there is no evidence of
an interaction. The presence of dust lanes in an apparent spiral pattern approximately centered on the compact radio core suggests a past merger, or accretion of a gas rich galaxy \citep{Moell92, Forbes96}. 
The optical nucleus has a LINER spectrum \citep{Ho97} and HST imaging  \citep{Masegosa2011} reveals extended H${\alpha}$ emission in a wide structure extending up to 2\arcsec\ W of the nucleus. The presence of two radio emitting bubbles located roughly symmetrically at $\sim 0.6$ kpc from the center and a ``ghost'' X-ray cavity suggest that the galaxy recently experienced stronger AGN activity, for which \citet{Machacek11} derive a duty cycle of $\sim 10$ Myr. The radio morphology is complex: \citet{FilhoFM2004} report VLBA observations at 2.3, 5 and 15 GHz which resolve the core into multiple blobs roughly aligned along the N--S axis. The nature of this structure is unclear,  possibilities include an AGN jet, or compact supernovae remnants \citep[e.g.][]{tarchi2000}.

\textbf{Photometric Analysis.}  One image, obtained with WFPC2/PC - F814W was analyzed for this galaxy (Figure~\ref{fig: NGC4636_WFPC2F814W}). We find a photocenter--AGN displacement $\Delta r \gtrsim 8.2\ \pm\ 2.5$ pc in the direction W-SW (PA = $253\ \pm\ 8^{\circ}$)
which is considered to be of intermediate significance. The isophote centers show systematic migrations of $\approx 2$ pixels in both $x$ and $y$ coordinates,
from W to E and S to N, respectively, as the SMA decreases from 300 to 200 pixels. These migrations are possibly the result of the dust lanes to the S and NE of the nucleus. 
The $y$-component of the displacement is not significant. The $x$-component is considered to be of intermediate significance (1.6 IQR) and is unlikely to be an artifact of extinction, which in this case, would bias the photocenter such as to increase the $x$- displacement. Of more concern is the determination of the AGN position. The central source is relatively weak and embedded in an asymmetric structure that affects the gaussian fits used to determine the AGN position. We therefore assign a larger uncertainty to the $x$-component of the AGN position. 

\textbf{Offset origin.} 
For $t_m = 5\ (0.4)$\,Gyr, we estimate a 25 (90)\% probability of finding a displacement larger than that observed, if due to phase II recoil oscillations. 
The probability of a recoil kick large enough ($\gtrsim$ 240 km s$^{-1}$; Table \ref{tab: param}) to initiate these oscillations is $\sim$45\% (L12).

As for NGC 4278, we use equations \ref{eq: tdamp} and \ref{eq: rms} and values from Table \ref{tab: param} to estimate the time since the onset of phase II, finding (t-t$_c$) $\lesssim$ 6.2 $\tau$ or 4 Gyr. This implies an upper limit to the time elapsed since the putative SBH-binary coalescence event of 2(t-t$_c$) or 8 Gyr.  

Although, as noted, there is ``fossil'' evidence for previous jet activity, the AGN does not currently produce a powerful radio jet, so it seems unlikely that the displacement is due to jet acceleration.
However, interactions with massive perturbers or an SBH binary on a highly eccentric orbit remain possibilities.


\subsubsection{NGC 4486 (M87, intermediate)}
NGC 4486 (M87) is the dominant galaxy of the Virgo Cluster. It hosts the best studied extragalactic jet, which exhibits prominent knots at optical and IR wavelengths yet, apart from this, the brightness profile is regular and featureless. The peculiarly large population of globular clusters, its size and location, and the presence of stellar streams support the hypothesis that the galaxy has experienced a number of minor mergers during its lifetime \citep{Janowiecki2010}. A remarkable feature is the unusually large core, which has a radius $r_{c} \sim 9\farcs41$ (CB05). This characteristic makes M87 a special case in our sample in that for all the other galaxies we restricted the isophotal analysis to radii larger than the core radius, whereas for M87 we analyzed isophotes within the core radius in order to be consistent with B10. 

\textbf{Photometric analysis.}  A total of six images obtained with ACS/HRC, WFPC2/PC and NICMOS2 were analyzed (see Table~\ref{tab: distanze}), including the combined ACS images (ACS/HRC/F606W and ACS/HRC/F814W) previously analyzed by B10. The results of the photometric analysis are shown in Figs.~\ref{fig: M87_ACS_F814W} and \ref{fig: NGC4486_ACSF606W}--\ref{fig: M87_WFPC2F814W} and the measured displacements relative to the AGN are plotted in Fig.~\ref{fig: multif}. In all images, we analyzed isophotes within the range $1\arcsec \leq r \leq 3\arcsec$, for consistency with B10. The displacements measured in different images vary in amplitude and significance. Notably, none of the three NICMOS2 images (F110W, F160W and F222W) exhibit significant displacements at the $\ge 3\sigma$ level. Furthermore, the $1\sigma$ uncertainties, when converted to linear distances, do not encompass the significant displacements recovered from the optical images. The displacements measured from the ACS and WFPC2 images also differ in amplitude but there is a tendency for the photocenter to be displaced NW of the AGN, roughly in the direction of the radio jet. 
The isophote centers are reasonably stable -- although there are deviations $\sim 1-2$ pixels  ($\le 0\farcs1$), systematic trends with SMA are not generally evident (Figs.~\ref{fig: M87_ACS_F814W} and \ref{fig: NGC4486_ACSF606W}--\ref{fig: M87_WFPC2F814W}). The most significant displacement is obtained from the ACS/HRC F814 image, for which the isophote centers cluster tightly around the mean photocenter (Fig.~\ref{fig: M87_ACS_F814W}). The measured displacement of the photocenter relative to the AGN is 
 $\Delta r  = 7.7 \pm 0.3$ pc NW, implying that the SBH is displaced by $\Delta r$ in the counter-jet direction. This result is in agreement with the displacement obtained by B10 from a similar analysis. The main differences between the method used by B10 and that employed in this paper are the iterative technique used here to generate the mask and the weight function used to derive the isophotal center: while B10 used a conventional uncertainty weight, here we weight for the light content of each isophotal annulus (see \textsection \ref{subsec: photo}). 
 
The weighted average displacement has a magnitude $4.3\pm0.2$ pc and direction NW (PA = $307\pm1^{\circ}$). 

\textbf{Offset origin.} The presence of a large number of bright globular clusters visible in all the images analyzed allows us to define a common reference frame which we can use to compare results from images obtained with different instrument/filter combinations. 
After selecting several globular clusters as reference points and matching their positions, we plotted the AGN and photocenter positions determined from the different images in the reference frame defined by the ACS image (Fig.~\ref{fig: M87_PC_AGN_match}). The AGN position as measured from the different images is fairly stable whereas the photocenters exhibit differences in position much larger than the scatter ($\sim 0.5$\,pixels) in the AGN position. This indicates that the differences in the displacements are due largely to differences in the positions of the photocenters. The origin of these discrepancies is unclear. Unlike other galaxies, there is no evidence of dust extinction or other asymmetries that may account for different results at different wavelengths.

We conducted several trials in order to determine the influence of (i) the degree of masking applied to the jet, (ii) the bright knot visible NW of the nucleus in the ACS image. 
The results were found to be independent of the degree of masking; no significant differences in the photocenter positions were found when the analysis was performed with no masking at all or with heavy jet/knot-masking. 

As a fraction of the rather large core radius, the weighted mean displacement is only $\approx 0.01\ r_c$ (Table~\ref{tab: probabilities}).
We estimate a 85 (99)\% probability of finding a displacement larger than that observed for $t_m = 5\ (0.4)$\,Gyr, assuming phase II recoil oscillations were triggered. 
The probability of obtaining the relatively large recoil kick  ($\gtrsim 325$\,km\,s$^{-1}$; Table~\ref{tab: param}) required for phase II oscillations is $\approx 33$\% (L12). 

The apparent close alignment of the offset with the jet direction favors jet acceleration but is also generally consistent with gravitational recoil. However, as there is no reason to expect
such an alignment if the displacement is due to gravitational interactions with massive perturbers, or orbital motion in a SBH-binary, these mechanisms are disfavored. A detailed discussion of the possible displacement mechanisms is given in \textsection 4 of B10.

\begin{figure}[t]
\centering
\includegraphics[ width=2.7 in, trim = 0cm 1cm 0cm 0cm, clip,angle=-90]{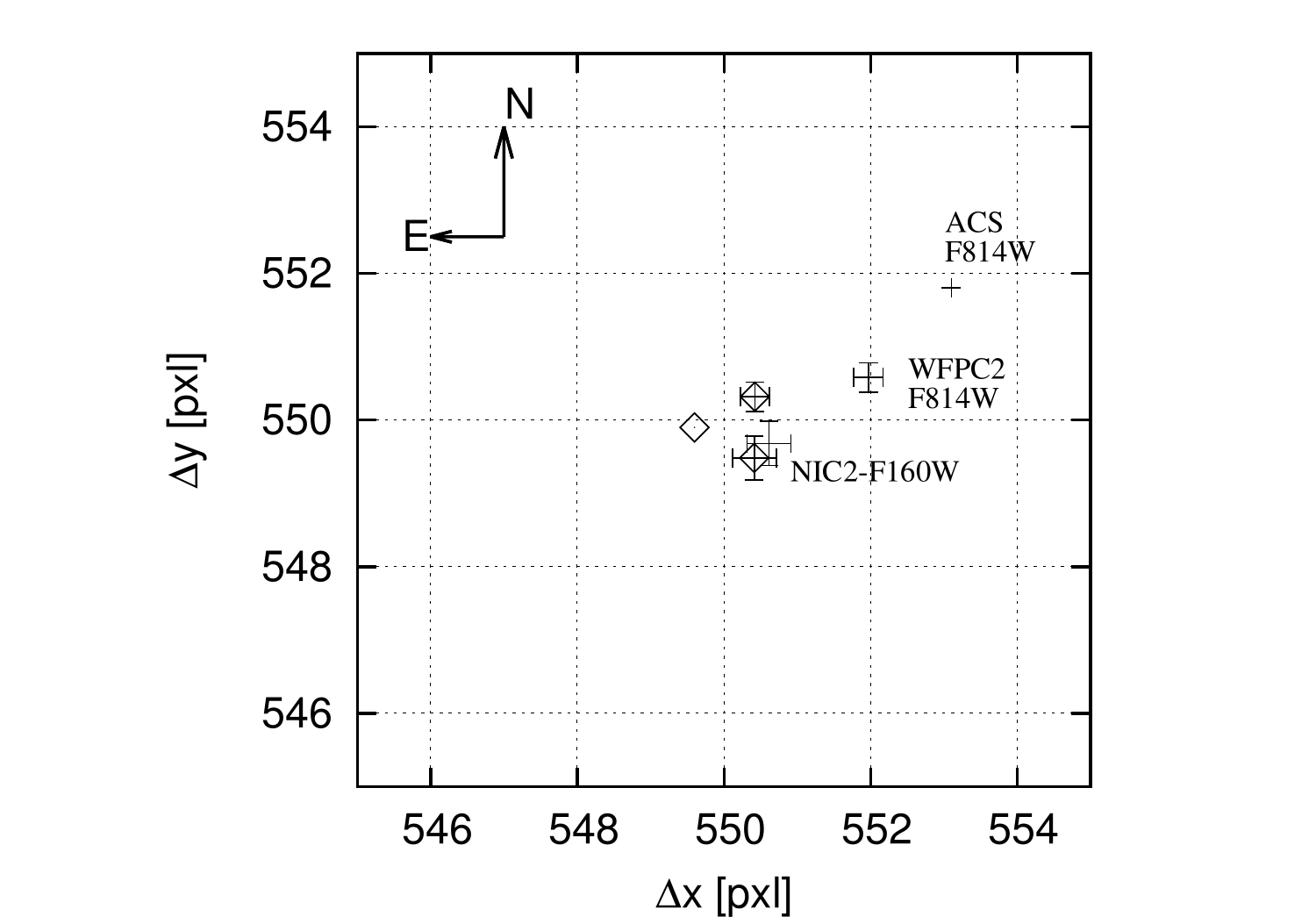}
\caption[fig: M87_PC_AGN_match]{M87 - Using a set of bright globular clusters visible in all the frames, we matched the AGN and isophotal center position with the coordinate system defined by the ACS frame. The result shows that while the AGN (diamonds) remains fairly stable, the isophotal center (crosses) has the larger scatter accounting for the discrepancies in the recovered offsets. The smaller error is associated with results from ACS, the larger with results from NICMOS. Results from only three images out of six are shown to make the image clear.}
\label{fig: M87_PC_AGN_match}
\end{figure}

\subsubsection{NGC 1399 (low)}

NGC 1399 is a regular giant E1 and is the central and dominant galaxy of the Fornax cluster \citep{Ferguson89}. It is associated with a low power radio source characterized by two nearly bi-symmetric jets extending $\sim 10$ kpc, approximately N and S of the core (PA $\approx 9^{\circ}$ W of N, \citealt{KilleenBE88}). \citet{Shurkin08} report the presence of X-ray cavities associated with the radio lobes. NGC 1399 has been the subject of a number of dynamical studies (e.g. \citet{hough2006} and references therein).
In particular, \citeauthor{hough2006} performed near-infrared adaptive optics assisted spectroscopic observations of the central 1\farcs5 of the galaxy and found evidence of a kinematically decoupled core and offset asymmetric isophotes that they considered to be consistent with the presence of a M31-like double nucleus.
The asymmetric isophotes, elongated in the direction E-SE, are localized in a region much smaller than that used for our isophotal analysis  ($r > 2\farcs5$). 

\textbf{Photometric analysis.} Four images, obtained with WFPC2/PC - F606W, F814W and WFC3/IR - F110W, F160W, were analyzed for this galaxy. From the F606W image we find a low significance displacement of amplitude 1.7 $\pm$ 0.4 pc, in almost the same N-NW direction as the radio jet axis, PA = $-8 \pm 16^{\circ}$. A $2\sigma$ offset in the same direction was recovered from the WFPC2/PC - F814W image, which is consistent within the errors with the F606W result. Offsets recovered from the WFC3 images are classified as non-significant, but, given their large error bars, they are consistent with the WFPC2 results (Fig.~\ref{fig: multif}). The error weighted offset, computed including results from both WFPC2 and WFC3, is 1.5 $\pm$ 0.4 pc with PA = -17 $\pm$ 16$^{\circ}$.
The isophote fitting analysis (Figs.~\ref{fig: NGC1399_9pF606} to \ref{fig: NGC1399_9pF160W}) reveals a remarkably regular photometric structure well modeled by concentric elliptical isophotes. For both images, the isophote center co-ordinates vary little with SMA and are very tightly clustered, with narrow ($\lesssim 0.5$ pixel) distributions in $x$ and $y$.

\textbf{Offset origin.} 
The weighted mean displacement is small compared to the core radius, $\approx 0.01\ r_c$ (Table~\ref{tab: probabilities}). A minimum kick velocity $\gtrsim 300$\,km\,s$^{-1}$ (Table~\ref{tab: param}),
is required to trigger phase II recoil oscillations; the probability of generating a kick of at least this magnitude is $\sim 33$\% (L12).
Assuming that phase II recoil oscillations were triggered, we estimate the probability of finding a displacement larger than that observed to be 11 (75)\%, for $t_m = 5\ (0.4)$\,Gyr. 

If the observed displacement is due to a recoiling SBH, we estimate that (t - t$_{c}$) $\lesssim$ 1.6 Gyr has elapsed since the oscillation damped below the core radius, using $\tau = 0.17$ Gyr and $\sigma = 346$ km s$^{-1}$ (Table \ref{tab: param}). 

The close alignment of the displacement with the radio axis is consistent with both recoil and jet acceleration as the cause of the displacement.
If the latter, eq.~\ref{eq: dr_norest} implies that, for a Bondi accretion rate $\dot M_{a} \sim 0.04$ M$_{\odot}$ yr$^{-1}$ \citep{LMAAQ2001} and a 10$^{9}$ M$_{\sun}$ black hole,
an asymmetry in the radio jet power in the range

\begin{align*}
3 \times 10^{-4} \lesssim f_\mathrm{jet} \lesssim 3 \times 10^{-3}
\end{align*}

\noindent that persists for $t \sim 10^{8}$ yr  would be needed to produce an offset 1.8 $\lesssim \Delta r \lesssim$ 18 pc. 

Other mechanisms (interactions with massive perturbers and a stalled SBH binary), are disfavored by the alignment between the displacement and jet direction.
The presence of an unresolved M31-like double nucleus as suggested by \citet{hough2006} might also explain the measured offset, although as in the case of a
binary SBH, the alignment with the radio source axis would have to be regarded as fortuitous.

\subsubsection{IC 4296 (low)}

The properties of this galaxy have been extensively studied in a series of papers by \citet{KBC86a, KBC86b} and \citet{KillBick88}. It is a fairly circular elliptical which is the brightest component in the small group A3565. It is associated with the extended low-to-intermediate luminosity radio source PKS 1333-33 \citep{Mills60}, which has two quasi-symmetric slightly curved jets with an average position angle  
$\sim 130^{\circ}$ and terminating in lobes  $\sim 200$\,kpc from the core \citep{KBC86a}. The curvature along the jets is consistent with the motion of the galaxy through the group inter-galactic medium \citep{KillBick88}. On the parsec scale, a jet and counter-jet emerge from the unresolved core in PA $\approx 140^{\circ}$,
consistent with the large-scale radio source, with the brighter jet on the NW side. The brightness asymmetry is attributed to Doppler boosting \citep{Pellegrini2003}. 
Isophotal studies suggest tidal interactions with the smaller companion IC 4299 \citep{Younis85}. Nevertheless, \citet{Efstathiou80} show that the isophotal ellipticity and position angle are constant at all radii from 2.4 to 48.8 arcsec (a region much larger than that studied in this work, $1\farcs5 < \Delta r < 6\farcs5$), which suggests that any distortions due to interactions with companion galaxies are negligible over this scale. 

\textbf{Photometric analysis.} Three images, obtained with ACS/HRC-F625W, WFPC2/PC-F814W and NICMOS2-F160W, were analyzed for this galaxy. 
In the ACS image a displacement of $\Delta r \geq 5.2 \pm 1$ pc was found. On
the basis of the IQR, that is considered to be of low significance. The displacement obtained from the NICMOS2 image is not considered significant, but is consistent with the ACS result, albeit within large errors.
For the WFPC2 image, the $y$-component of the displacement is consistent with the ACS result but the $x$ component exhibits a marginally significant difference 
and in fact indicates a displacement in the opposite direction (E). However, neither component of the WFPC2 displacement is classified as significant.
These results are summarized in Fig. \ref{fig: multif}.

The discrepancy in the photocenter positions appears to be linked to systematic migrations in the isophote center co-ordinates determined from the fits in the NICMOS2 and WFPC2 images:
there are systematic migrations northwards in the NICMOS2 image (Fig. \ref{fig: IC4296_NIC2}) and eastwards in the WFPC2 image (Fig.~\ref{fig: IC4296_WFPC2}) resulting in
distributions with large IQRs in the $y$ and $x$ co-ordinates, respectively. In contrast, the co-ordinates extracted from the ACS image (Fig.~\ref{fig: IC4296_ACS}) show little variation with SMA and cluster quite symmetrically around the photocenter. 

This galaxy exhibits a warped dusty disk, which crosses the nucleus in a roughly E-W direction and is clearly visible in all three images (Figs.~\ref{fig: IC4296_ACS}, \ref{fig: IC4296_WFPC2} and 
\ref{fig: IC4296_NIC2}). This feature is too small ($r \lesssim 0\farcs9$) to affect the photocenter position, which is calculated from isophote fits over radii $1\farcs5 \leq r \leq 6\farcs5$.
However, it does create large gradients in the background brightness around the nucleus which affect the gaussian fits to the central peak, especially when large fitting-boxes are used. 
The effects are particularly noticeable for the WFPC2 image, where the AGN position systematically shifts South-East as the size of the fitting-box is increased. 
This was mitigated by reducing the size of the fitting-box until stable values were found for the AGN co-ordinates. 

The average error-weighted offset obtained for this galaxy from the ACS, NICMOS2 and WFPC2 images is $\Delta r \geq 3.8 \pm 0.7$ pc in PA = -22 $\pm$ 7$^{\circ}$ E of N.

\textbf{Offset origin.} The status of IC 4296 as the brightest member of its group and the fact that it has a kinematically peculiar core \citep{KBC86a} suggest that this galaxy has experienced minor mergers during the past few Gyr. 

The weighted mean displacement is a small fraction of the core radius, $\approx 0.01\ r_c$ (Table~\ref{tab: probabilities}). The probability of generating a recoil kick 
large enough to trigger phase II  recoil oscillations  ($\gtrsim 200$\,km\,s$^{-1}$; Table~\ref{tab: param}),
is  $\sim 45$\% (L12) and given that phase II oscillations were triggered, we estimate a 37 (97)\% probability of finding a displacement larger than that observed, for $t_m = 5\ (0.4)$\,Gyr. 

Using eqs.~\ref{eq: tdamp} and \ref{eq: rms} with $\sigma=333$ km s$^{-1}$, $r_\mathrm{rms} \ge 3.8$ pc and $r_{c} = 347$ pc, we find $(t-t_{c}) \lesssim 9\ \tau \approx 5.8$ Gyr. In this case, the low density of the large core is responsible for the long damping time. 

It has been pointed out by \citet{KBC86a} that on kpc scales the radio jets exhibit transverse oscillations with wavelengths 
$\sim 10$\,kpc and an amplitude comparable with the jet width. Similar oscillation wavelengths are expected from precession due to orbital motion in a $\sim pc$ scale SBH binary (see Section~\ref{subsec: disp_mech}). However, \citet{KillBick88} found that, in detail, the oscillation pattern is not consistent with precession and favor instead helical Kelvin-Helmholtz instabilities. 

If we disregard the discrepant WFPC2 result, the direction of the weighted mean displacement derived from the NICMOS2 and ACS images is PA = $132 \pm 8^{\circ}$. In projection, therefore,
the displacement is remarkably well-aligned with the radio source axis, suggesting jet acceleration as a plausible candidate for the displacement mechanism.
\citet{Pellegrini2003} infer a viewing angle $\theta = 63.5 \pm 3.5\arcdeg$ for the parsec-scale jets. Assuming that the 
displacement is in the same direction in space as the jet, the de-projected amplitude is $\Delta r_\mathrm{jet} = 4.3 \pm 3$ pc. 

Given that the galaxy is characterized by a low density core, taking the age of the radio source to be $t \sim 10^{8}$ yr \citep{KillBick88}, $\dot M_{a} = \dot m_\mathrm{Bondi} \sim 0.02$ M$_{\odot}$ yr$^{-1}$ \citep{Pellegrini2003}, $M_{\bullet}\sim 10^{9}$ M$_{\odot}$ from the $M-\sigma$ relation and $3 \lesssim \Delta r \lesssim 30$ pc, we find from eq.~\ref{eq: dr_norest} that
\begin{align}
10^{-3} \lesssim f_\mathrm{jet} \lesssim 9.7 \times 10^{-3}.
\end{align}

Therefore, an asymmetry in jet power amounting to $\lesssim 1 \%$ of the accretion luminosity could account for the observed offset if it persists for $\sim10^{8}$ yr.

The close alignment with the radio jet axis argues against the displacement being caused by gravitational interactions with massive perturbers.

\subsubsection{NGC 5419 (low, double nucleus)}

This galaxy is the dominant member of the cluster Abell S753 \citep{AbellCO89}. A low surface brightness X-ray halo is centered on the galaxy and extends over a radius $r \approx 16 '$ (190 h$^{-1}$ kpc). An unusual diffuse radio source, PKS B1400-33, is associated with the cluster, but is offset relative to the X-ray emission and NGC 5419 \citep{SubraEtAl03}. The galaxy is
located just beyond the NW edge of the extended radio emission and itself is associated with a bright, compact radio source. 
The low surface brightness and steep spectral index of the diffuse source suggest that it is a relic, with a spectral age of $5 \times 10^{8}$ yr. However, it is unclear if the relic 
was created by an earlier episode of powerful activity in NGC 5419, or whether it is a relic lobe injected into the cluster by a previously active double radio source located outside the cluster (the scenario favored by \citeauthor{SubraEtAl03}, although an optical counterpart has not been firmly identified). If the relic originated from NGC 5419, the spatial offset may be due to proper motion of the galaxy or to buoyancy of the synchrotron plasma \citep[e.g.,][]{McNamara01}.

\textbf{Photometric analysis.}
The galaxy has a double nucleus \citep{lauer, CapettiB06}, which is well-resolved in the WFPC2 planetary camera image (WFPC2/PC-F555W). The two nuclei are separated by $0\farcs27$, with the fainter (secondary) nucleus located almost at the south of the brighter (primary) nucleus. The photocenter is offset by $\Delta r_{1} \gtrsim 7.5 \pm 1.7$ pc W-SW (PA $=252\pm12^{\circ}$) from the primary and $\Delta r_{2} \gtrsim 62 \pm 2$ pc N-NW (PA = $346\pm2^{\circ}$) from the secondary nucleus.

A periodic pattern is visible in the residual image obtained after subtraction of the photometric model (middle panel in Fig.~\ref{fig: NGC5419_WF}). This is likely to be due to the pronounced ``boxiness'' of the isophotes, which could not be perfectly reproduced even though higher harmonics were included in the model. The isophote center co-ordinates cluster fairly tightly around the photocenter, which is closest to the primary nucleus, and generally show little variation with SMA (Fig.~\ref{fig: NGC5419_WF}), with the exception of a small migration ($\approx$ 0\farcs05) eastward of the photocenter in the inner 170 pixels (8\farcs5).  

\textbf{Offset origin.} The projected separation between the two nuclei ($\Delta r \approx 75$ pc) is slightly smaller than the estimated SBH influence radius ($r_{h} \approx 90$ pc) and
much smaller than the core radius ($r_{c} \approx 500$ pc; Table~\ref{tab: distanze}). Hence, it is possible that the 
secondary nucleus is a second SBH, or even the nucleus of a satellite galaxy, on a slowly-evolving orbit in the low-density core \citep{FabioAndDavid11}. 

The displacement of the photocenter relative to the primary nucleus is a small fraction of the core radius, $\approx 0.02r_c$ (Table~\ref{tab: probabilities}). 
The probability of generating a recoil kick velocity large enough to trigger phase II recoil oscillations ($v \gtrsim 300$\,km\,s$^{-1}$, Table~\ref{tab: param}),
is  $\sim 33$\% (L12) and given that phase II oscillations occur, we estimate a 46 (97)\% probability of finding a displacement larger than that observed, for 
$t_m = 5\ (0.4)$\,Gyr.  
Using eqs.~\ref{eq: tdamp} and \ref{eq: rms} with $\sigma=351$ km s$^{-1}$, $r_\mathrm{rms} \ge 7$ pc and $r_{c} = 500$ pc, we find $(t-t_{c}) \lesssim 9\ \tau \approx 10$ Gyr, a factor ten larger than the estimated spectral age of the diffuse radio source. Therefore, if an SBH binary coalescence event is responsible for the displacement, it seems likely that it pre-dates the AGN phase that created the radio relic (if indeed NGC 5419 is the origin of the relic).

\subsection{Notes on other galaxies}
In the following galaxies, photocenter -- SBH displacements were not considered significant.

\subsubsection{NGC 4168}
NGC 4168 is an E2 in the Virgo cluster. Broad H$\alpha$ has been detected in its nuclear optical spectrum, resulting in classification as a Seyfert 1.9 \citep{Ho97}.
 VLBA observations at 8.4 GHz show an unresolved radio source with no evidence of jets \citep{AndersonUH2004}. 
Although the galaxy appears very regular in the WFPC2-WF/F606W image, the photometric analysis reveals systematic variations in the isophote centers with SMA that are much larger in amplitude than the
displacement between the mean photocenter and the AGN position (Fig.~\ref{fig: NGC4168_9p}). In particular, there is an abrupt shift to the E in the $x$ co-ordinate and a systematic migration to the S in $y$, for SMA $\gtrsim$ 260 pixels (13\arcsec). The measured displacements in $x$ and $y$ are $< 0.8$ IQR and therefore considered ``null'' results.

\subsubsection{NGC 4261} 
\label{app: NGC 4261}

NGC 4261 is a massive elliptical (E2-3) in the outskirts of the Virgo cluster. \citet{Nolt93} identifies this galaxy as the dominant member of a group of 33 galaxies, the Virgo W cloud. It is associated with a bright FR\,I radio source, 3C 270, which has twin jets  close to the plane of the sky, $\theta = 63 \pm 3^{\circ}$, and aligned on the axis W-E, PA  $= 87 \pm 8^{\circ}$, with the western radio component slightly brighter than the eastern \citep{JonesWMP00, Worrall10}. Deep imaging reveals the presence of tidal tails to the NW and a tidal fan SE \citep{Tal09} while \citet{GiordanoCT05} find anisotropies in the globular cluster distribution. 

\citet{Cappellari07} report the presence of a kpc-scale kinematically decoupled core hosting a 100 pc scale disk of dust, cool molecular and atomic gas \citep{Jaffe93, JaffeMcNa94}. \citet{JaffeetAl96} $\&$ \citet{ferrarese1996} studied the central region of this galaxy in detail using HST/WFPC2/PC1 V (filter F547M), R (F675W) and I (F791W) images. After correcting the R images for line emission, they determined that the central dusty disk is characterized by a rotation axis which closely matches the axis of the jets but it is not coaxial with the semi-major axis of the galaxy. The disk is not centered on the AGN, nor on the photocenter, which they find is itself displaced relative to the AGN by $\Delta r = 3.3 \pm 1.3$ pc NE. 

In this study, we chose to analyze a NICMOS2 F160W image (with a pixel size of 0\farcs05), since the nuclear source is bright and dust extinction and contamination due to emission from ionized gas are minimized. The galaxy exhibits boxy isophotes which are difficult to model accurately. Even though the third and forth harmonics were included in the isophote fits, the residual image still shows a spiral pattern suggesting that a more complex model perhaps involving multiple components would be necessary to successfully model the surface brightness distribution (Fig.~\ref{fig: NGC4261_9p}). 
Nevertheless, the isophote centers are relatively stable, except for a slow migration to the N for SMA $\gtrsim 120$ pixels (6$^{\prime\prime}$).
The mean photocenter displacement relative to the AGN is $\Delta r = 3.3 \pm 2.2$ pc W-NW (PA = 297 $\pm$ 38$^{\circ}$). The $x$ component of the displacement is large compared to the IQR. 
However, because of the large uncertainty in positions derived from NICMOS2 images (0.23 pixels), we do not consider this offset significant.

\subsubsection{NGC 4373}

This galaxy is located in the outskirts of the Hydra-Centaurus supercluster and is associated with a compact radio source \citep{Sadler1989}. 
It forms an isolated pair with IC 3290 \citep{SadlerS84}, which is located 2$^{\prime}$ W-SW of NGC 4373. \citet{GTR2007} suggest that there is an ongoing strong interaction with the companion.
Our photometric analysis of the WFPC2/PC-F814W image reveals boxy isophotes which produce cross-shaped residuals, even though higher harmonics were included in the isophote fits (Fig.~\ref{fig: NGC4373_W2}). The  $y$ coordinate of the isophote center position has a narrow distribution centered on the mean photocenter and shows only a slight drift northward, for SMA $>$ 170 pixel (8\farcs5).
The $x$ co-ordinate exhibits a systematic migration eastwards from the photocenter, resulting in a broader, skewed distribution.
The mean photocenter is consistent with the AGN position within the uncertainties.

	
\subsubsection{NGC 4552}
NGC 4552 is a typical giant elliptical in the Virgo cluster which hosts a very low-luminosity AGN. HST imaging and spectroscopy reveal a variable UV point source that has optical emission lines
characteristic of AGN \citep{Cappellari99, Maoz2005}. Interestingly, a second, transient UV source of comparable luminosity, but offset from the first by $0\farcs14$, was present in 1991, but was not detected in 1993 \citet{Renzini95, Cappellari99, Maoz2005}.
Radio observations at 8.4 \citep{Filho2000} and 5 GHz \citep{Nagar2002} show the presence of a flat spectrum radio core. \citet{Nagar2002} also report variability at the $\sim$ 20\% level on a $\sim$ 1
year timescale and find evidence of pc-scale jets, or extensions,  along the East-West axis. NGC 4552 exhibits several morphological features indicative of past mergers, including  
three concentric shells at 5$^\prime$, 7$^\prime$ and 10$^\prime$ from the nucleus \citep{Malin79}. Traces of a recent minor merger with a gas rich satellite galaxy have been identified in the presence of dust patches in the inner region \citep{DokkumFranx1995}, and H${\alpha}$ emission extending up to $\sim2.5$ kpc from the center \citep{TrinchieriSerego1991, MacchettoEtAl1996}. 

The photometric analysis of the WFPC2/PC-F814W image shows that the $y$-coordinate of the isophote center migrates systematically southward relative to the photocenter for SMA $>$ 200 pixels (10\arcsec) with a maximum shift $\lesssim$ 0\farcs2 (Fig.~\ref{fig: NGC4552_WFPC2F814W}). There is a smaller westward shift in the $x$ coordinate.
The inner $0\farcs3$ of the residual image shows evidence of an arc-like dust lane partly encircling the central source with an opening toward the S. This feature was previously noted by \citet{Carollo97} and \citet{Cappellari99} (see Fig.~12 in the latter) and is the cause of the increased uncertainty on the $y$ co-ordinate of the AGN position. We did not find a significant ($>3\sigma$) displacement in this object.


\subsubsection{NGC 4636}
NGC 4636 is a giant elliptical in the outskirts of the Virgo cluster. Its nucleus has a LINER emission line spectrum, in which a weak broad H$\alpha$ line has been detected \citep{Ho97}. 
It has a jet-like radio source extending $\approx 1.5$ kpc either side of the nucleus in position angle $\theta \approx 33^{\circ}$\citep{BD85}. NGC 4636 is also associated with a luminous, extended X-ray source
with a peculiar morphology featuring two symmetric, arm-like extensions emerging from the bright central region and extending up to $\approx 8$ kpc along the NE-SW axis. Other fainter arm-like features emerge from the nucleus and \citet{Jones02} suggest that these features trace cyclic nuclear outburst events occurring  on a timescale  $\sim10^{7}$ yr. 
According to \citet{Temi03}, the observed FIR fluxes (at 60, 90 and 180 $\mu$m) suggest the accretion of dust in a recent (within $\sim10^{8}$ yr) merger with a gas-rich galaxy.

Our isophotal modeling of the WFPC2/PC-F814W image indicates the presence of faint dust lanes inspiraling toward the nucleus (Fig.~\ref{fig: NGC4636_WFPC2F814W}). 
There is a systematic shift in the isophote centers to the north and west relative to the photocenter for SMA $>$ 170 pixels (8\farcs5), which results in broad distributions in the $x$ and $y$
co-ordinates (Fig.~\ref{fig: NGC4636_WFPC2F814W}). The photocenter position is consistent with the AGN position in the $y$ direction, but exhibits a $>3\sigma$ offset in $x$.
Nevertheless, as the IQR of the isophote center coordinate distribution exceeds the displacement we consider this a null result. 

\subsubsection{NGC 4696 (double nucleus)}

NGC 4696 is the dominant galaxy of the Centaurus cluster. There is an extensive literature on the multi-wavewelength properties of this galaxy, which has a secondary nucleus \citep{LaineEtAl03}
offset by $\approx 0\farcs3$ to the SW of the primary (\citealt{LaineEtAl03} and Fig.~\ref{fig: NGC4696_ACS}). The double nucleus is partially surrounded by a complex of dust lanes and associated emission line filaments, the most prominent of which is in the form of a spiral leading towards the center. These features have been interpreted as evidence of a recent minor merger with a
gas-rich galaxy (\citealt{SparkMG89, FarageEtAl10}; alternatively the emission-line gas may have been ``drawn out'' by rising radio bubbles \citealt{Canning2011}). The galaxy is associated with the moderately powerful radio source PKS 1246-410, which extends $\approx 10$ kpc along an approximately E-W axis, with the ends of both arms bending southwards. The radio source wraps around the region of brightest X-ray emission, suggesting interactions between the radio plasma and the hot thermal gas \citep{Taylor06}. VLBA observations also obtained by \citet{Taylor06} reveal a relatively weak core and a one-sided jet extending over $\approx 25$\,pc in PA = $-150^{\circ}$. 

Subtraction of our photometric model from the ACS/WFC-F814W image (Fig.~\ref{fig: NGC4696_ACS}) clearly shows the dust features around the nucleus. The positive residuals probably trace
the emission-line filaments. The isophote centers exhibit large excursions, resulting in broad distributions in both $x$ and (especially) $y$ coordinates.  
Although we measure a $3\sigma$ offset in $x$ between the brightest nucleus and the mean photocenter, we consider this a null result because of the large IQR.

\subsubsection{IC 1459}

IC 1459 is a giant elliptical galaxy (E3--4) and a member of a loose group (identified as  ``group 15" in  \citealt{HG82}) of ten galaxies, which are mainly spirals. Despite its relatively unremarkable appearance to casual inspection, this galaxy turns out to be notably peculiar (see \citealt{Forb95} for a detailed analysis): it has one of the faster counter rotating kinematically distinct cores \citep{FranxILL88}, faint stellar spiral arms \citep{Malin85}, outer shells \citep{Forb95} and irregular dust lanes or patches near the nucleus \citep{Forb94}. 
It has a hard X-ray point source \citep{GonzalezMMG09} and a strong, compact radio core at 5 GHz \citep{Slee94}.

We analyzed a WFPC2/PC-F814W image (Fig.~\ref{fig: IC1459_F814W}). After subtraction of the photometric model, the residual image shows a periodic pattern, indicating that the isophotes deviate from perfect ellipses. Systematic variations in the isophote center co-ordinates are seen in both $x$ and $y$, with increasing  scatter also present for SMA $>$ 230 pixels (11\farcs5). We did not find a significant ($>3\sigma$) displacement in this galaxy.

\subsubsection{IC 4931}

This relatively little studied galaxy is the brightest member of the group A3656 \citep{PostmanL95}. Its radio source is relatively weak, with a total radio flux at 5 GHz of 0.9 mJy (see Table \ref{tab: radio} to compare with other galaxies). It is also detected at 1.4 GHz with a flux density of 22 mJy \citep{Brown2011}. 
Apart from the optical nucleus and the radio emission, there is no supporting evidence indicating the presence of an AGN such as an X-ray point source, or line emission.

Our analysis of the WFPC2/PC-F814 image (Fig.~\ref{fig: IC4931_F814W}) shows  large systematic variations in the isophote center co-ordinates, with the $x$ coordinate in particular
migrating up to 0\farcs4 W relative the photocenter for  $SMA \gtrsim$ 100 pixels (5\arcsec). The co-ordinate distributions are among the broadest  of the whole sample, spanning a range of 0\farcs5 in $x$.
We measured a $\sim 3\sigma$ offset between the photocenter and the AGN in the $x$ co-ordinate (E), but this is considered a null result because of the large IQR.

\section{C: Images}
\label{app: images}

\begin{figure*}[h]
\begin{center}$
\begin{array}{ccc}
\includegraphics[trim=3.75cm 1cm 3cm 0cm, clip=true, scale=0.48]{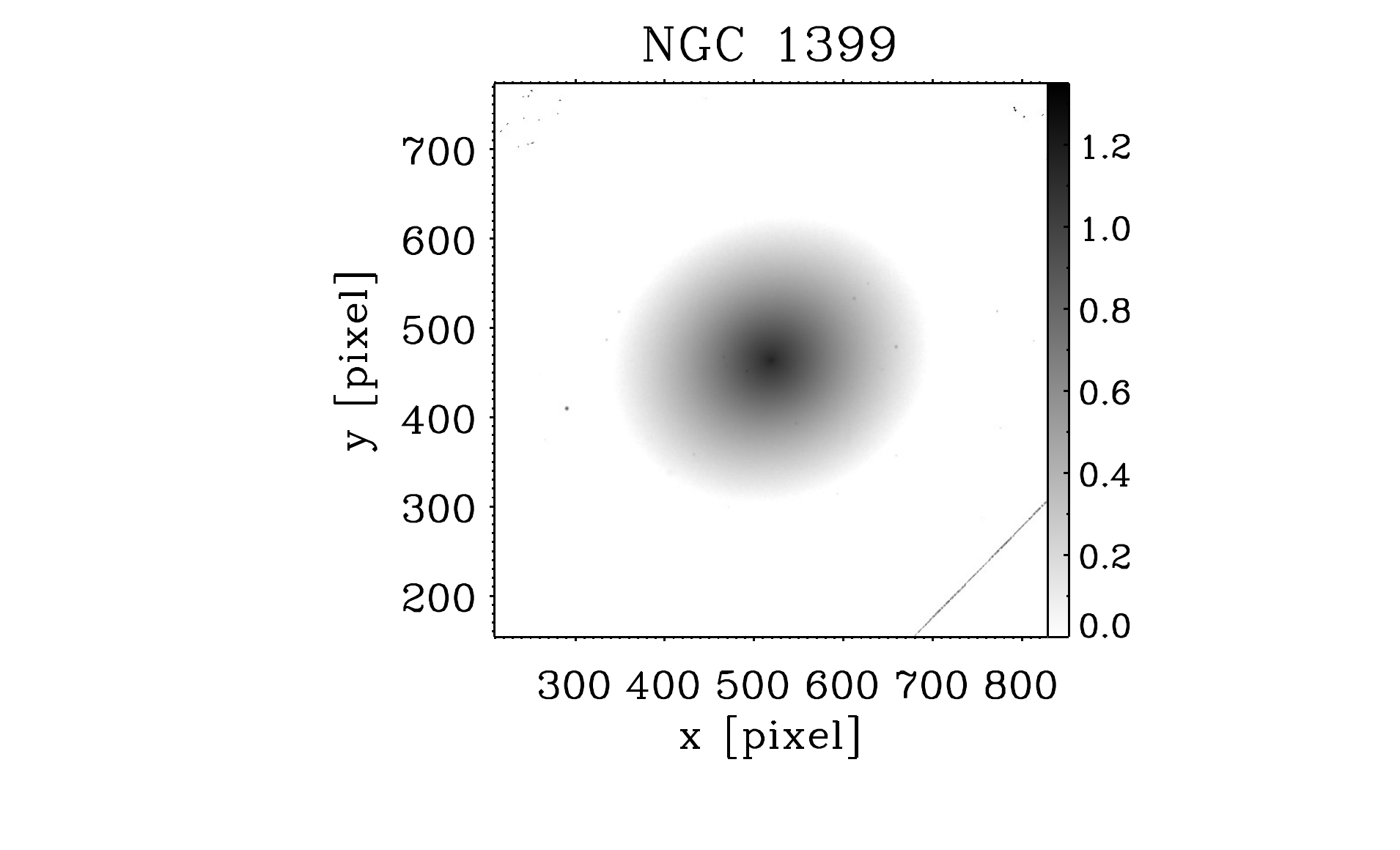} & \includegraphics[trim= 4.cm 1cm 3cm 0cm, clip=true, scale=0.48]{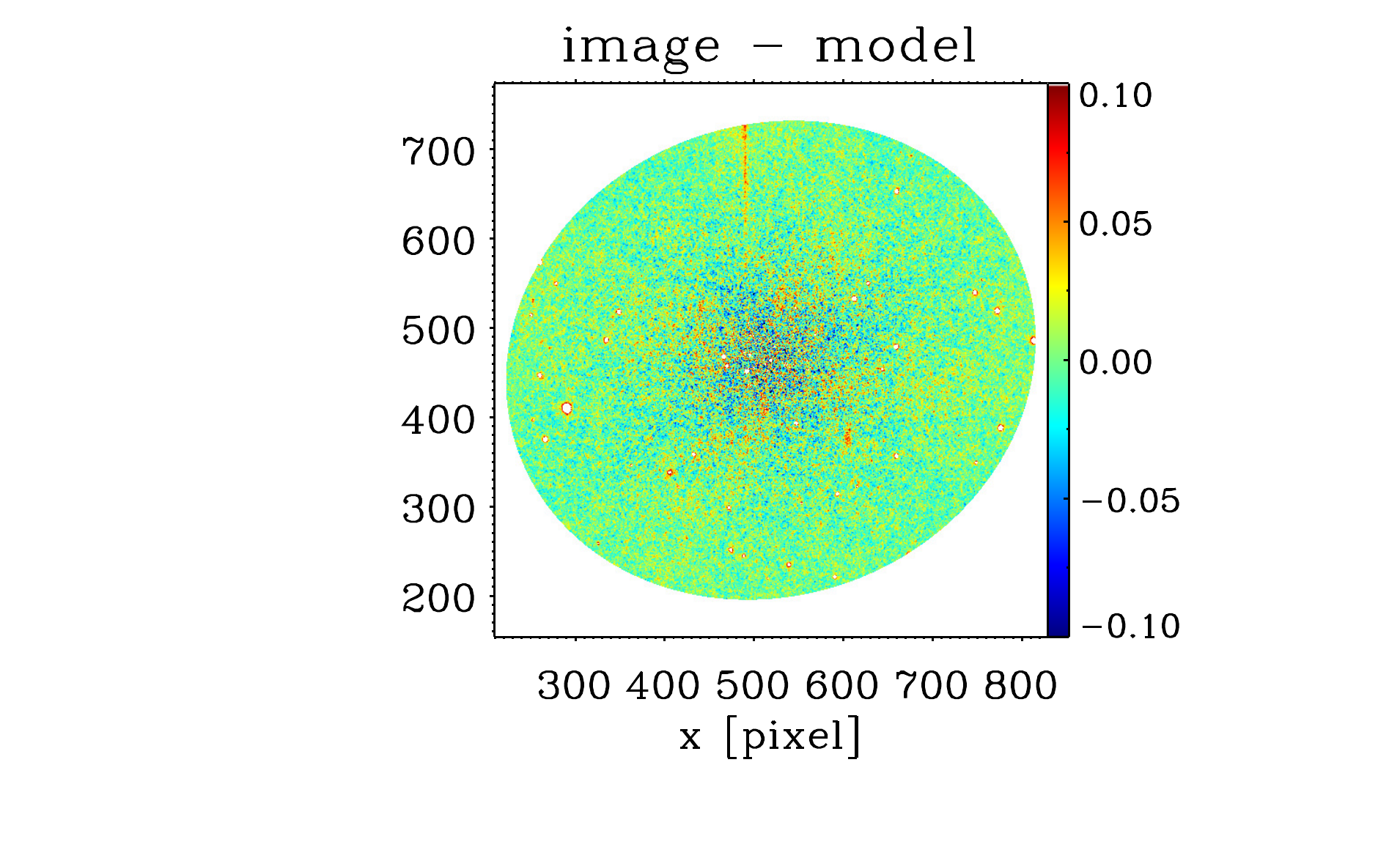}	& \includegraphics[trim= 4.cm 1cm 3cm 0cm, clip=true, scale=0.48]{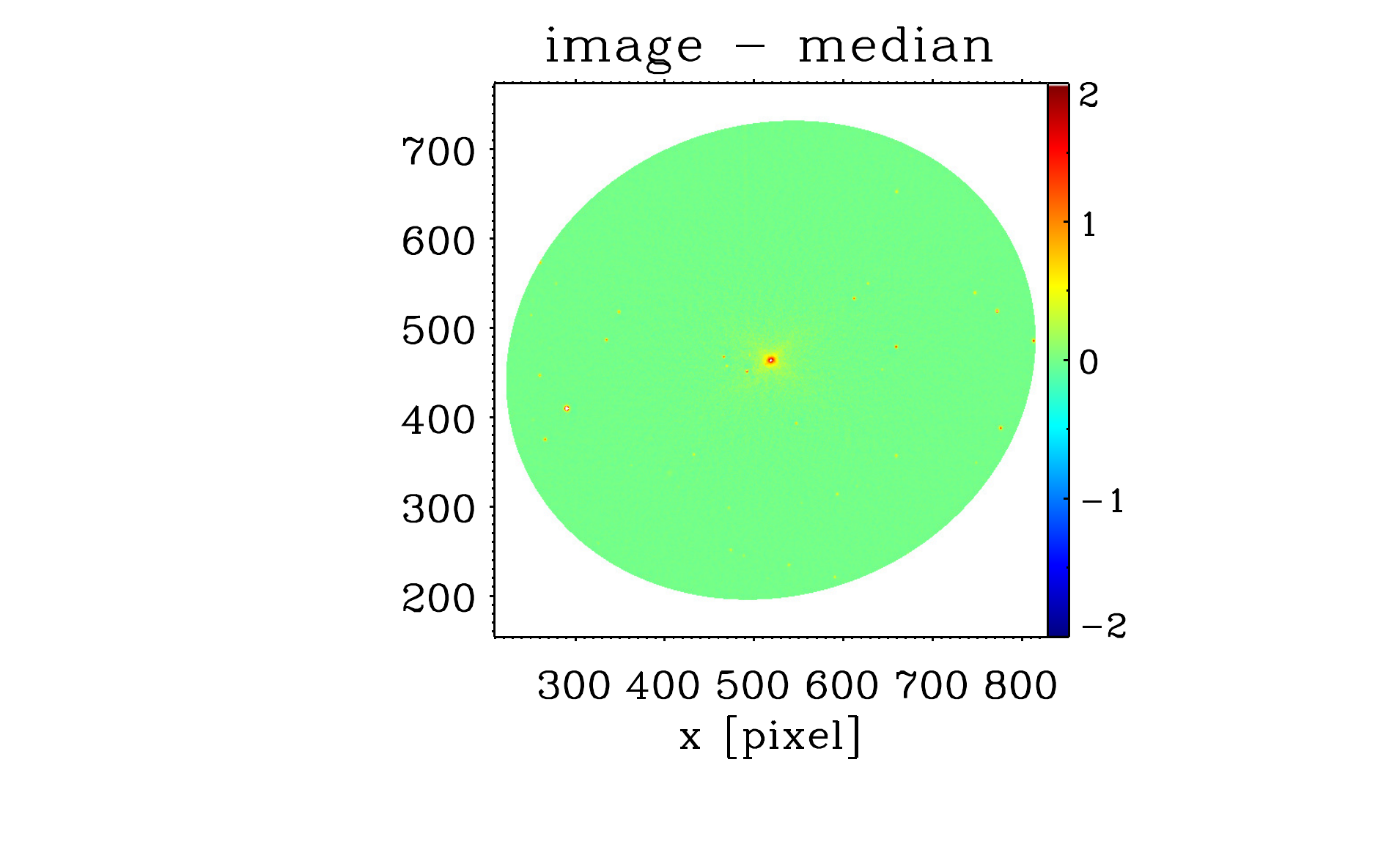} \\

\includegraphics[trim=0.7cm 0cm 0cm 0cm, clip=true, scale=0.46]{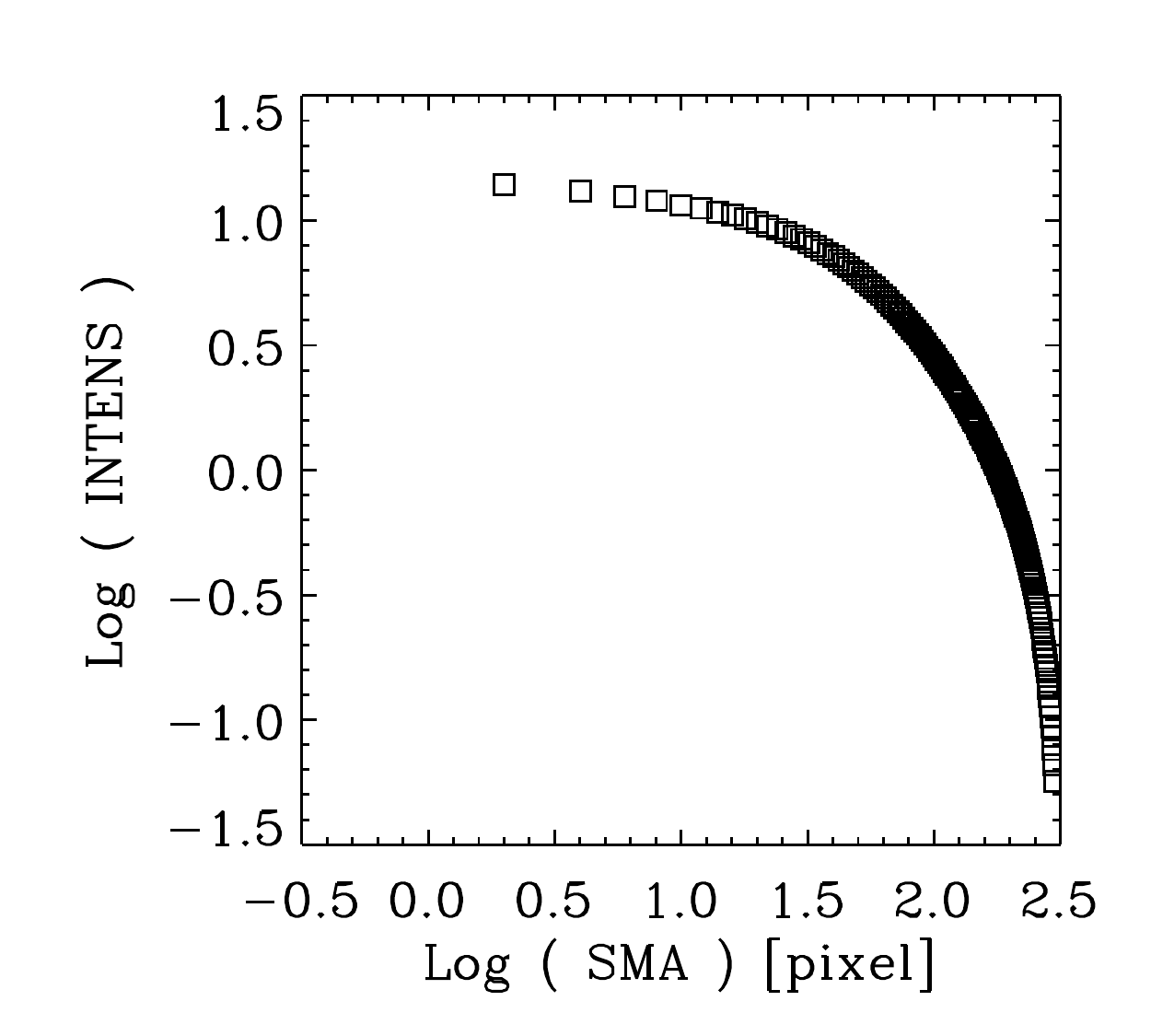}	    &  \includegraphics[trim=0.6cm 0cm 0cm 0cm, clip=true, scale=0.46]{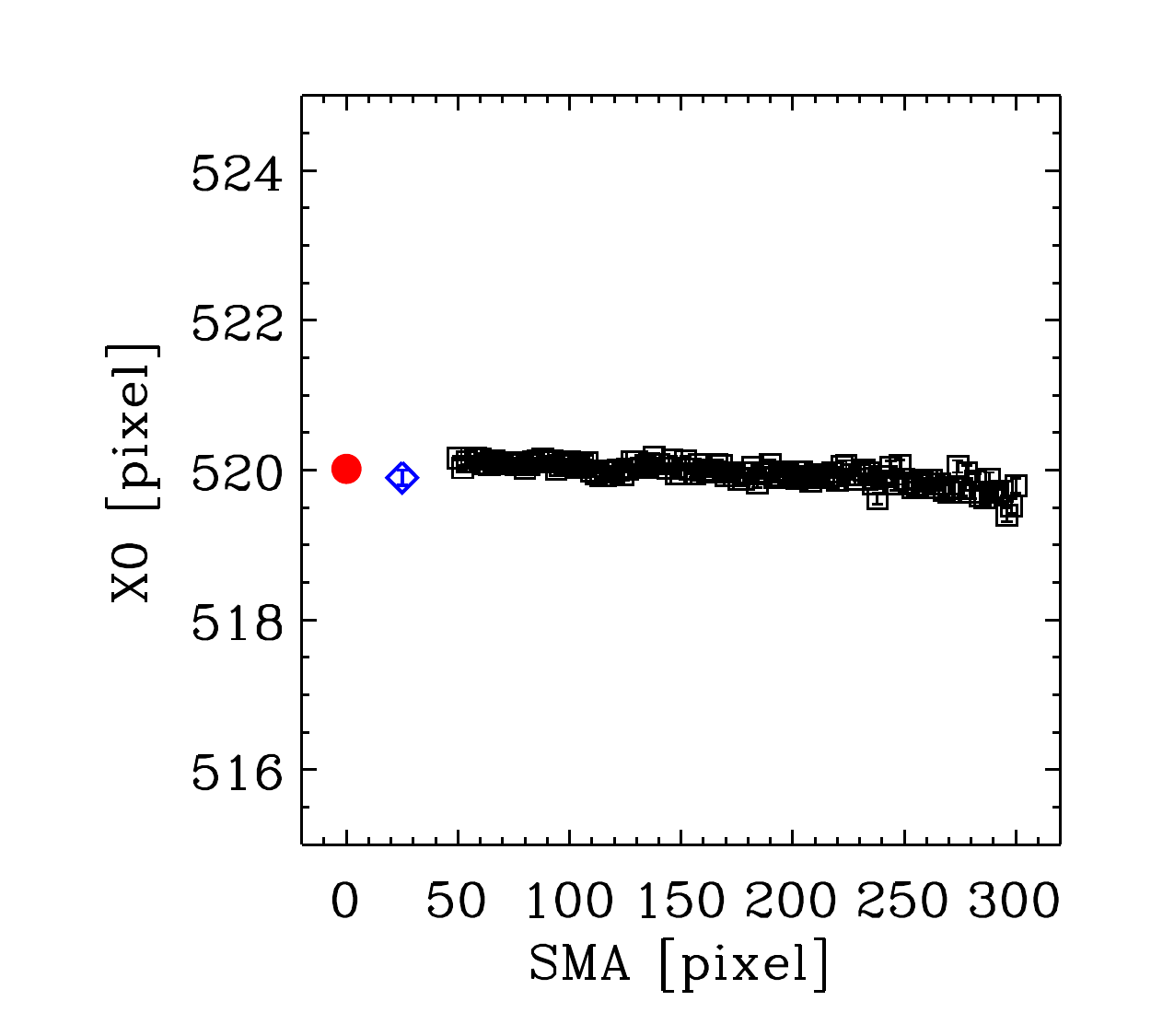}  &  \includegraphics[trim=0.6cm 0cm 0cm 0cm, clip=true, scale=0.46]{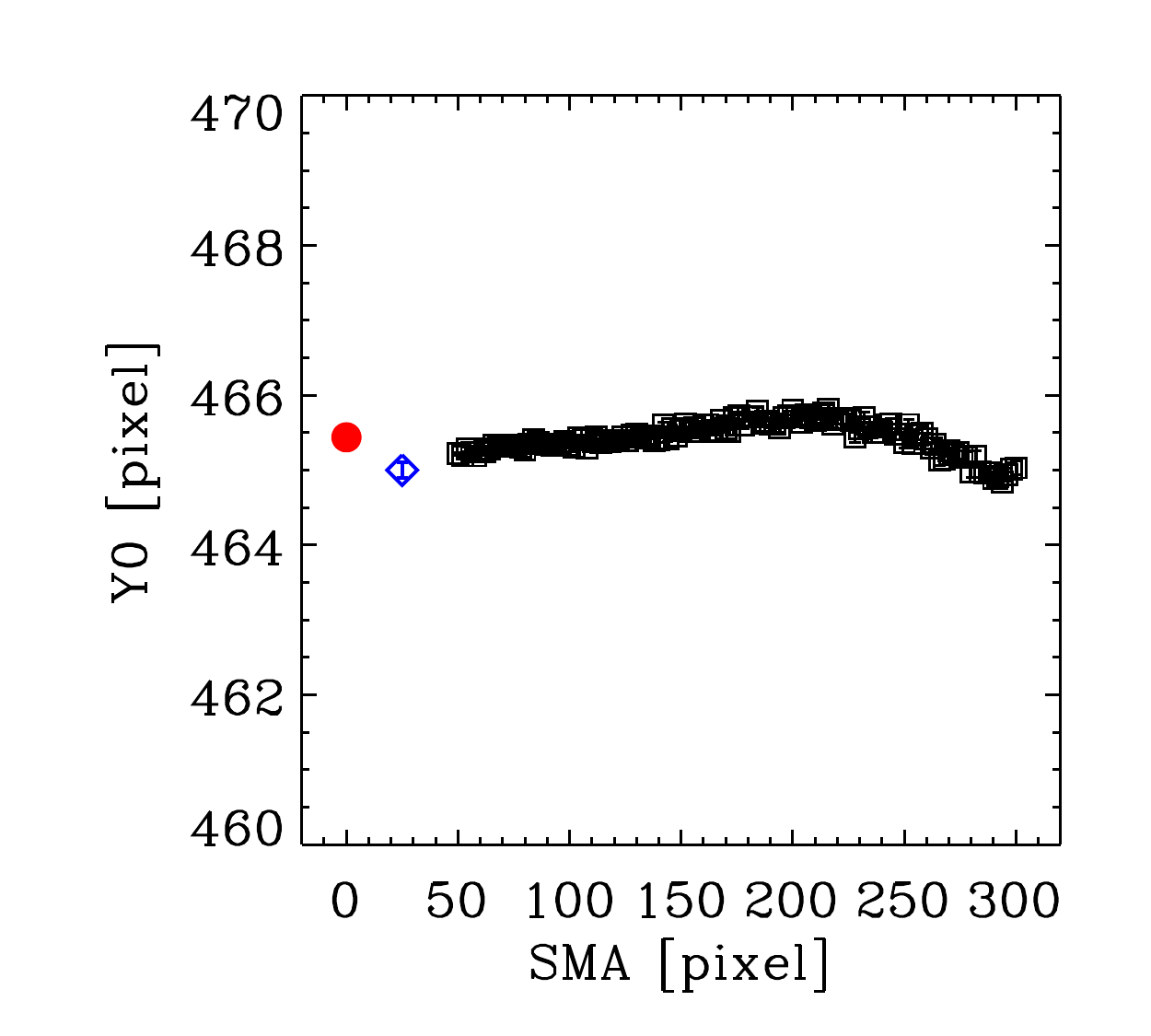} \\	

\includegraphics[trim=0.65cm 0cm 0cm 0cm, clip=true, scale=0.46]{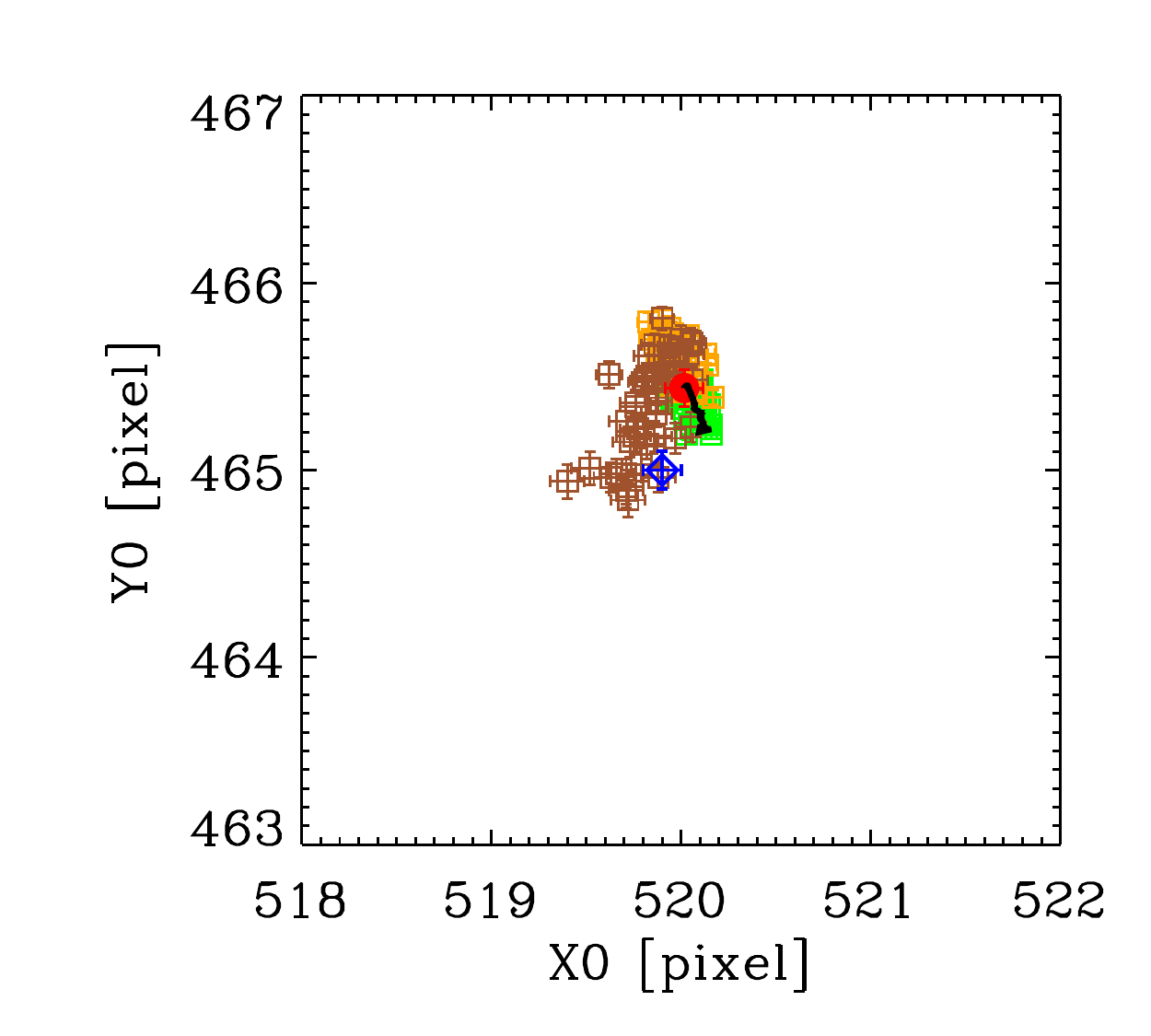}	&  \includegraphics[trim=0.6cm 0cm 0cm 0cm, clip=true, scale=0.46]{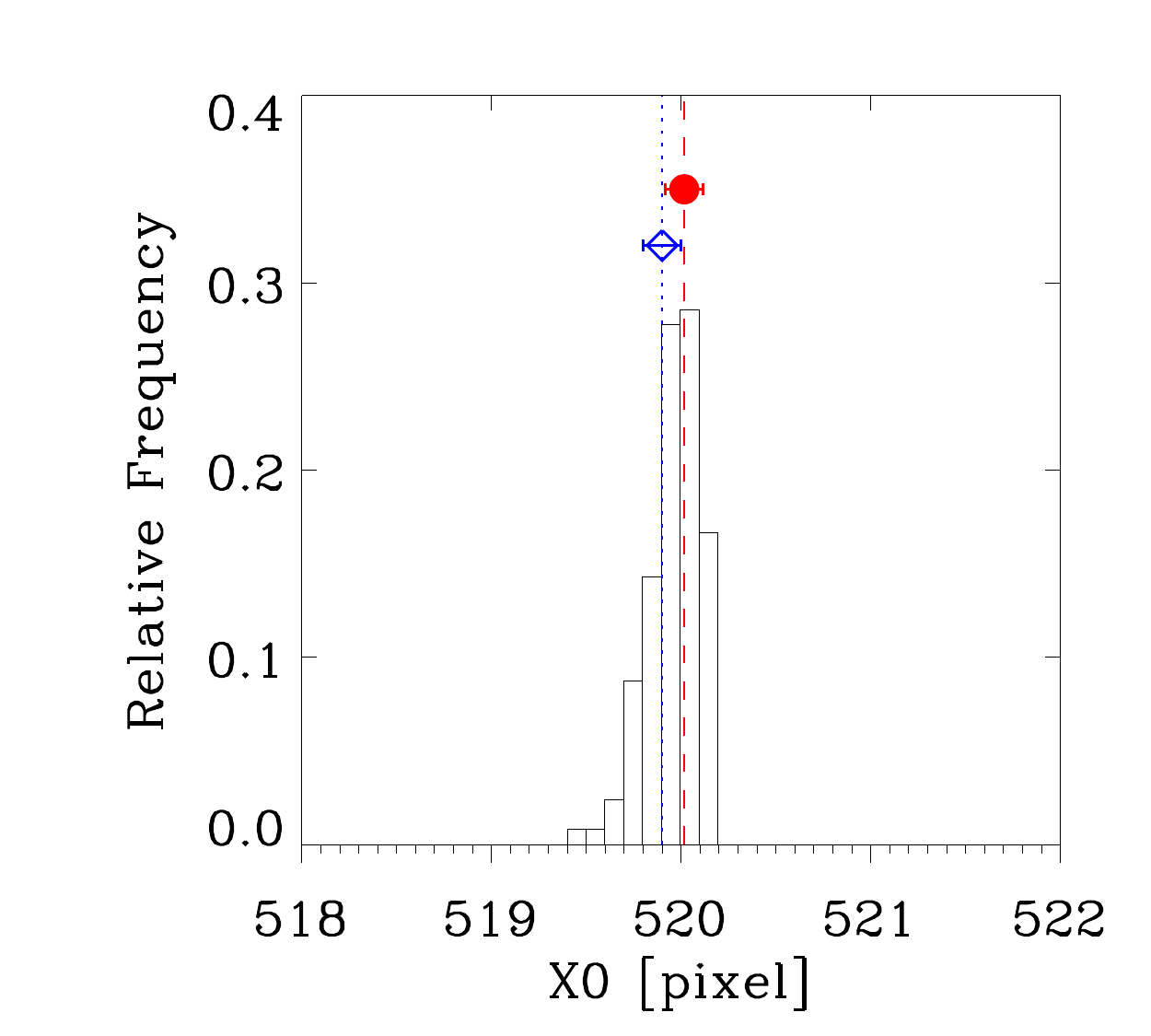}	& \includegraphics[trim=0.6cm 0cm 0cm 0cm, clip=true, scale=0.46]{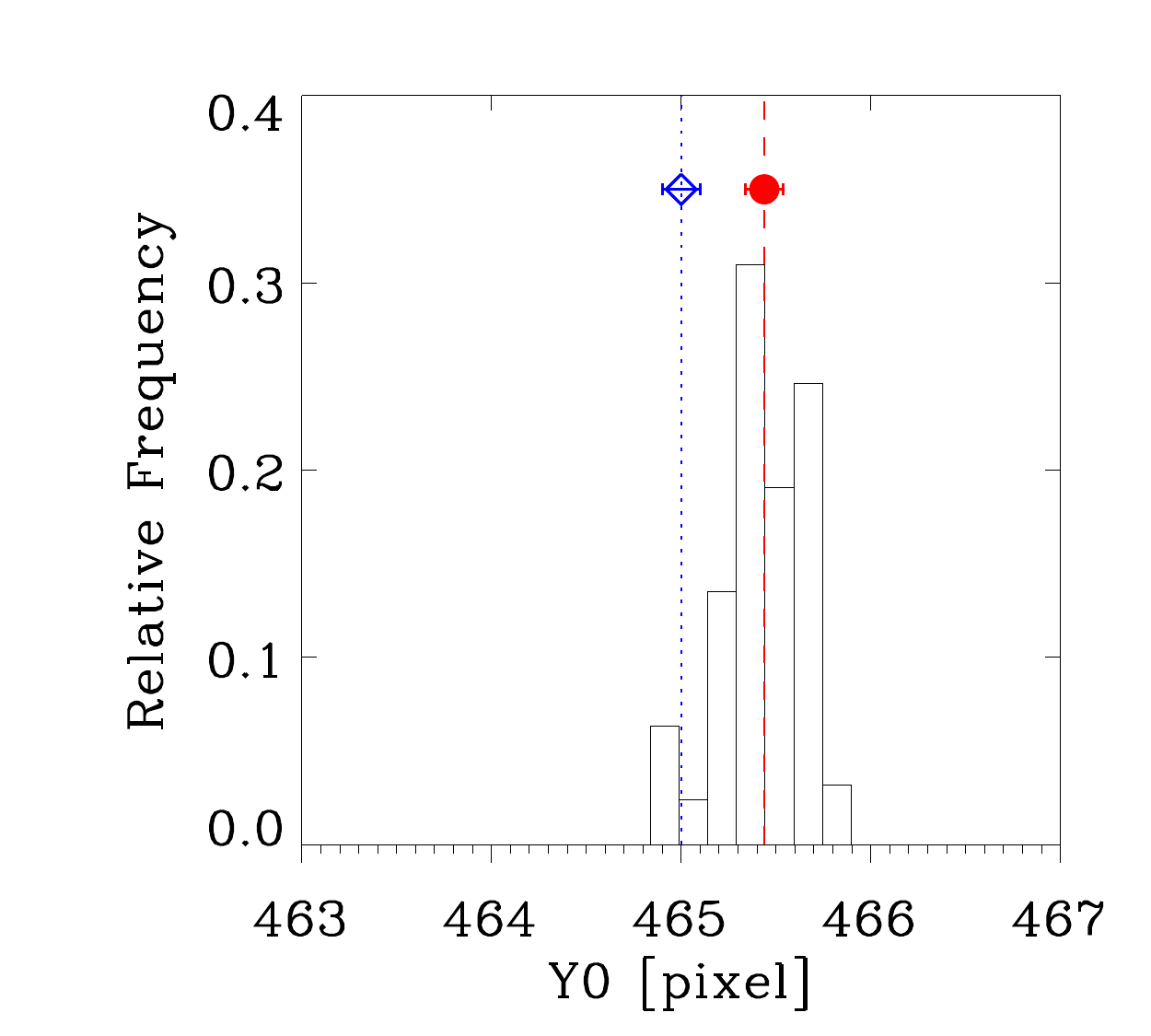}\\
\end{array}$
\end{center}
\caption[NGC 1399]{As in Fig.\ref{fig: NGC4373_W2} for galaxy NGC 1399, WFPC2/PC - F606W, scale=$0\farcs05$/pxl.}
\label{fig: NGC1399_9pF606}
\end{figure*} 

\begin{figure*}[h]
\begin{center}$
\begin{array}{ccc}
\includegraphics[trim=3.75cm 1cm 3cm 0cm, clip=true, scale=0.48]{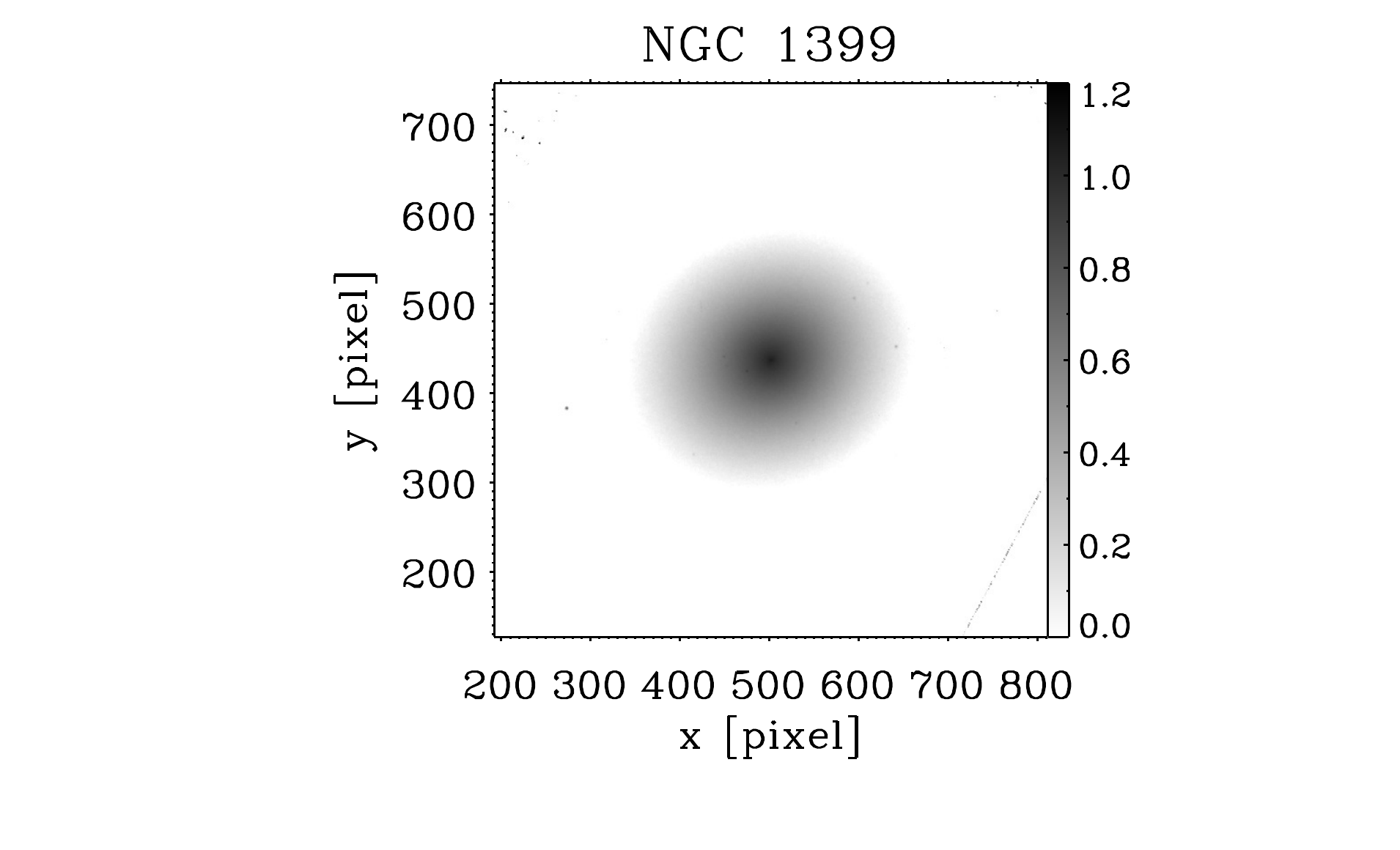} & \includegraphics[trim= 4.cm 1cm 3cm 0cm, clip=true, scale=0.48]{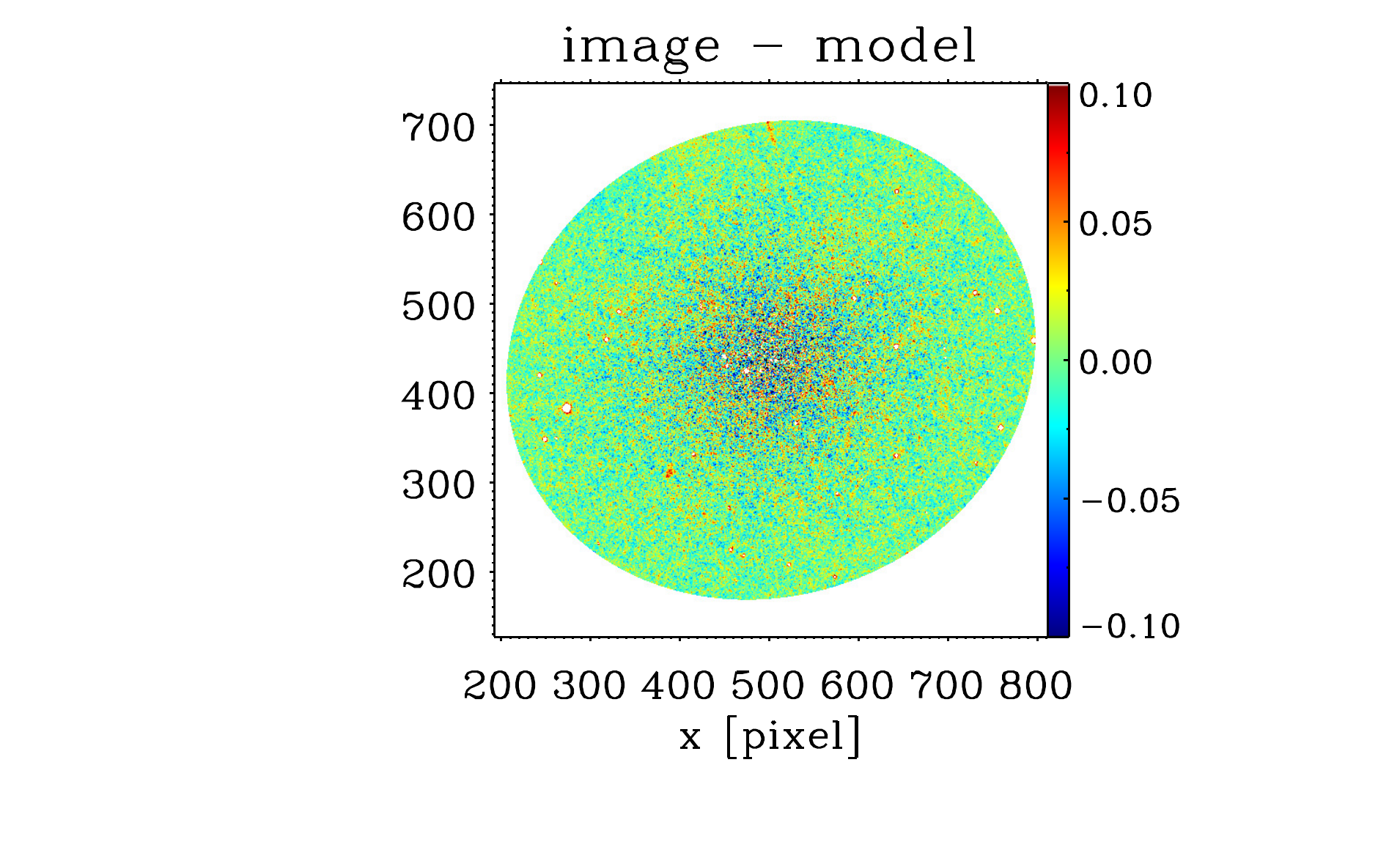}	& \includegraphics[trim= 4.cm 1cm 3cm 0cm, clip=true, scale=0.48]{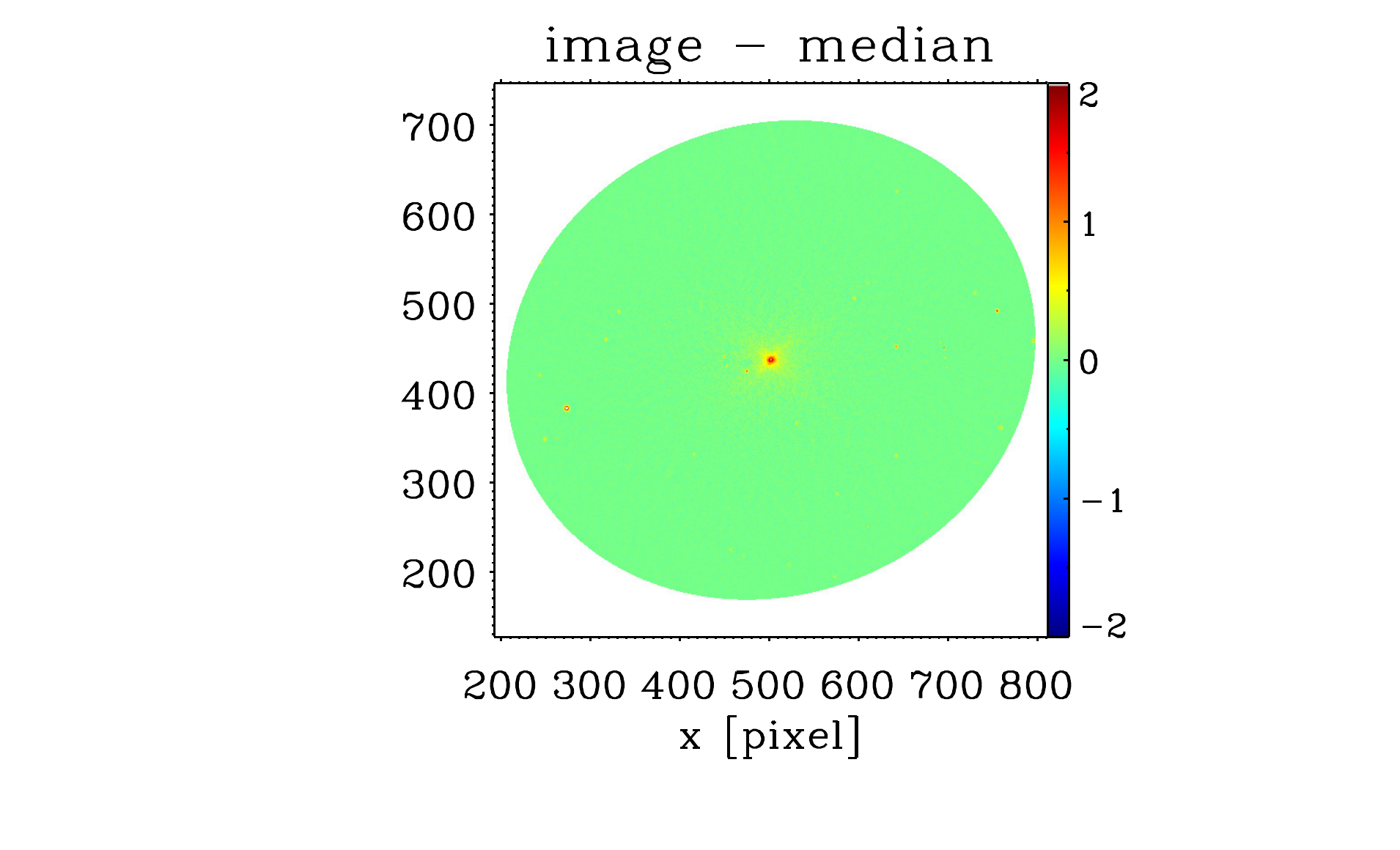} \\

\includegraphics[trim=0.7cm 0cm 0cm 0cm, clip=true, scale=0.46]{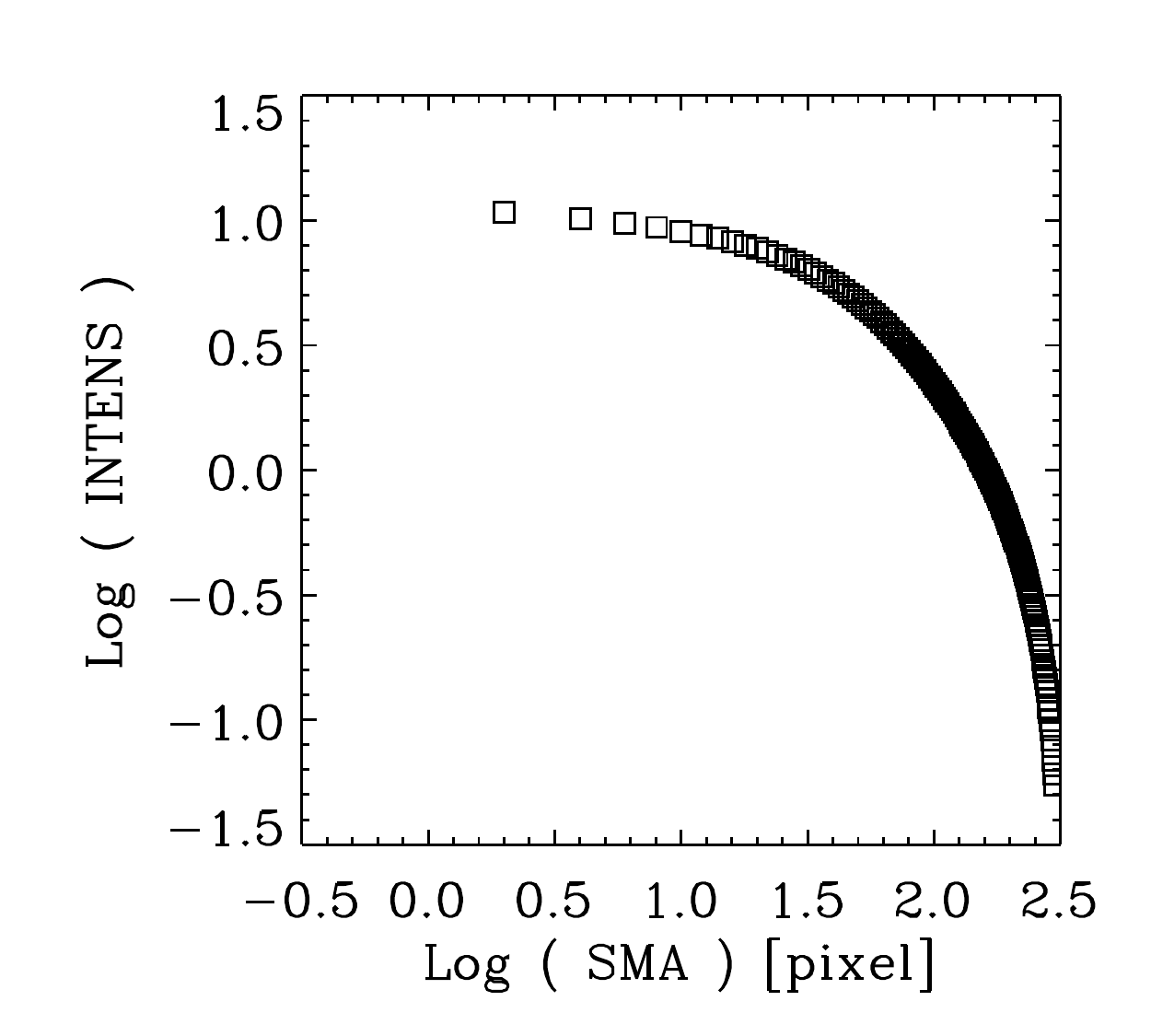}	    &  \includegraphics[trim=0.6cm 0cm 0cm 0cm, clip=true, scale=0.46]{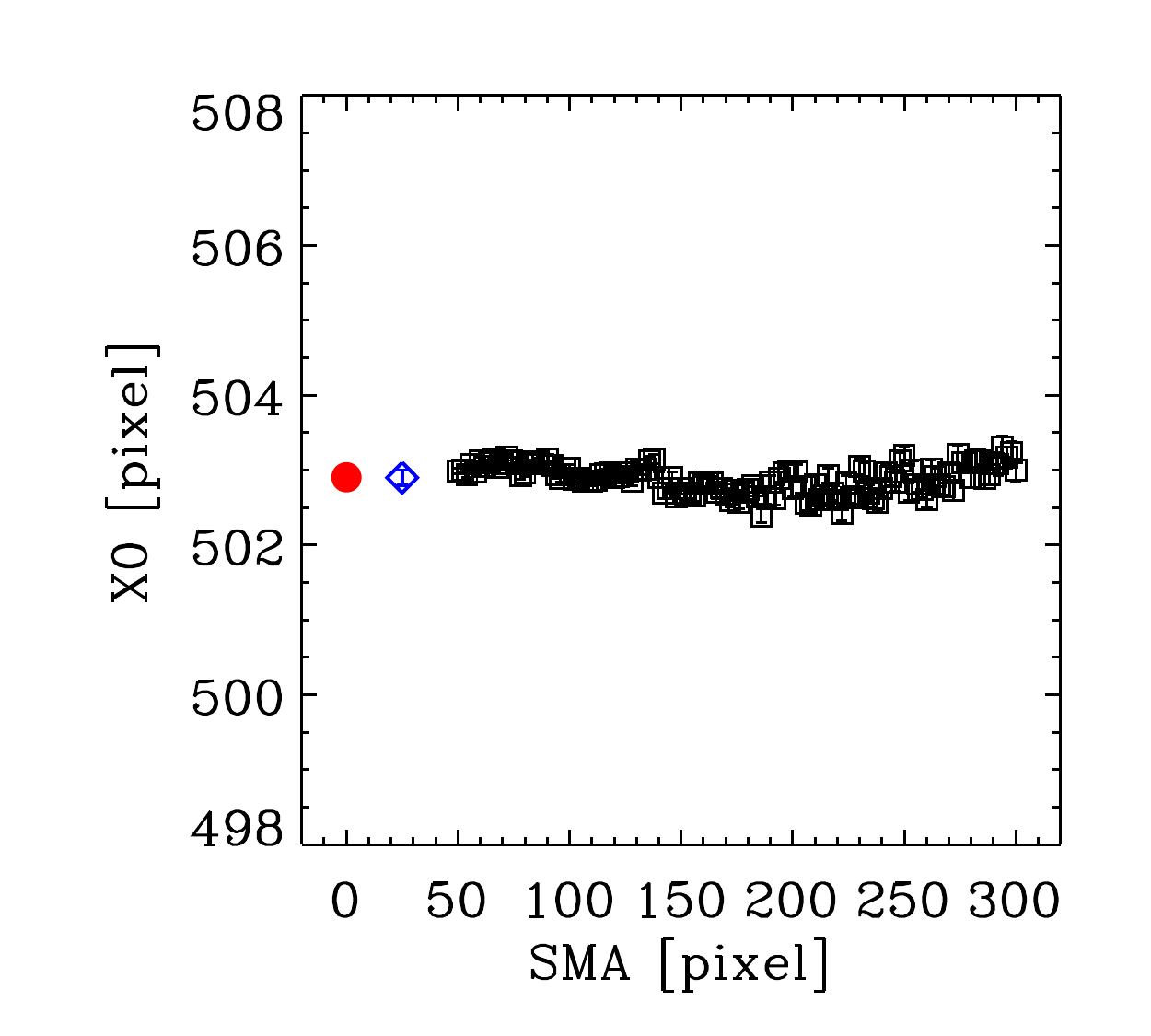}  &  \includegraphics[trim=0.6cm 0cm 0cm 0cm, clip=true, scale=0.46]{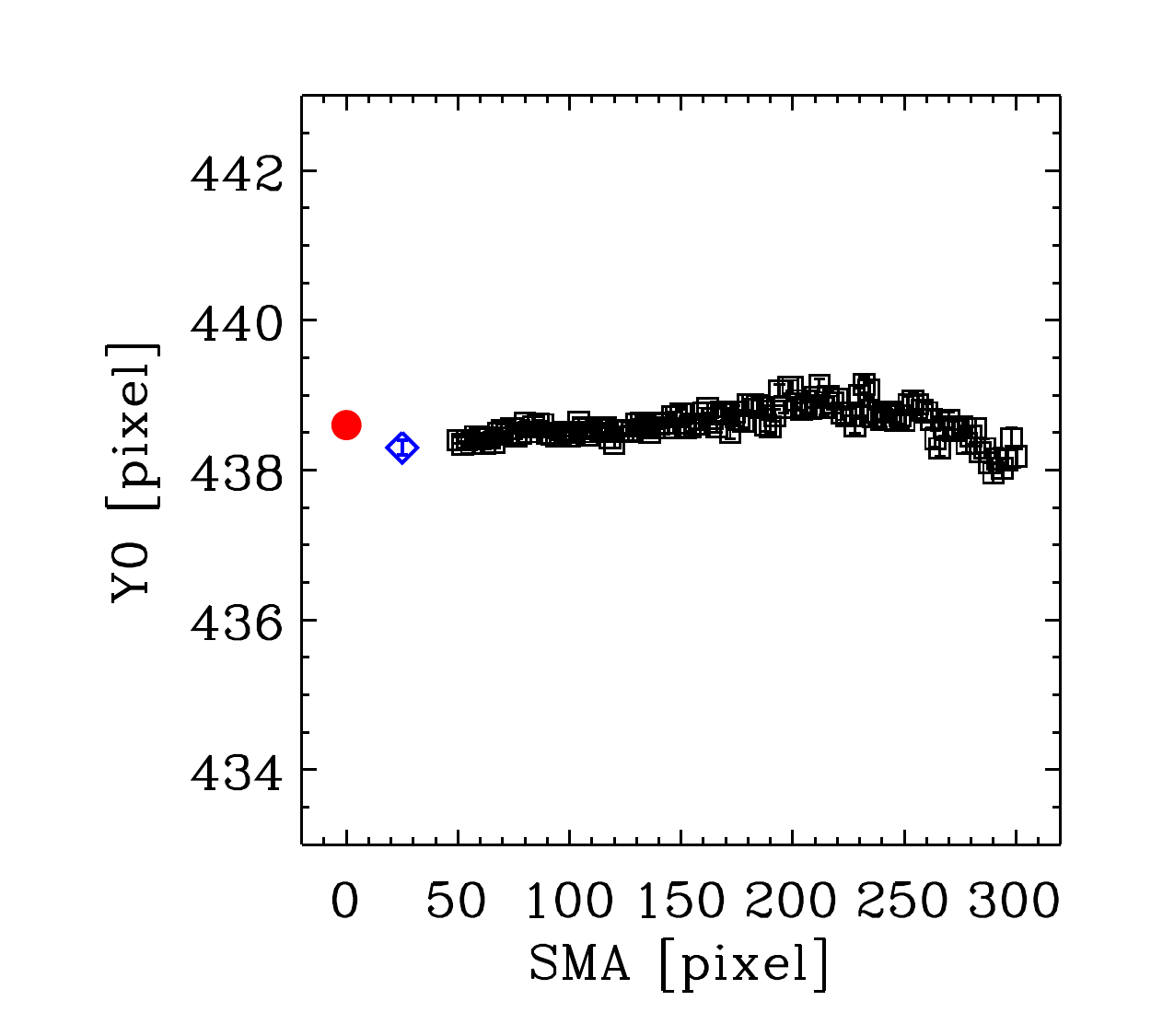} \\	

 \includegraphics[trim=0.65cm 0cm 0cm 0cm, clip=true, scale=0.46]{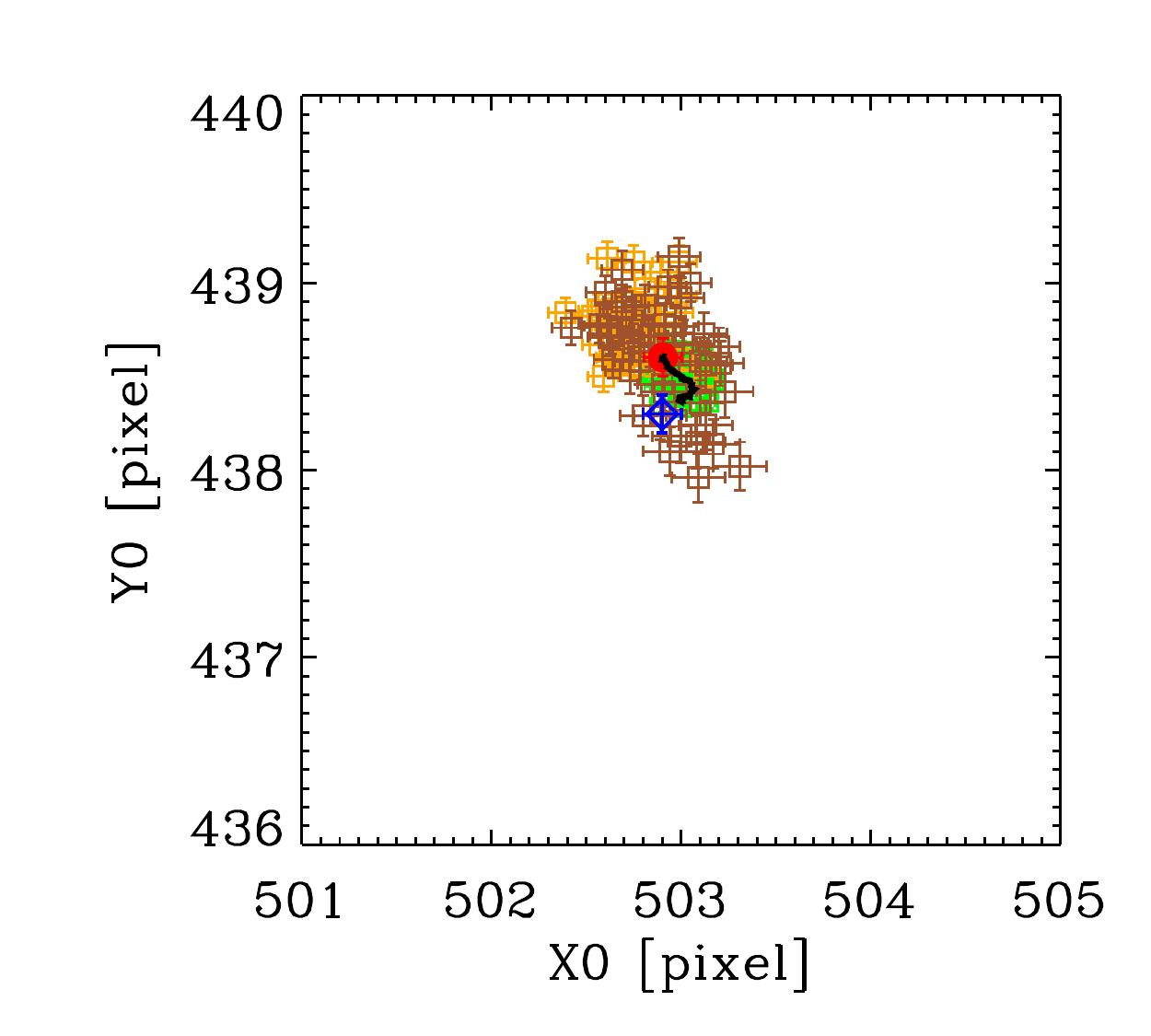}	&  \includegraphics[trim=0.6cm 0cm 0cm 0cm, clip=true, scale=0.46]{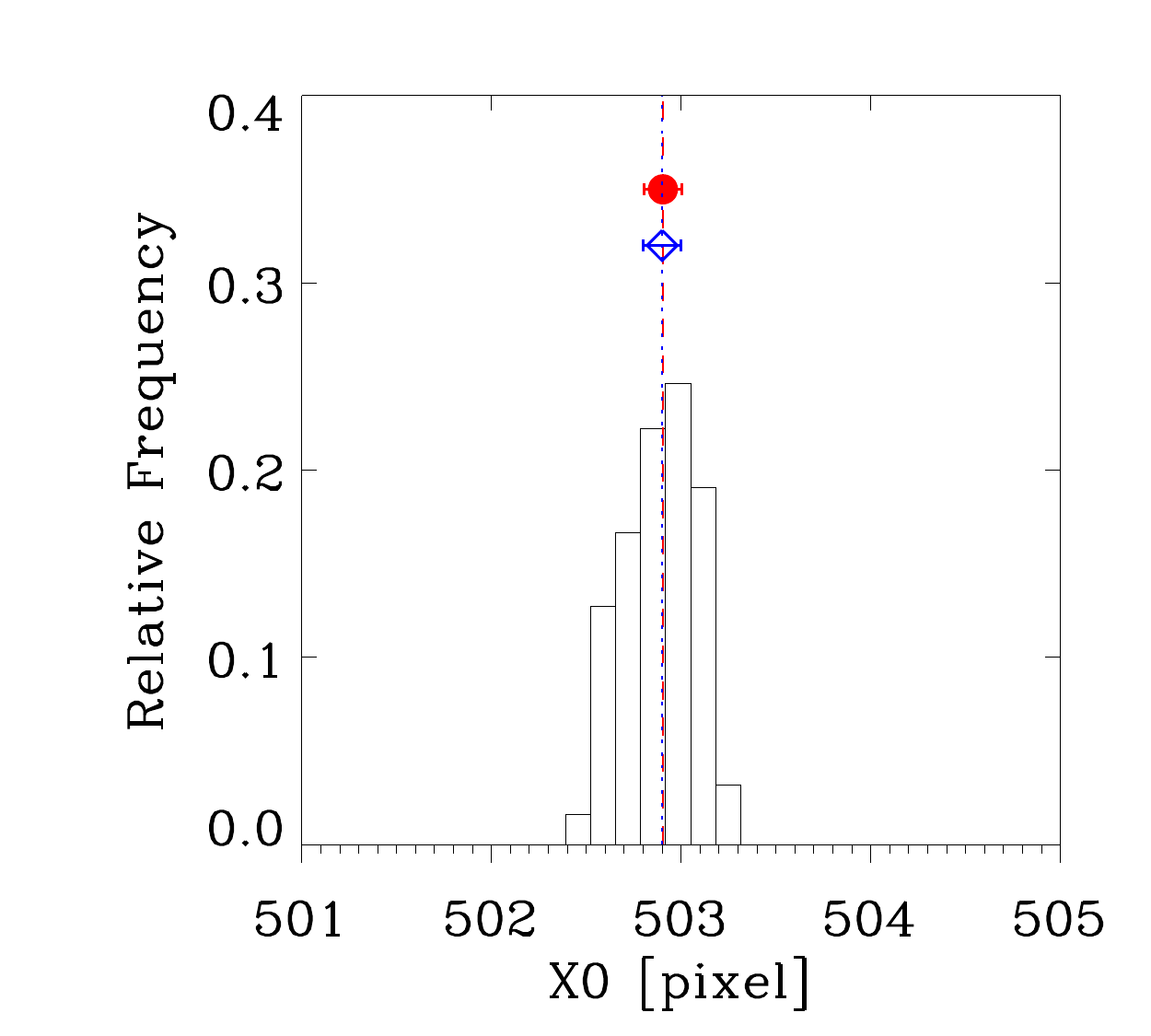}	& \includegraphics[trim=0.6cm 0cm 0cm 0cm, clip=true, scale=0.46]{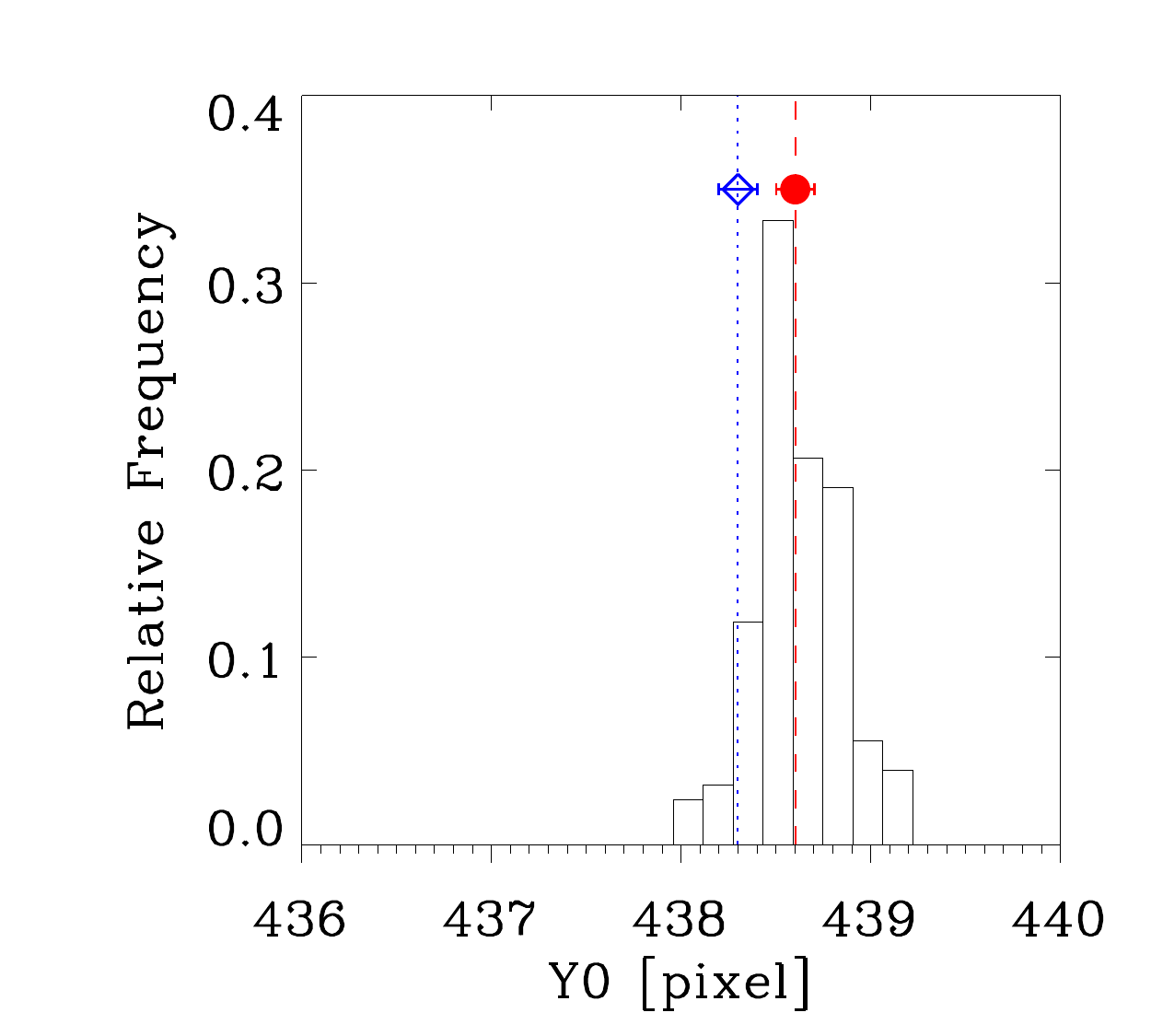}\\
\end{array}$
\end{center}
\caption[NGC 1399]{As in Fig.\ref{fig: NGC4373_W2} for galaxy NGC 1399, WFPC2/PC - F814W, scale=$0\farcs05$/pxl.}
\label{fig: NGC1399_9pF814}
\end{figure*} 

\begin{figure*}[h]
\begin{center}$
\begin{array}{ccc}
\includegraphics[trim=3.75cm 1cm 3cm 0cm, clip=true, scale=0.48]{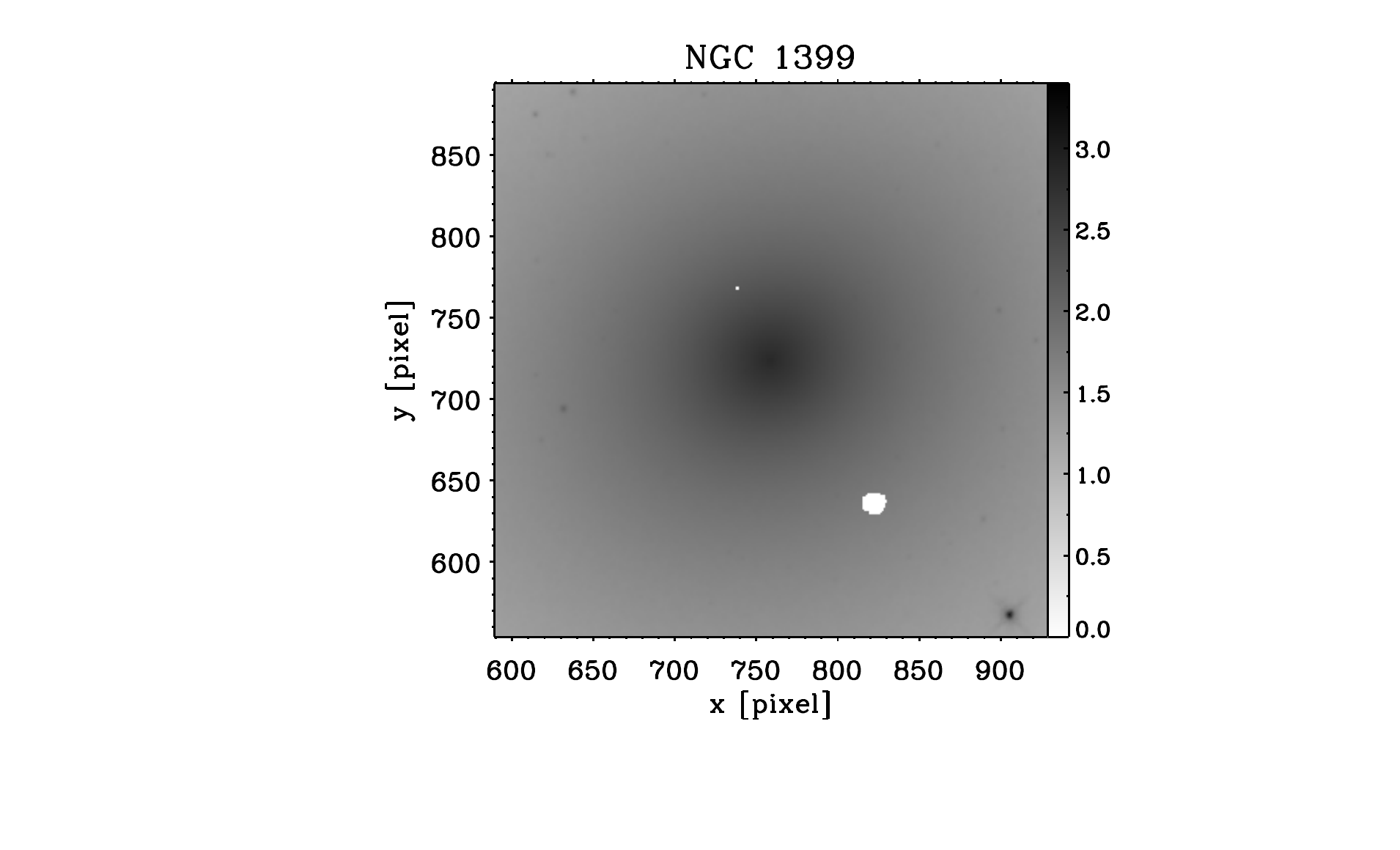} & \includegraphics[trim= 4.cm 1cm 3cm 0cm, clip=true, scale=0.48]{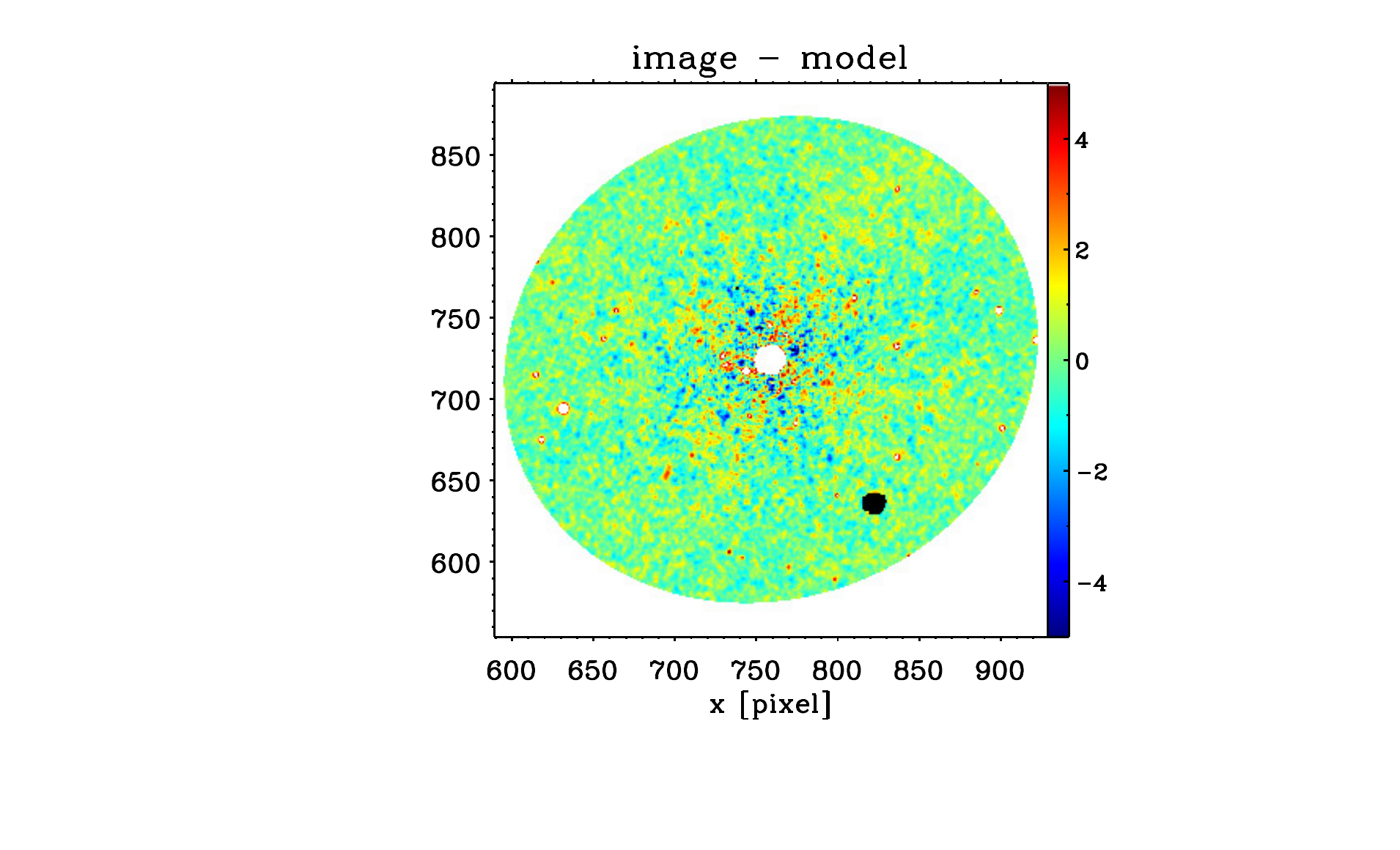}	& \includegraphics[trim= 4.cm 1cm 3cm 0cm, clip=true, scale=0.48]{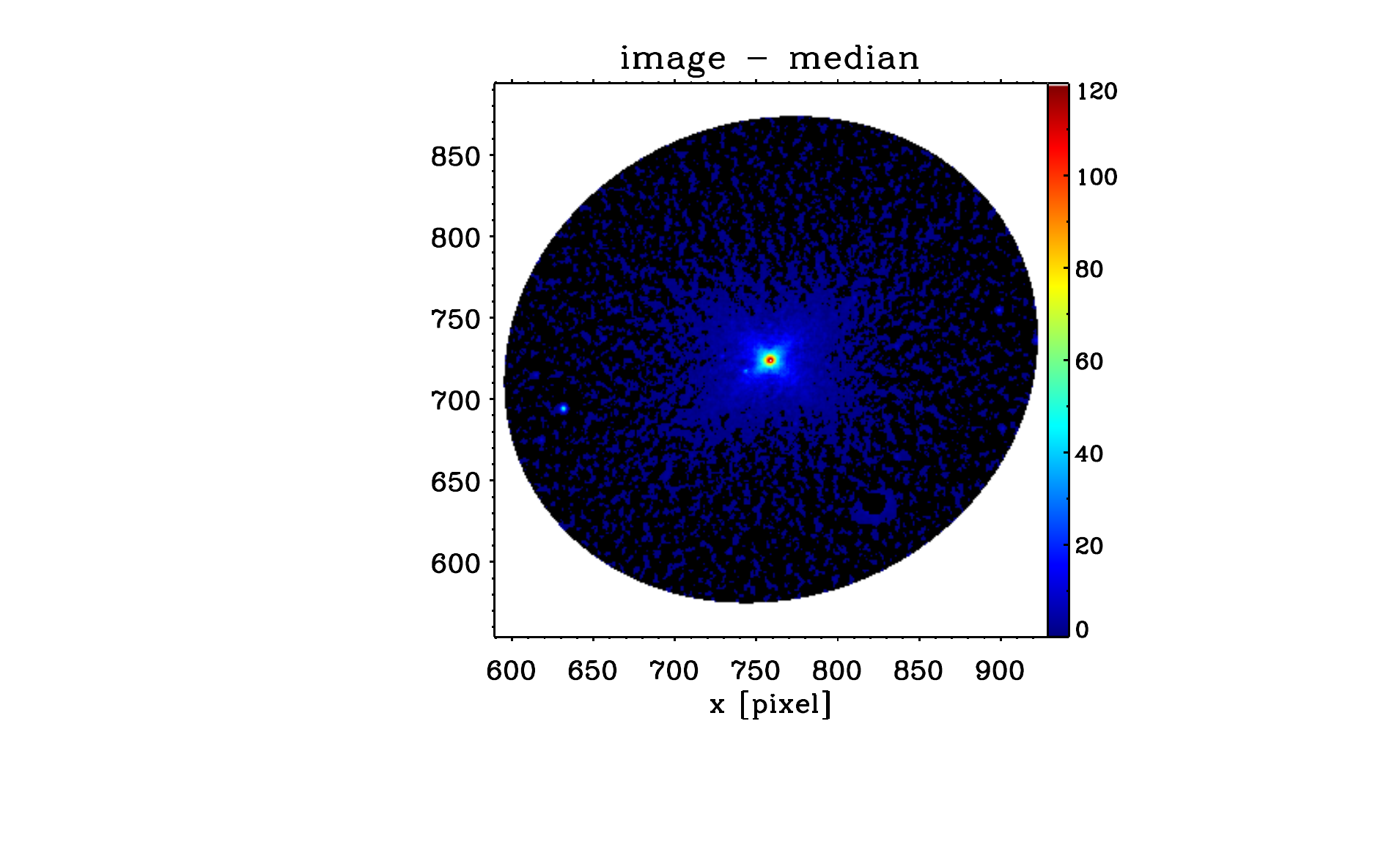} \\

\includegraphics[trim=0.7cm 0cm 0cm 0cm, clip=true, scale=0.46]{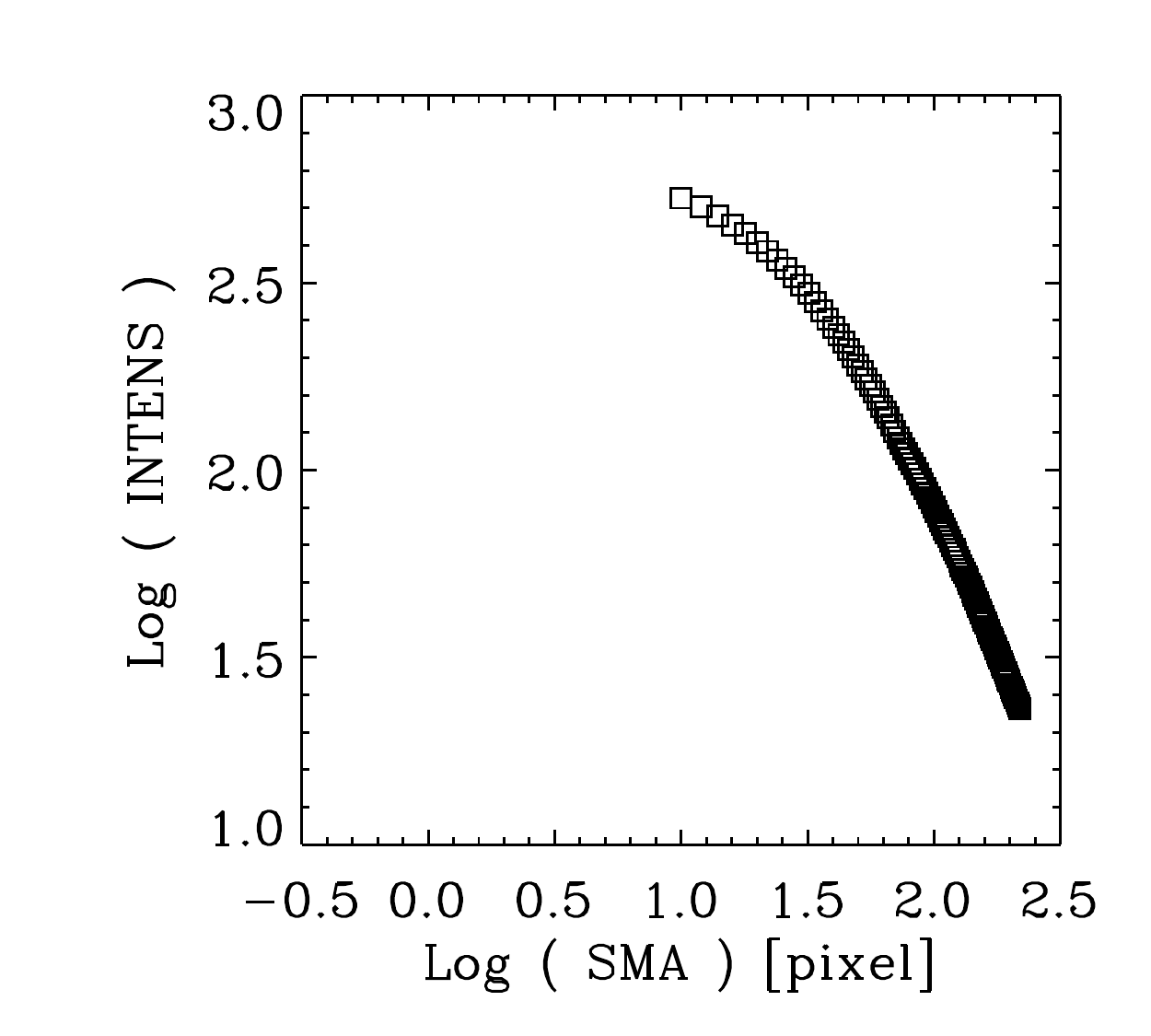}	    &  \includegraphics[trim=0.6cm 0cm 0cm 0cm, clip=true, scale=0.46]{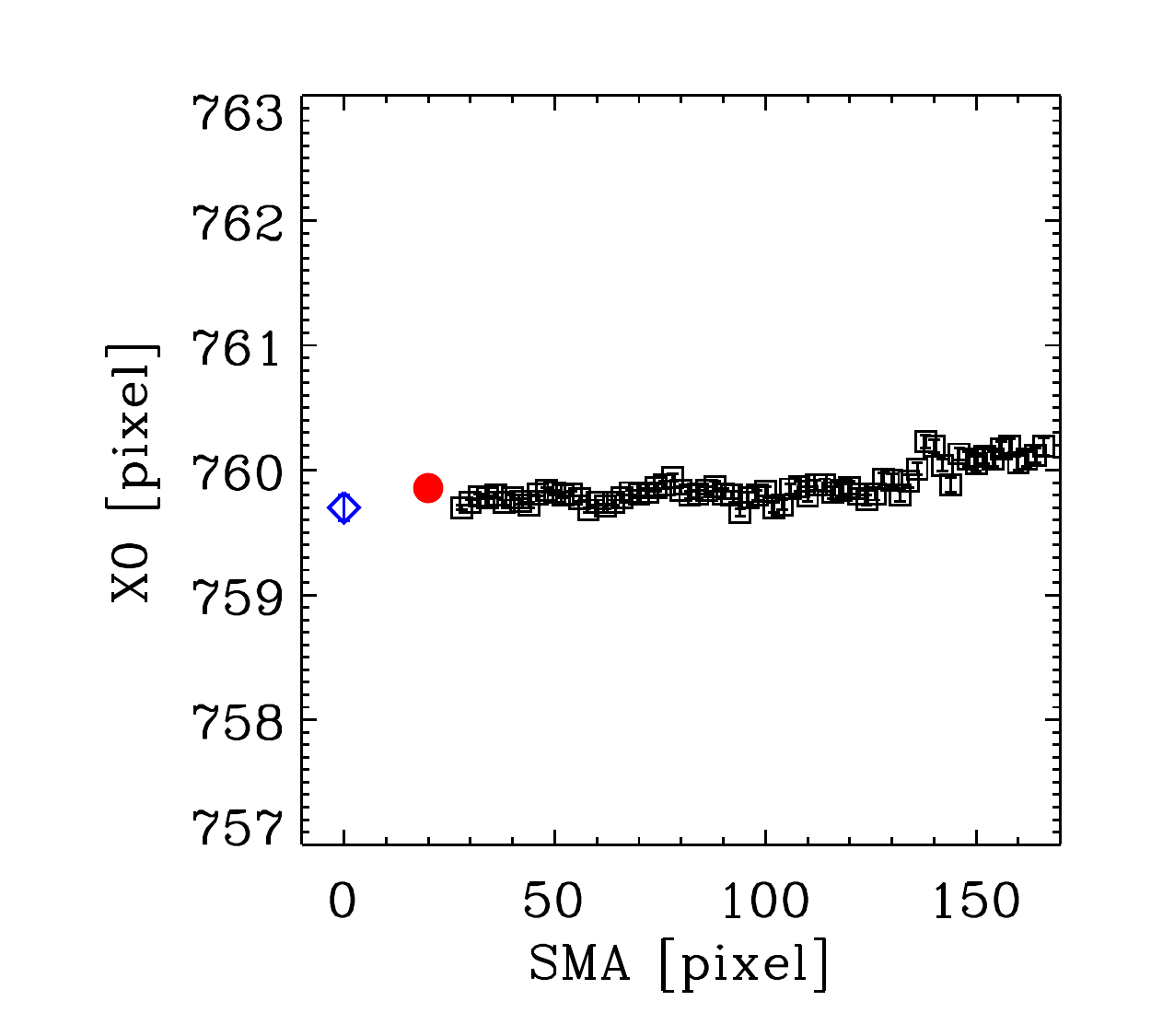}  &  \includegraphics[trim=0.6cm 0cm 0cm 0cm, clip=true, scale=0.46]{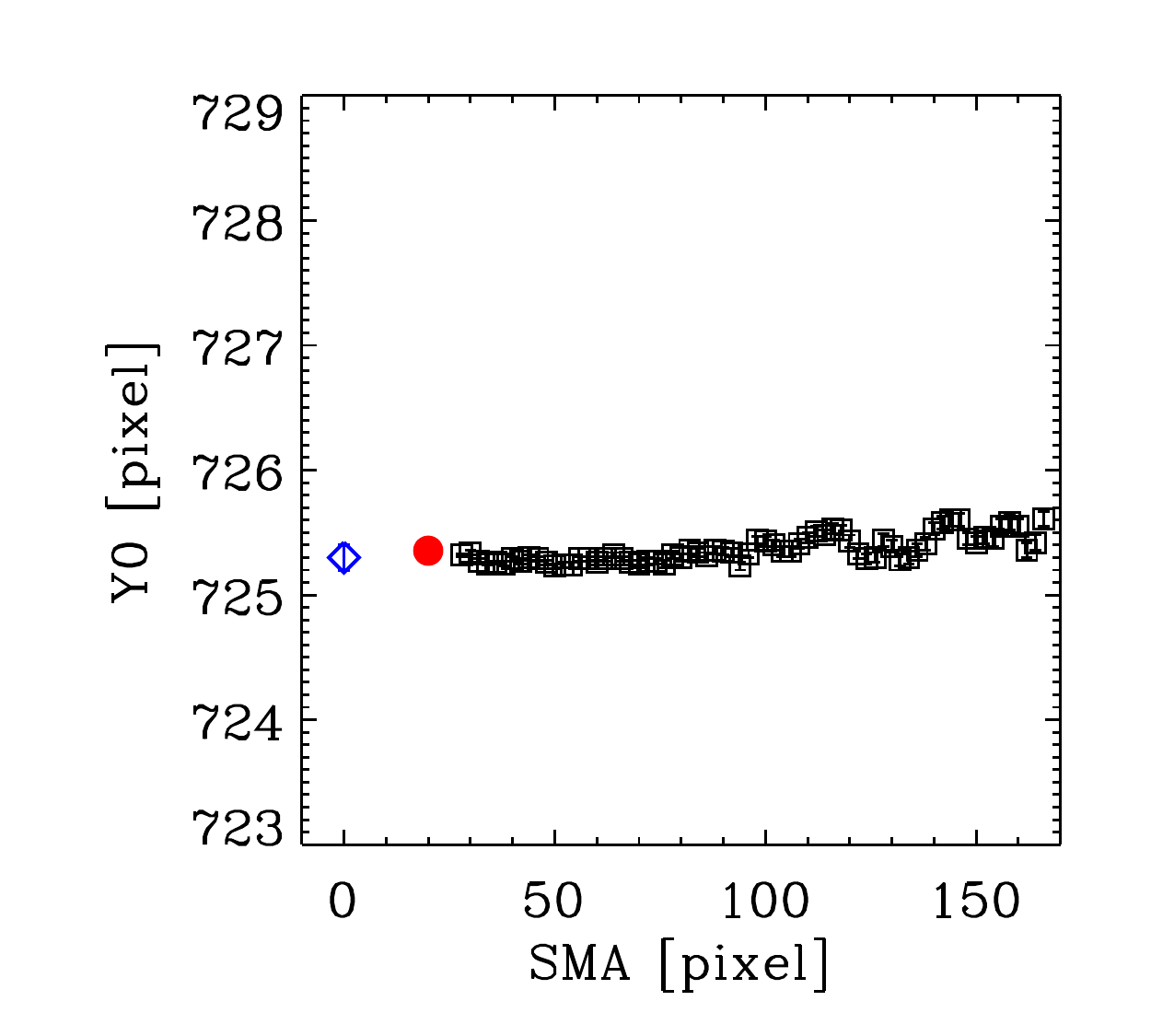} \\	

\includegraphics[trim=0.65cm 0cm 0cm 0cm, clip=true, scale=0.46]{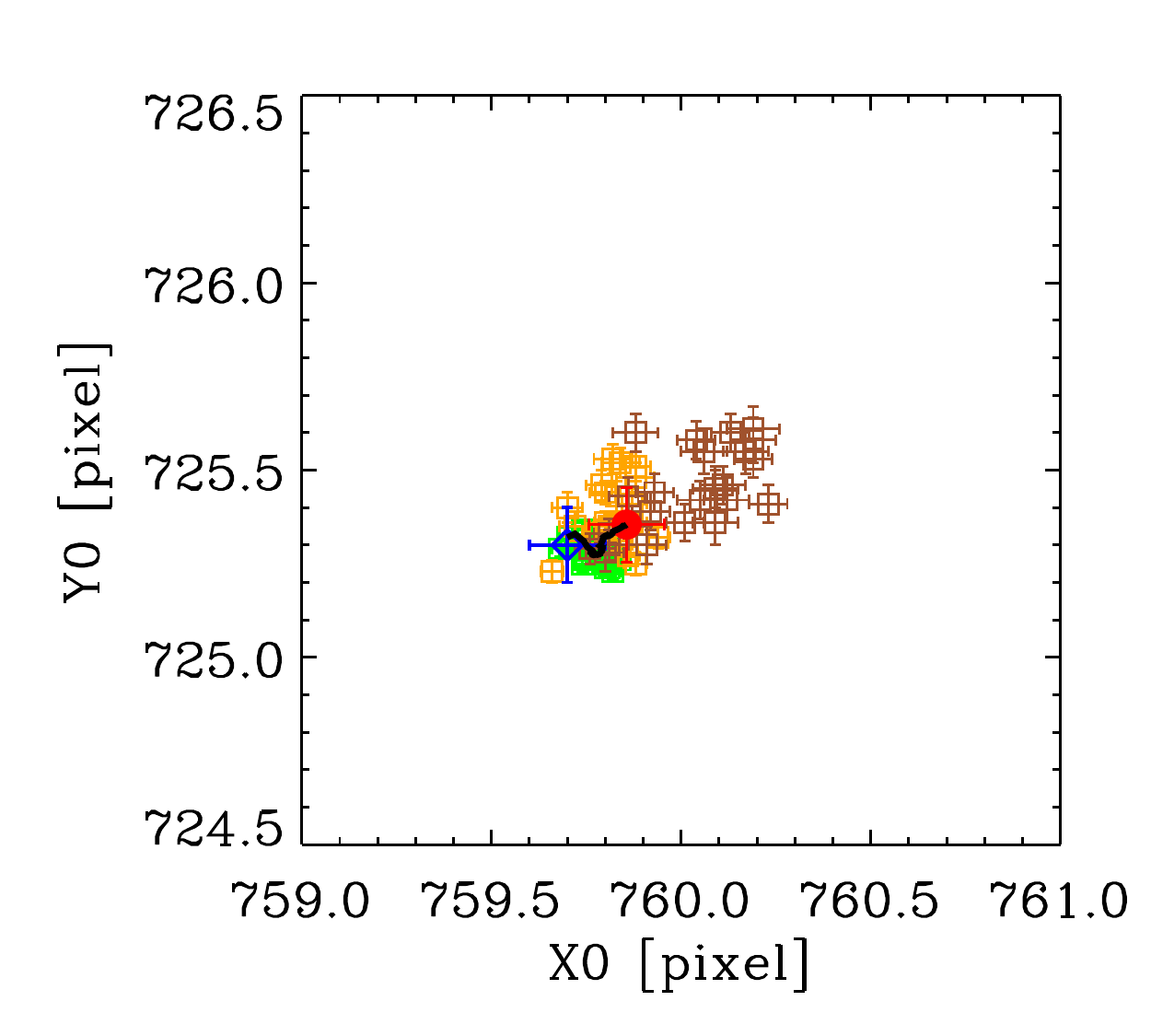}	&  \includegraphics[trim=0.6cm 0cm 0cm 0cm, clip=true, scale=0.46]{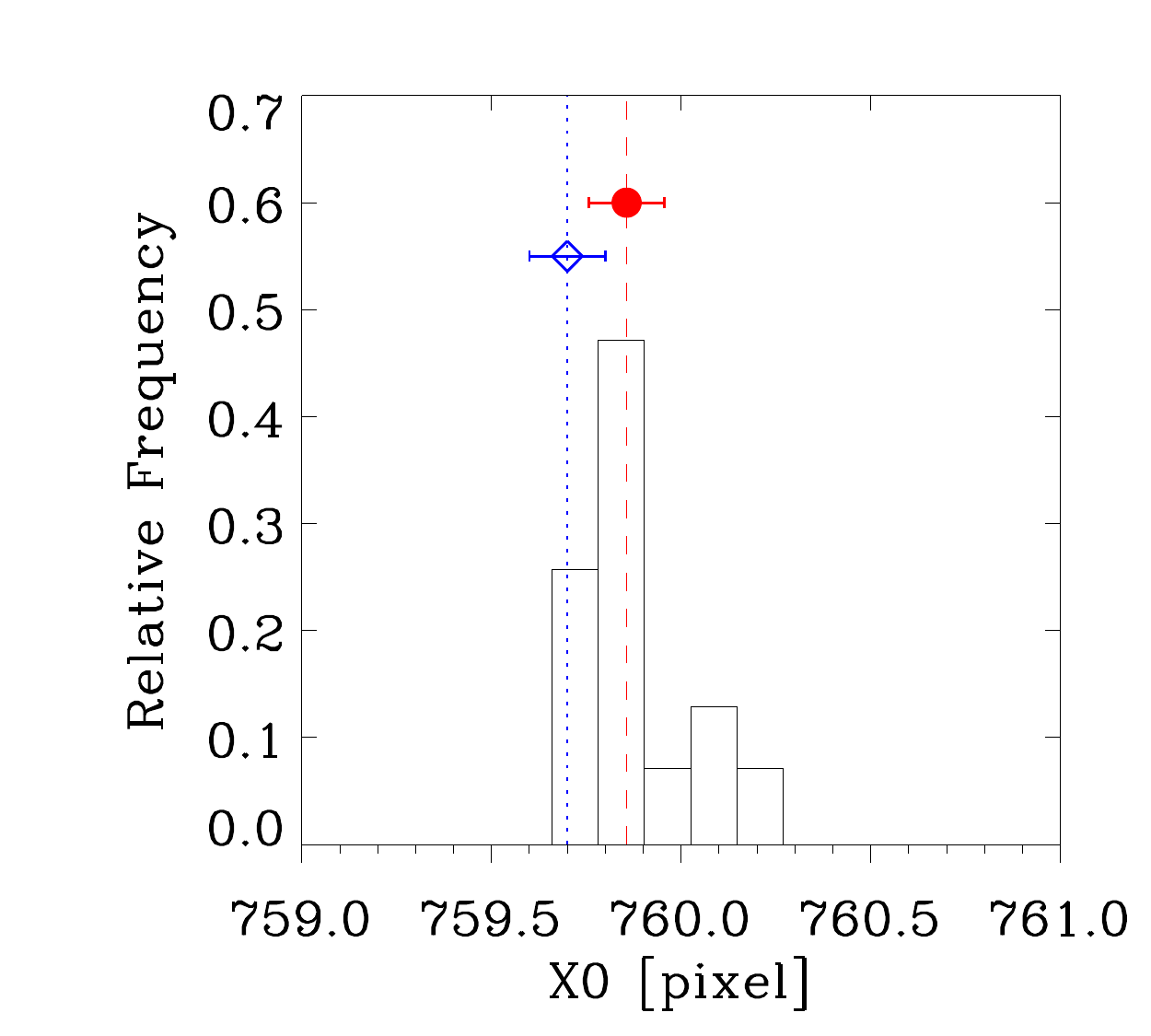}	& \includegraphics[trim=0.6cm 0cm 0cm 0cm, clip=true, scale=0.46]{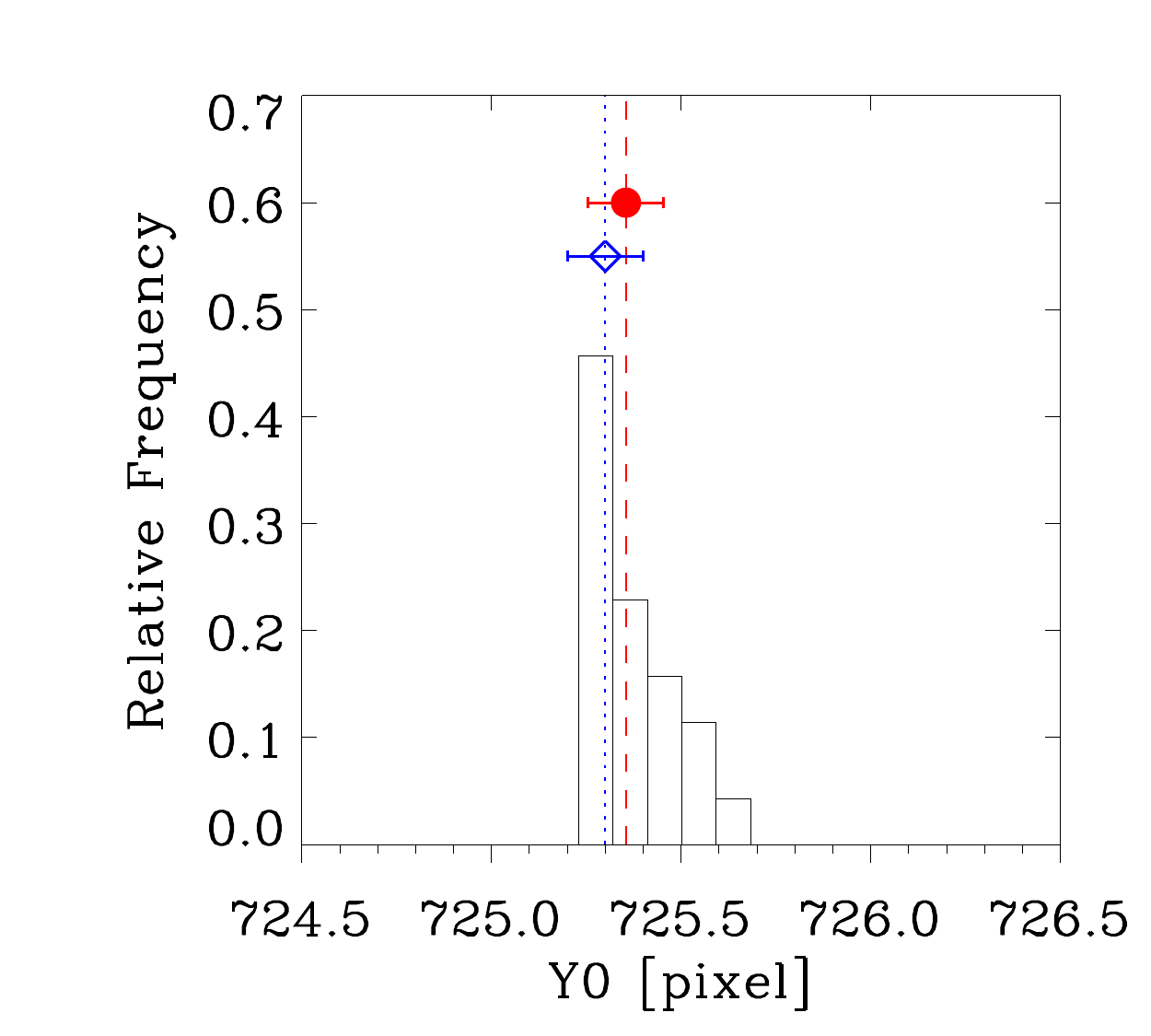}\\
\end{array}$
\end{center}
\caption[NGC 1399]{As in Fig.\ref{fig: NGC4373_W2} for galaxy NGC 1399, WFC3/IR - F110W, scale=$0\farcs09$/pxl.}
\label{fig: NGC1399_9pF110W}
\end{figure*} 

\begin{figure*}[h]
\begin{center}$
\begin{array}{ccc}
\includegraphics[trim=3.75cm 1cm 3cm 0cm, clip=true, scale=0.48]{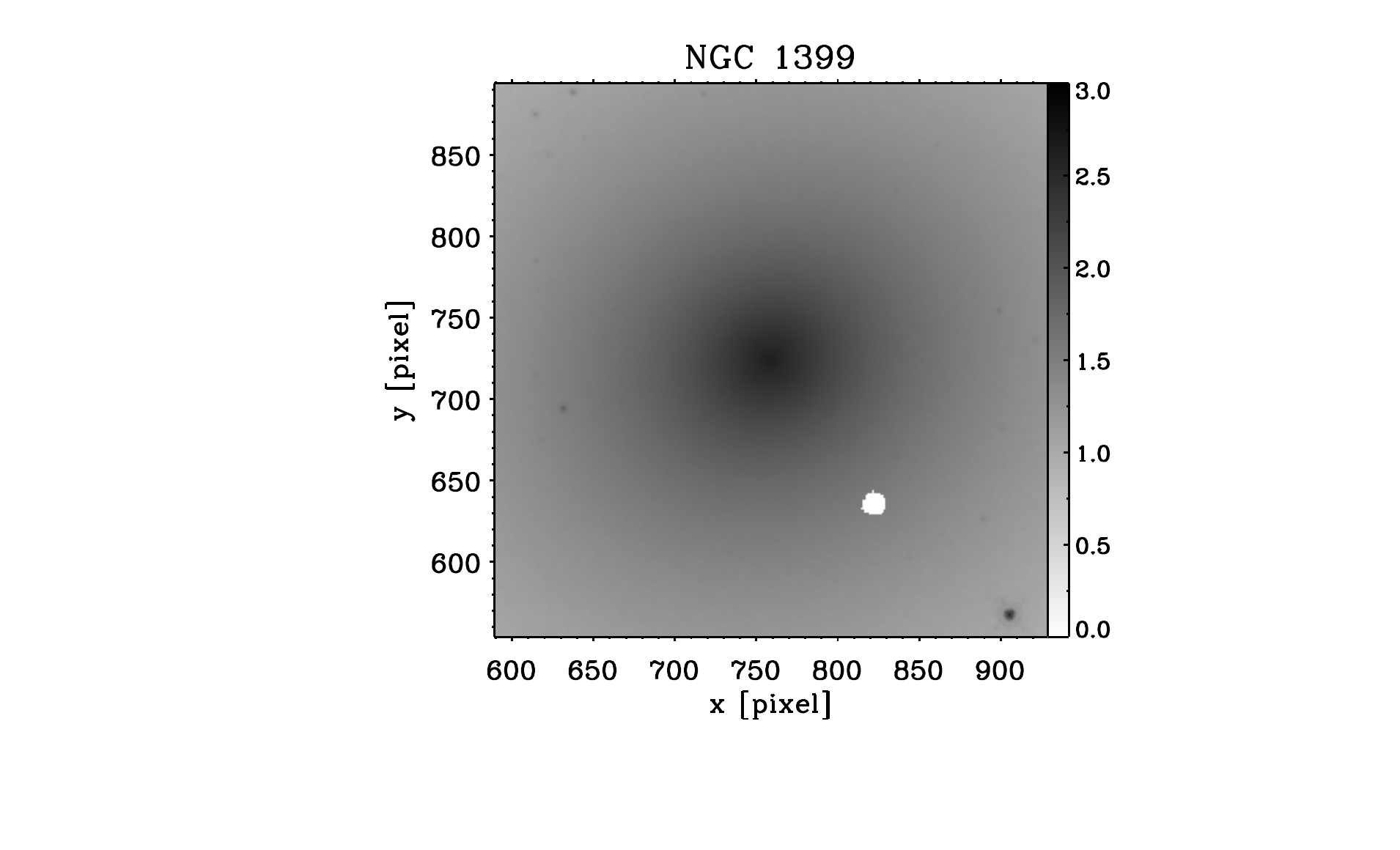} & \includegraphics[trim= 4.cm 1cm 3cm 0cm, clip=true, scale=0.48]{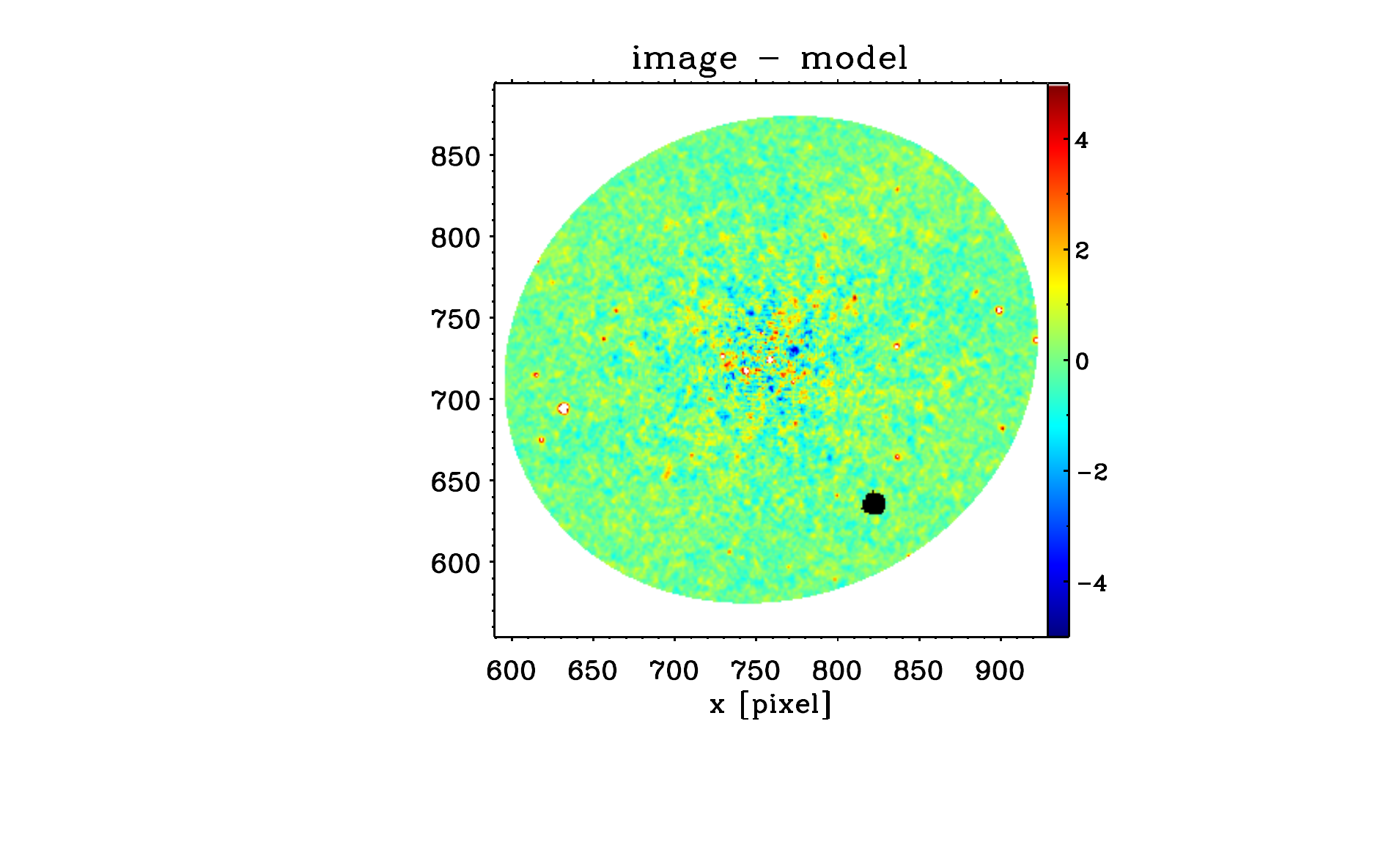}	& \includegraphics[trim= 4.cm 1cm 3cm 0cm, clip=true, scale=0.48]{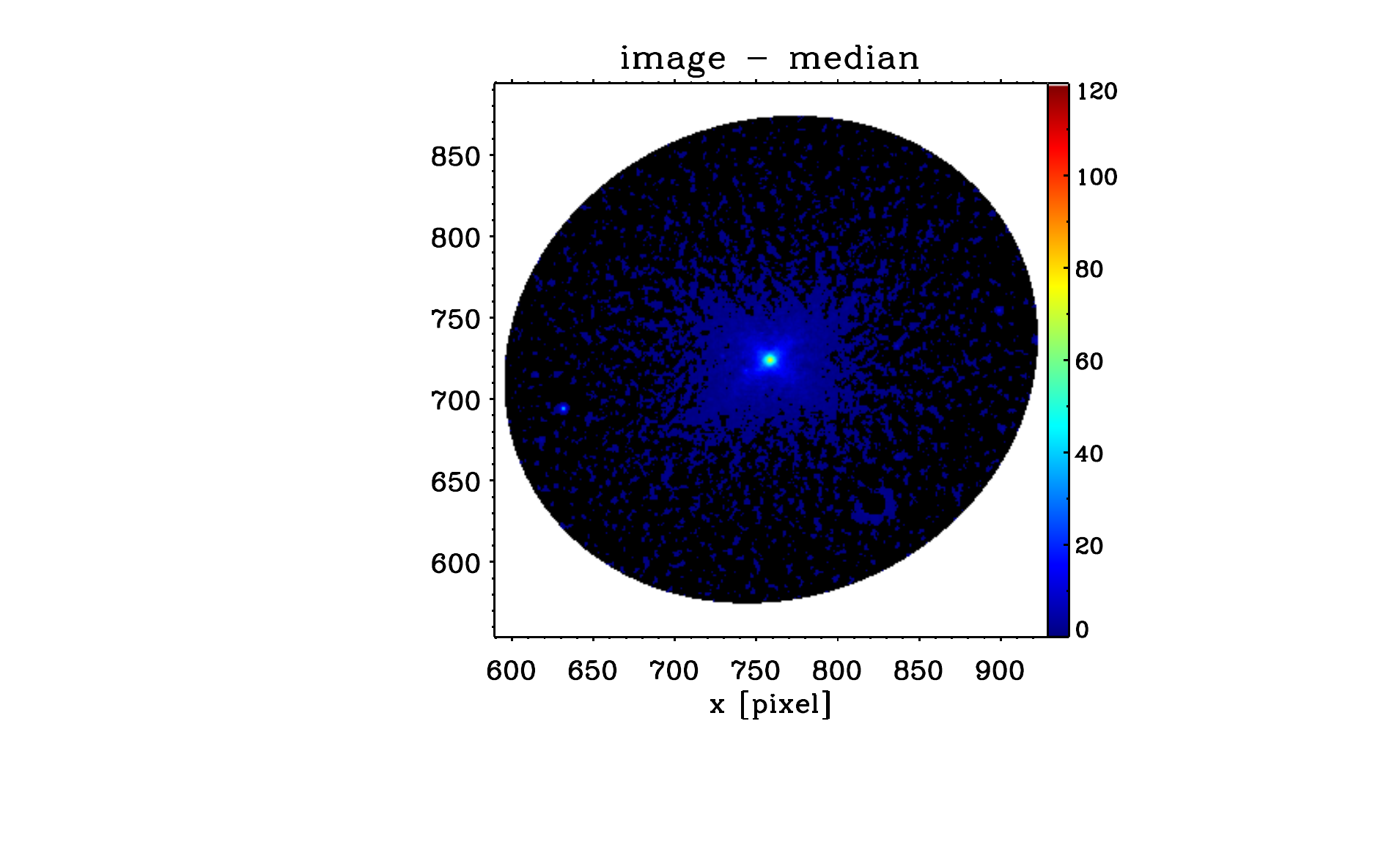} \\

\includegraphics[trim=0.7cm 0cm 0cm 0cm, clip=true, scale=0.46]{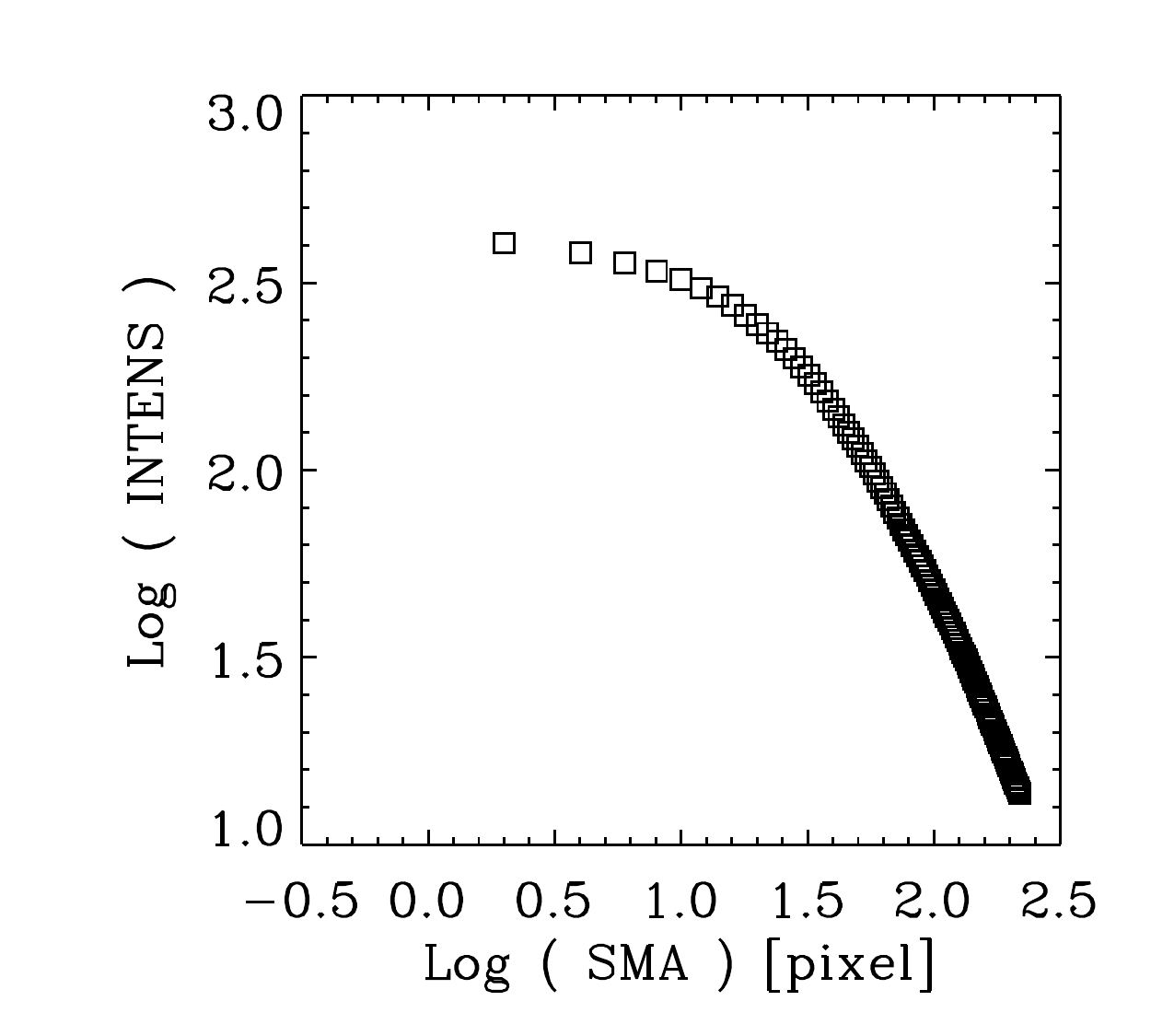}	    &  \includegraphics[trim=0.6cm 0cm 0cm 0cm, clip=true, scale=0.46]{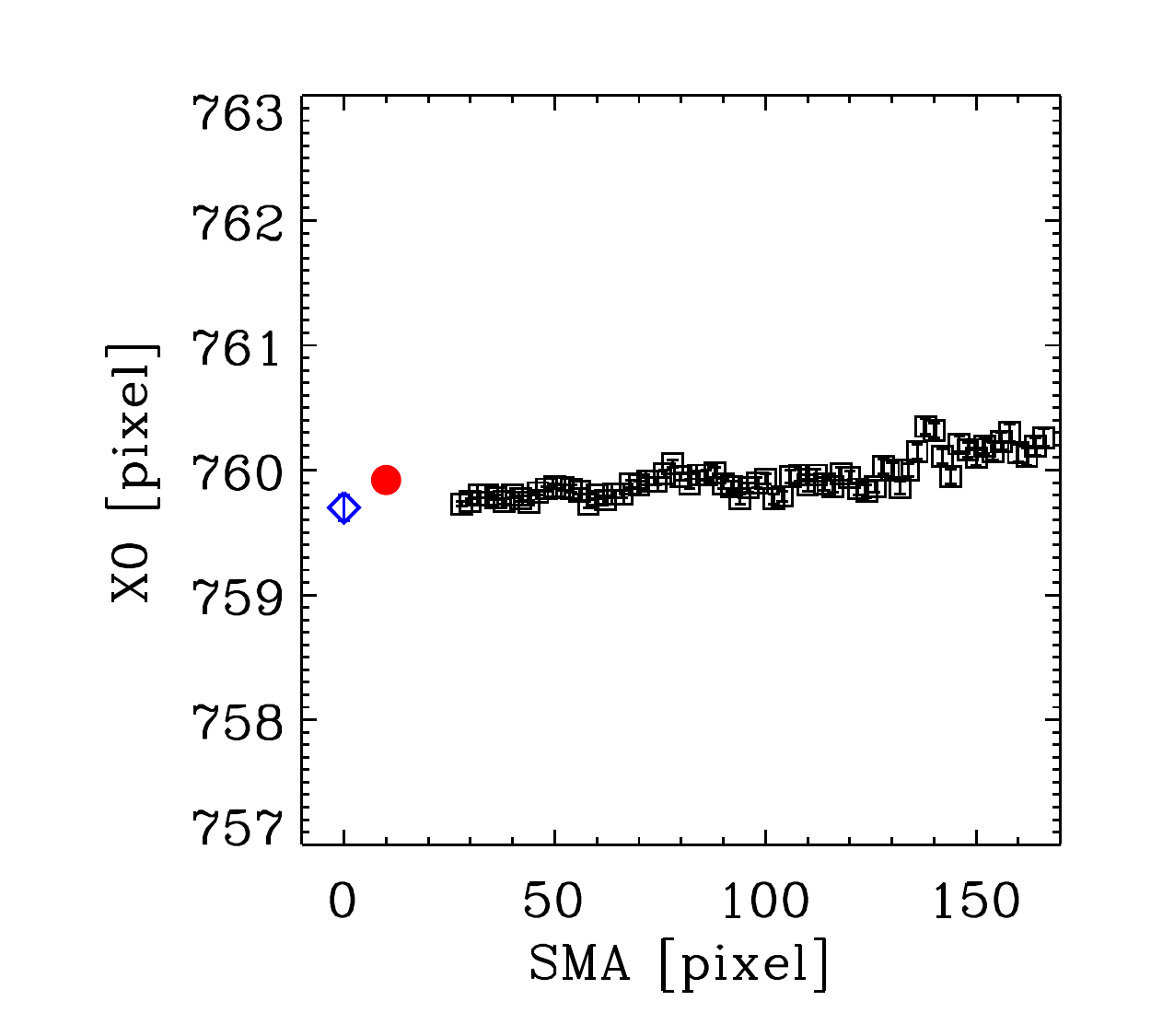}  &  \includegraphics[trim=0.6cm 0cm 0cm 0cm, clip=true, scale=0.46]{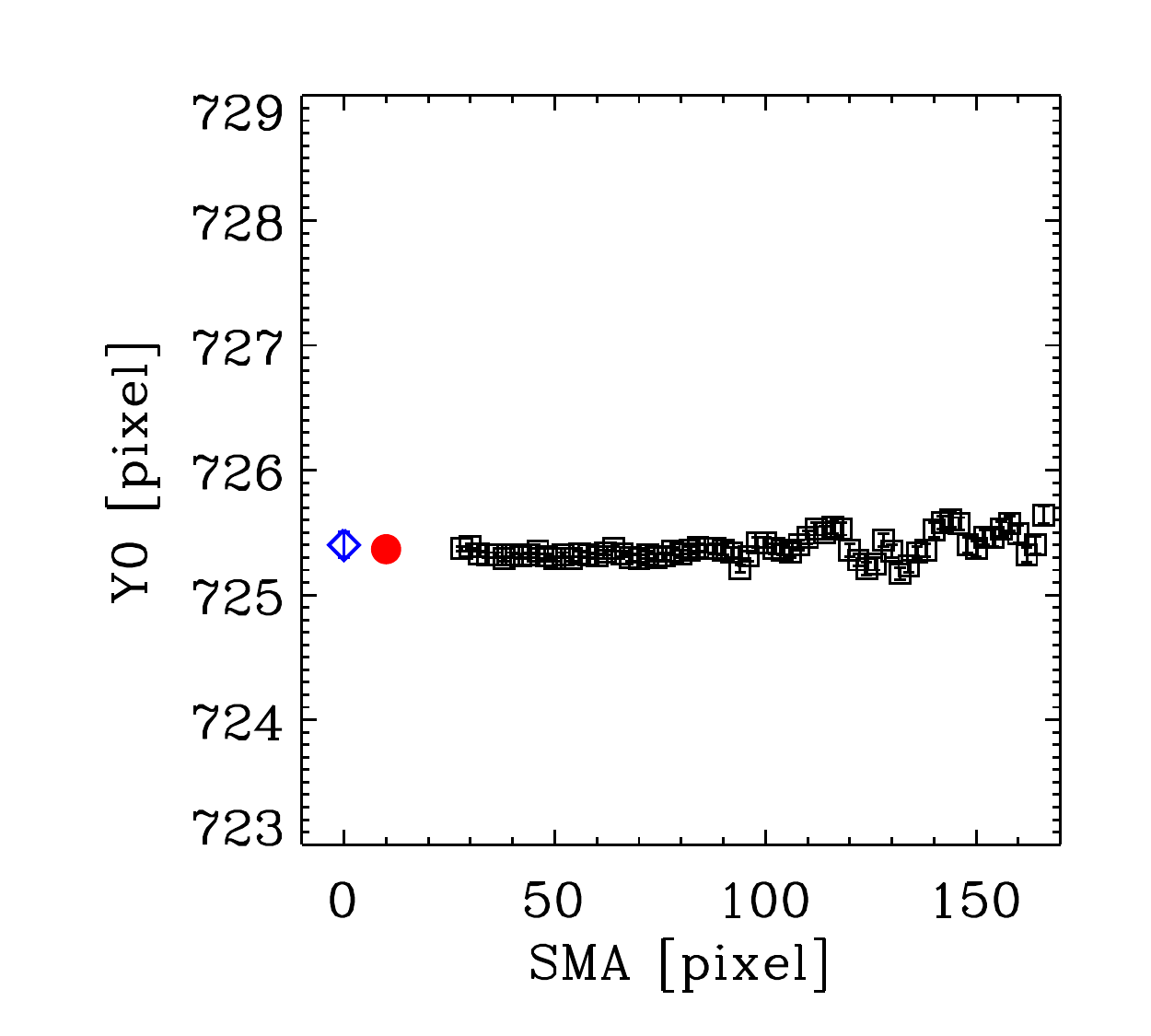} \\	

 \includegraphics[trim=0.65cm 0cm 0cm 0cm, clip=true, scale=0.46]{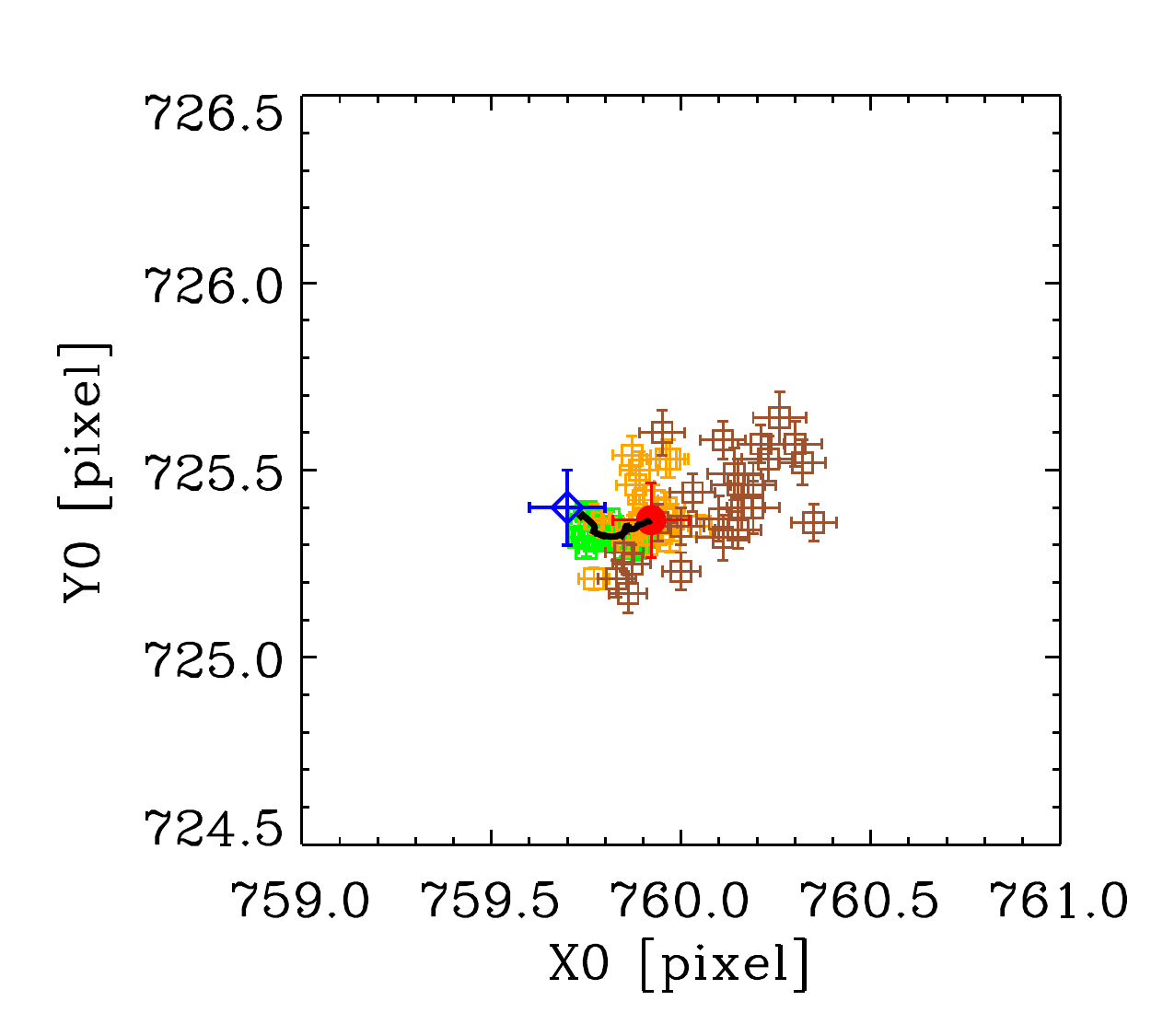}	&  \includegraphics[trim=0.6cm 0cm 0cm 0cm, clip=true, scale=0.46]{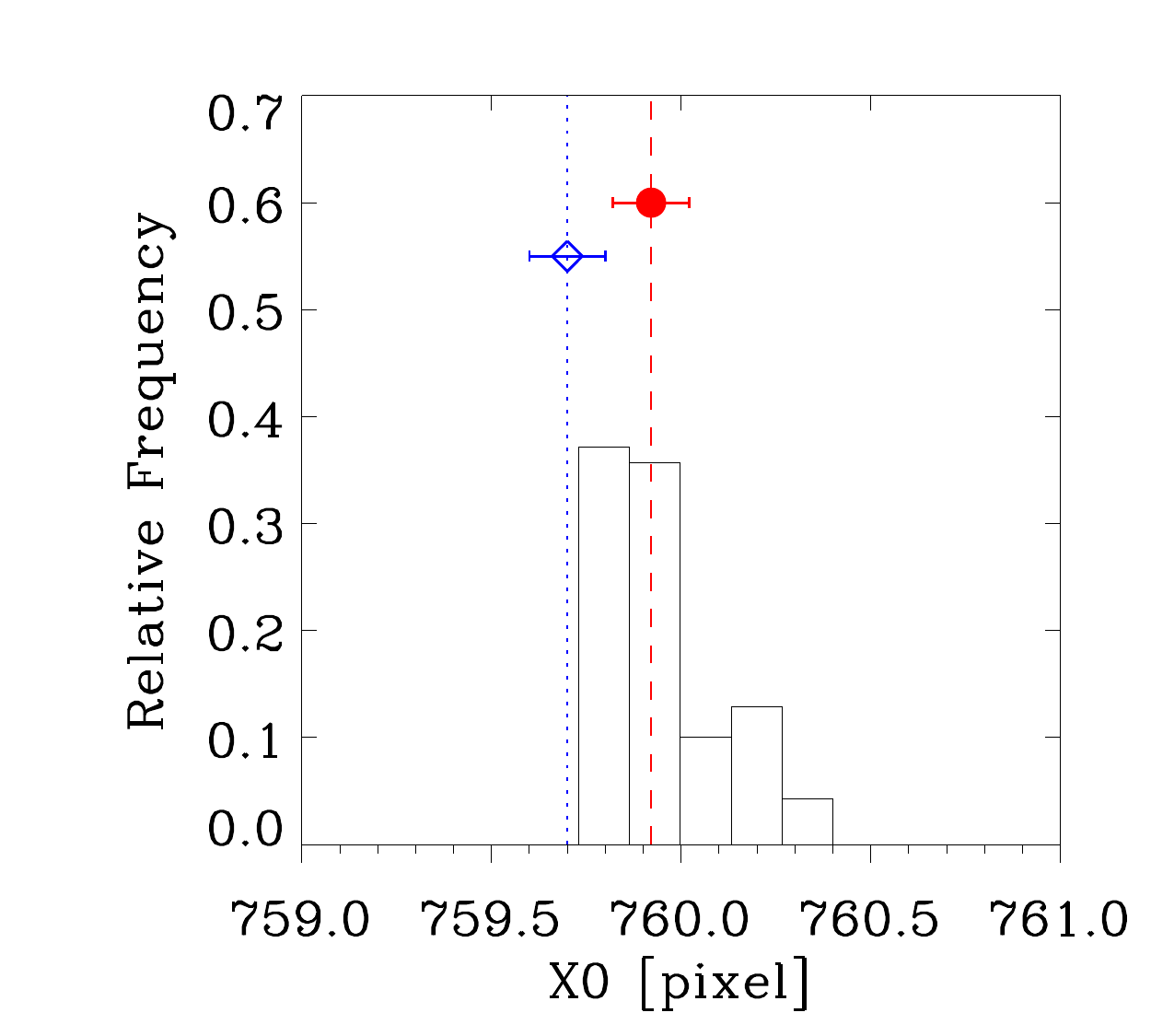}	& \includegraphics[trim=0.6cm 0cm 0cm 0cm, clip=true, scale=0.46]{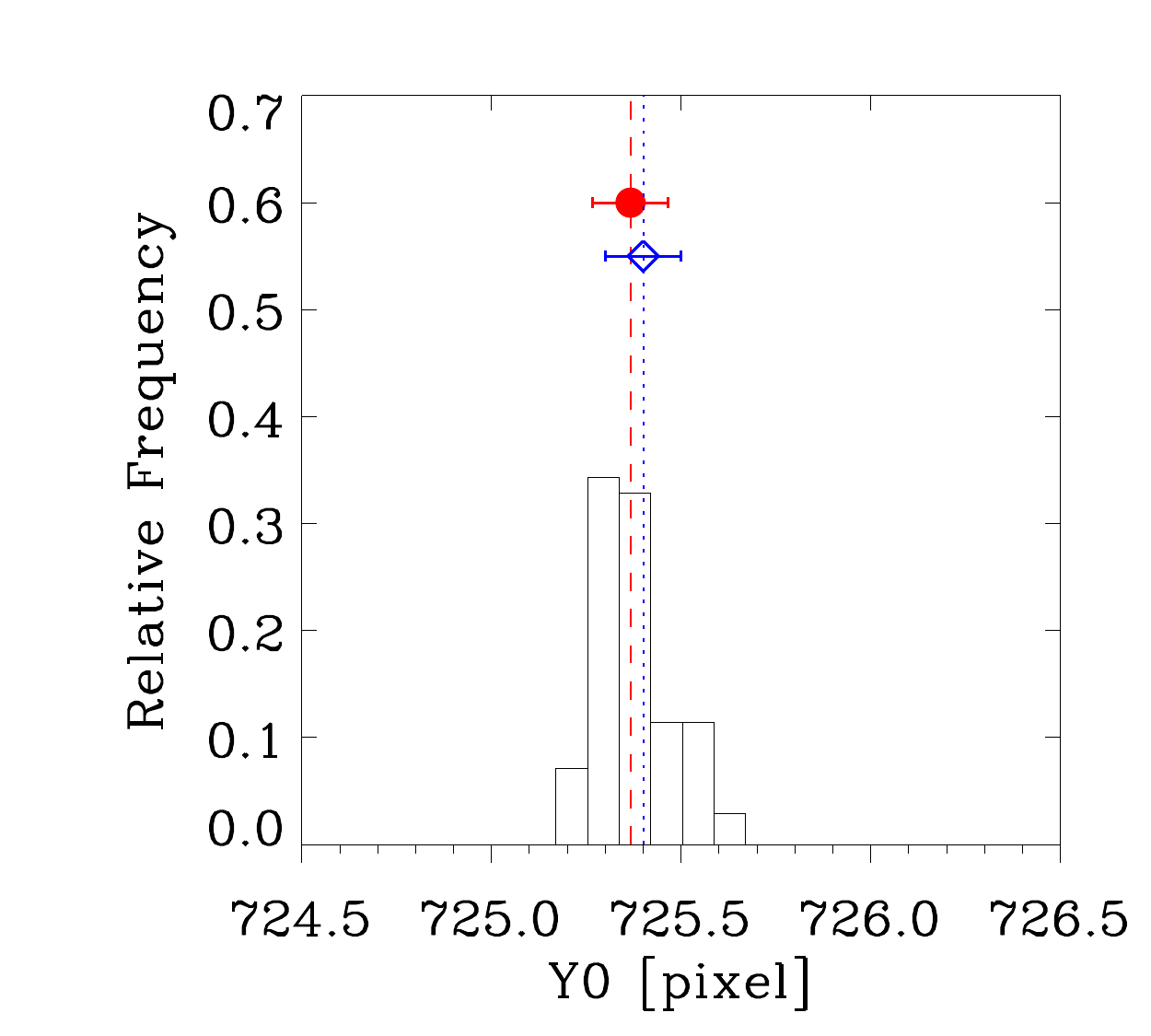}\\
\end{array}$
\end{center}
\caption[NGC 1399]{As in Fig.\ref{fig: NGC4373_W2} for galaxy NGC 1399, WFC3/IR - F160W, scale=$0\farcs09$/pxl.}
\label{fig: NGC1399_9pF160W}
\end{figure*} 

\begin{figure*}[h]
\begin{center}$
\begin{array}{ccc}
\includegraphics[trim=3.75cm 1cm 3cm 0cm, clip=true, scale=0.48]{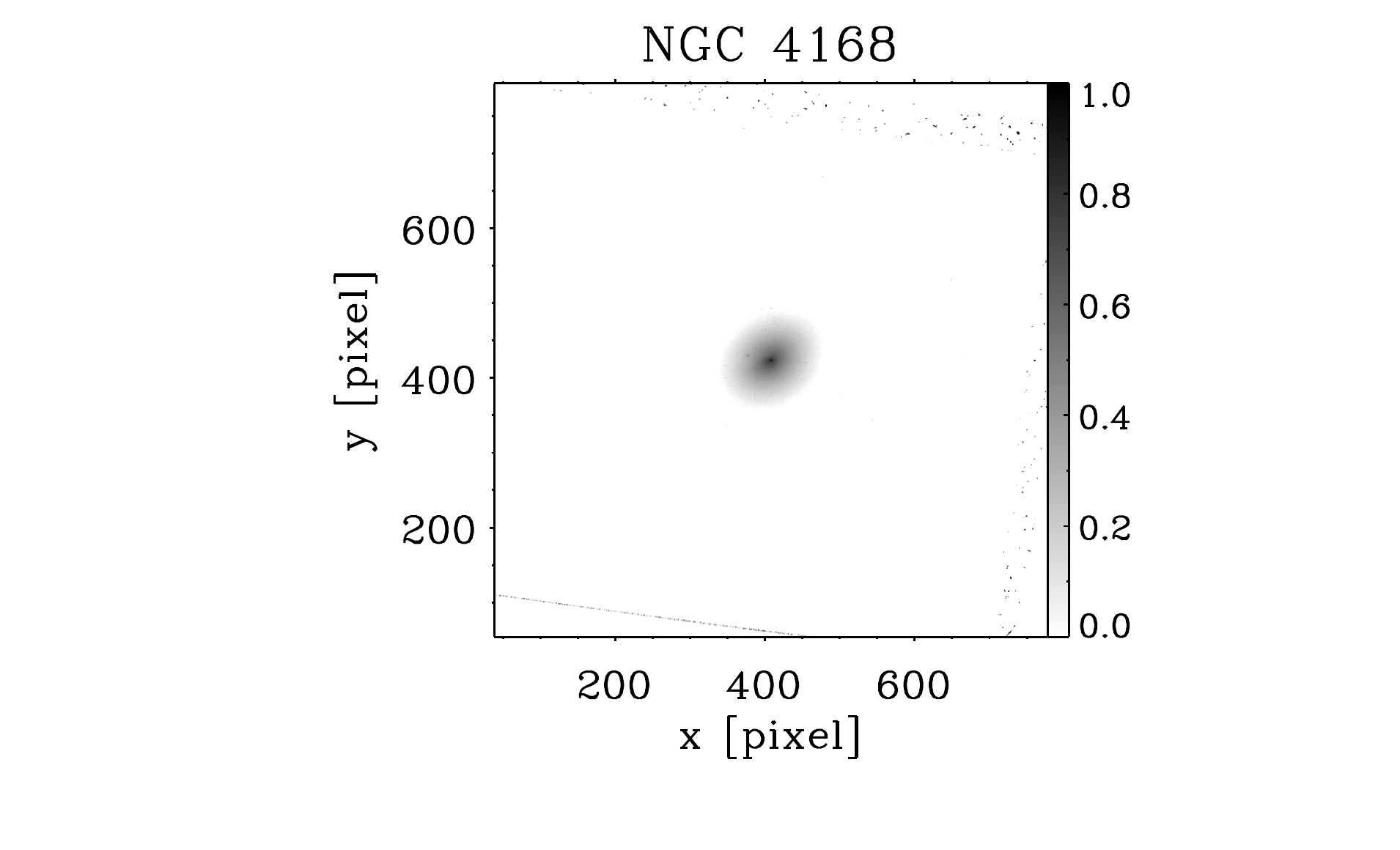} & \includegraphics[trim= 4.cm 1cm 3cm 0cm, clip=true, scale=0.48]{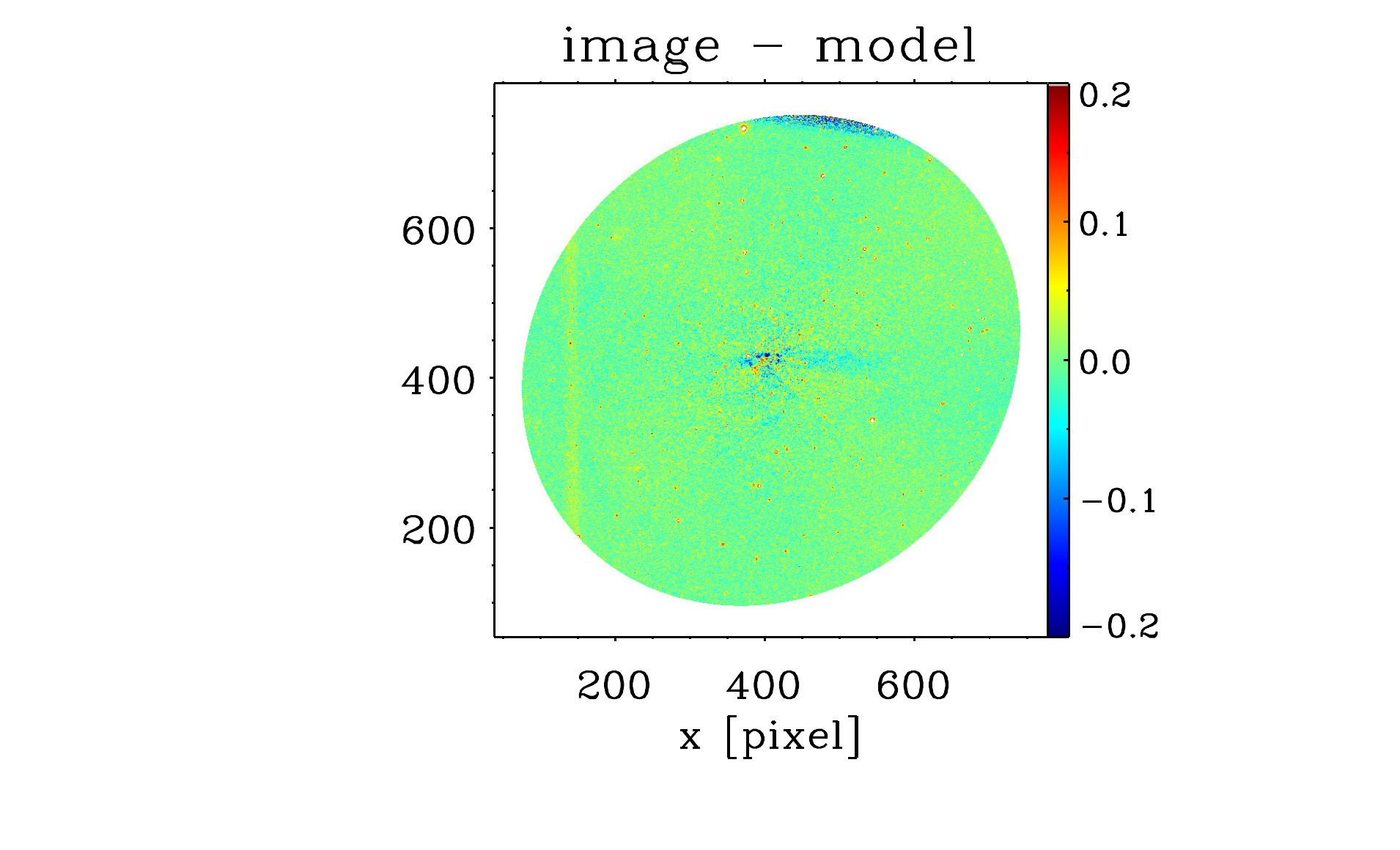}	& \includegraphics[trim= 4.cm 1cm 3cm 0cm, clip=true, scale=0.48]{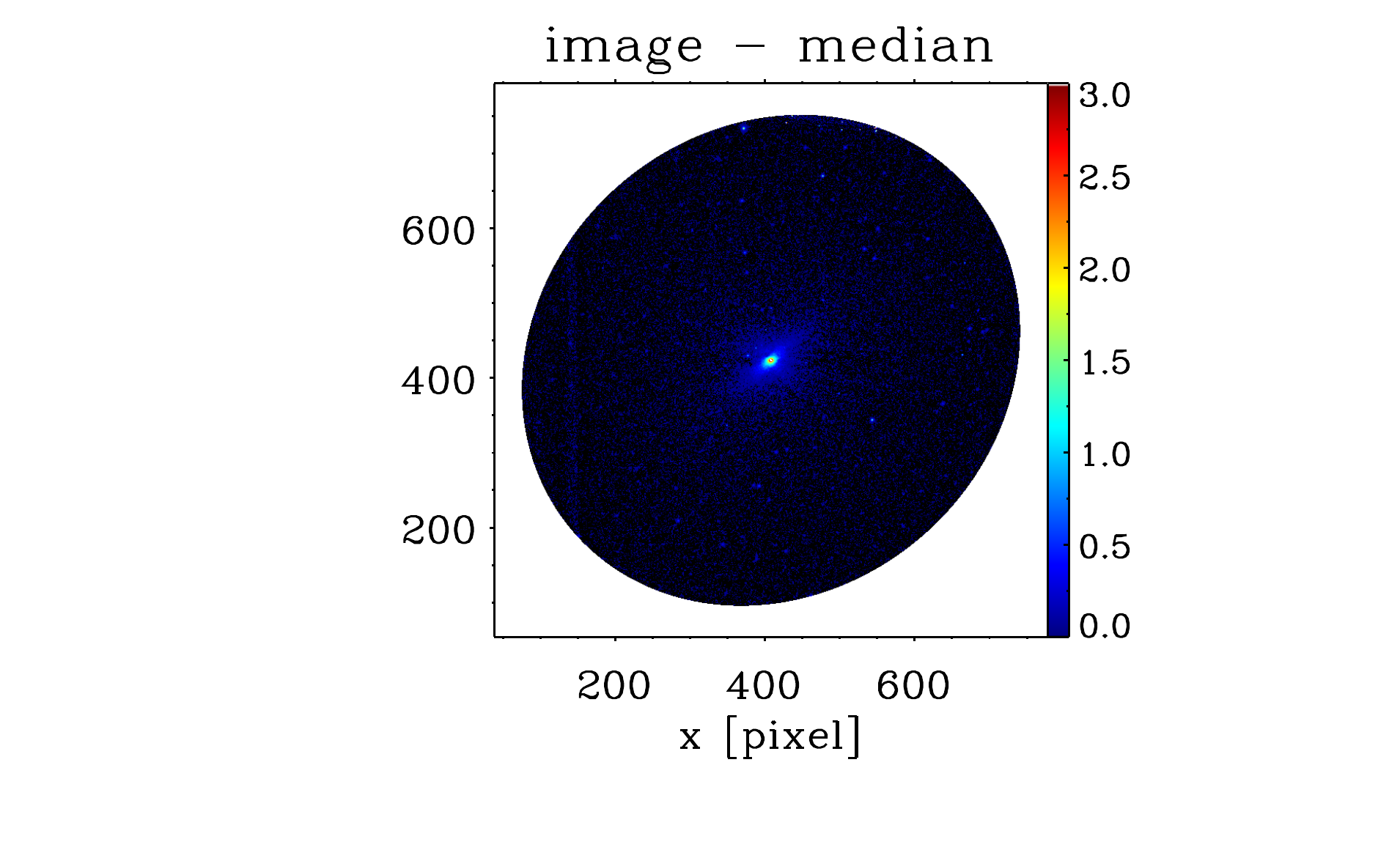} \\
\includegraphics[trim=0.7cm 0cm 0cm 0cm, clip=true, scale=0.46]{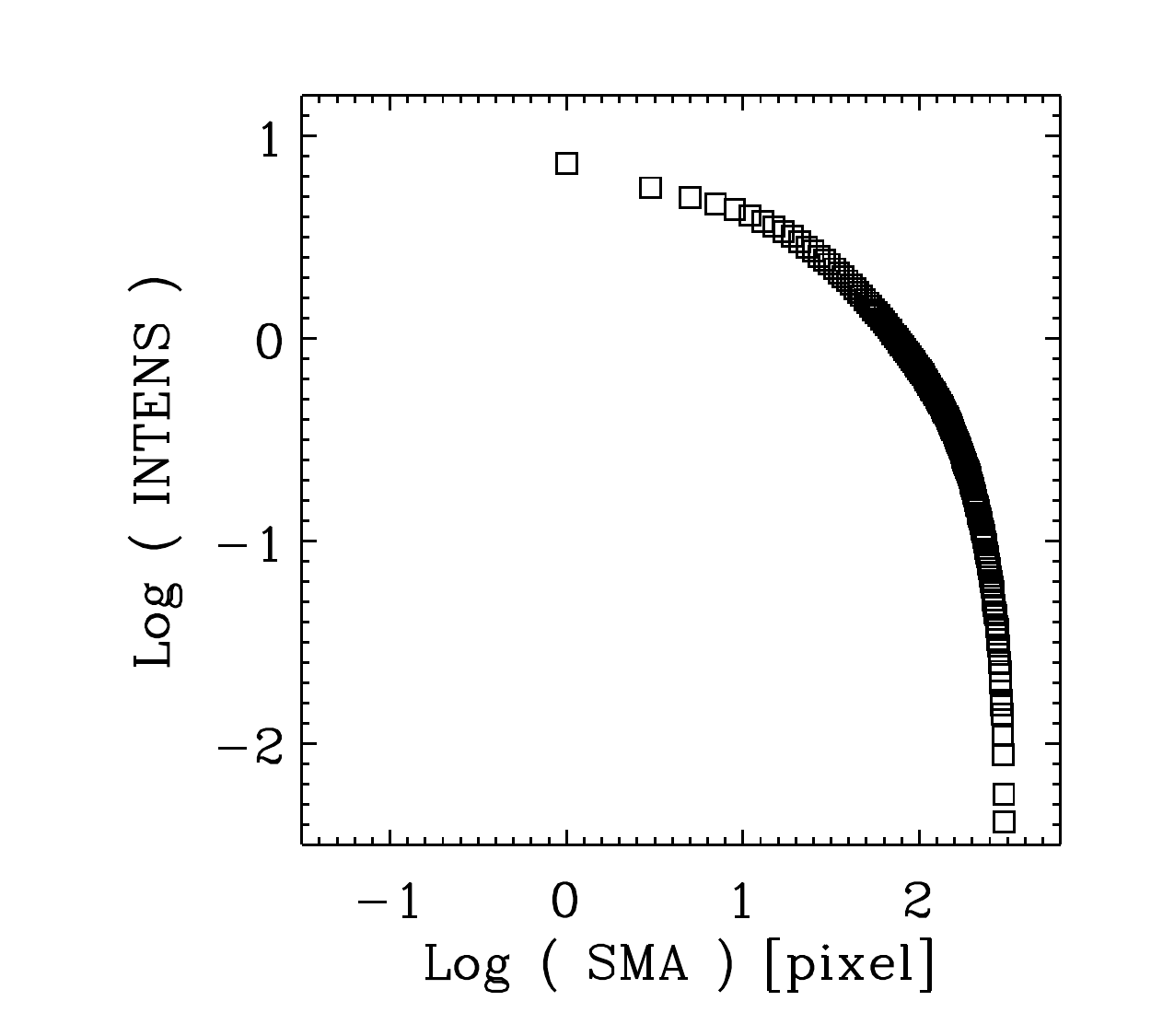}	    &  \includegraphics[trim=0.6cm 0cm 0cm 0cm, clip=true, scale=0.46]{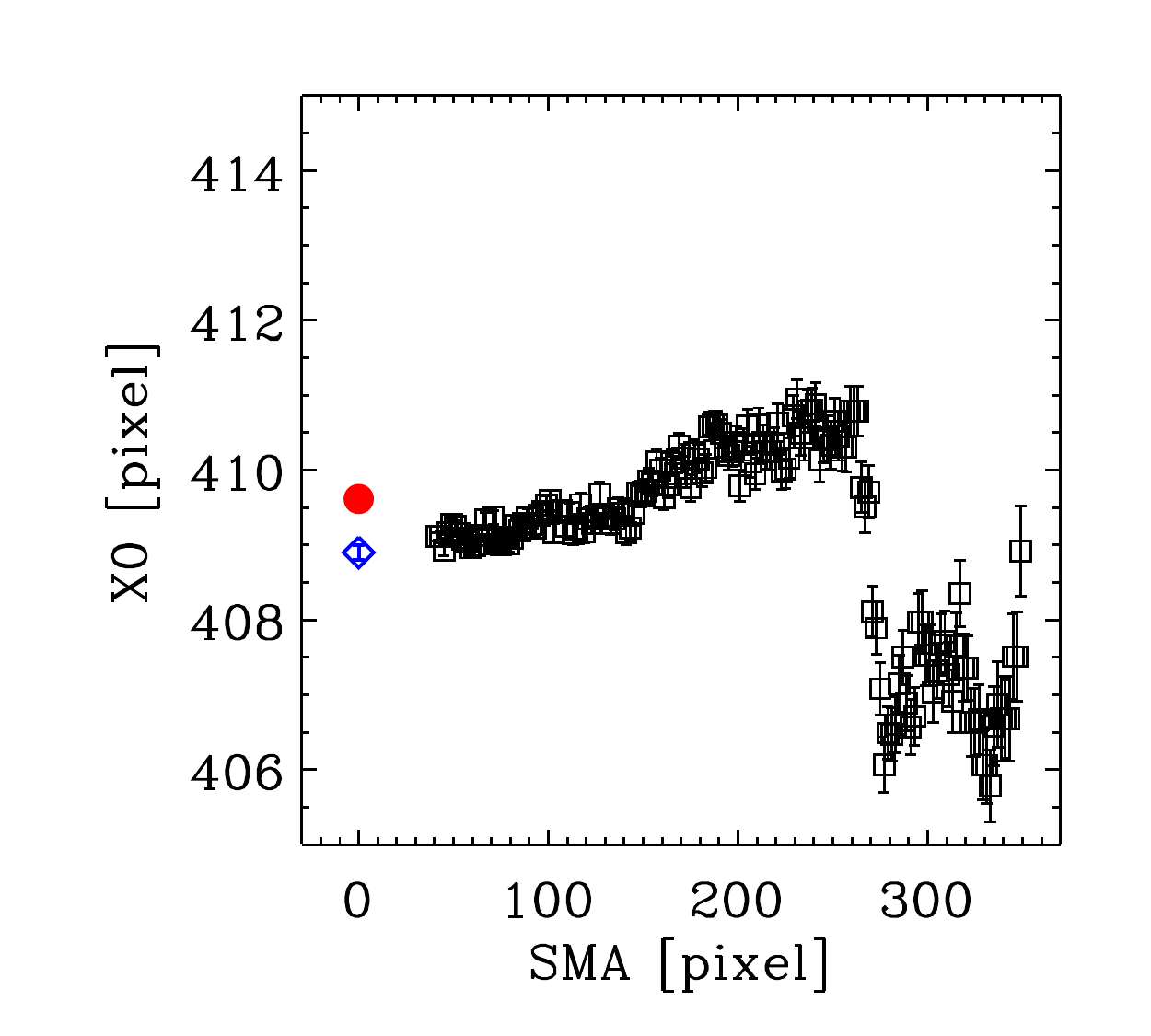}  &  \includegraphics[trim=0.6cm 0cm 0cm 0cm, clip=true, scale=0.46]{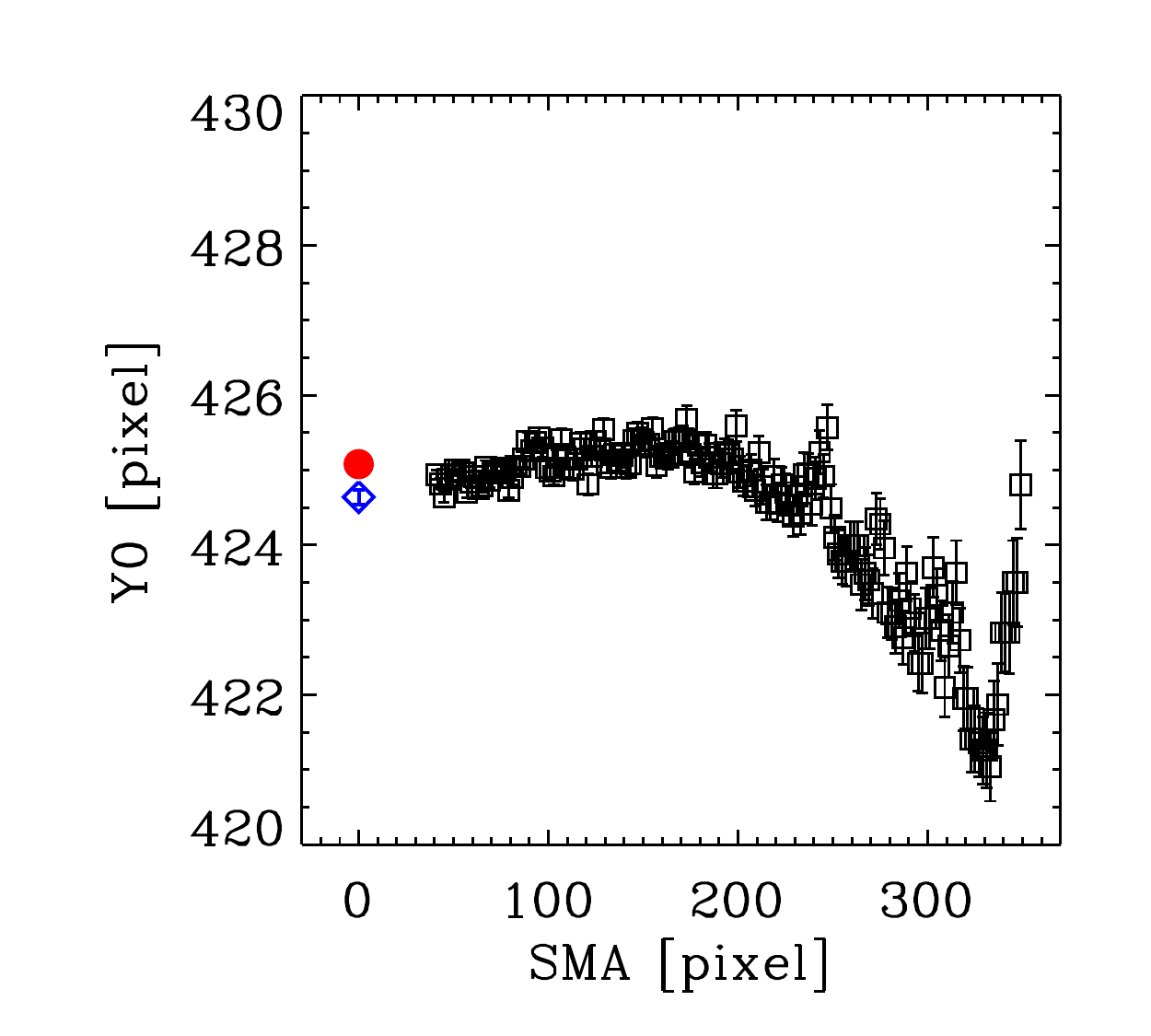} \\	
 \includegraphics[trim=0.65cm 0cm 0cm 0cm, clip=true, scale=0.46]{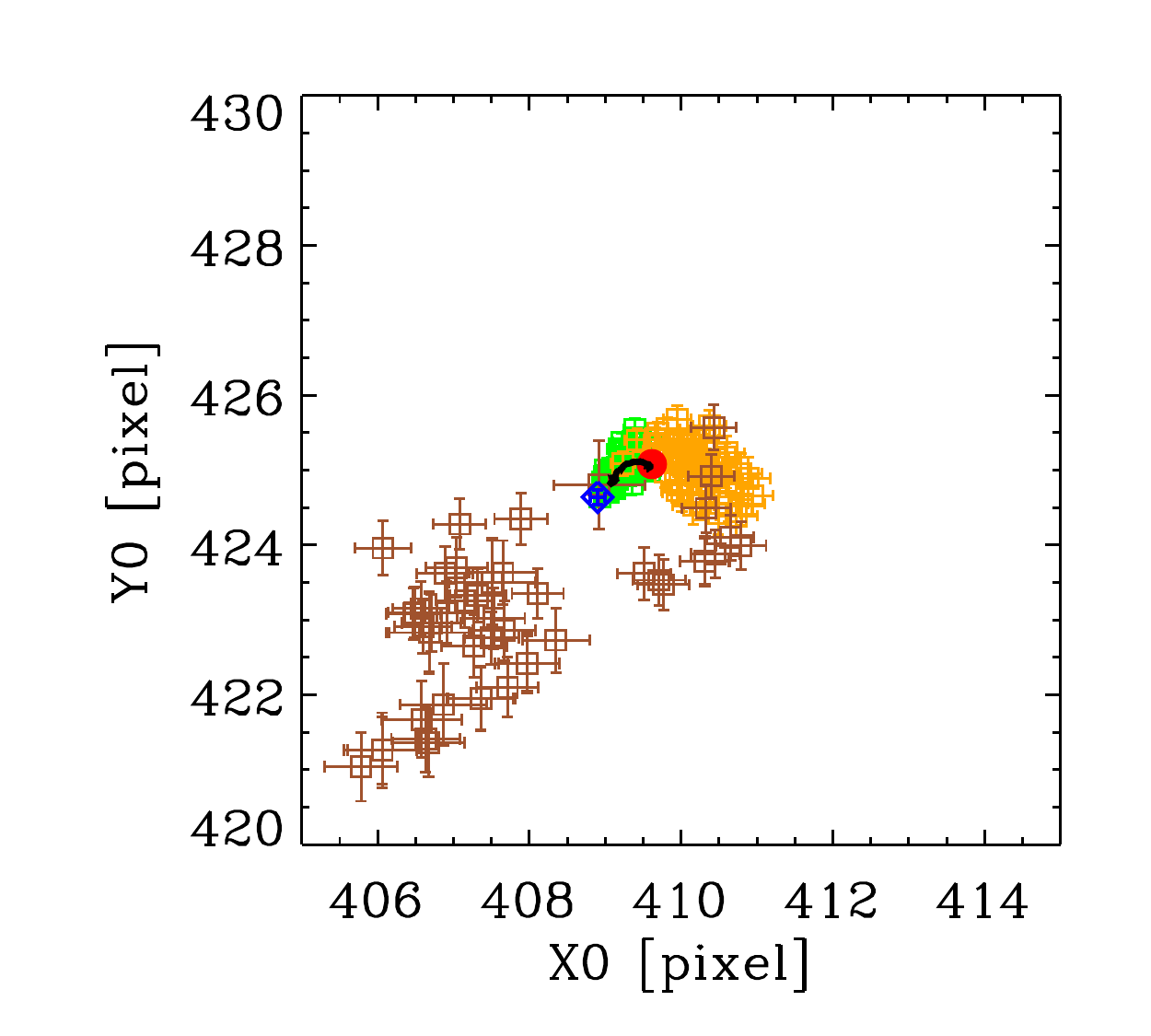}	&  \includegraphics[trim=0.6cm 0cm 0cm 0cm, clip=true, scale=0.46]{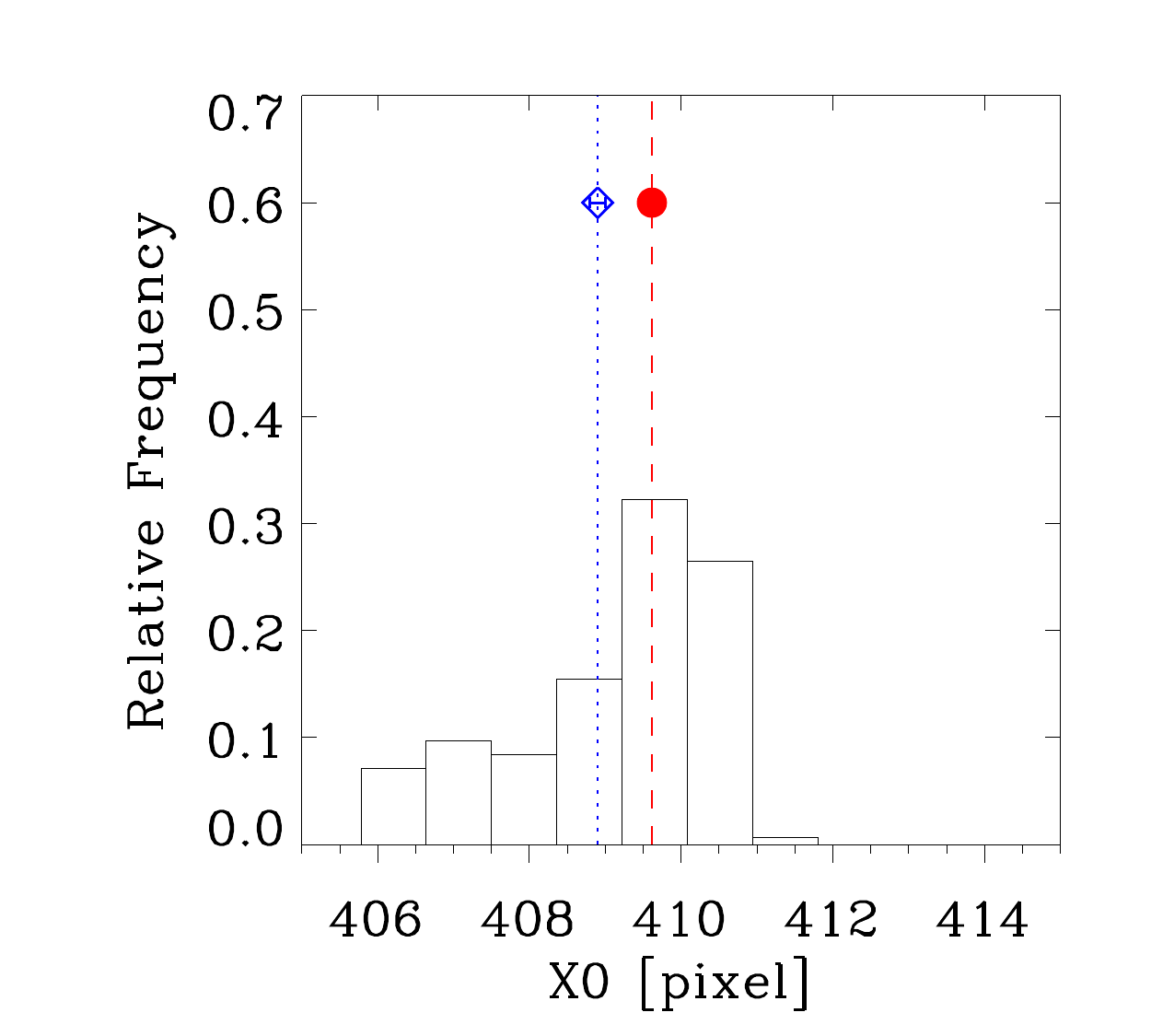}	& \includegraphics[trim=0.6cm 0cm 0cm 0cm, clip=true, scale=0.46]{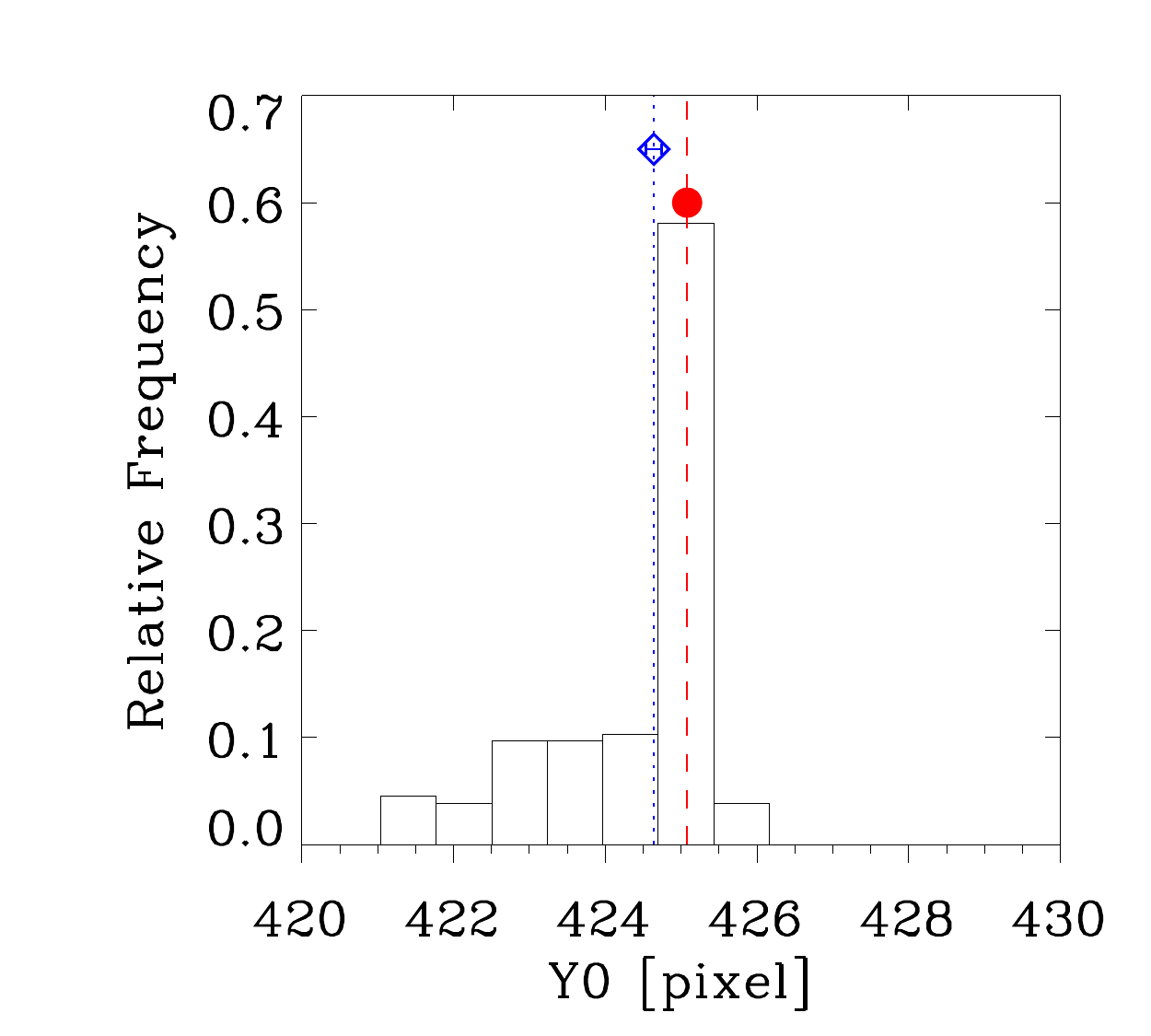}\\
\end{array}$
\end{center}
\caption[NGC 4168]{As in Fig.\ref{fig: NGC4373_W2} for galaxy NGC 4168, WFPC2/PC - F702W, scale=$0\farcs05$/pxl.}
\label{fig: NGC4168_9p}
\end{figure*} 

\begin{figure*}[hp]
\begin{center}$
\begin{array}{ccc}
\includegraphics[trim=3.75cm 1cm 3cm 0cm, clip=true, scale=0.48]{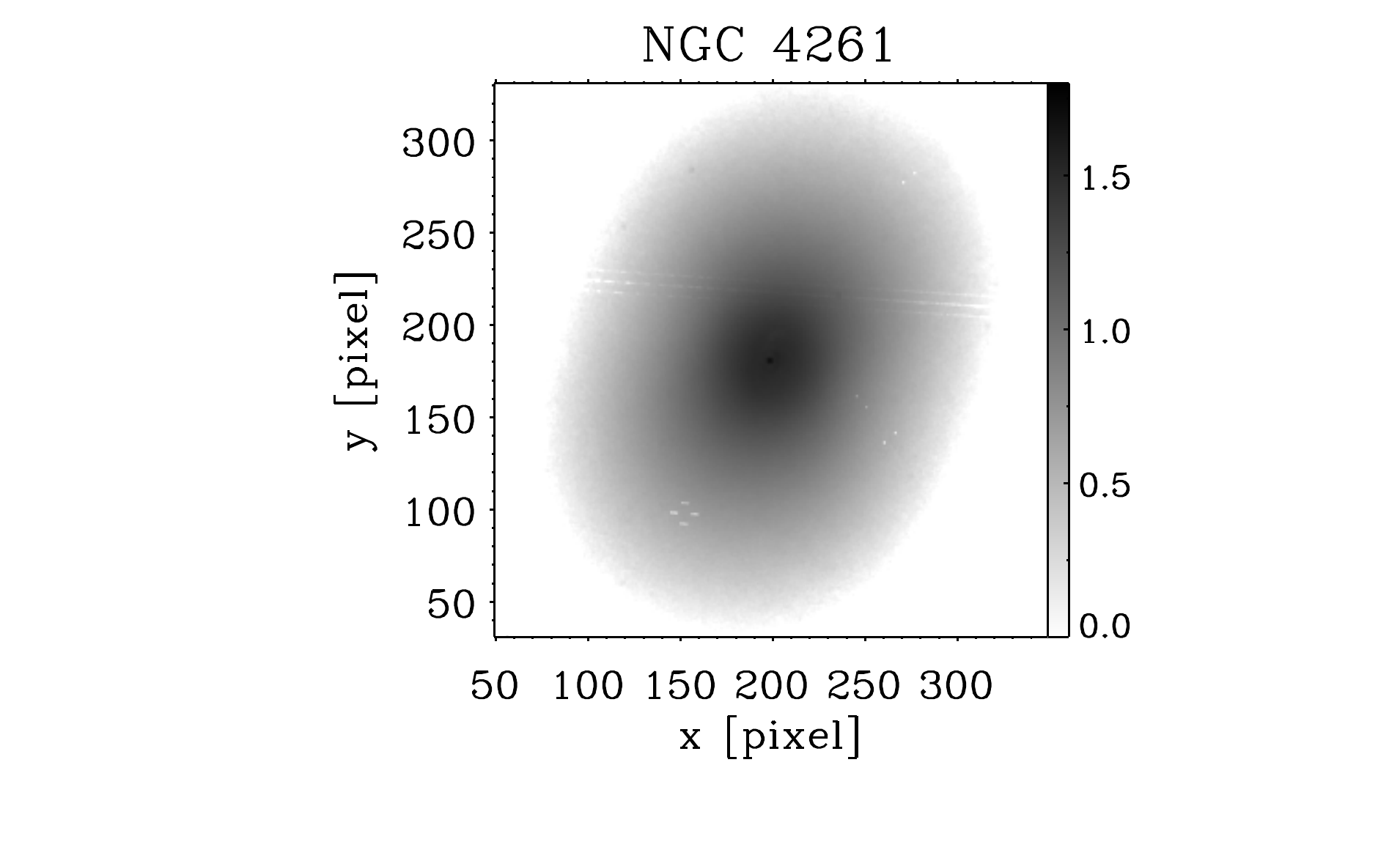} & \includegraphics[trim= 4cm 1cm 3cm 0cm, clip=true, scale=0.48]{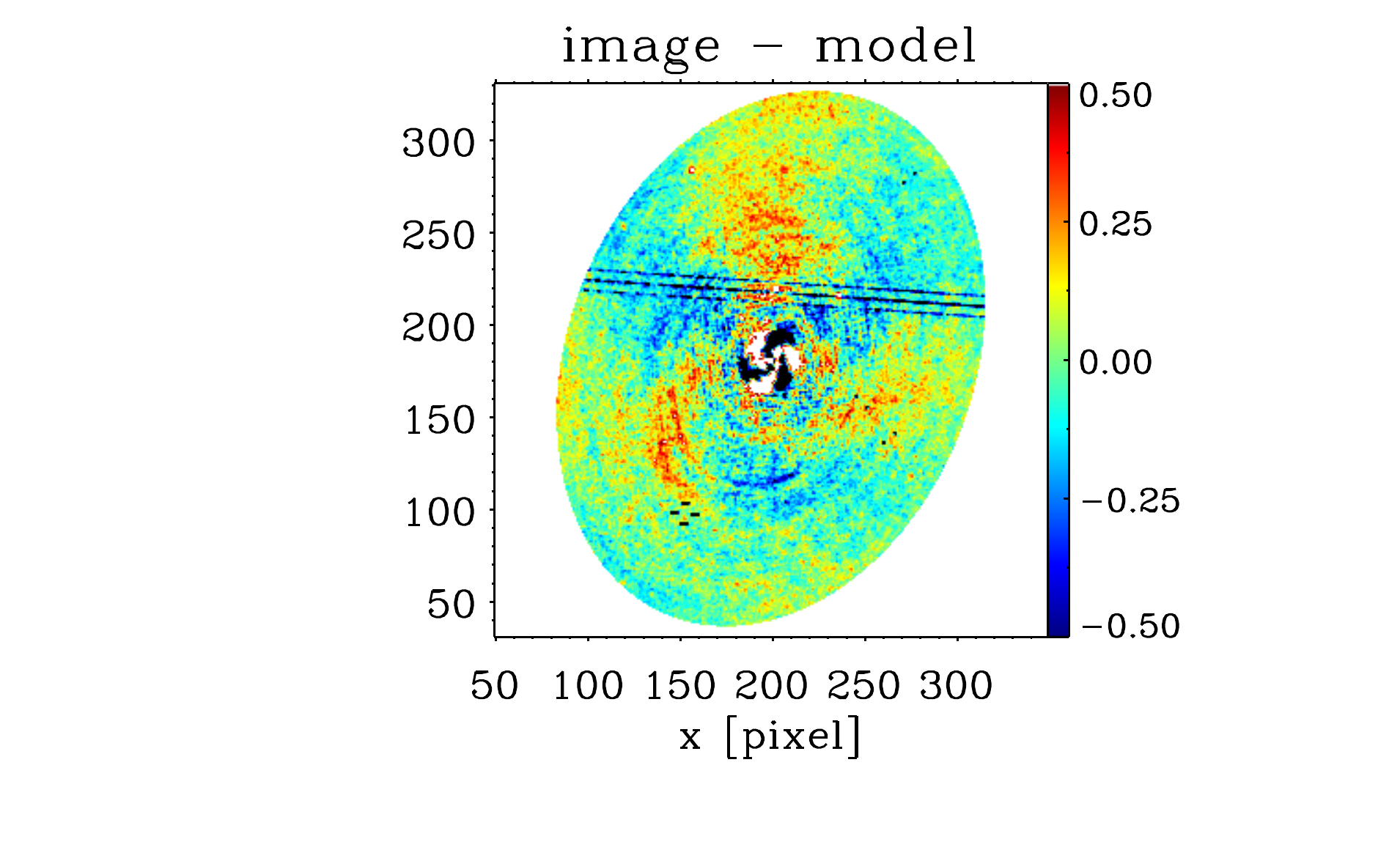}	& \includegraphics[trim= 4cm 1cm 3cm 0cm, clip=true, scale=0.48]{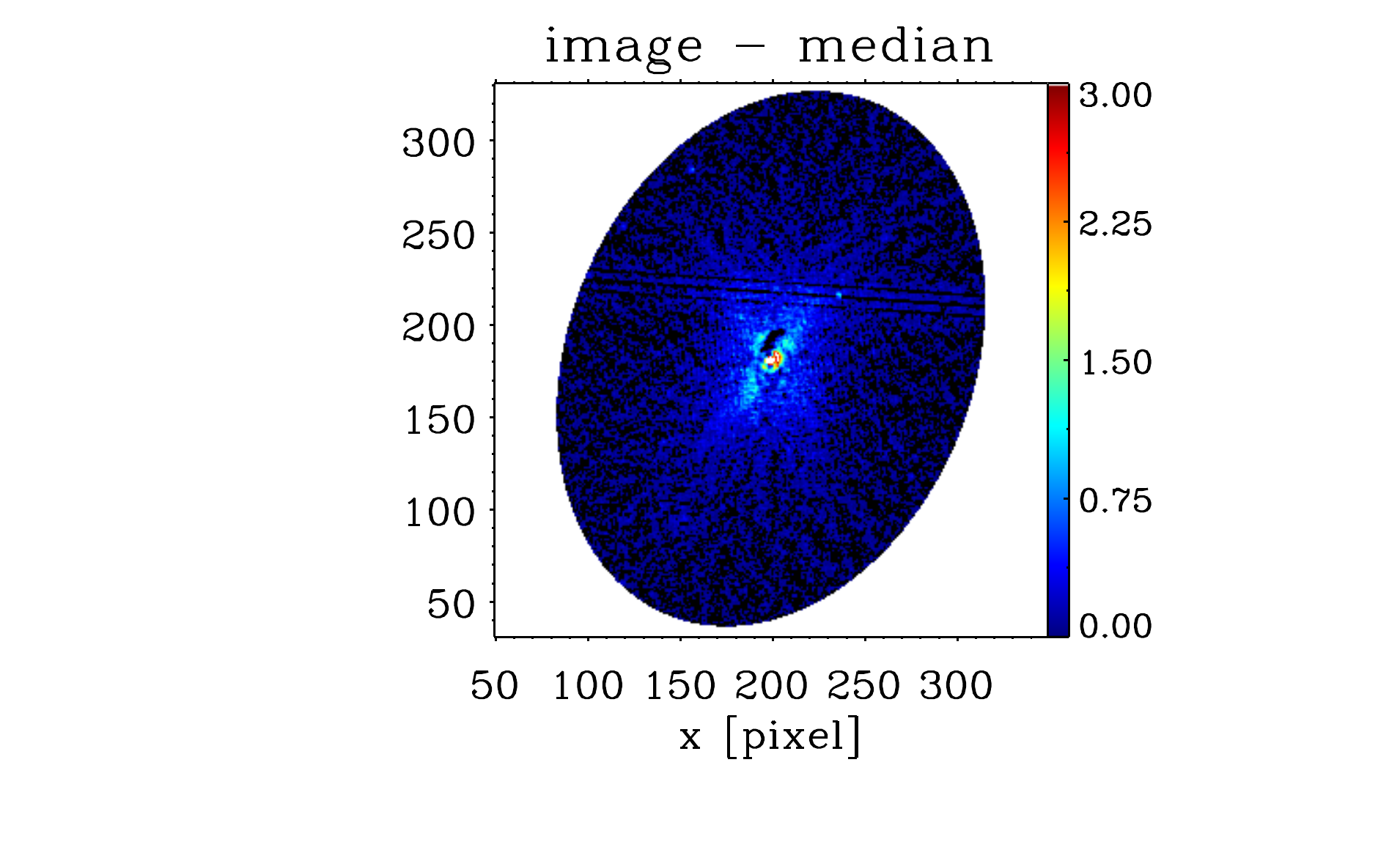} \\
\includegraphics[trim=0.7cm 0cm 0cm 0cm, clip=true, scale=0.46]{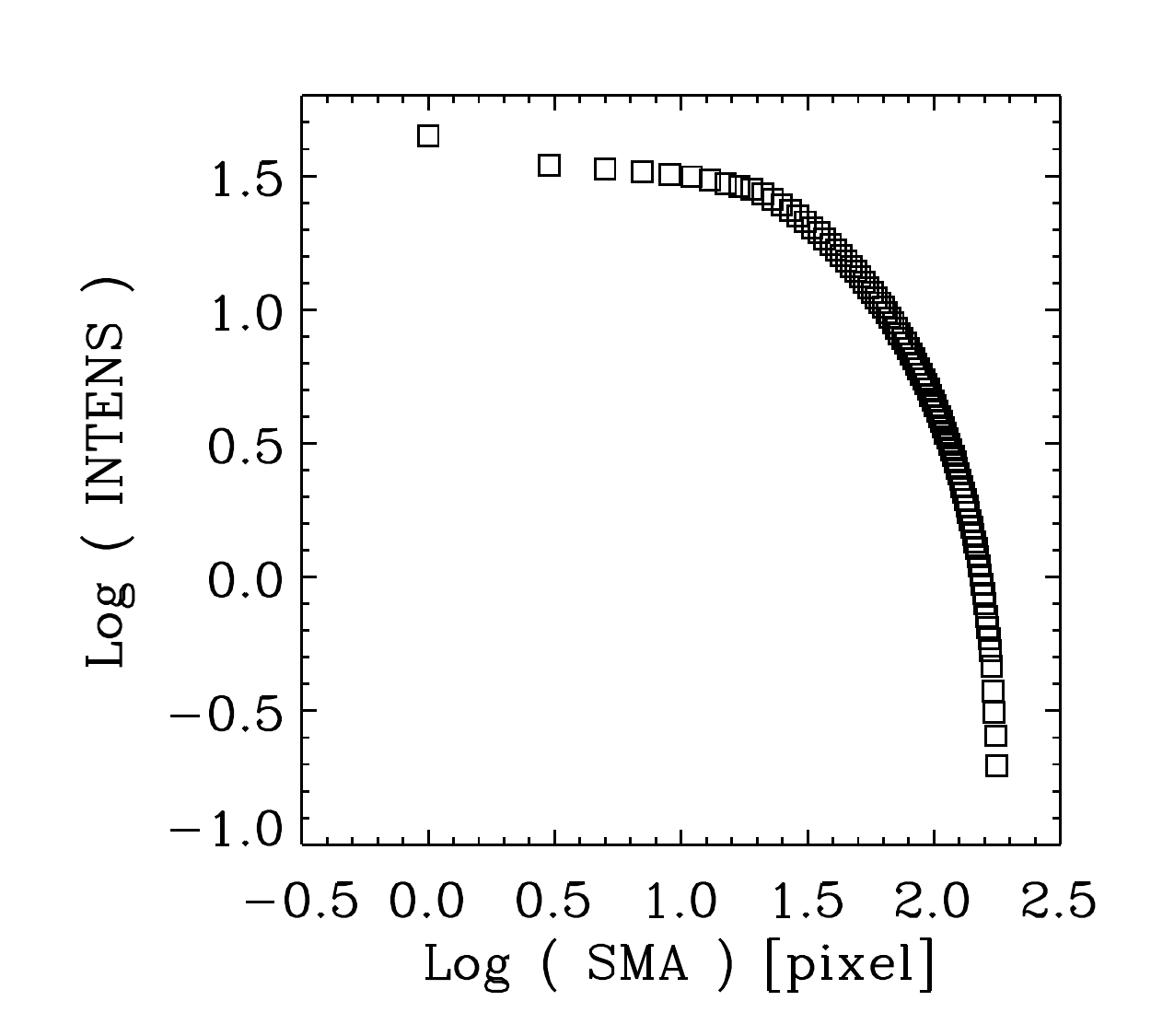}	    &  \includegraphics[trim=0.6cm 0cm 0cm 0cm, clip=true, scale=0.46]{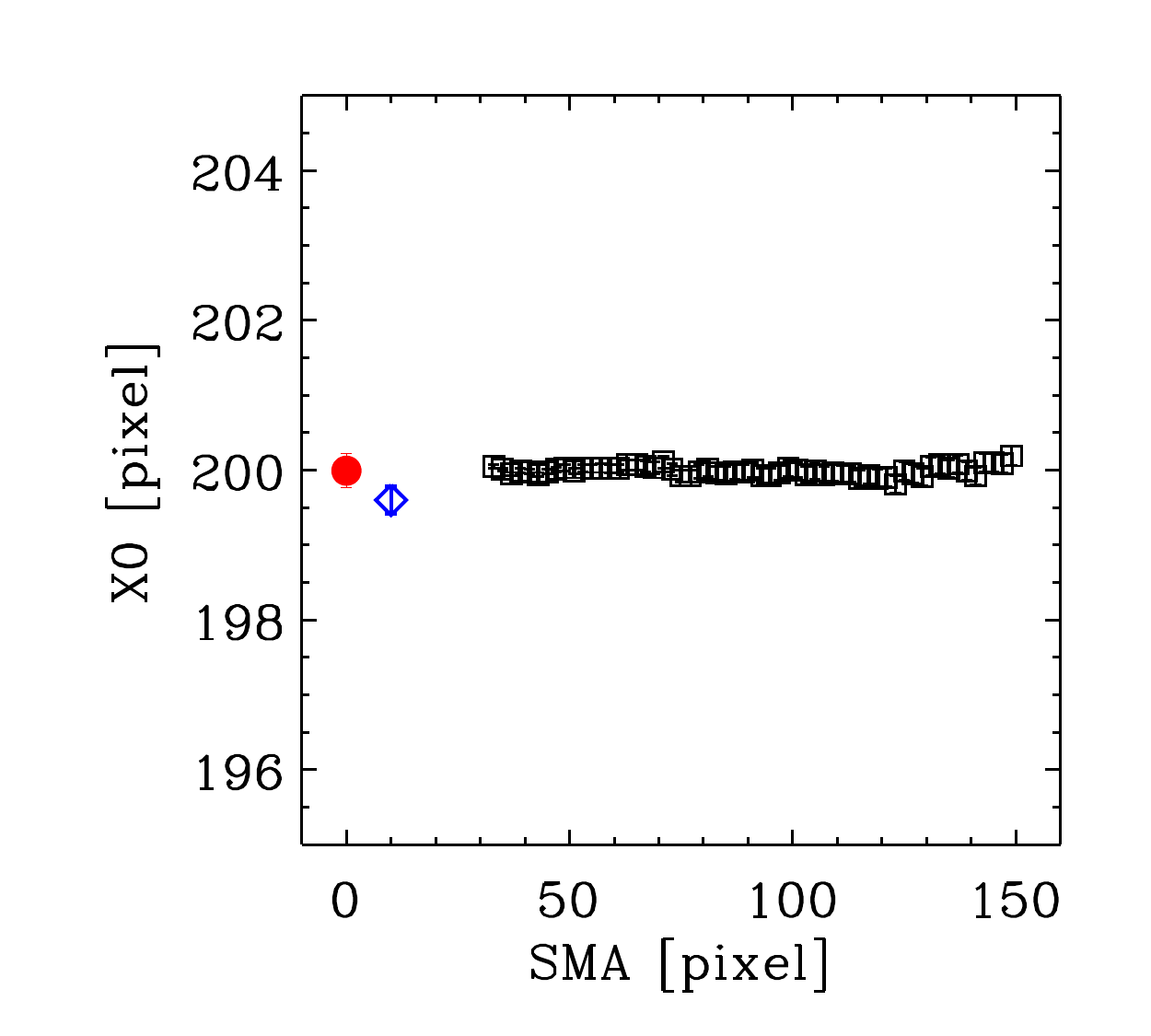}  &  \includegraphics[trim=0.6cm 0cm 0cm 0cm, clip=true, scale=0.46]{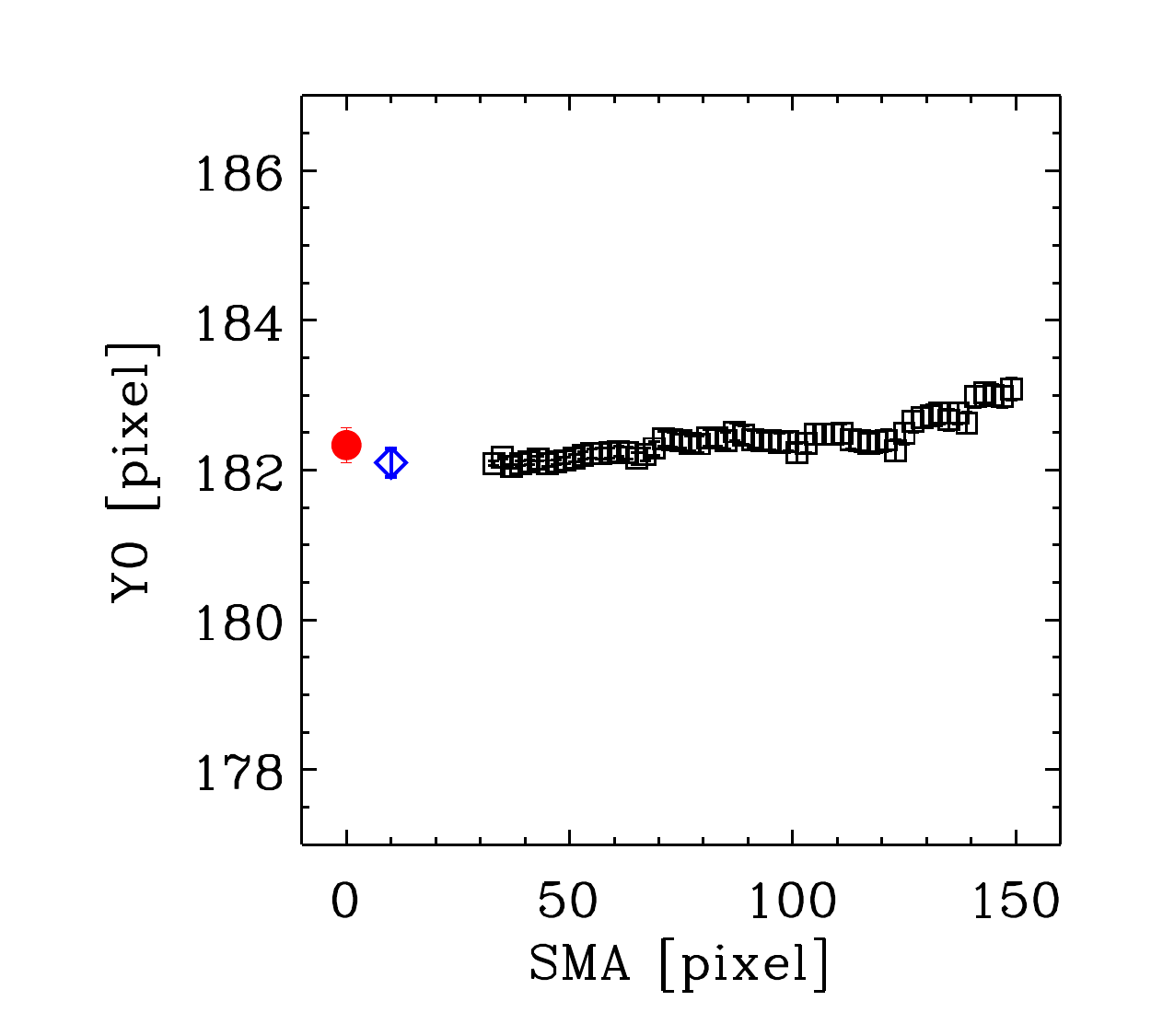} \\	
 \includegraphics[trim=0.65cm 0cm 0cm 0cm, clip=true, scale=0.46]{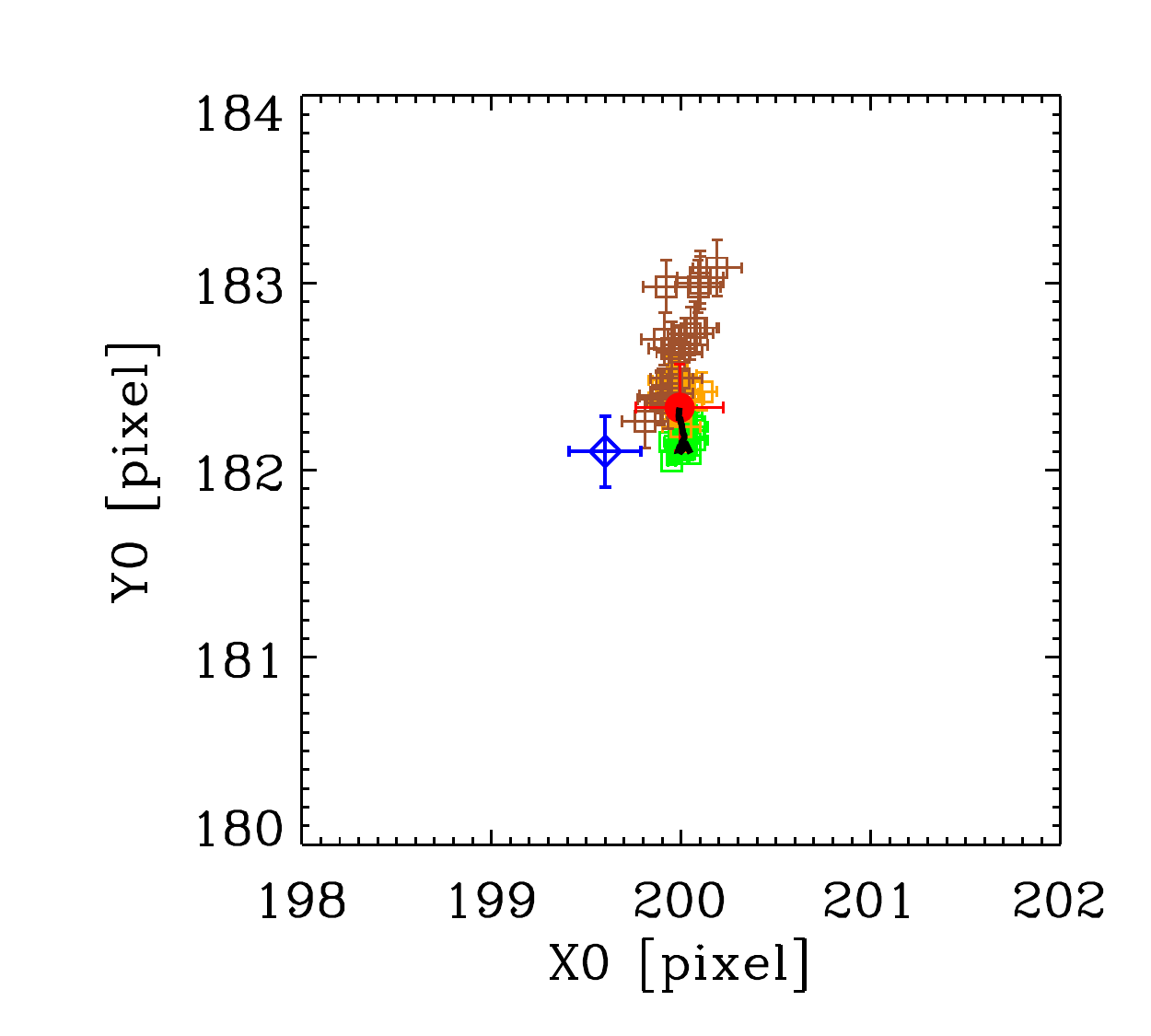}	&  \includegraphics[trim=0.6cm 0cm 0cm 0cm, clip=true, scale=0.46]{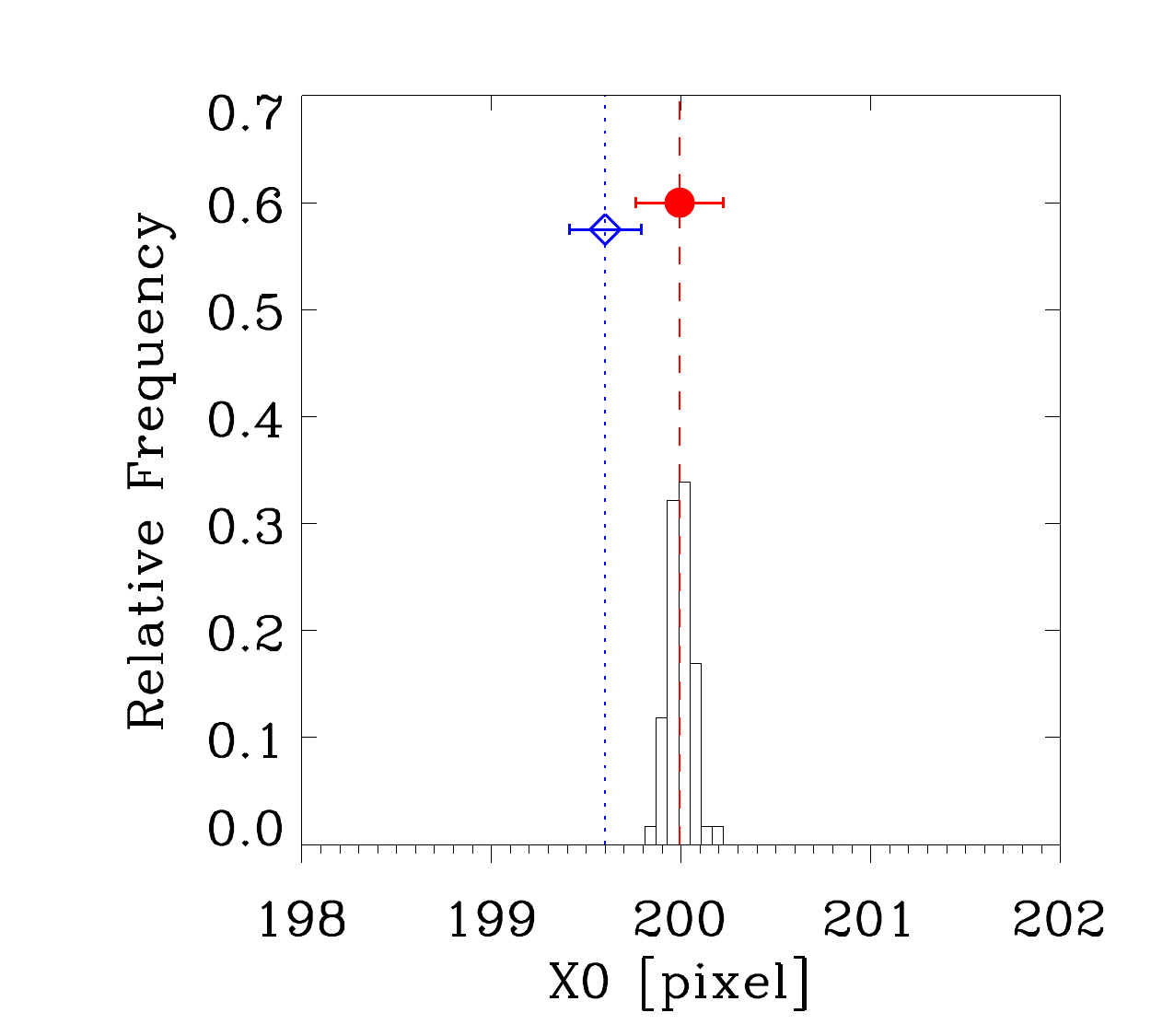}	& \includegraphics[trim=0.6cm 0cm 0cm 0cm, clip=true, scale=0.46]{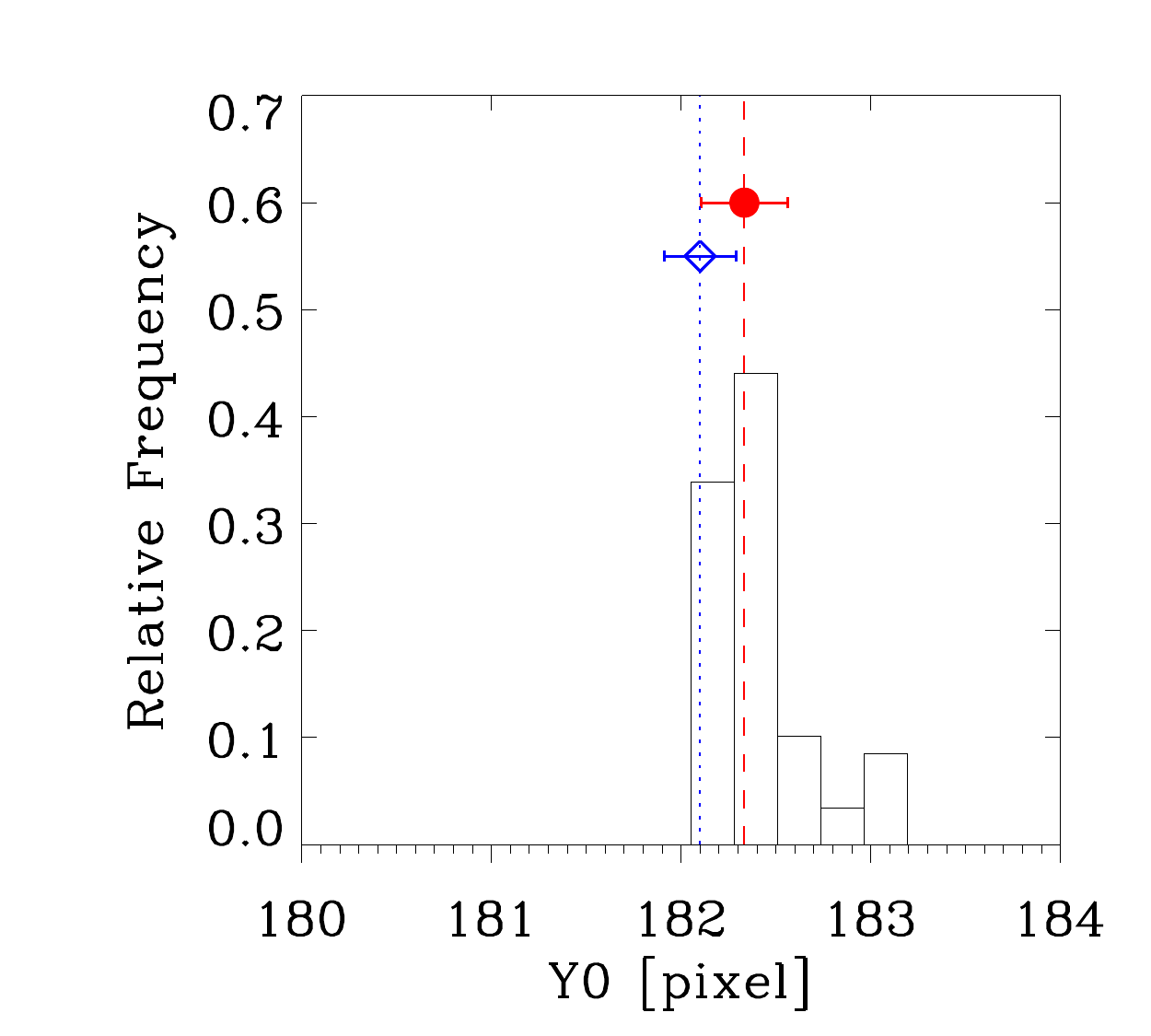}\\
\end{array}$
\end{center}
\caption[]{As in Fig.\ref{fig: NGC4373_W2} for galaxy NGC 4261, NICMOS2 - F160W, scale=$0\farcs05$/pxl.}
\label{fig: NGC4261_9p}
\end{figure*} 

\begin{figure*}[h]
\begin{center}$
\begin{array}{ccc}
\includegraphics[trim=3.75cm 1cm 3cm 0cm, clip=true, scale=0.48]{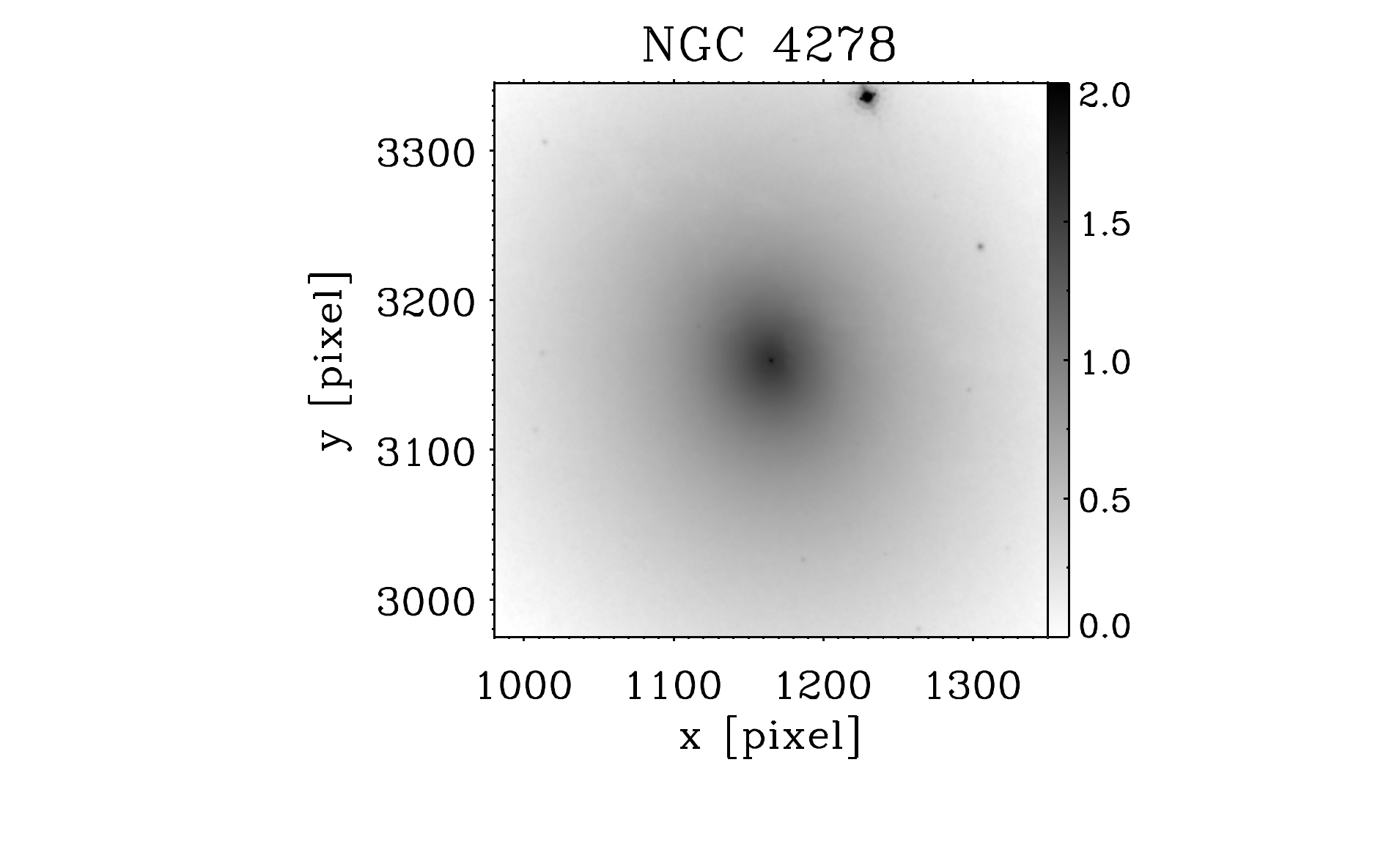} & \includegraphics[trim= 4.cm 1cm 3cm 0cm, clip=true, scale=0.48]{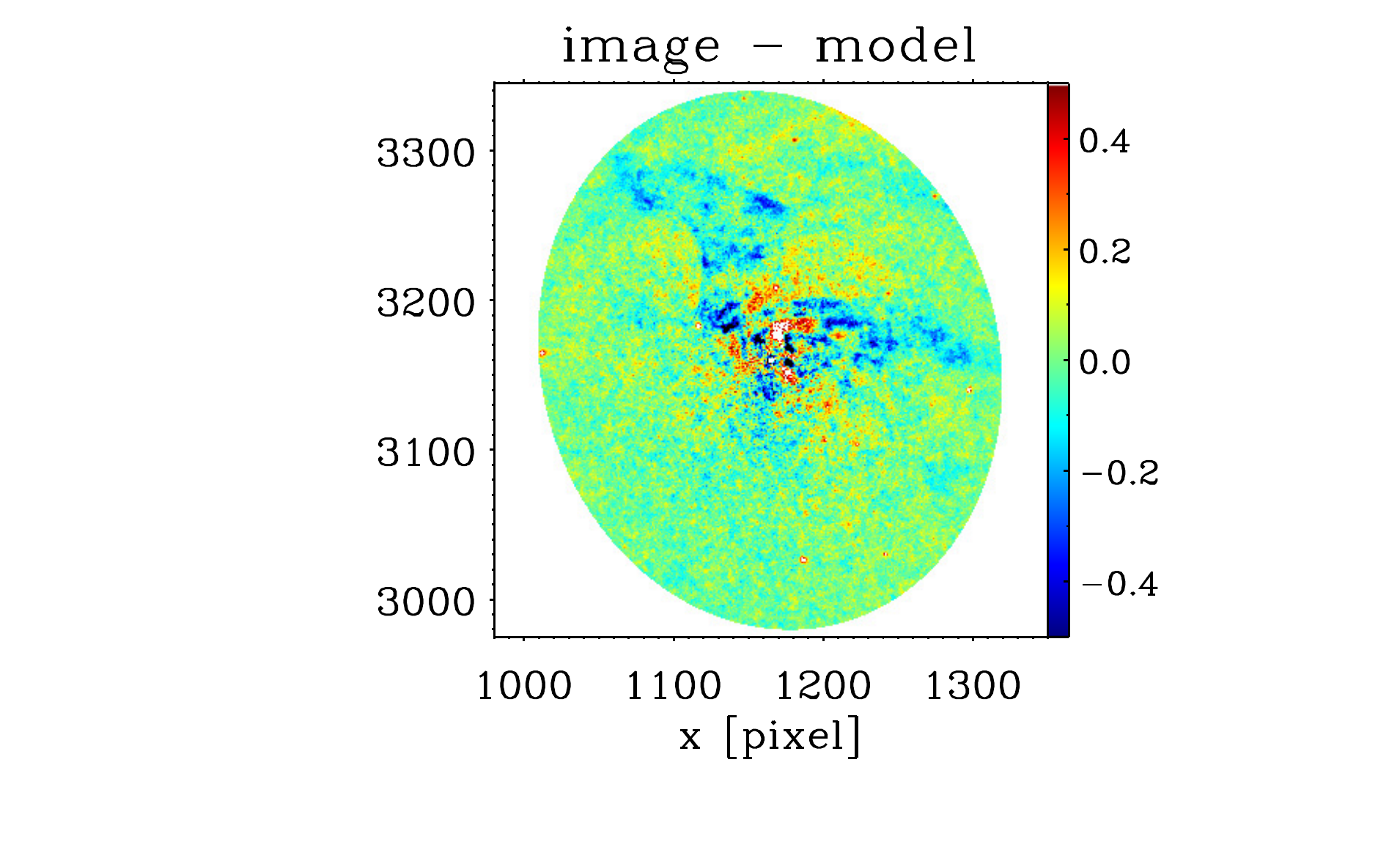}	& \includegraphics[trim= 4.cm 1cm 3cm 0cm, clip=true, scale=0.48]{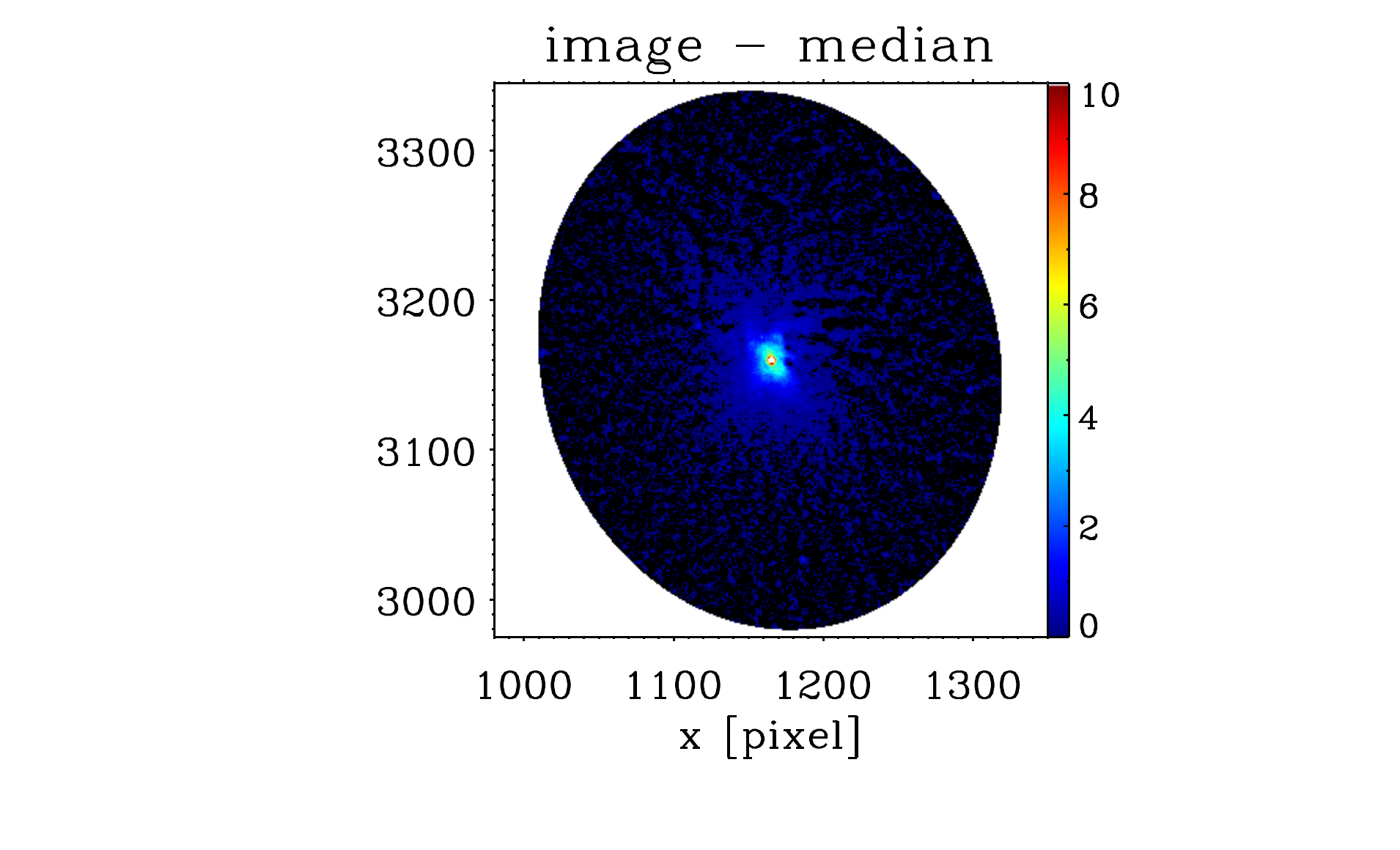} \\
\includegraphics[trim=0.7cm 0cm 0cm 0cm, clip=true, scale=0.46]{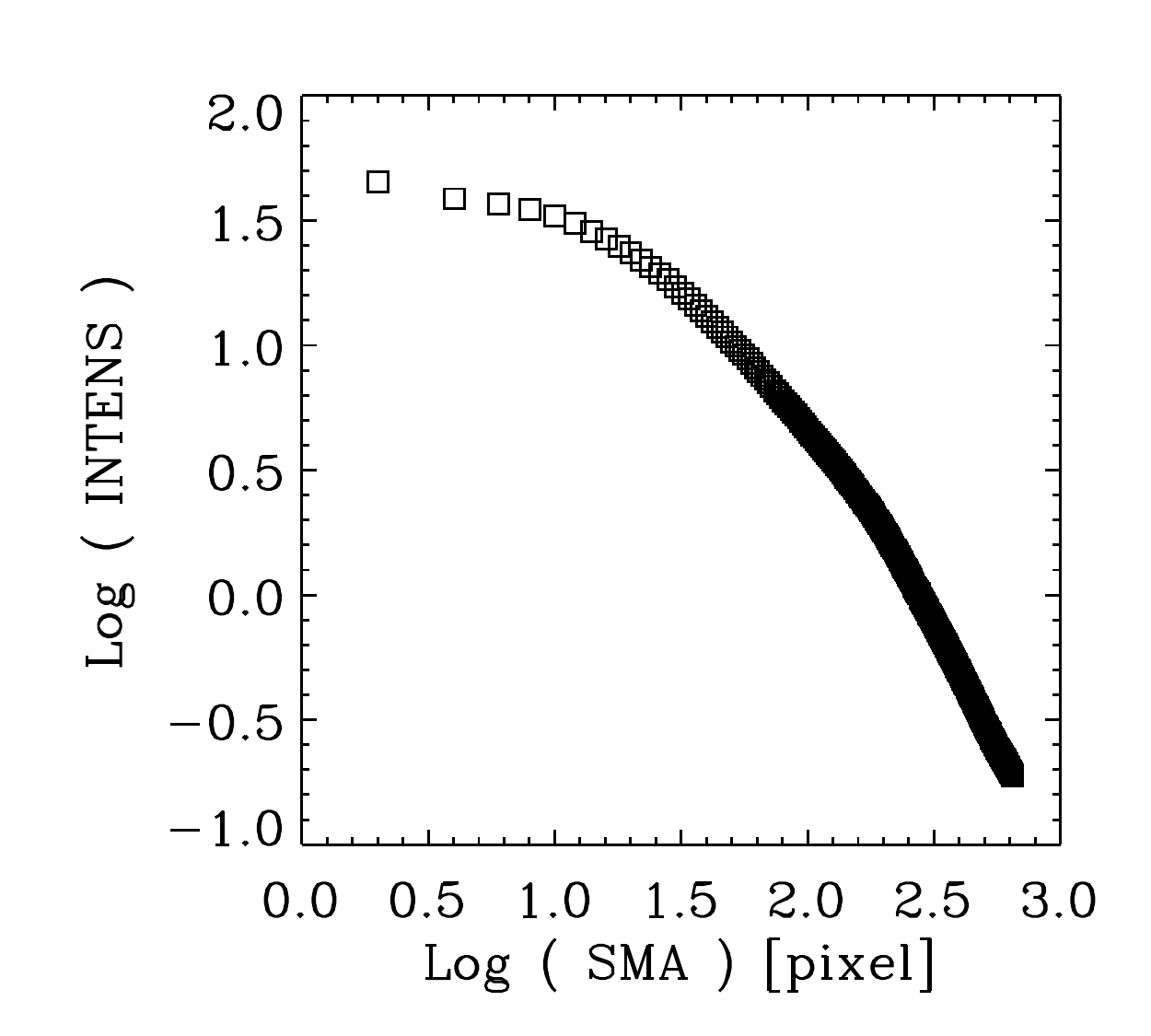}	    &  \includegraphics[trim=0.6cm 0cm 0cm 0cm, clip=true, scale=0.46]{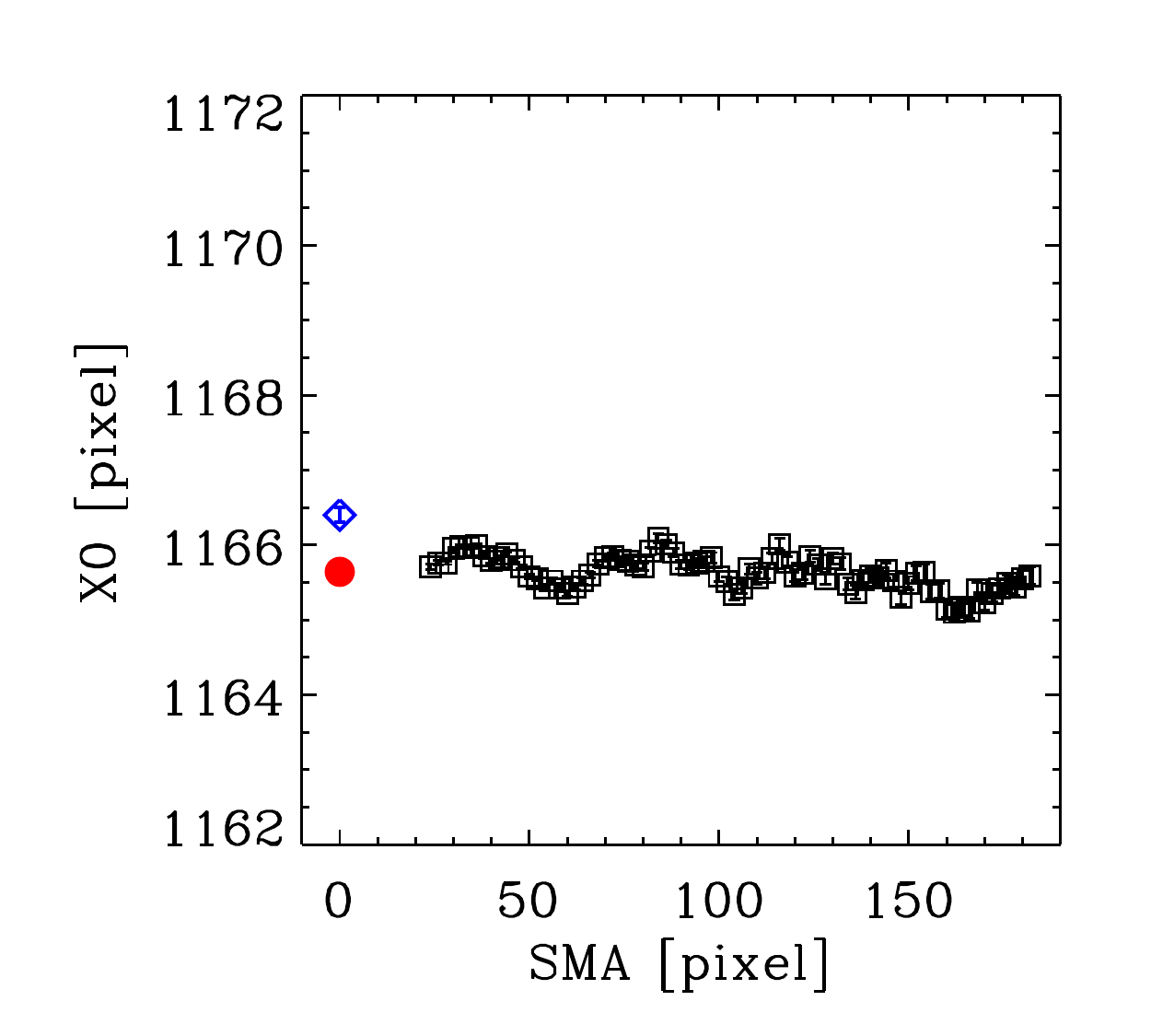}  &  \includegraphics[trim=0.6cm 0cm 0cm 0cm, clip=true, scale=0.46]{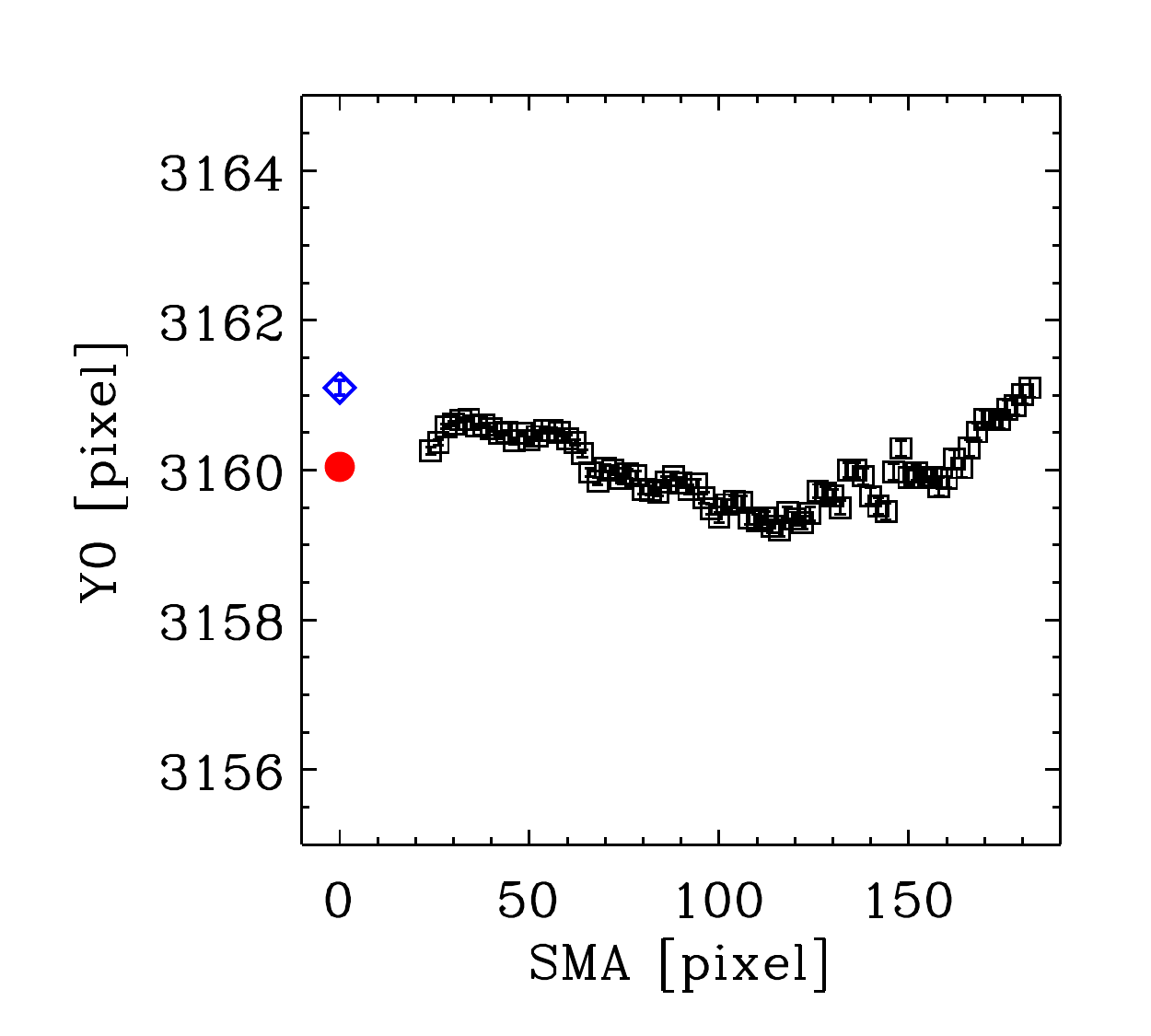} \\	
 \includegraphics[trim=0.65cm 0cm 0cm 0cm, clip=true, scale=0.46]{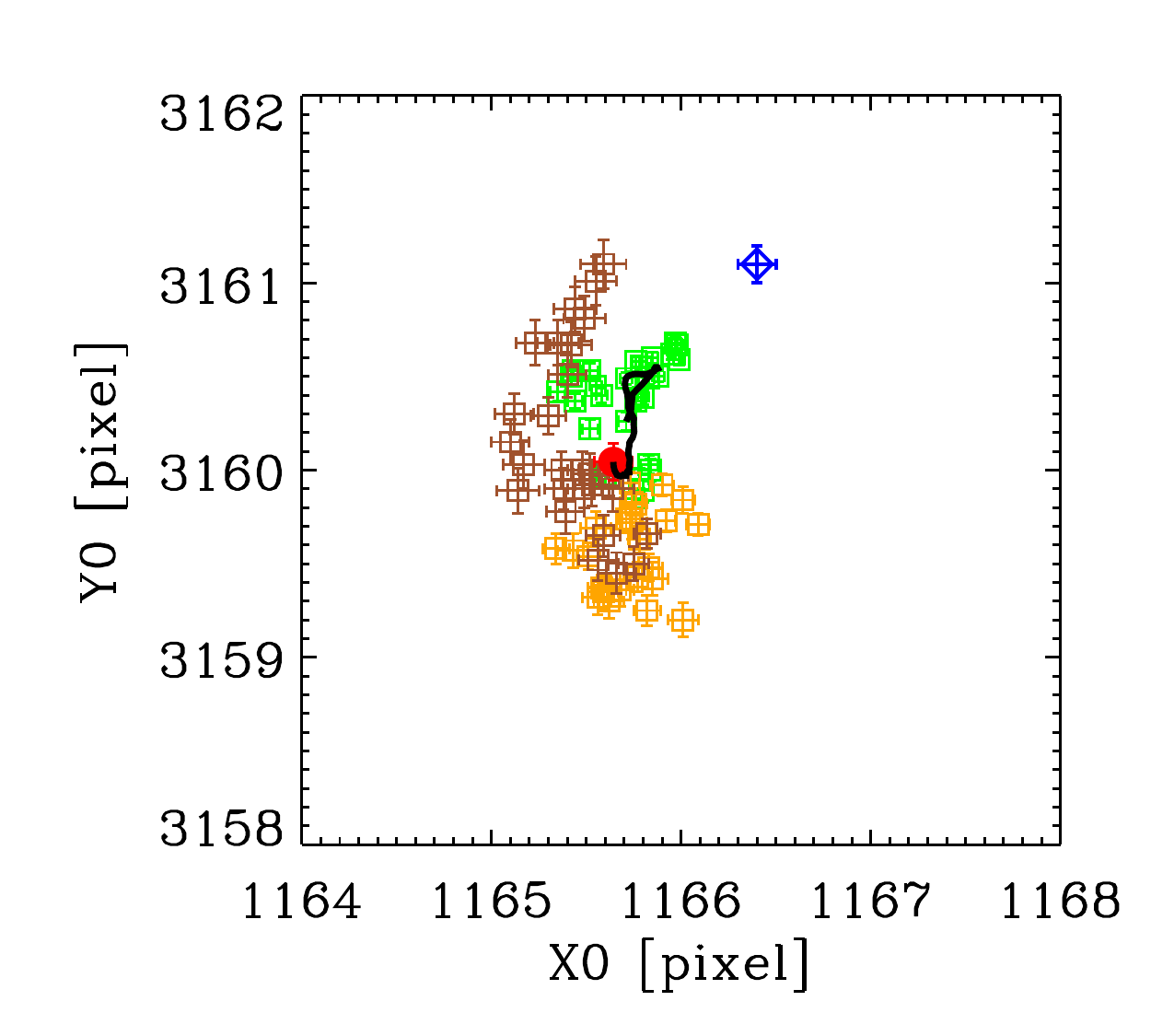}	&  \includegraphics[trim=0.6cm 0cm 0cm 0cm, clip=true, scale=0.46]{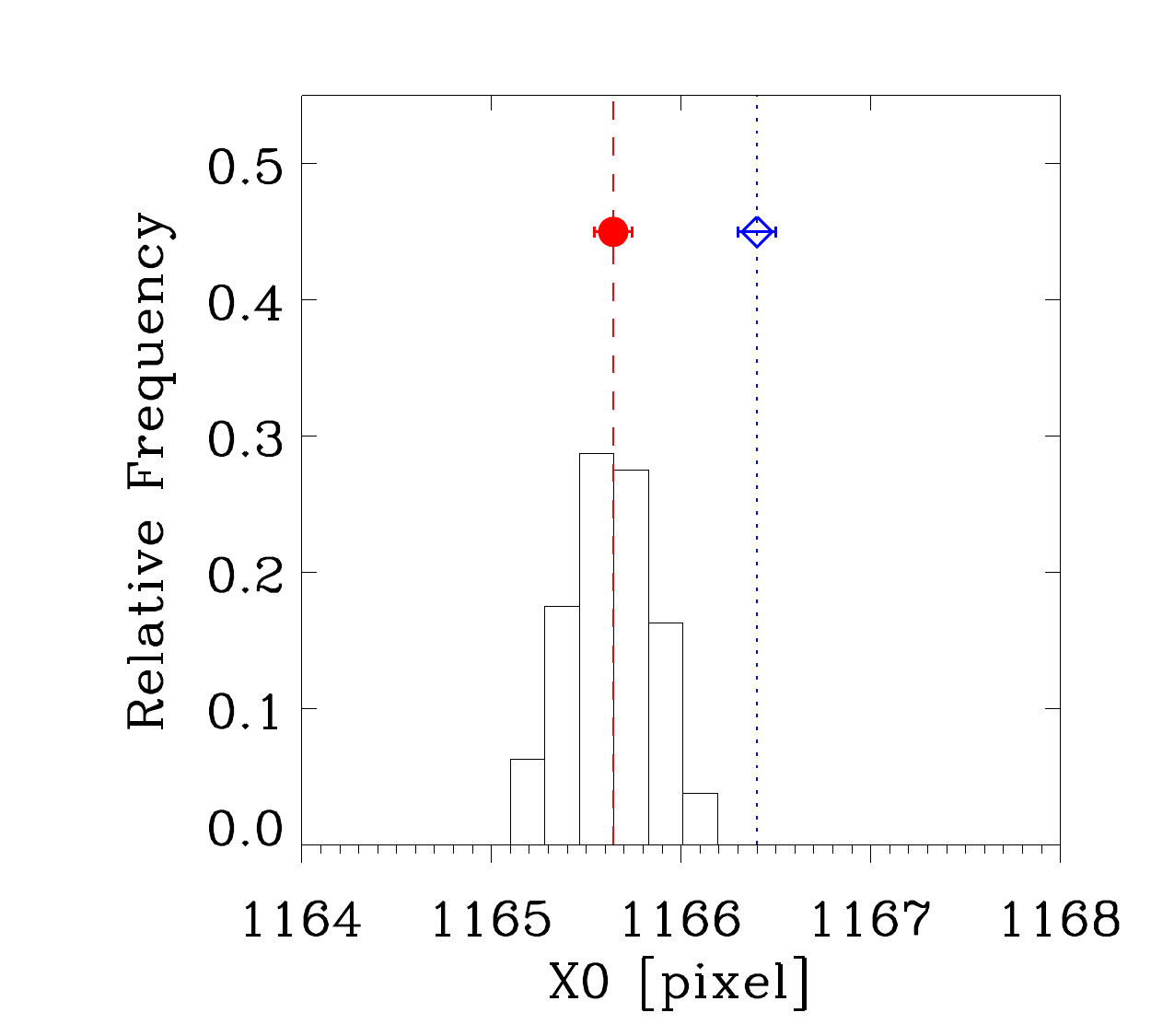}	& \includegraphics[trim=0.6cm 0cm 0cm 0cm, clip=true, scale=0.46]{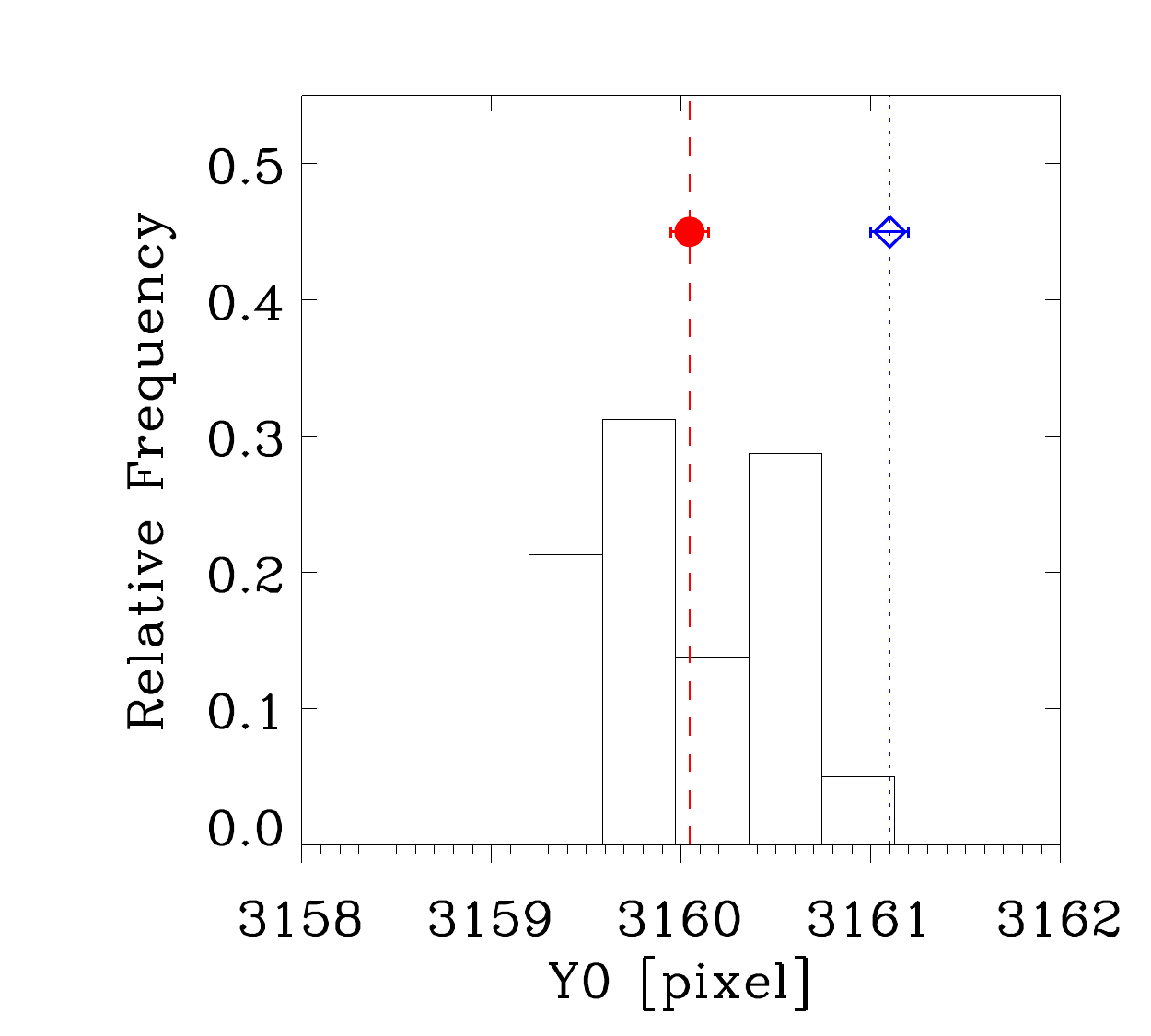}\\
\end{array}$
\end{center}
\caption[NGC 4278]{As in Fig.\ref{fig: NGC4373_W2} for galaxy NGC 4278, ACS/WFC - F850LP, scale=$0\farcs05$/pxl.}
\label{fig: NGC4278_9p_850}
\end{figure*} 

\begin{figure*}[h]
\begin{center}$
\begin{array}{ccc}
\includegraphics[trim=3.75cm 1cm 3cm 0cm, clip=true, scale=0.48]{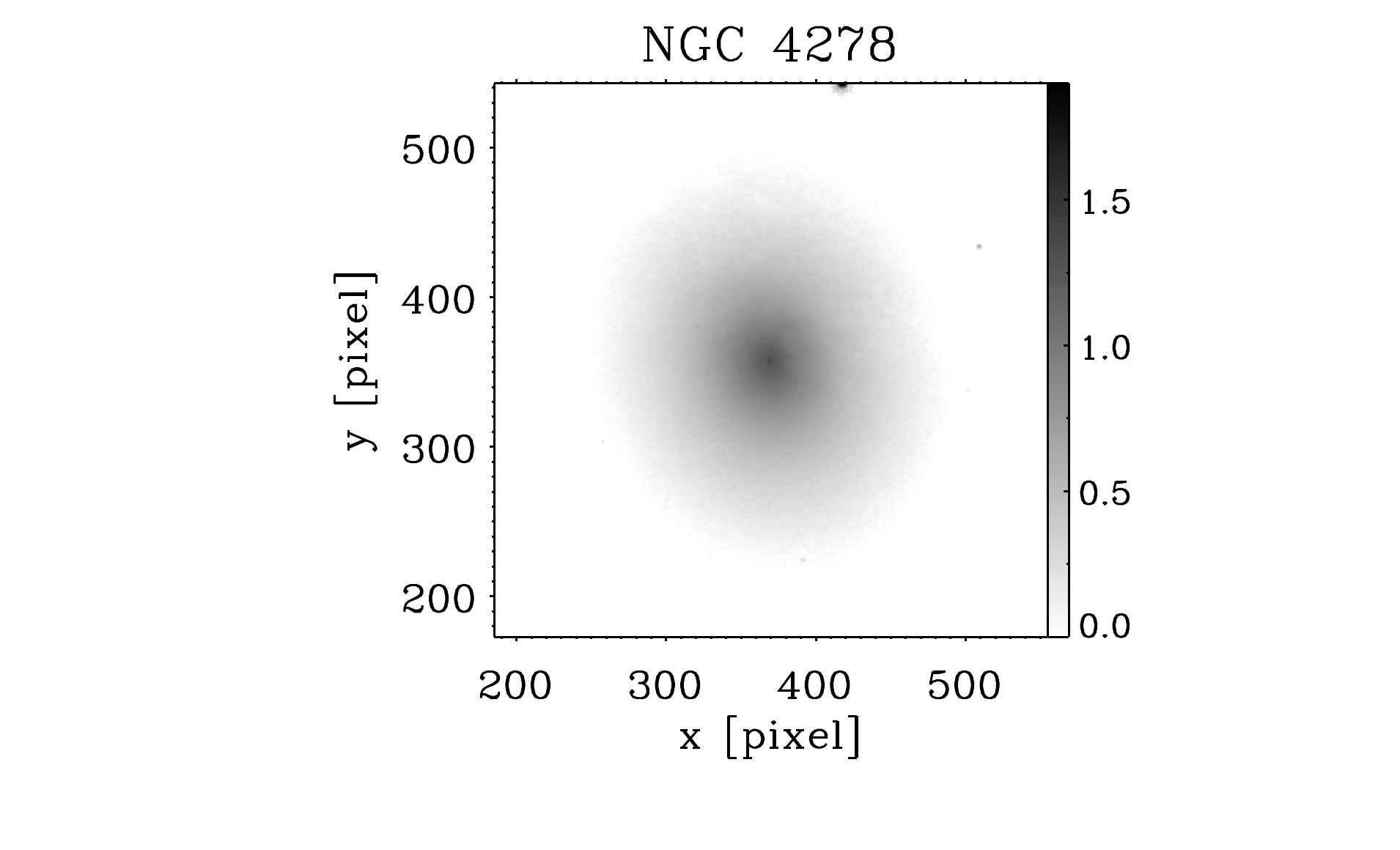} & \includegraphics[trim= 4.cm 1cm 3cm 0cm, clip=true, scale=0.48]{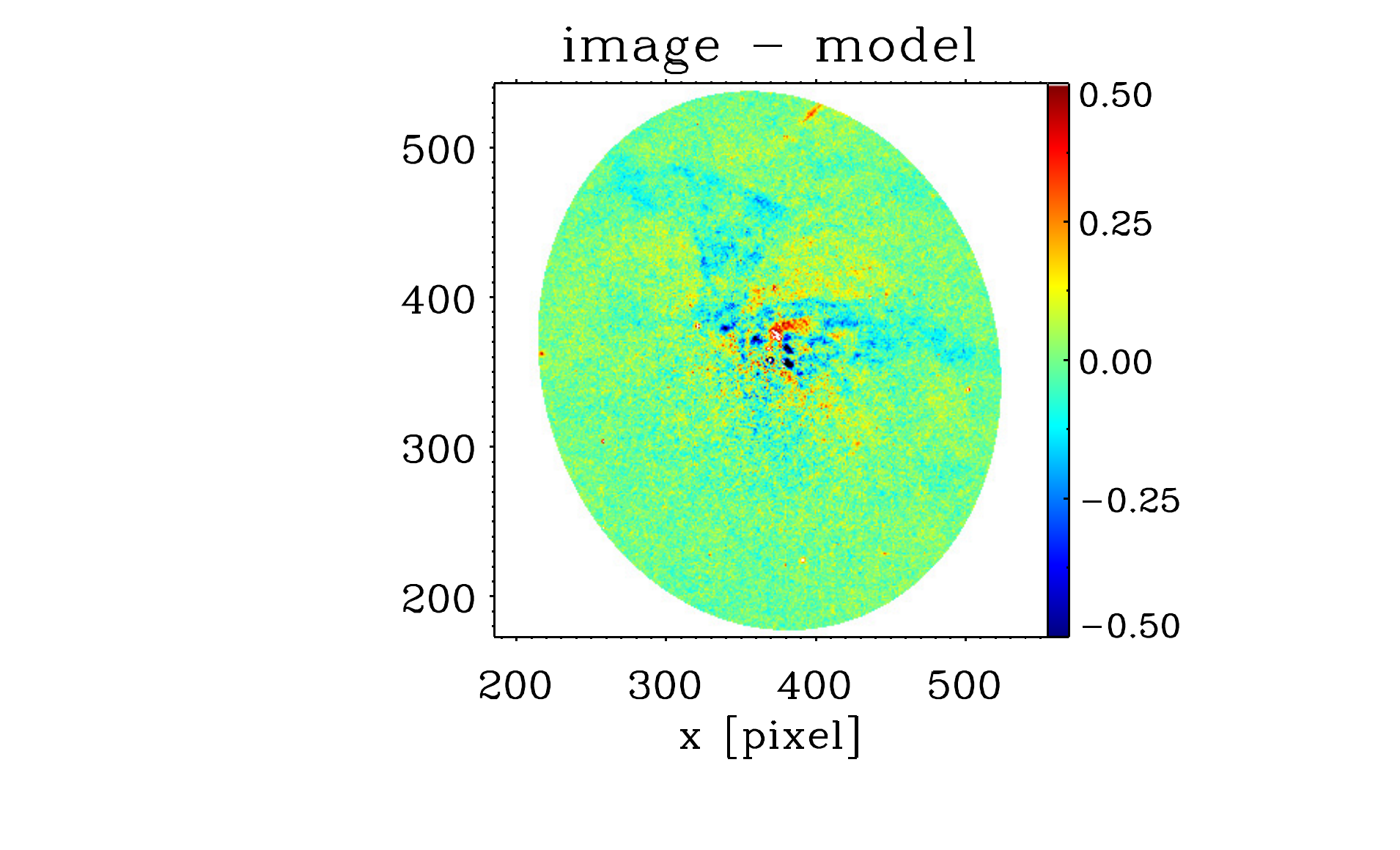}	& \includegraphics[trim= 4.cm 1cm 3cm 0cm, clip=true, scale=0.48]{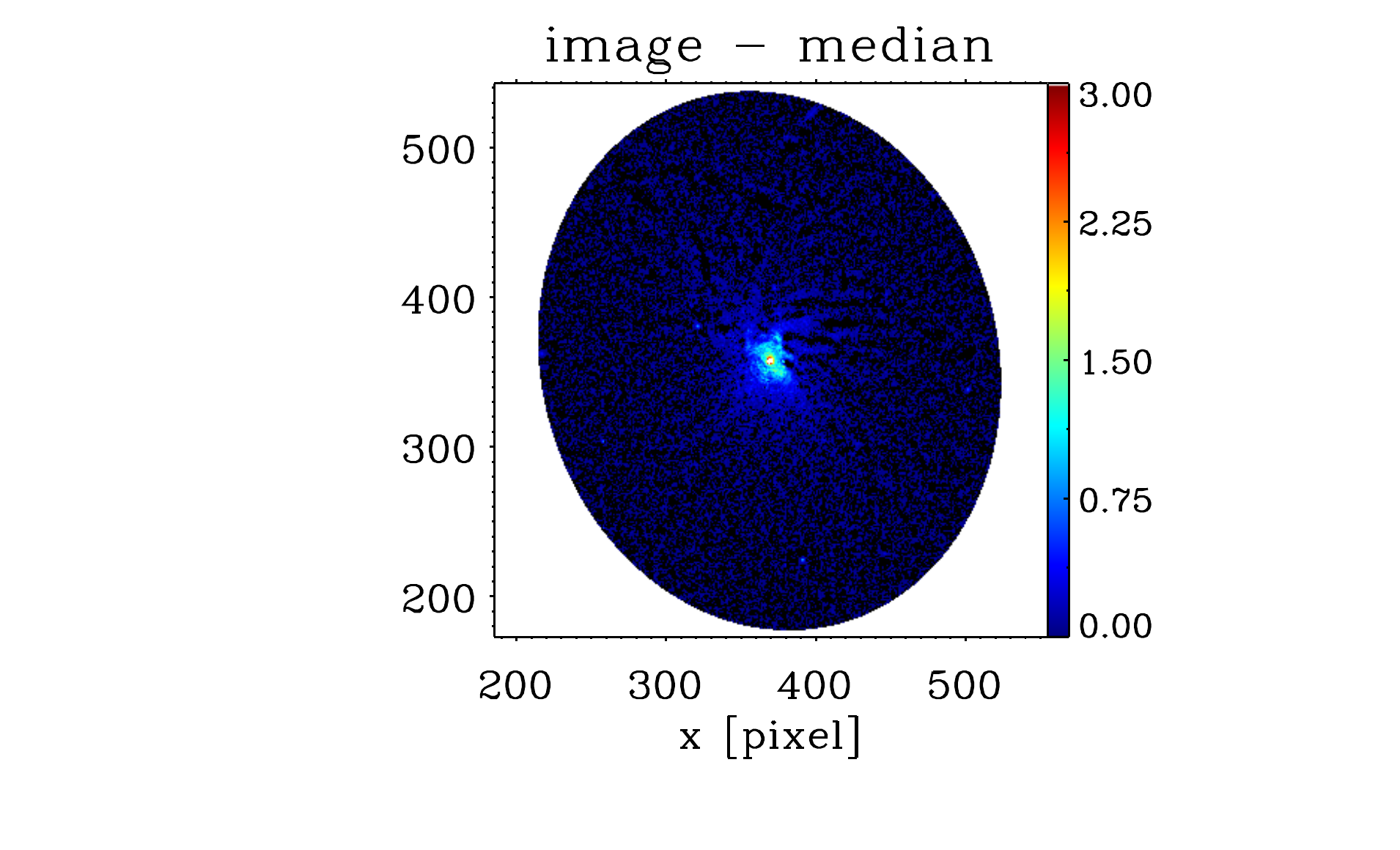} \\
\includegraphics[trim=0.7cm 0cm 0cm 0cm, clip=true, scale=0.46]{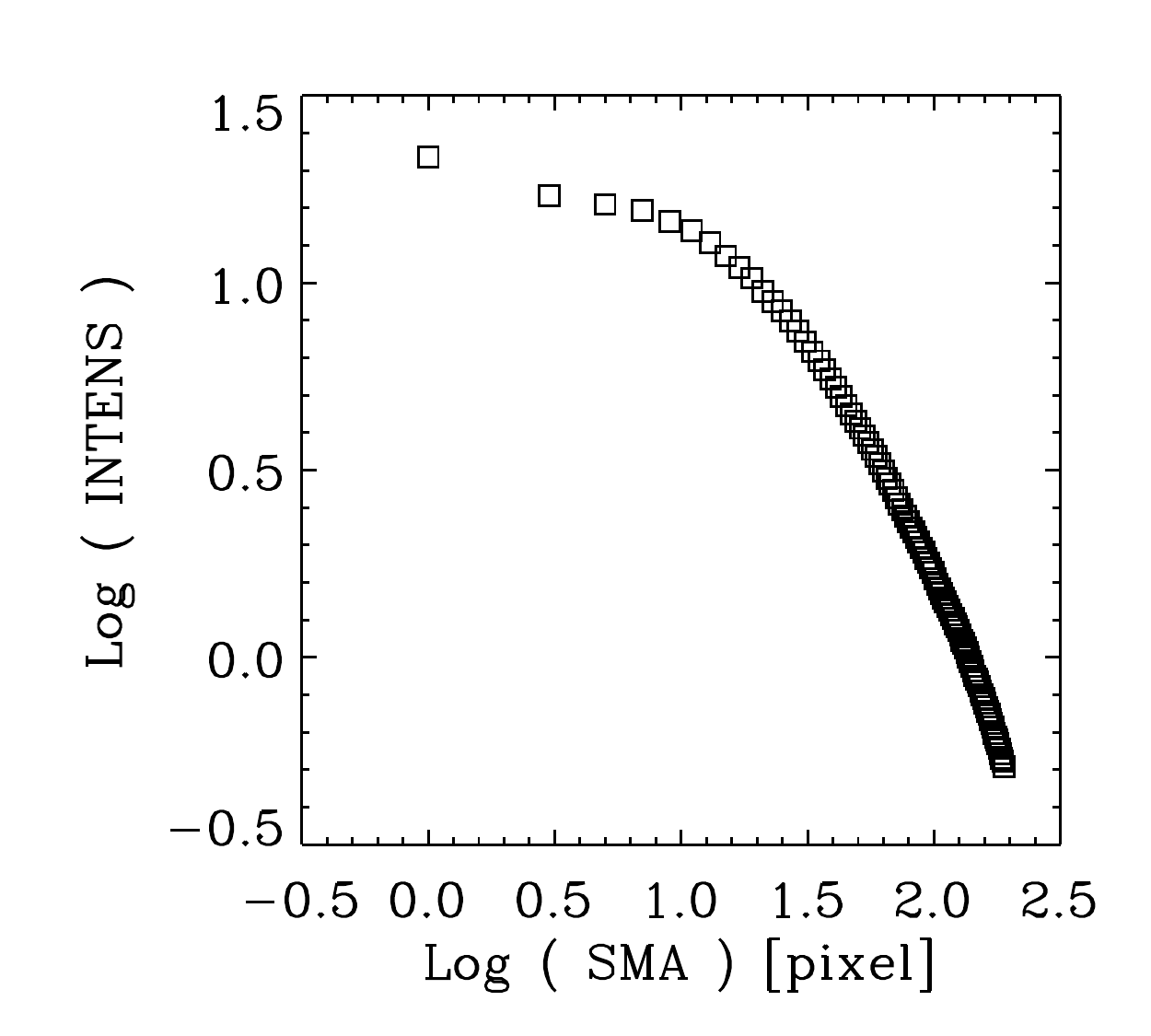}	    &  \includegraphics[trim=0.6cm 0cm 0cm 0cm, clip=true, scale=0.46]{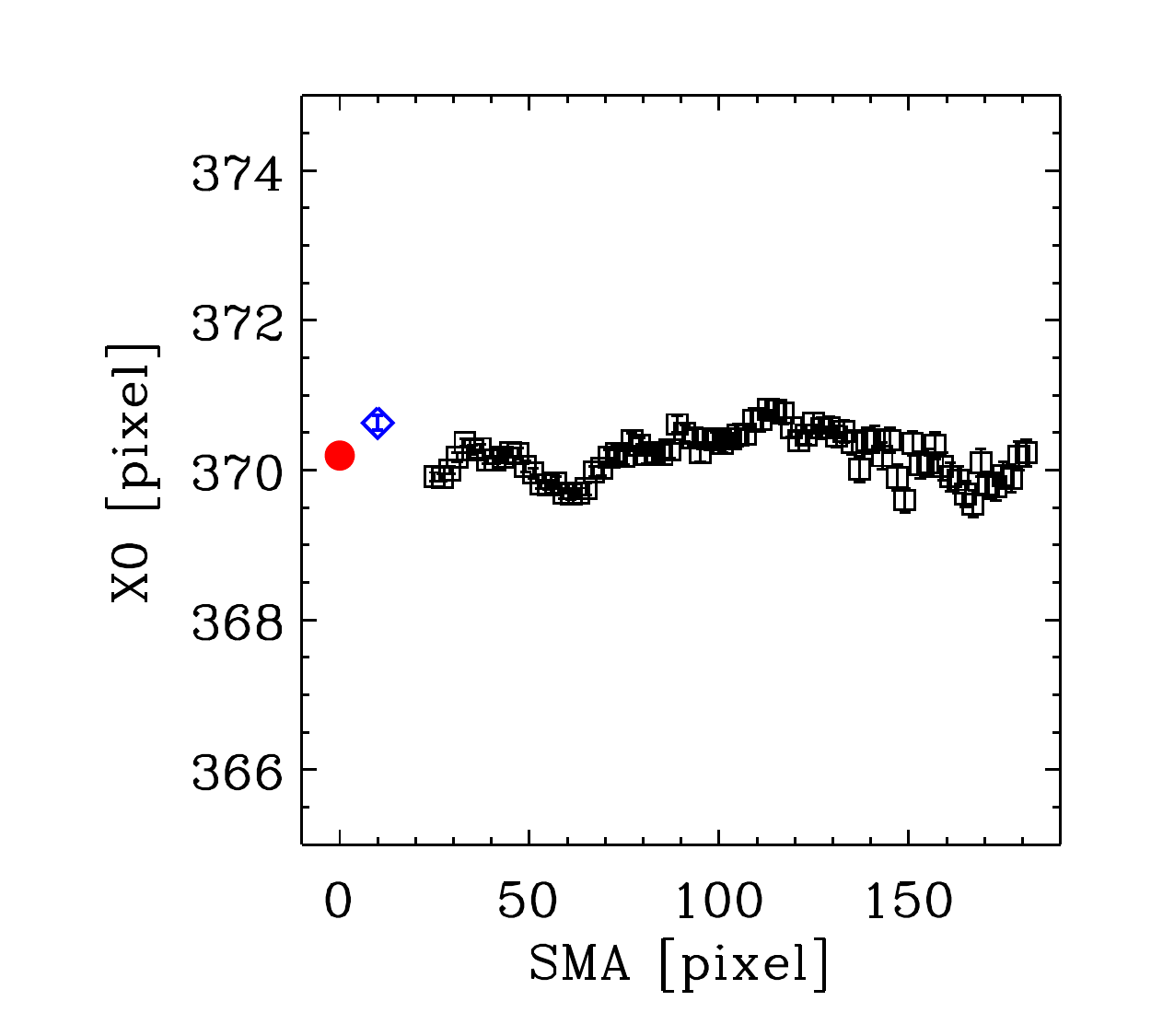}  &  \includegraphics[trim=0.6cm 0cm 0cm 0cm, clip=true, scale=0.46]{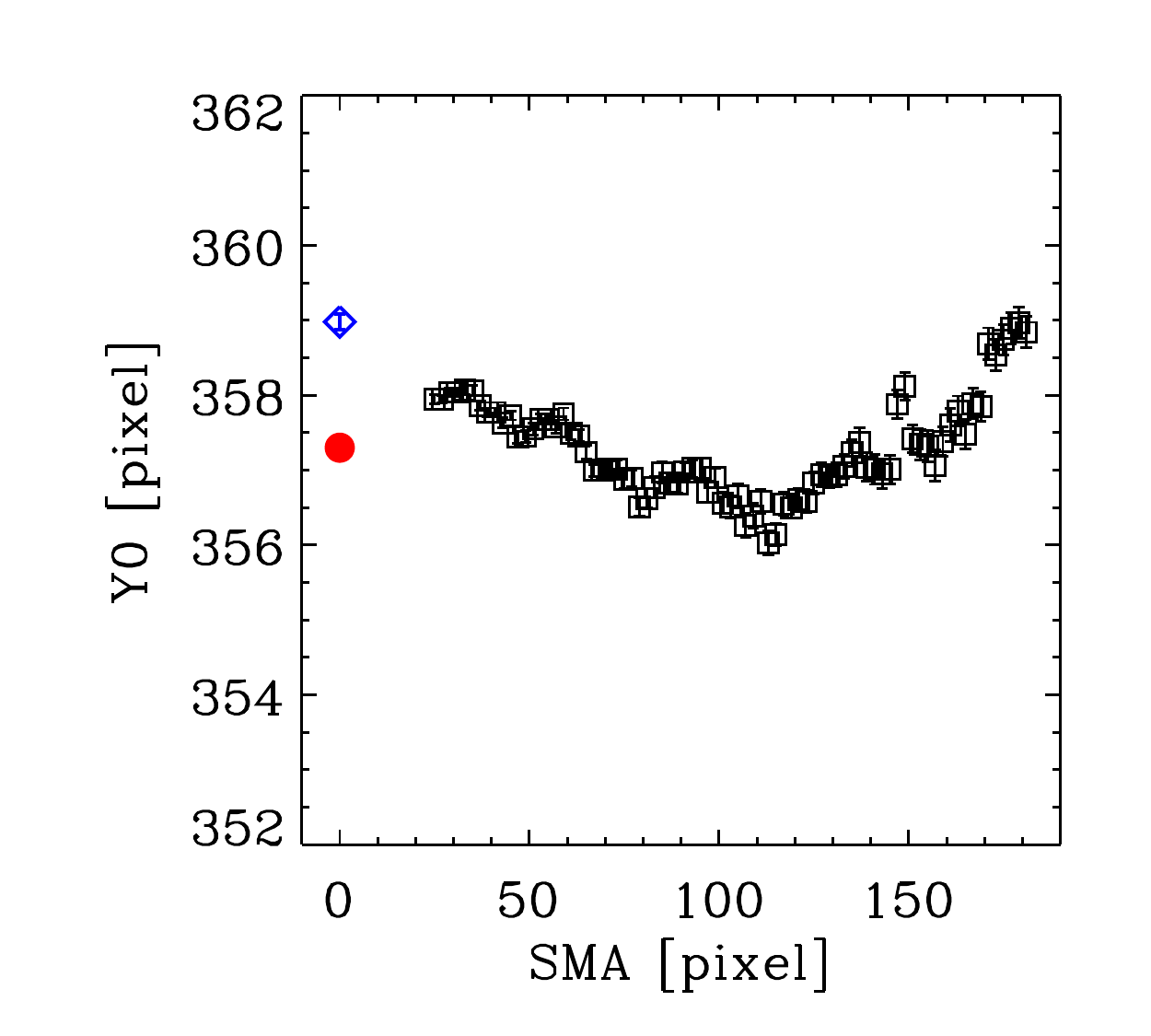} \\	
 \includegraphics[trim=0.65cm 0cm 0cm 0cm, clip=true, scale=0.46]{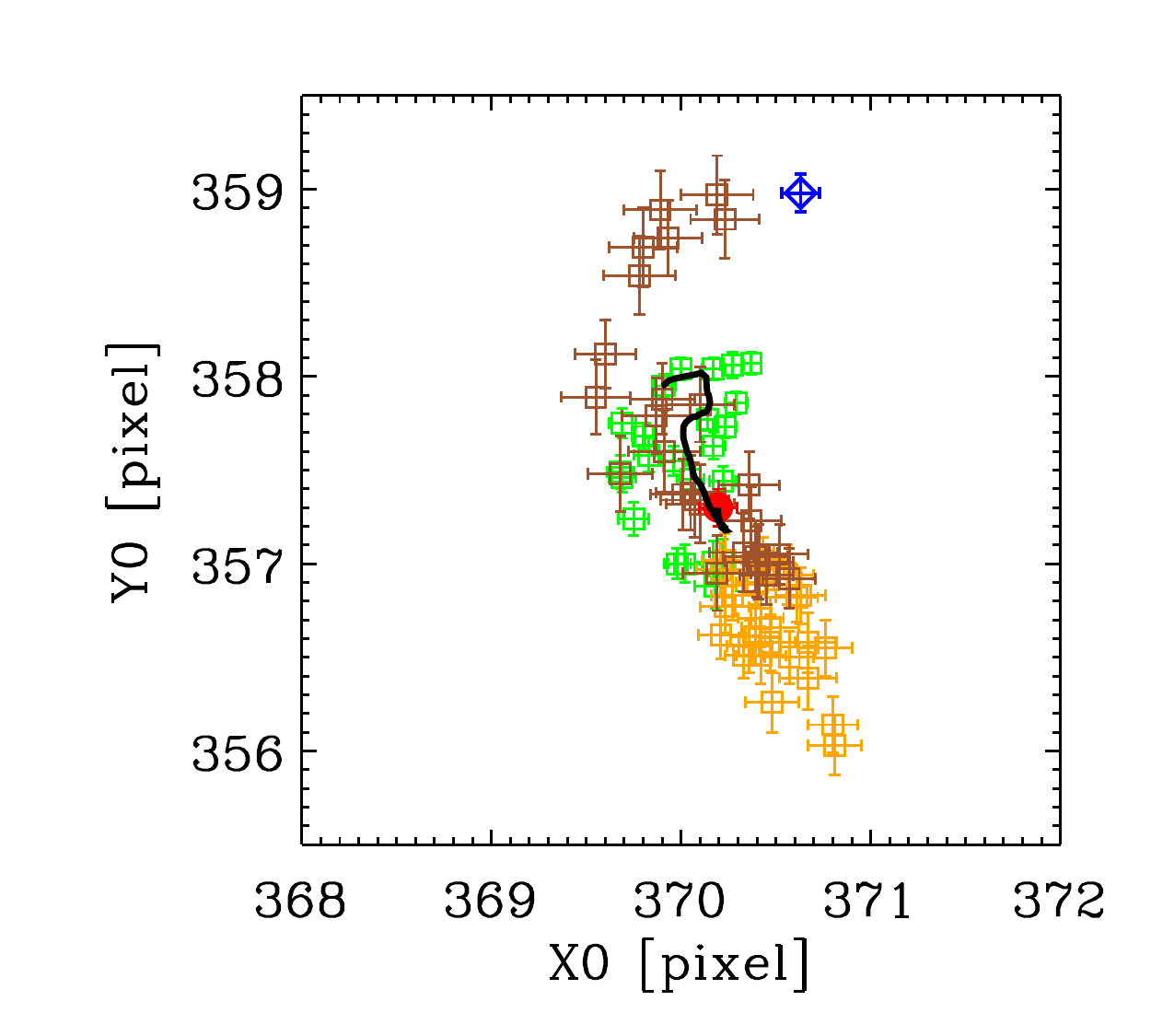}	&  \includegraphics[trim=0.6cm 0cm 0cm 0cm, clip=true, scale=0.46]{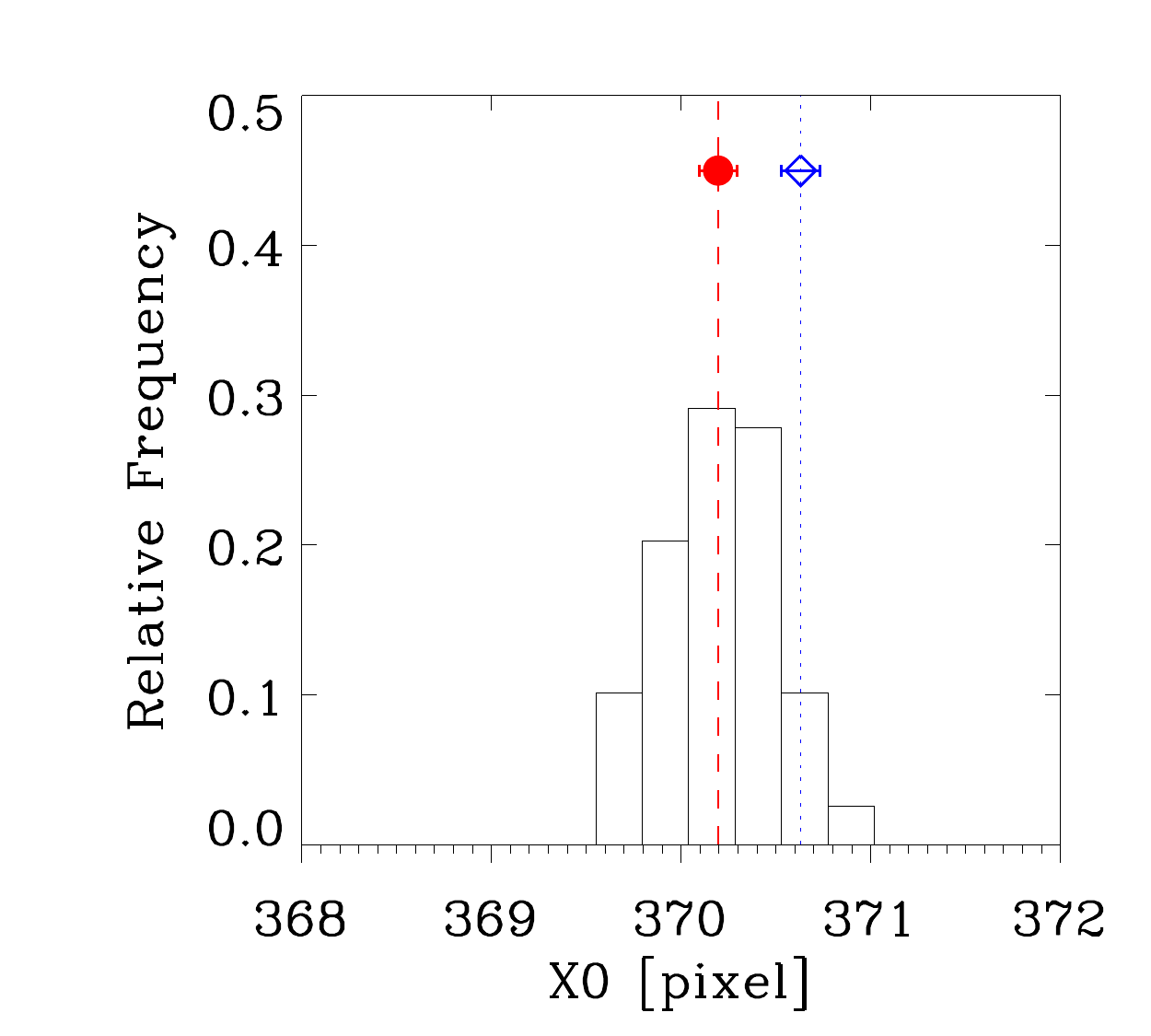}	& \includegraphics[trim=0.6cm 0cm 0cm 0cm, clip=true, scale=0.46]{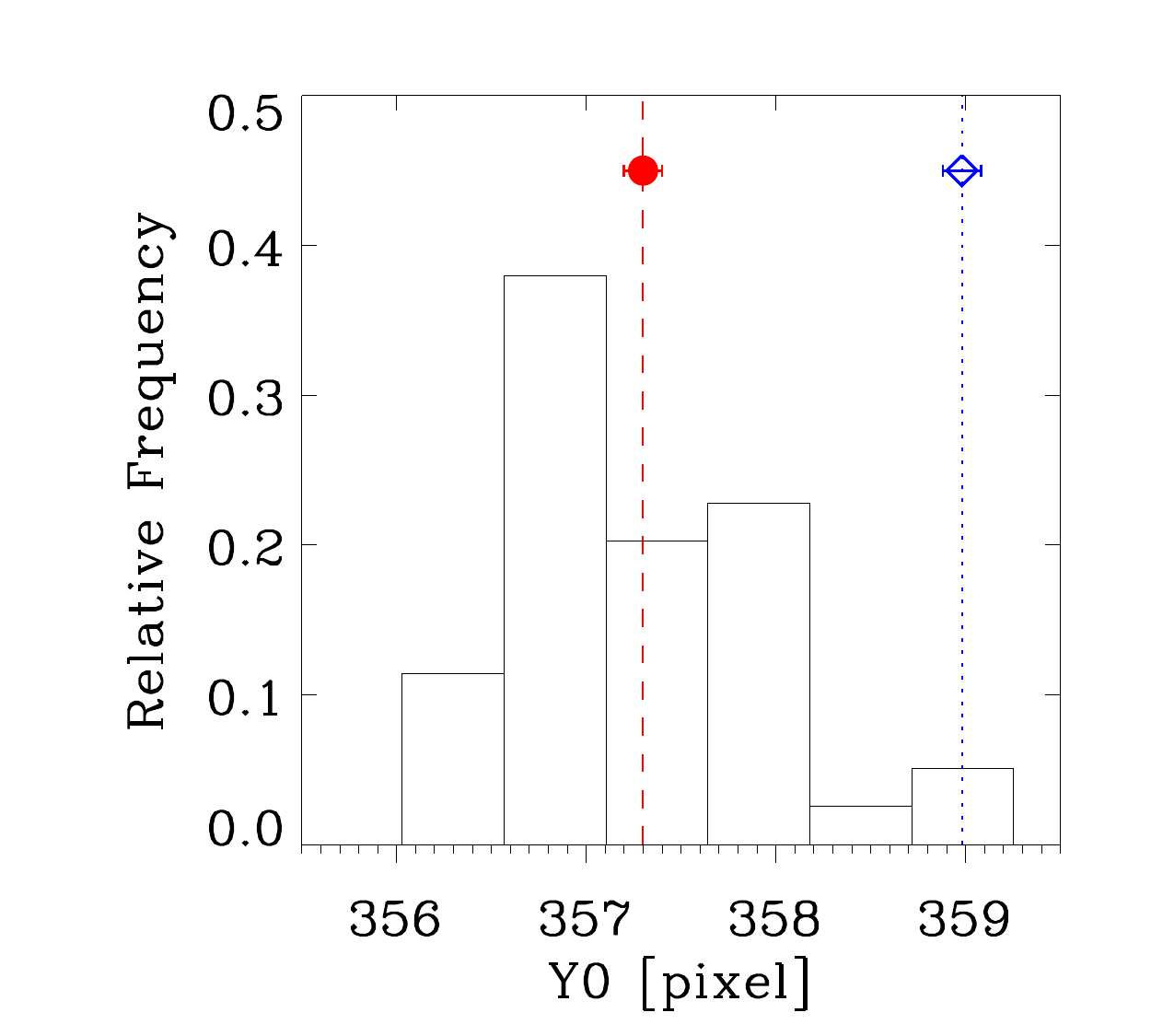}\\
\end{array}$
\end{center}
\caption[NGC 4278]{As in Fig.\ref{fig: NGC4373_W2} for galaxy NGC 4278, WFPC2/PC - F814W, scale=$0\farcs05$/pxl.}
\label{fig: NGC4278_9p_814}
\end{figure*} 

\begin{figure*}[h]
\begin{center}$
\begin{array}{ccc}
\includegraphics[trim=3.75cm 1cm 3cm 0cm, clip=true, scale=0.48]{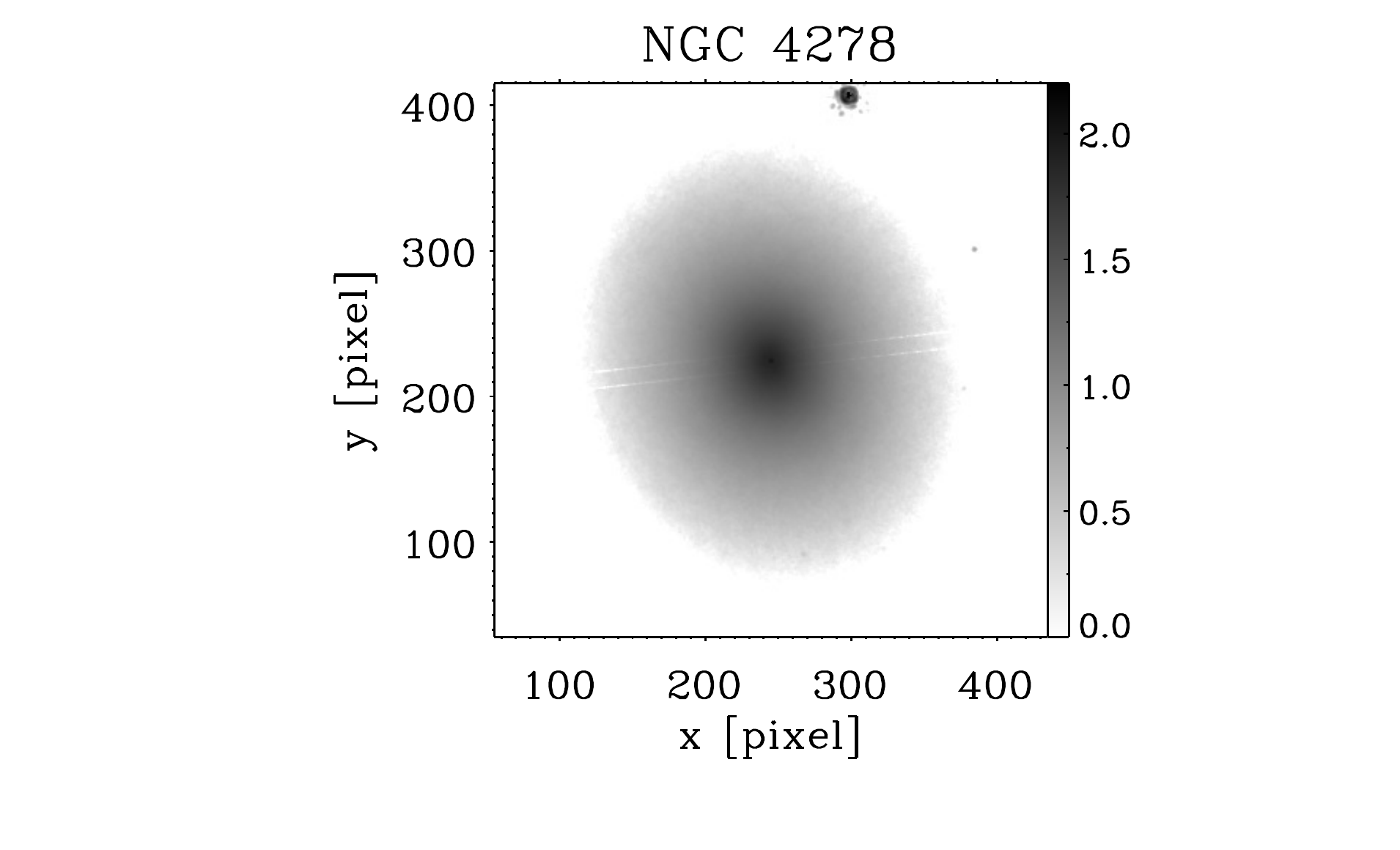} & \includegraphics[trim= 4.cm 1cm 3cm 0cm, clip=true, scale=0.48]{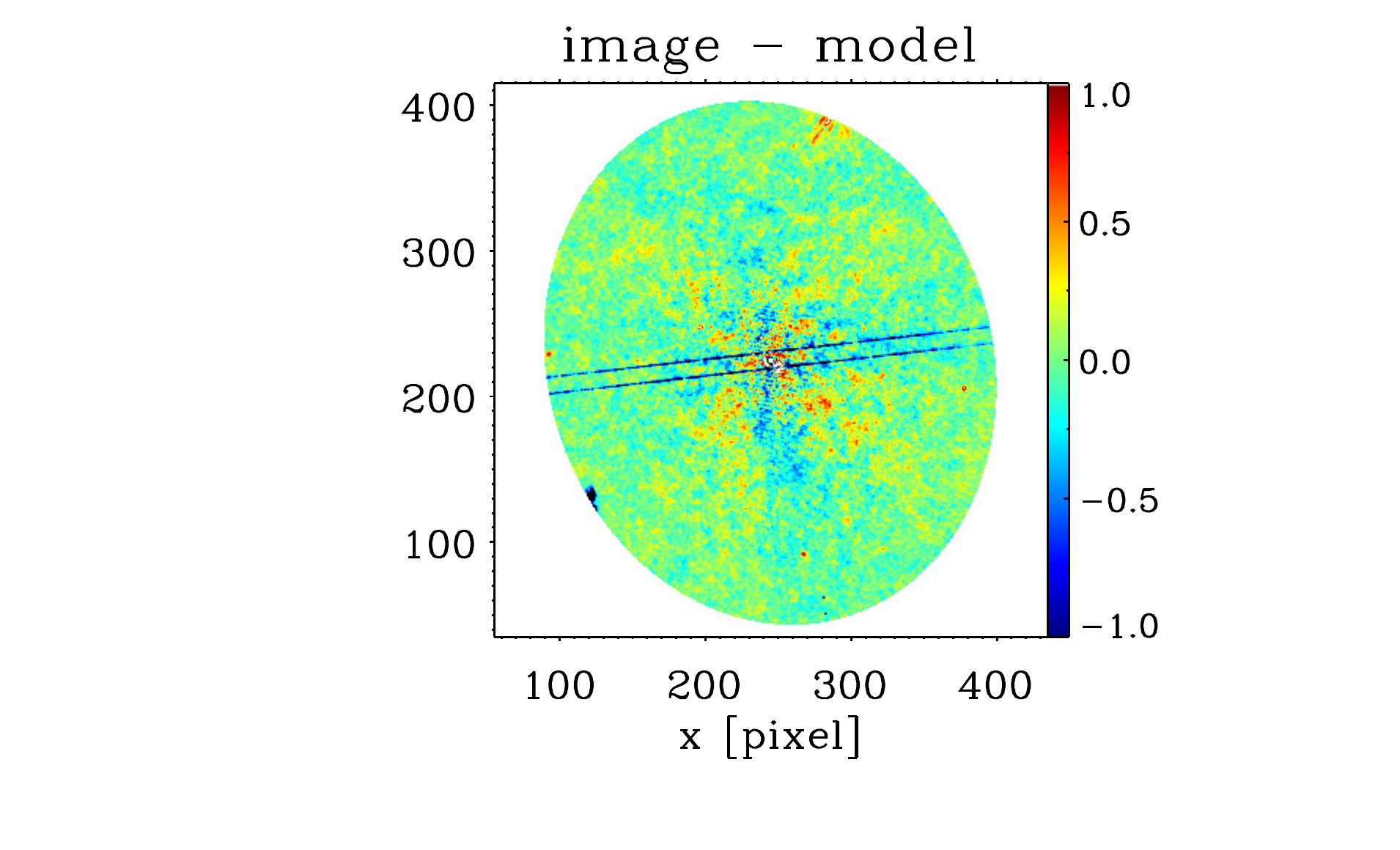}	& \includegraphics[trim= 4.cm 1cm 3cm 0cm, clip=true, scale=0.48]{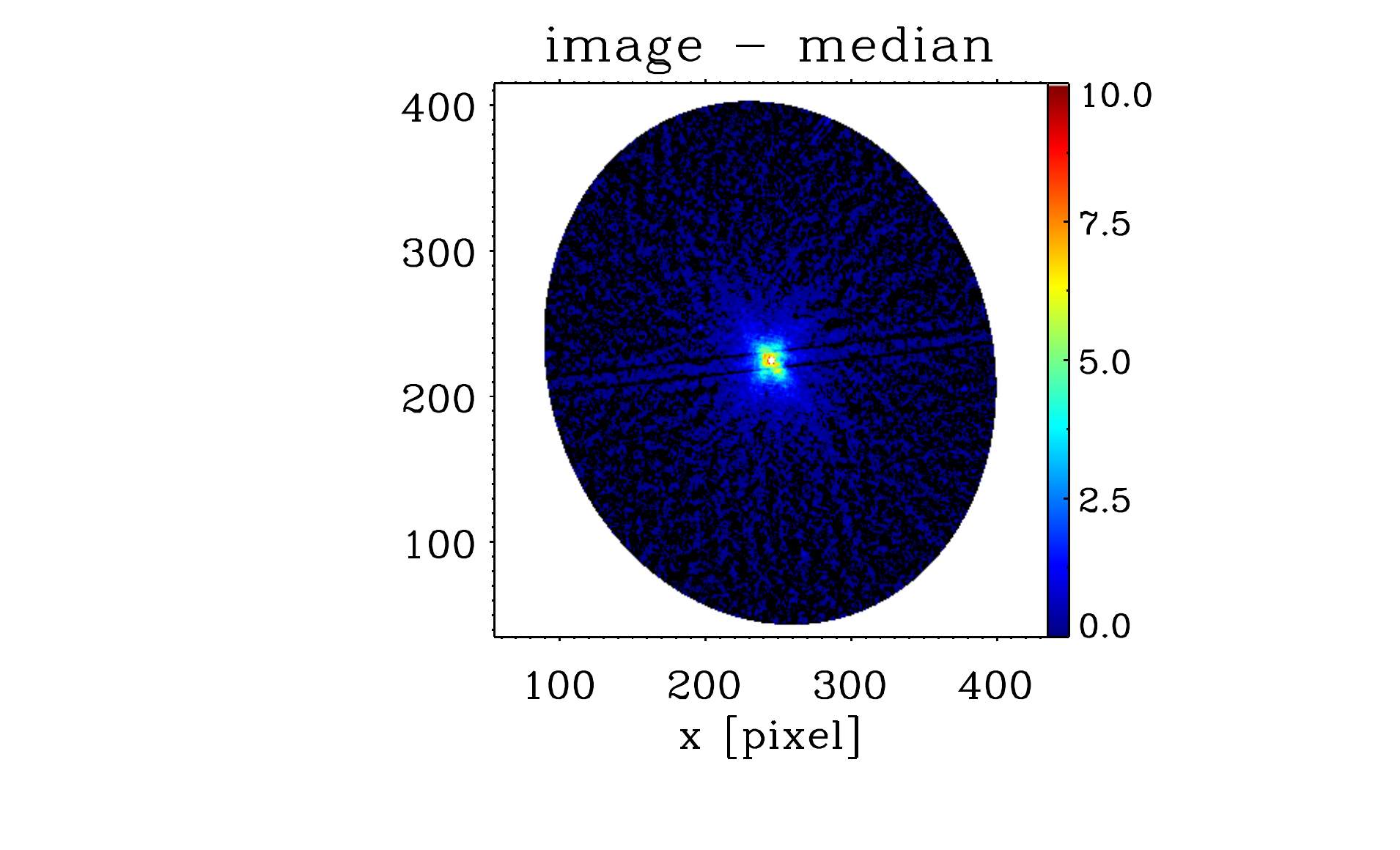} \\
\includegraphics[trim=0.7cm 0cm 0cm 0cm, clip=true, scale=0.46]{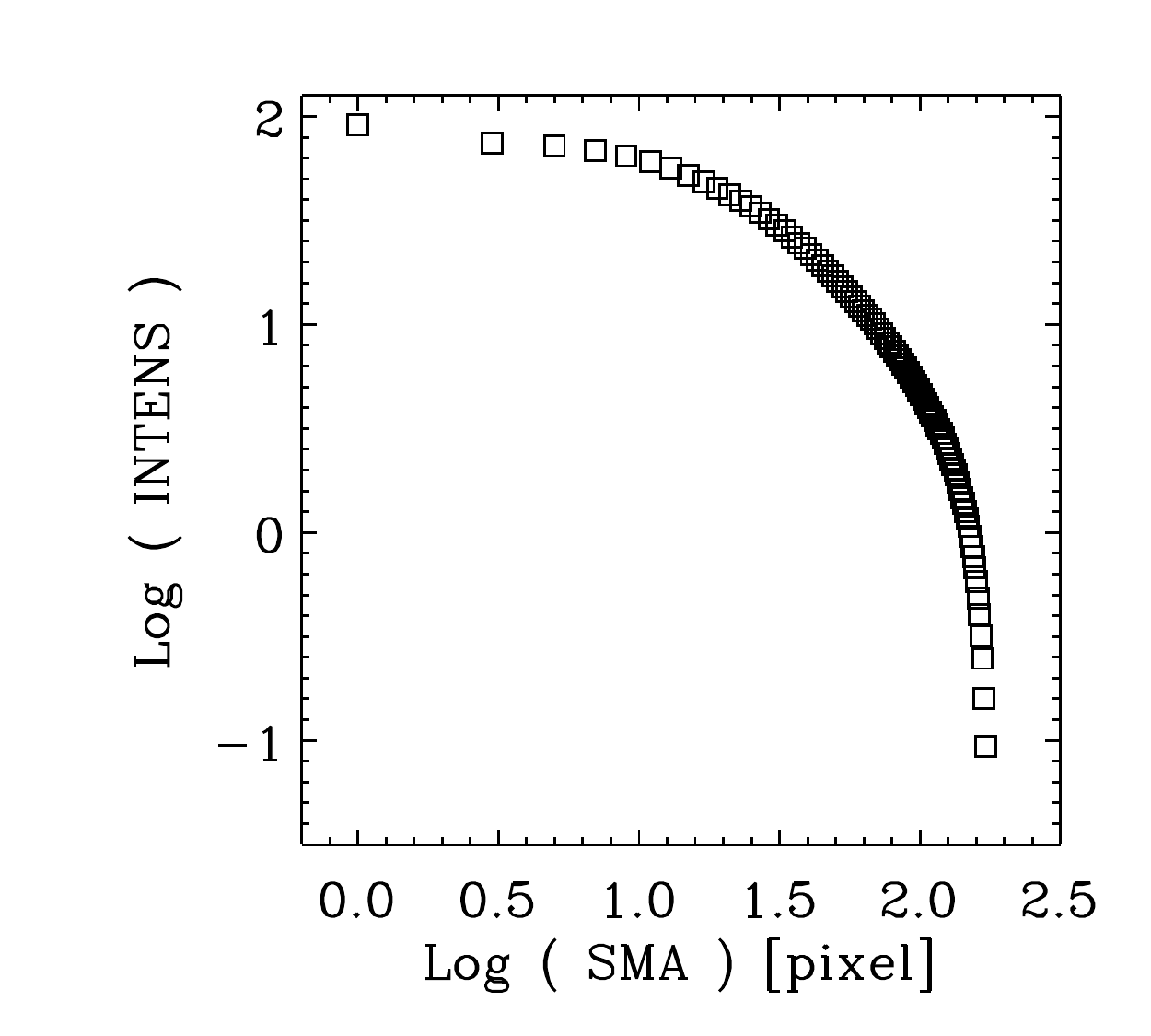}	    &  \includegraphics[trim=0.6cm 0cm 0cm 0cm, clip=true, scale=0.46]{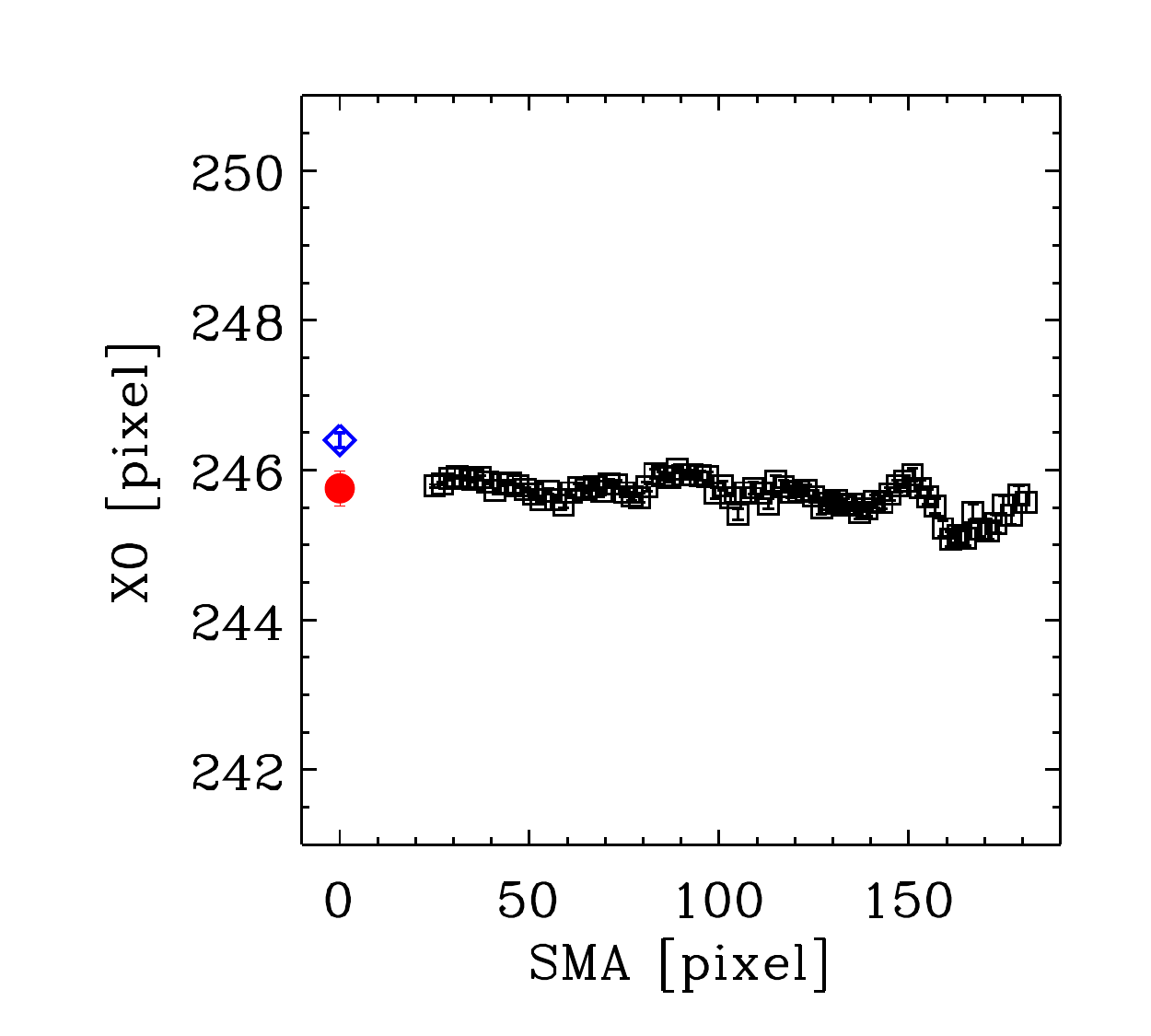}  &  \includegraphics[trim=0.6cm 0cm 0cm 0cm, clip=true, scale=0.46]{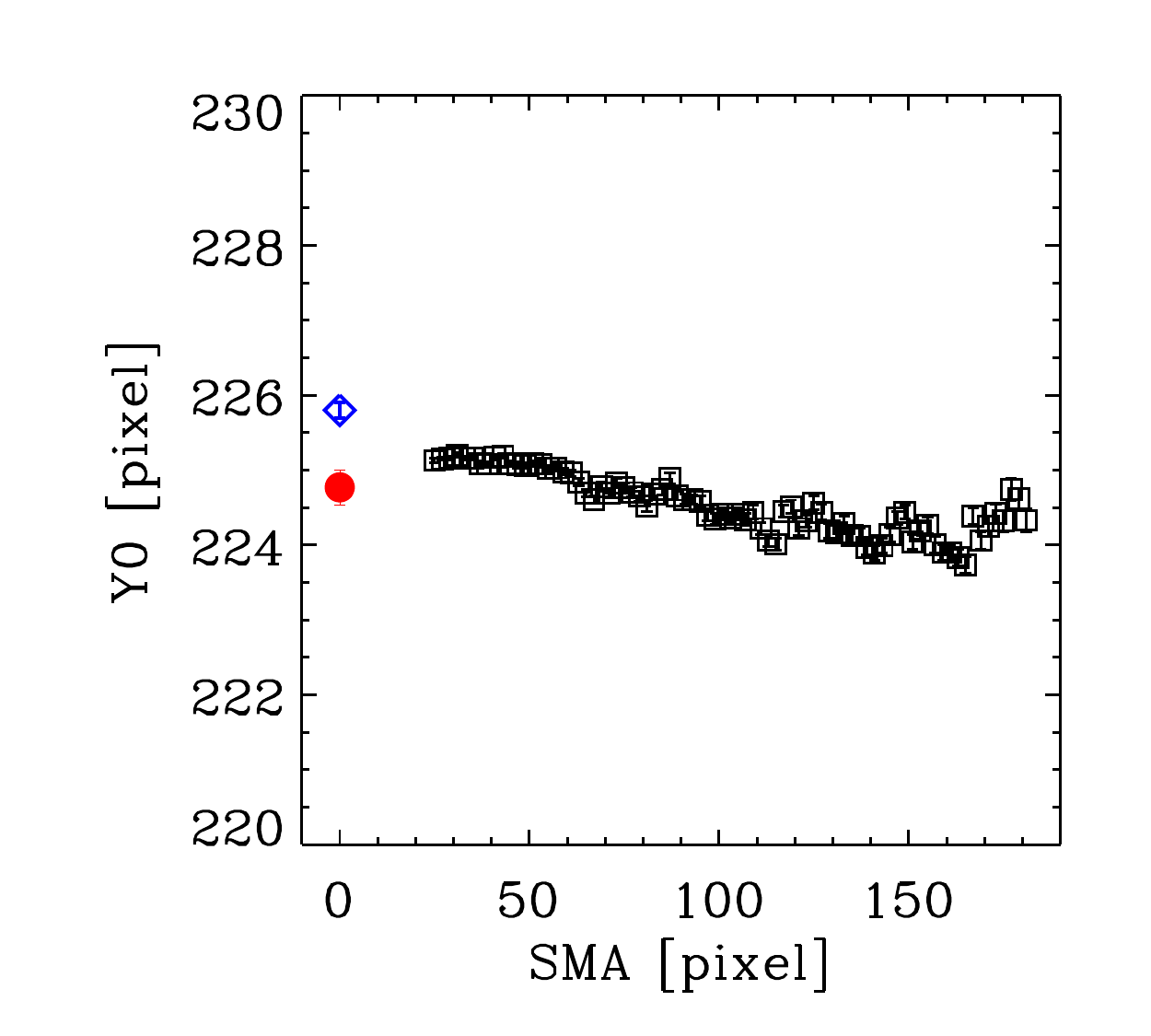} \\	
 \includegraphics[trim=0.65cm 0cm 0cm 0cm, clip=true, scale=0.46]{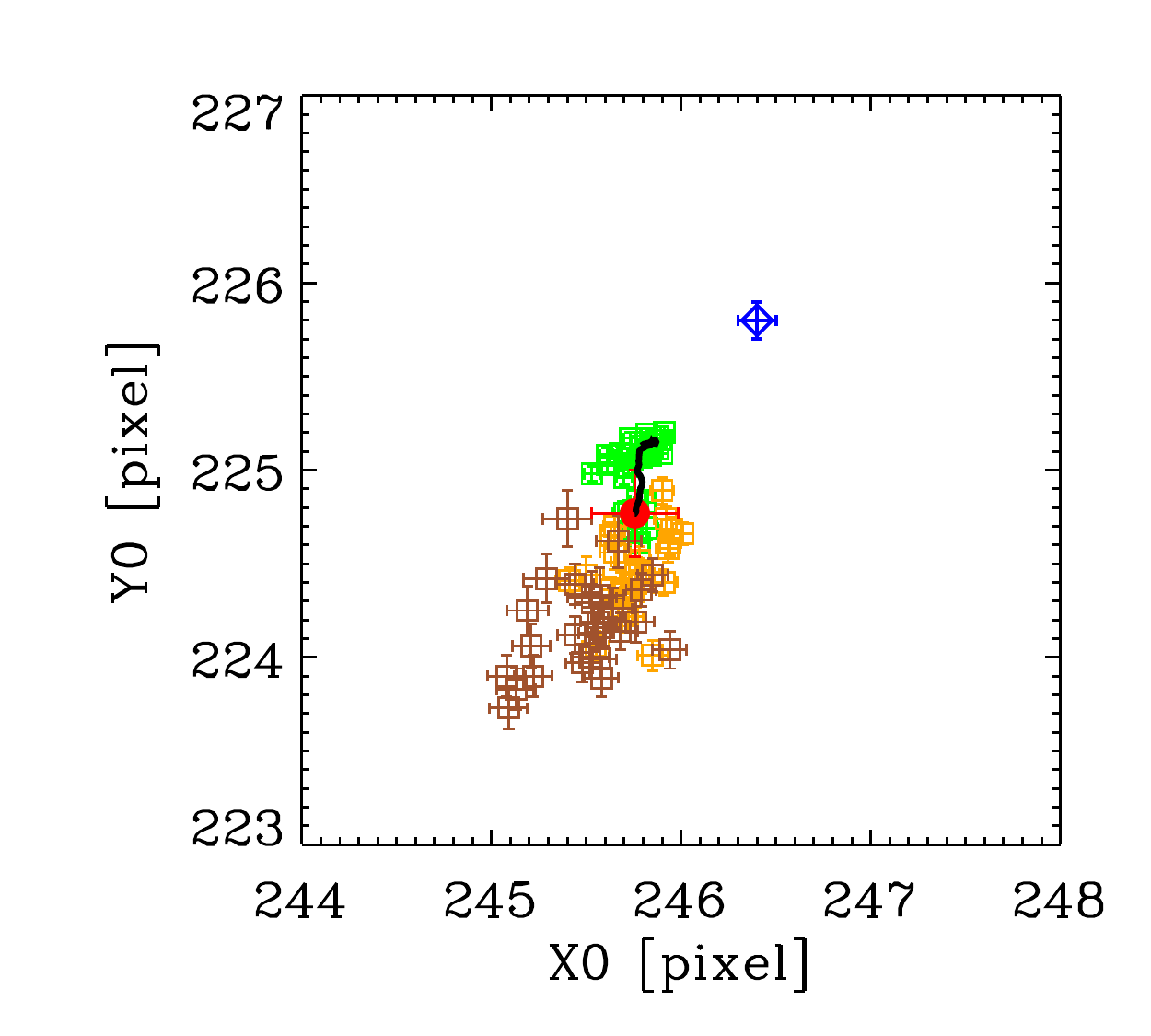}	&  \includegraphics[trim=0.6cm 0cm 0cm 0cm, clip=true, scale=0.46]{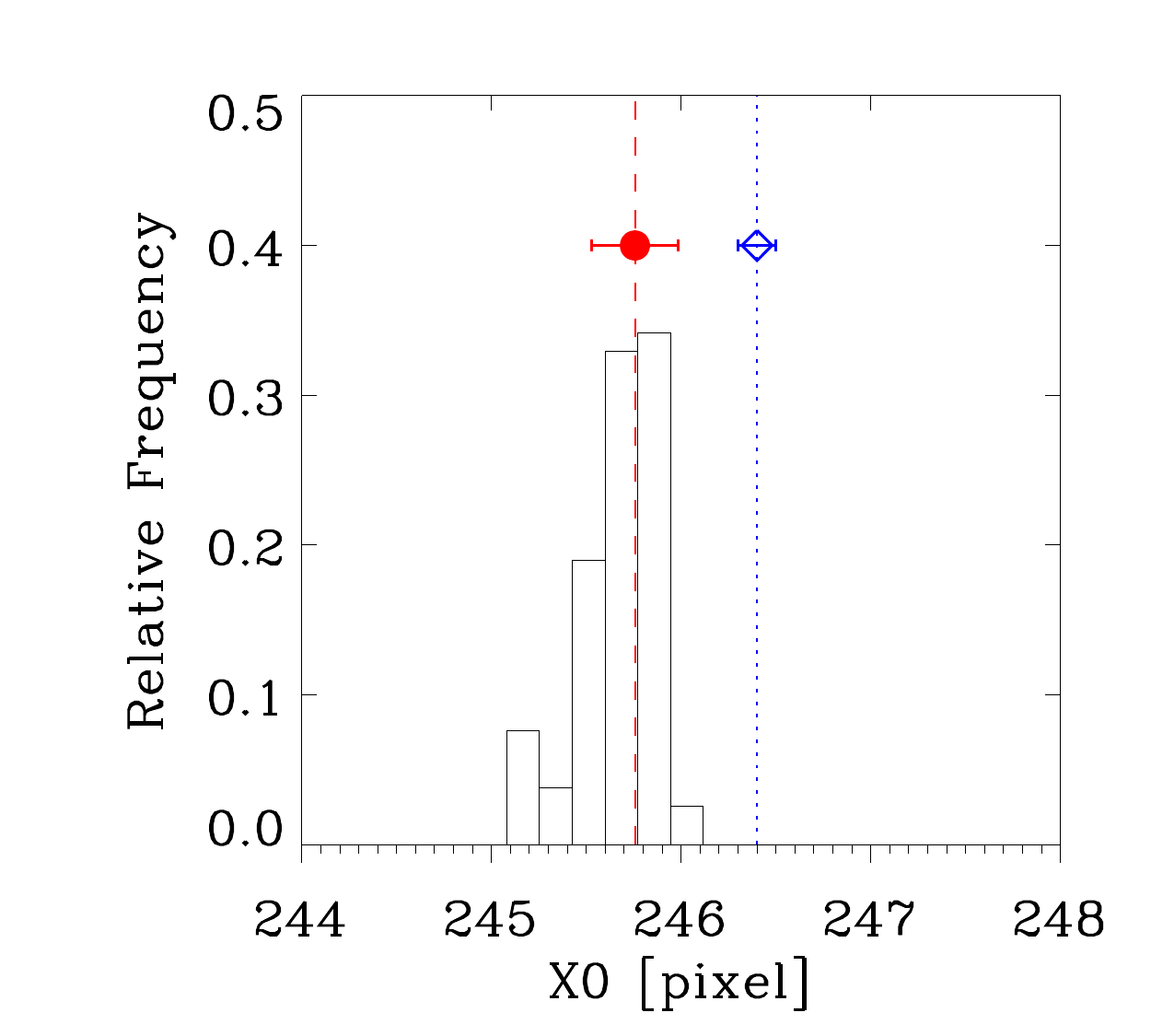}	& \includegraphics[trim=0.6cm 0cm 0cm 0cm, clip=true, scale=0.46]{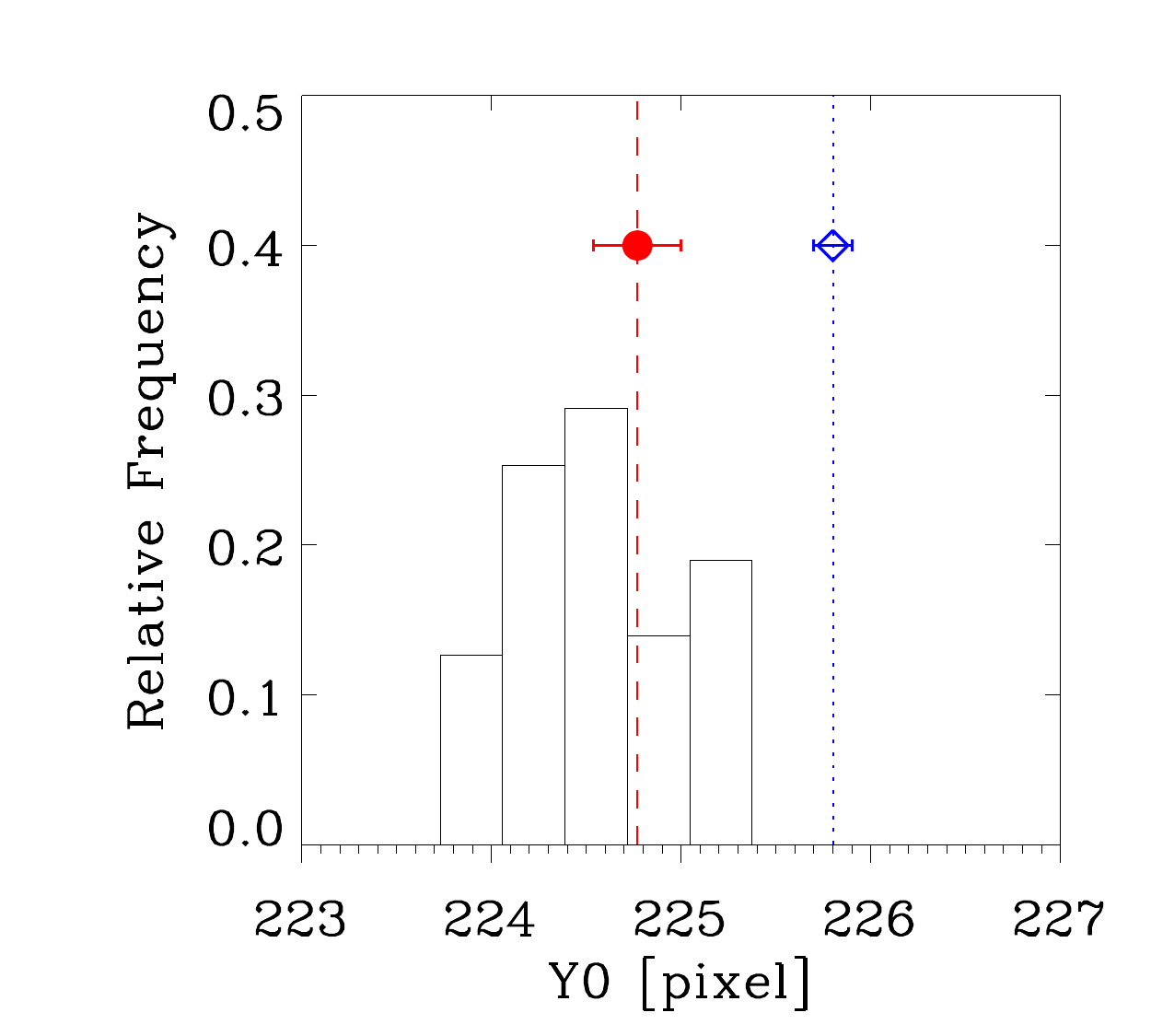}\\
\end{array}$
\end{center}
\caption[NGC 4278]{ As in Fig.\ref{fig: NGC4373_W2} for galaxy NGC 4278, NICMOS2 - F160W, scale=$0\farcs05$/pxl.}
\label{fig: NGC4278_NIC2}
\end{figure*} 

\begin{figure*}[h]
\begin{center}$
\begin{array}{ccc}
\includegraphics[trim=3.75cm 1cm 3cm 0cm, clip=true, scale=0.48]{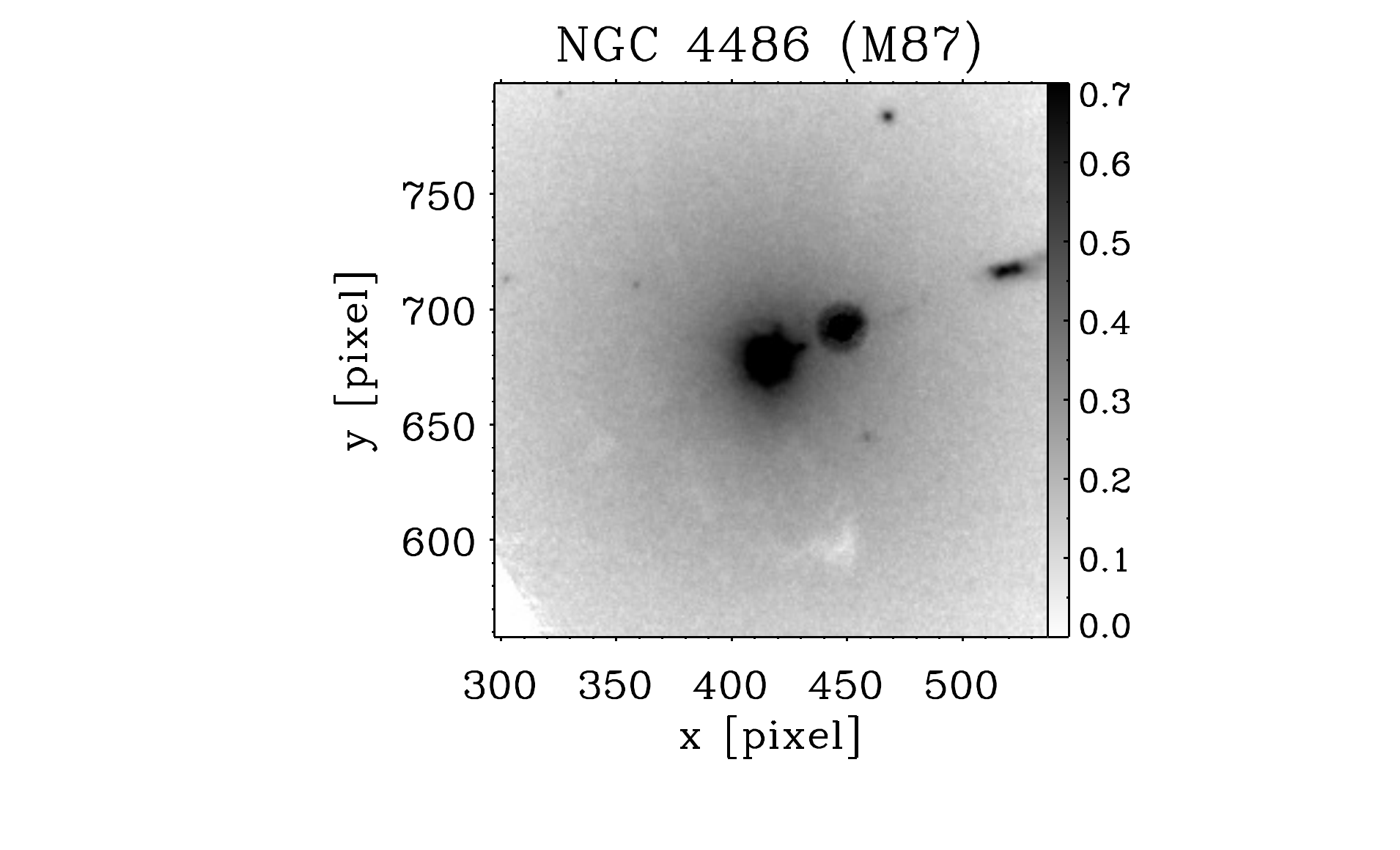} & \includegraphics[trim= 4.cm 1cm 3cm 0cm, clip=true, scale=0.48]{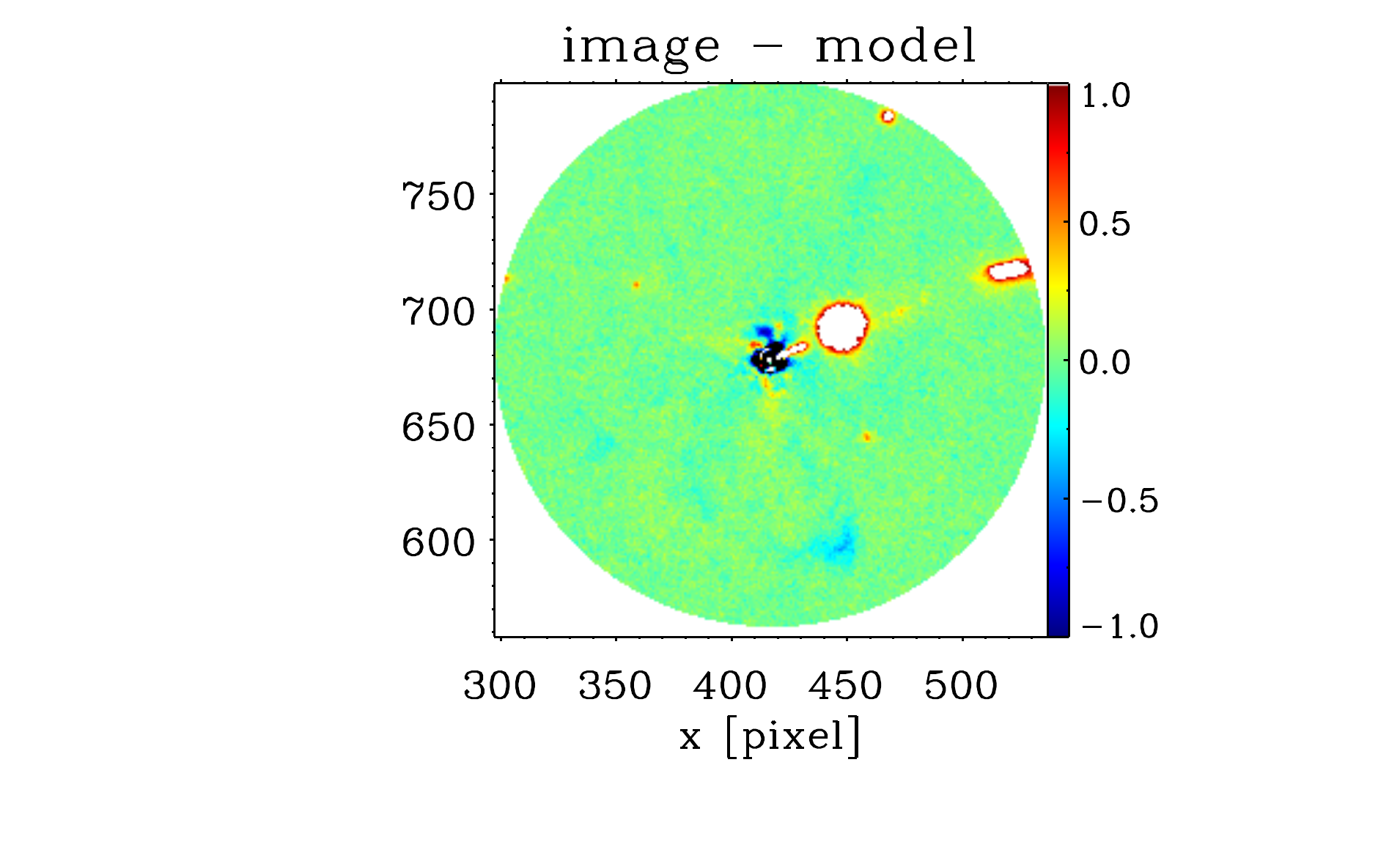}	& \includegraphics[trim= 4.cm 1cm 3cm 0cm, clip=true, scale=0.48]{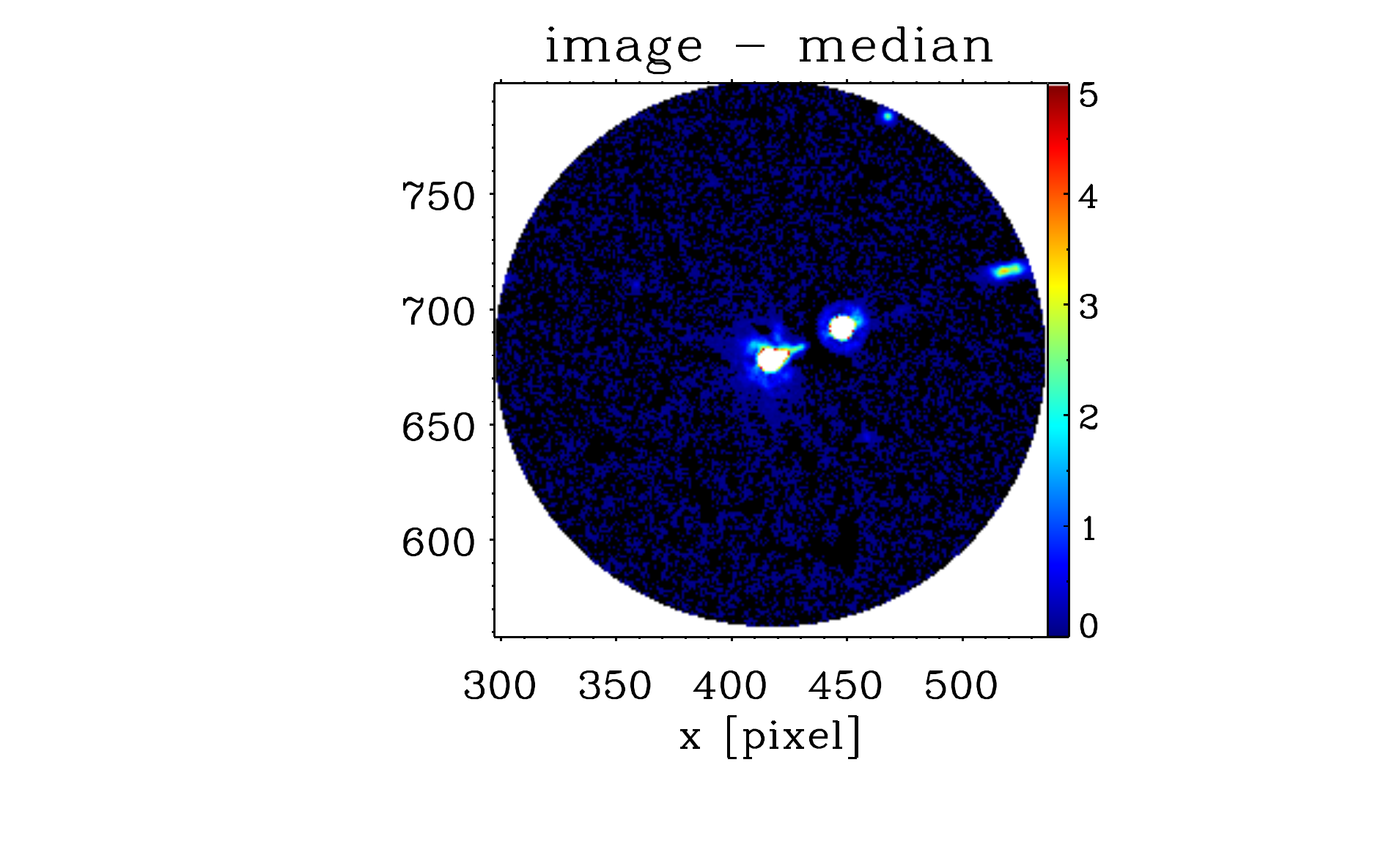} \\
\includegraphics[trim=0.7cm 0cm 0cm 0cm, clip=true, scale=0.46]{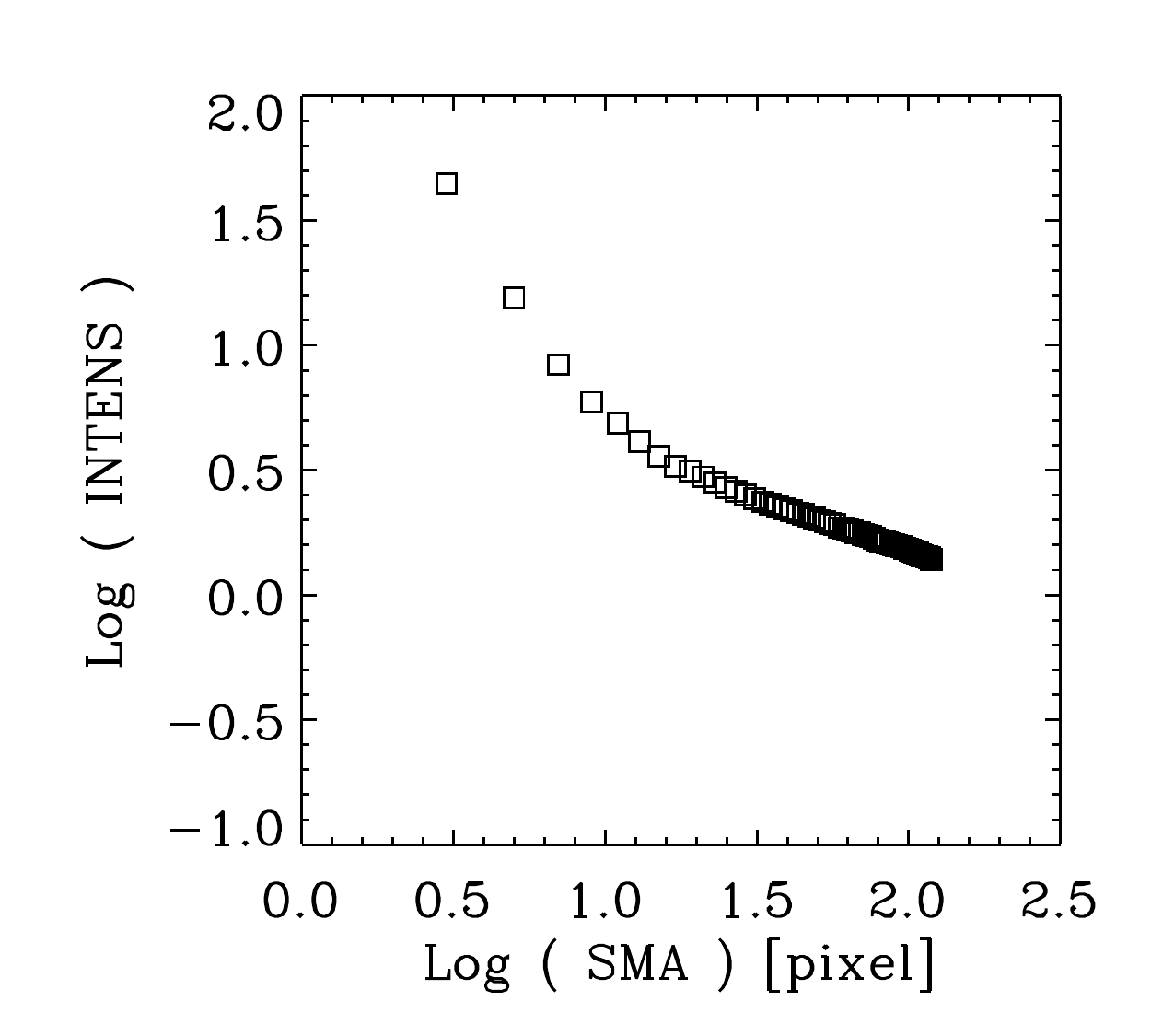}	    &  \includegraphics[trim=0.6cm 0cm 0cm 0cm, clip=true, scale=0.46]{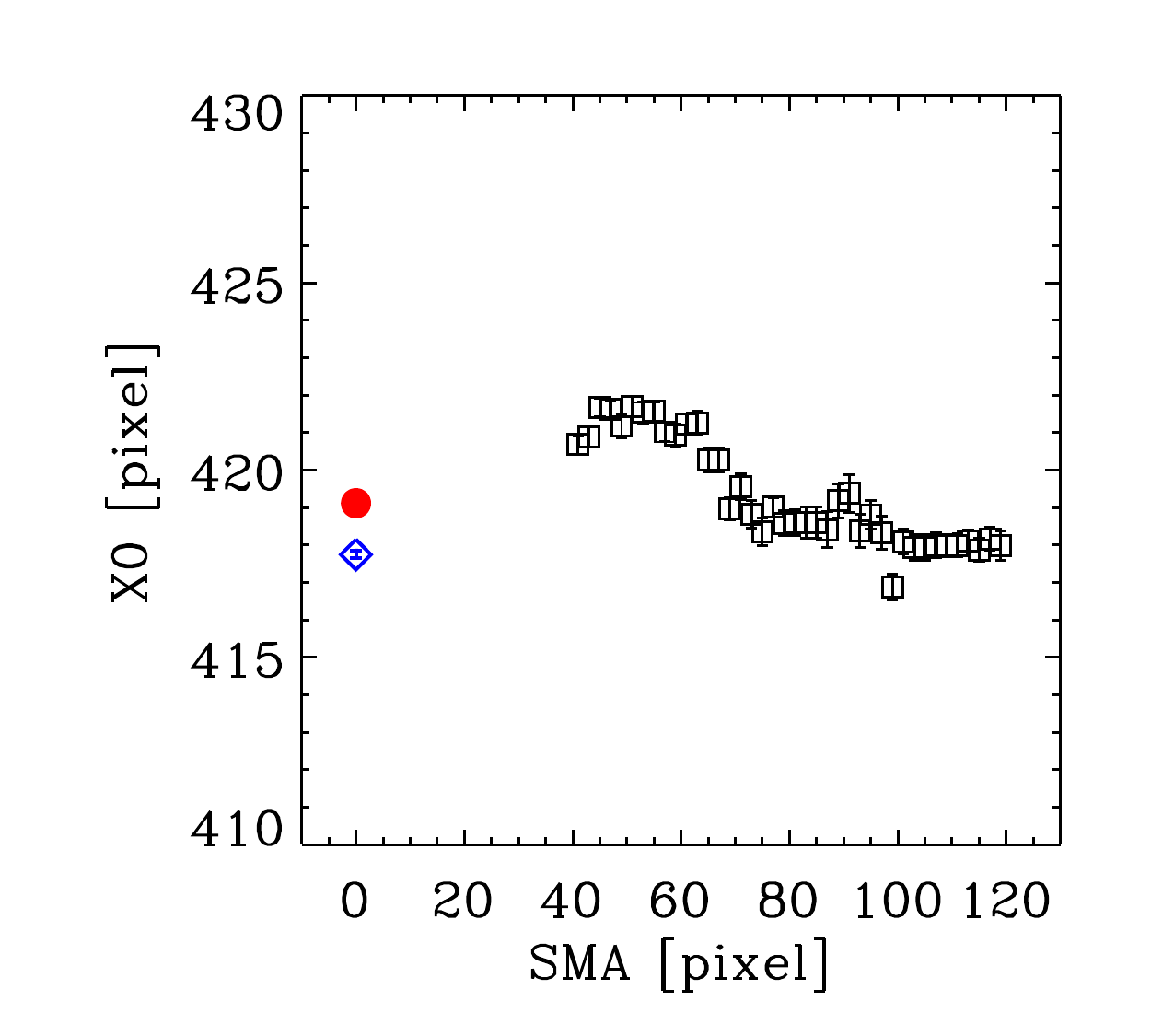}  &  \includegraphics[trim=0.6cm 0cm 0cm 0cm, clip=true, scale=0.46]{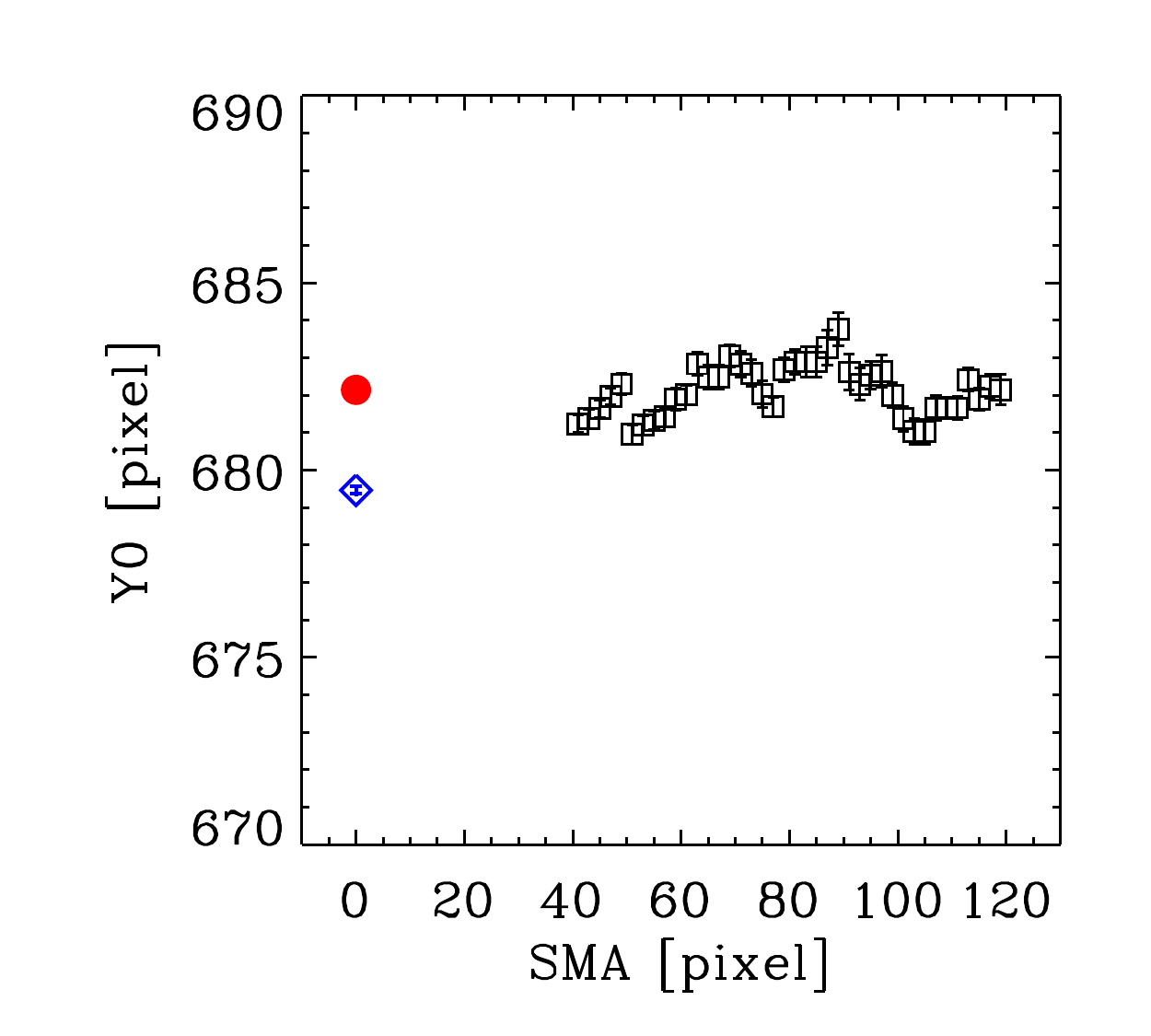} \\	
 \includegraphics[trim=0.65cm 0cm 0cm 0cm, clip=true, scale=0.46]{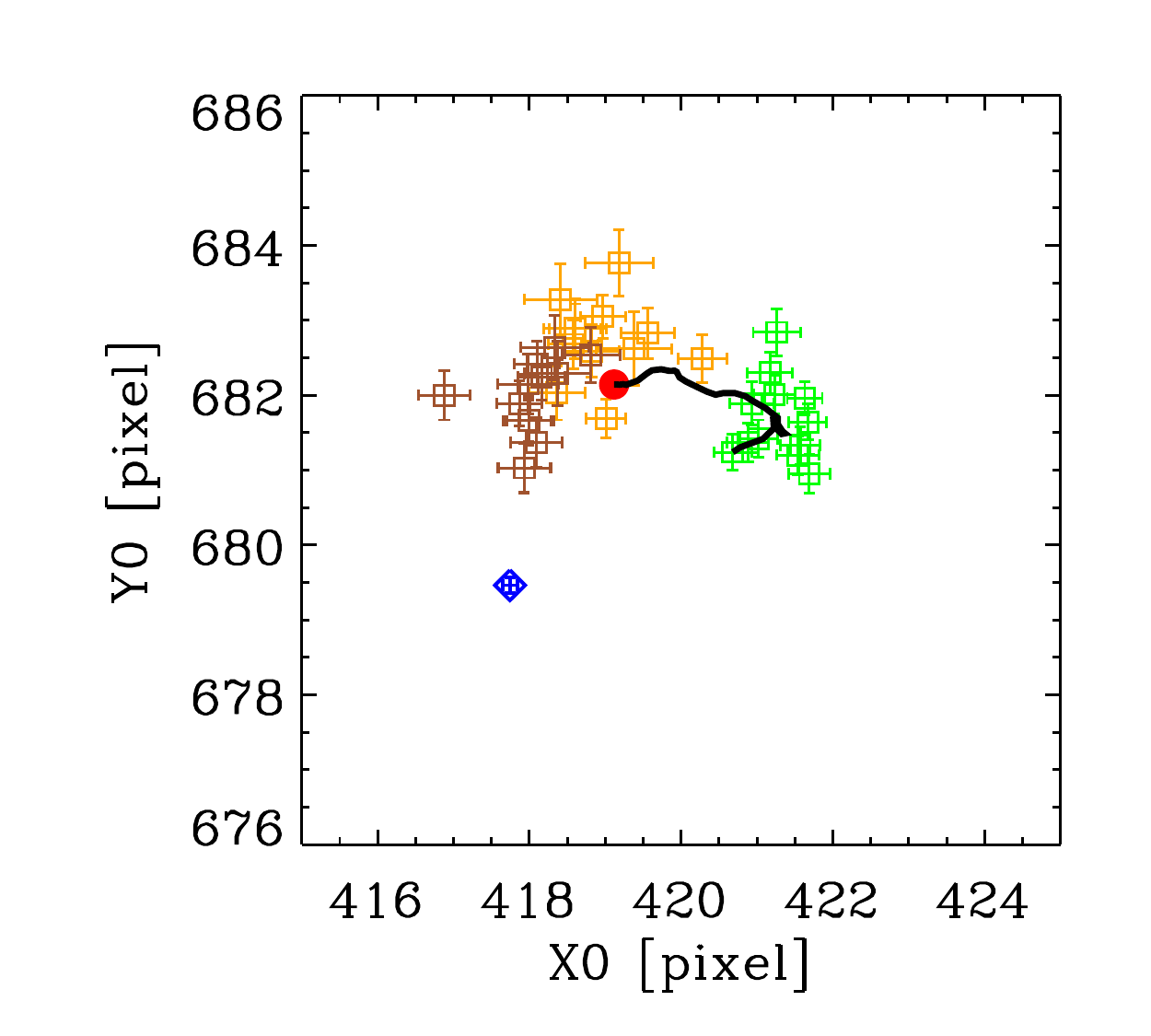}	&  \includegraphics[trim=0.6cm 0cm 0cm 0cm, clip=true, scale=0.46]{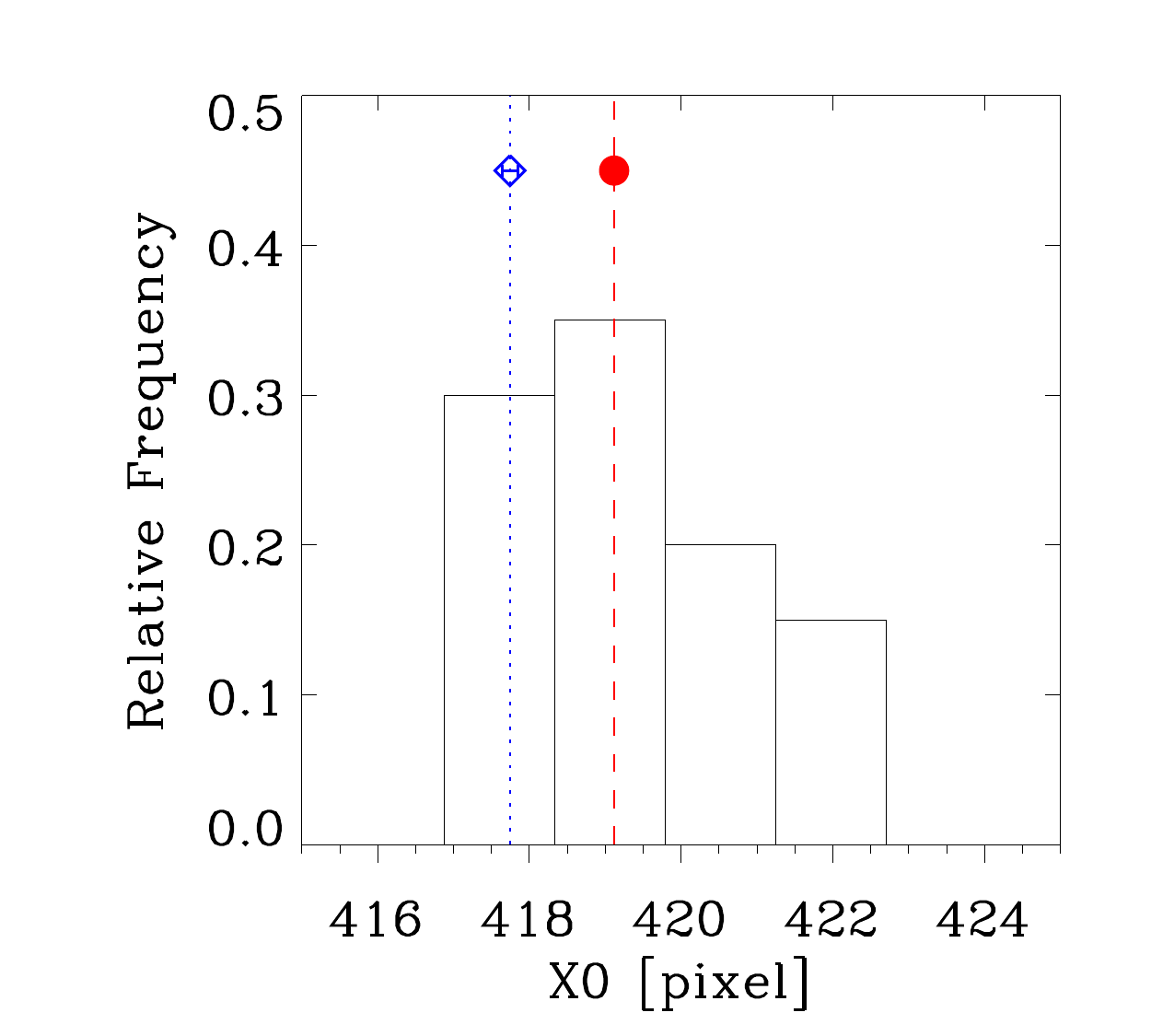}	& \includegraphics[trim=0.6cm 0cm 0cm 0cm, clip=true, scale=0.46]{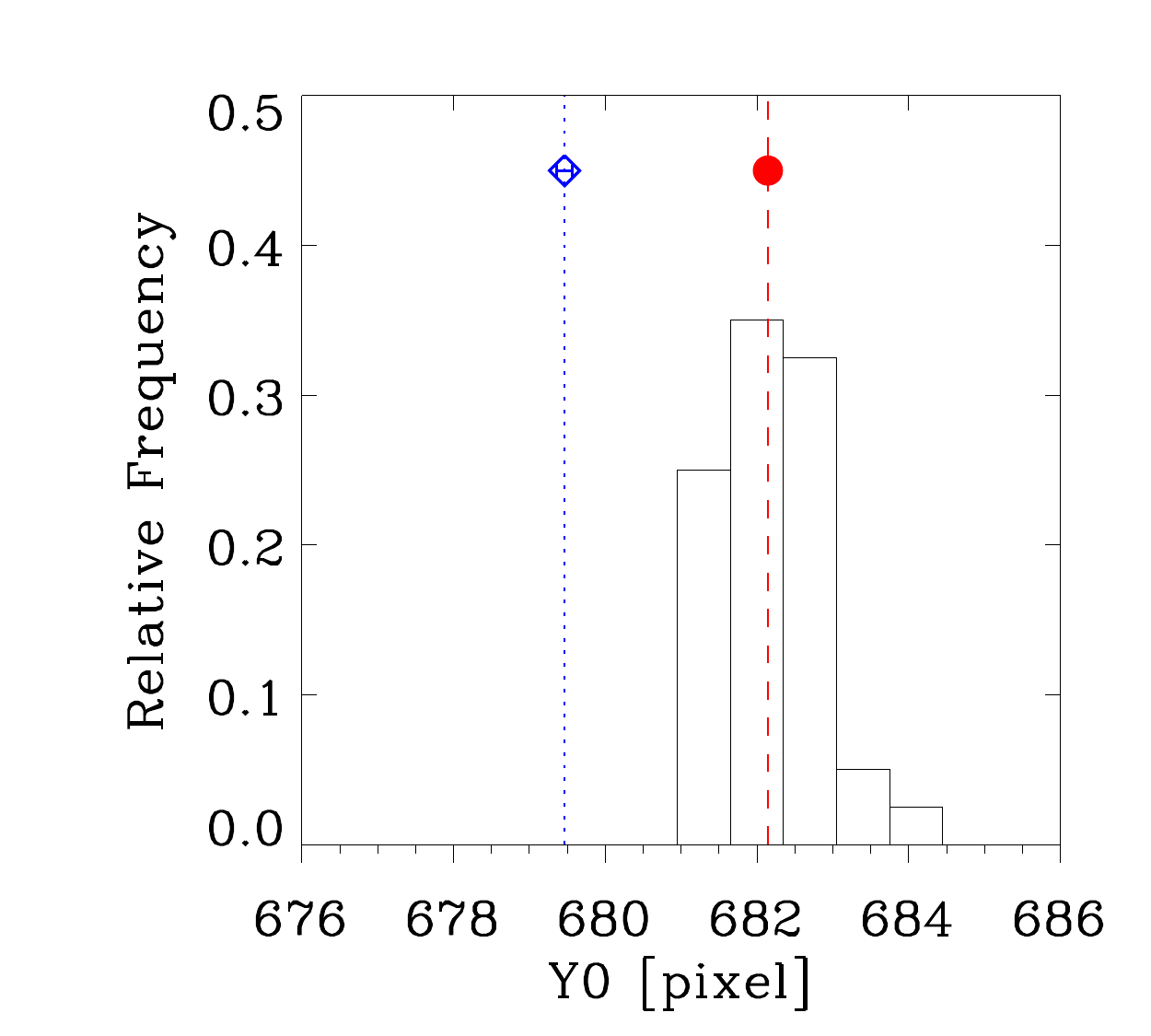}\\
\end{array}$
\end{center}
\caption[M87 (ACS)]{As in Fig.\ref{fig: NGC4373_W2} for galaxy NGC 4486 (M 87), ACS/HRC - F606W, scale=0\farcs025/pxl.}
\label{fig: NGC4486_ACSF606W}
\end{figure*} 

\begin{figure*}[h]
\begin{center}$
\begin{array}{ccc}
\includegraphics[trim=3.75cm 1cm 3cm 0cm, clip=true, scale=0.48]{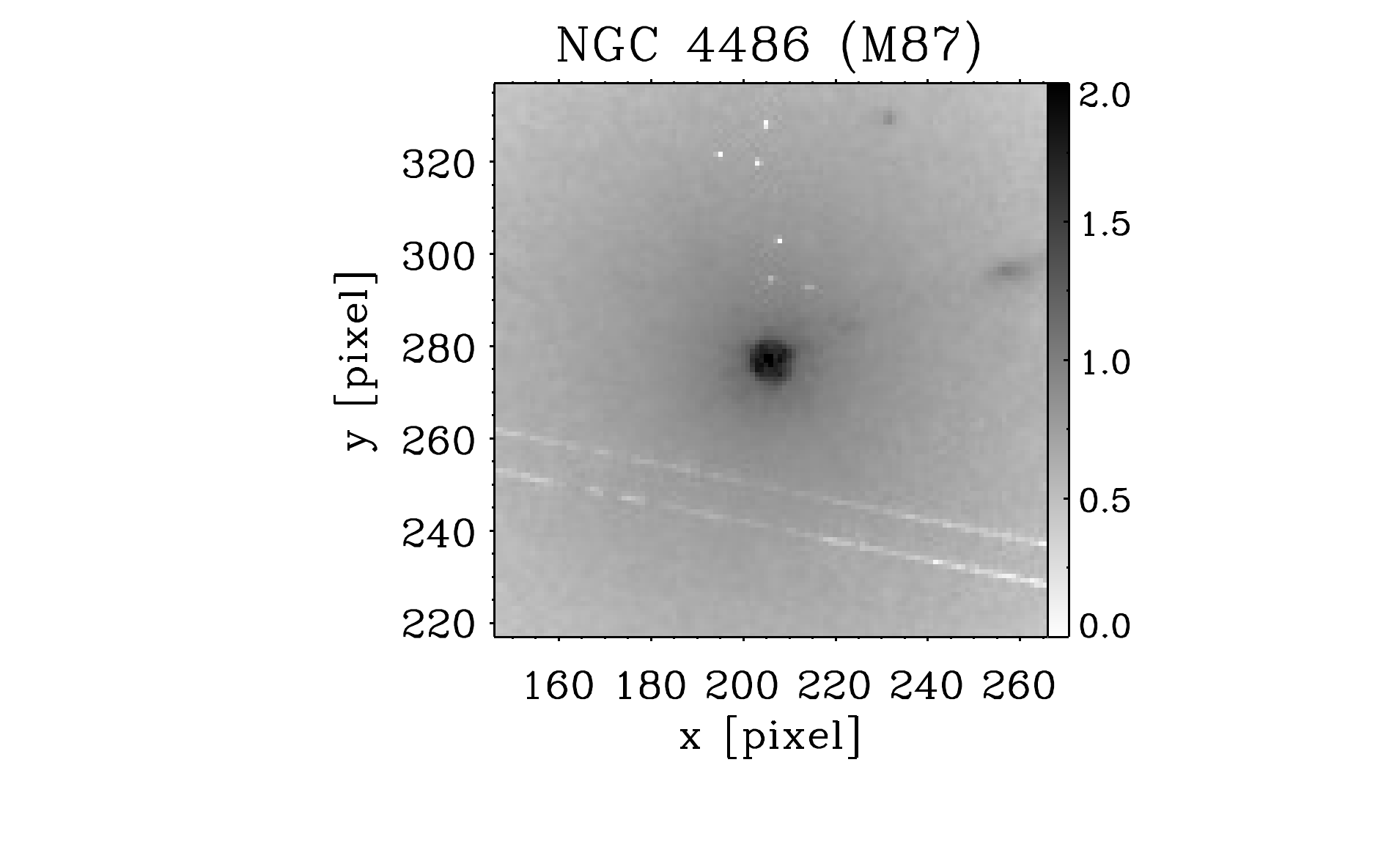} & \includegraphics[trim= 4.cm 1cm 3cm 0cm, clip=true, scale=0.48]{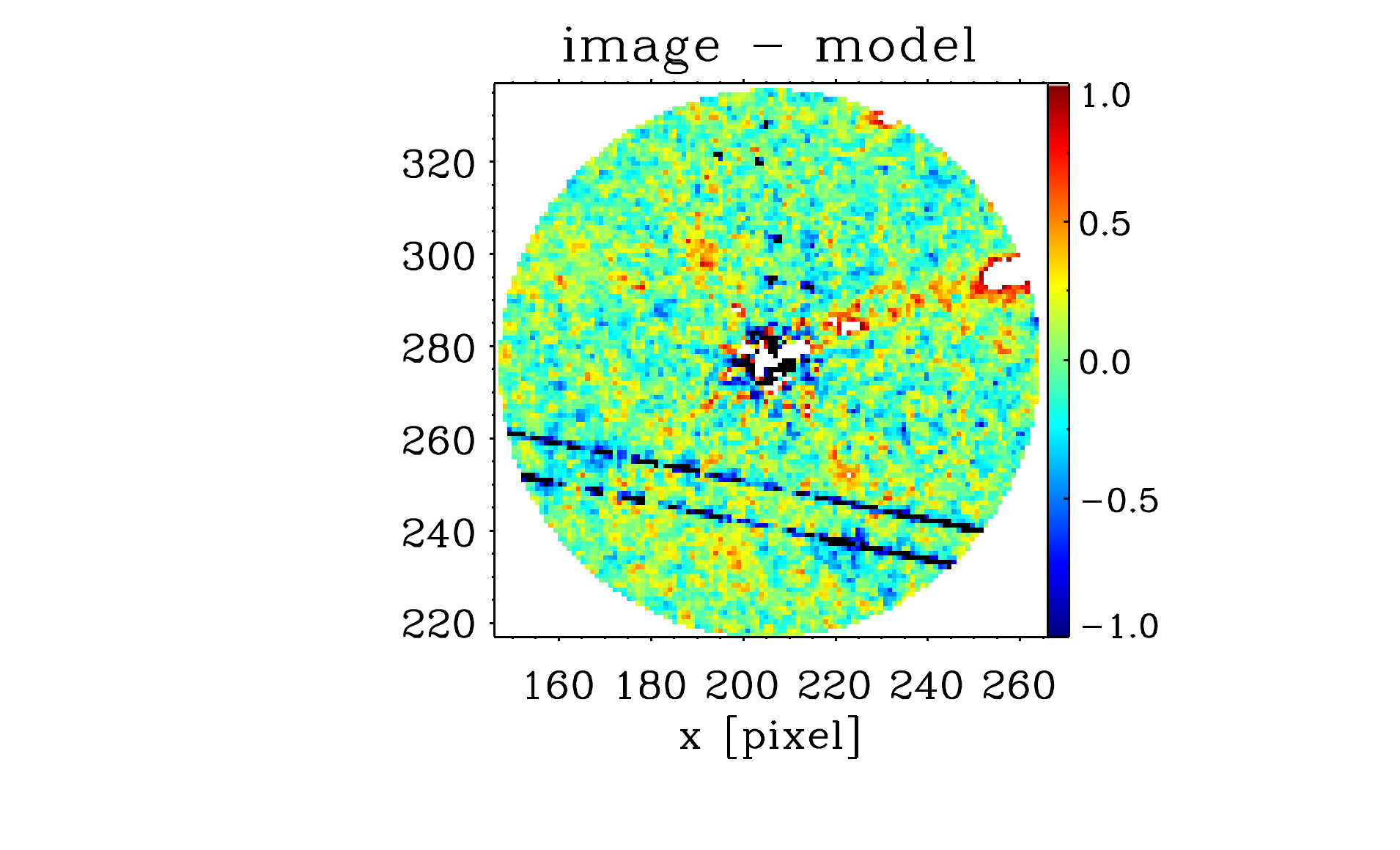}	& \includegraphics[trim= 4.cm 1cm 3cm 0cm, clip=true, scale=0.48]{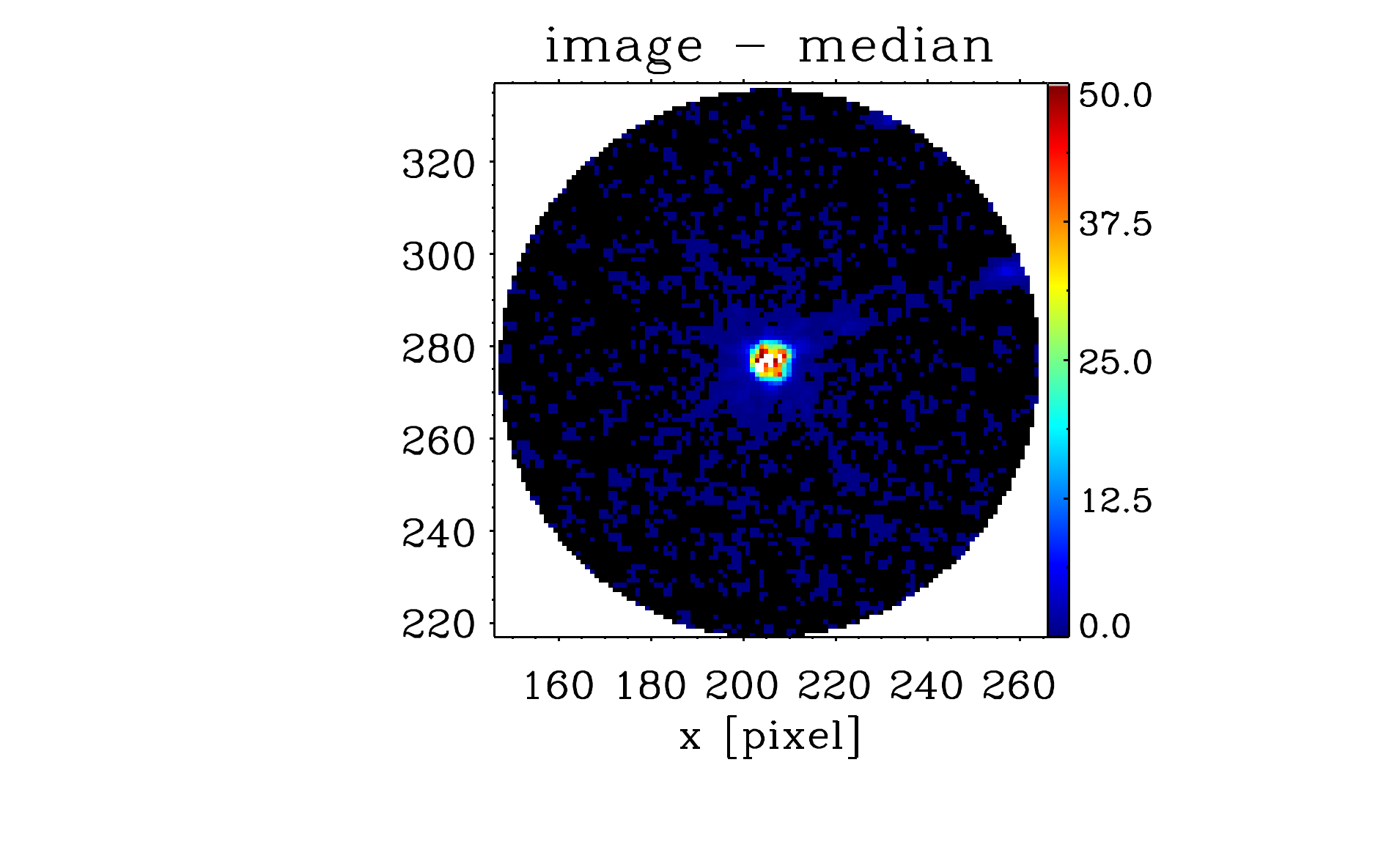} \\
\includegraphics[trim=0.7cm 0cm 0cm 0cm, clip=true, scale=0.46]{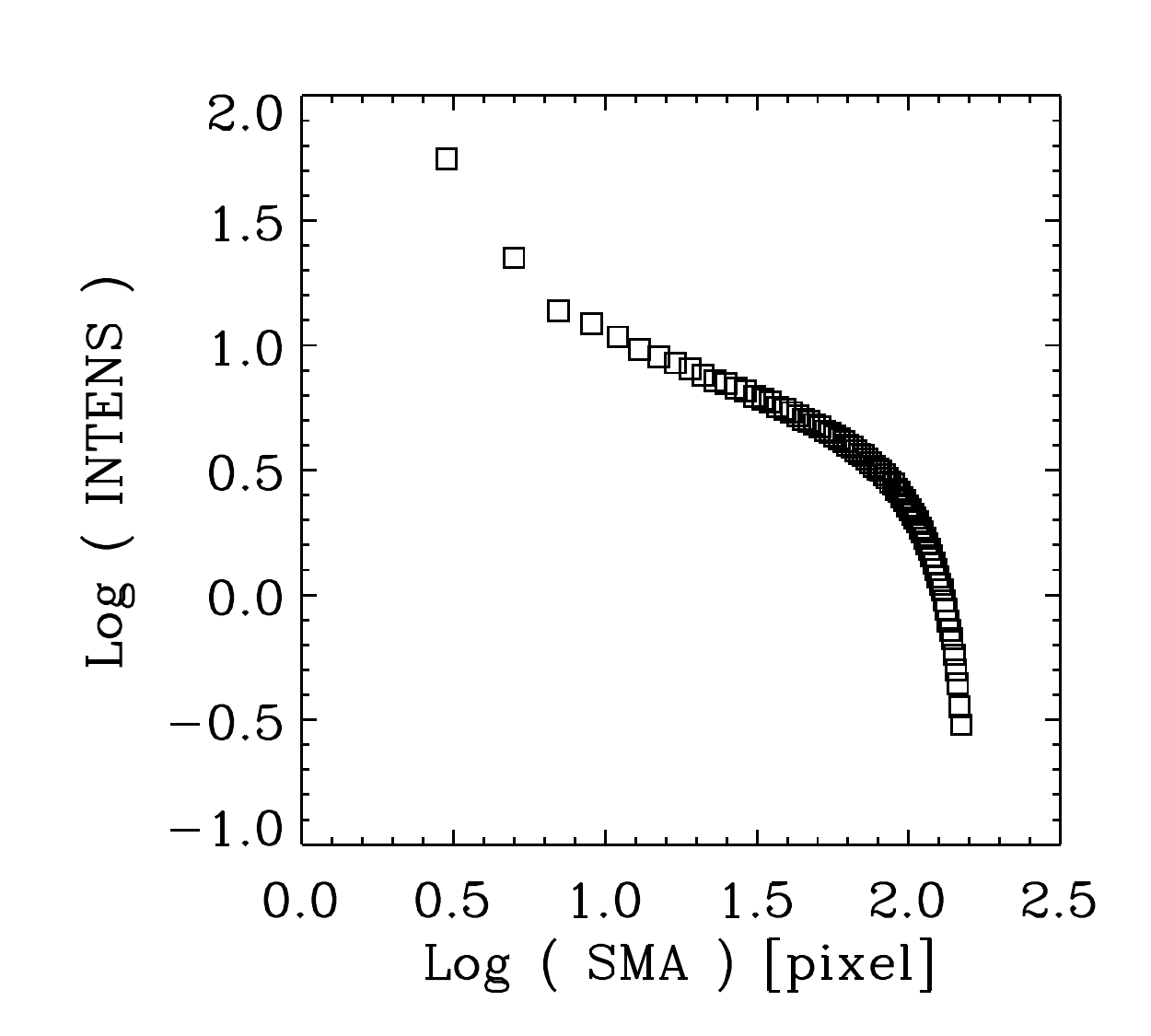}	    &  \includegraphics[trim=0.6cm 0cm 0cm 0cm, clip=true, scale=0.46]{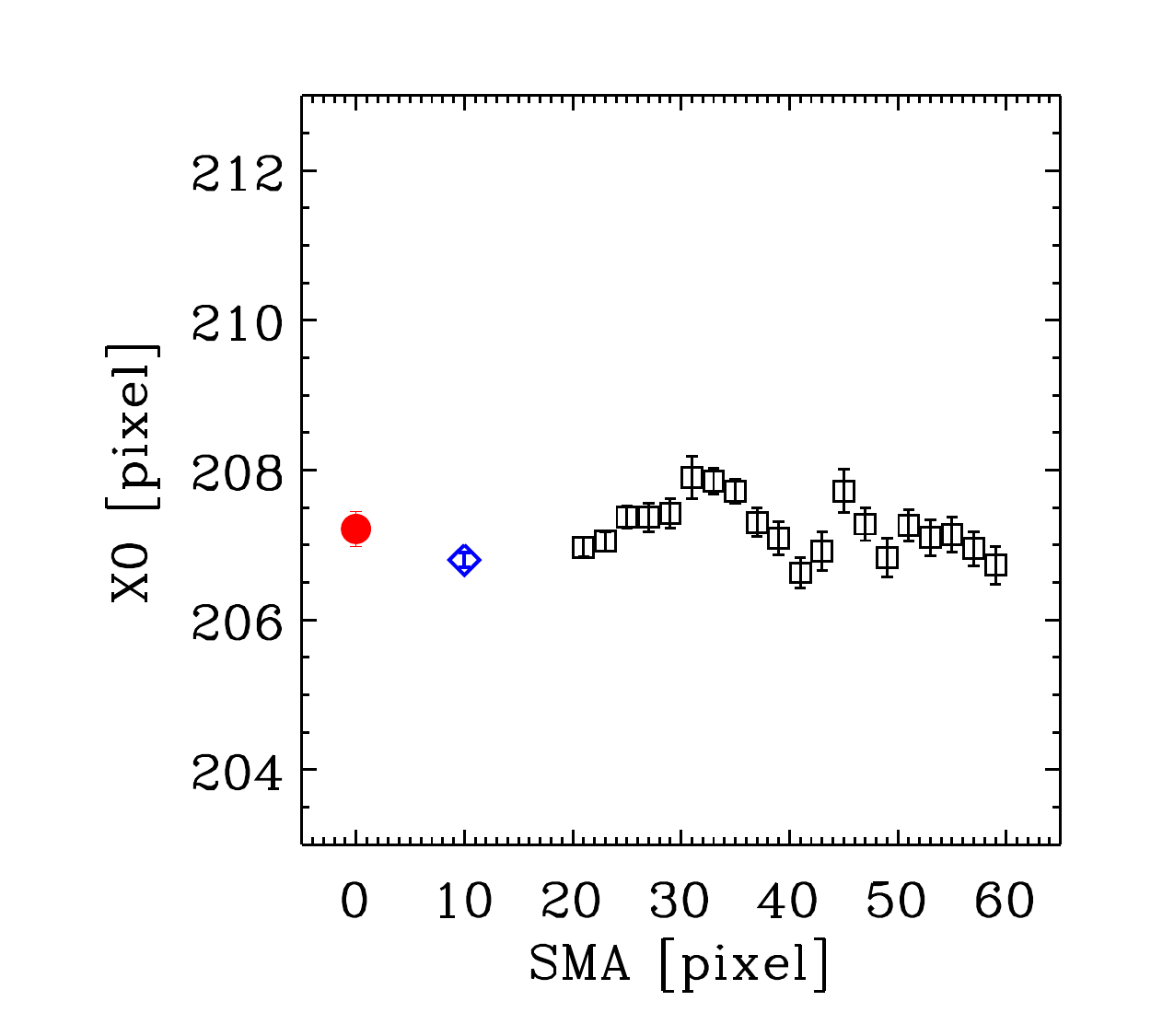}  &  \includegraphics[trim=0.6cm 0cm 0cm 0cm, clip=true, scale=0.46]{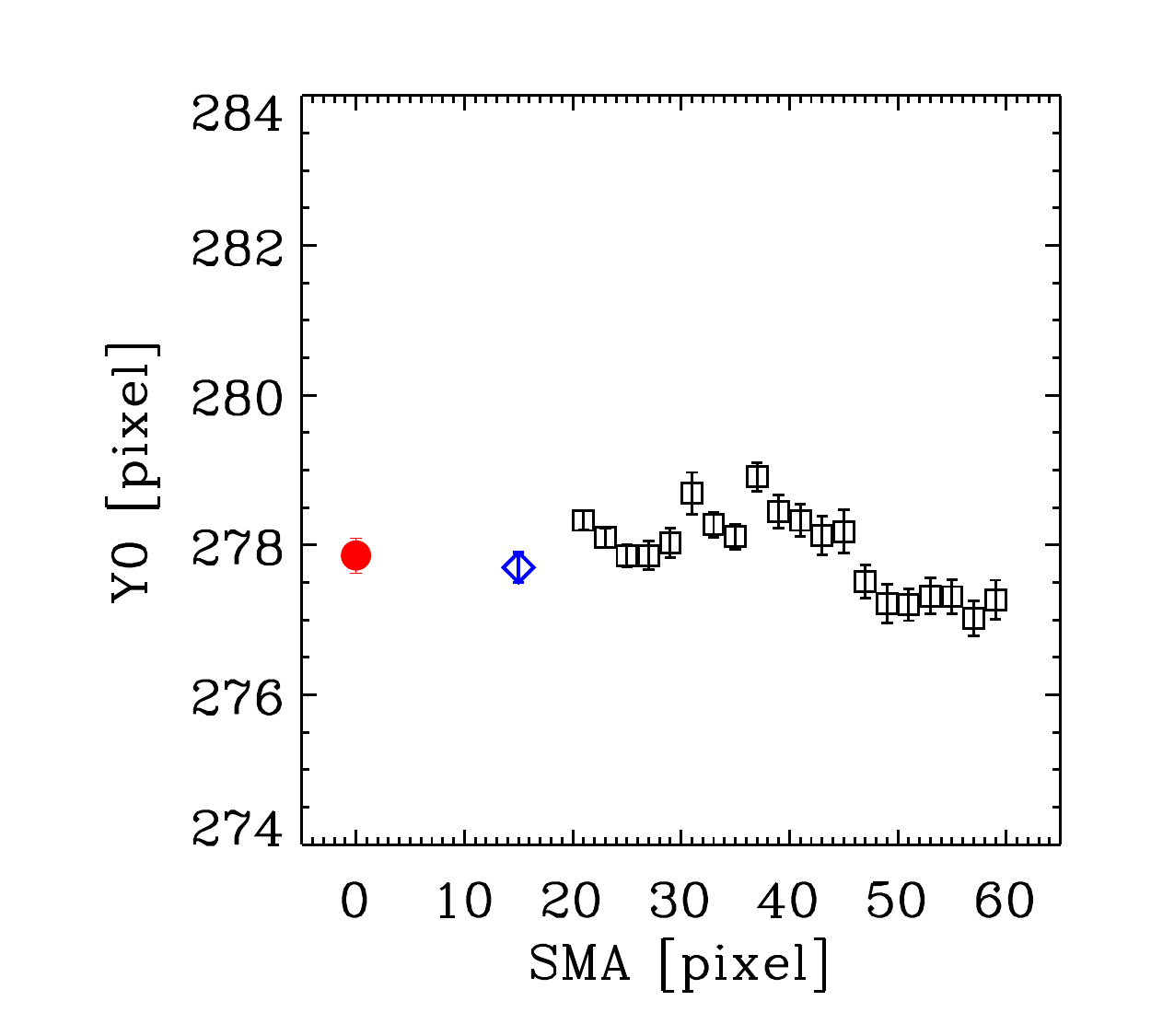} \\	
 \includegraphics[trim=0.65cm 0cm 0cm 0cm, clip=true, scale=0.46]{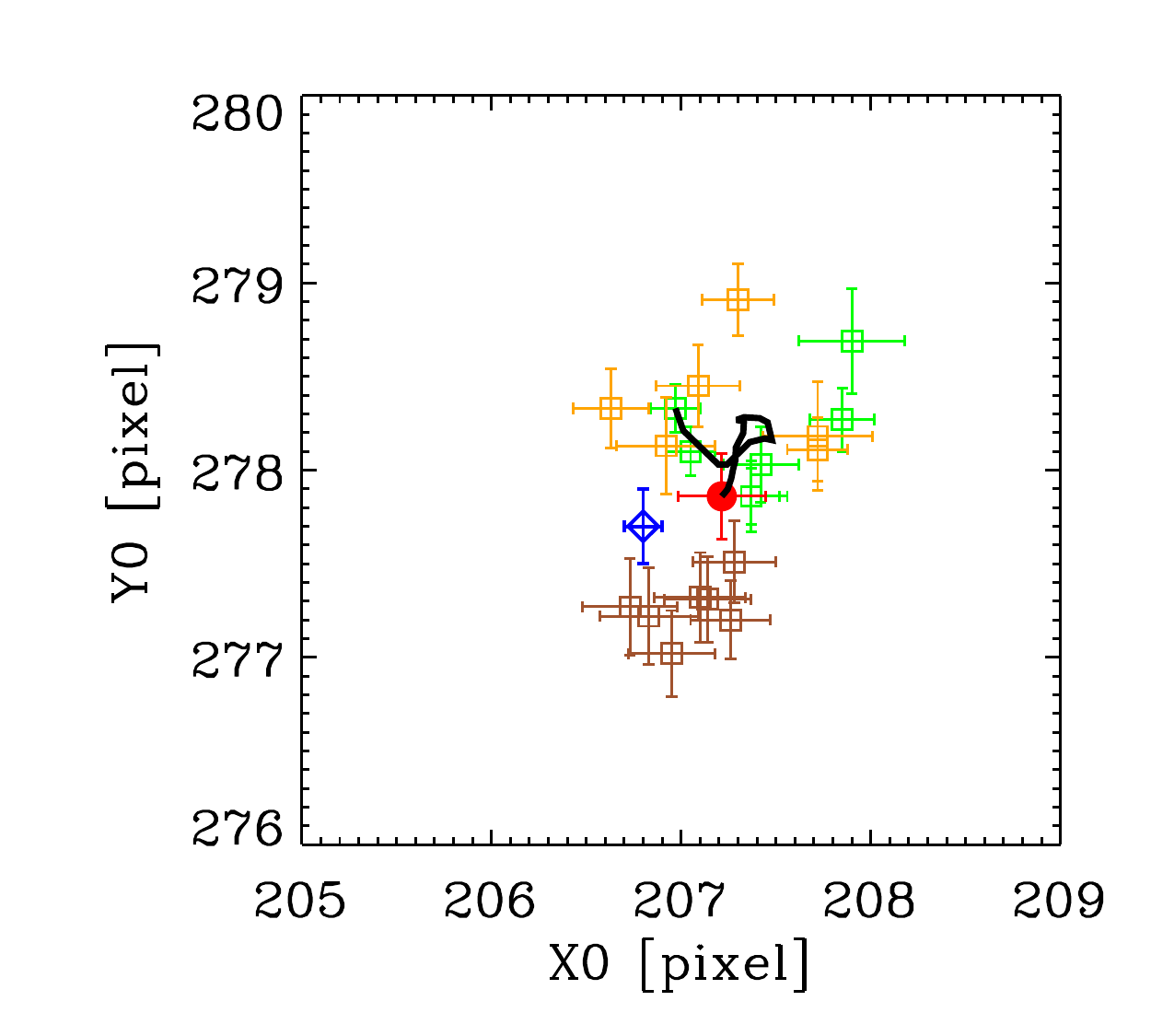}	&  \includegraphics[trim=0.6cm 0cm 0cm 0cm, clip=true, scale=0.46]{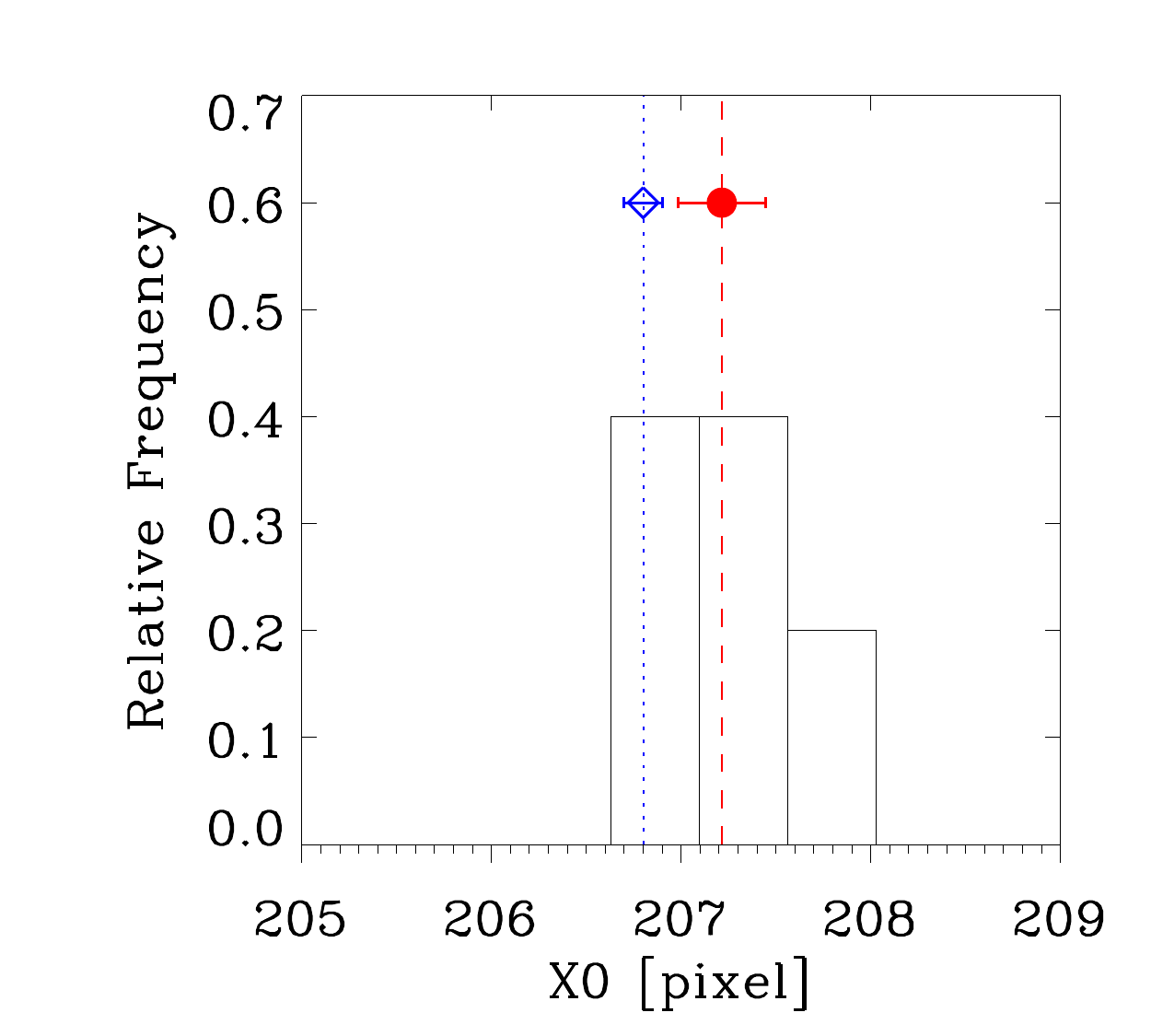}	& \includegraphics[trim=0.6cm 0cm 0cm 0cm, clip=true, scale=0.46]{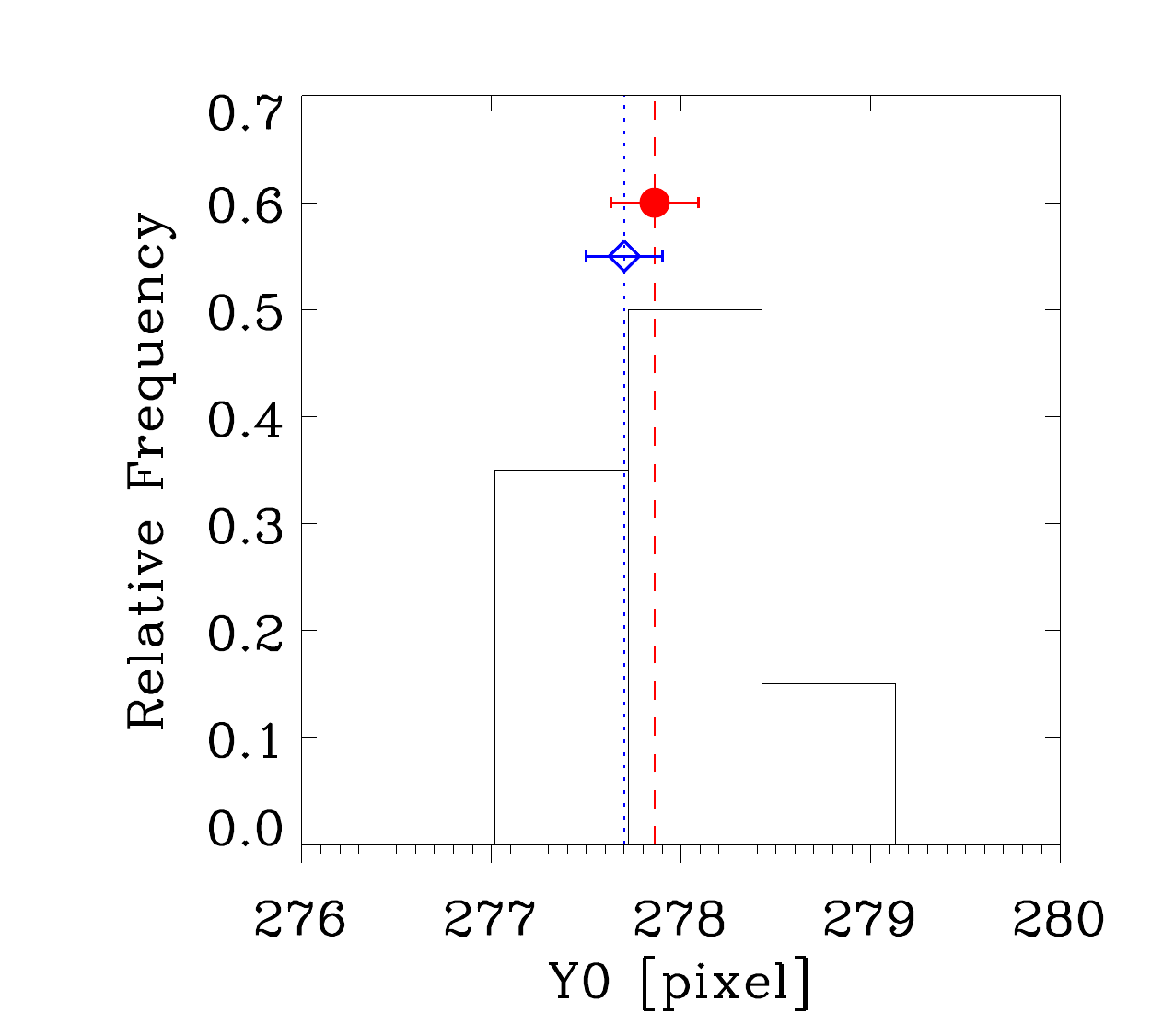}\\
\end{array}$
\end{center}
\caption[M87 (ACS)]{As in Fig.\ref{fig: NGC4373_W2} for galaxy NGC 4486 (M 87), NICMOS2 - F110W, scale=$0\farcs05$/pxl.}
\label{fig: M87_NIC2F110W}
\end{figure*} 

\begin{figure*}[h]
\begin{center}$
\begin{array}{ccc}
\includegraphics[trim=3.75cm 1cm 3cm 0cm, clip=true, scale=0.48]{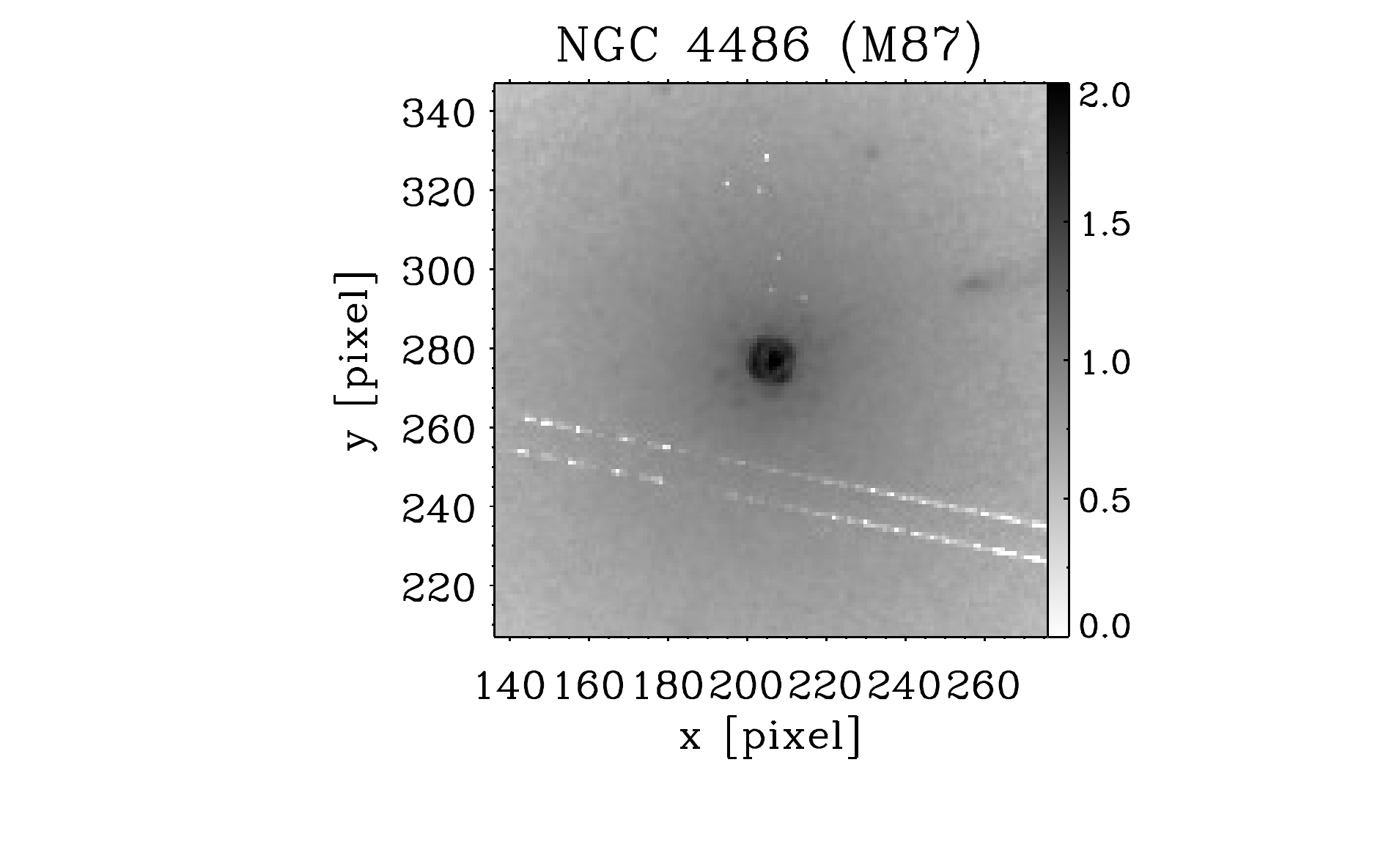} & \includegraphics[trim= 4.cm 1cm 3cm 0cm, clip=true, scale=0.48]{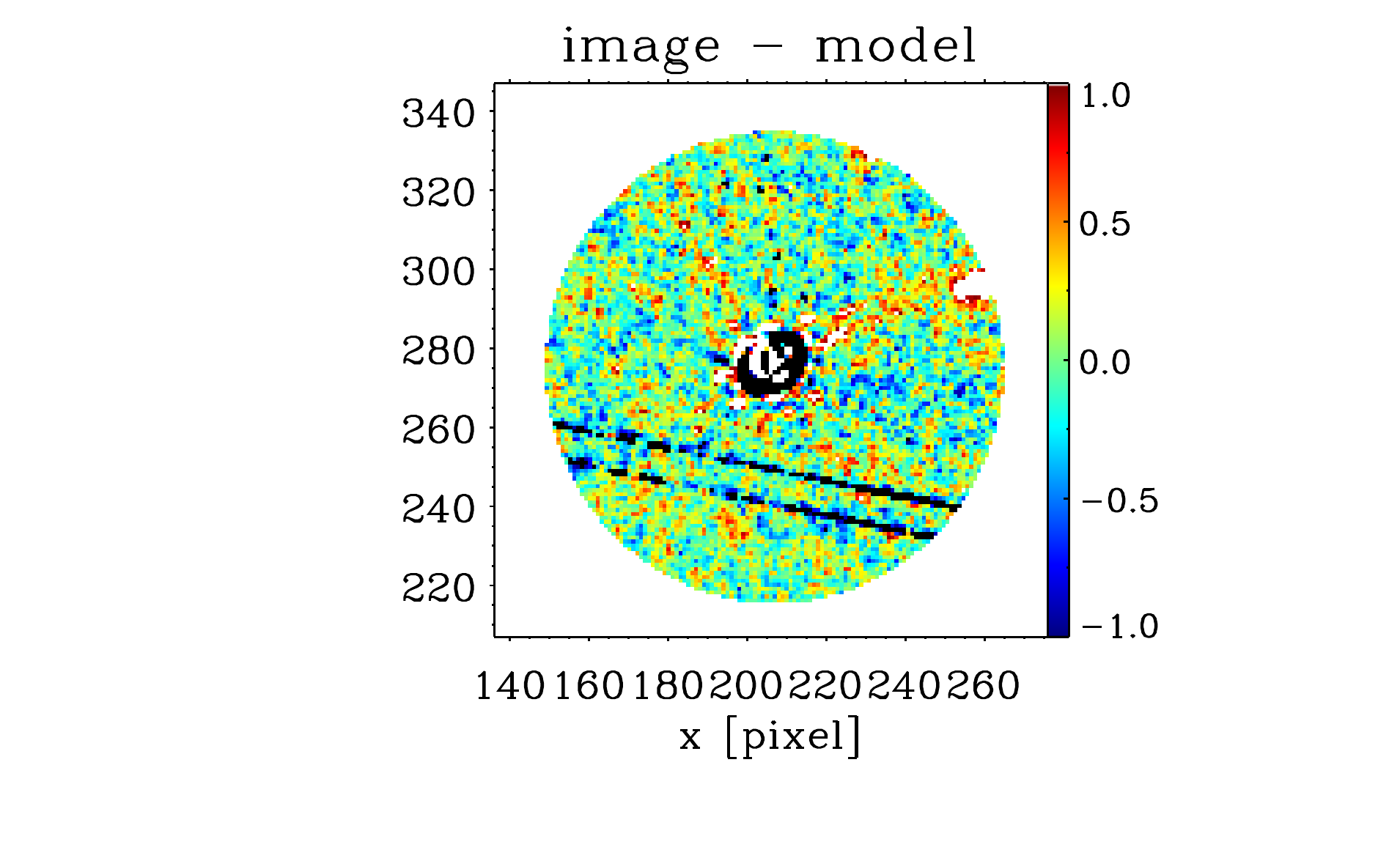}	& \includegraphics[trim= 4.cm 1cm 3cm 0cm, clip=true, scale=0.48]{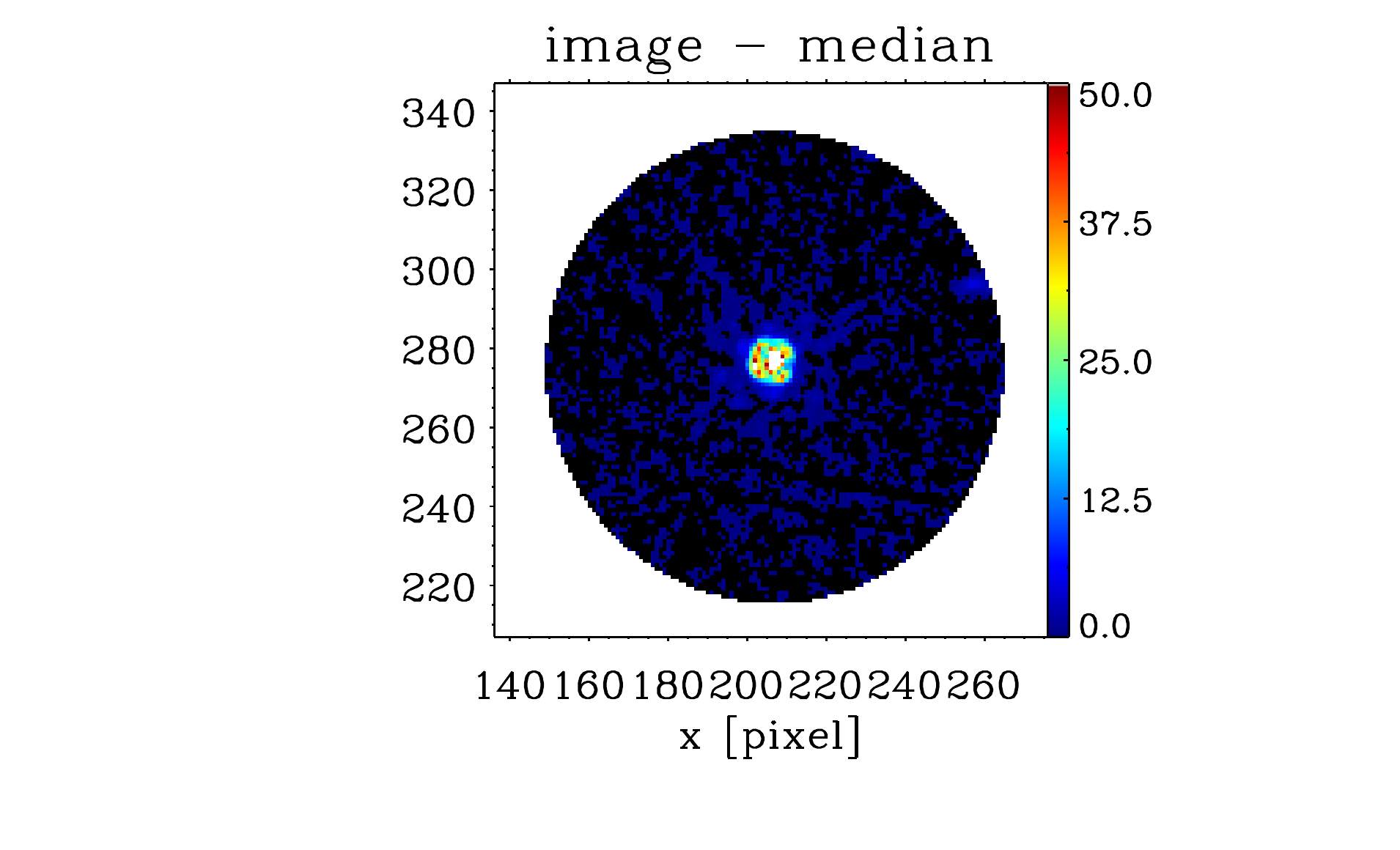} \\
\includegraphics[trim=0.7cm 0cm 0cm 0cm, clip=true, scale=0.46]{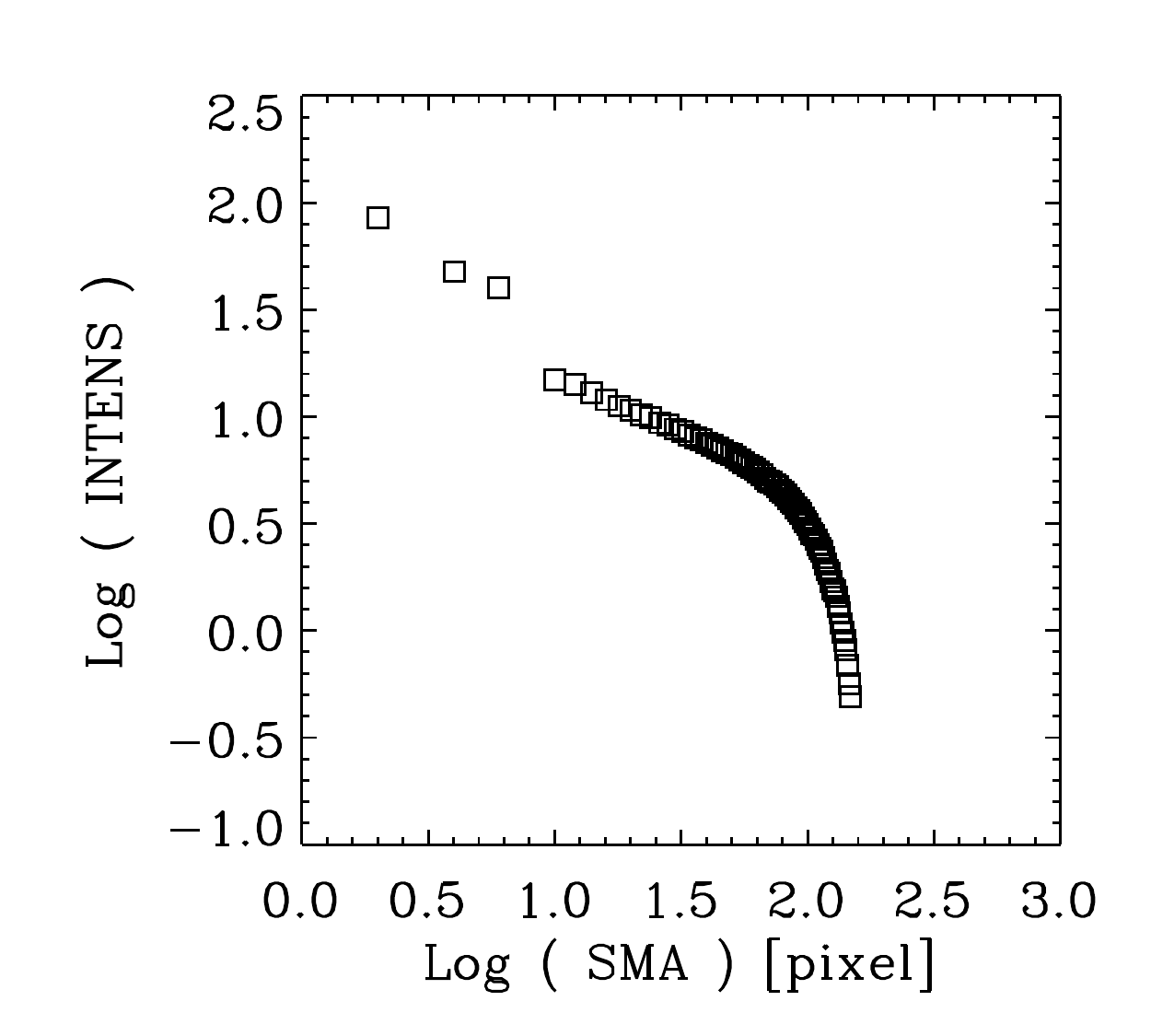}	    &  \includegraphics[trim=0.6cm 0cm 0cm 0cm, clip=true, scale=0.46]{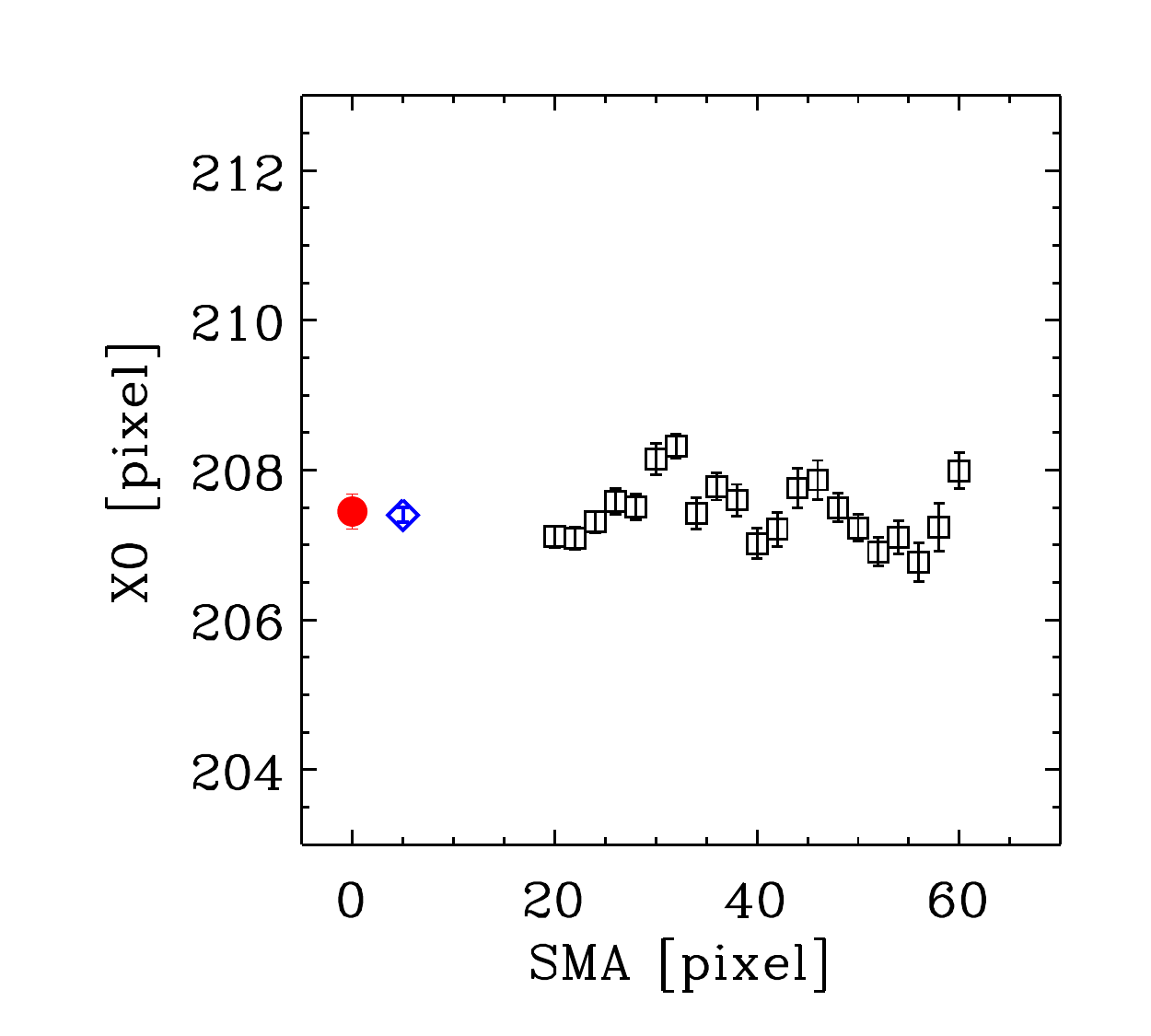}  &  \includegraphics[trim=0.6cm 0cm 0cm 0cm, clip=true, scale=0.46]{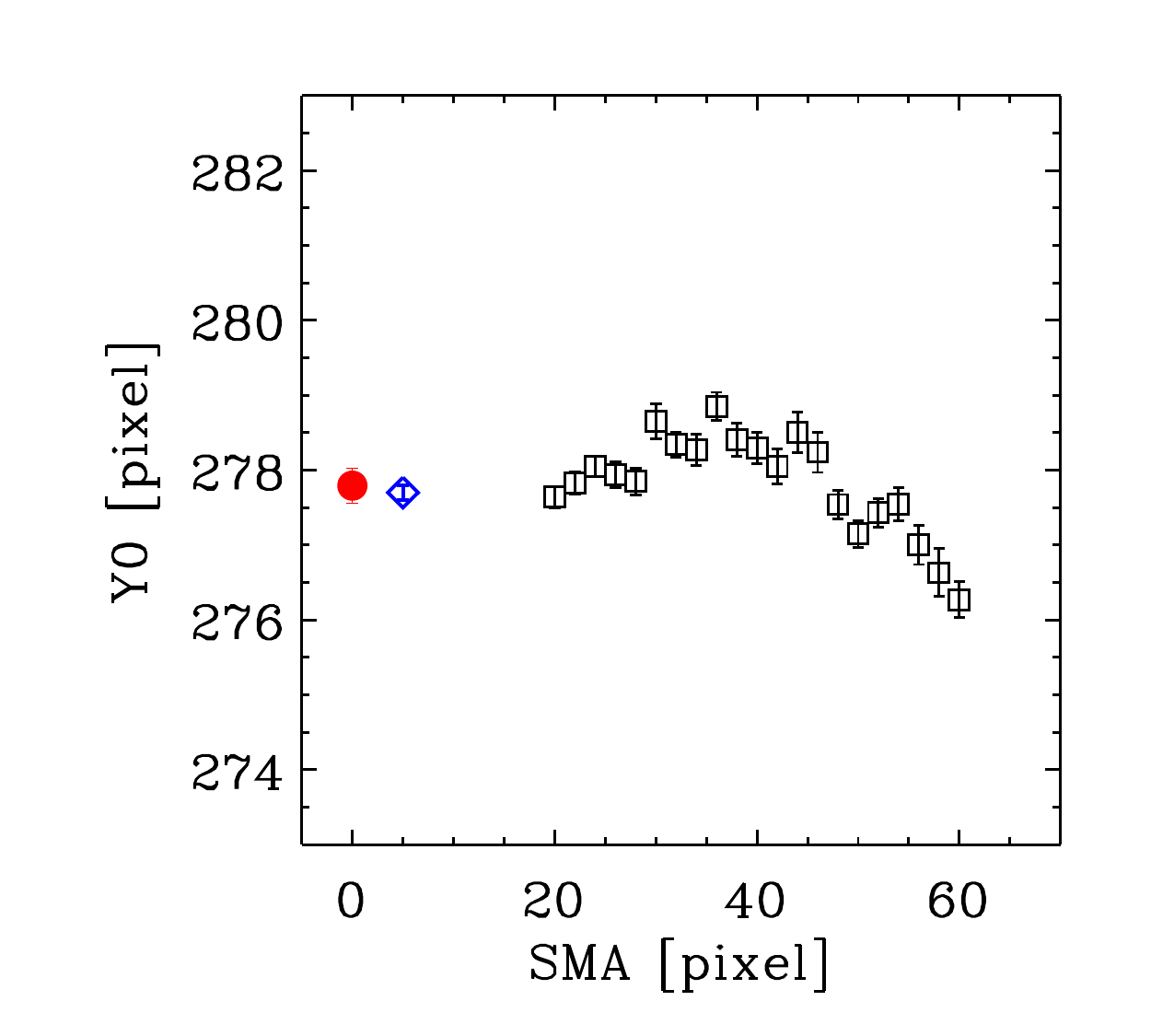} \\	
 \includegraphics[trim=0.65cm 0cm 0cm 0cm, clip=true, scale=0.46]{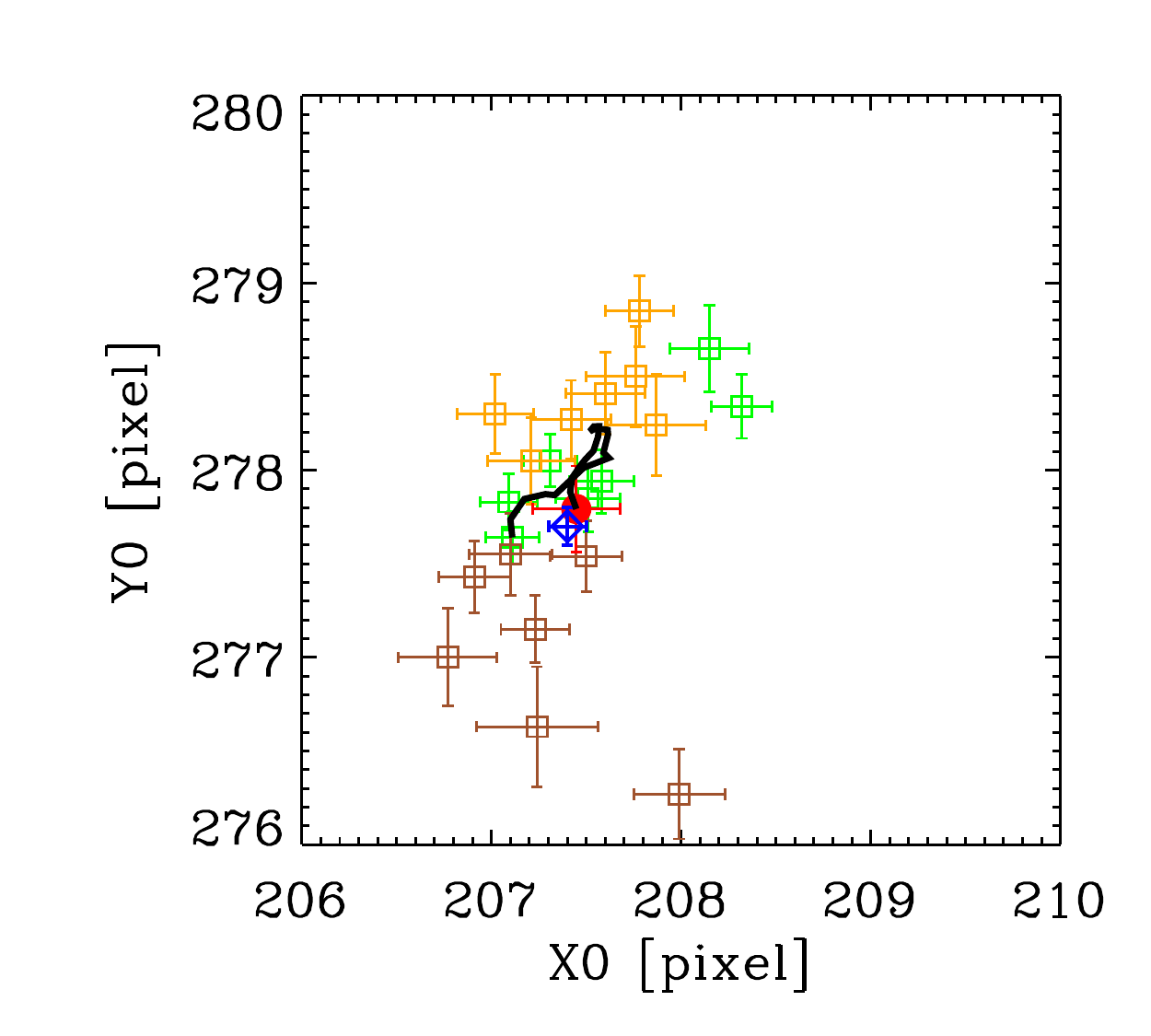}	&  \includegraphics[trim=0.6cm 0cm 0cm 0cm, clip=true, scale=0.46]{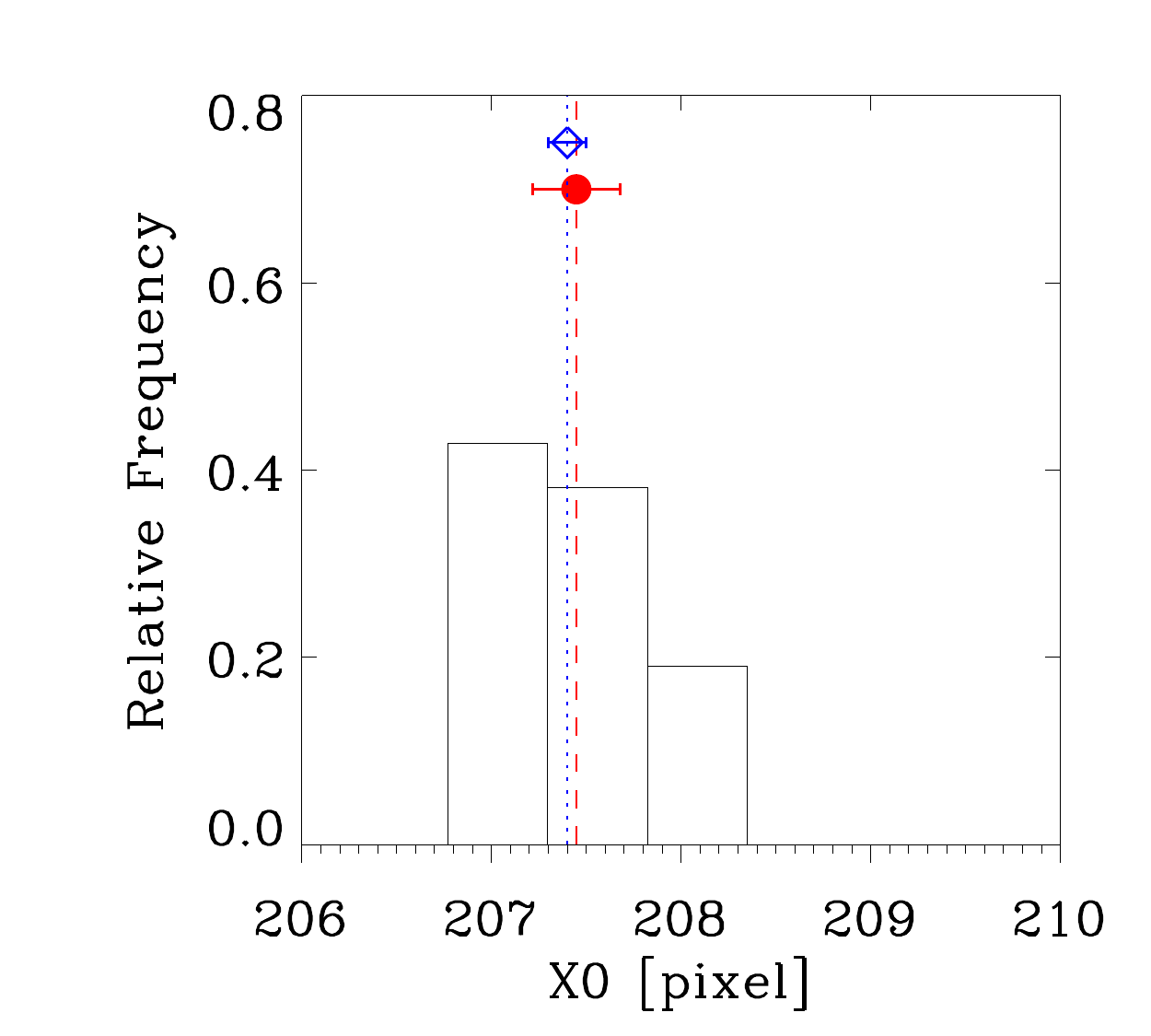}	& \includegraphics[trim=0.6cm 0cm 0cm 0cm, clip=true, scale=0.46]{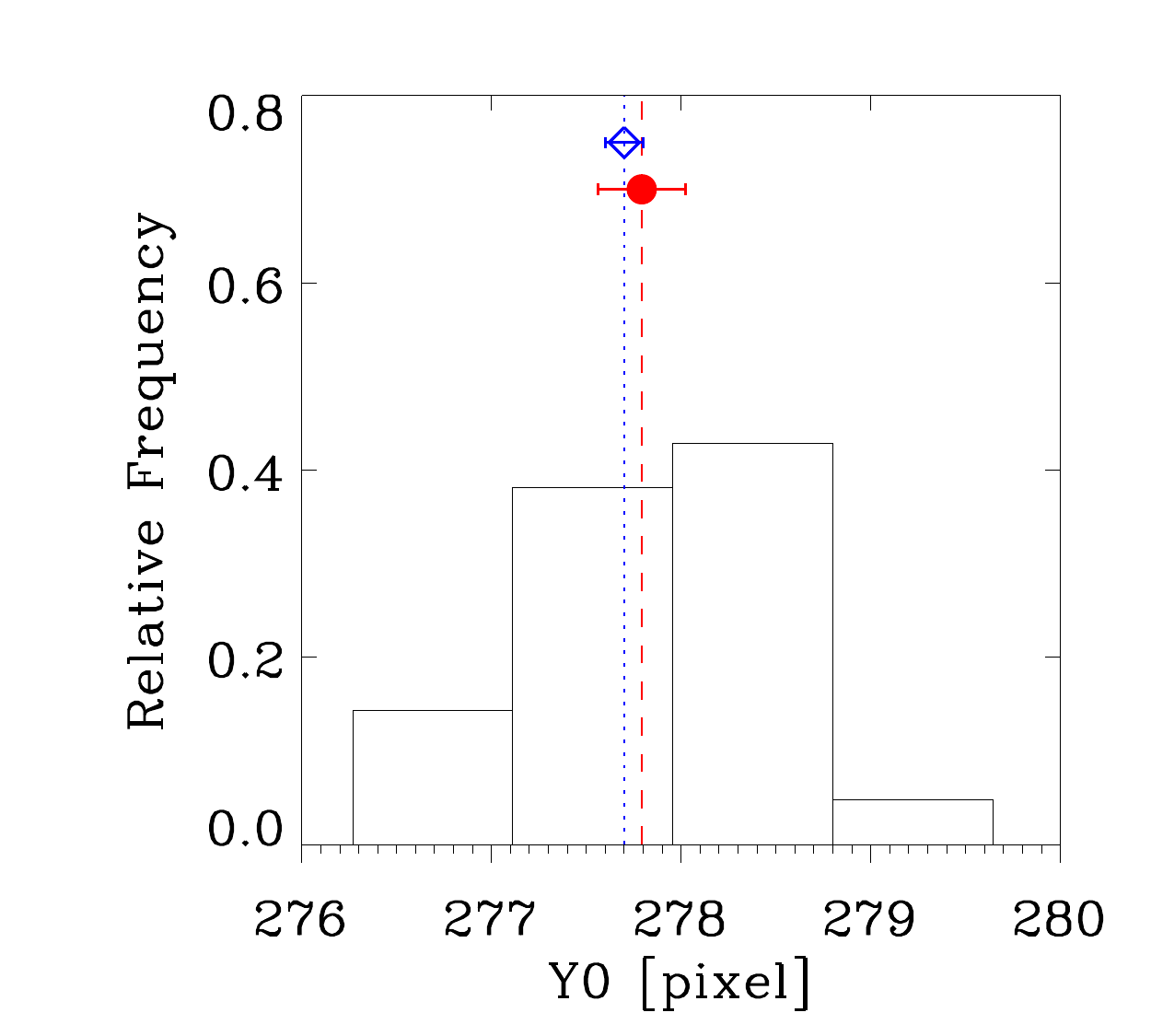}\\
\end{array}$
\end{center}
\caption[M87 (NICMOS2-F160W)]{As in Fig.\ref{fig: NGC4373_W2} for galaxy NGC 4486 (M 87), NICMOS2 - F160W, scale=$0\farcs05$/pxl.}
\label{fig: M87_NIC2F160W}
\end{figure*} 

\begin{figure*}[h]
\begin{center}$
\begin{array}{ccc}
\includegraphics[trim=3.75cm 1cm 3cm 0cm, clip=true, scale=0.48]{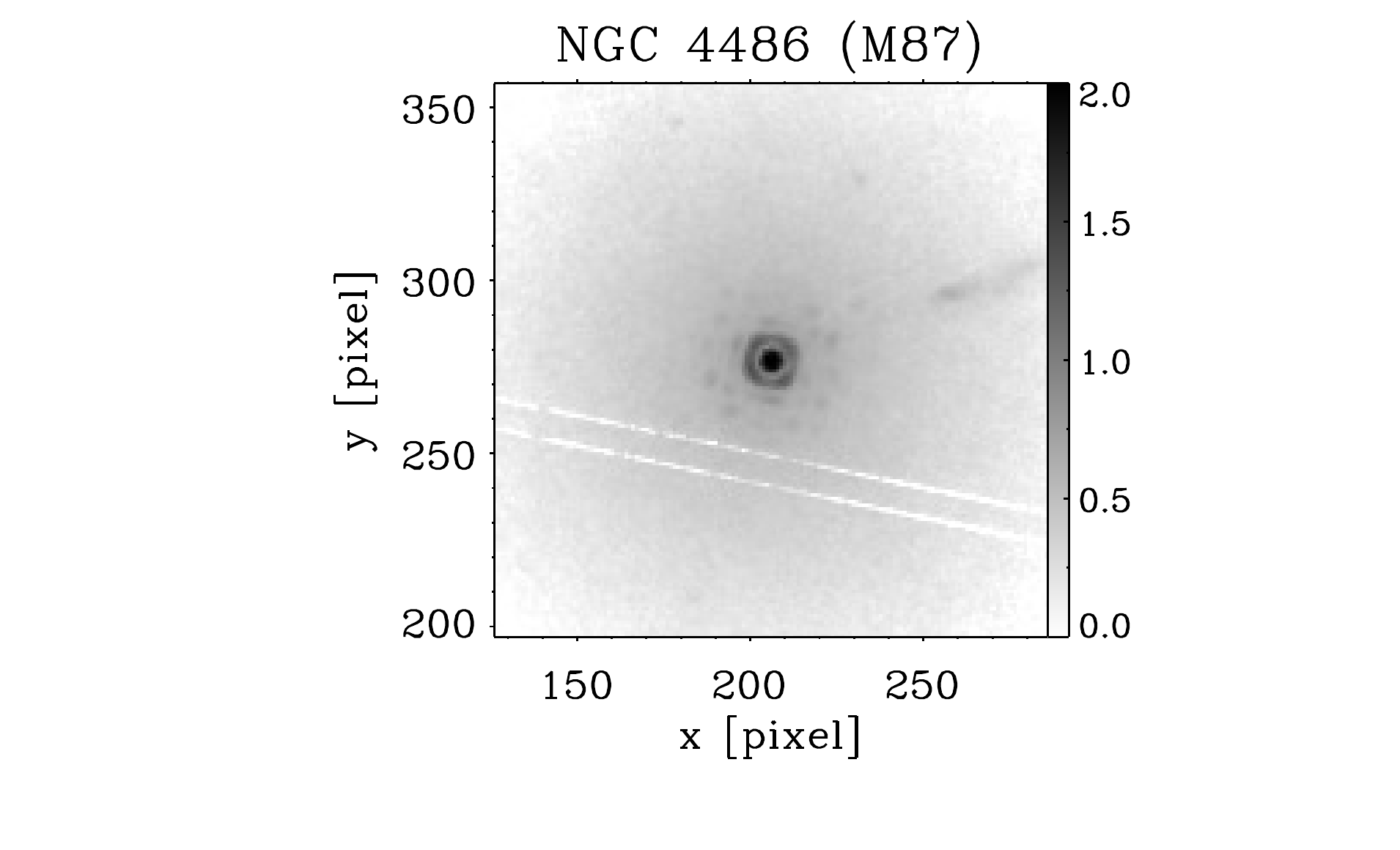} & \includegraphics[trim= 4.cm 1cm 3cm 0cm, clip=true, scale=0.48]{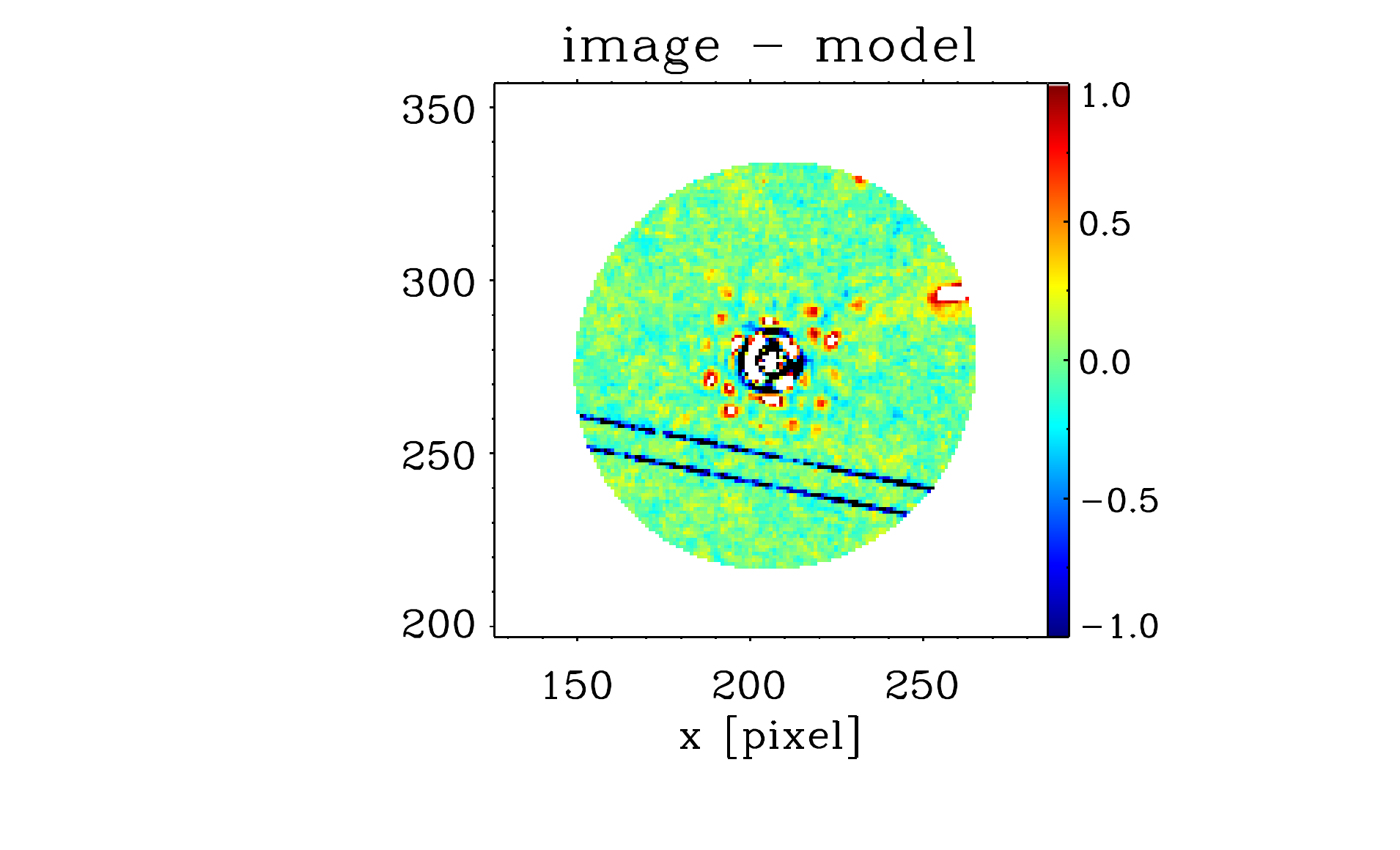}	& \includegraphics[trim= 4.cm 1cm 3cm 0cm, clip=true, scale=0.48]{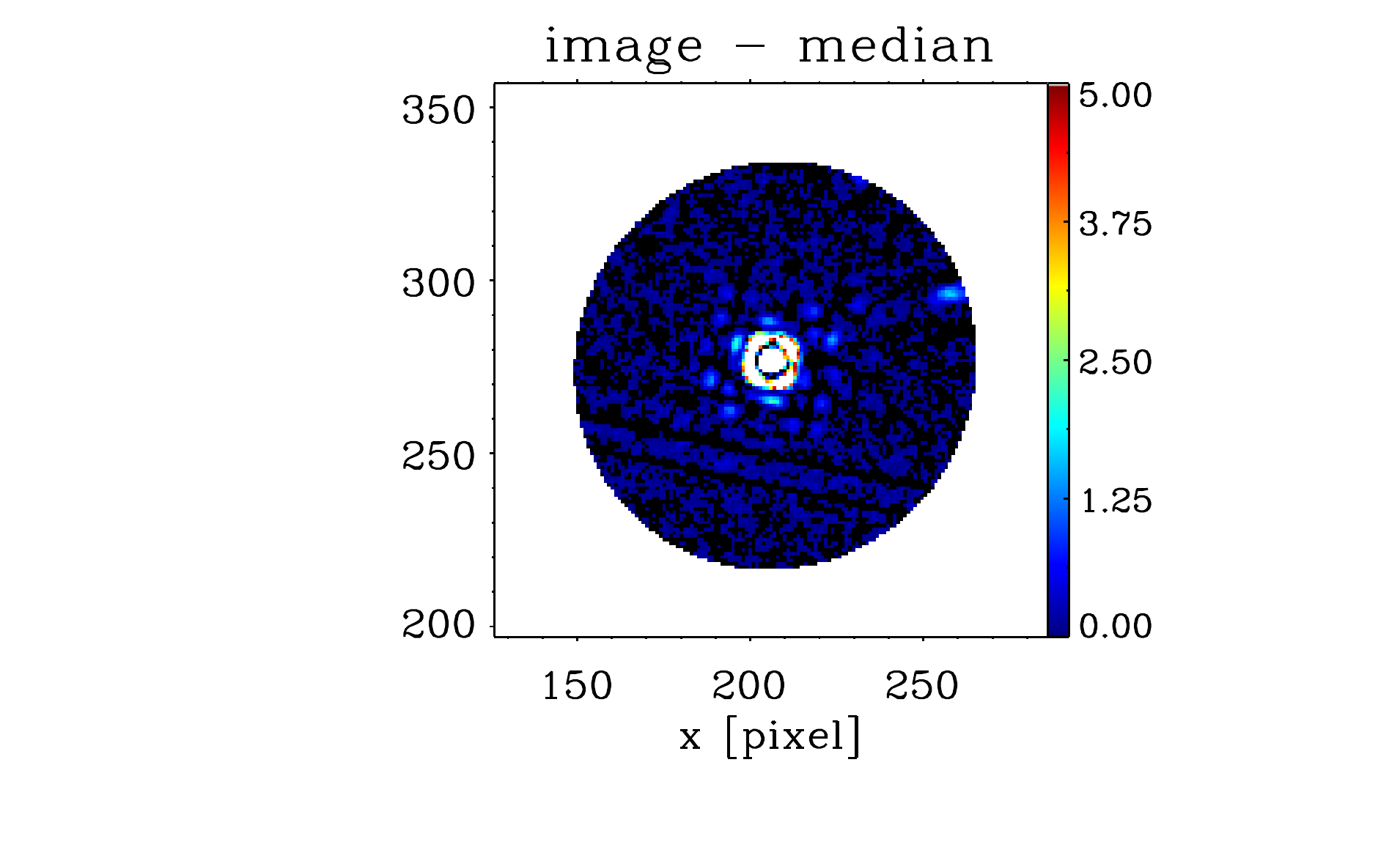} \\
\includegraphics[trim=0.7cm 0cm 0cm 0cm, clip=true, scale=0.46]{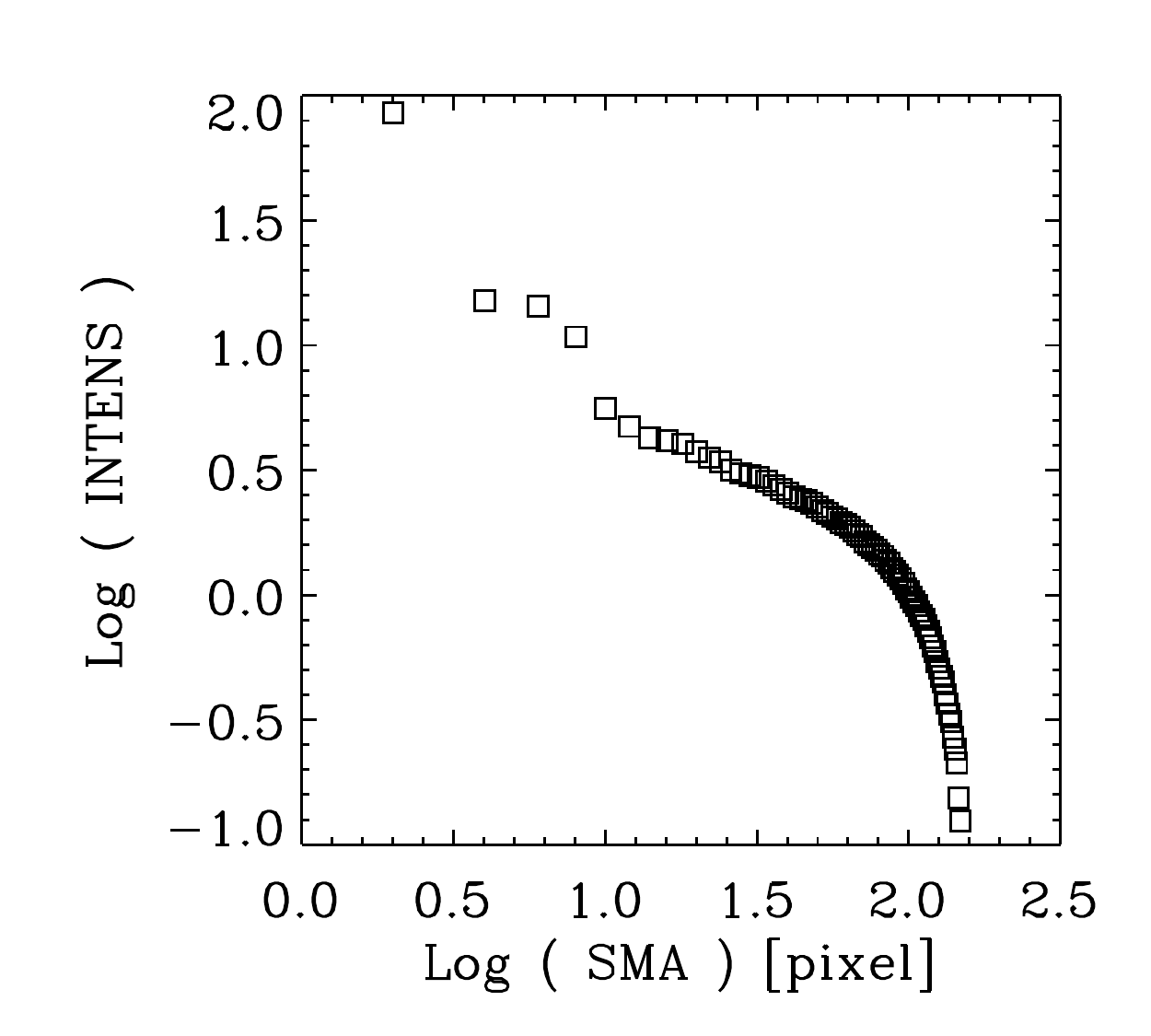}	    &  \includegraphics[trim=0.6cm 0cm 0cm 0cm, clip=true, scale=0.46]{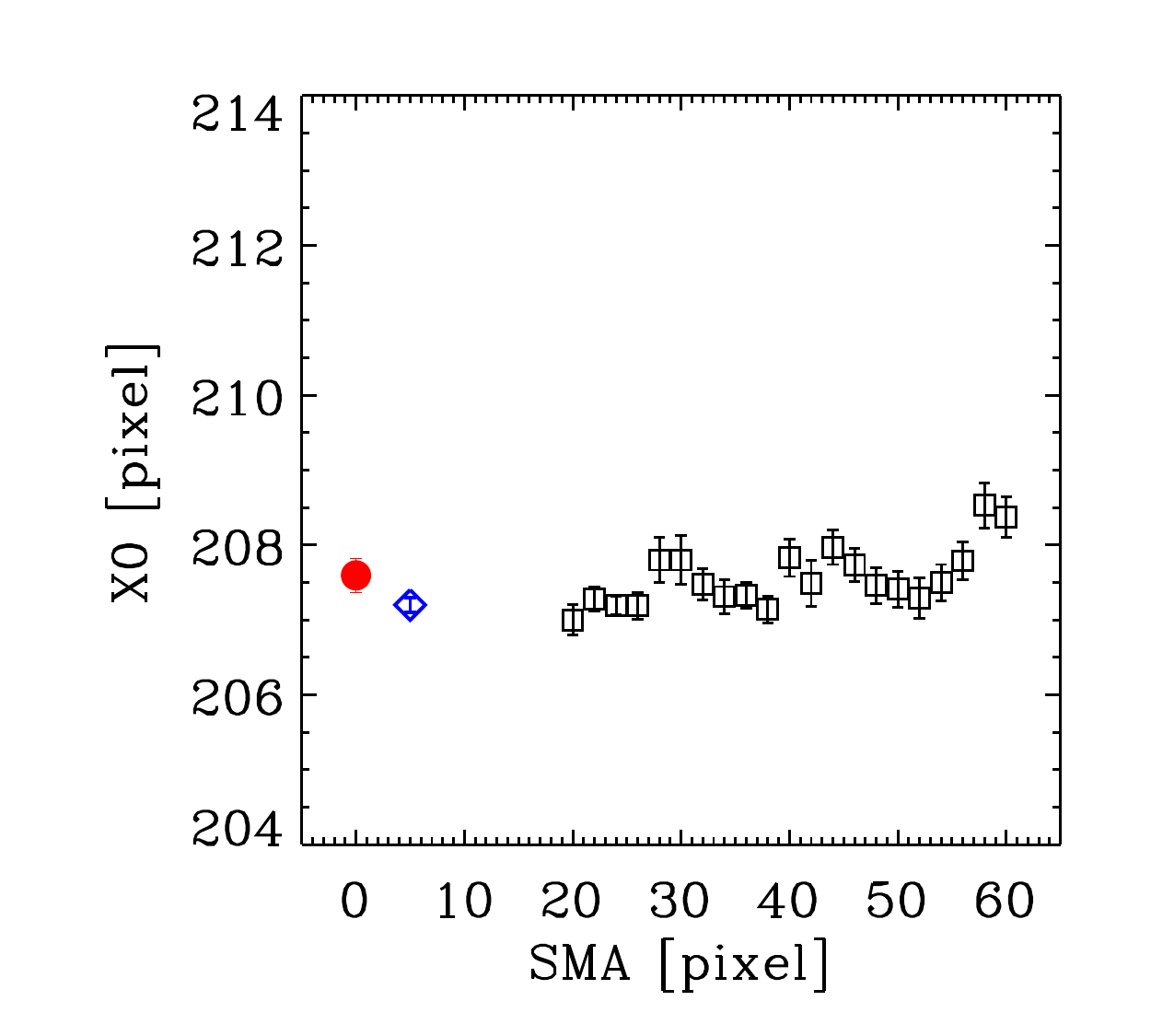}  &  \includegraphics[trim=0.6cm 0cm 0cm 0cm, clip=true, scale=0.46]{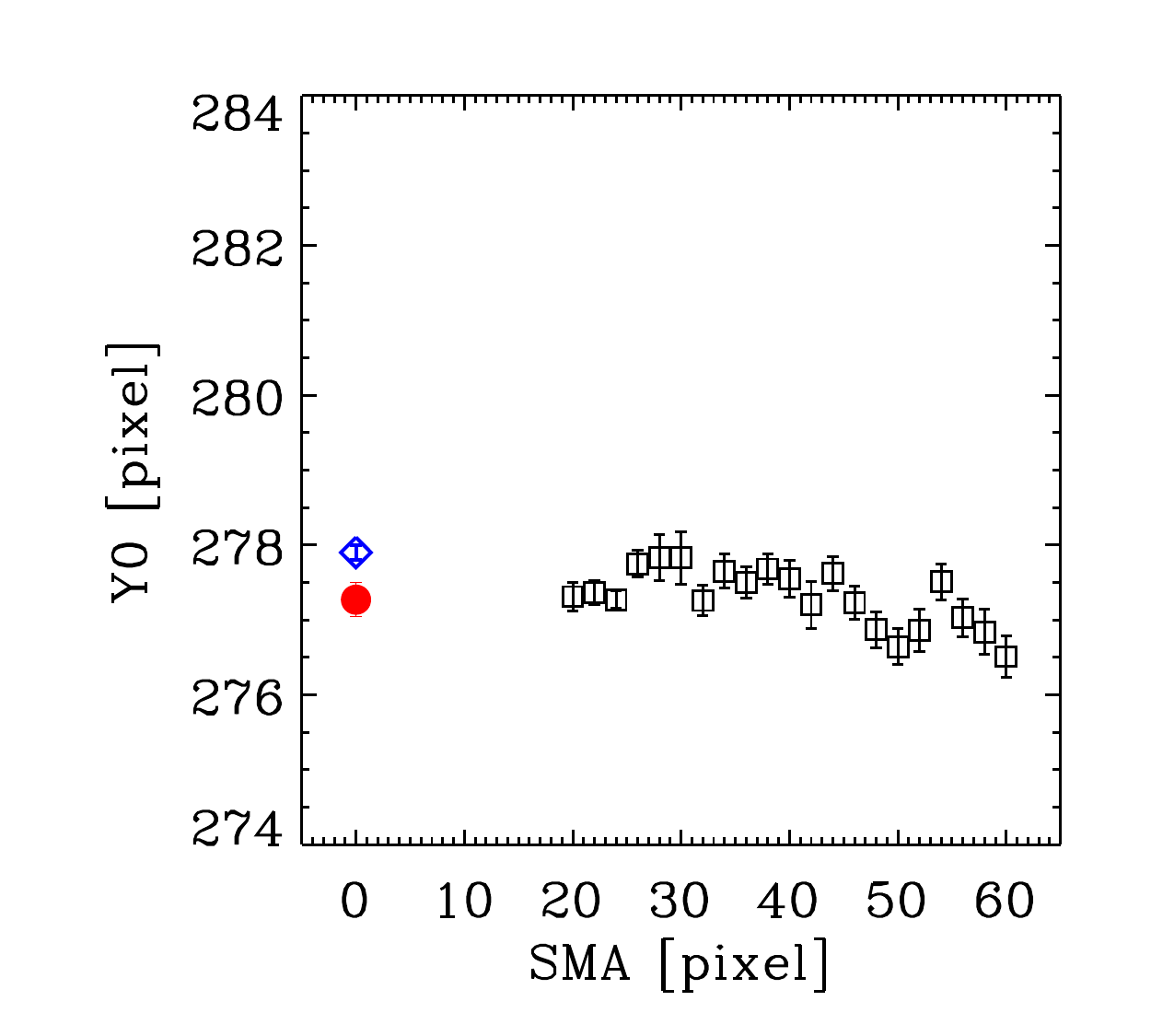} \\	
 \includegraphics[trim=0.65cm 0cm 0cm 0cm, clip=true, scale=0.46]{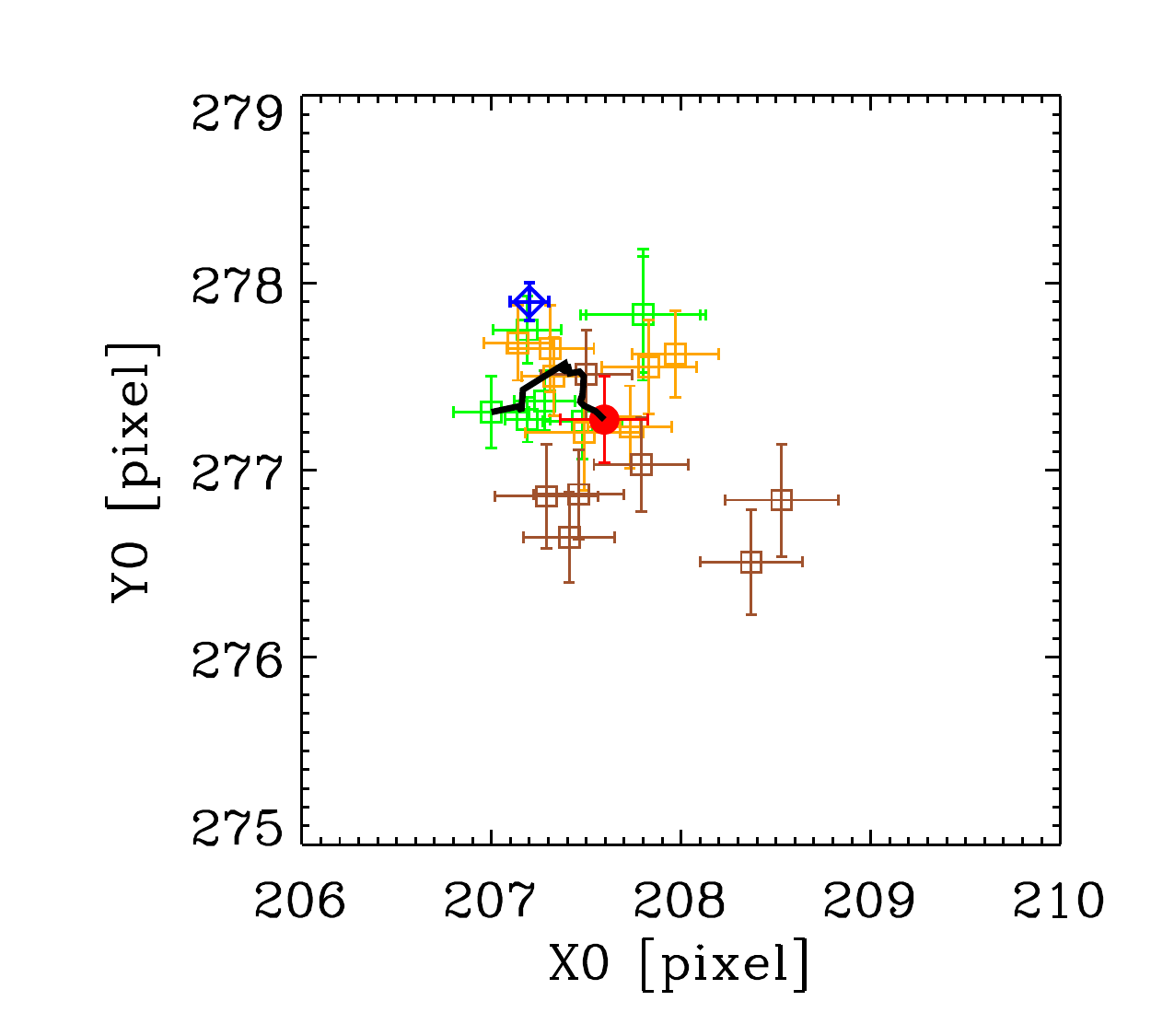}	&  \includegraphics[trim=0.6cm 0cm 0cm 0cm, clip=true, scale=0.46]{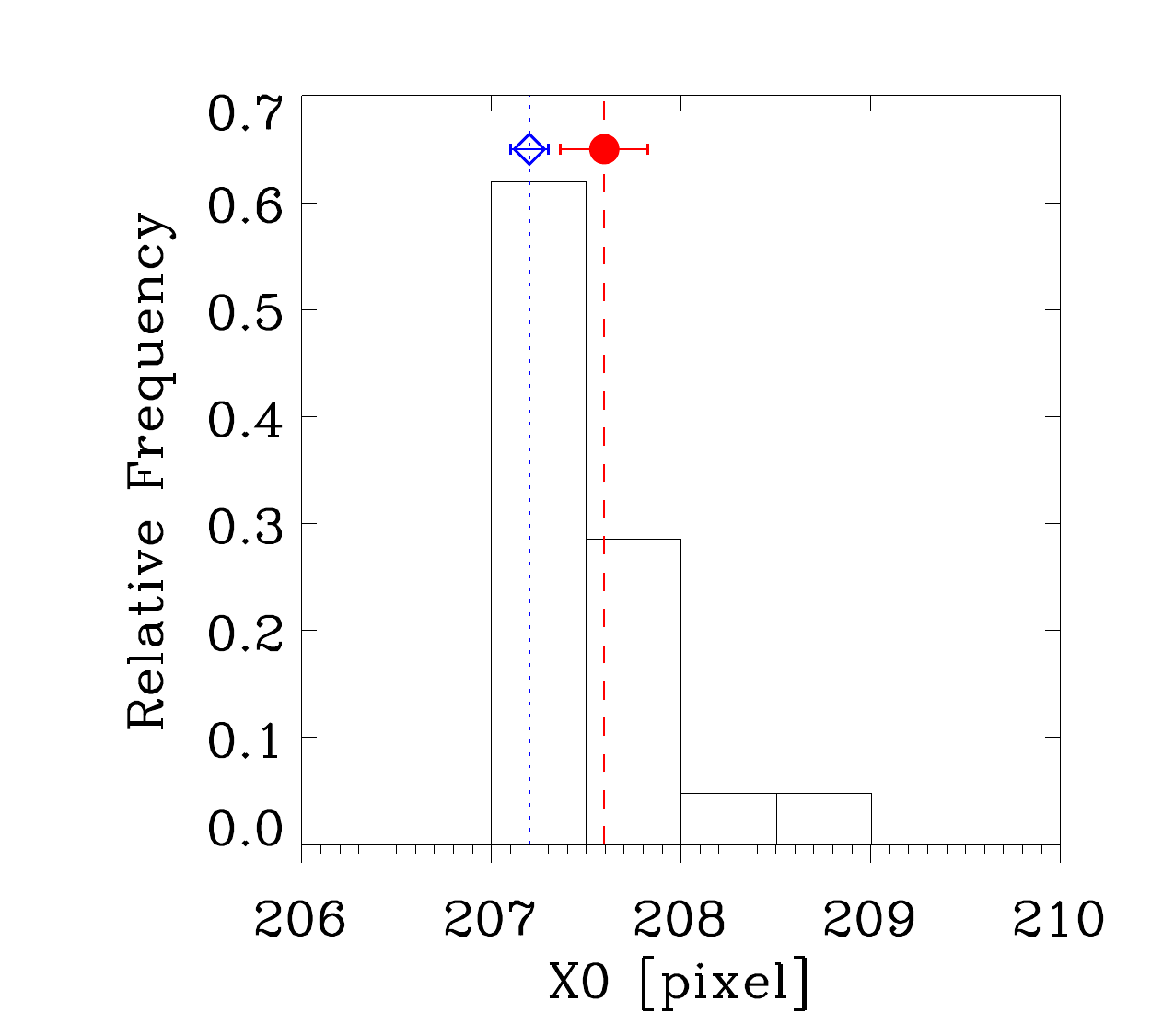}	& \includegraphics[trim=0.6cm 0cm 0cm 0cm, clip=true, scale=0.46]{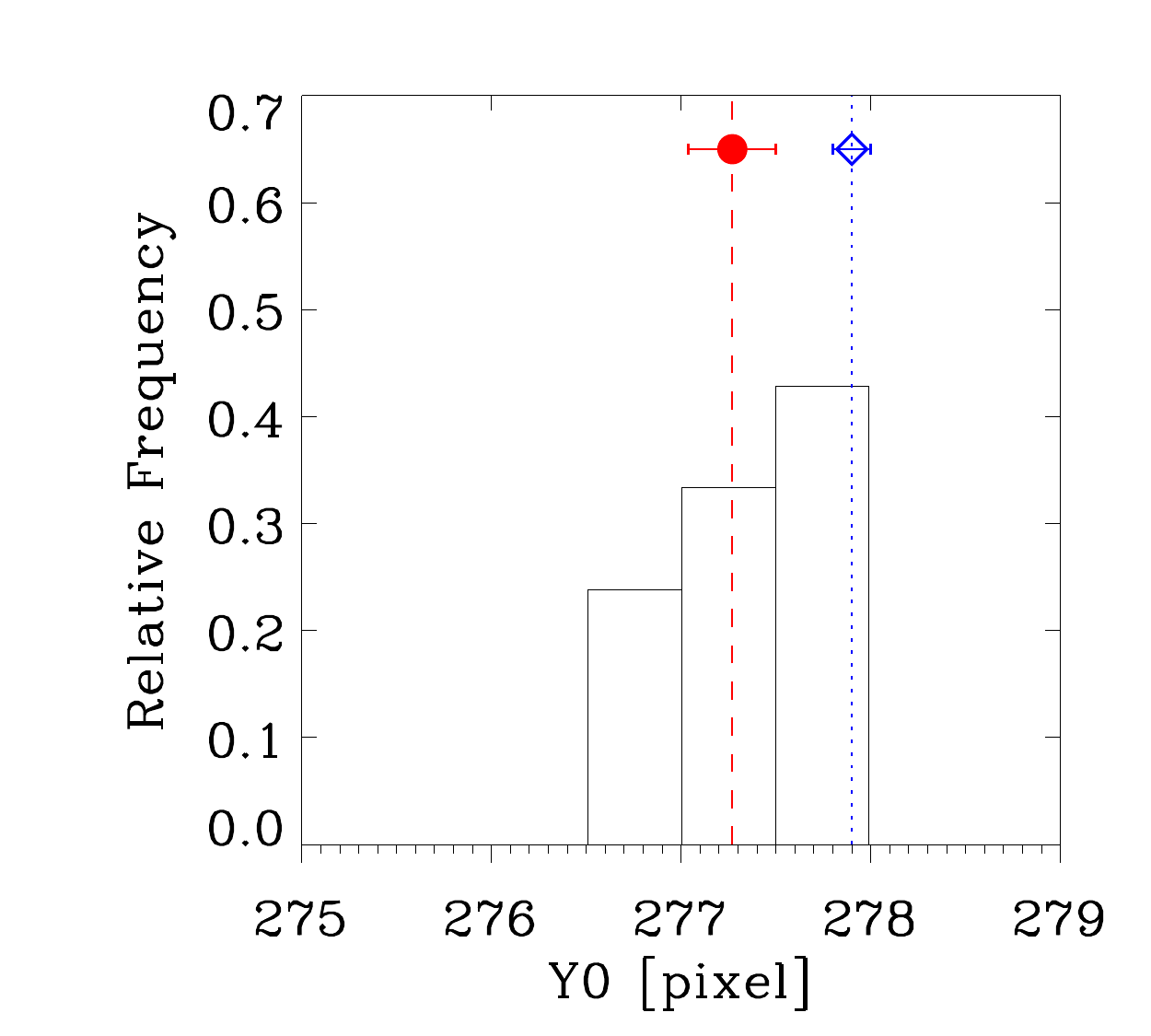}\\
\end{array}$
\end{center}
\caption[M87 (NICMOS2-F222M)]{As in Fig.\ref{fig: NGC4373_W2} for galaxy NGC 4486 (M 87), NICMOS2 - F222M, scale=$0\farcs05$/pxl.}
\label{fig: M87_NIC2F222W}
\end{figure*} 

\begin{figure*}[h]
\begin{center}$
\begin{array}{ccc}
\includegraphics[trim=3.75cm 1cm 3cm 0cm, clip=true, scale=0.48]{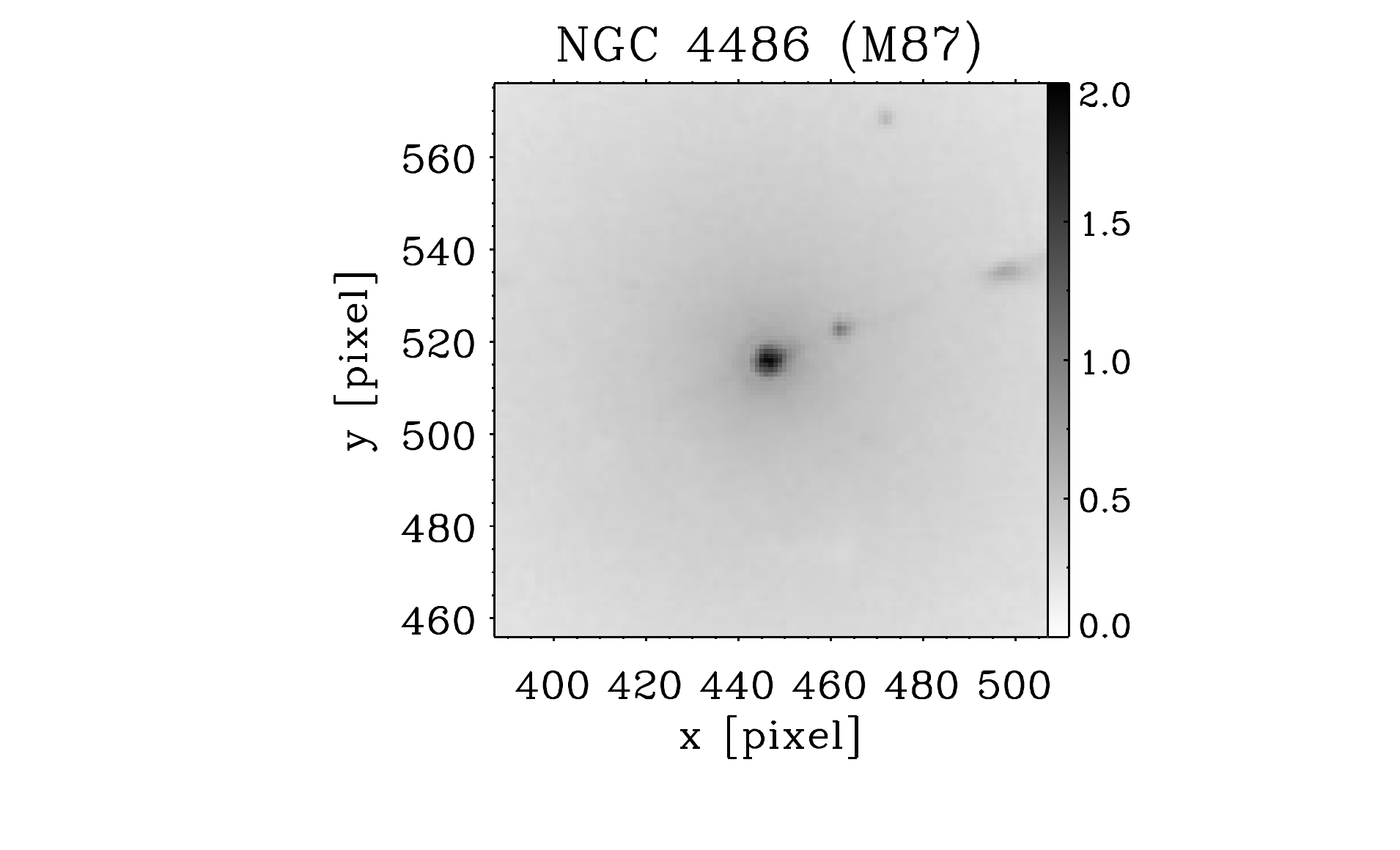} & \includegraphics[trim= 4.cm 1cm 3cm 0cm, clip=true, scale=0.48]{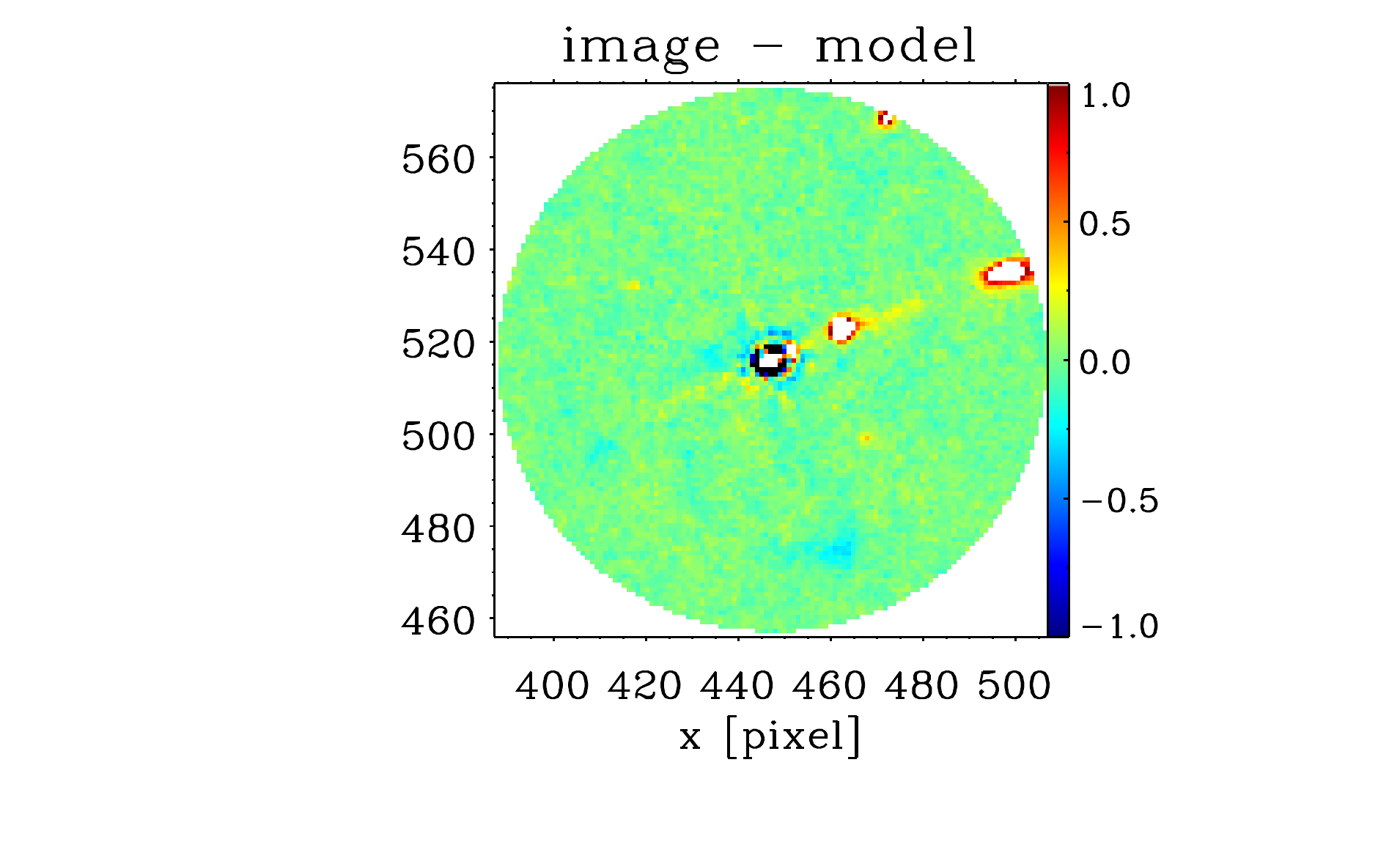}	& \includegraphics[trim= 4.cm 1cm 3cm 0cm, clip=true, scale=0.48]{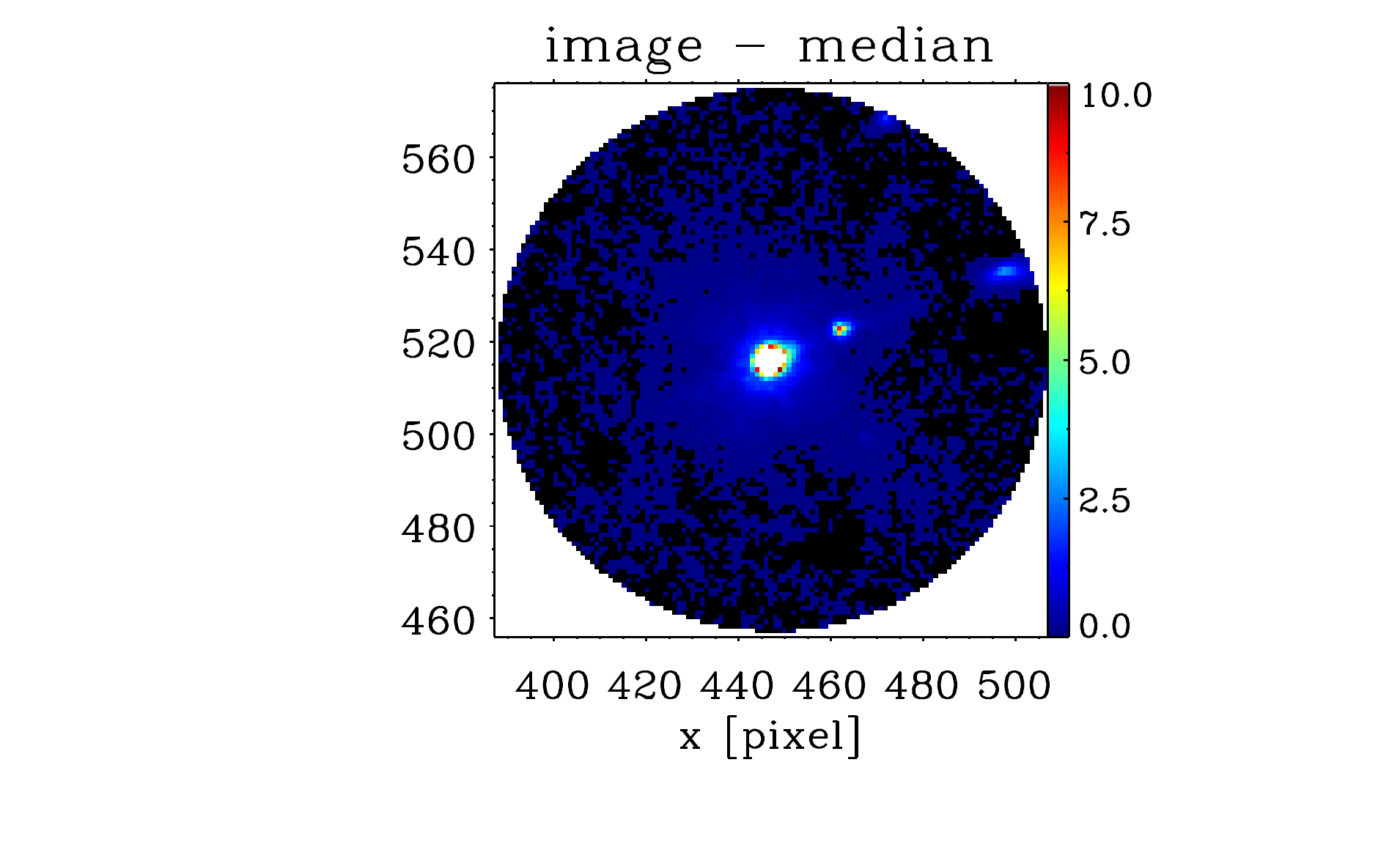} \\
\includegraphics[trim=0.7cm 0cm 0cm 0cm, clip=true, scale=0.46]{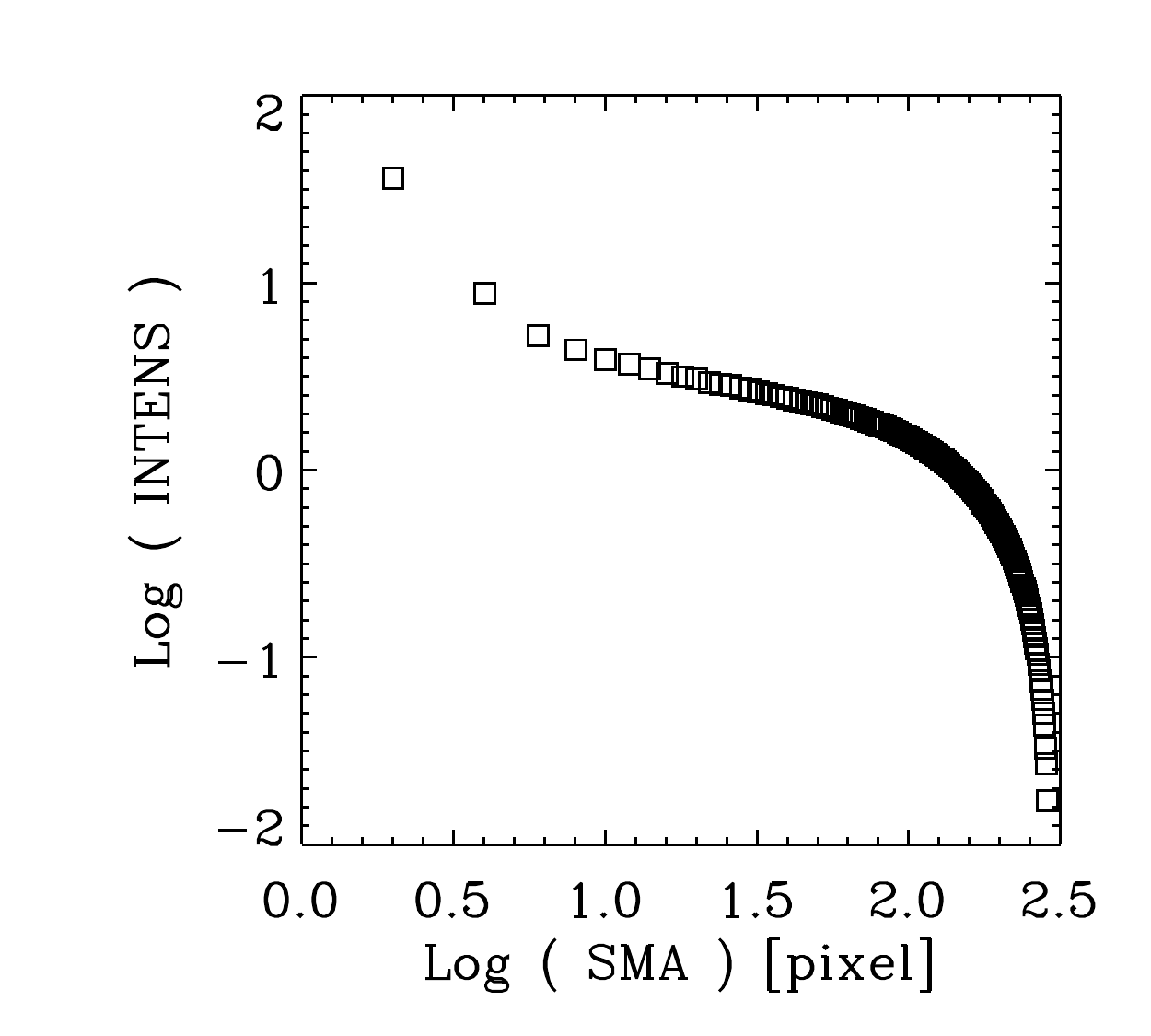}	    &  \includegraphics[trim=0.6cm 0cm 0cm 0cm, clip=true, scale=0.46]{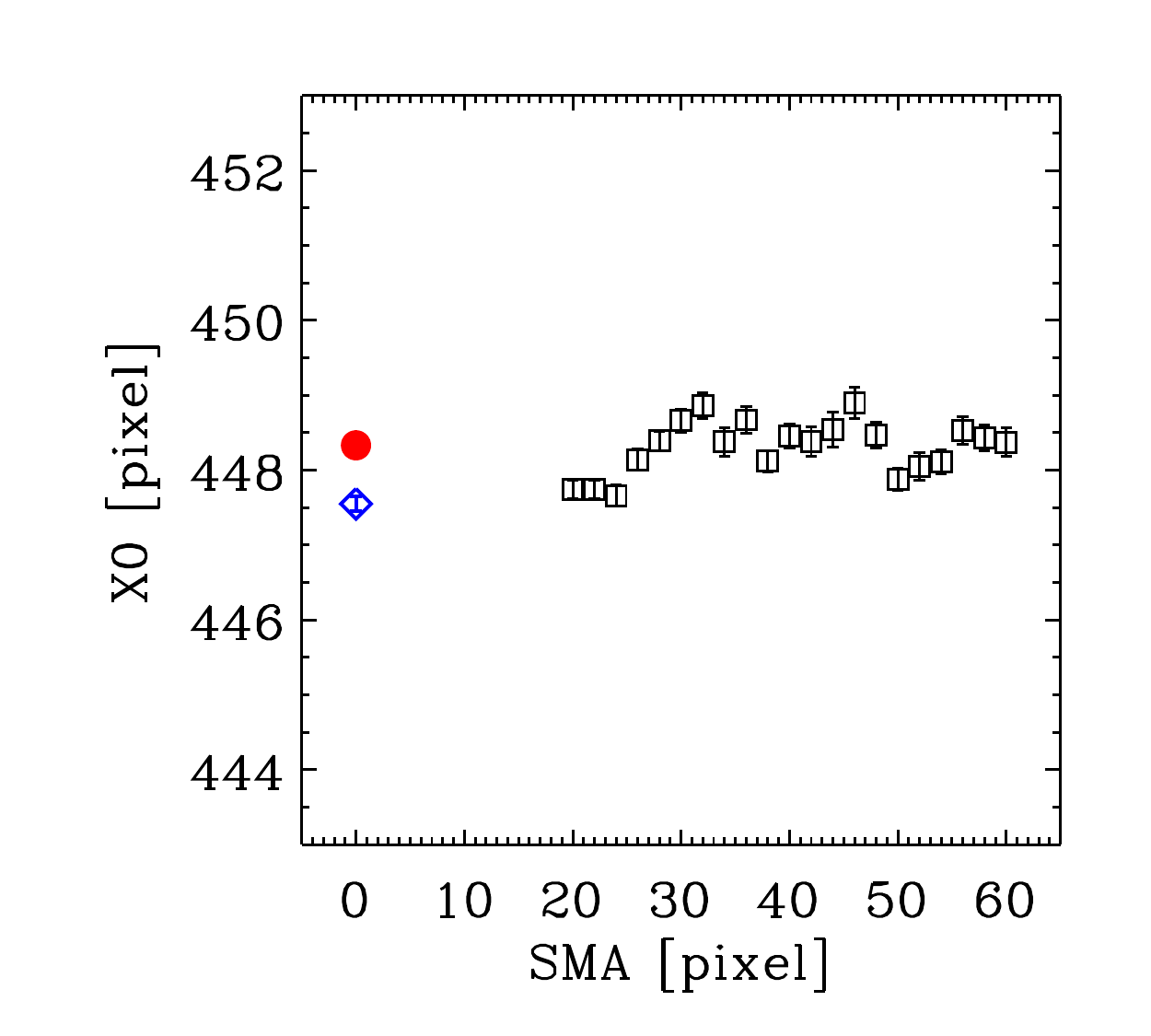}  &  \includegraphics[trim=0.6cm 0cm 0cm 0cm, clip=true, scale=0.46]{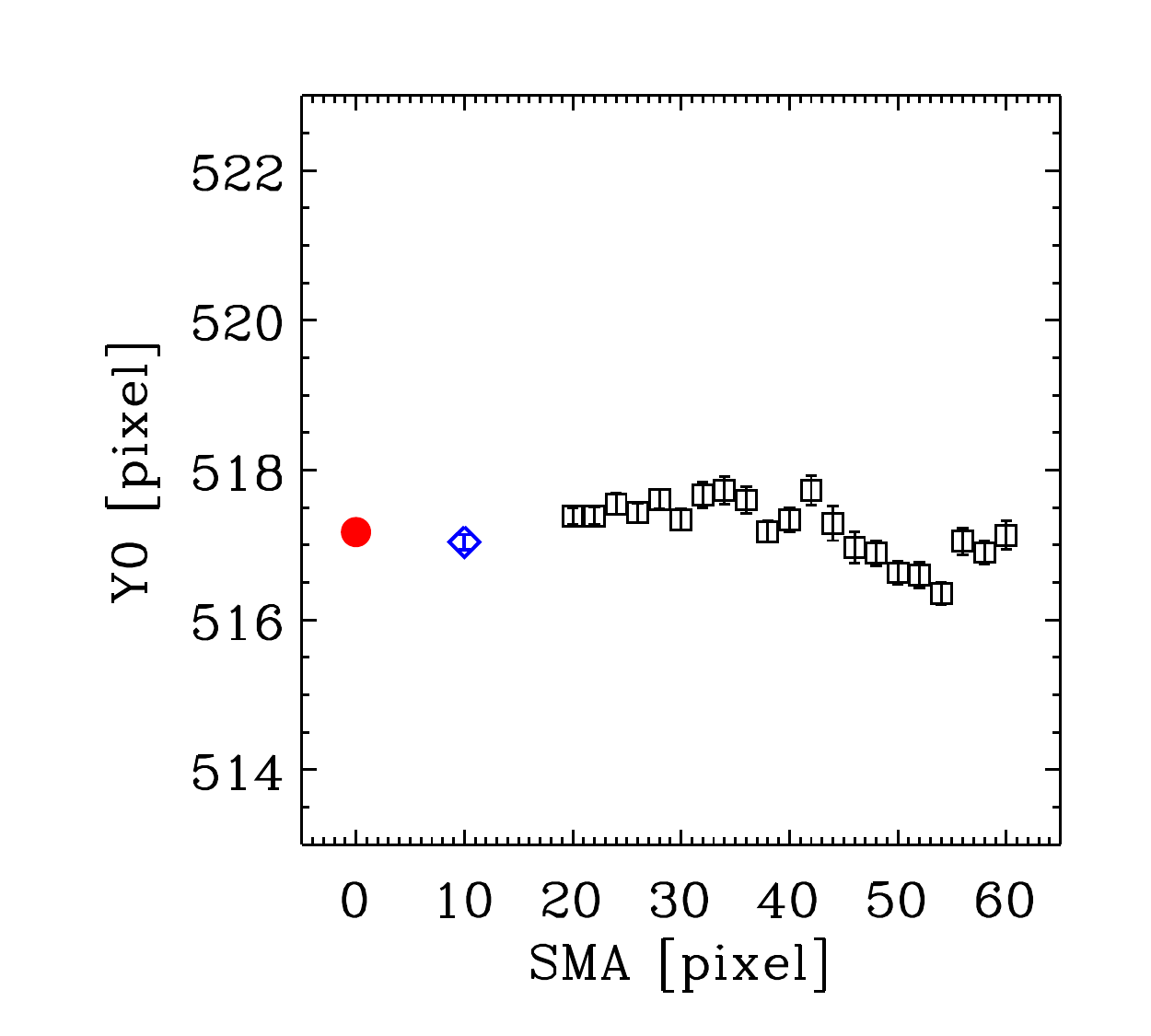} \\	
 \includegraphics[trim=0.65cm 0cm 0cm 0cm, clip=true, scale=0.46]{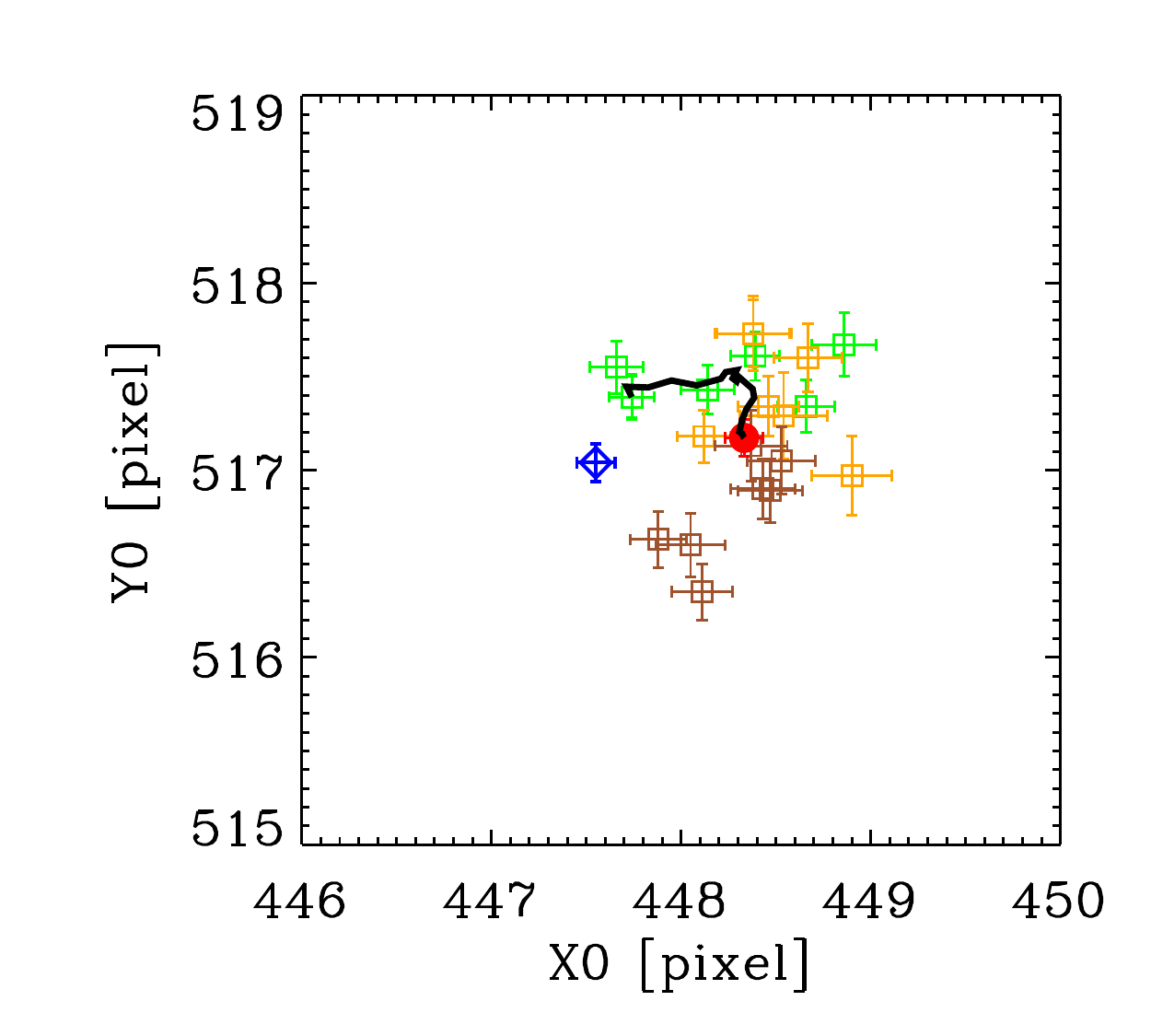}	&  \includegraphics[trim=0.6cm 0cm 0cm 0cm, clip=true, scale=0.46]{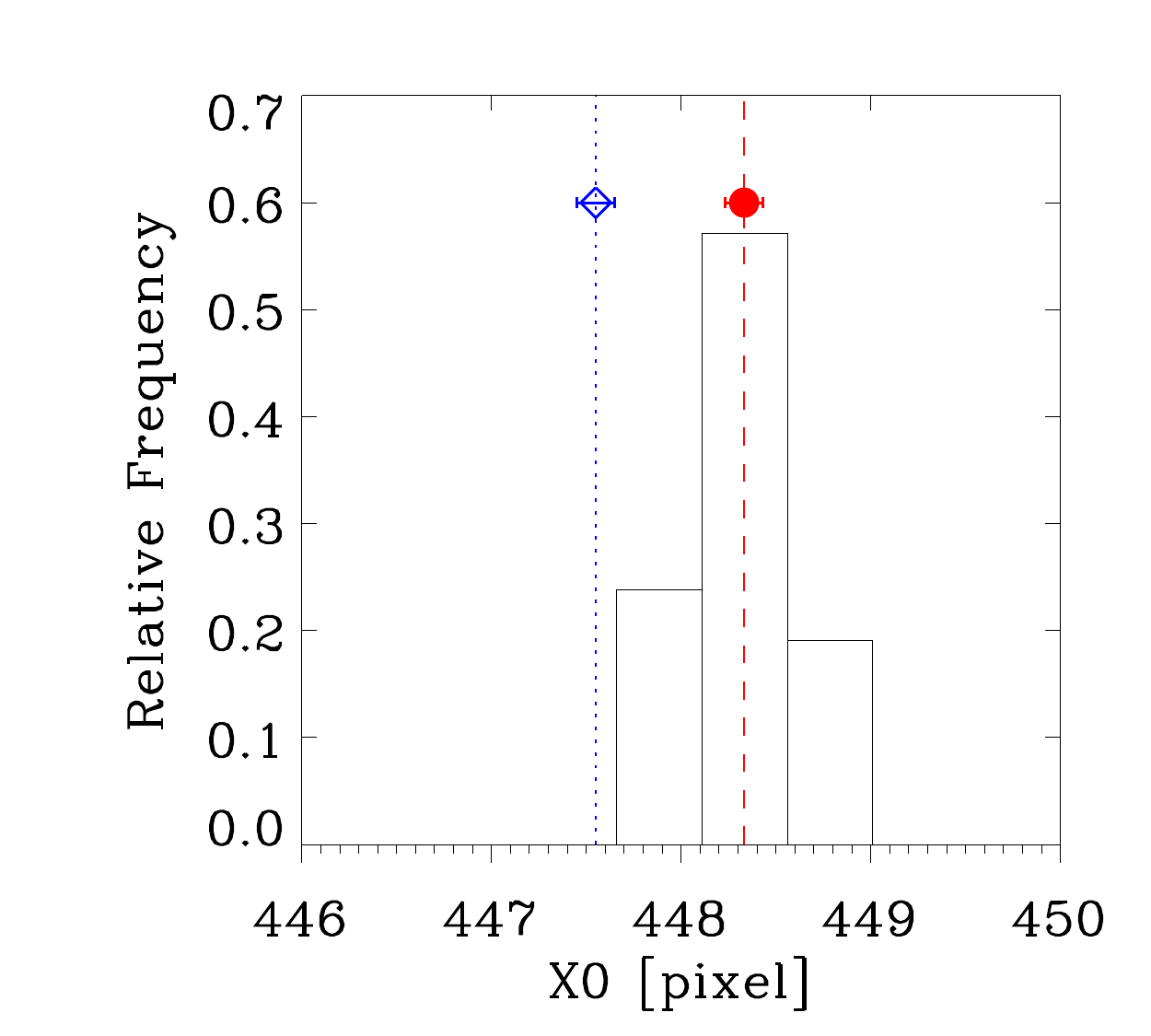}	& \includegraphics[trim=0.6cm 0cm 0cm 0cm, clip=true, scale=0.46]{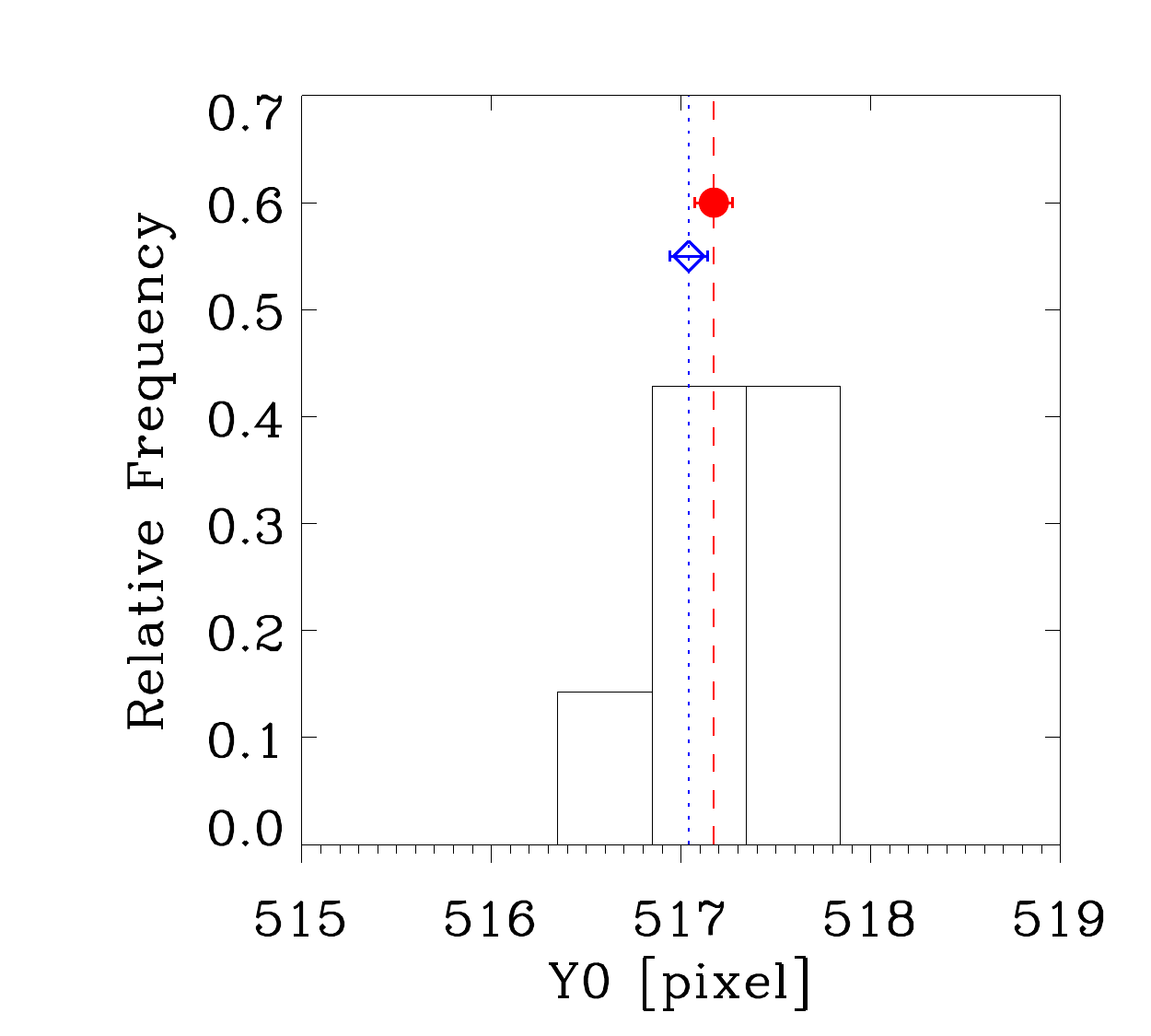}\\
\end{array}$
\end{center}
\caption[M87 (WFPC2_F814W)]{As in Fig.\ref{fig: NGC4373_W2} for galaxy NGC 4486 (M 87), WFPC2/PC - F814W, scale=$0\farcs05$/pxl.}
\label{fig: M87_WFPC2F814W}
\end{figure*} 

\begin{figure*}[h]
\begin{center}$
\begin{array}{ccc}
\includegraphics[trim=3.75cm 1cm 3cm 0cm, clip=true, scale=0.48]{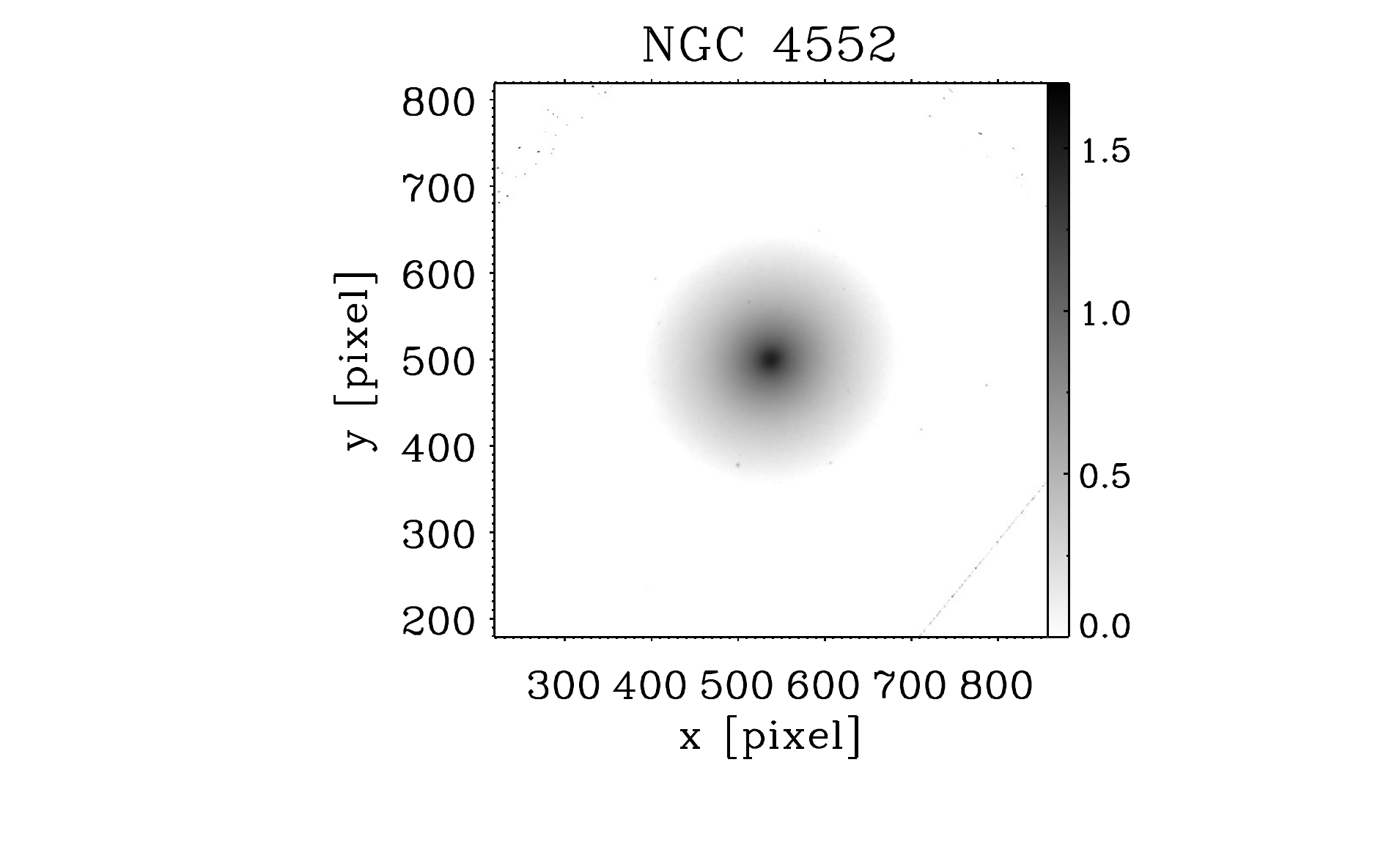} & \includegraphics[trim= 4.cm 1cm 3cm 0cm, clip=true, scale=0.48]{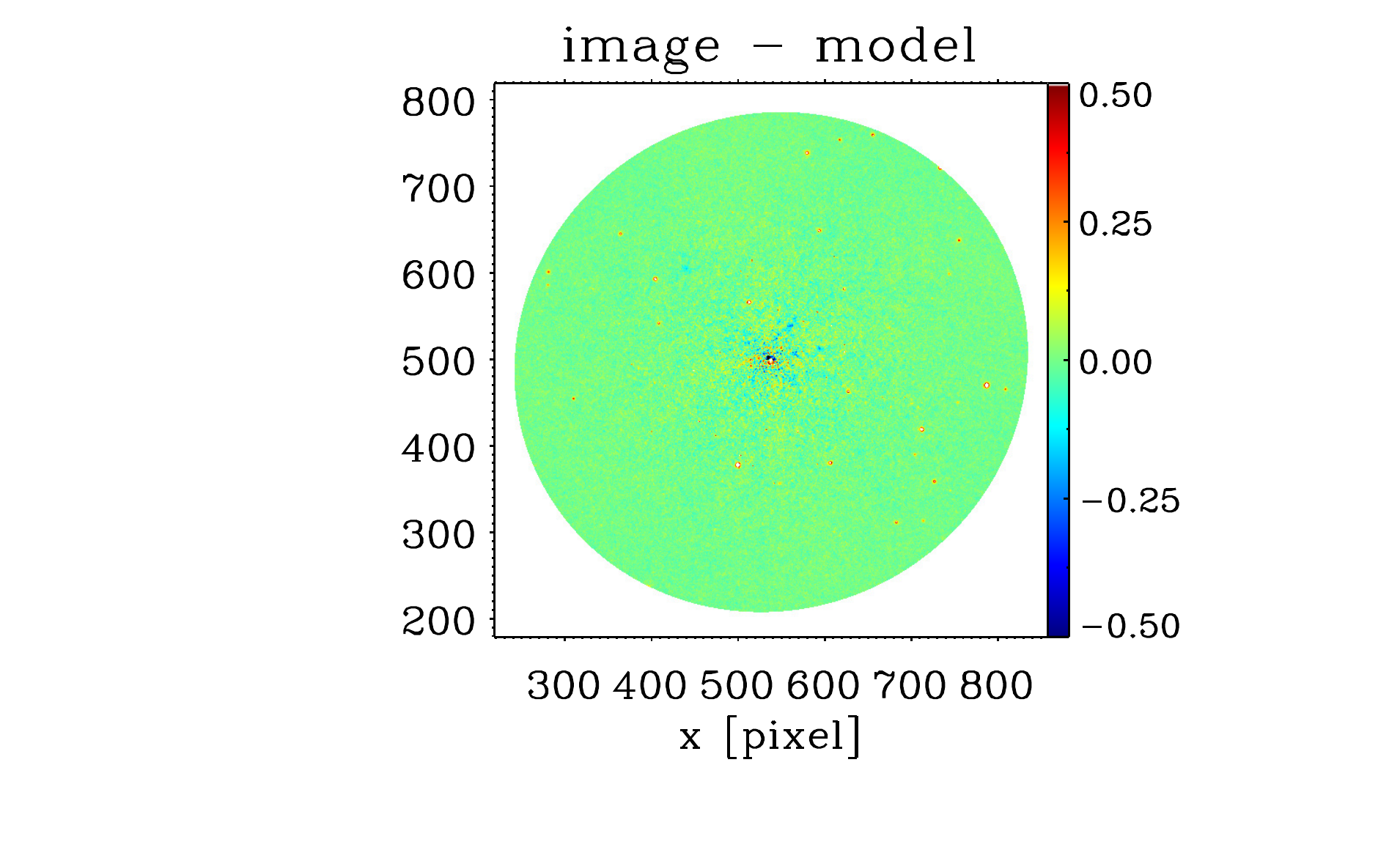}	& \includegraphics[trim= 4.cm 1cm 3cm 0cm, clip=true, scale=0.48]{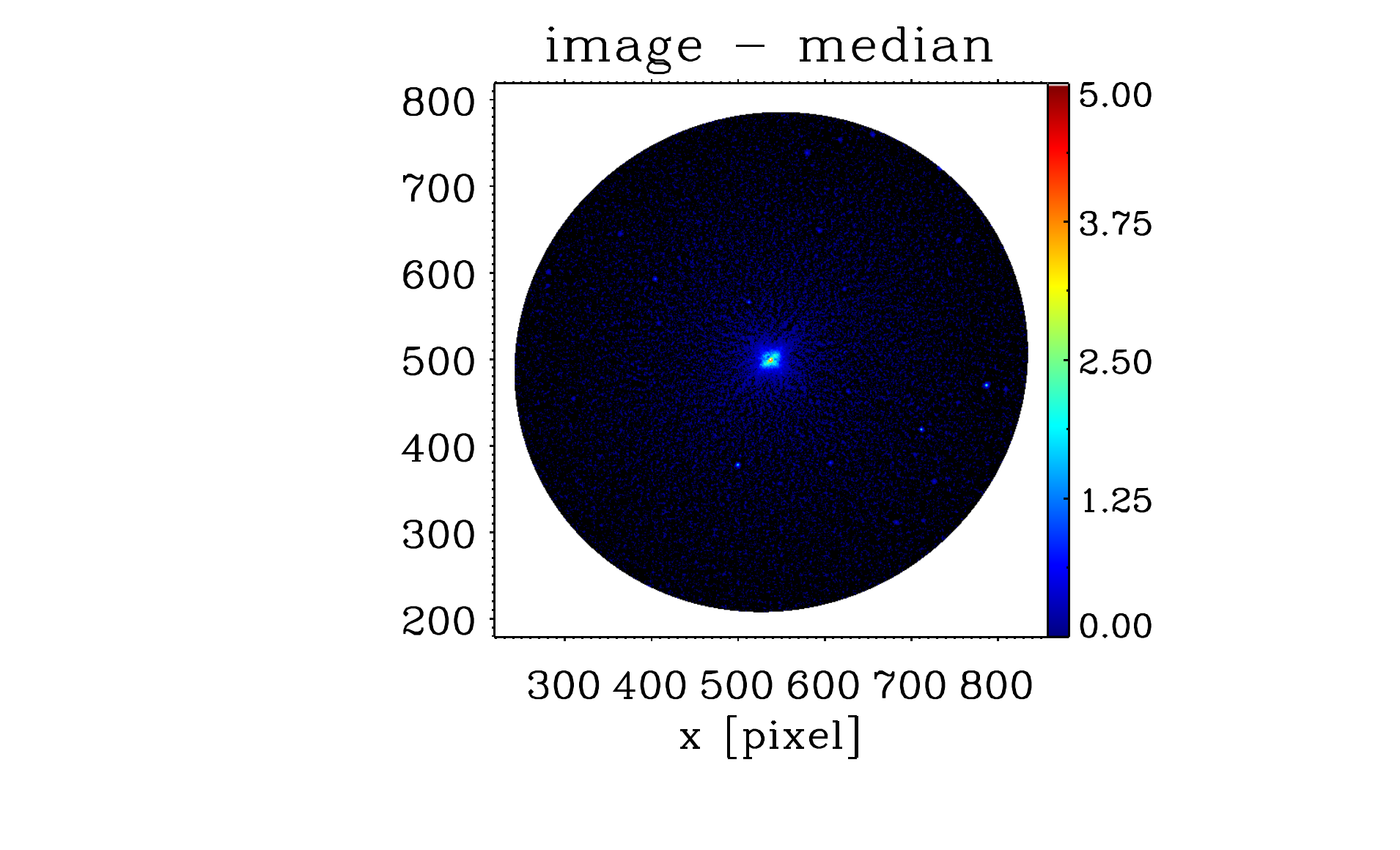} \\
\includegraphics[trim=0.7cm 0cm 0cm 0cm, clip=true, scale=0.46]{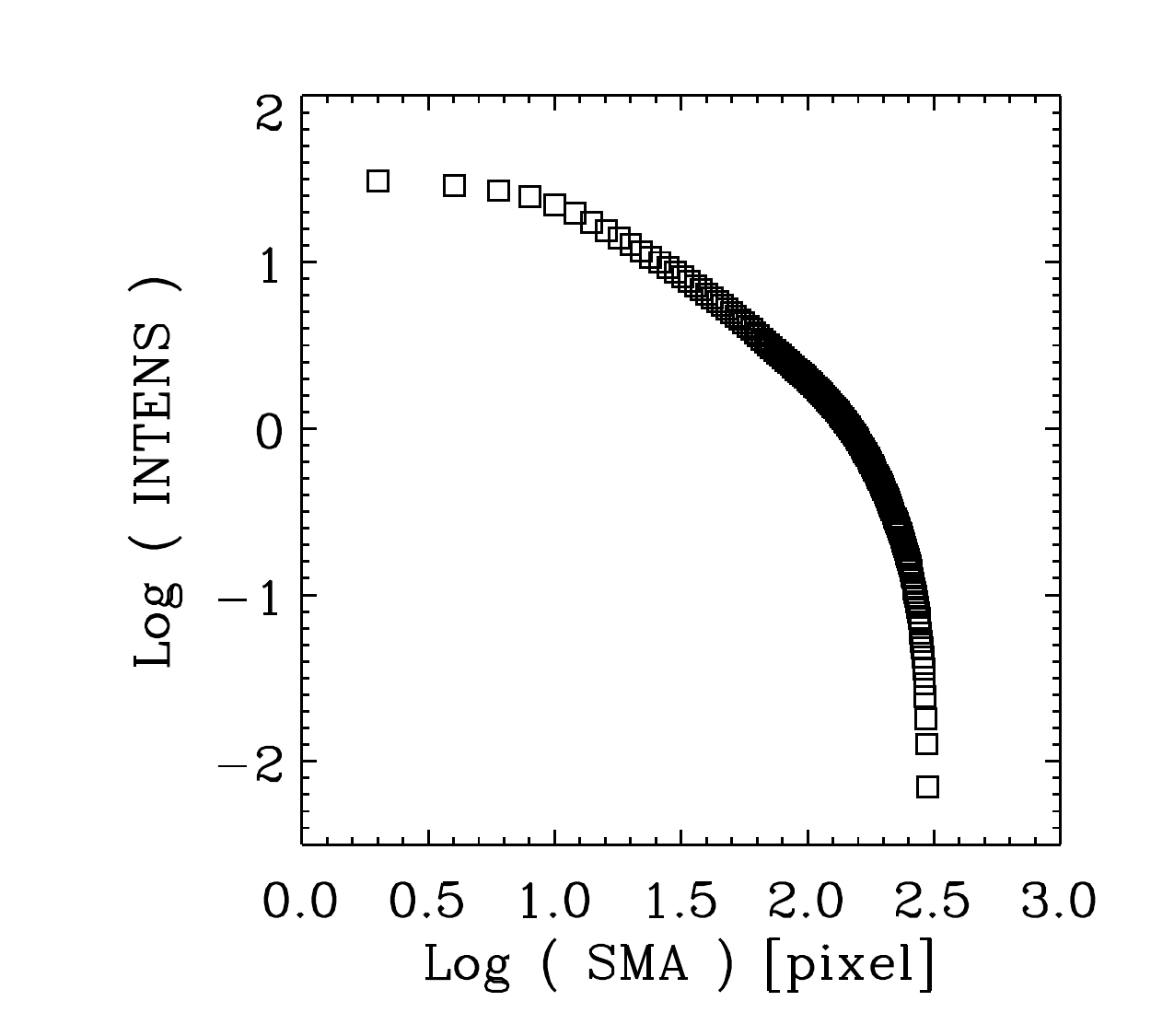}	    &  \includegraphics[trim=0.6cm 0cm 0cm 0cm, clip=true, scale=0.46]{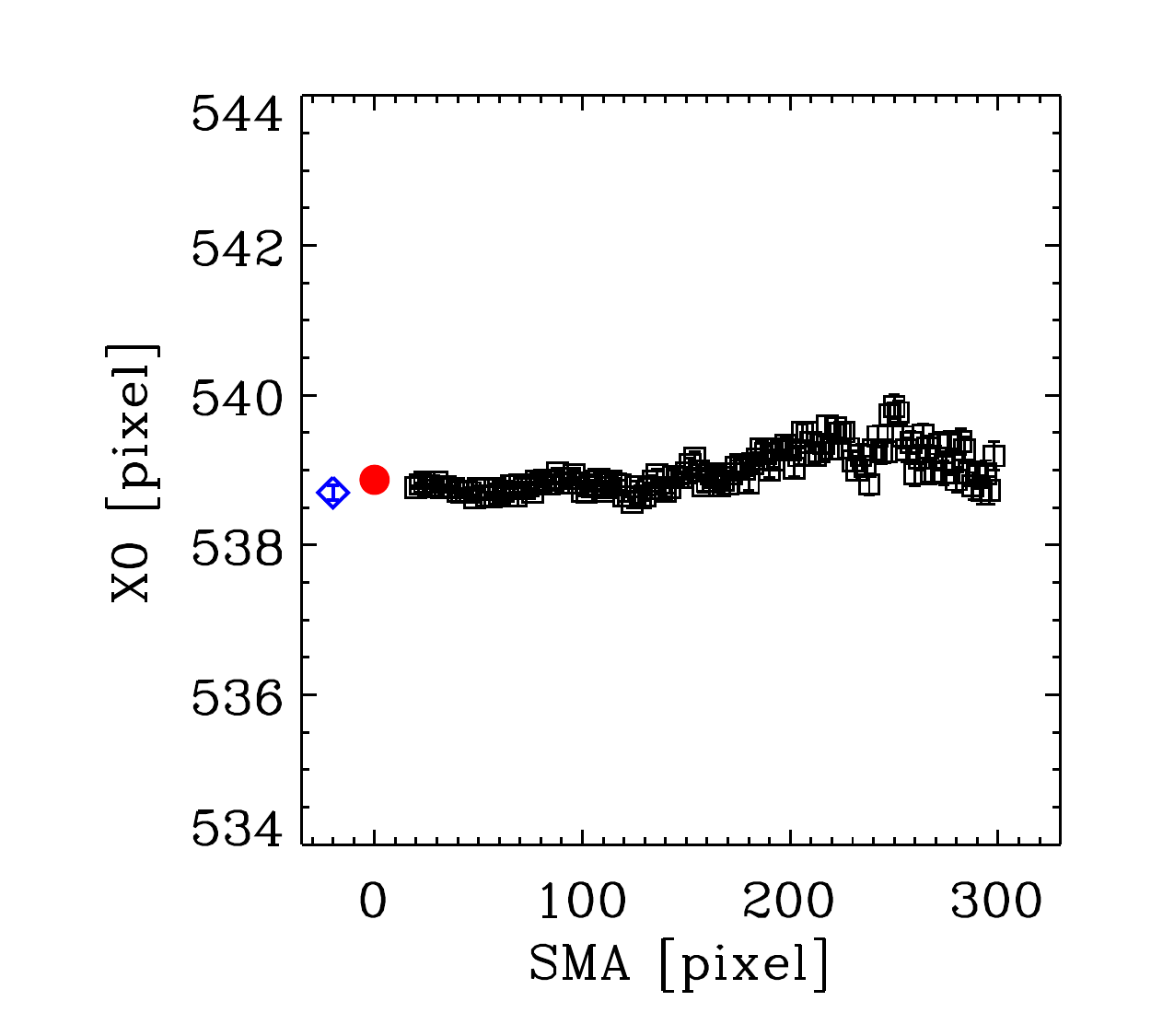}  &  \includegraphics[trim=0.6cm 0cm 0cm 0cm, clip=true, scale=0.46]{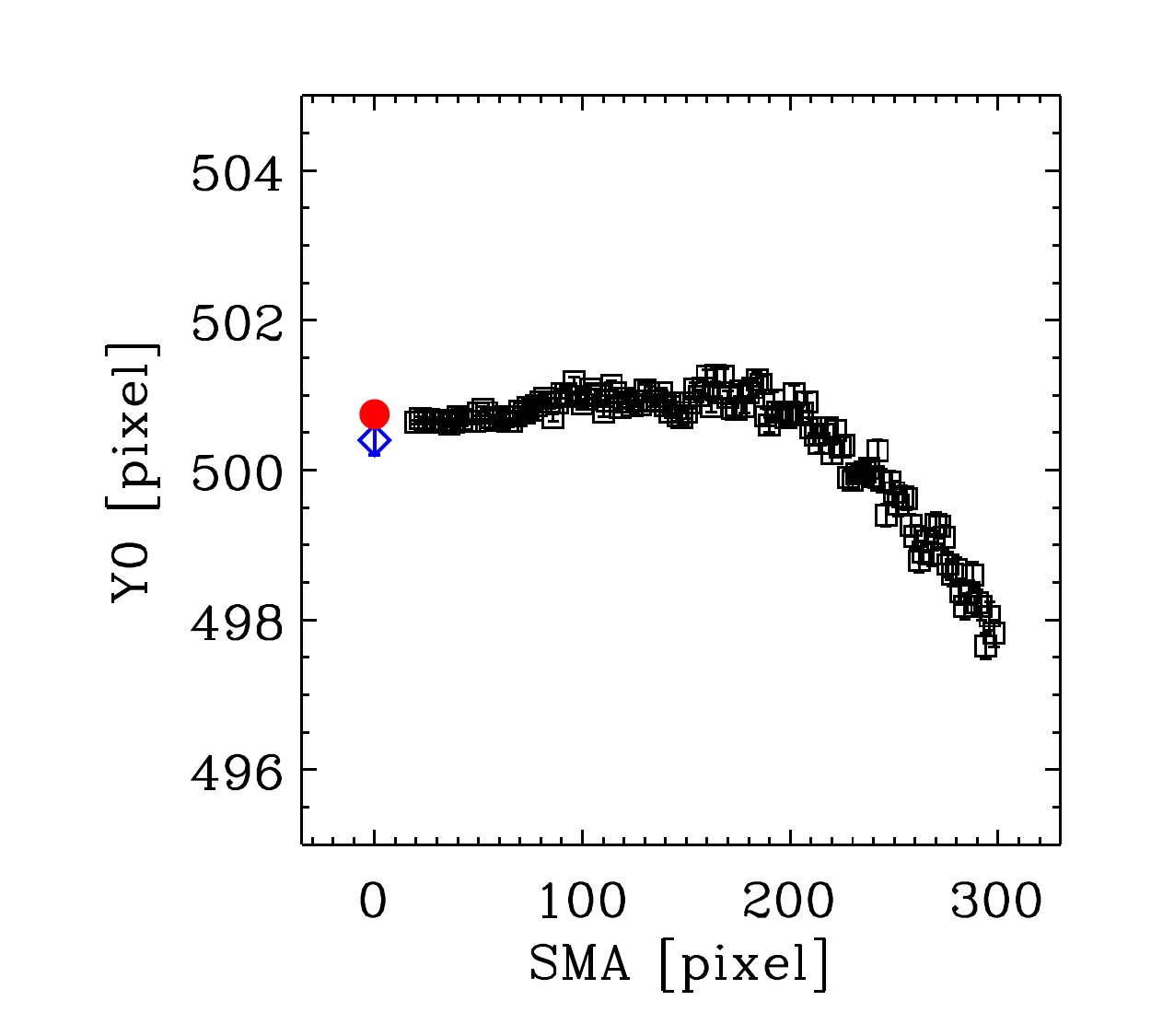} \\	
 \includegraphics[trim=0.65cm 0cm 0cm 0cm, clip=true, scale=0.46]{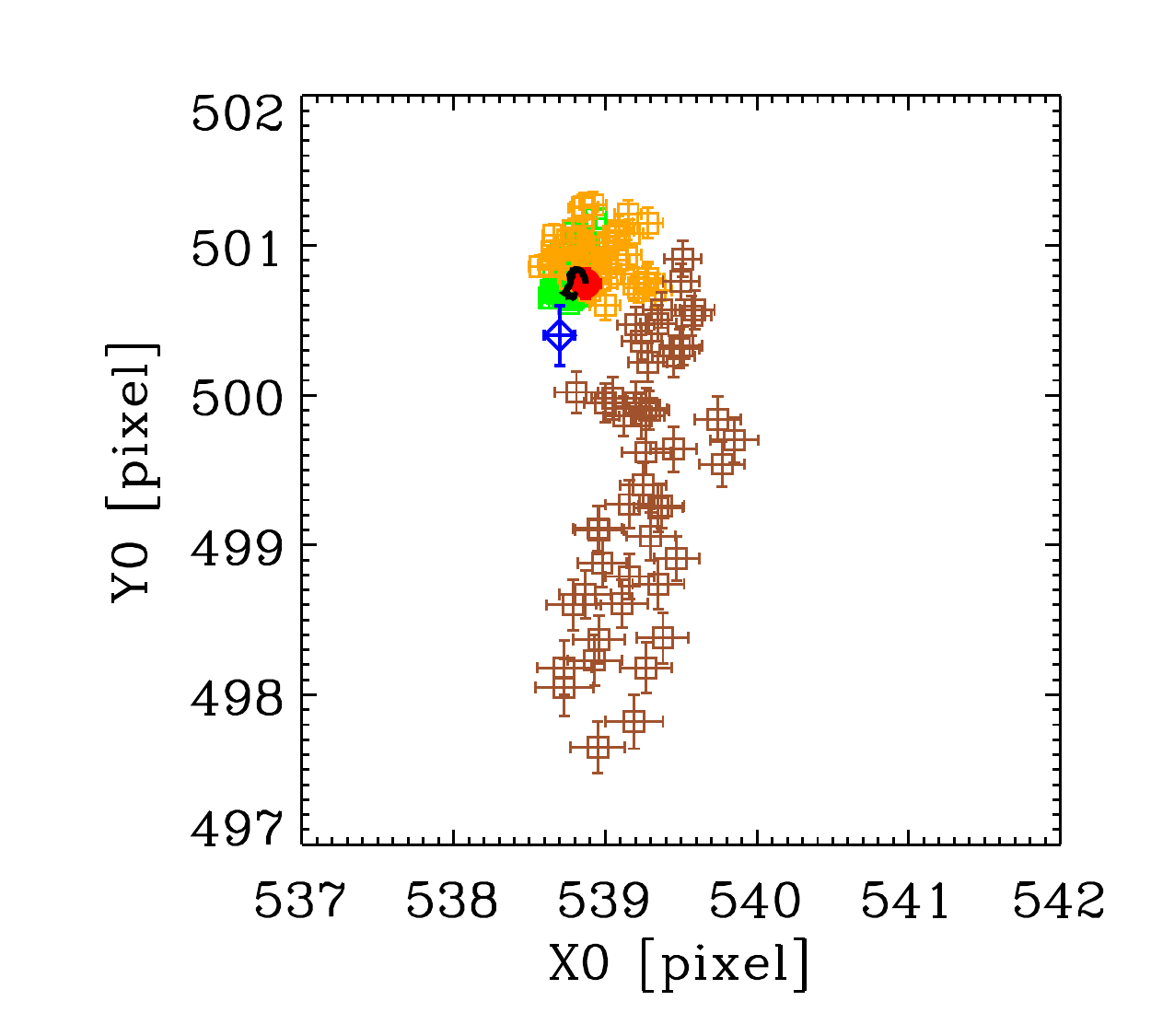}	&  \includegraphics[trim=0.6cm 0cm 0cm 0cm, clip=true, scale=0.46]{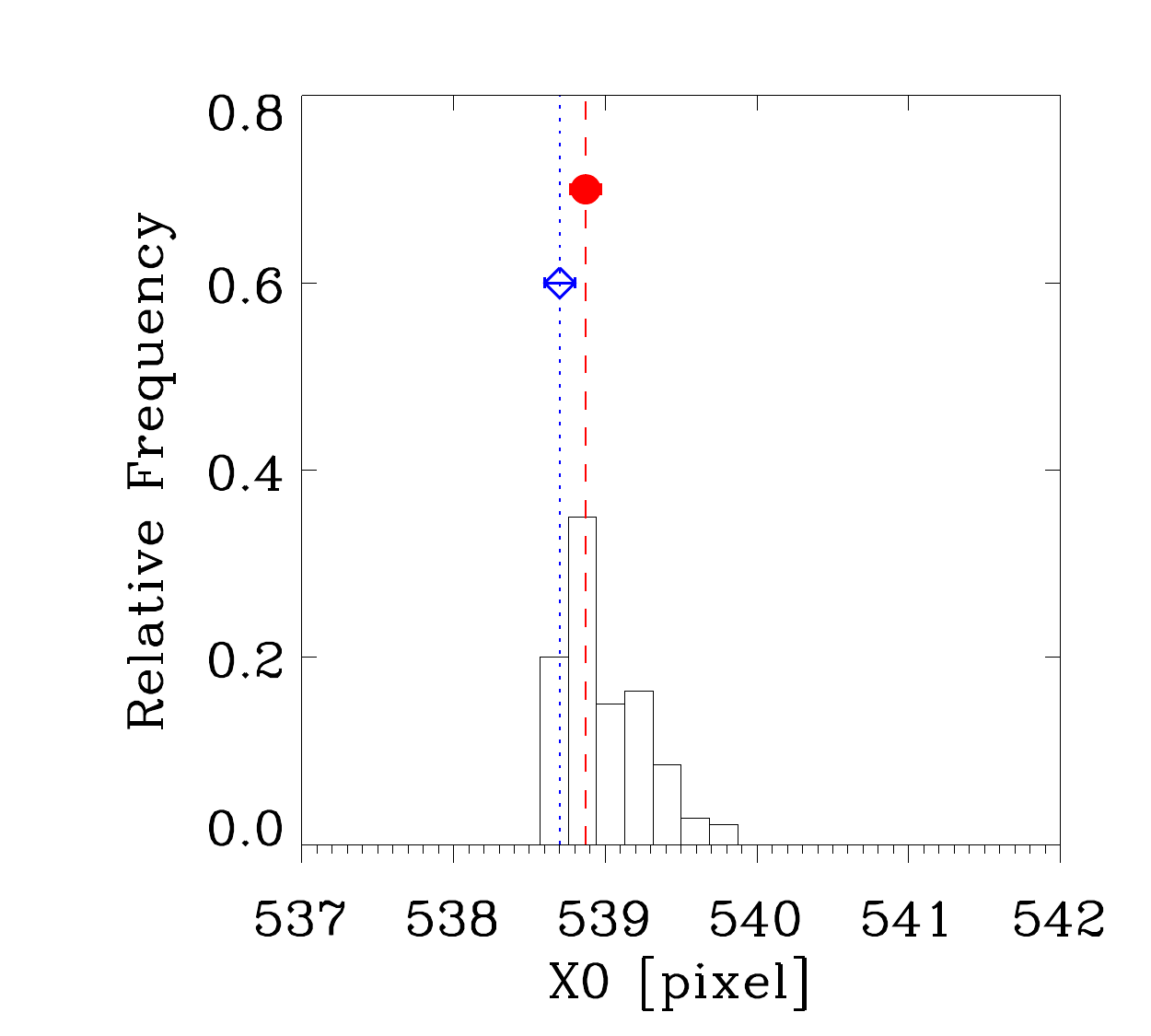}	& \includegraphics[trim=0.6cm 0cm 0cm 0cm, clip=true, scale=0.46]{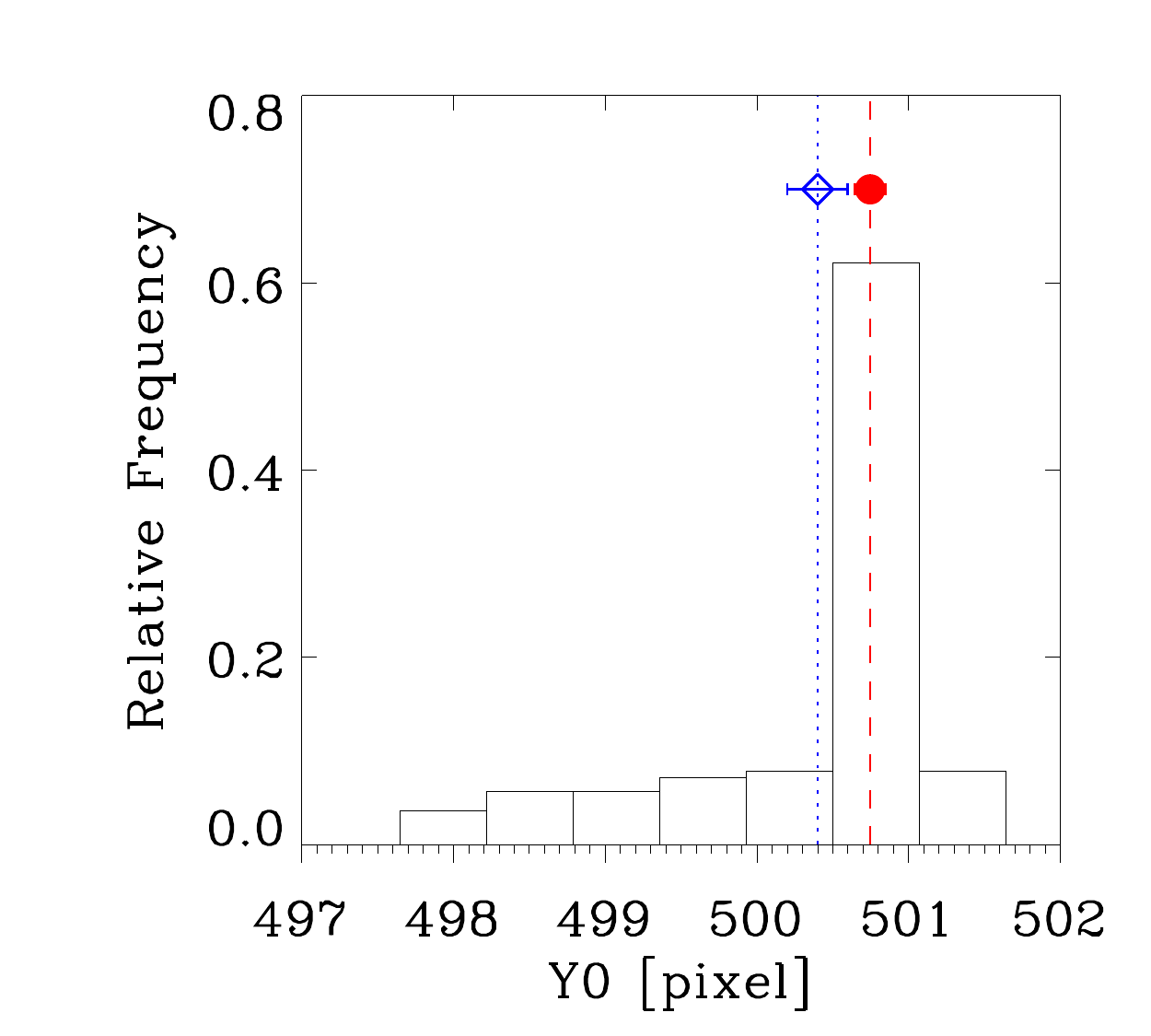}\\
\end{array}$
\end{center}
\caption[NGC 4552 (WFPC2_F814W)]{As in Fig.\ref{fig: NGC4373_W2} for galaxy NGC 4552, WFPC2/PC - F814W, scale=$0\farcs05$/pxl.}
\label{fig: NGC4552_WFPC2F814W}
\end{figure*} 

\begin{figure*}[h]
\begin{center}$
\begin{array}{ccc}
\includegraphics[trim=3.75cm 1cm 3cm 0cm, clip=true, scale=0.48]{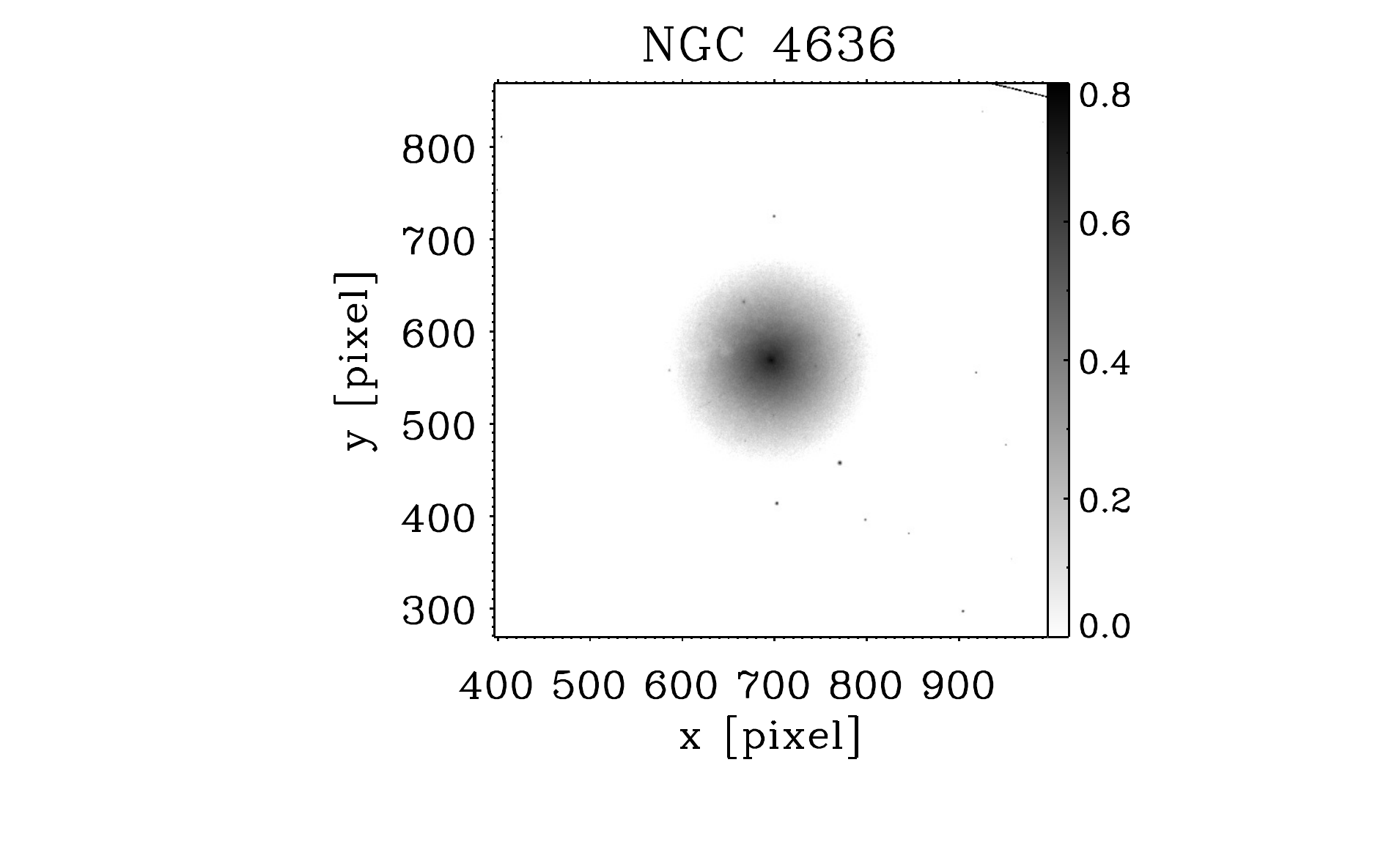} & \includegraphics[trim= 4.cm 1cm 3cm 0cm, clip=true, scale=0.48]{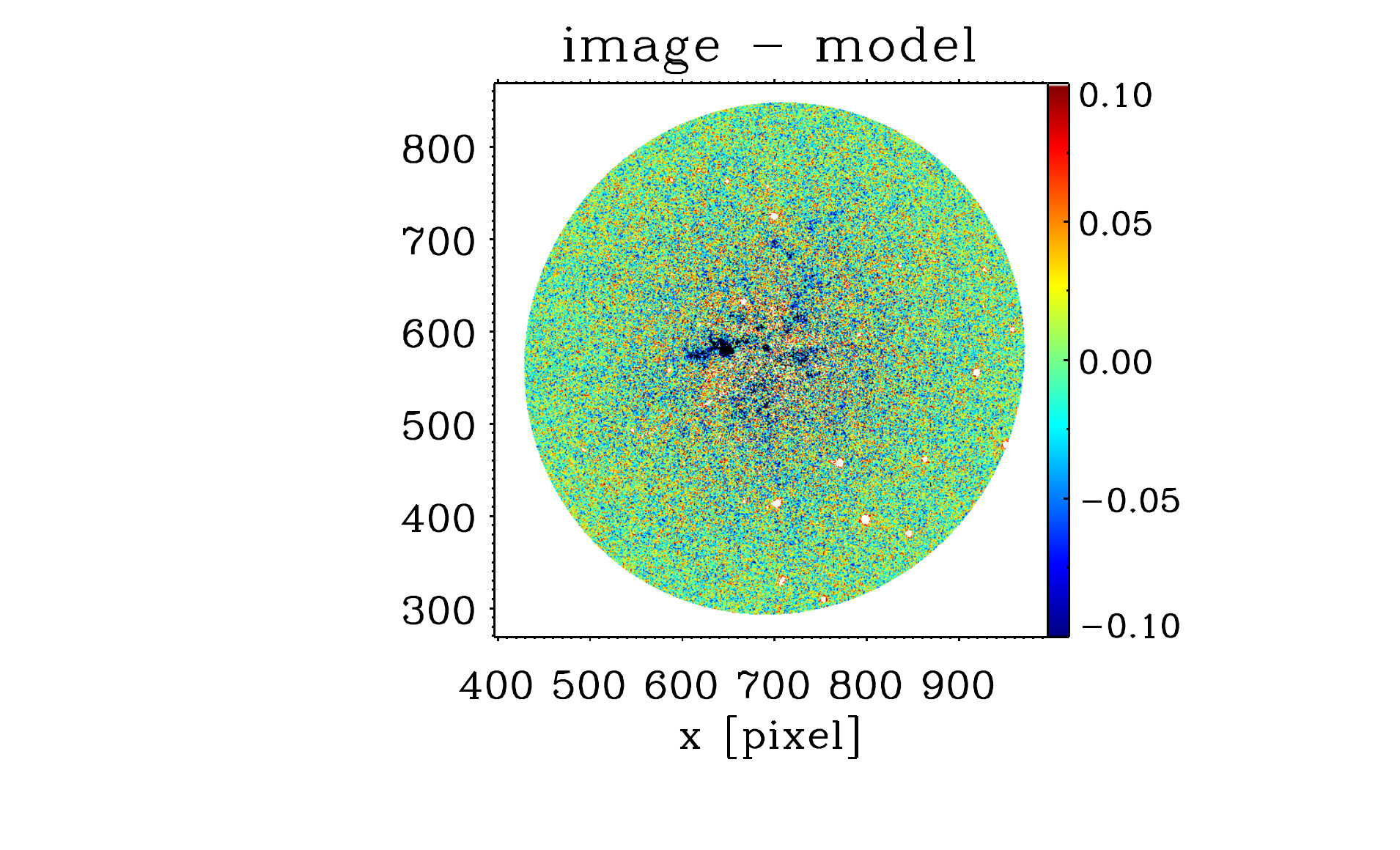}	& \includegraphics[trim= 4.cm 1cm 3cm 0cm, clip=true, scale=0.48]{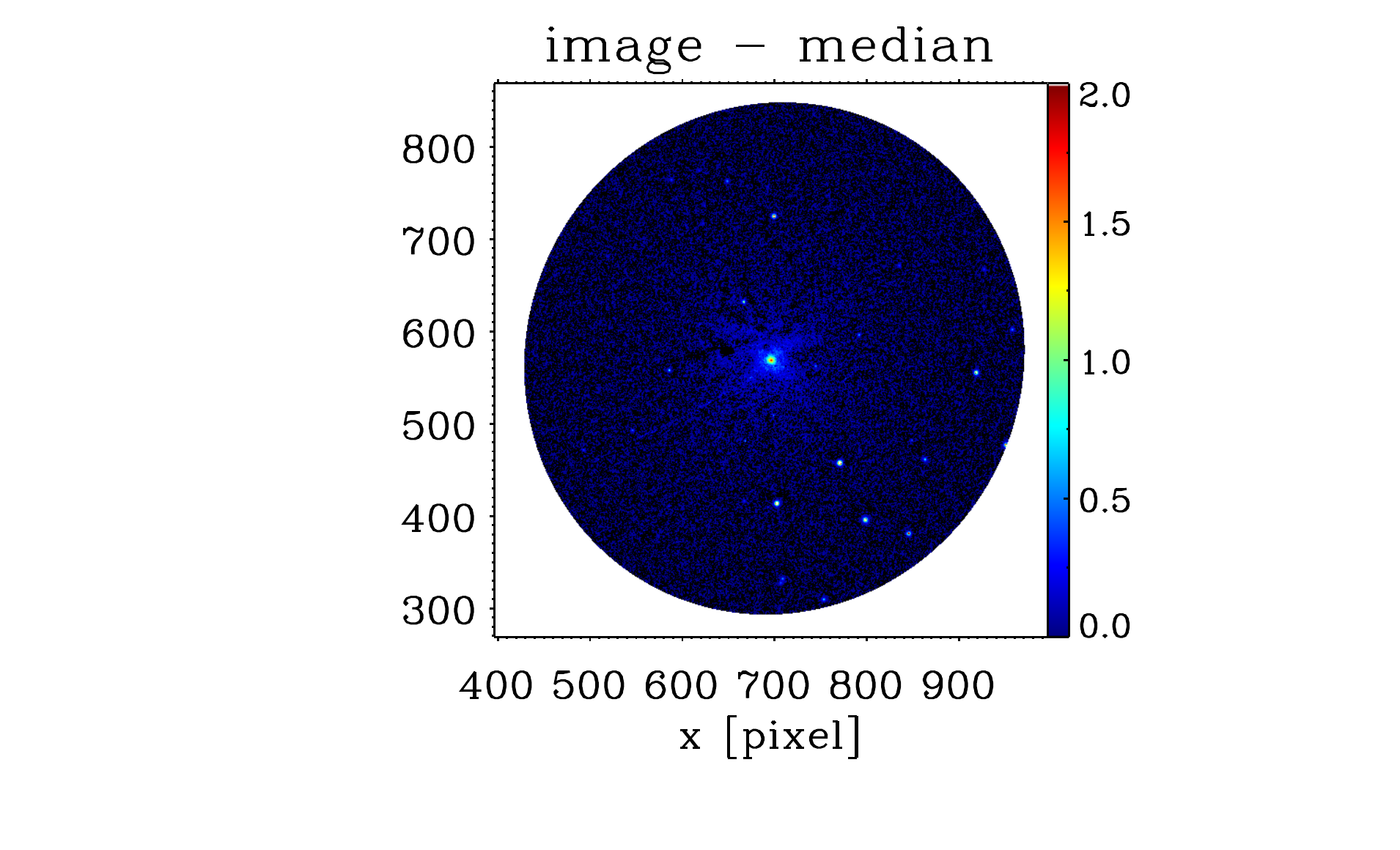} \\
\includegraphics[trim=0.7cm 0cm 0cm 0cm, clip=true, scale=0.46]{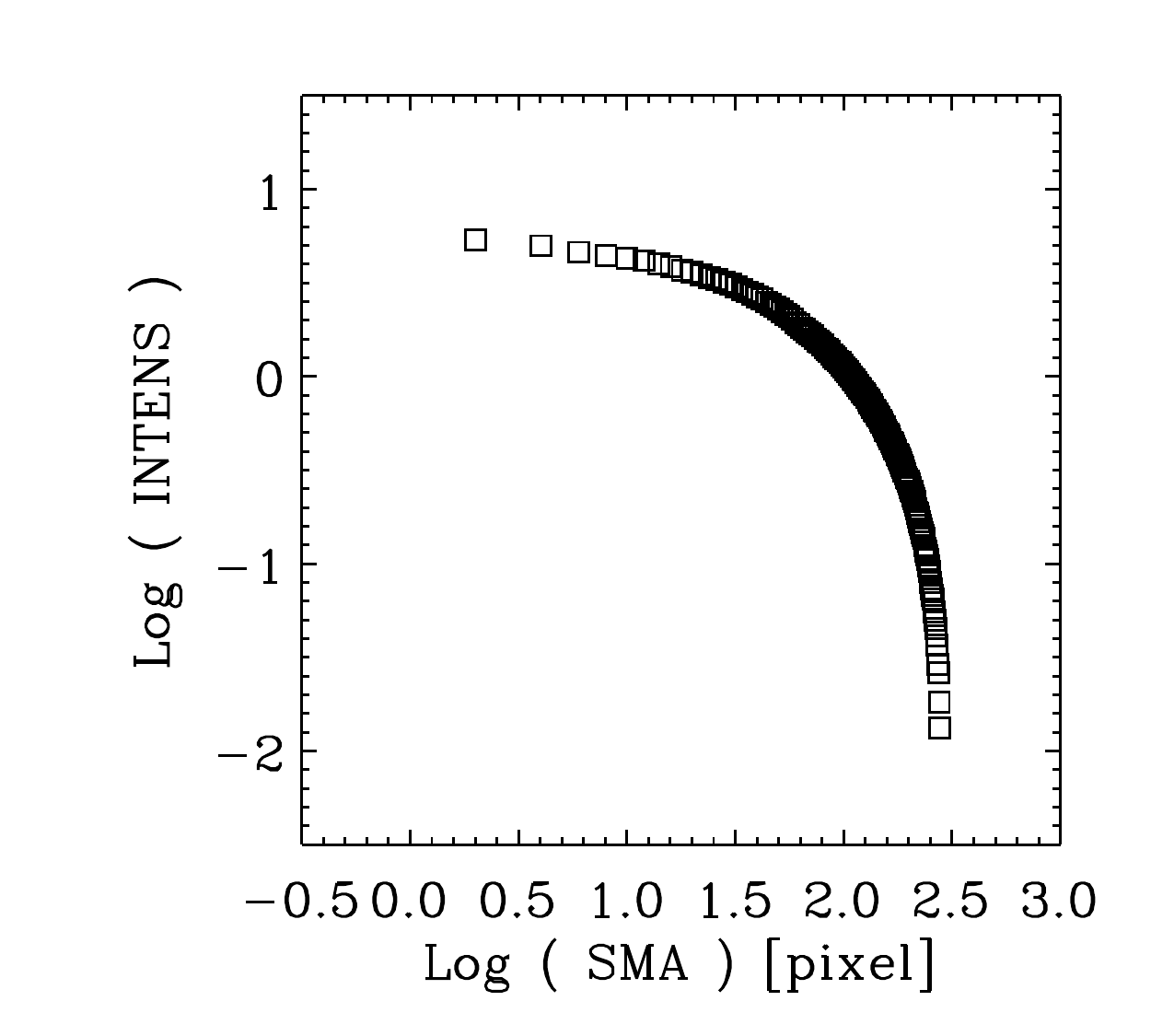}	    &  \includegraphics[trim=0.6cm 0cm 0cm 0cm, clip=true, scale=0.46]{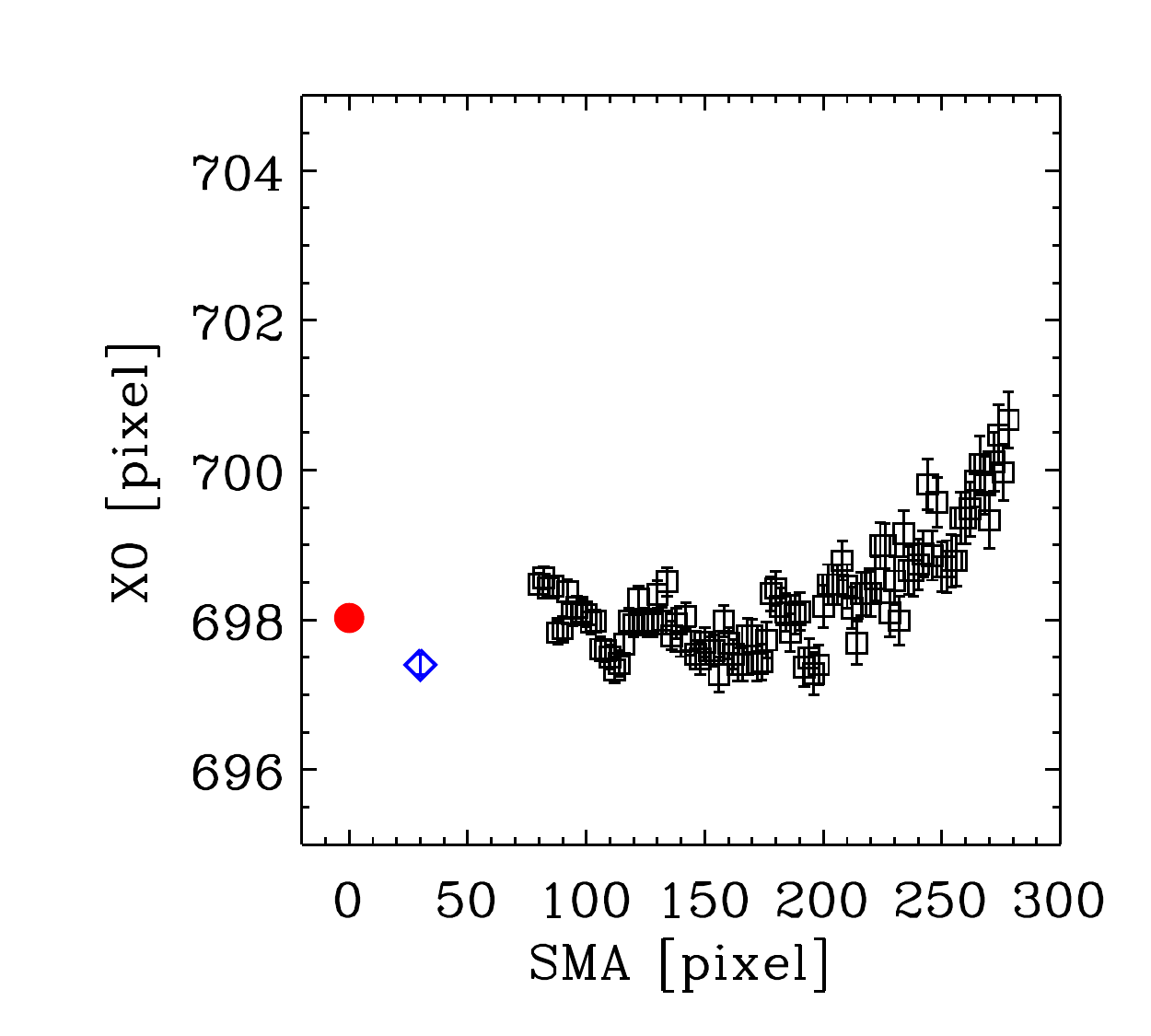}  &  \includegraphics[trim=0.6cm 0cm 0cm 0cm, clip=true, scale=0.46]{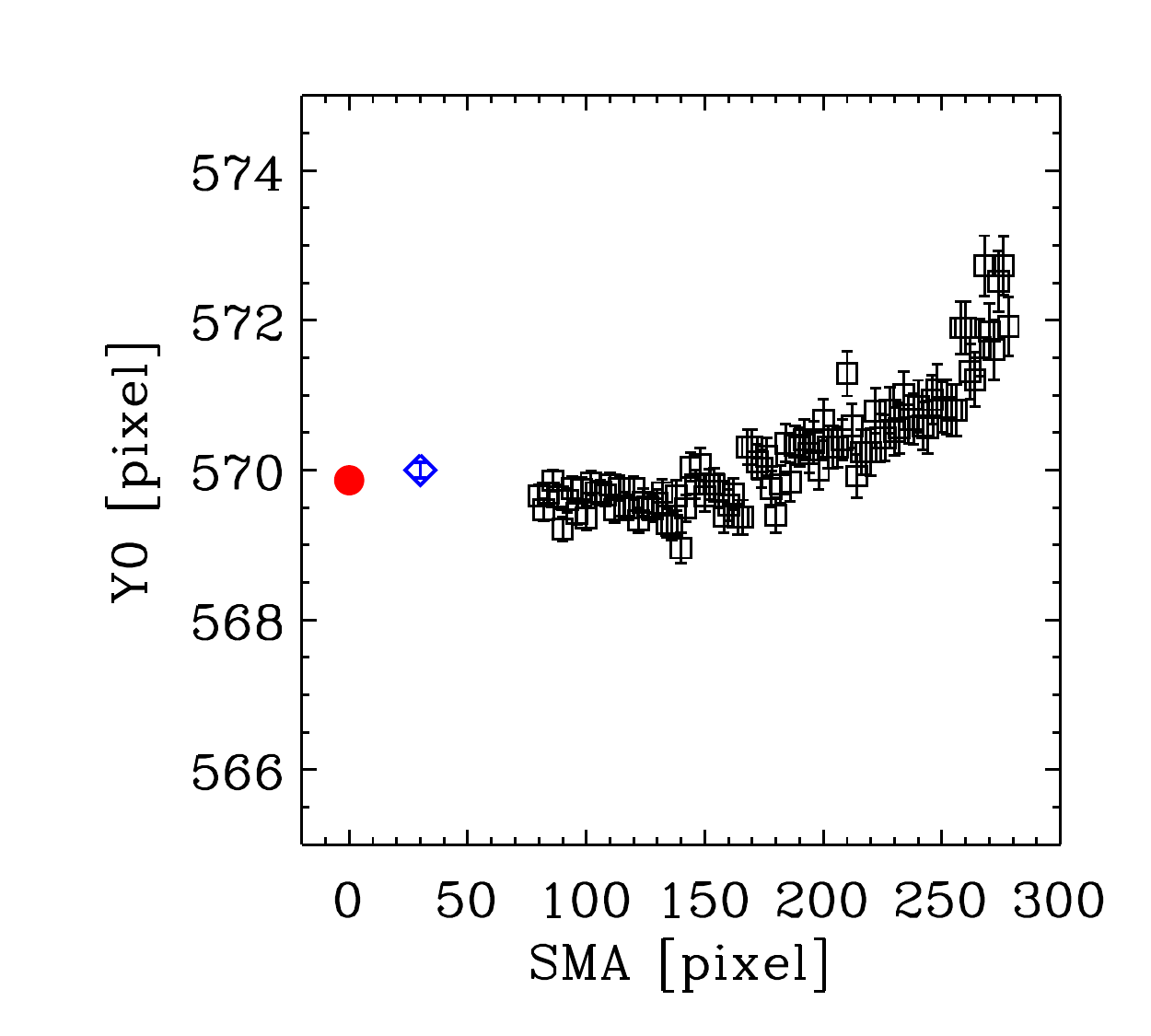} \\	
 \includegraphics[trim=0.65cm 0cm 0cm 0cm, clip=true, scale=0.46]{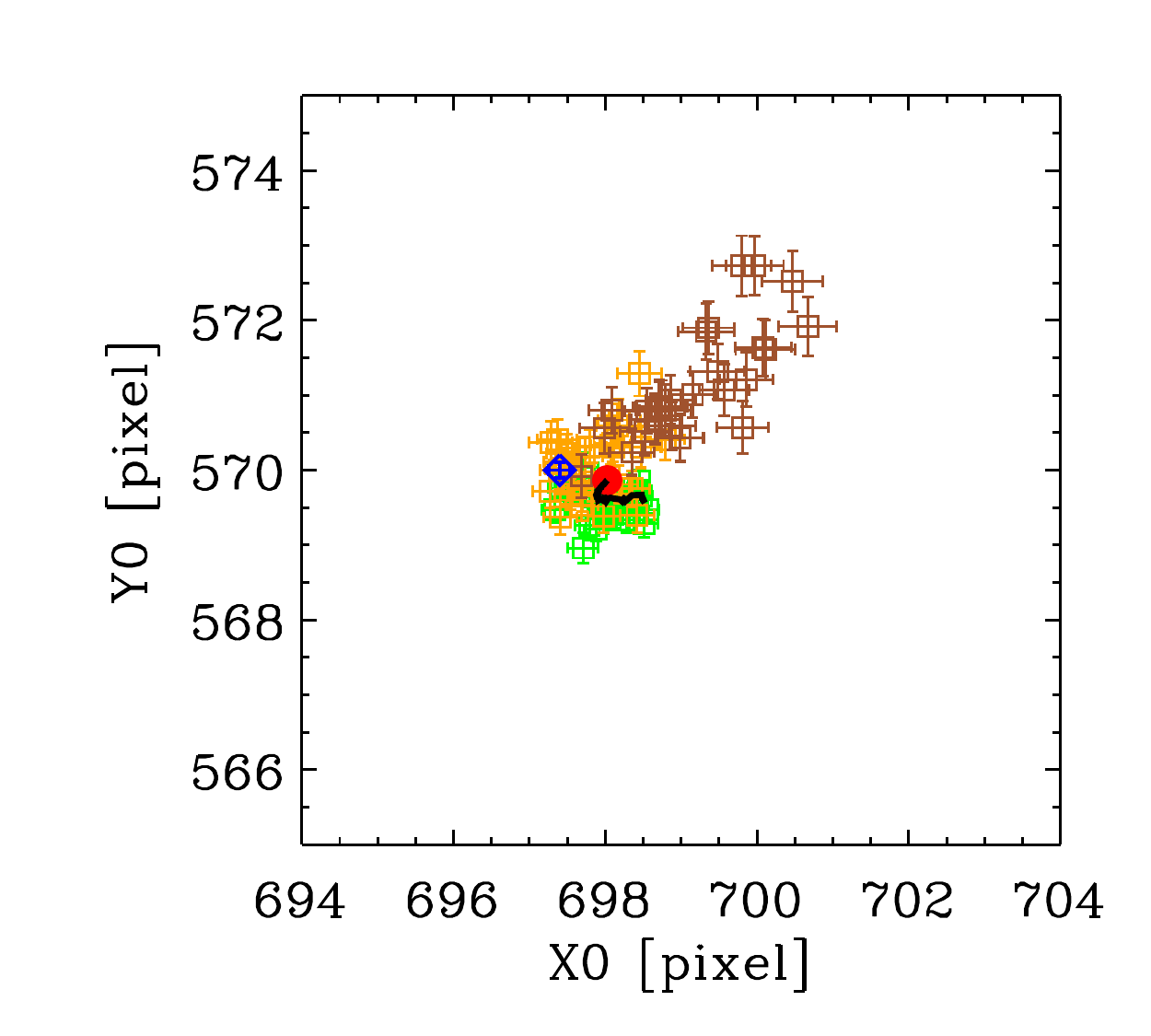}	&  \includegraphics[trim=0.6cm 0cm 0cm 0cm, clip=true, scale=0.46]{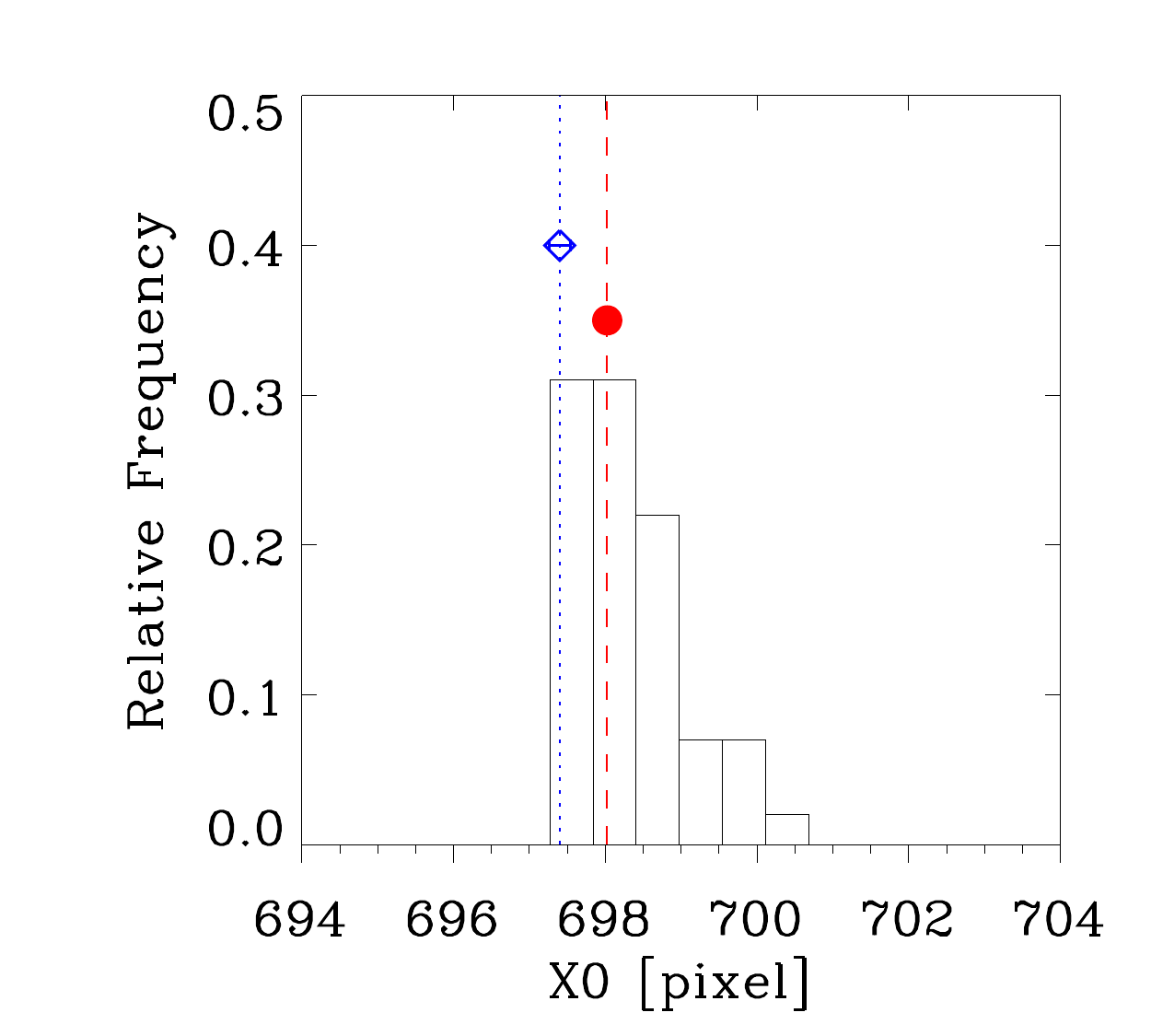}	& \includegraphics[trim=0.6cm 0cm 0cm 0cm, clip=true, scale=0.46]{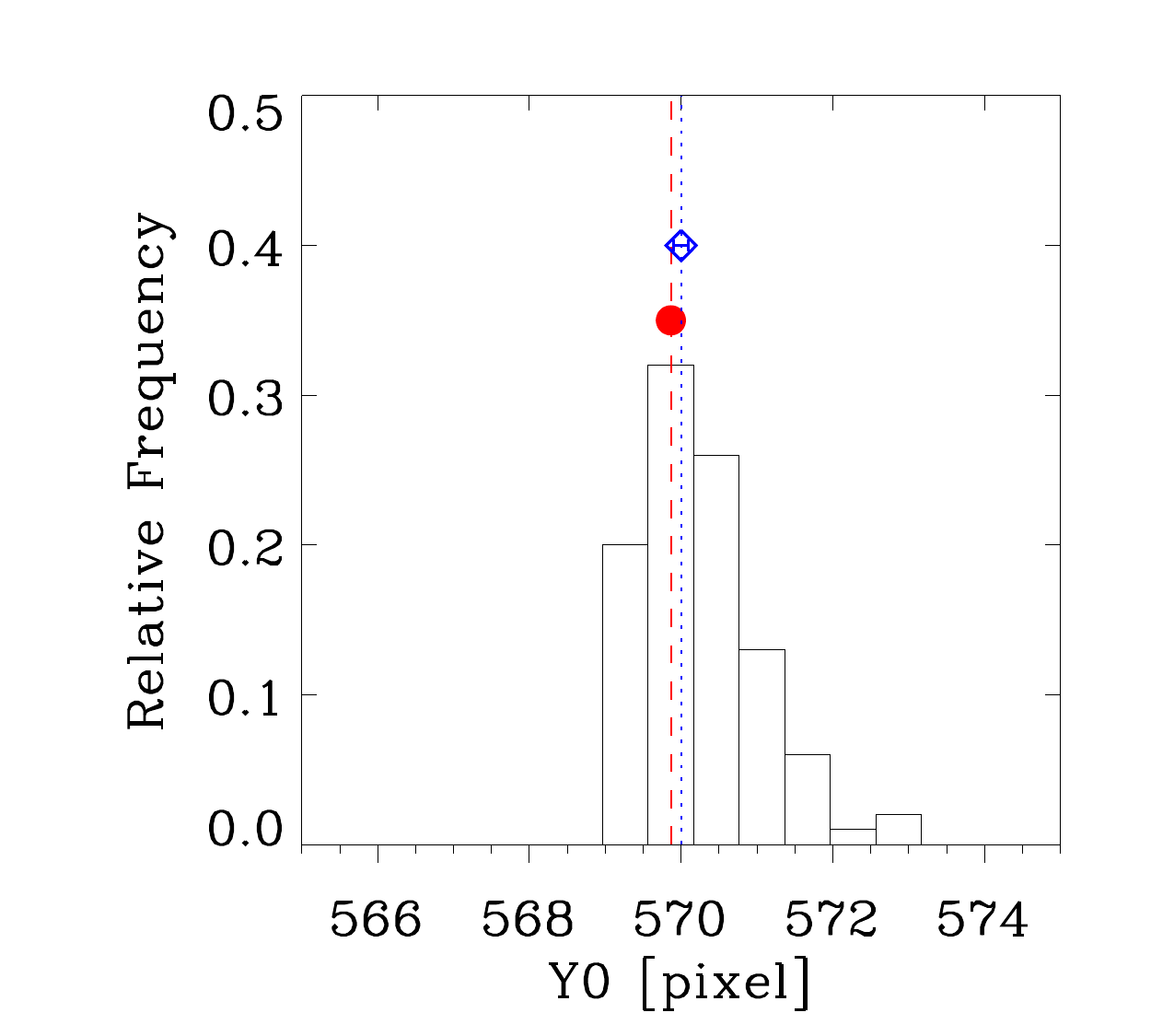}\\
\end{array}$
\end{center}
\caption[NGC 4636 (WFPC2_F814W)]{As in Fig.\ref{fig: NGC4373_W2} for galaxy NGC 4636, WFPC2/PC - F814W, scale=$0\farcs05$/pxl.}
\label{fig: NGC4636_WFPC2F814W}
\end{figure*} 

\cleardoublepage
\begin{figure*}[h]
\begin{center}$
\begin{array}{ccc}
\includegraphics[trim=3.75cm 1cm 3cm 0cm, clip=true, scale=0.48]{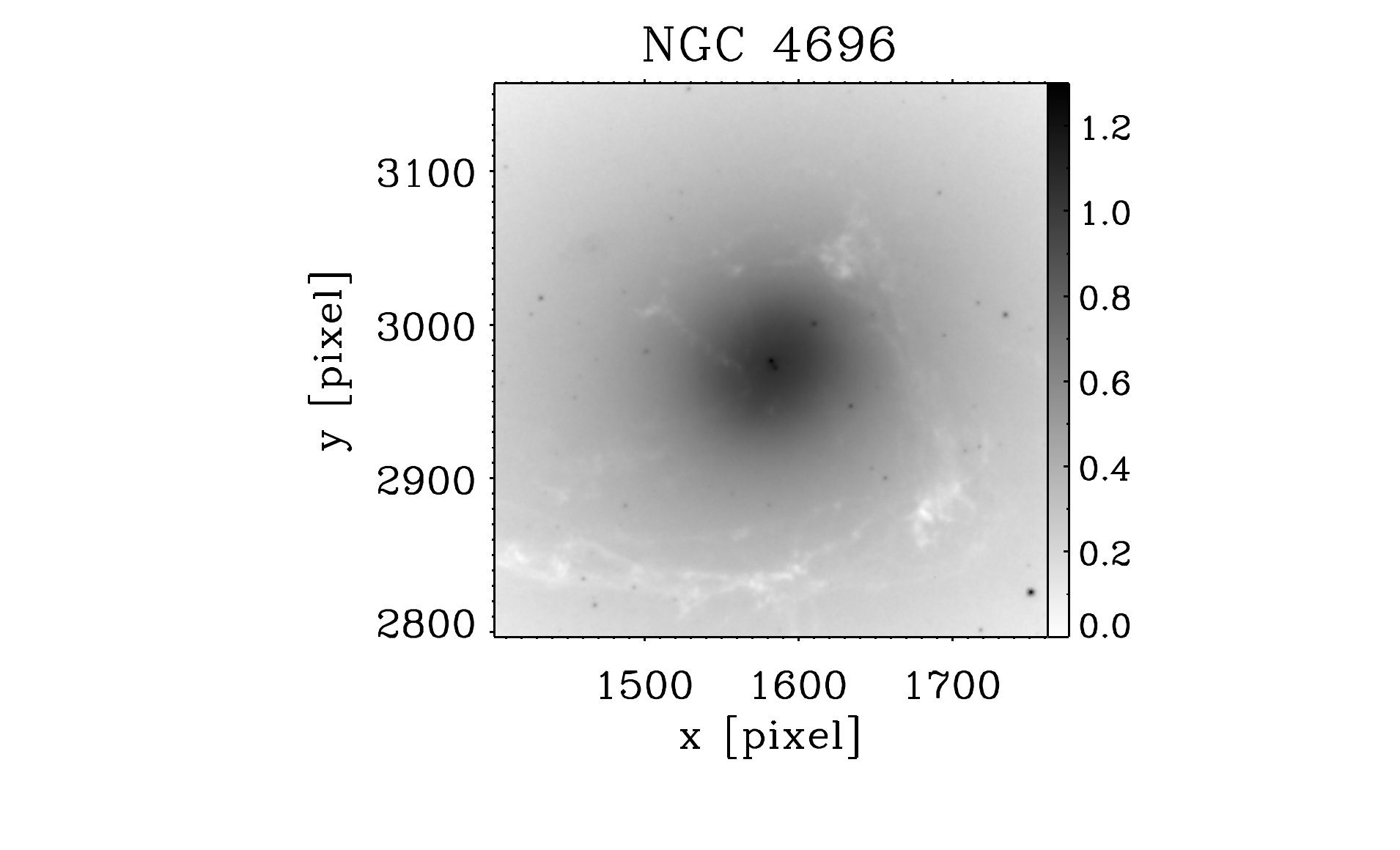} & \includegraphics[trim= 4.cm 1cm 3cm 0cm, clip=true, scale=0.48]{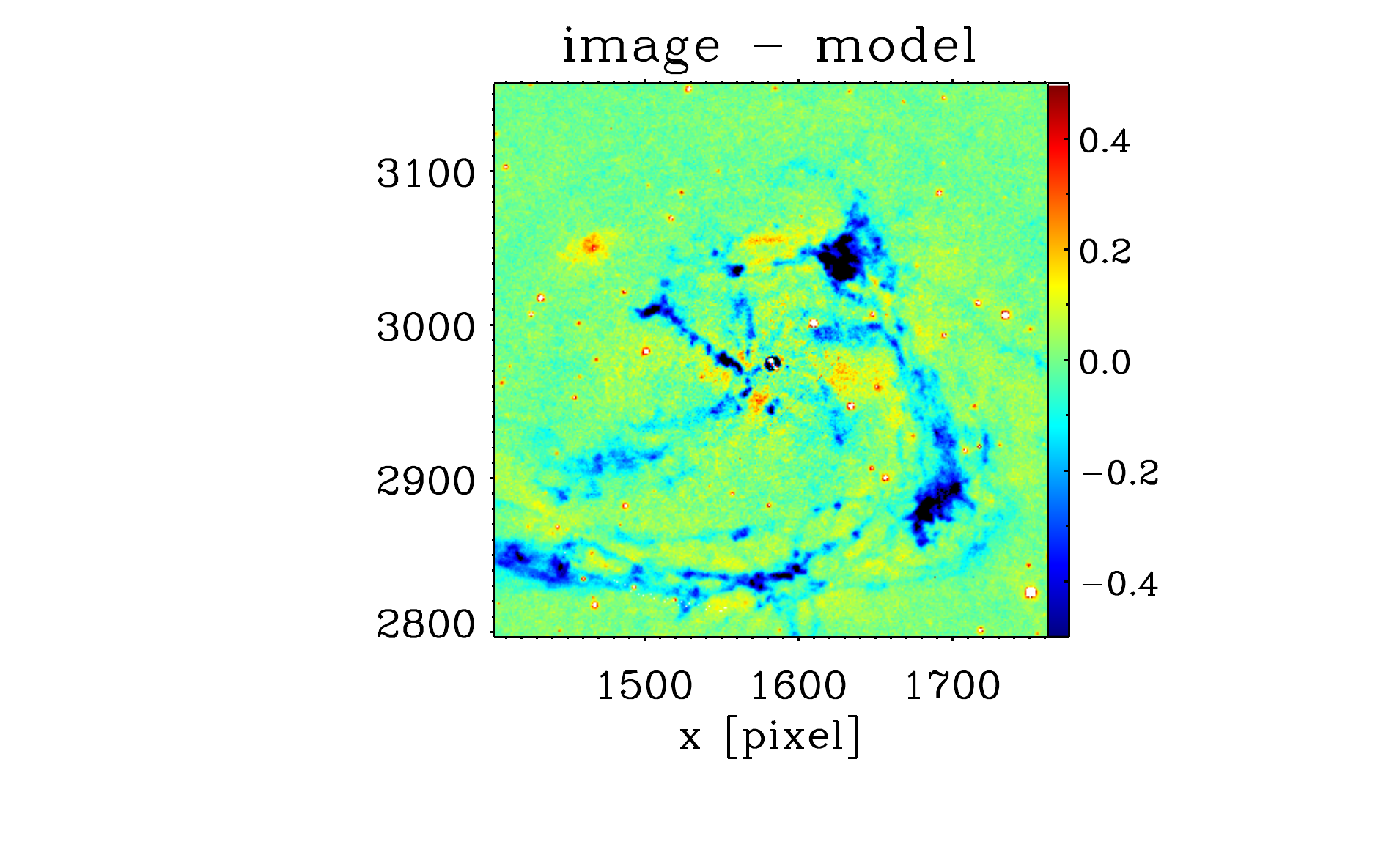}	& \includegraphics[trim= 4.cm 1cm 3cm 0cm, clip=true, scale=0.48]{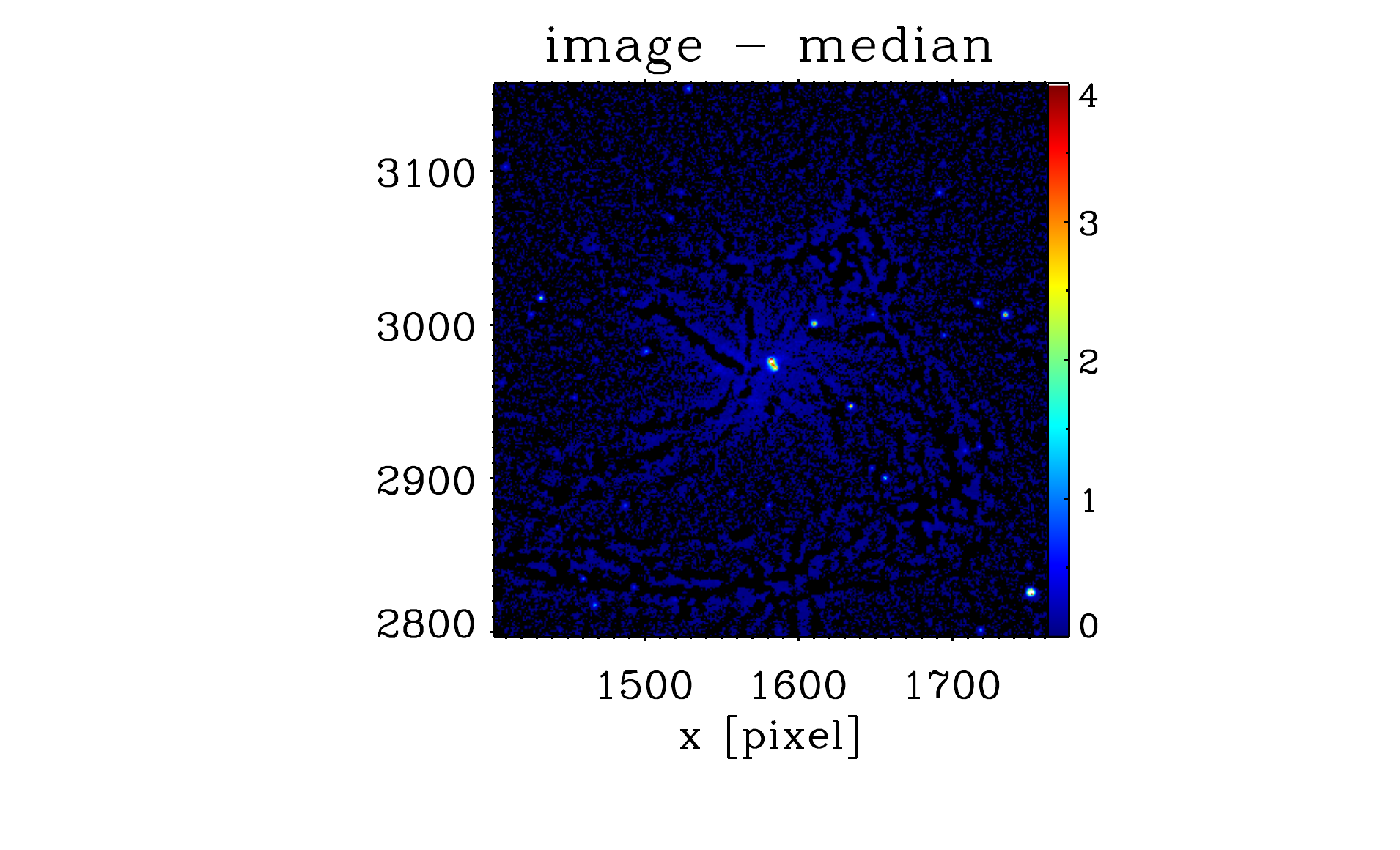} \\
\includegraphics[trim=0.7cm 0cm 0cm 0cm, clip=true, scale=0.46]{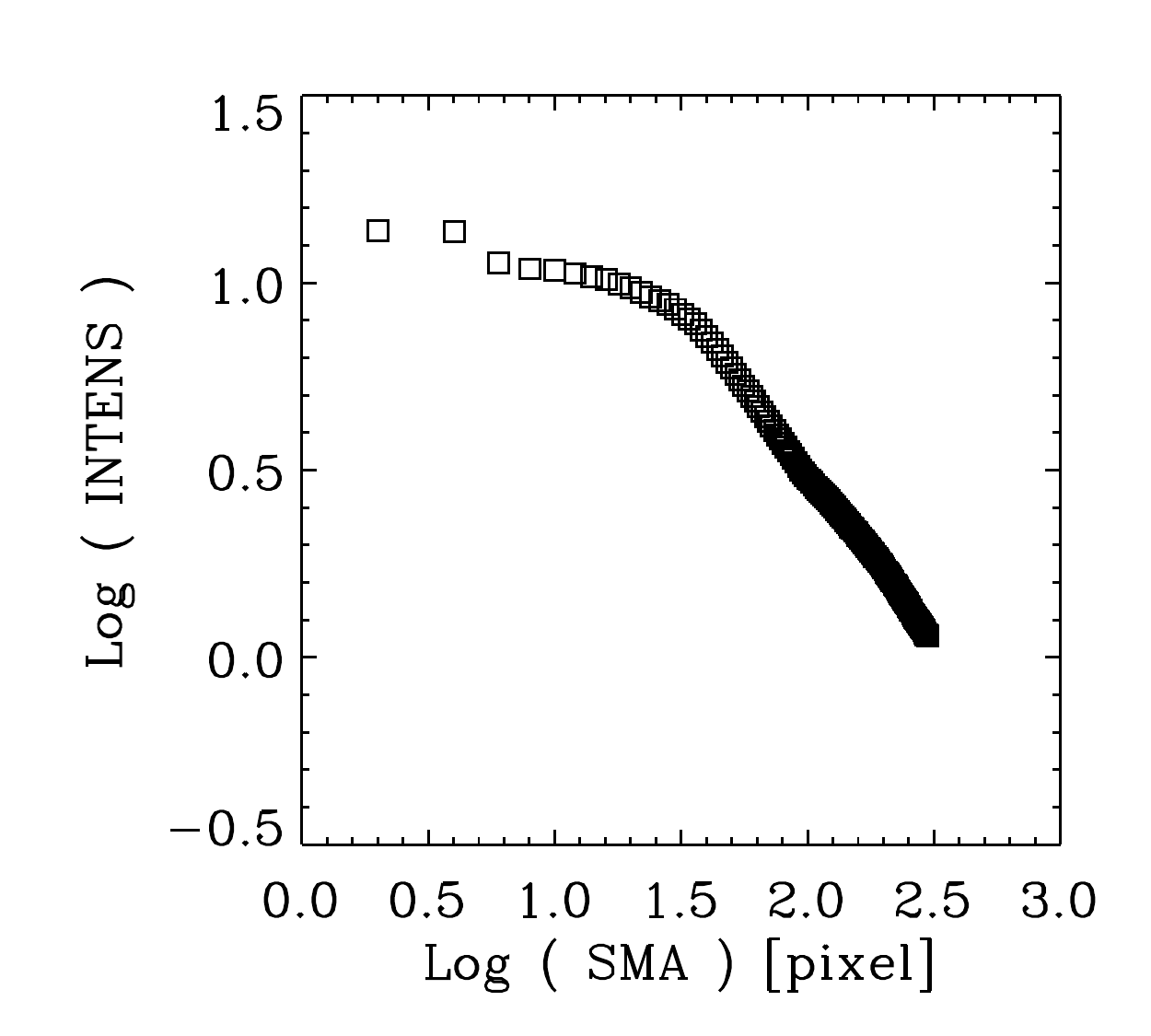}	    &  \includegraphics[trim=0.6cm 0cm 0cm 0cm, clip=true, scale=0.46]{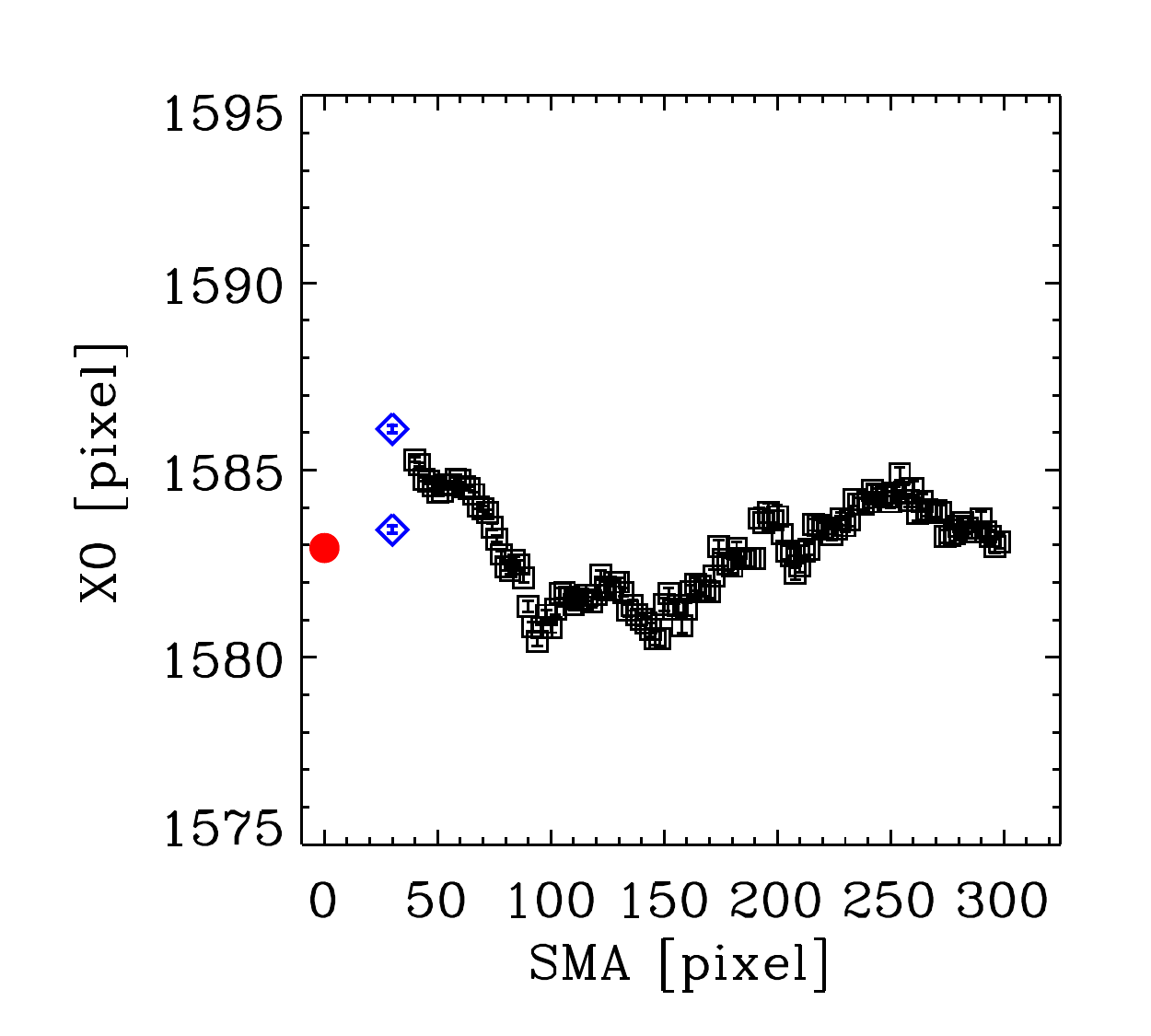}  &  \includegraphics[trim=0.6cm 0cm 0cm 0cm, clip=true, scale=0.46]{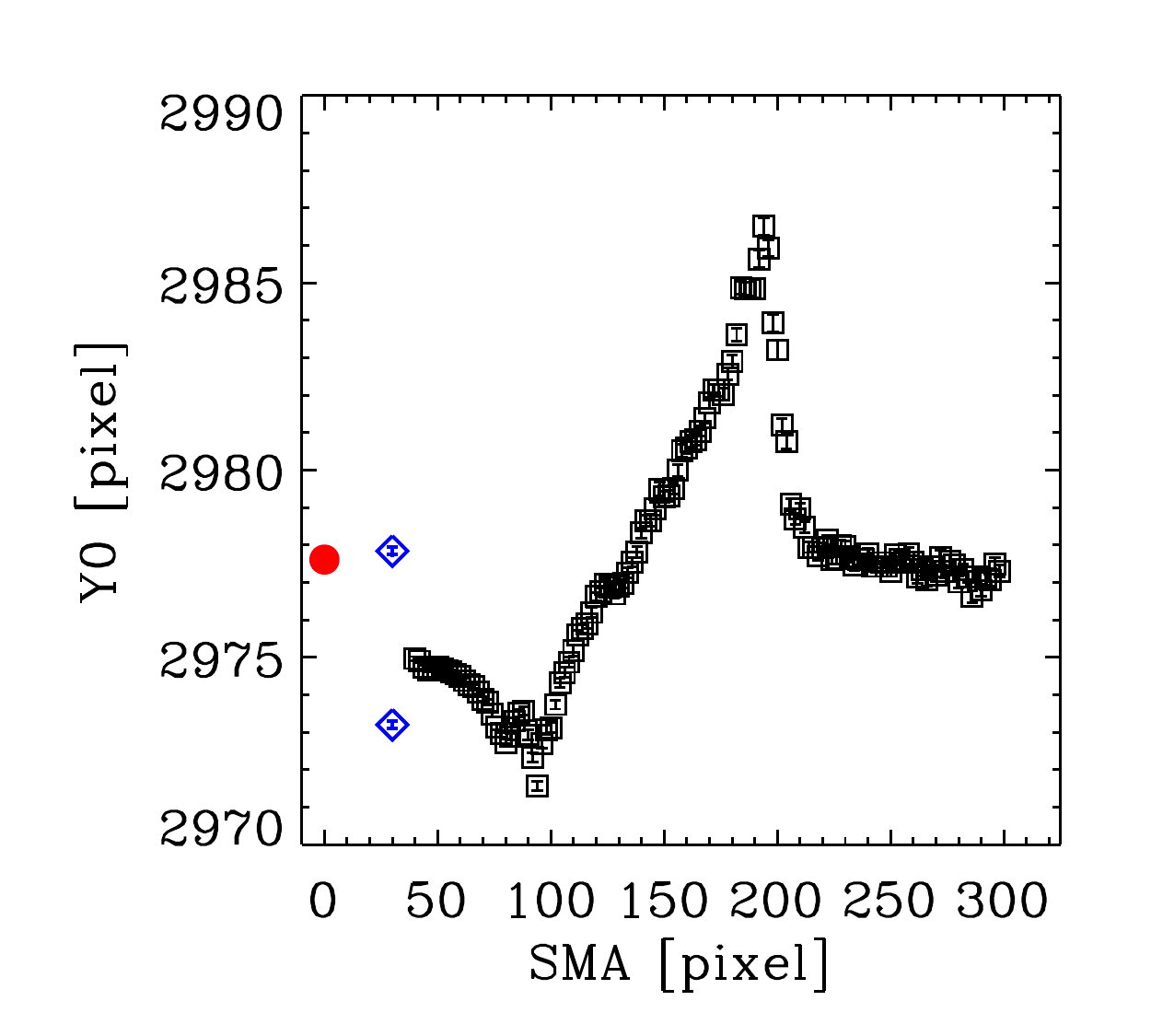} \\	
 \includegraphics[trim=0.65cm 0cm 0cm 0cm, clip=true, scale=0.46]{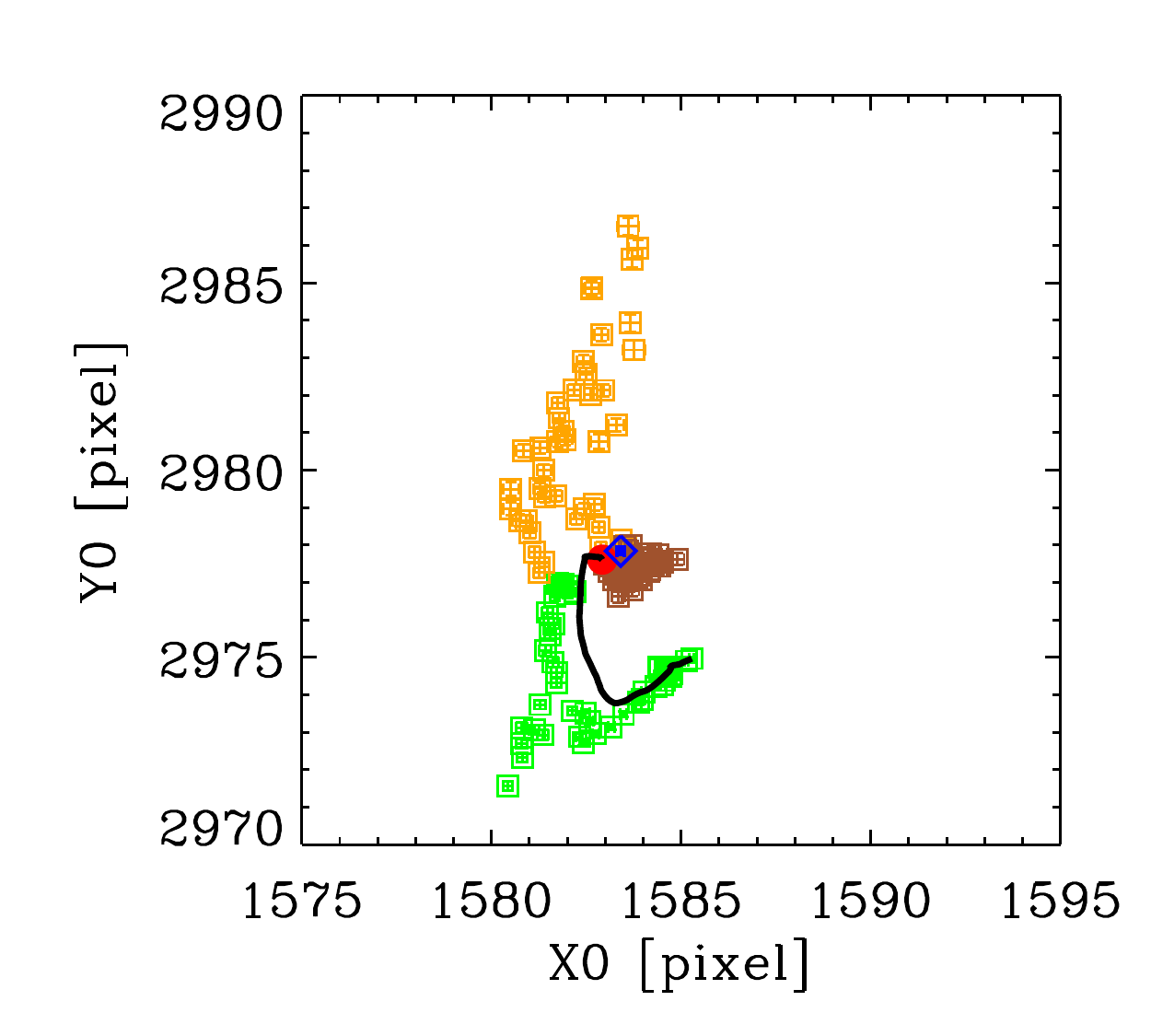}	&  \includegraphics[trim=0.6cm 0cm 0cm 0cm, clip=true, scale=0.46]{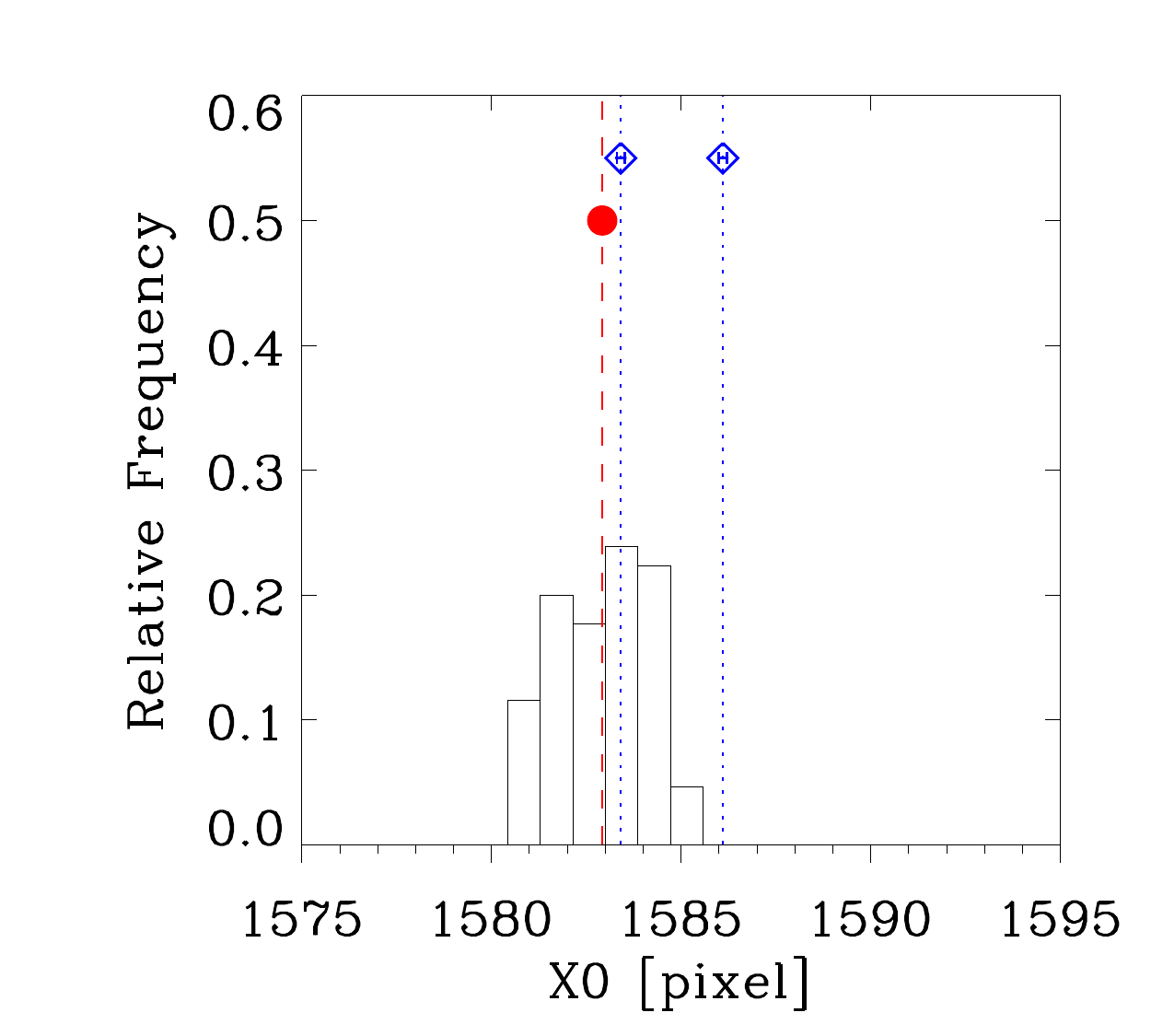}	& \includegraphics[trim=0.6cm 0cm 0cm 0cm, clip=true, scale=0.46]{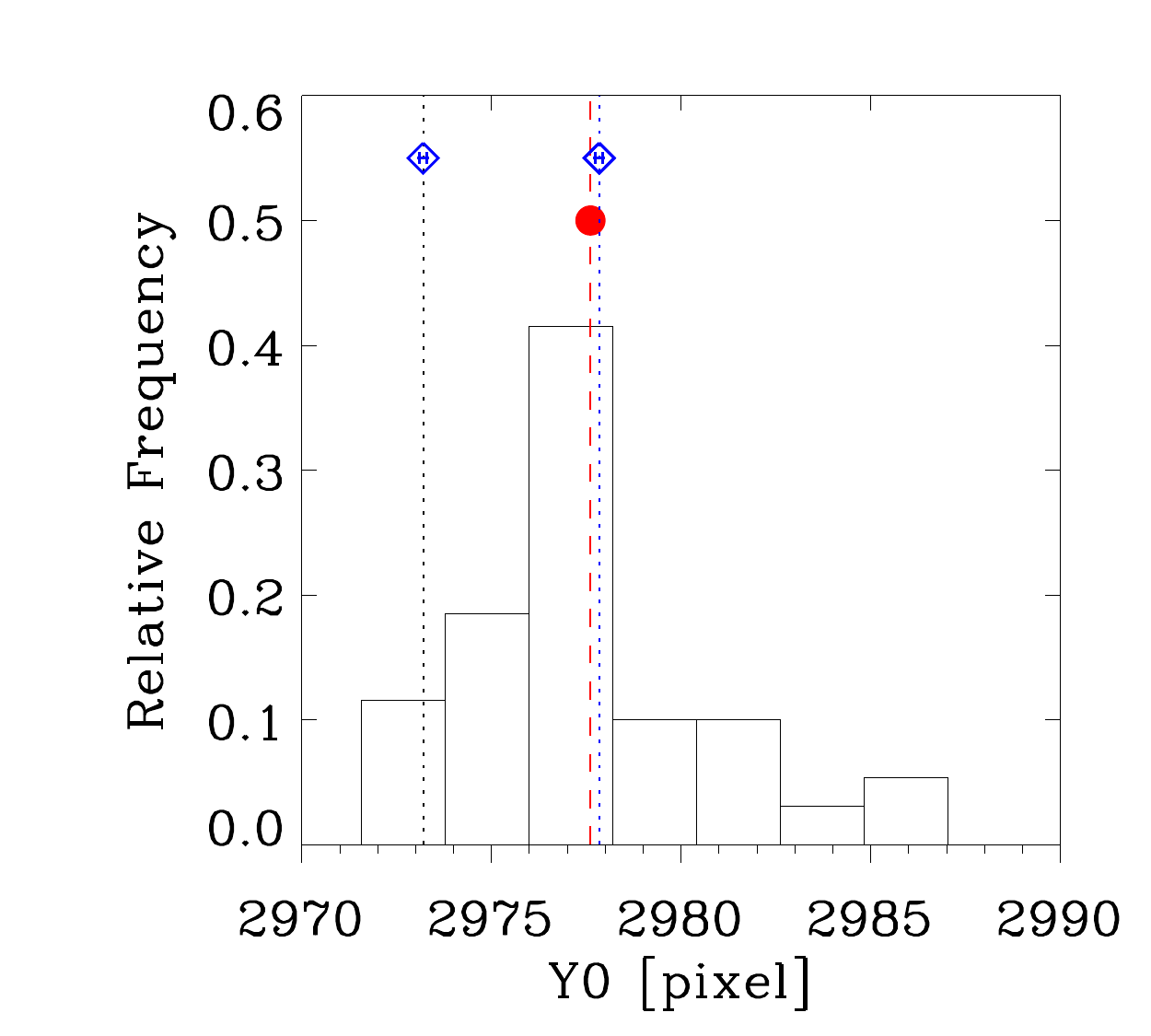}\\
\end{array}$
\end{center}
\caption[NGC 4696 (ACS_F814W)]{As in Fig.\ref{fig: NGC4373_W2} for galaxy NGC 4696, ACS/WFC - F814W, scale=$0\farcs05$/pxl.}
\label{fig: NGC4696_ACS}
\end{figure*}  

\begin{figure*}[h]
\begin{center}$
\begin{array}{ccc}
\includegraphics[trim=3.75cm 1cm 3cm 0cm, clip=true, scale=0.48]{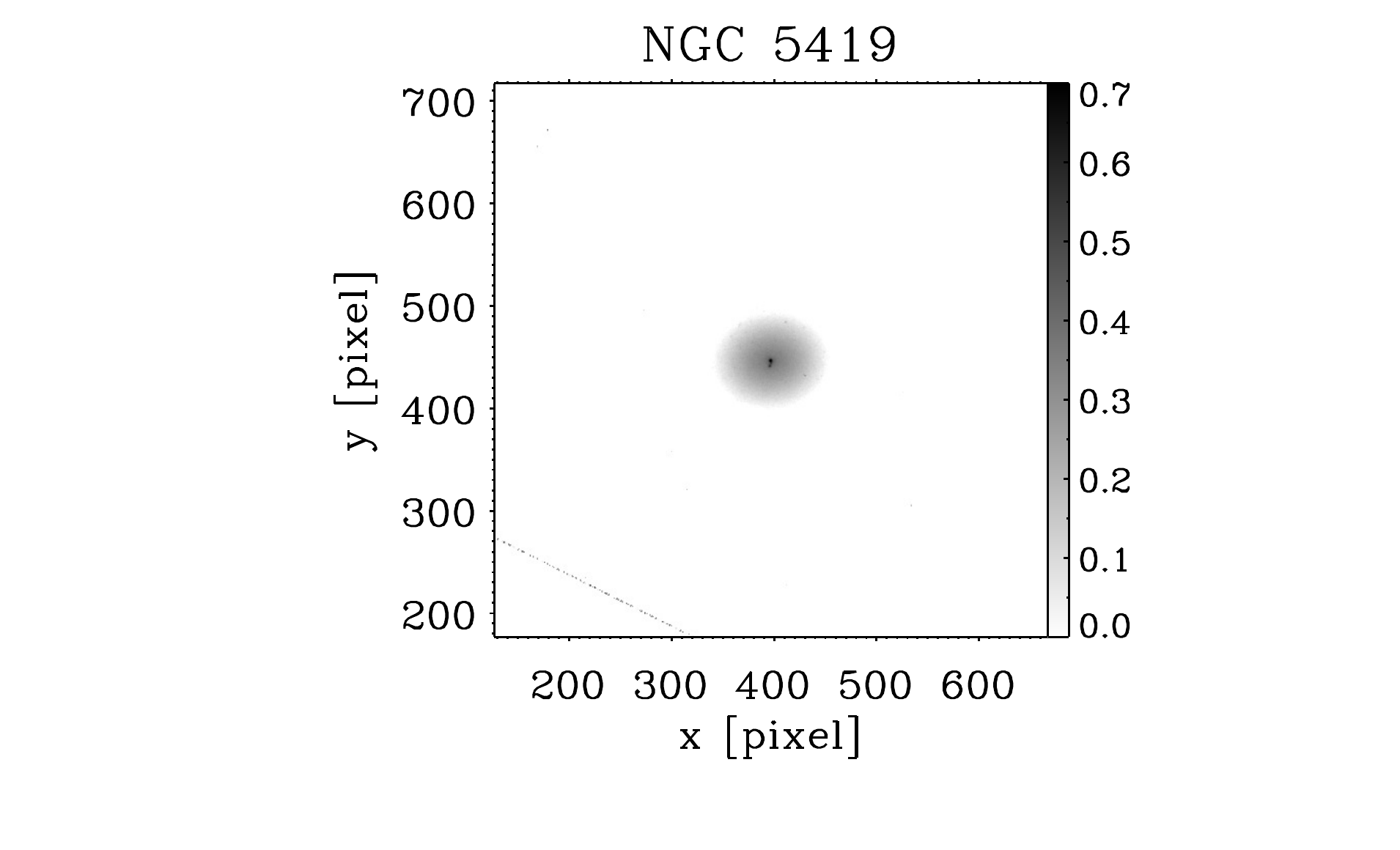} & \includegraphics[trim= 4.cm 1cm 3cm 0cm, clip=true, scale=0.48]{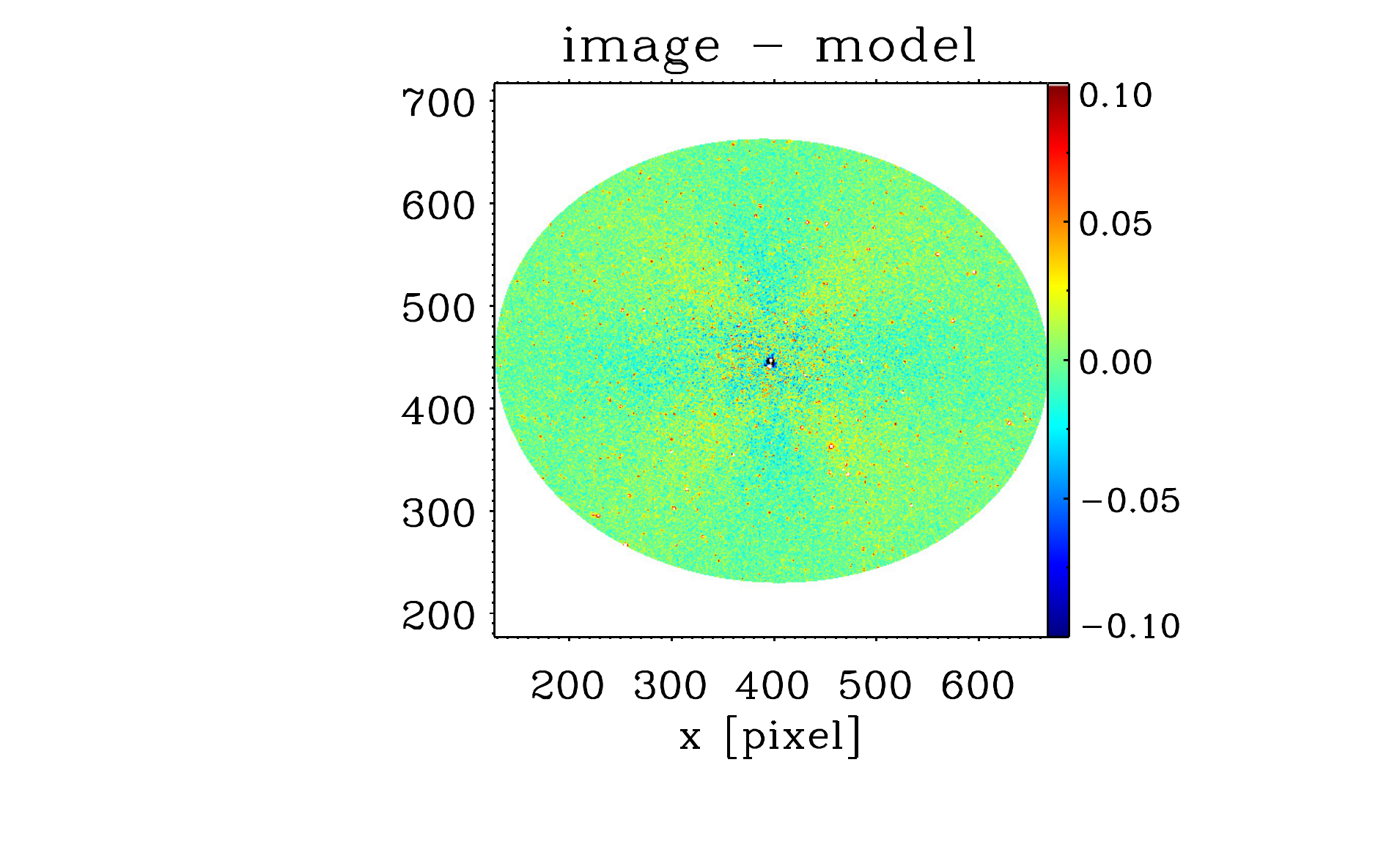}	& \includegraphics[trim= 4.cm 1cm 3cm 0cm, clip=true, scale=0.48]{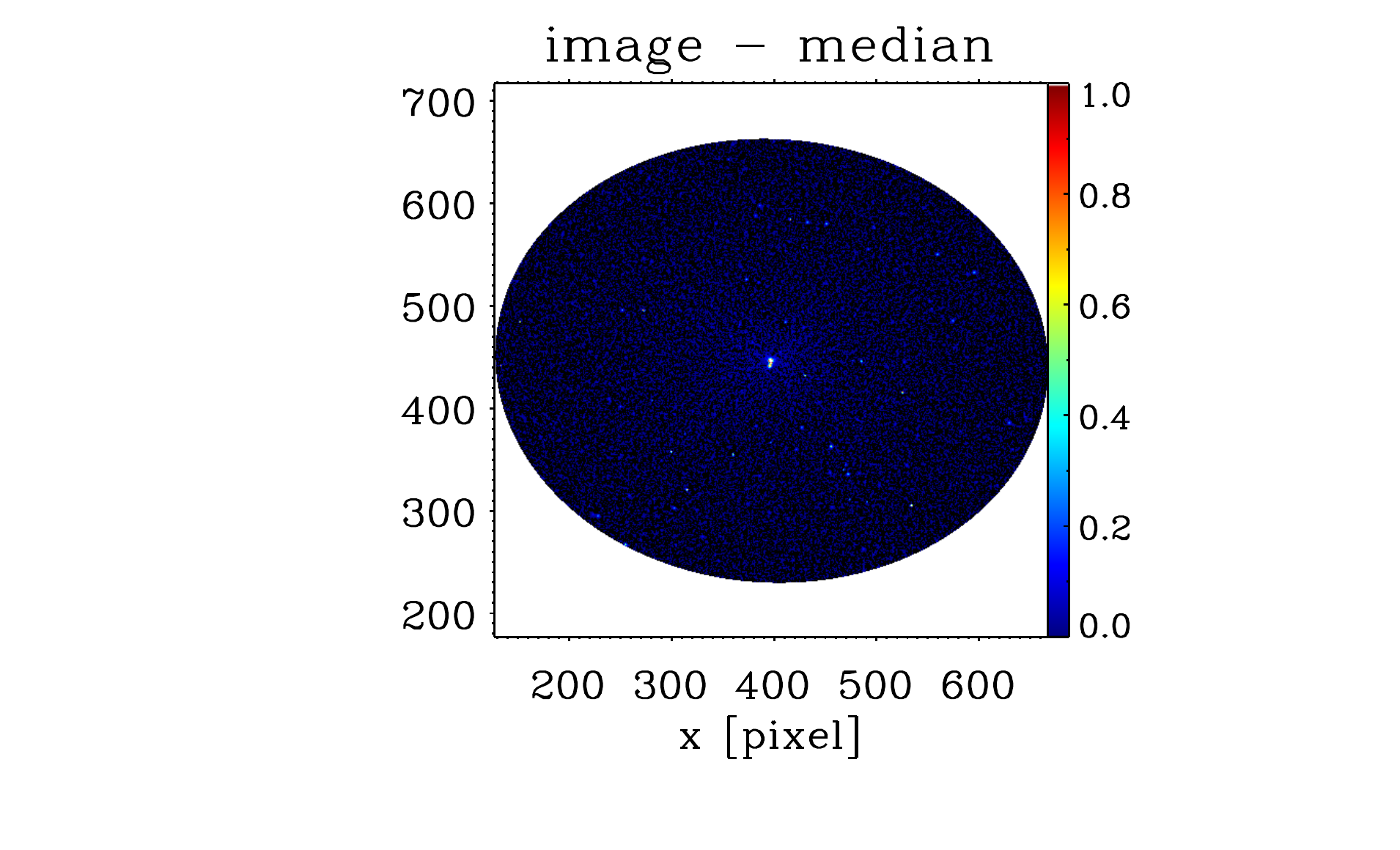} \\
\includegraphics[trim=0.7cm 0cm 0cm 0cm, clip=true, scale=0.46]{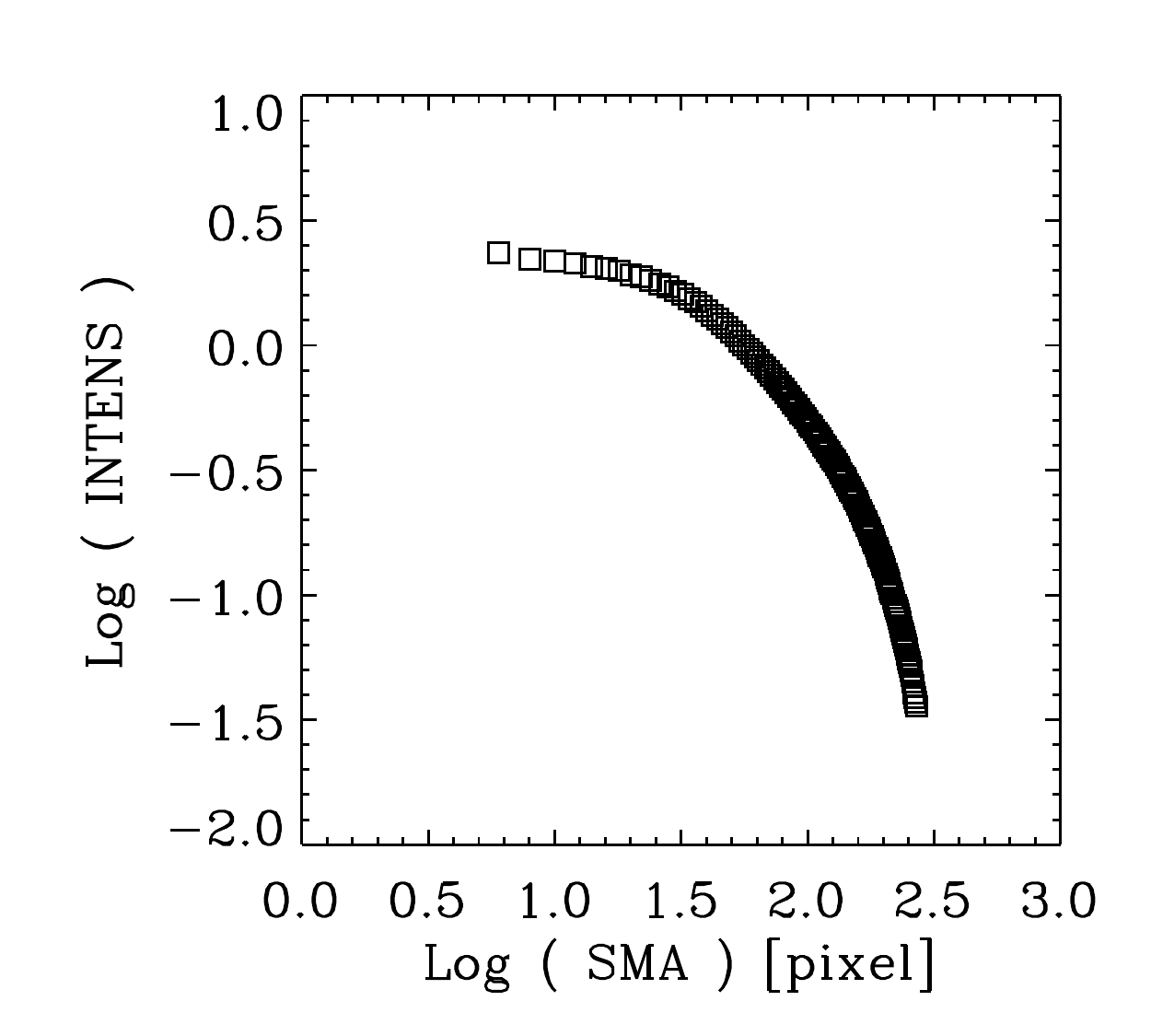}	    &  \includegraphics[trim=0.6cm 0cm 0cm 0cm, clip=true, scale=0.46]{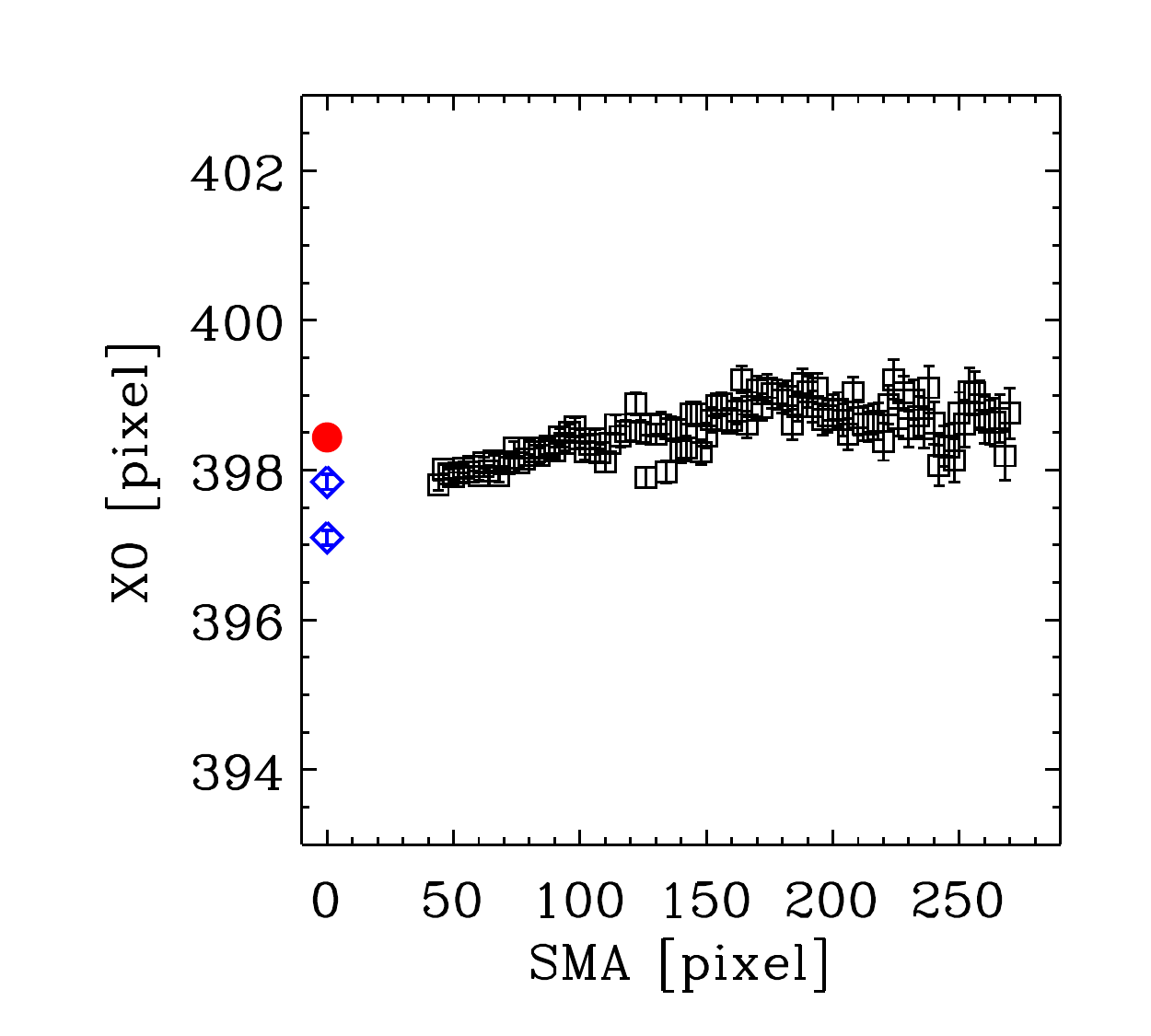}  &  \includegraphics[trim=0.6cm 0cm 0cm 0cm, clip=true, scale=0.46]{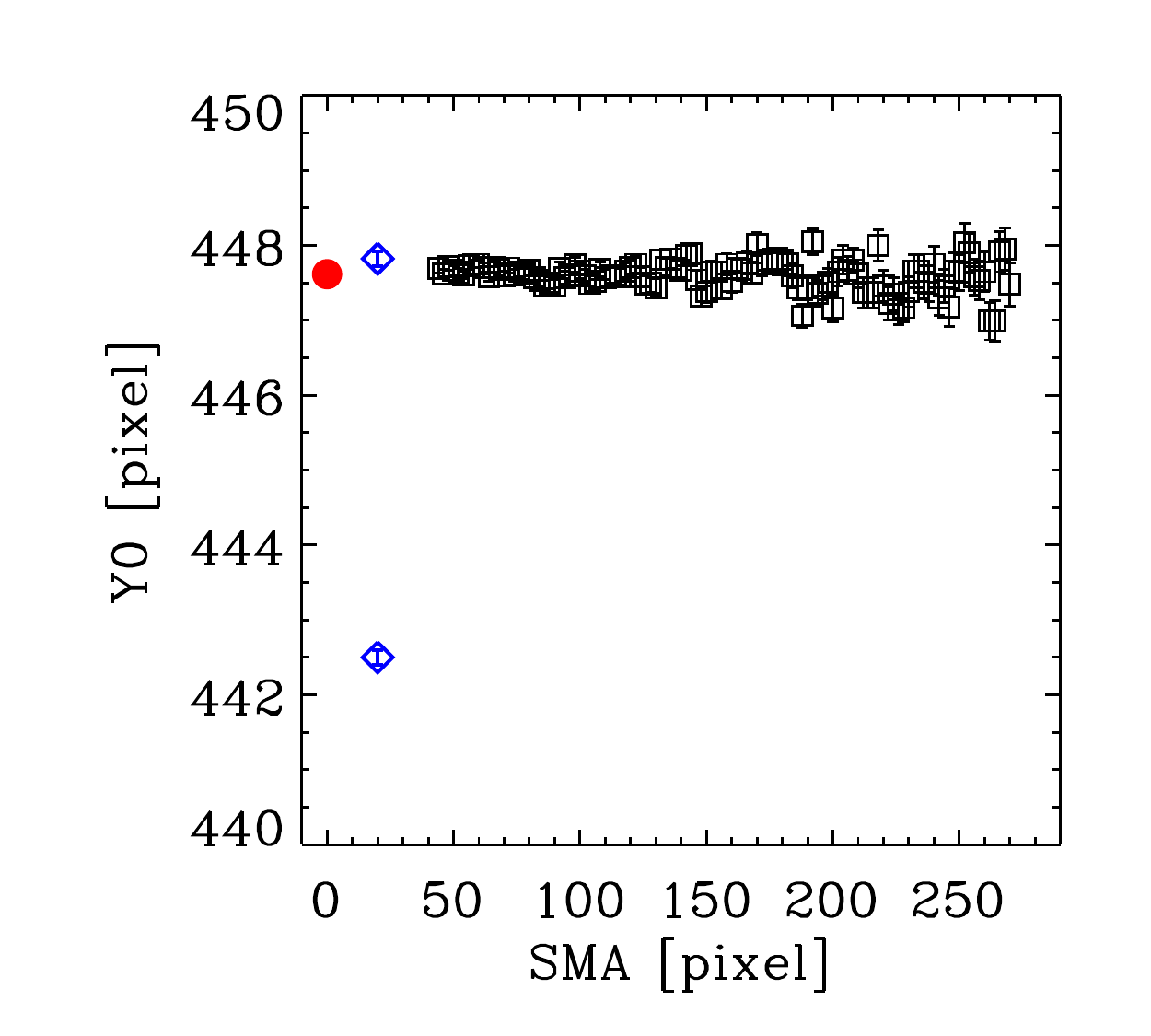} \\	
 \includegraphics[trim=0.65cm 0cm 0cm 0cm, clip=true, scale=0.46]{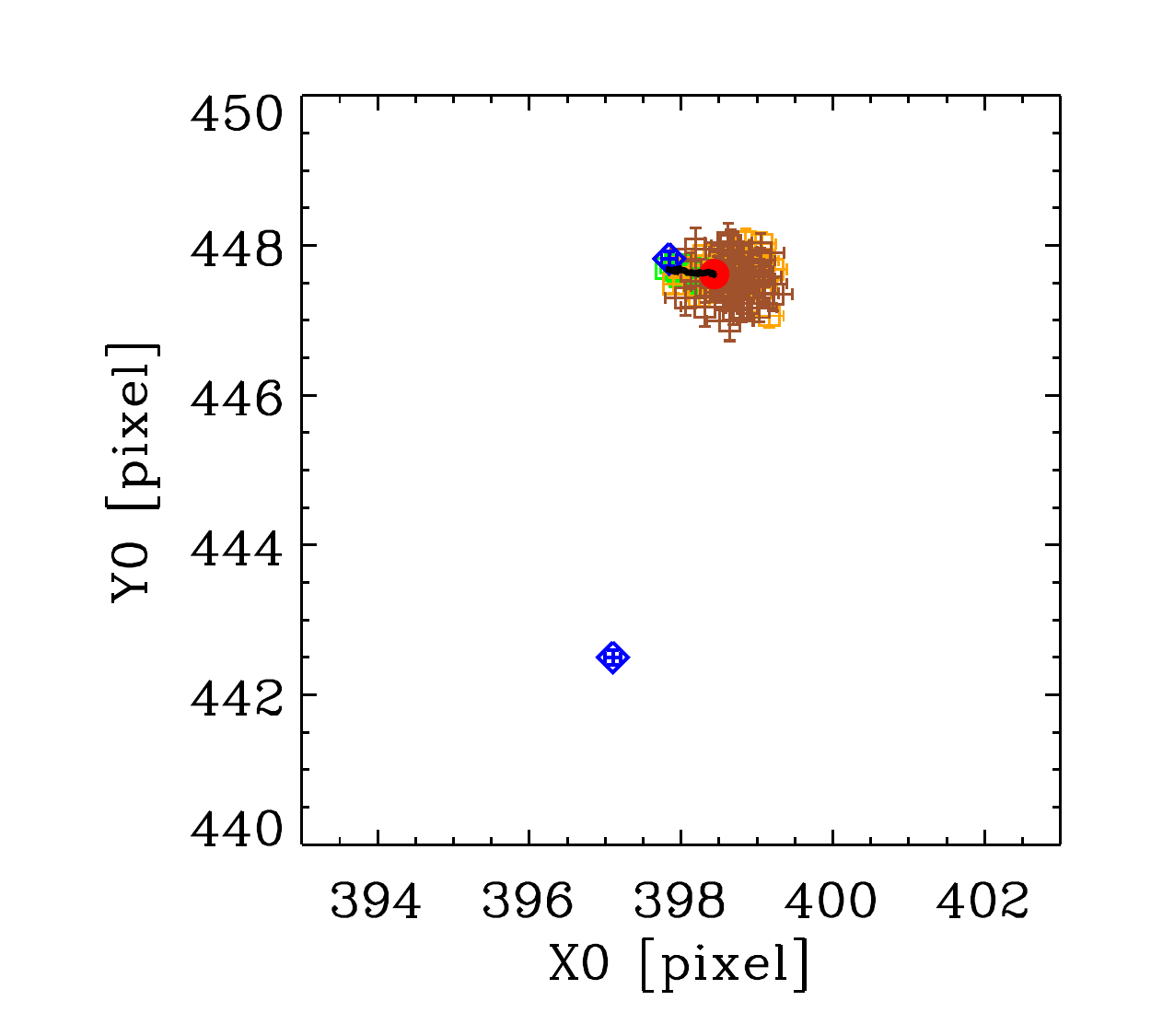}	&  \includegraphics[trim=0.6cm 0cm 0cm 0cm, clip=true, scale=0.46]{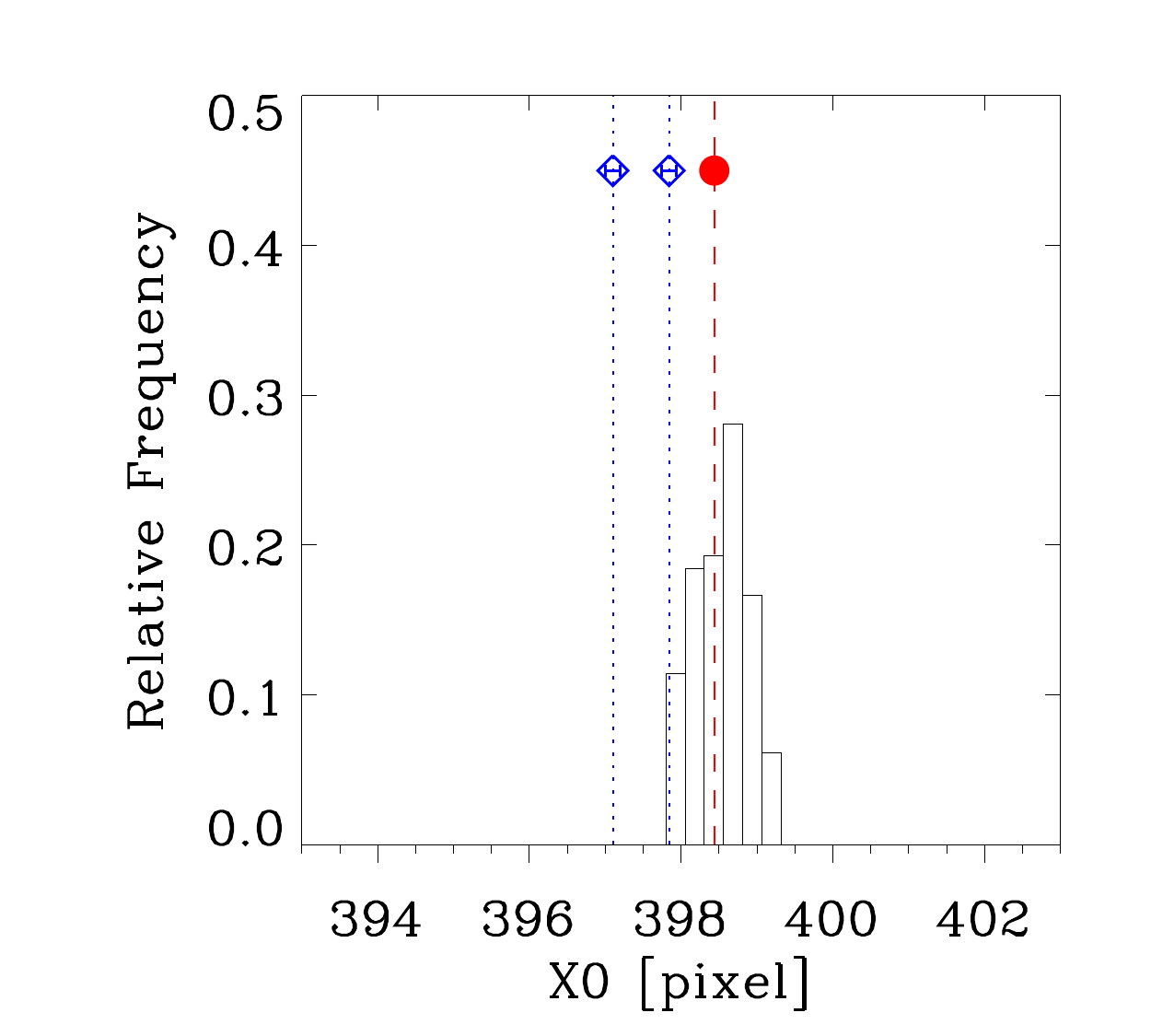}	& \includegraphics[trim=0.6cm 0cm 0cm 0cm, clip=true, scale=0.46]{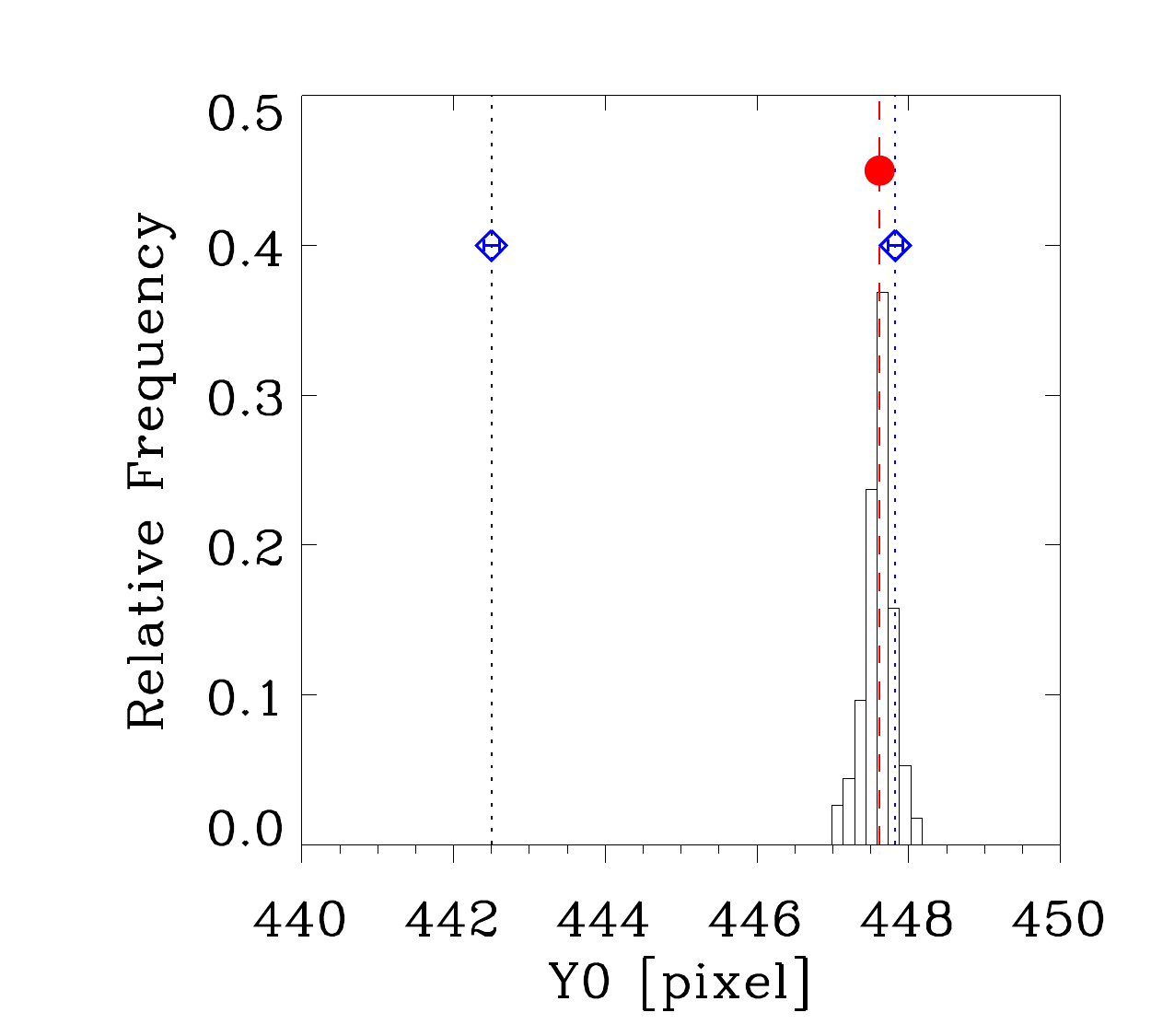}\\
\end{array}$
\end{center}
\caption[NGC5419 (WFPC2WF)]{As in Fig.\ref{fig: NGC4373_W2} for galaxy NGC 5419, WFPC2/PC - F555W, scale=$0\farcs05$/pxl.}
\label{fig: NGC5419_WF}
\end{figure*} 

\begin{figure*}[h]
\begin{center}$
\begin{array}{ccc}
\includegraphics[trim=3.75cm 1cm 3cm 0cm, clip=true, scale=0.48]{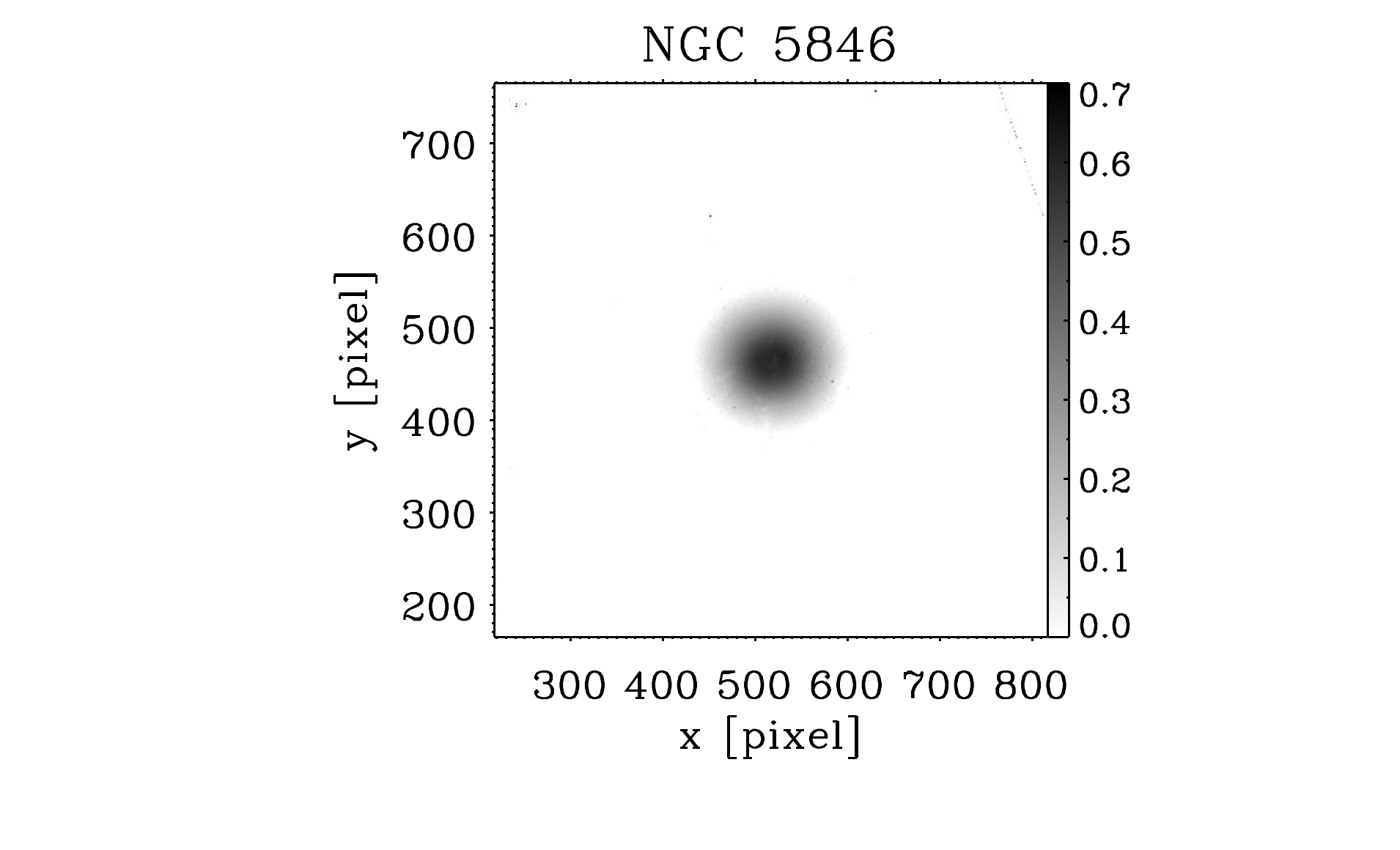} & \includegraphics[trim= 4.cm 1cm 3cm 0cm, clip=true, scale=0.48]{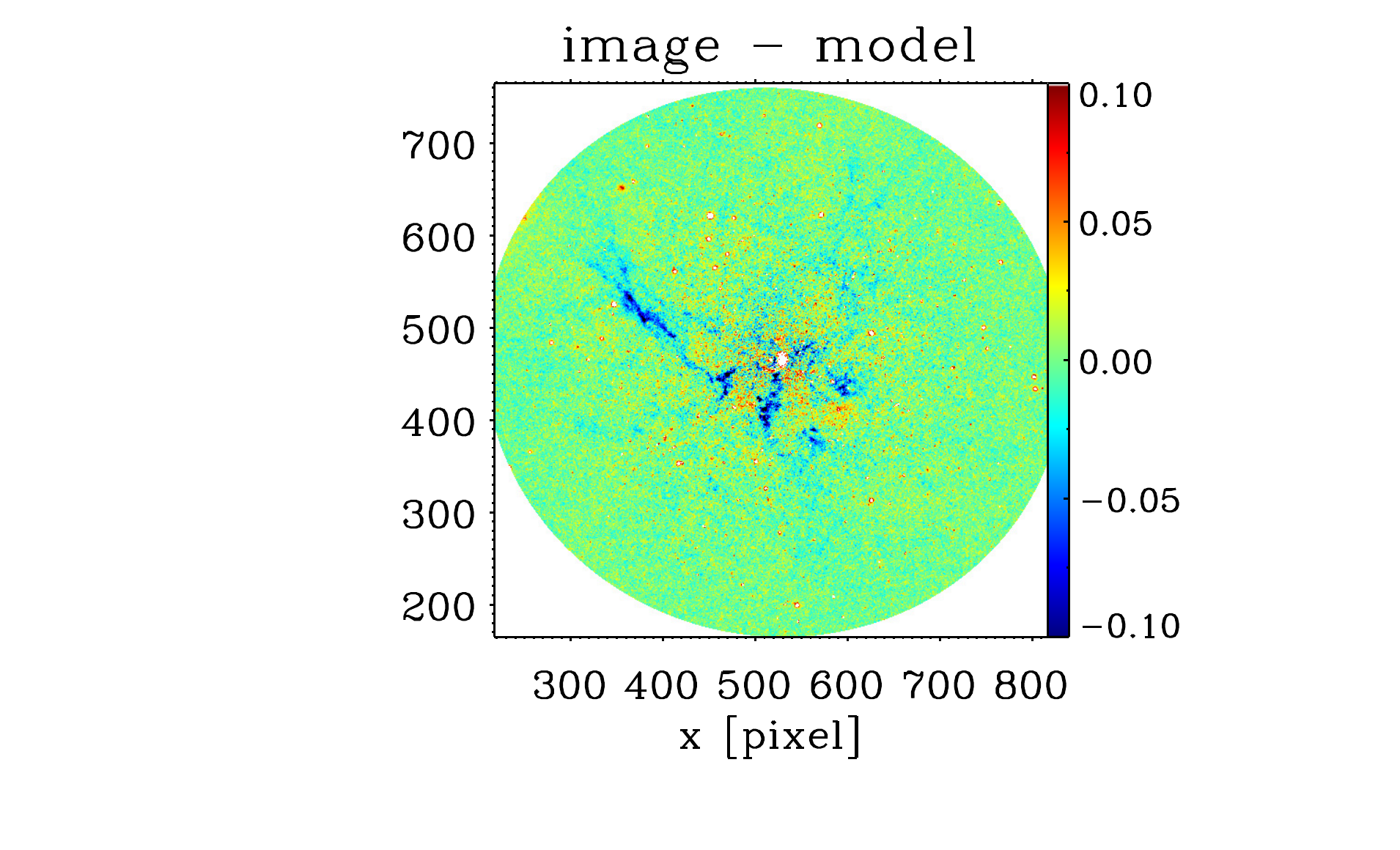}	& \includegraphics[trim= 4.cm 1cm 3cm 0cm, clip=true, scale=0.48]{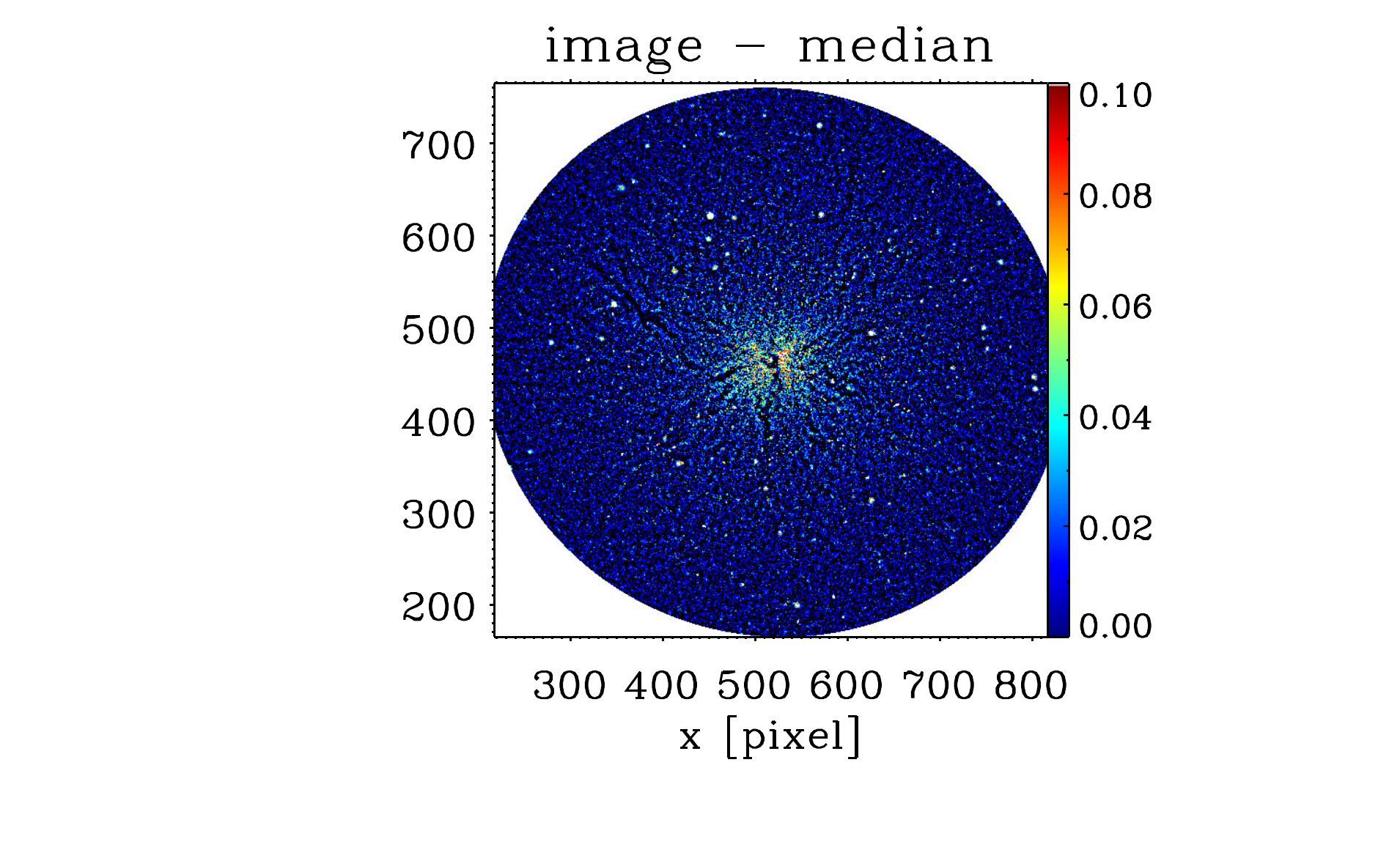} \\
\includegraphics[trim=0.7cm 0cm 0cm 0cm, clip=true, scale=0.46]{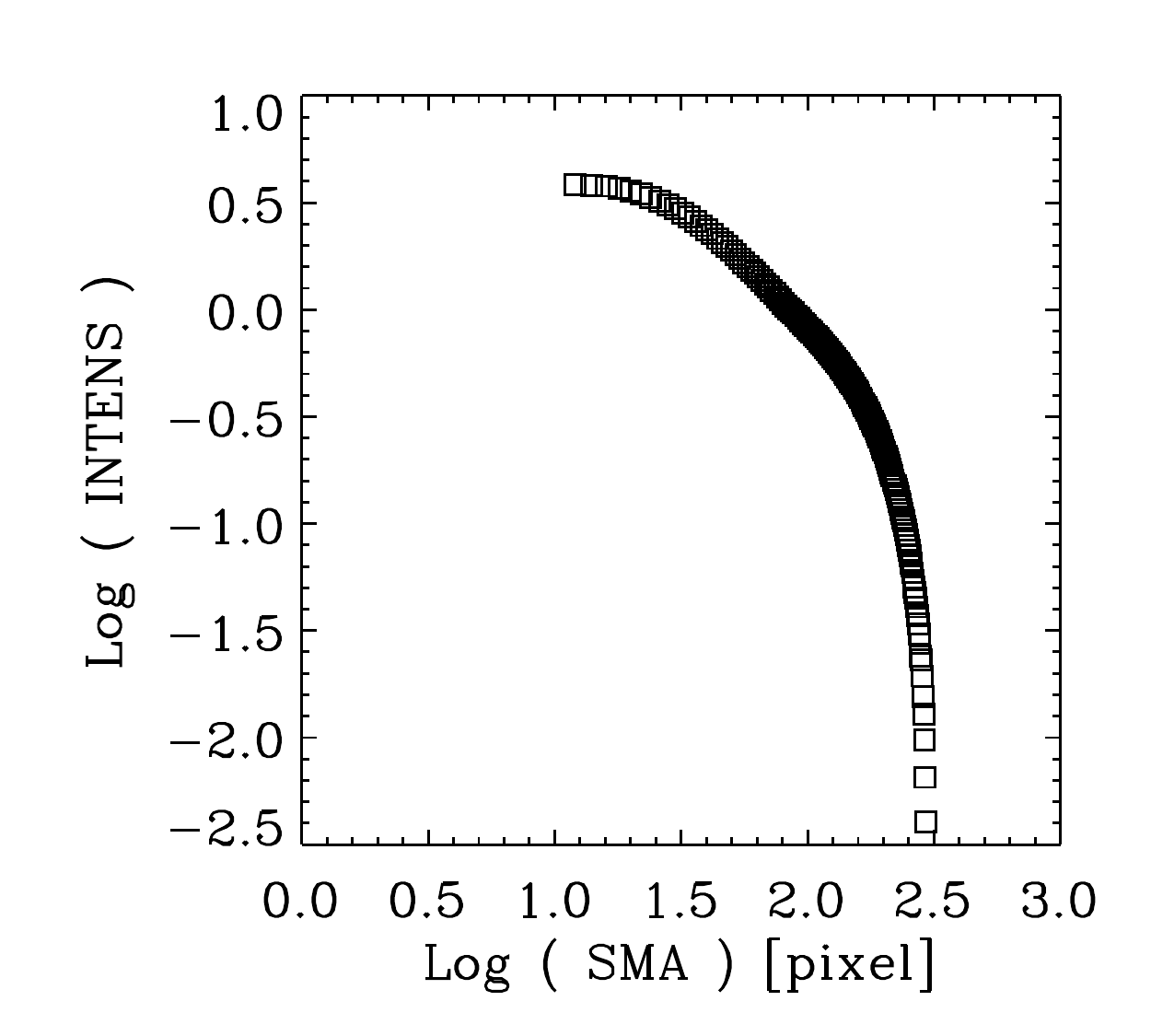}	    &  \includegraphics[trim=0.6cm 0cm 0cm 0cm, clip=true, scale=0.46]{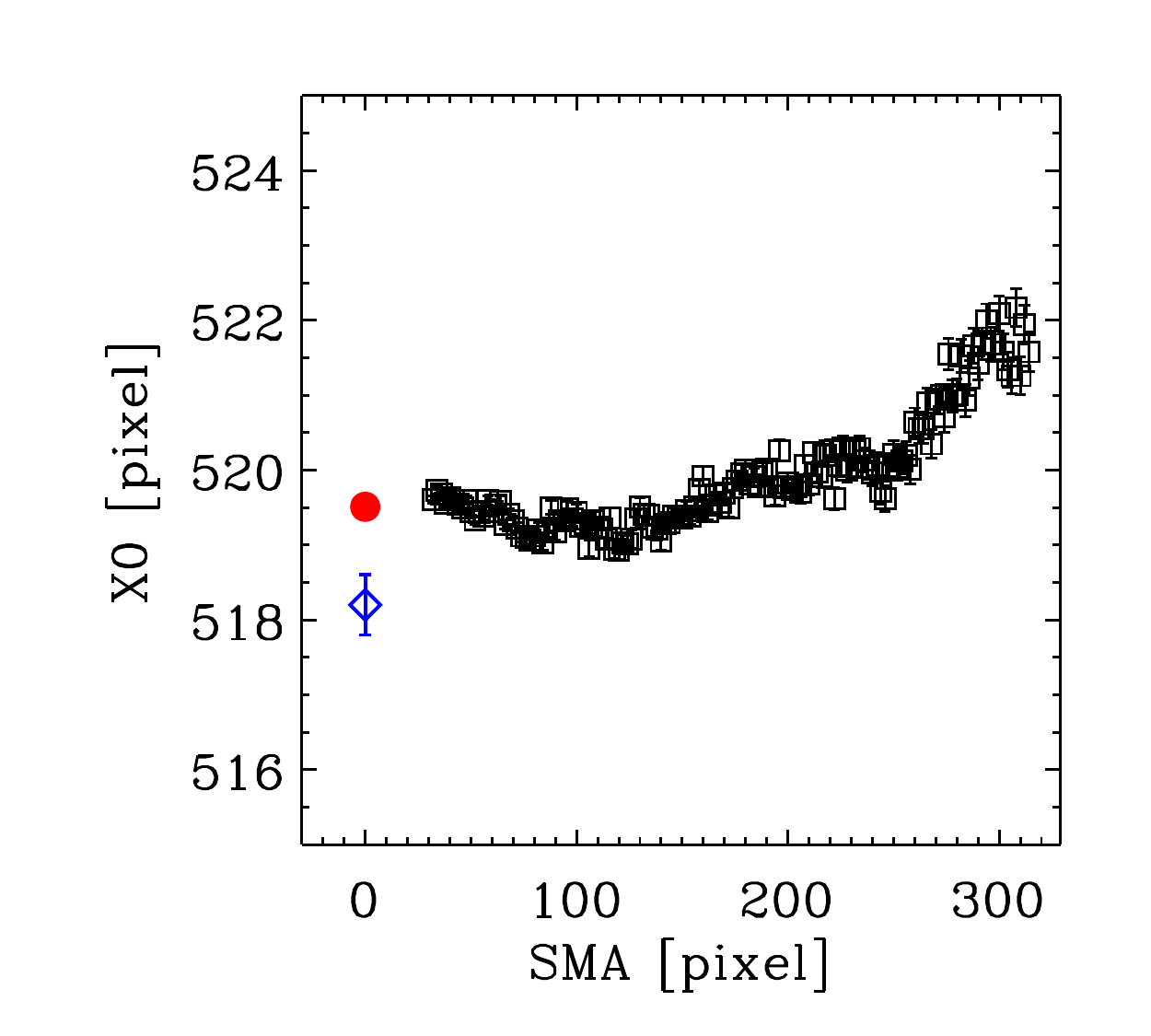}  &  \includegraphics[trim=0.6cm 0cm 0cm 0cm, clip=true, scale=0.46]{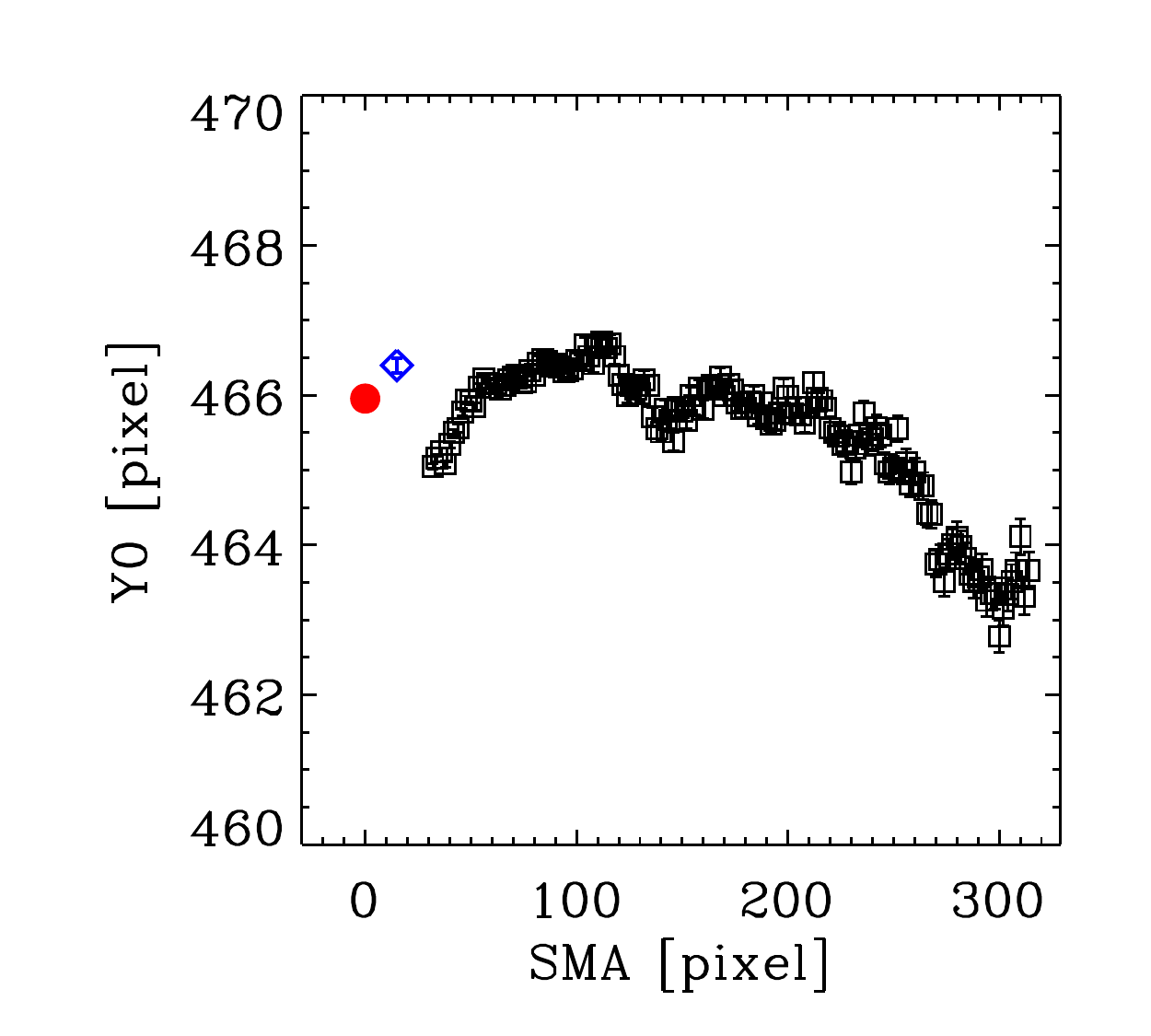} \\	
 \includegraphics[trim=0.65cm 0cm 0cm 0cm, clip=true, scale=0.46]{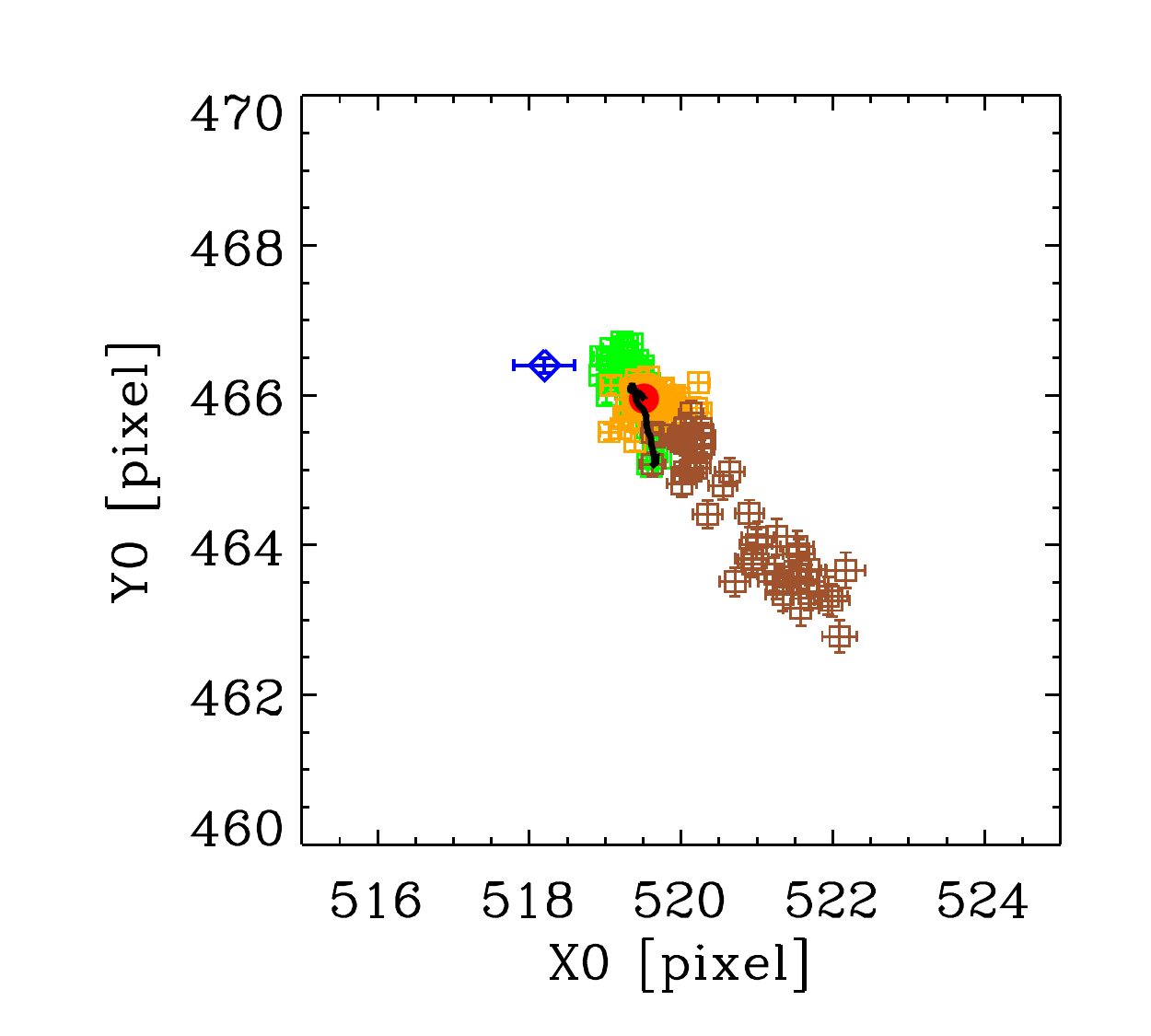}	&  \includegraphics[trim=0.6cm 0cm 0cm 0cm, clip=true, scale=0.46]{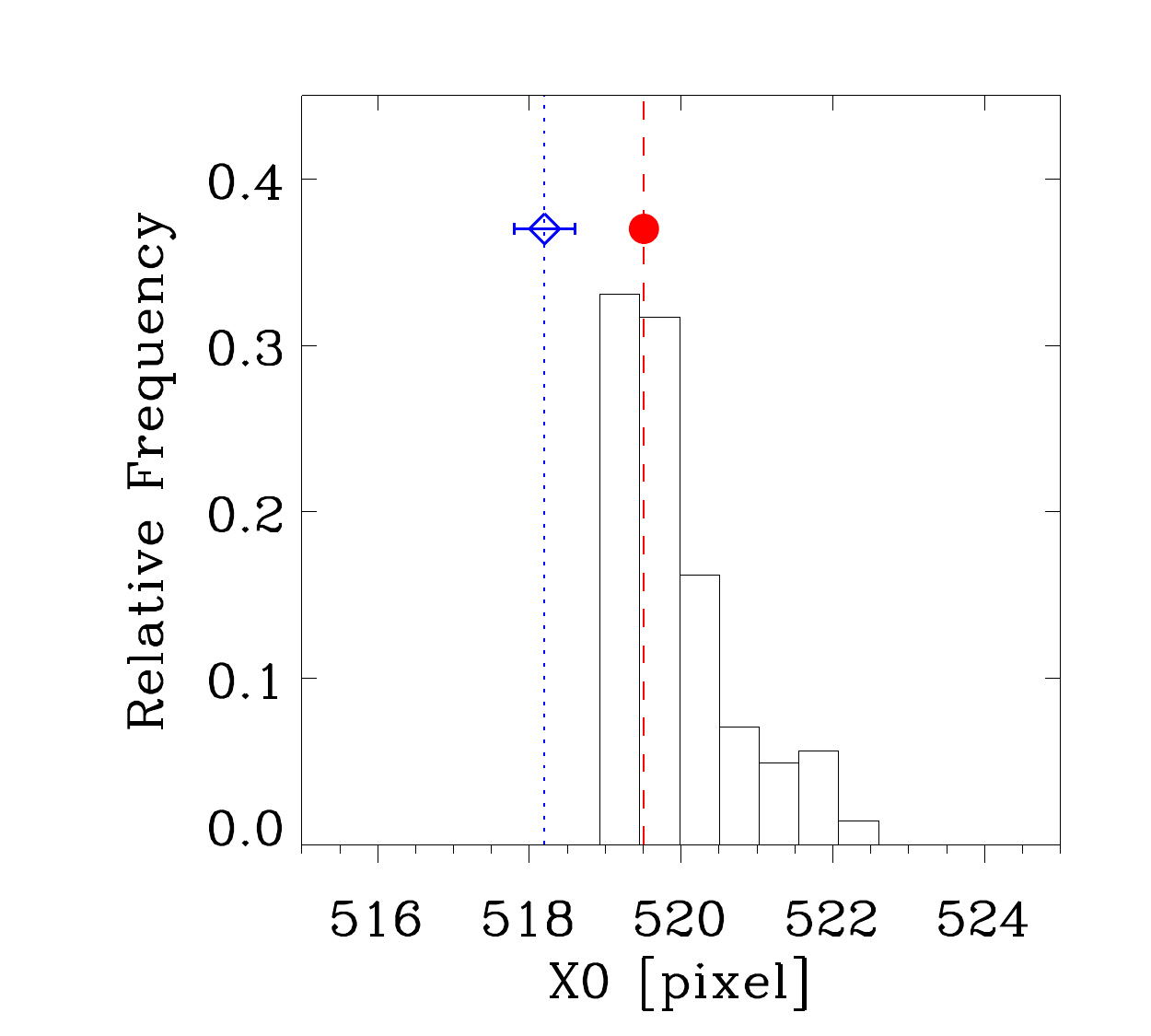}	& \includegraphics[trim=0.6cm 0cm 0cm 0cm, clip=true, scale=0.46]{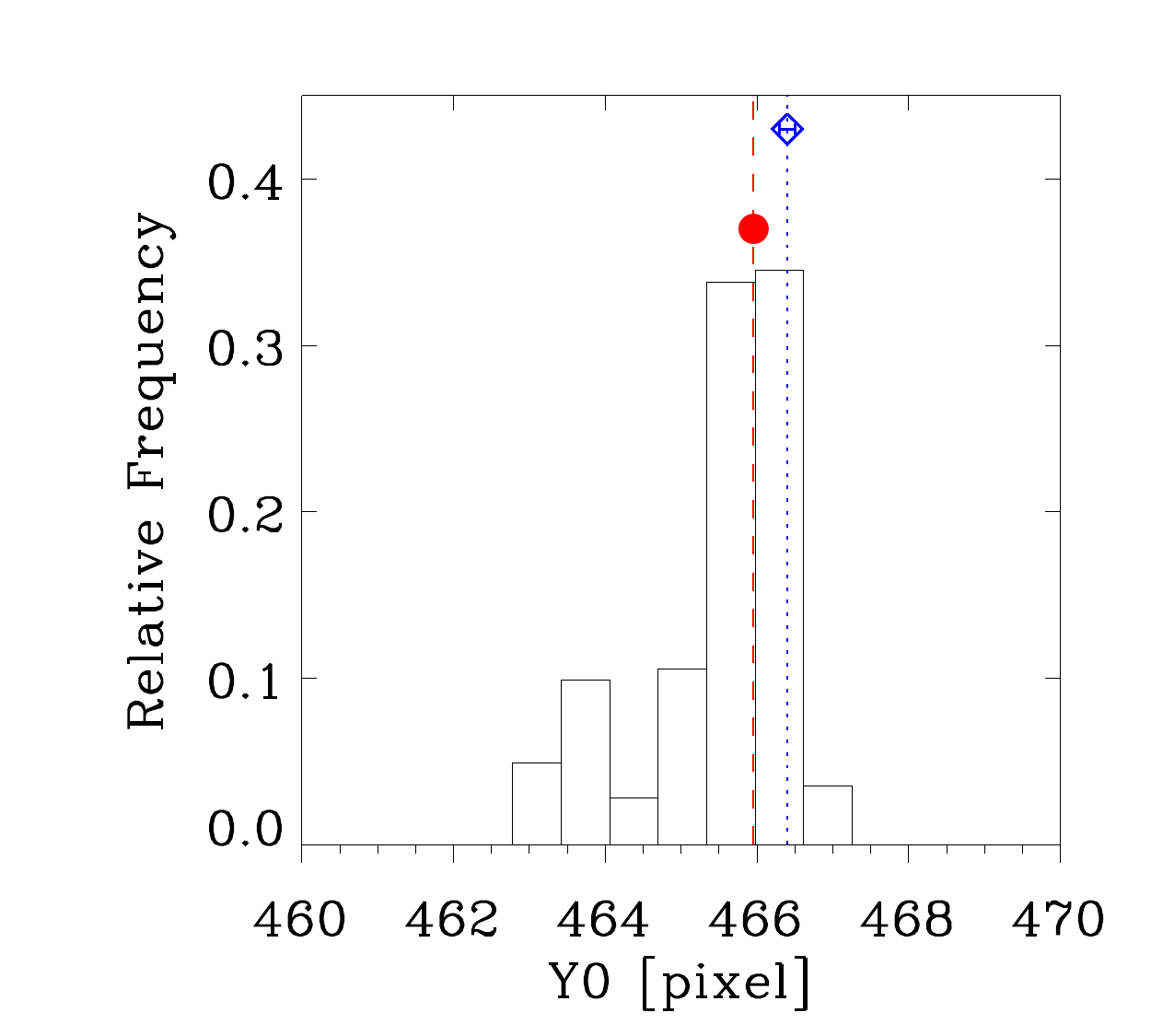}\\
\end{array}$
\end{center}
\caption[NGC 5846 (WFPC2_F814W)]{As in Fig.\ref{fig: NGC4373_W2} for galaxy NGC 5846, WFPC2/PC - F814W, scale=$0\farcs05$/pxl.}
\label{fig: NGC5846_WFPC2F814W}
\end{figure*} 

\begin{figure*}[h]
\begin{center}$
\begin{array}{ccc}
\includegraphics[trim=3.75cm 1cm 3cm 0cm, clip=true, scale=0.48]{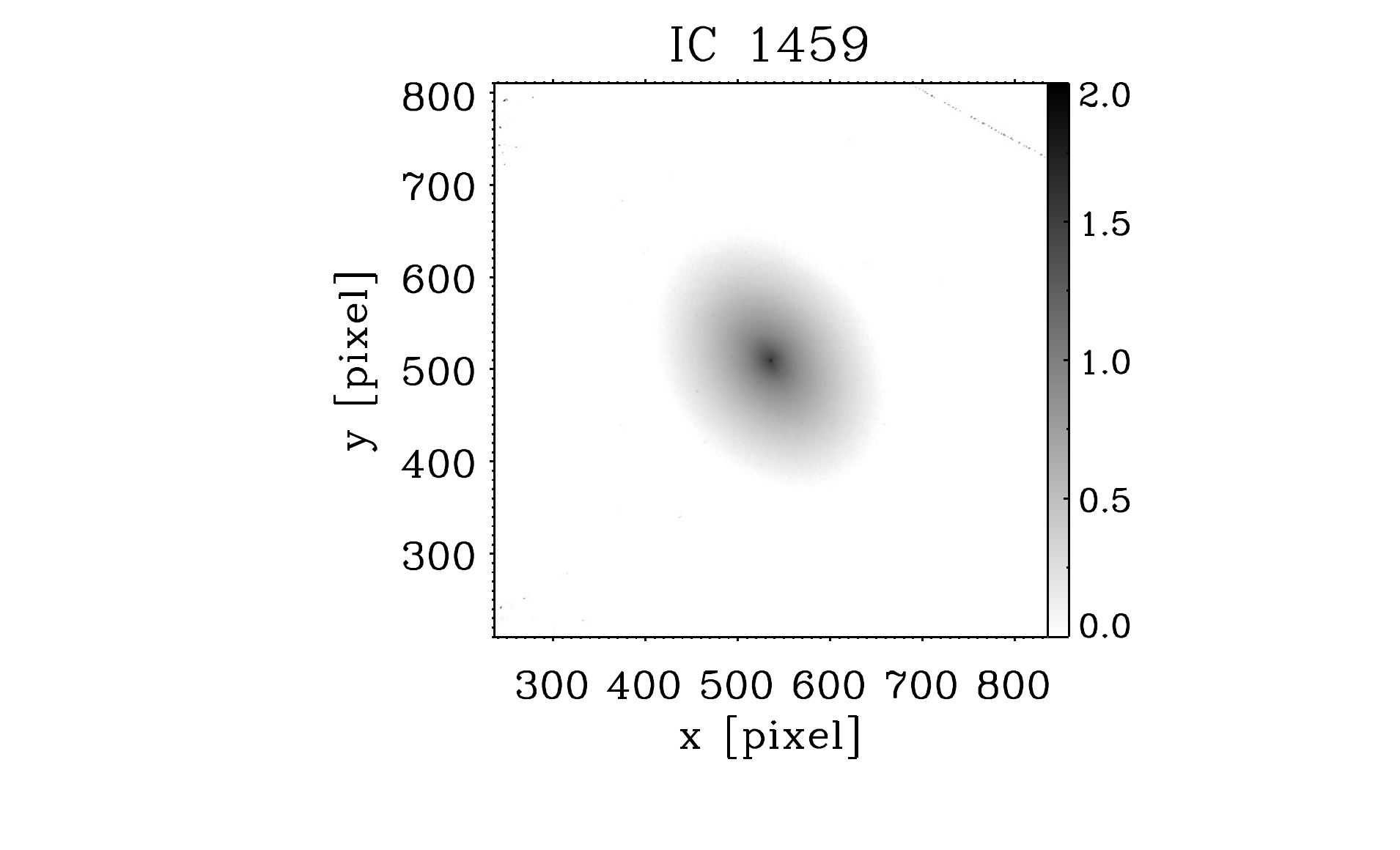} & \includegraphics[trim= 4.cm 1cm 3cm 0cm, clip=true, scale=0.48]{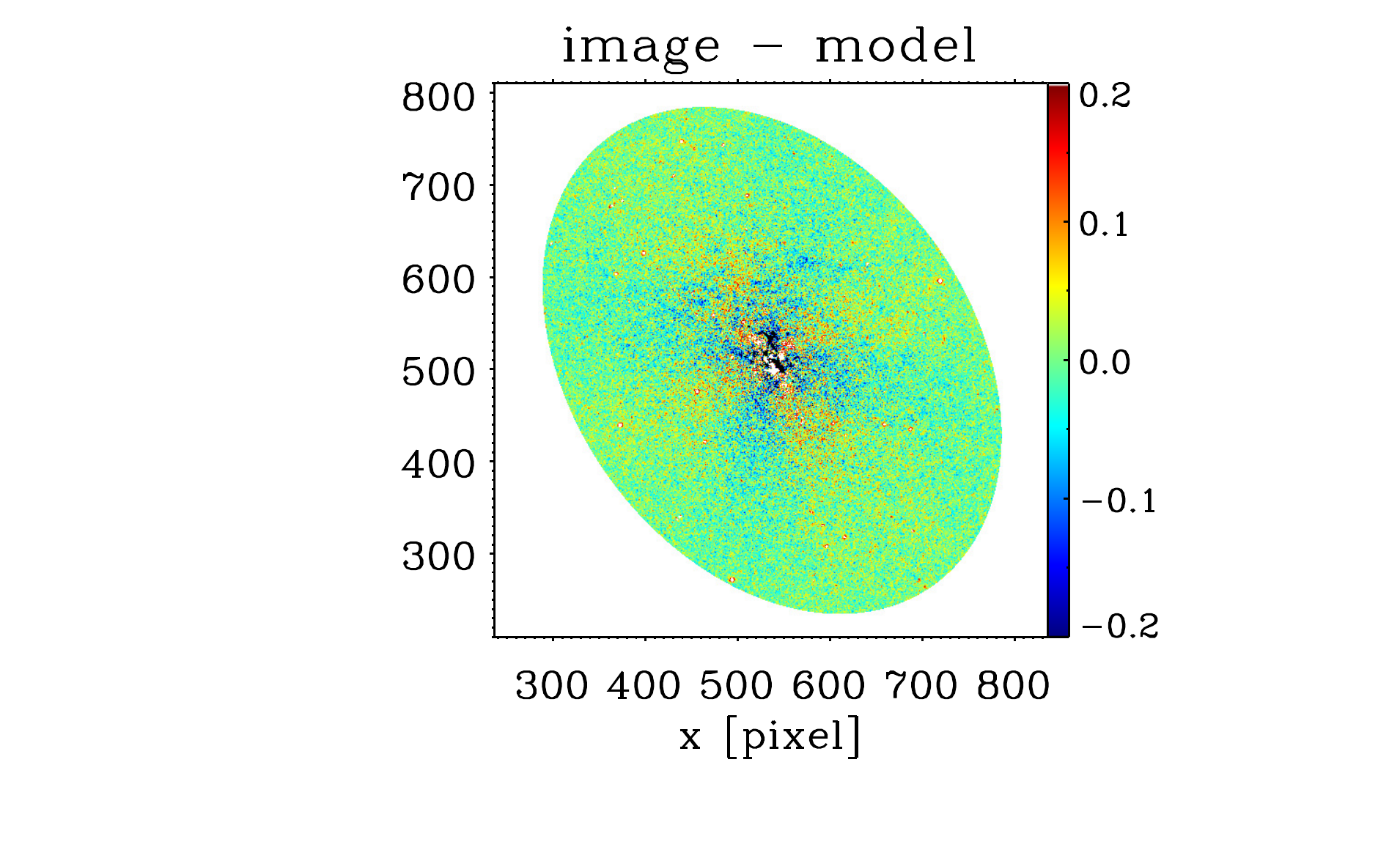}	& \includegraphics[trim= 4.cm 1cm 3cm 0cm, clip=true, scale=0.48]{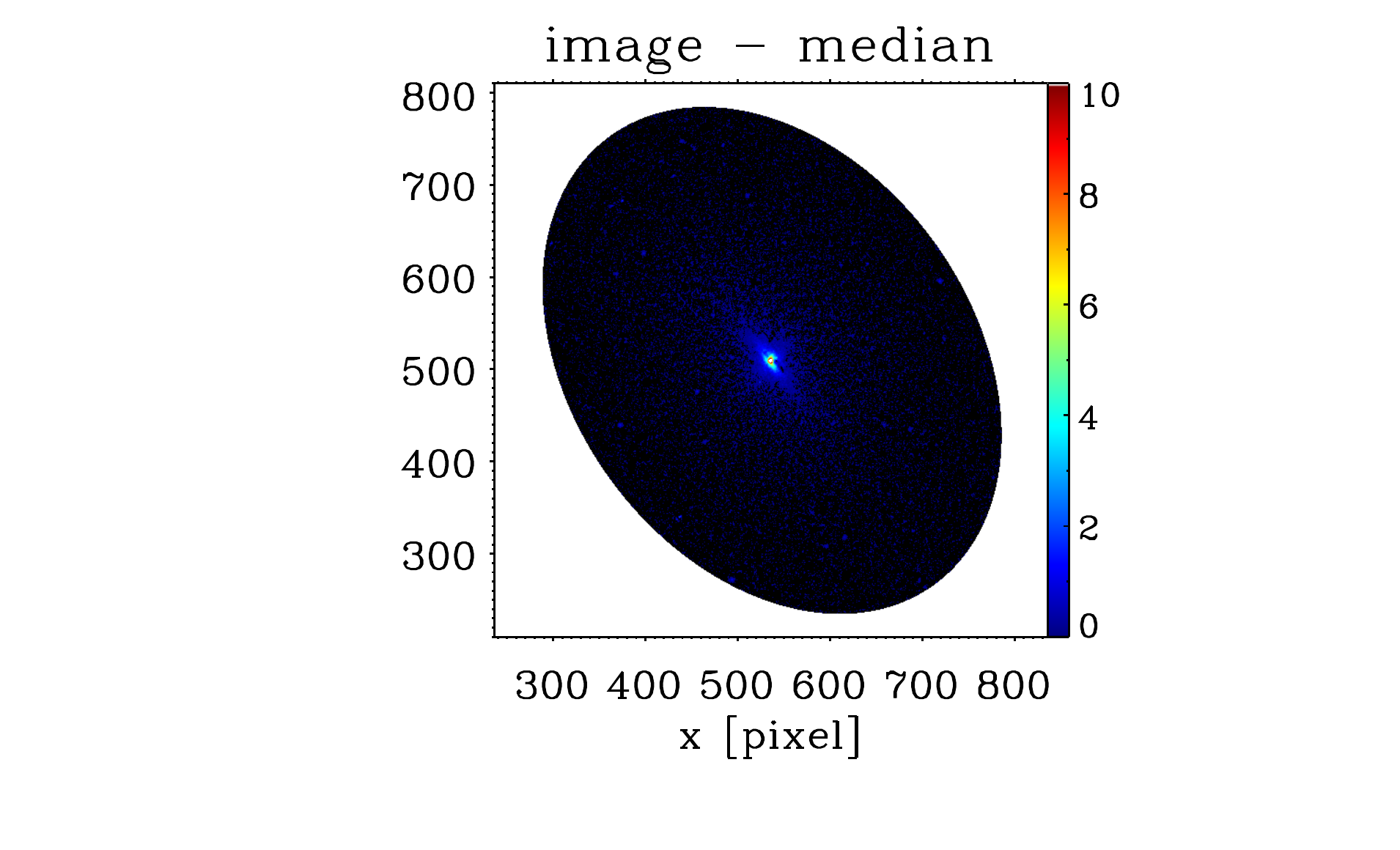} \\
\includegraphics[trim=0.7cm 0cm 0cm 0cm, clip=true, scale=0.46]{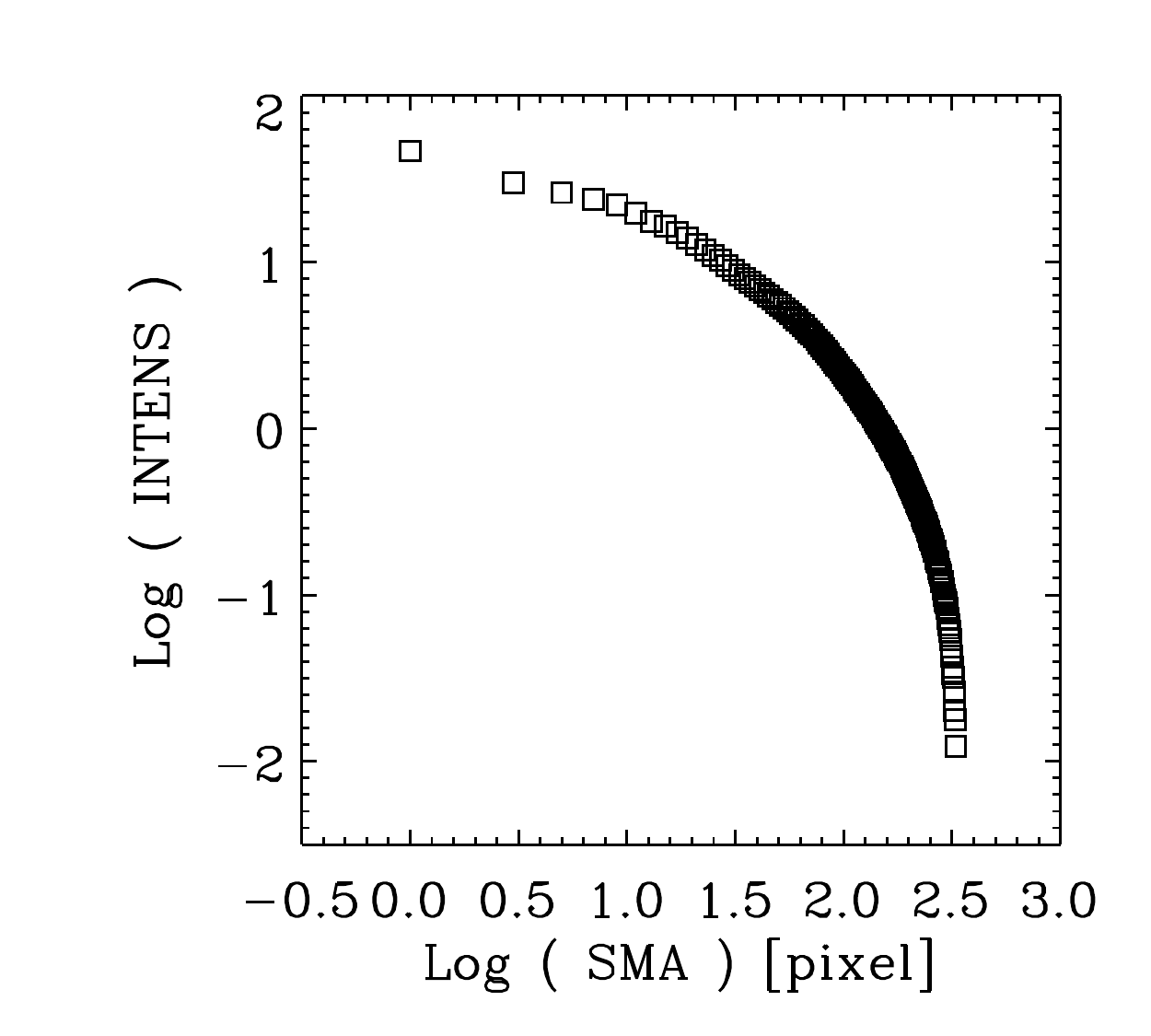}	    &  \includegraphics[trim=0.6cm 0cm 0cm 0cm, clip=true, scale=0.46]{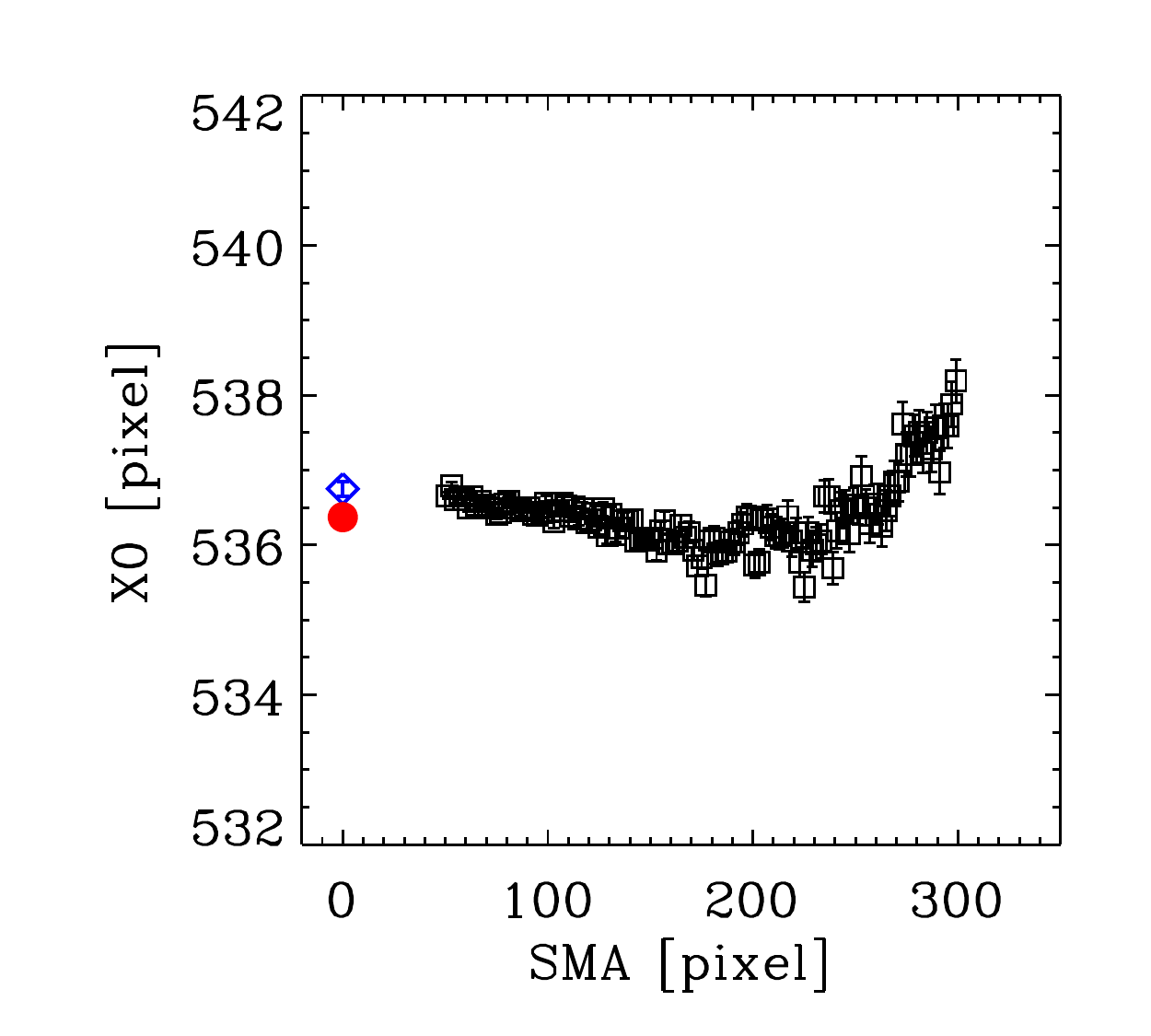}  &  \includegraphics[trim=0.6cm 0cm 0cm 0cm, clip=true, scale=0.46]{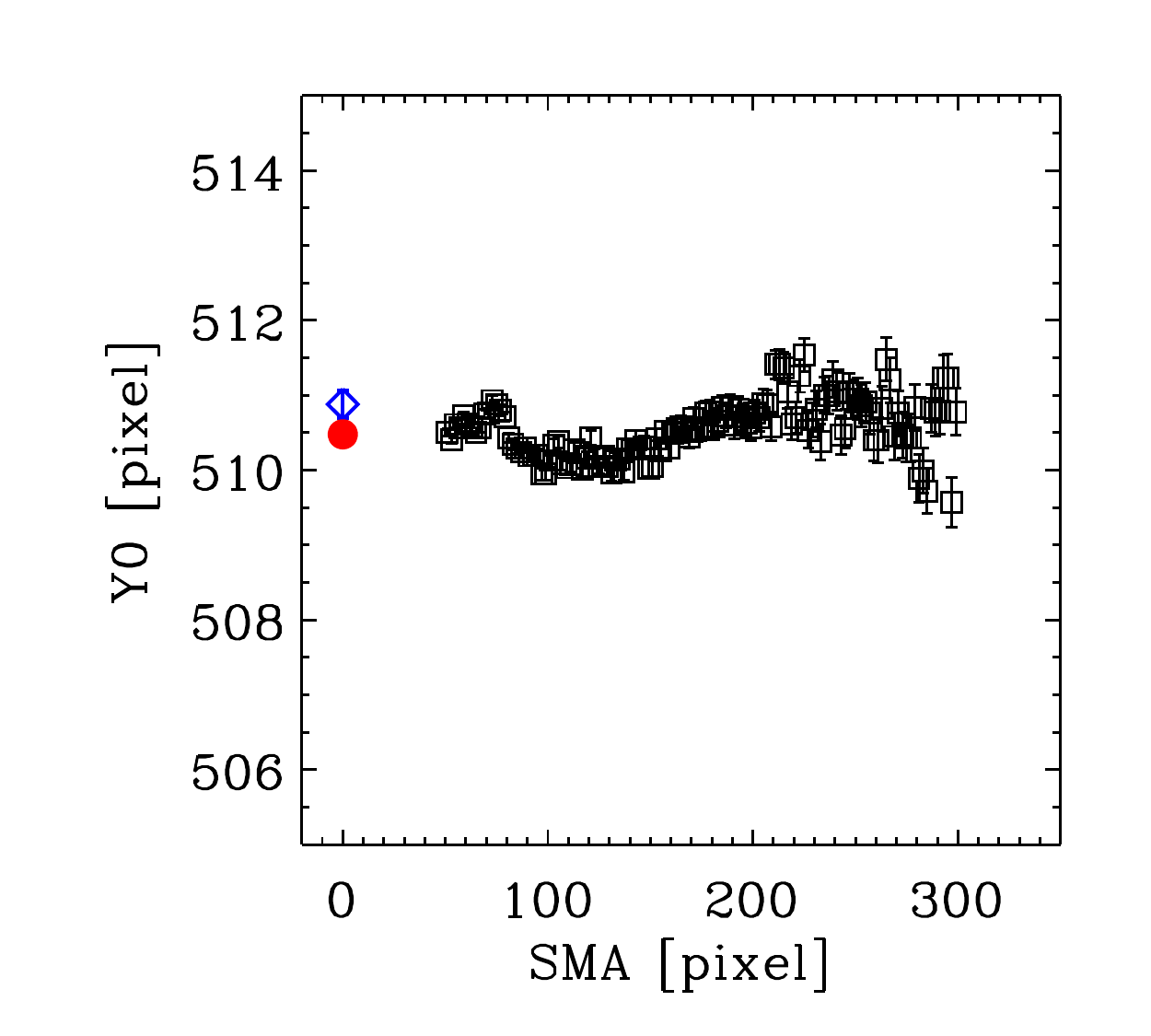} \\	
 \includegraphics[trim=0.65cm 0cm 0cm 0cm, clip=true, scale=0.46]{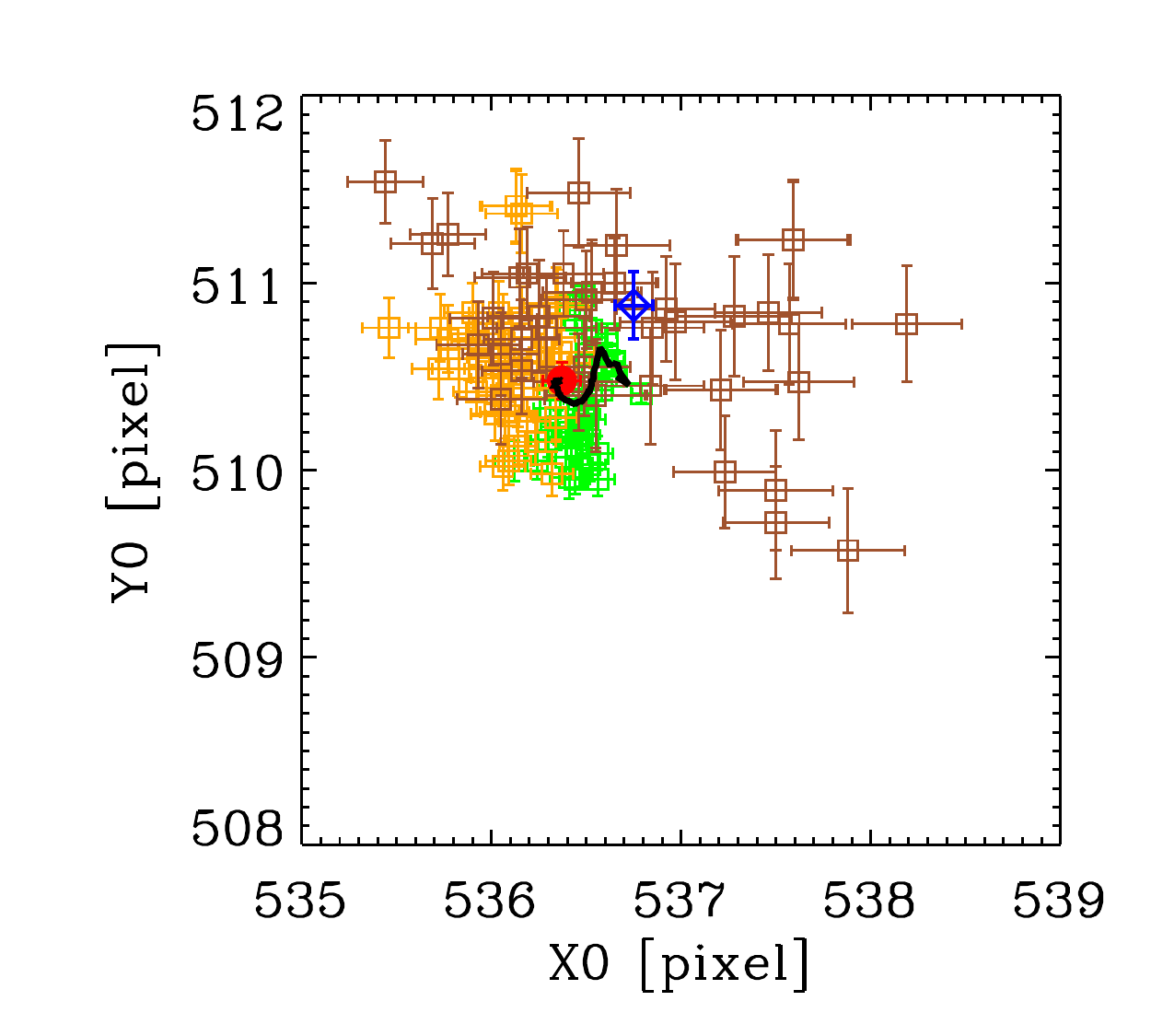}	&  \includegraphics[trim=0.6cm 0cm 0cm 0cm, clip=true, scale=0.46]{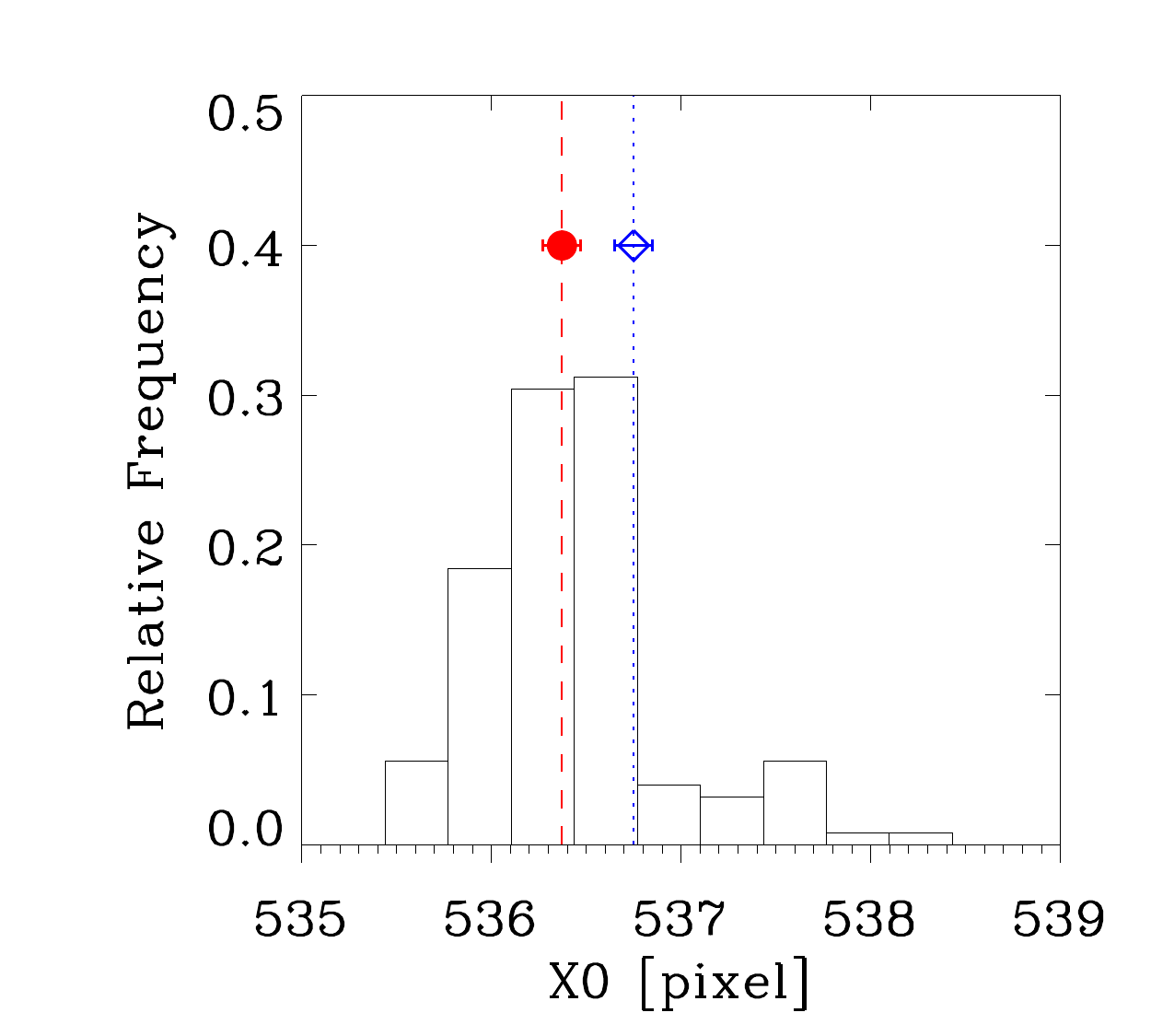}	& \includegraphics[trim=0.6cm 0cm 0cm 0cm, clip=true, scale=0.46]{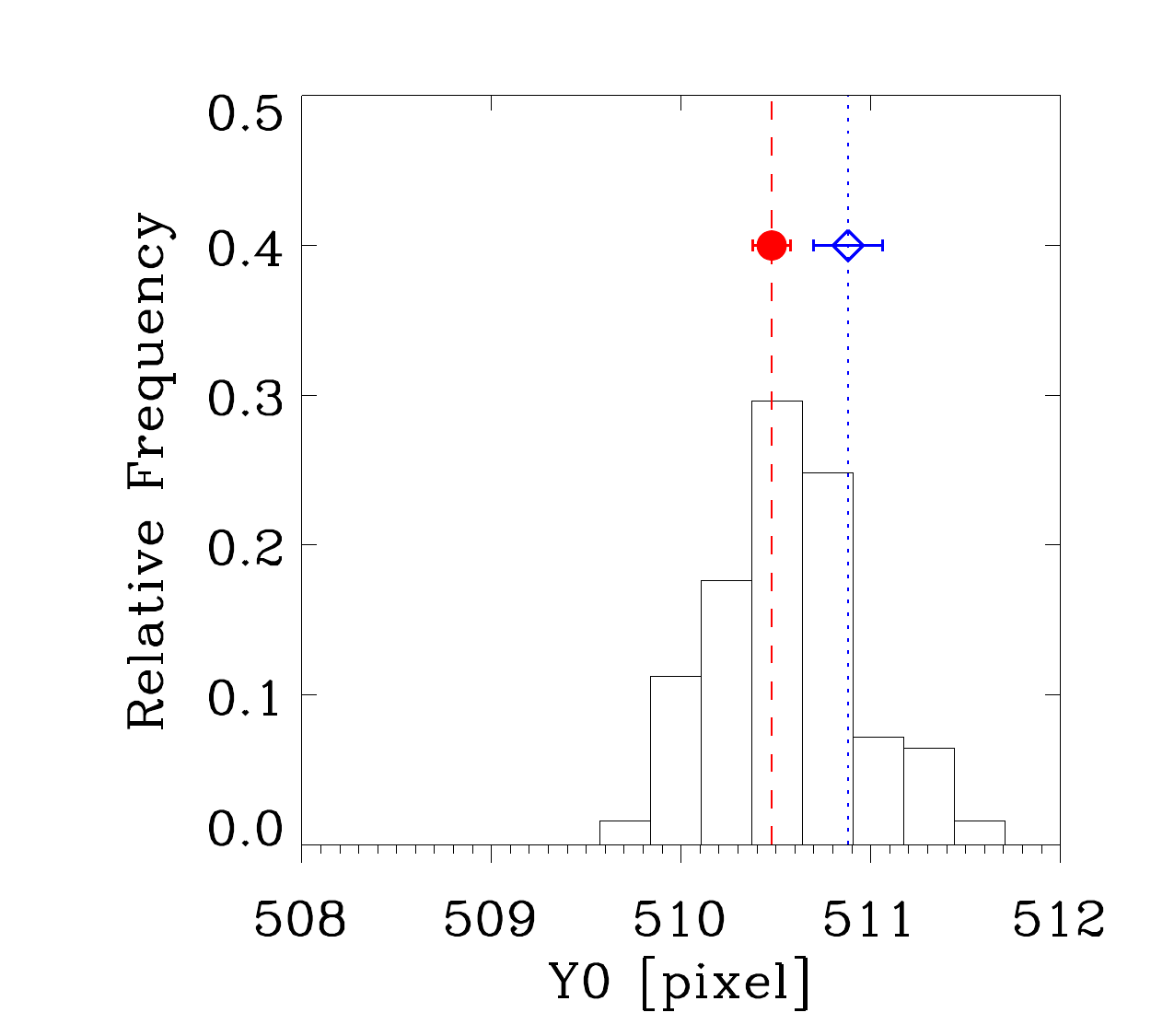}\\
\end{array}$
\end{center}
\caption[IC 1459 (WFPC2_F814W)]{As in Fig.\ref{fig: NGC4373_W2} for galaxy IC 1459, WFPC2/PC - F814W, scale=$0\farcs05$/pxl.}
\label{fig: IC1459_F814W}
\end{figure*} 

\cleardoublepage 
\begin{figure*}[h]
\begin{center}$
\begin{array}{ccc}
\includegraphics[trim=3.75cm 1cm 3cm 0cm, clip=true, scale=0.48]{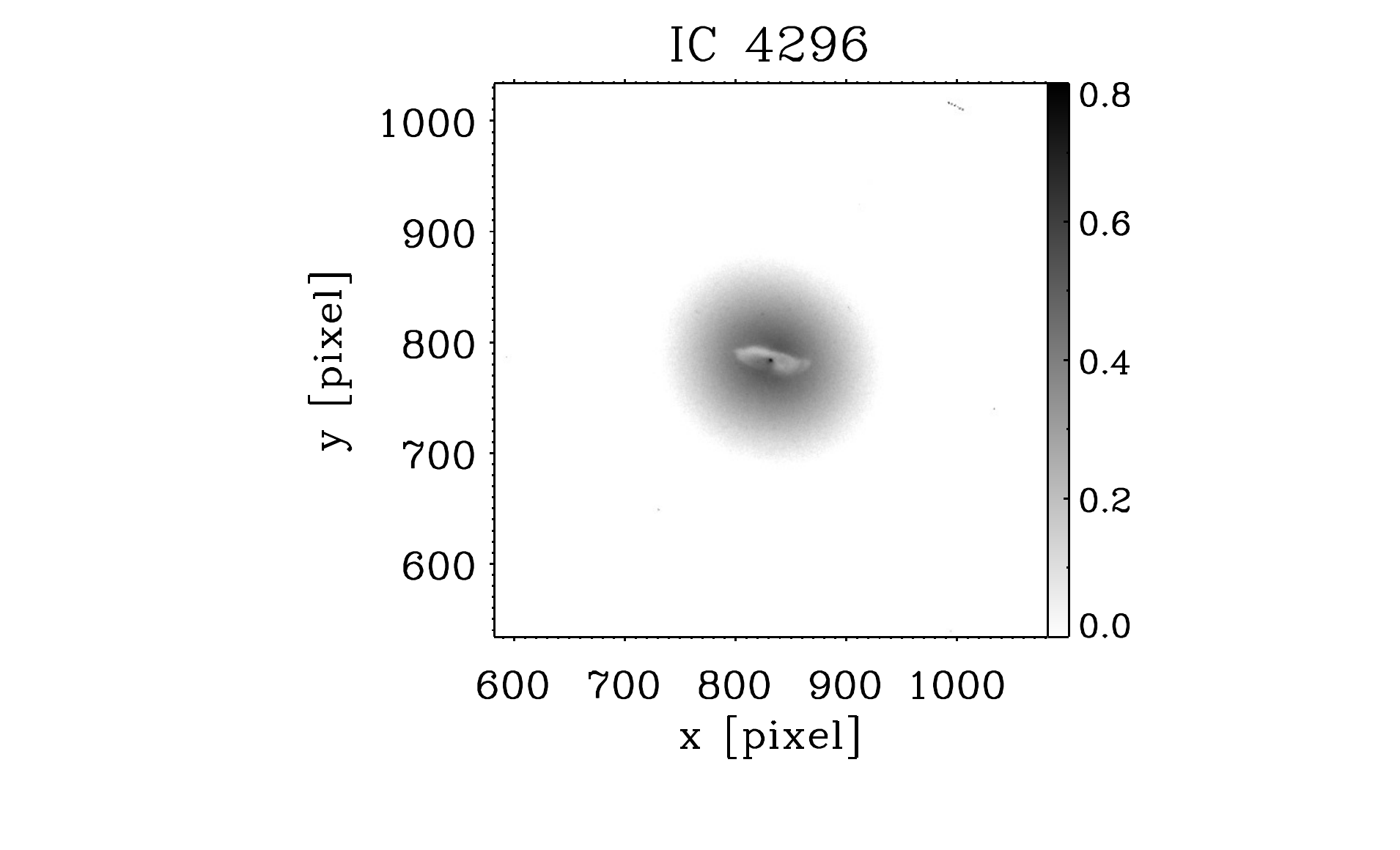} & \includegraphics[trim= 4.cm 1cm 3cm 0cm, clip=true, scale=0.48]{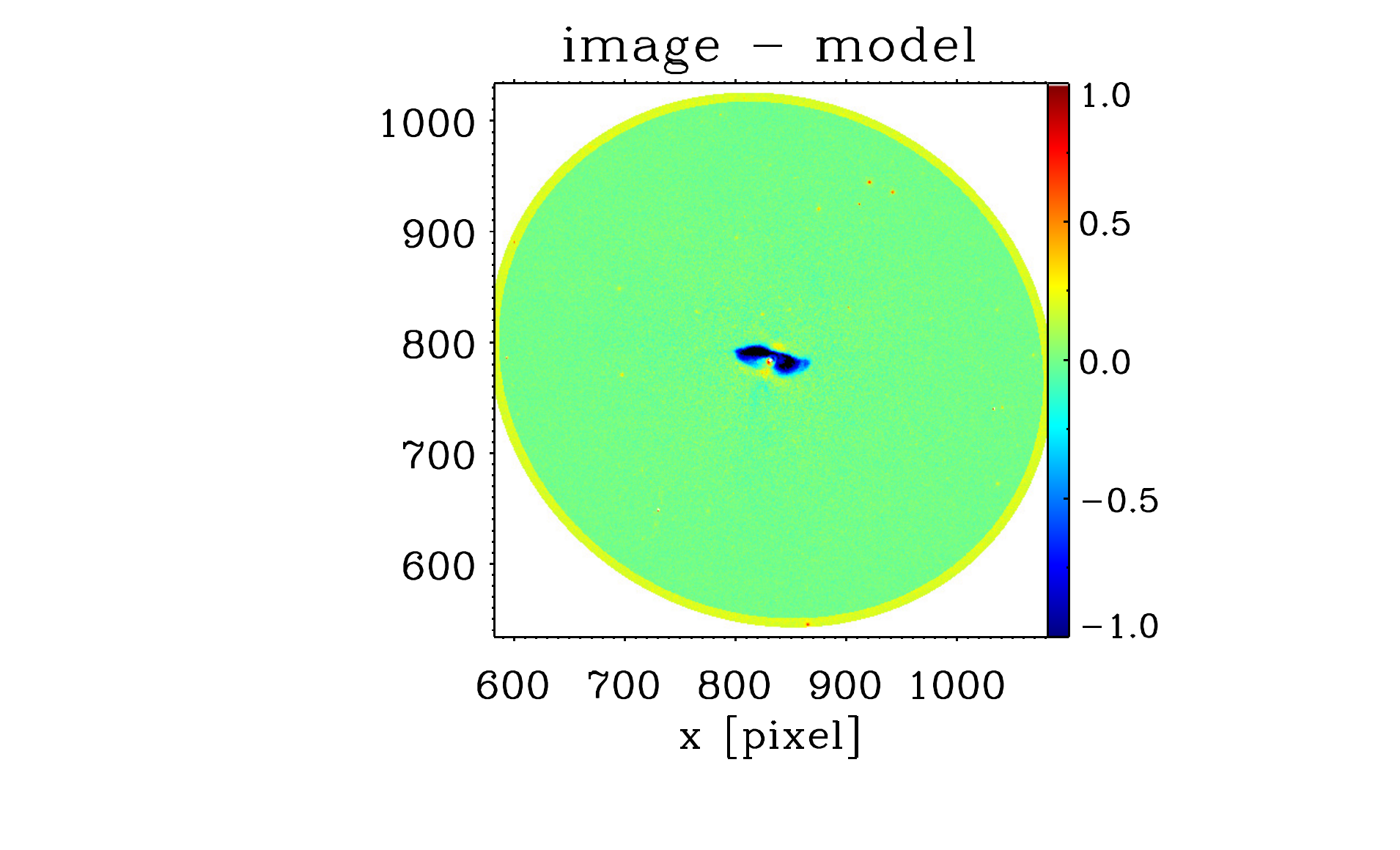}	& \includegraphics[trim= 4.cm 1cm 3cm 0cm, clip=true, scale=0.48]{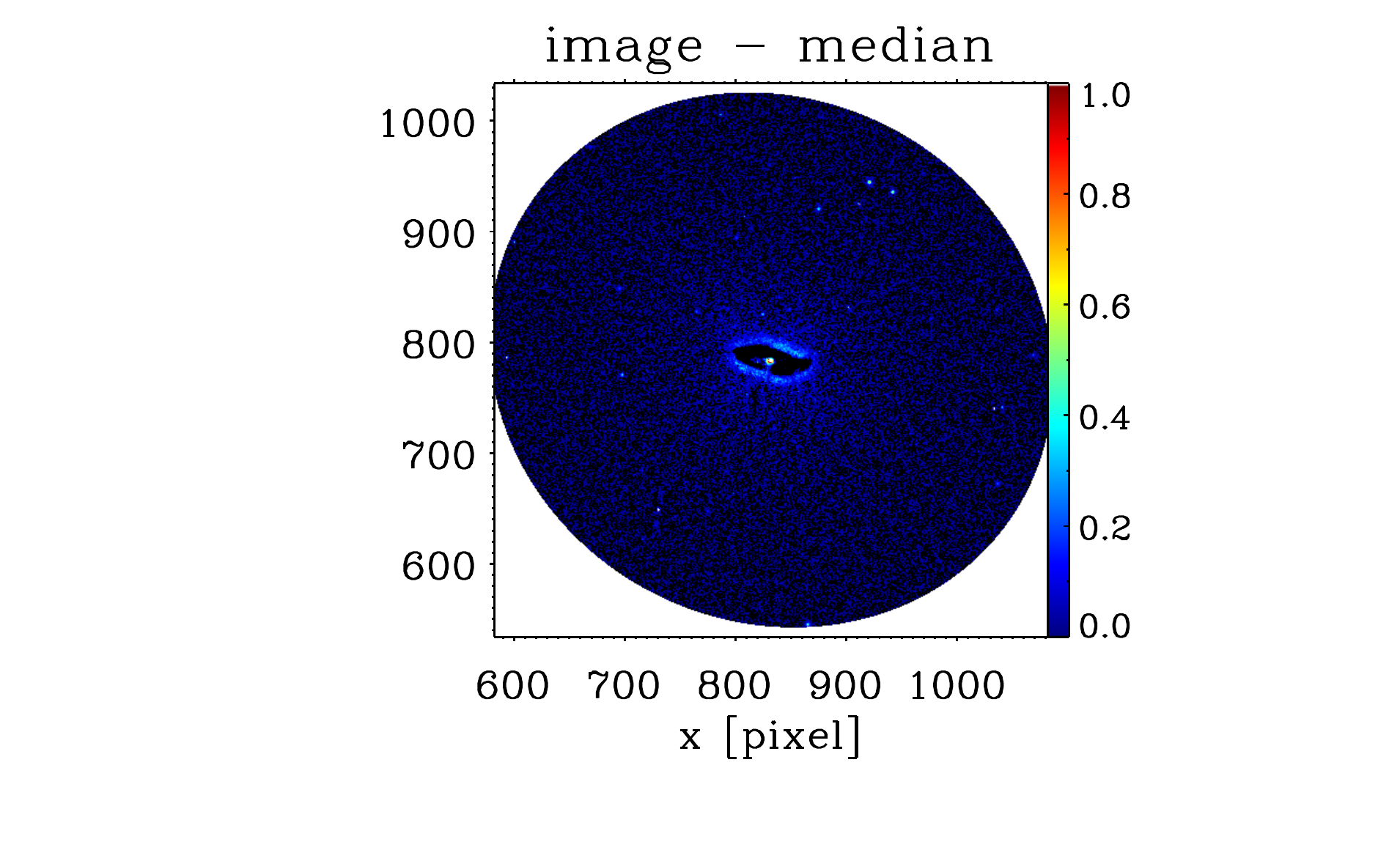} \\
\includegraphics[trim=0.7cm 0cm 0cm 0cm, clip=true, scale=0.46]{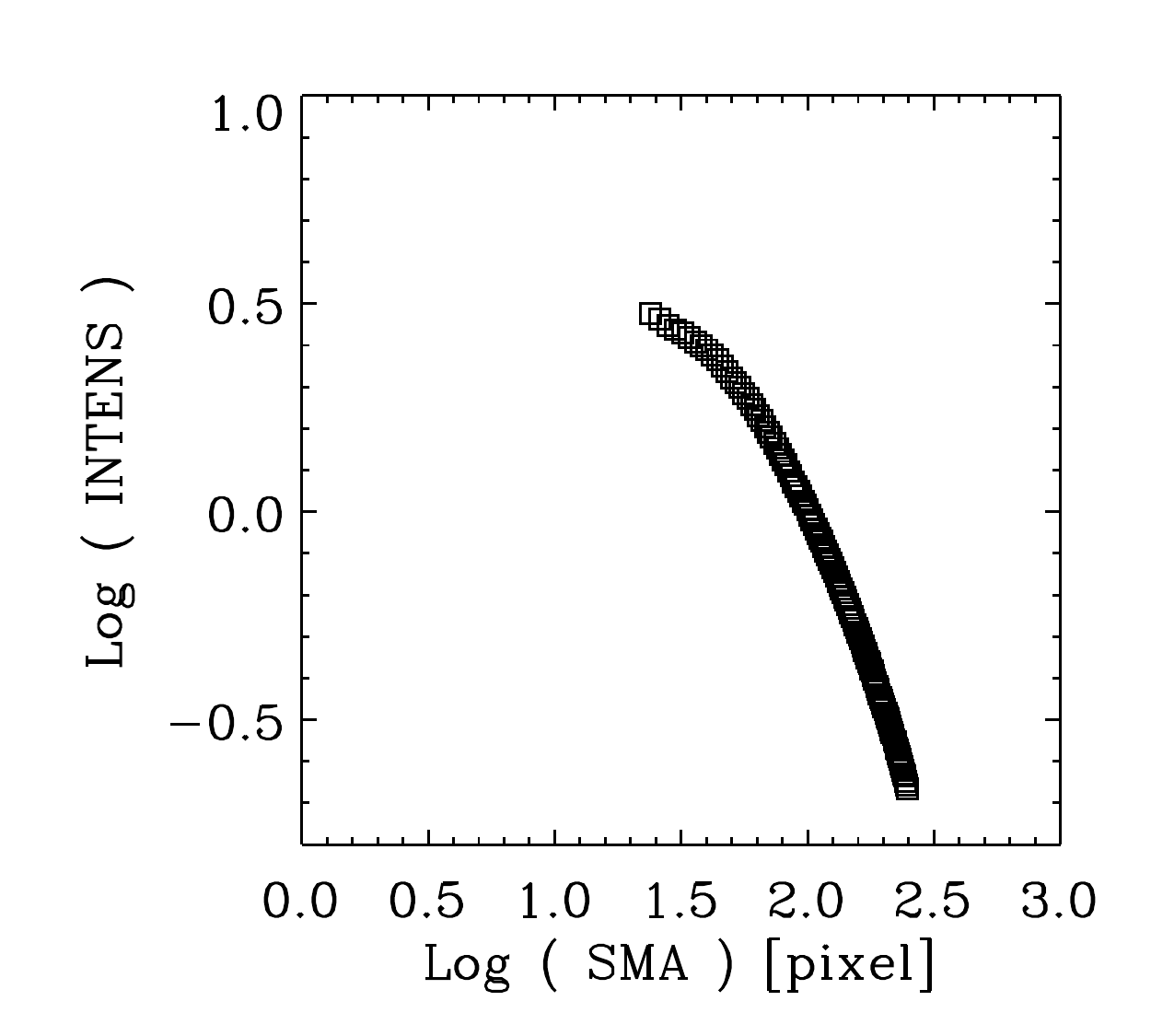}	    &  \includegraphics[trim=0.6cm 0cm 0cm 0cm, clip=true, scale=0.46]{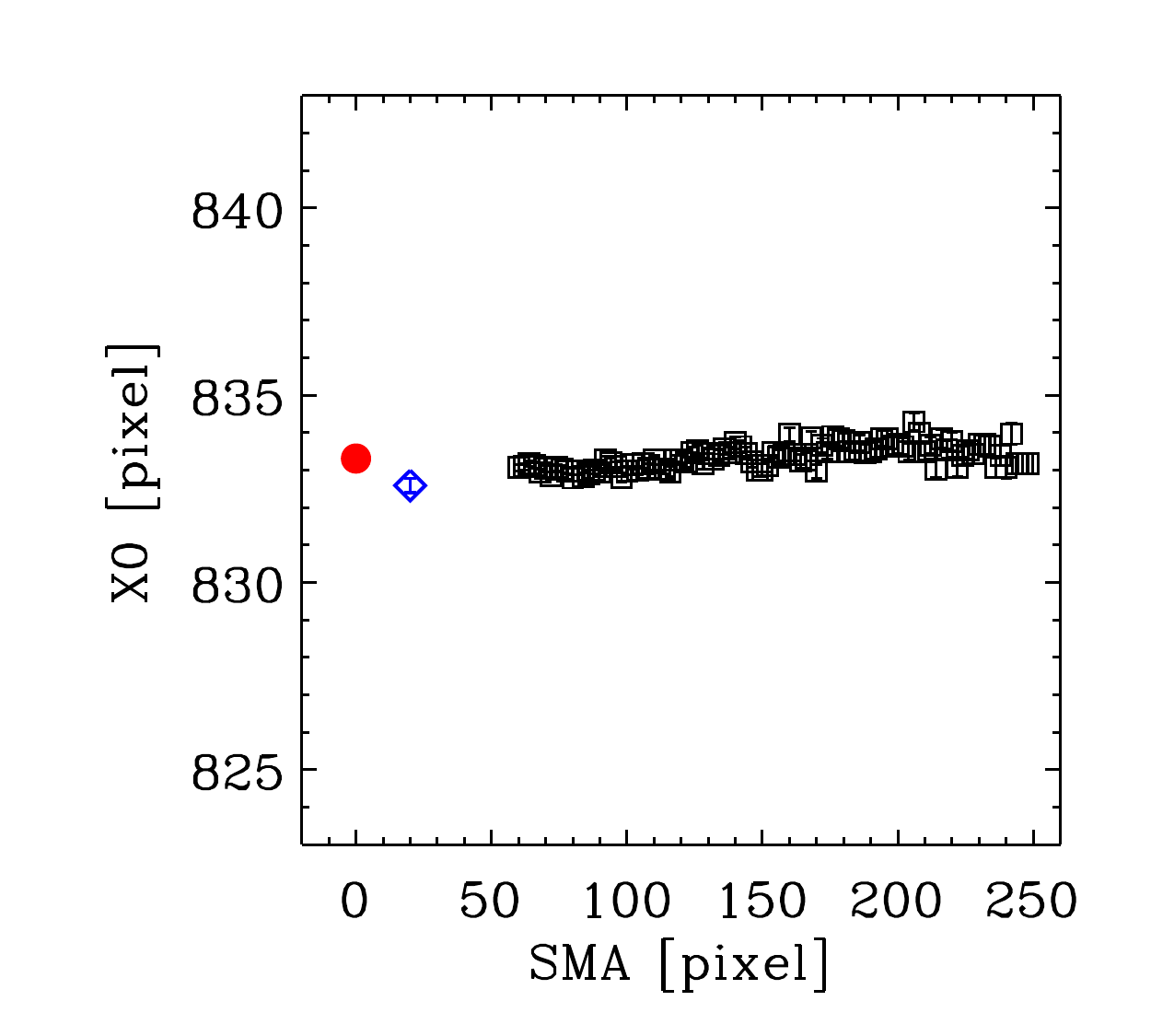}  &  \includegraphics[trim=0.6cm 0cm 0cm 0cm, clip=true, scale=0.46]{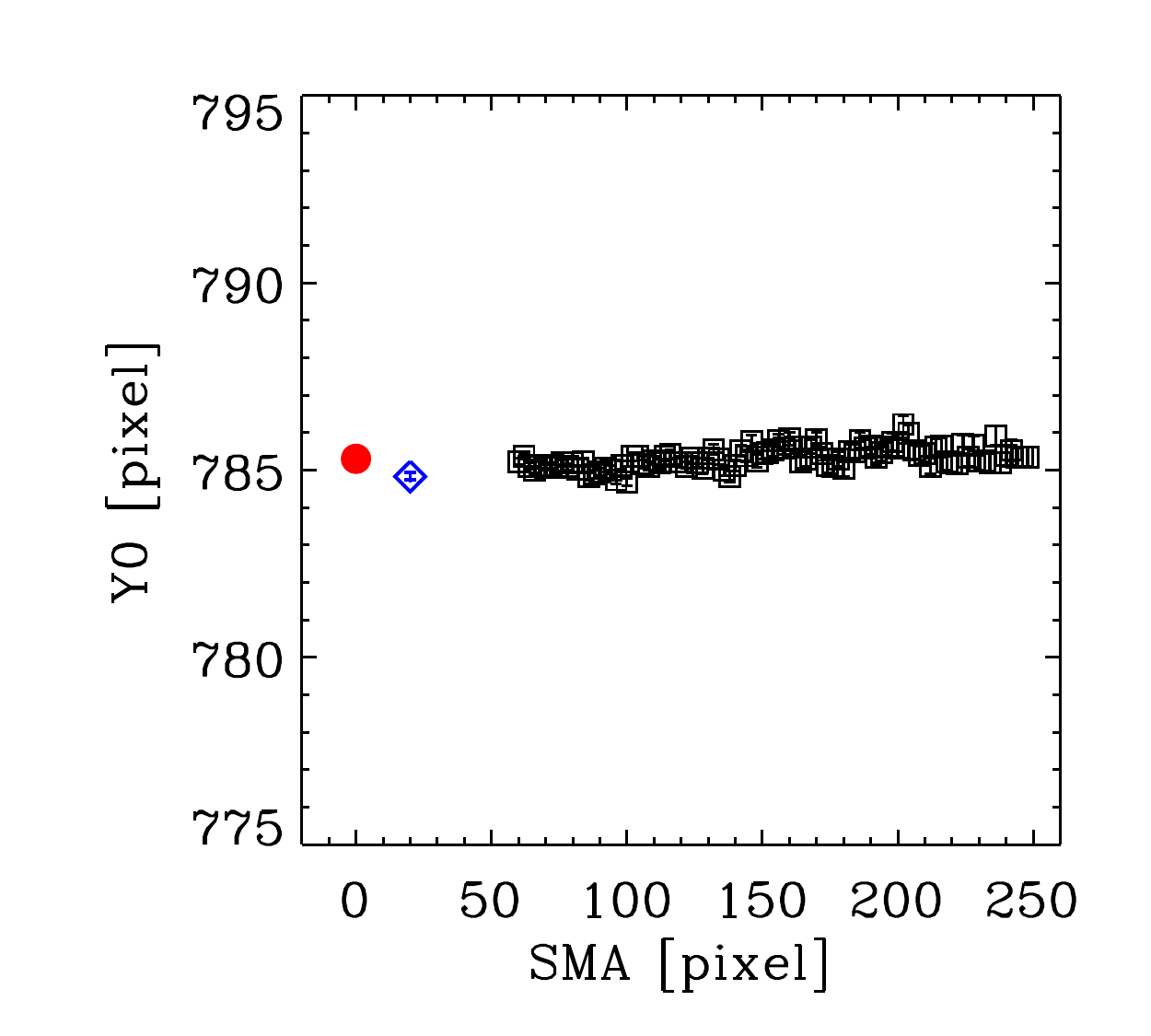} \\	
 \includegraphics[trim=0.65cm 0cm 0cm 0cm, clip=true, scale=0.46]{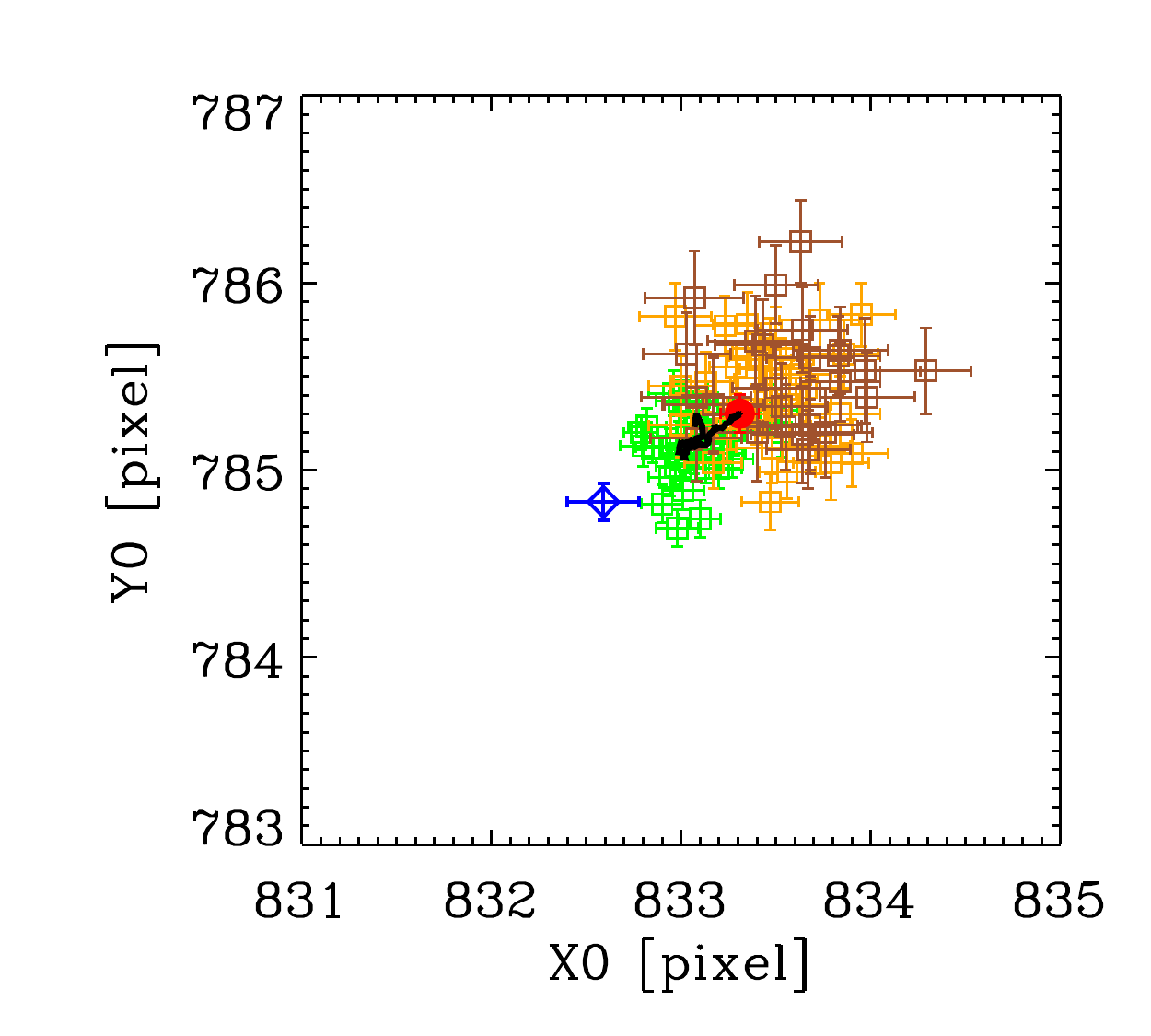}	&  \includegraphics[trim=0.6cm 0cm 0cm 0cm, clip=true, scale=0.46]{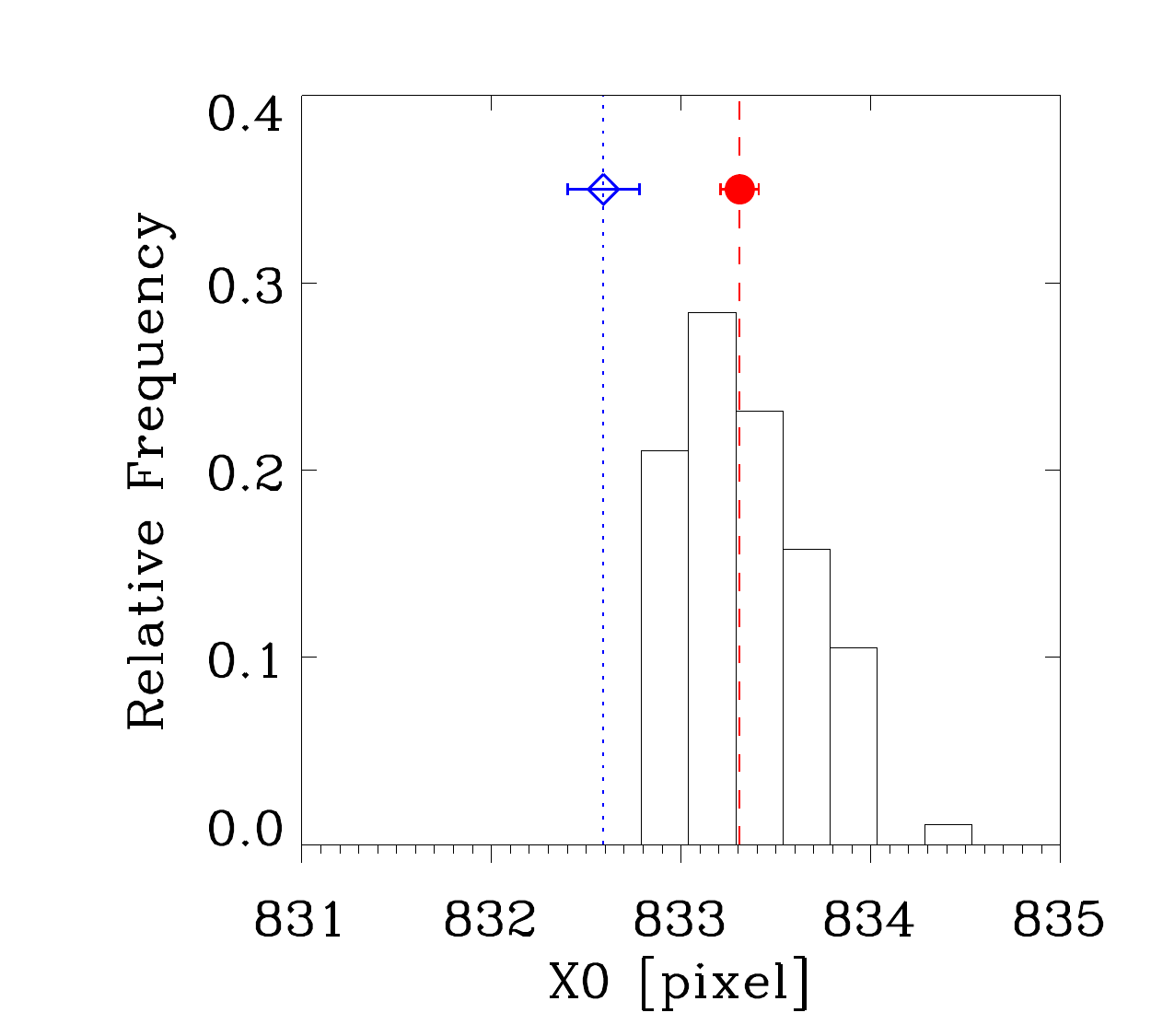}	& \includegraphics[trim=0.6cm 0cm 0cm 0cm, clip=true, scale=0.46]{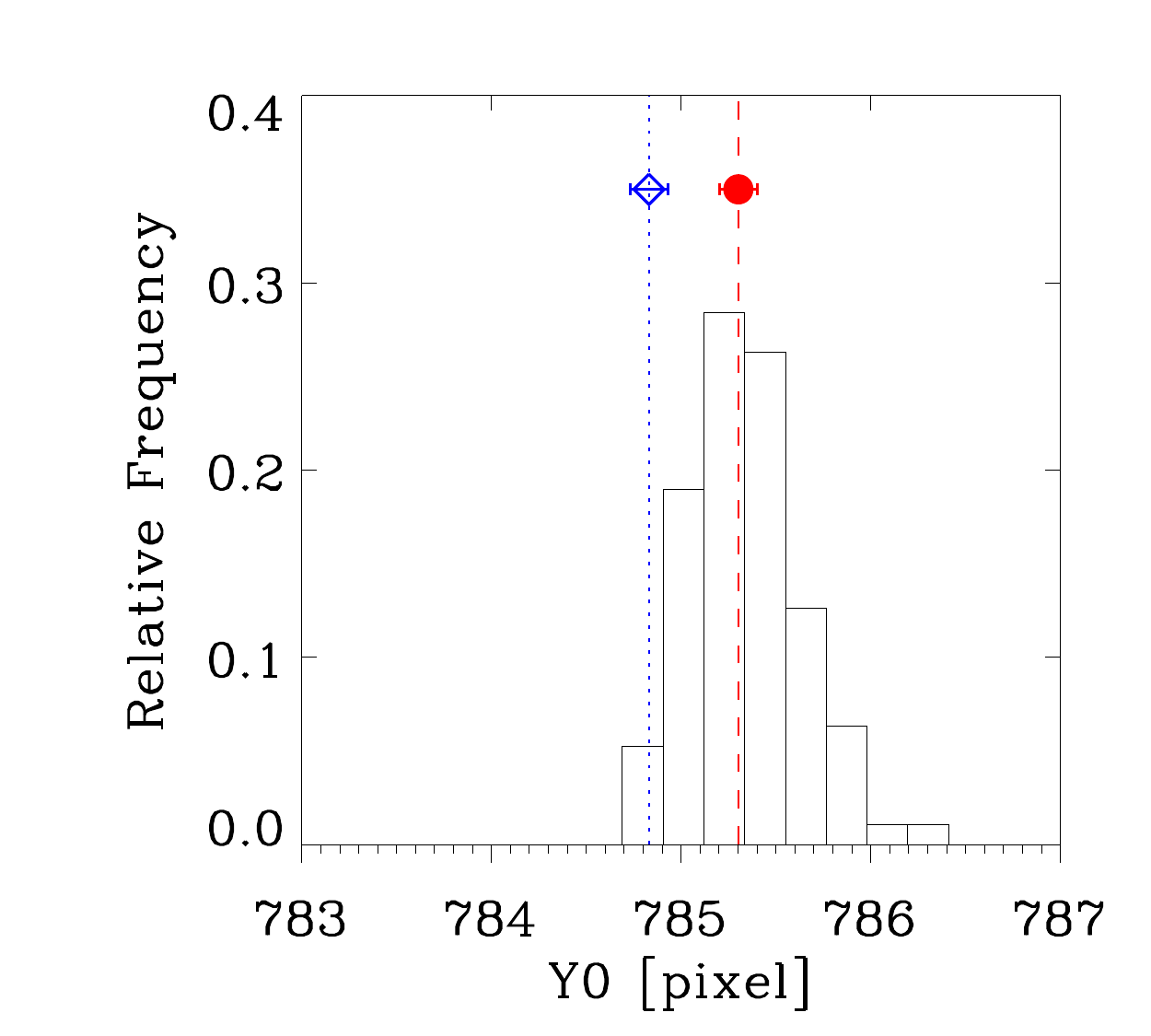}\\
\end{array}$
\end{center}
\caption[IC 4296 (ACS)]{As in Fig.\ref{fig: NGC4373_W2} for galaxy IC 4296, ACS/HRC - F625W, scale=0\farcs025/pxl.}
\label{fig: IC4296_ACS}
\end{figure*} 

\begin{figure*}[h]
\begin{center}$
\begin{array}{ccc}
\includegraphics[trim=3.75cm 1cm 3cm 0cm, clip=true, scale=0.48]{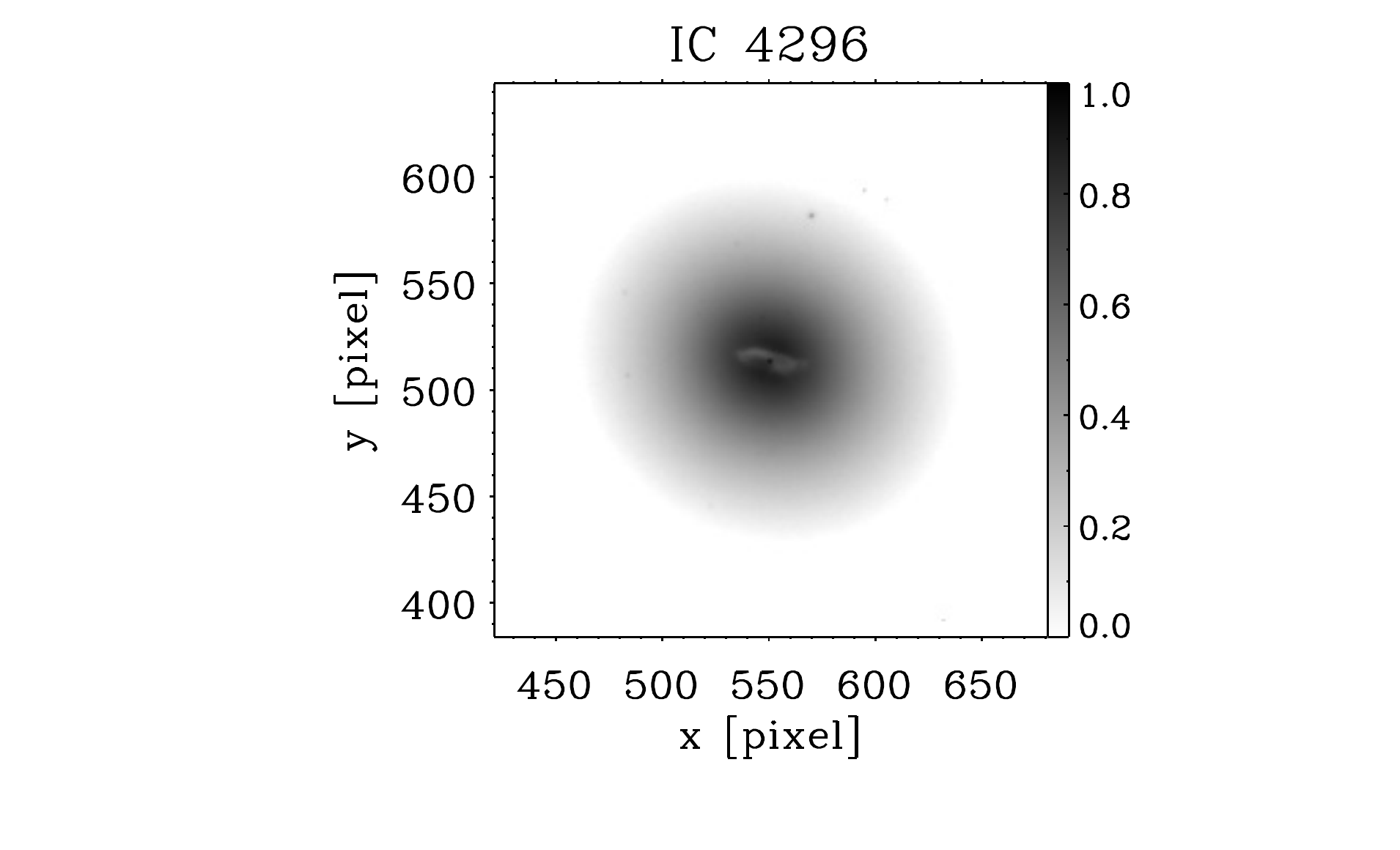} & \includegraphics[trim= 4.cm 1cm 3cm 0cm, clip=true, scale=0.48]{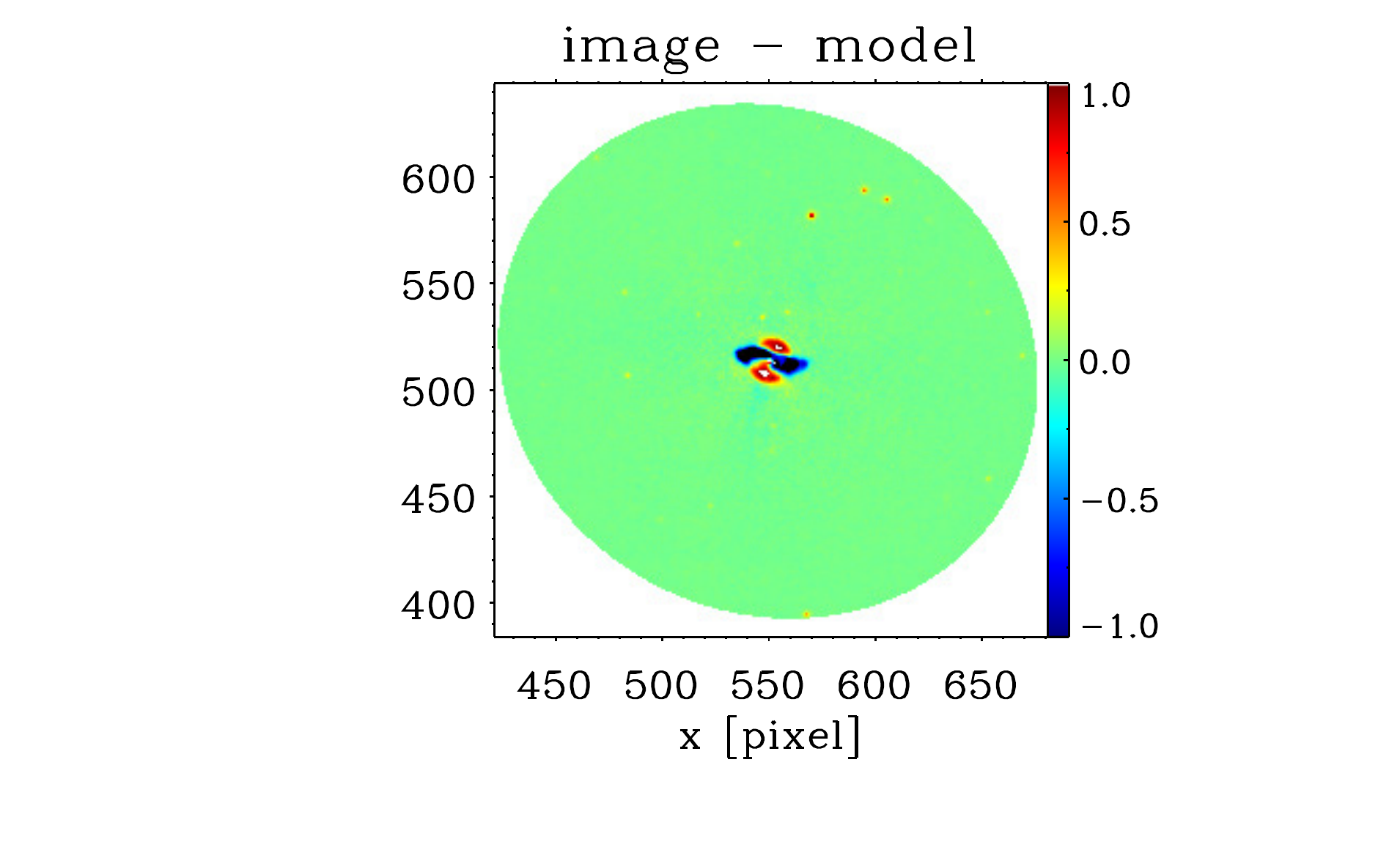}	& \includegraphics[trim= 4.cm 1cm 3cm 0cm, clip=true, scale=0.48]{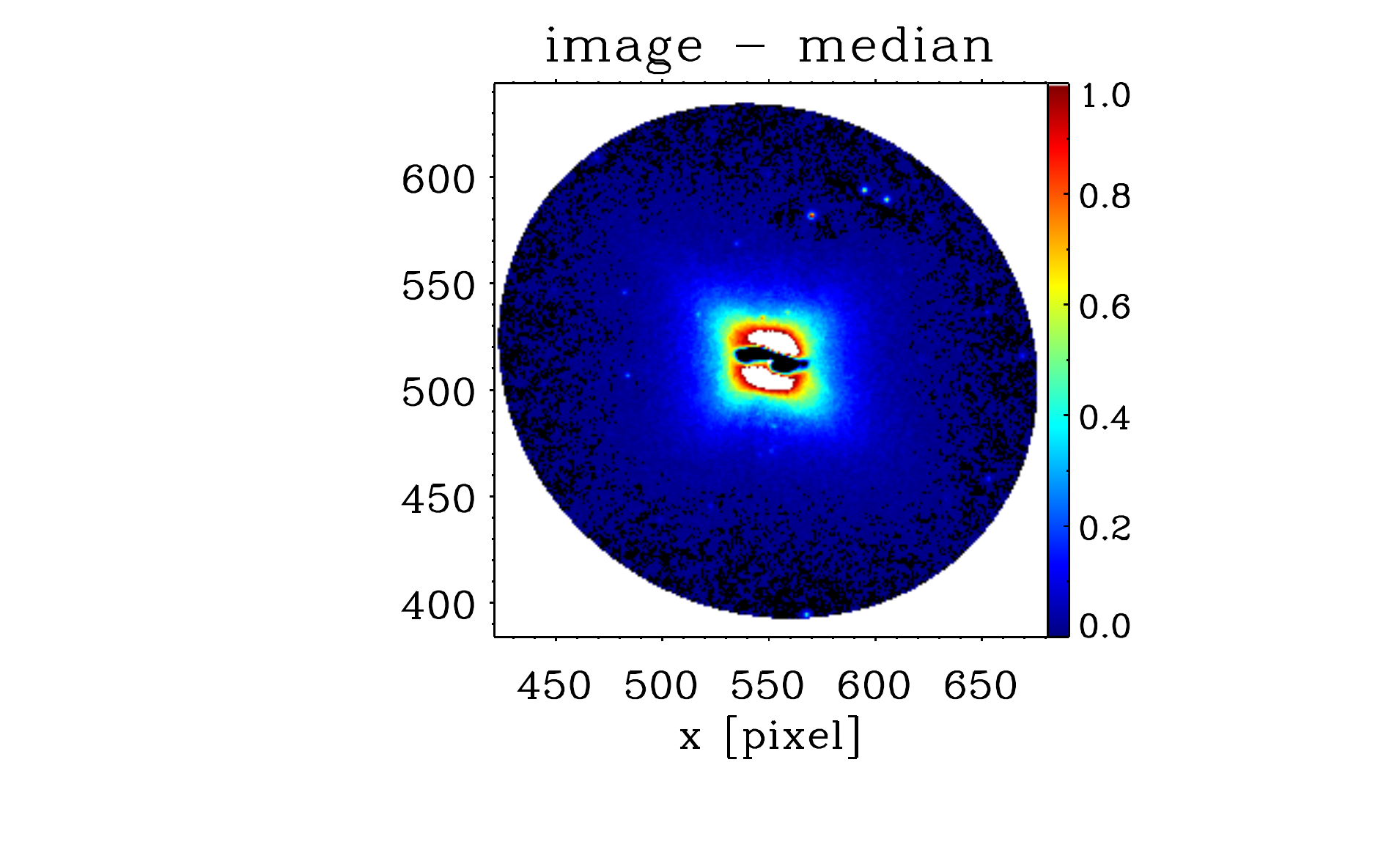} \\
\includegraphics[trim=0.7cm 0cm 0cm 0cm, clip=true, scale=0.46]{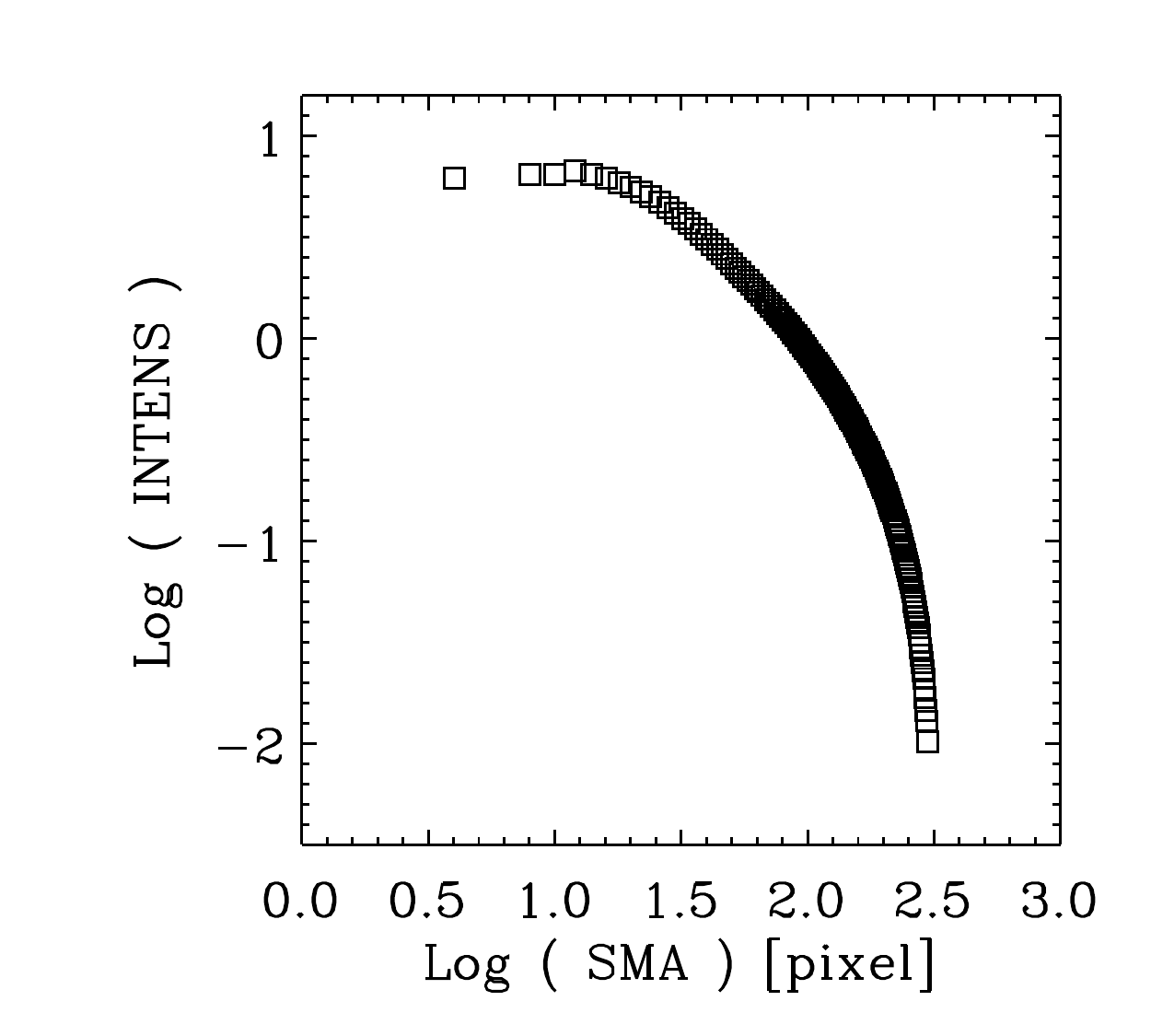}	    &  \includegraphics[trim=0.6cm 0cm 0cm 0cm, clip=true, scale=0.46]{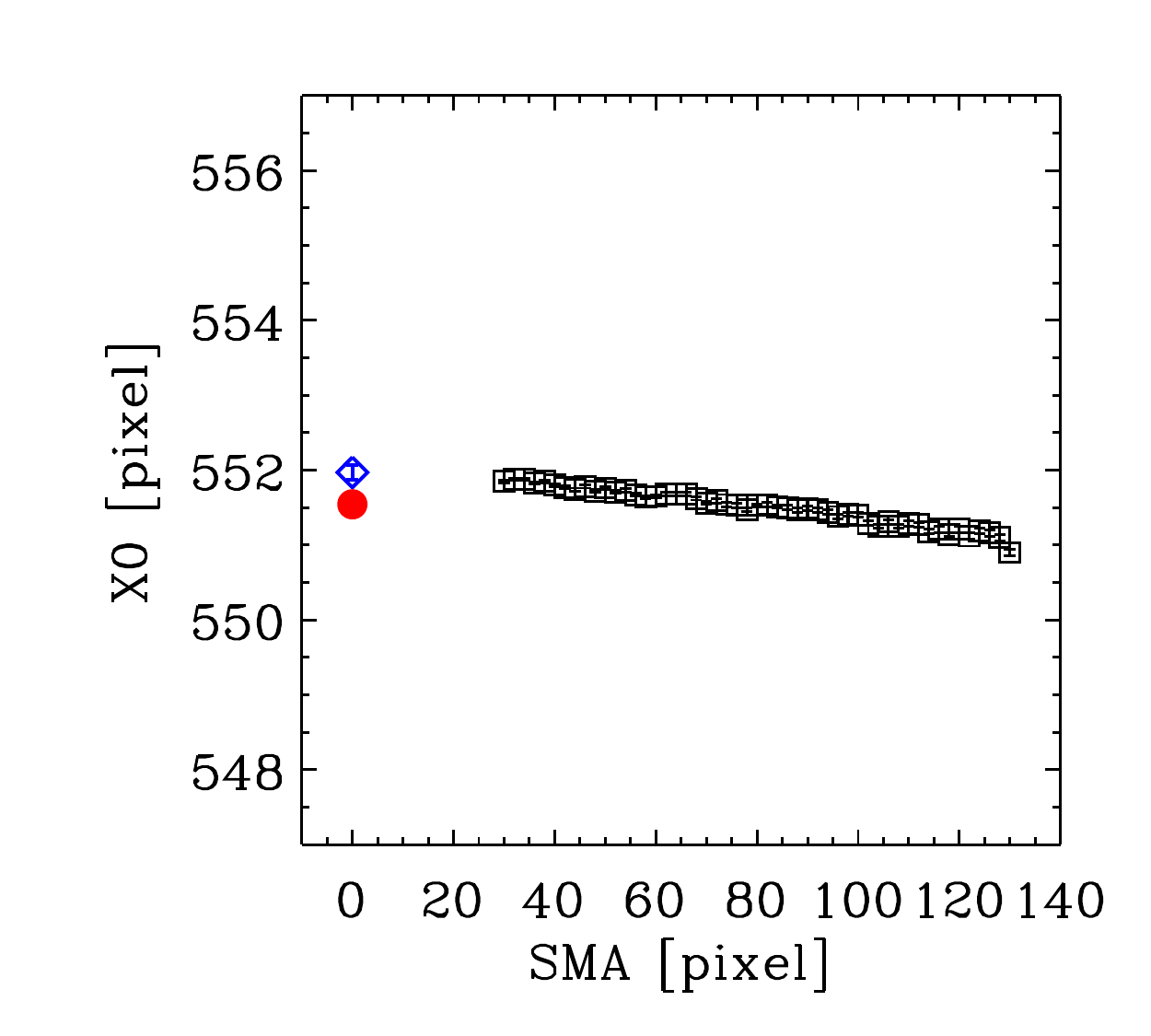}  &  \includegraphics[trim=0.6cm 0cm 0cm 0cm, clip=true, scale=0.46]{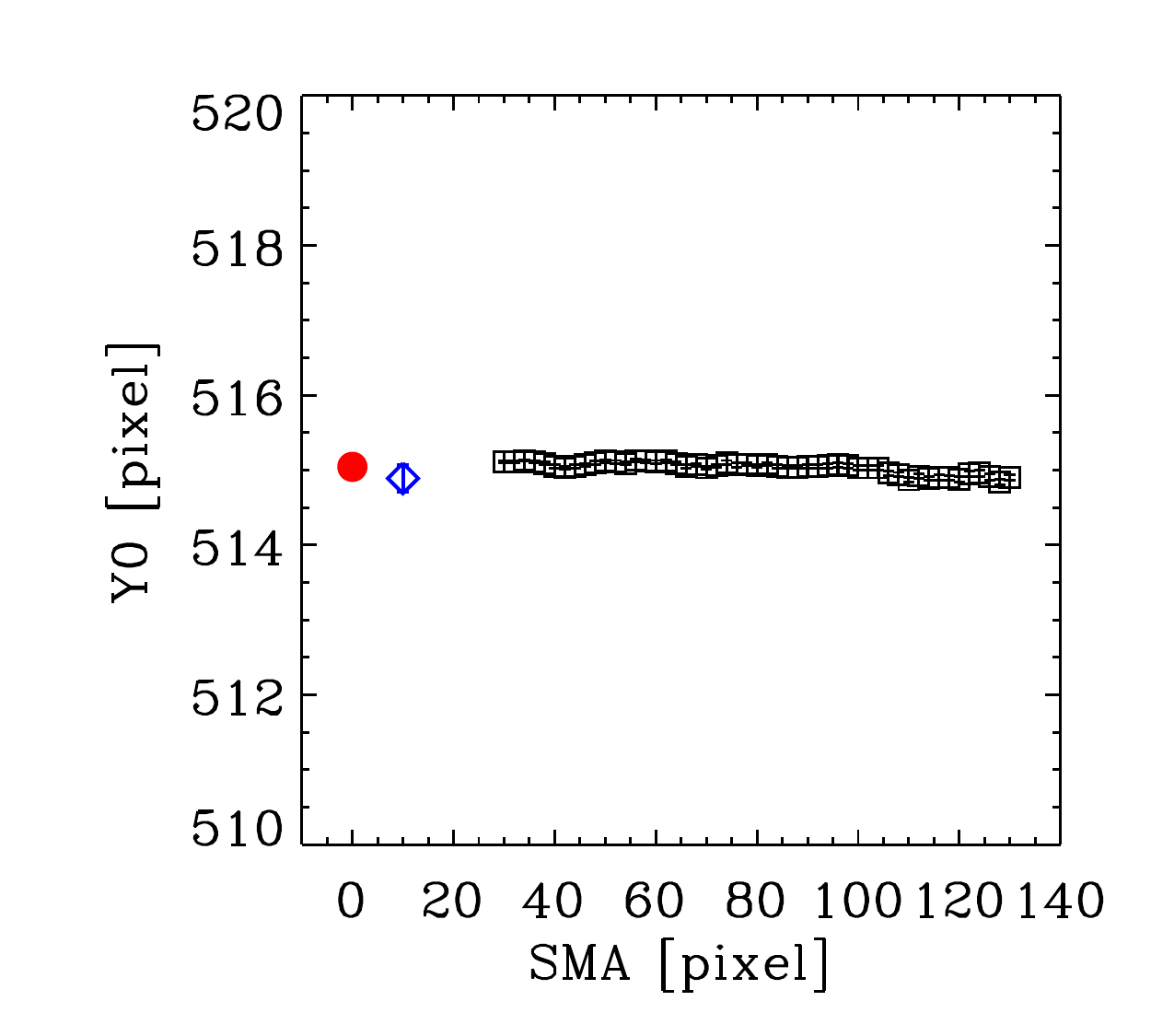} \\	
 \includegraphics[trim=0.65cm 0cm 0cm 0cm, clip=true, scale=0.46]{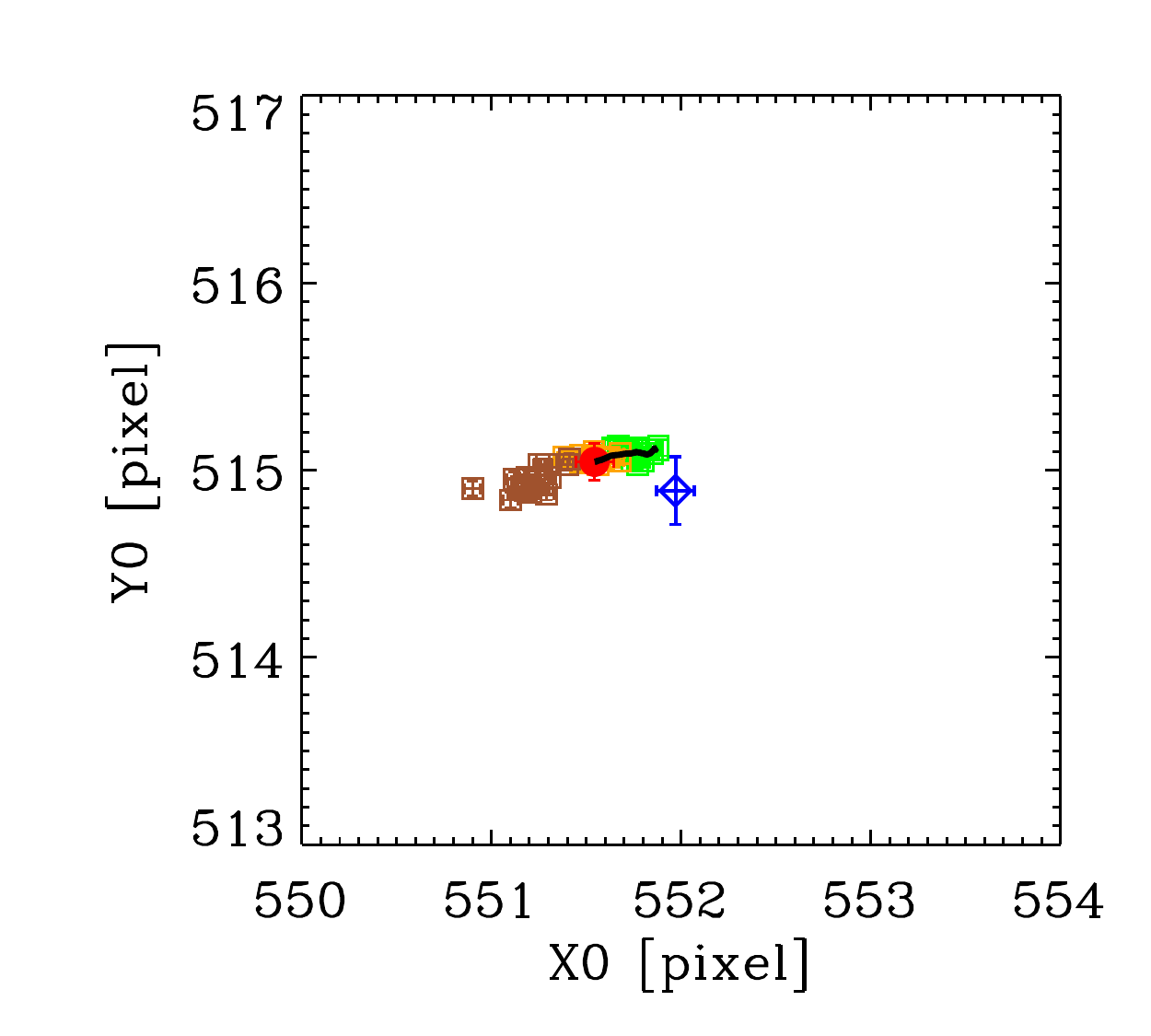}	&  \includegraphics[trim=0.6cm 0cm 0cm 0cm, clip=true, scale=0.46]{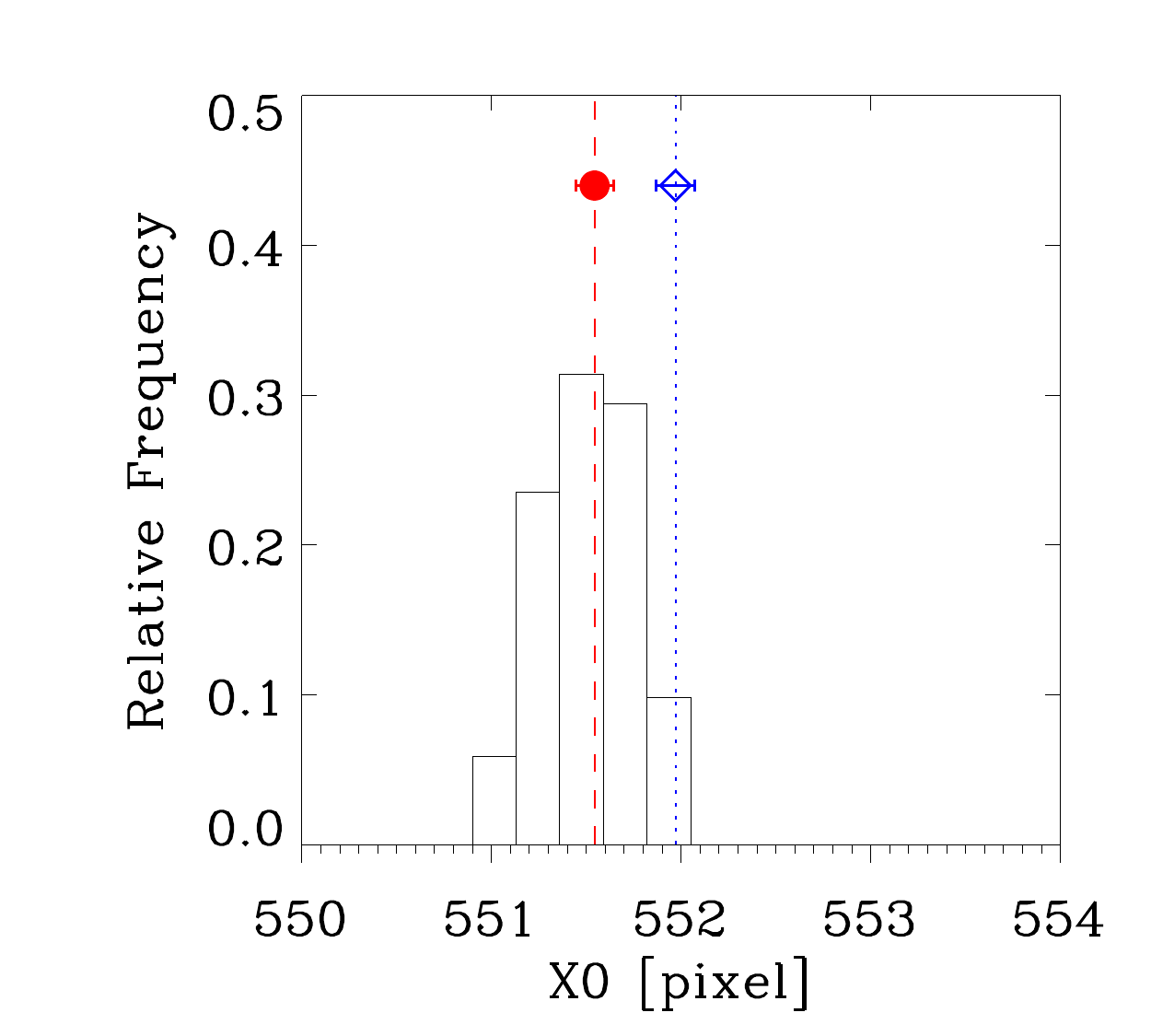}	& \includegraphics[trim=0.6cm 0cm 0cm 0cm, clip=true, scale=0.46]{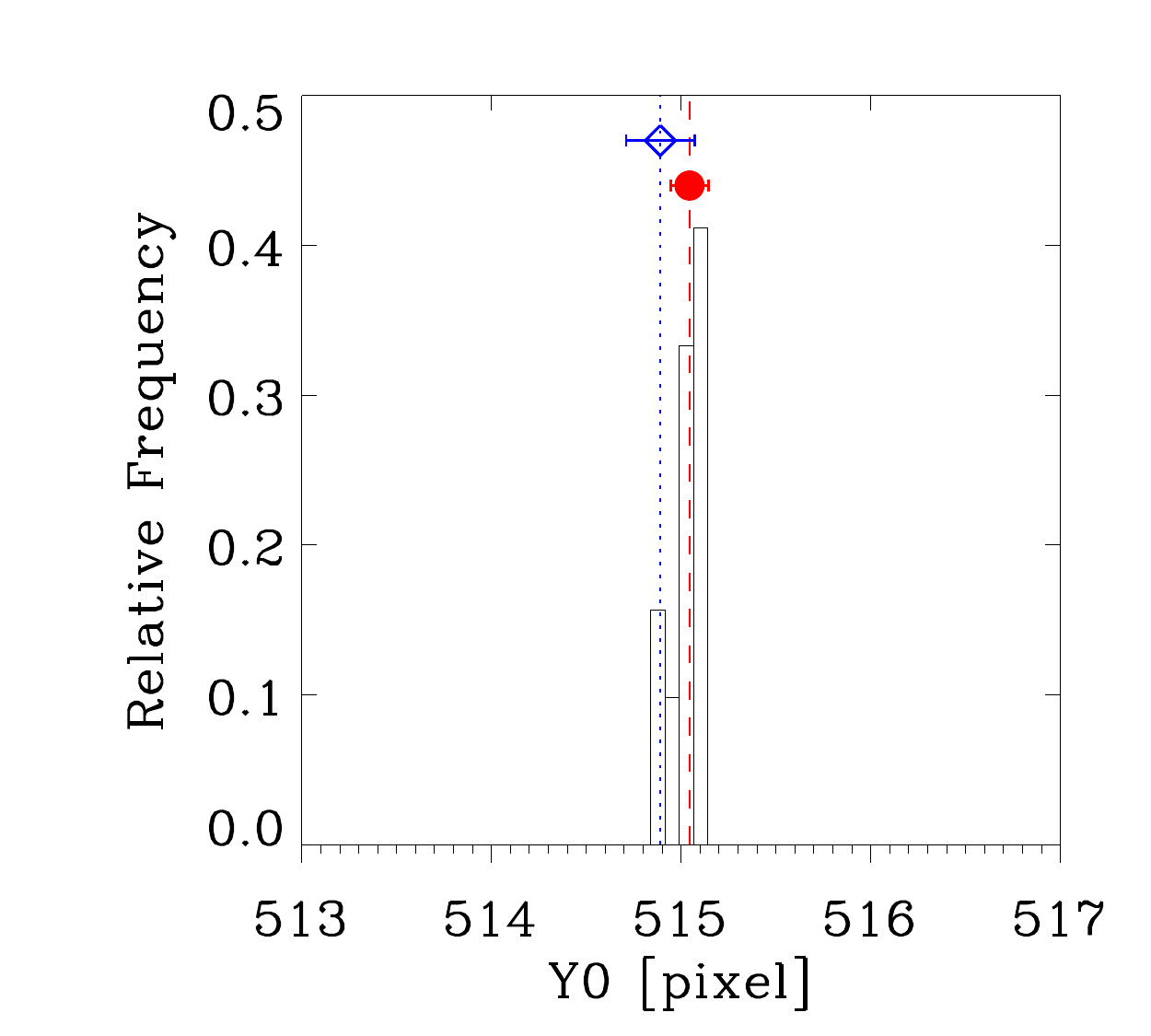}\\
\end{array}$
\end{center}
\caption[IC 4296 (WFPC2)]{As in Fig.\ref{fig: NGC4373_W2} for galaxy IC 4296, WFPC2/PC - F814W, scale=$0\farcs05$/pxl.}
\label{fig: IC4296_WFPC2}
\end{figure*} 

\begin{figure*}[h]
\begin{center}$
\begin{array}{ccc}
\includegraphics[trim=3.75cm 1cm 3cm 0cm, clip=true, scale=0.48]{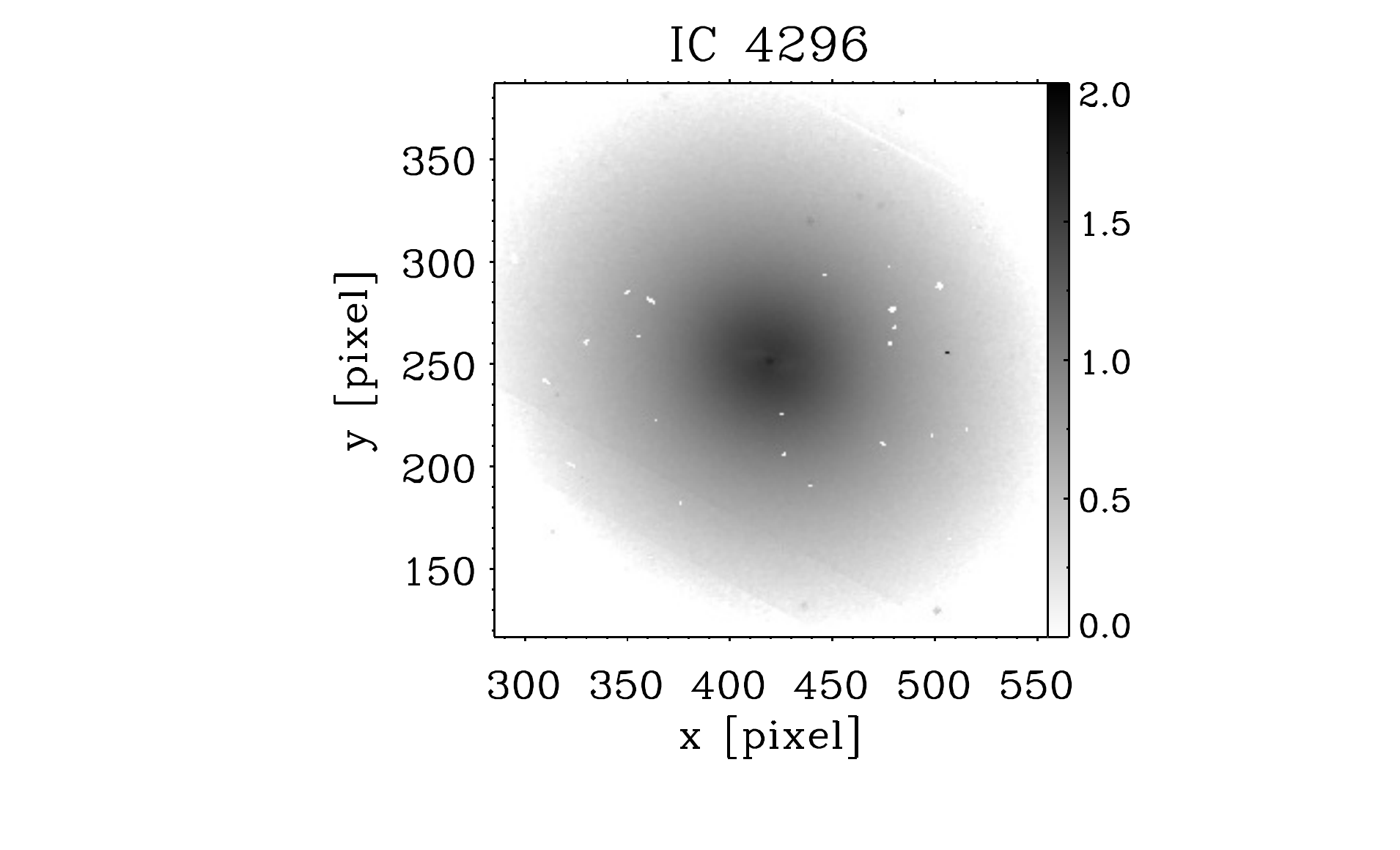} & \includegraphics[trim= 4.cm 1cm 3cm 0cm, clip=true, scale=0.48]{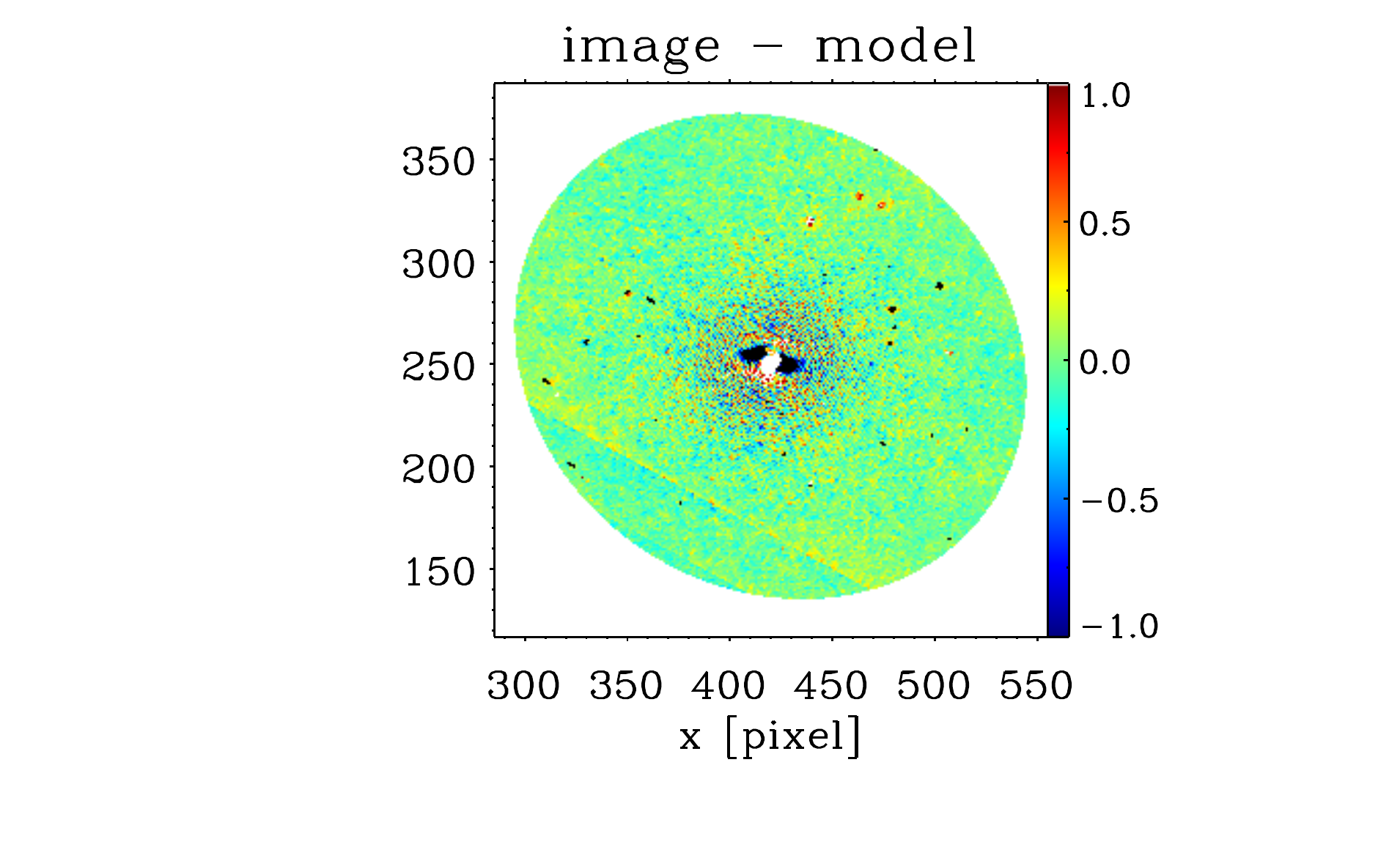}	& \includegraphics[trim= 4.cm 1cm 3cm 0cm, clip=true, scale=0.48]{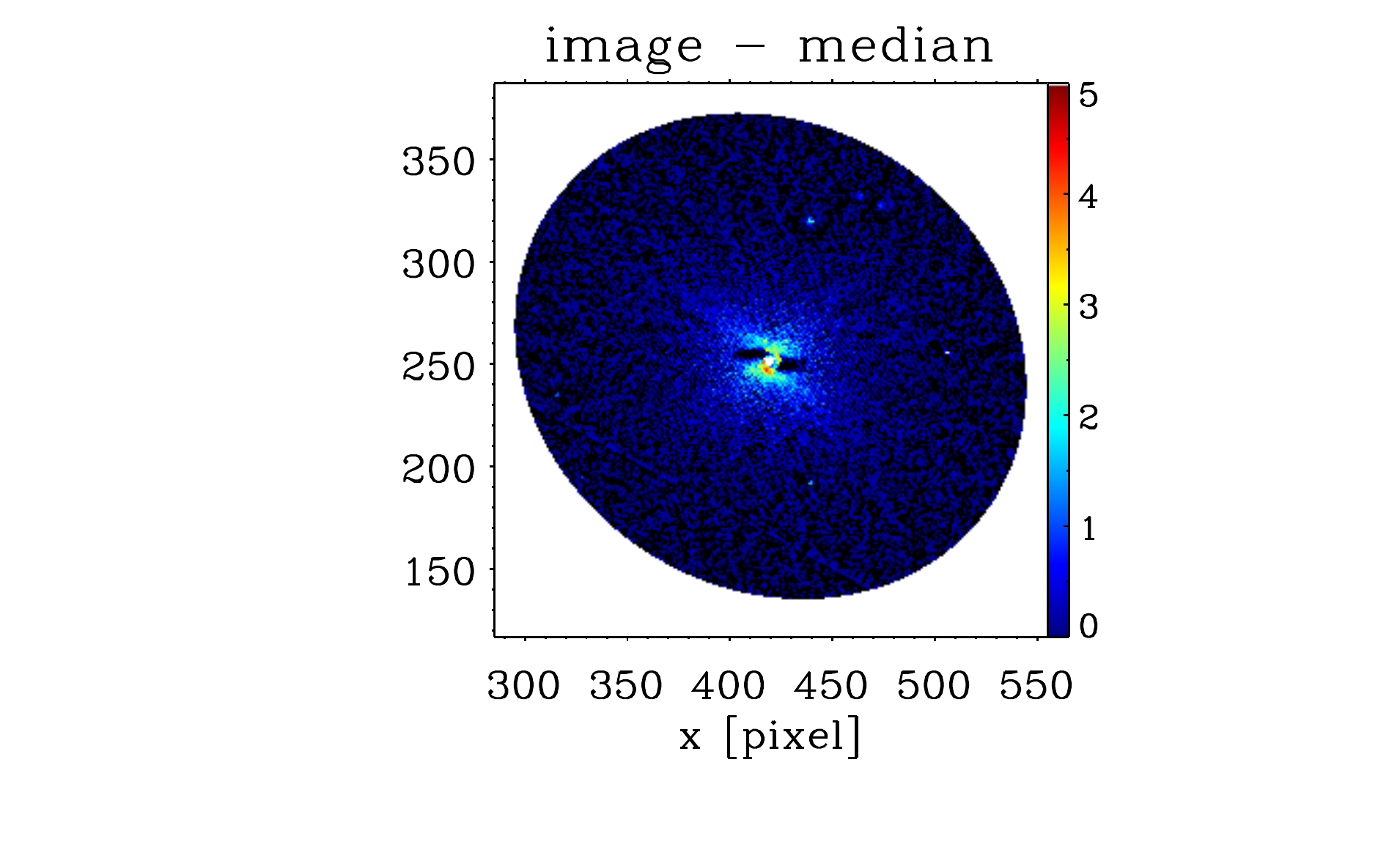} \\
\includegraphics[trim=0.7cm 0cm 0cm 0cm, clip=true, scale=0.46]{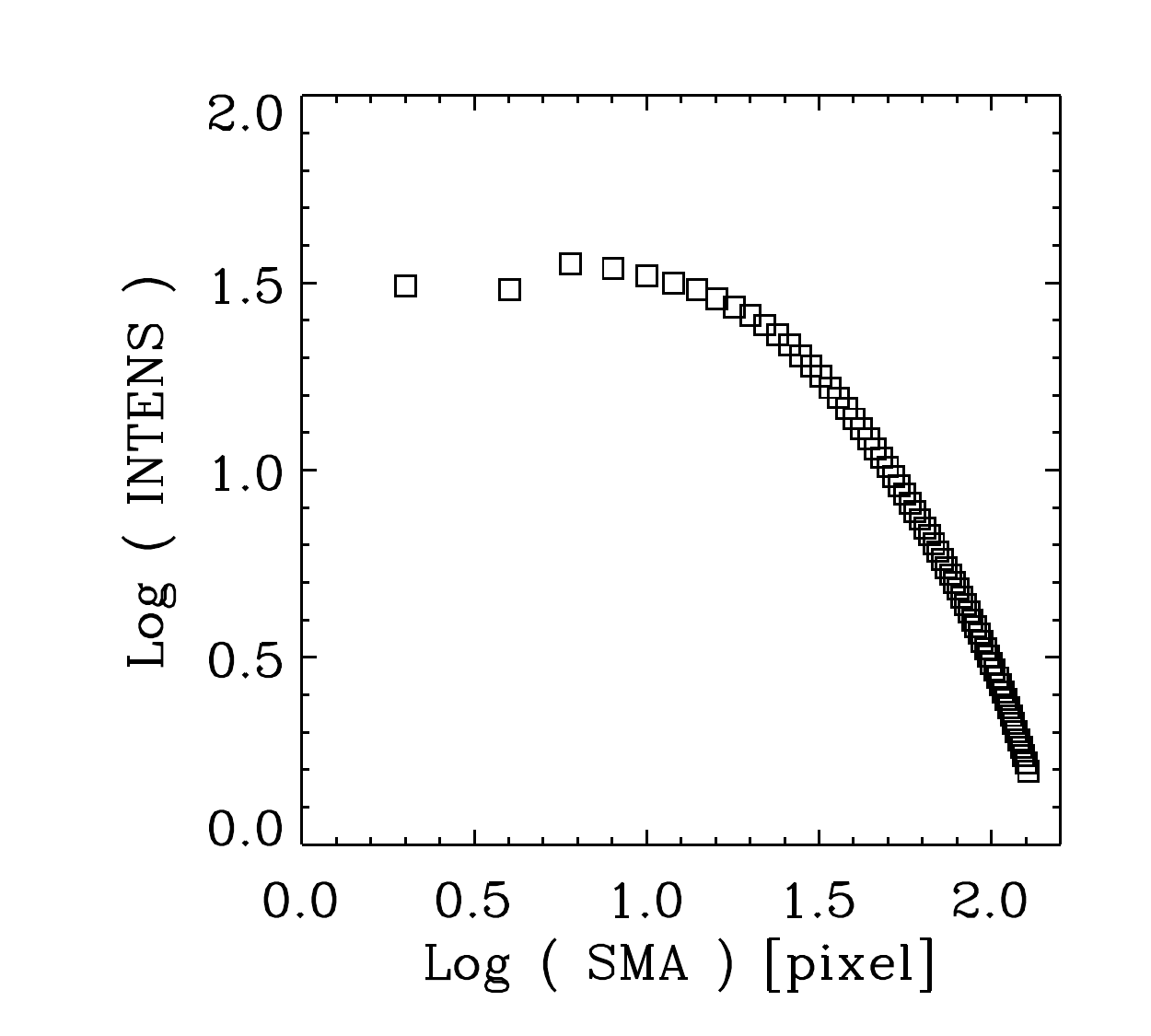}	    &  \includegraphics[trim=0.6cm 0cm 0cm 0cm, clip=true, scale=0.46]{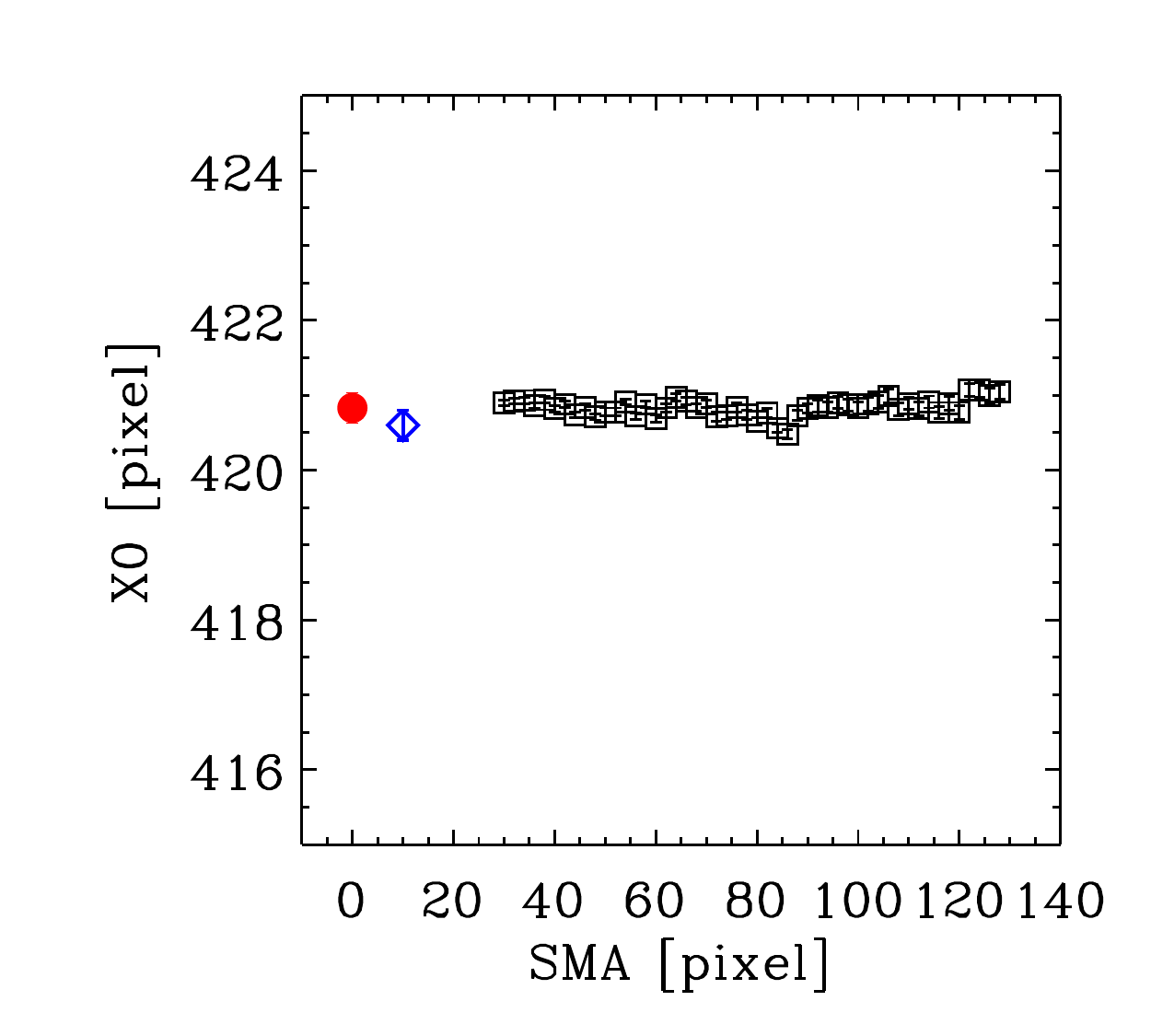}  &  \includegraphics[trim=0.1cm 0cm 0cm 0cm, clip=true, scale=0.46]{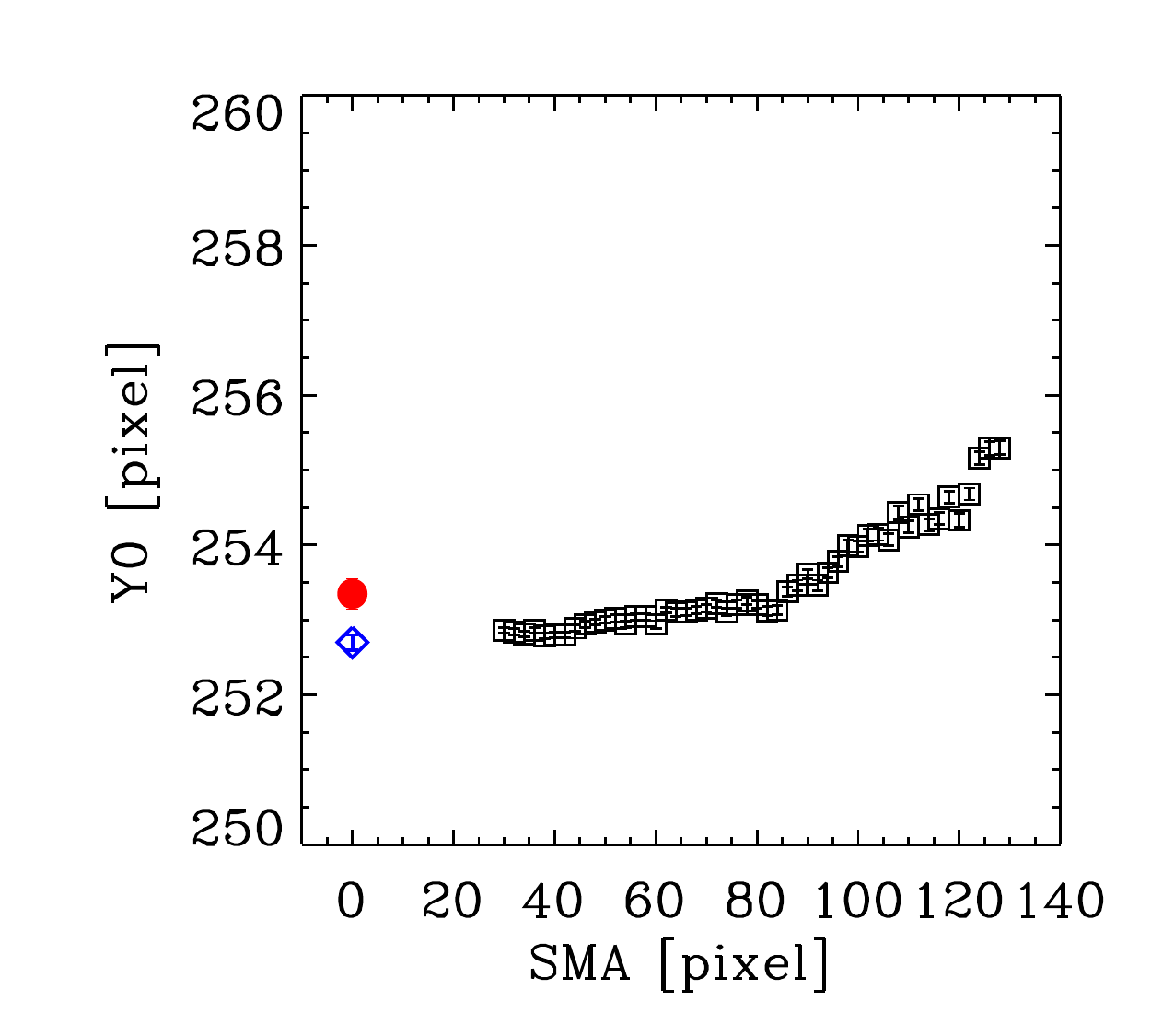} \\	
 \includegraphics[trim=0.1cm 0cm 0cm 0cm, clip=true, scale=0.46]{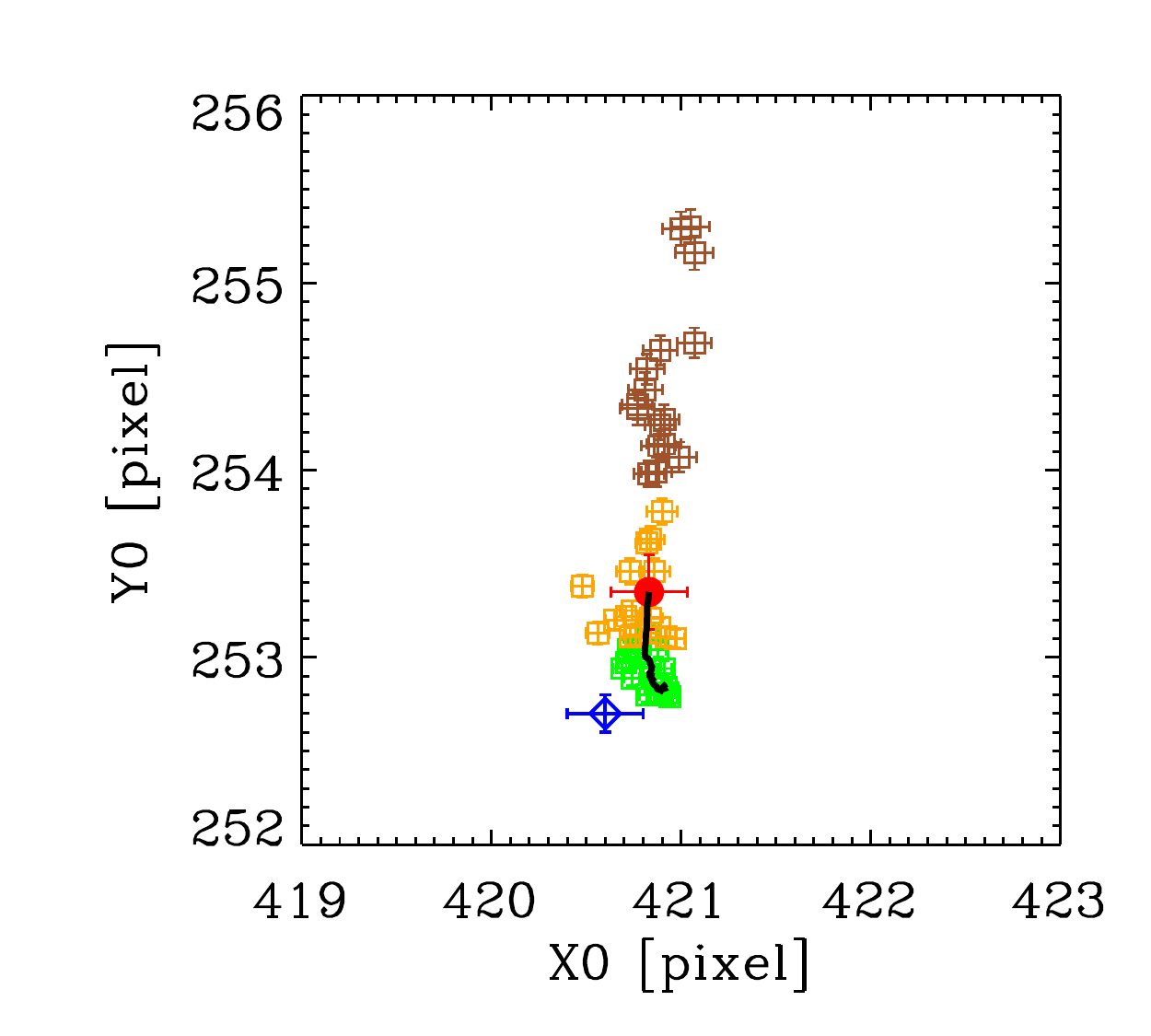}	&  \includegraphics[trim=0.6cm 0cm 0cm 0cm, clip=true, scale=0.46]{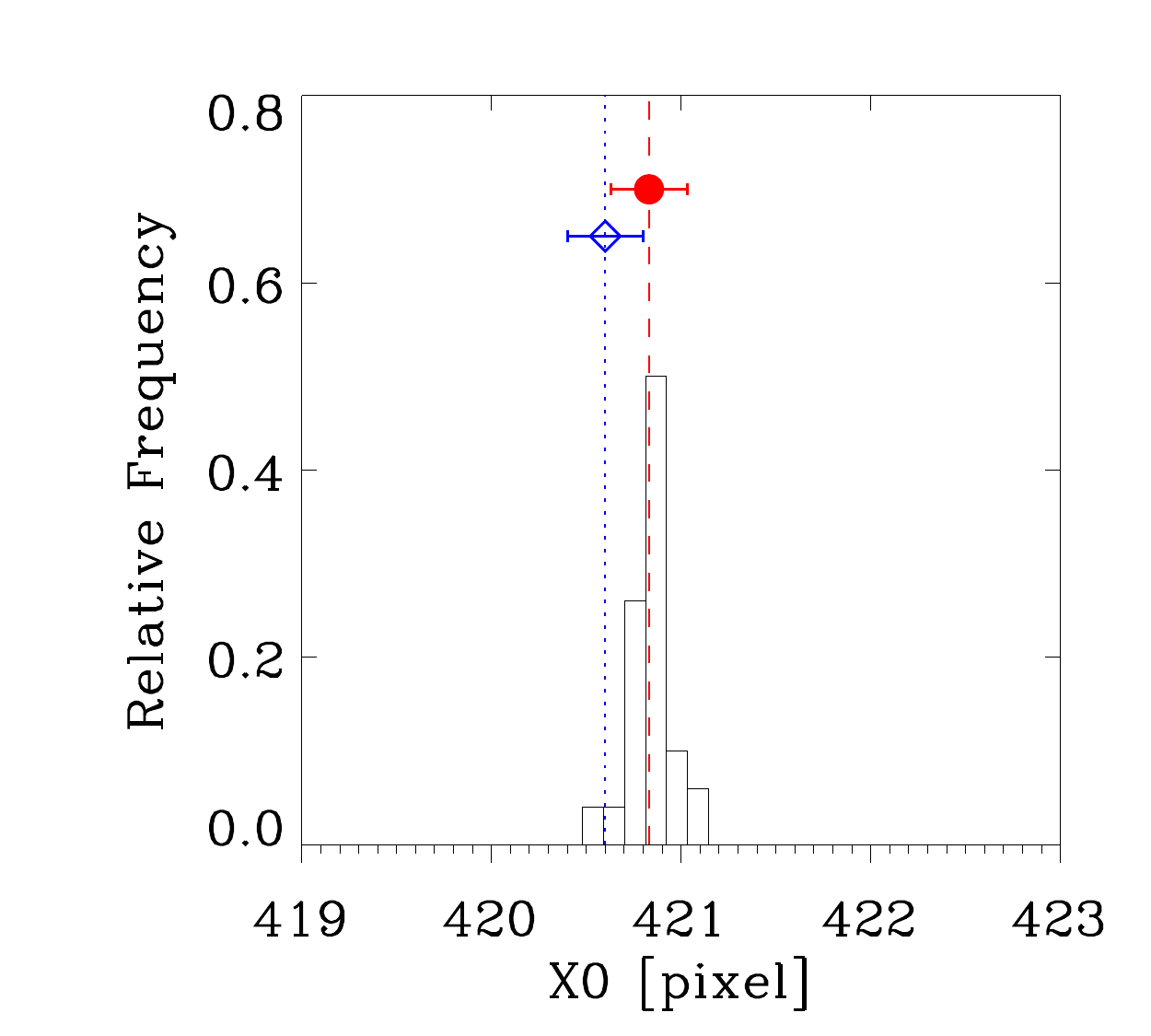}	& \includegraphics[trim=0.6cm 0cm 0cm 0cm, clip=true, scale=0.46]{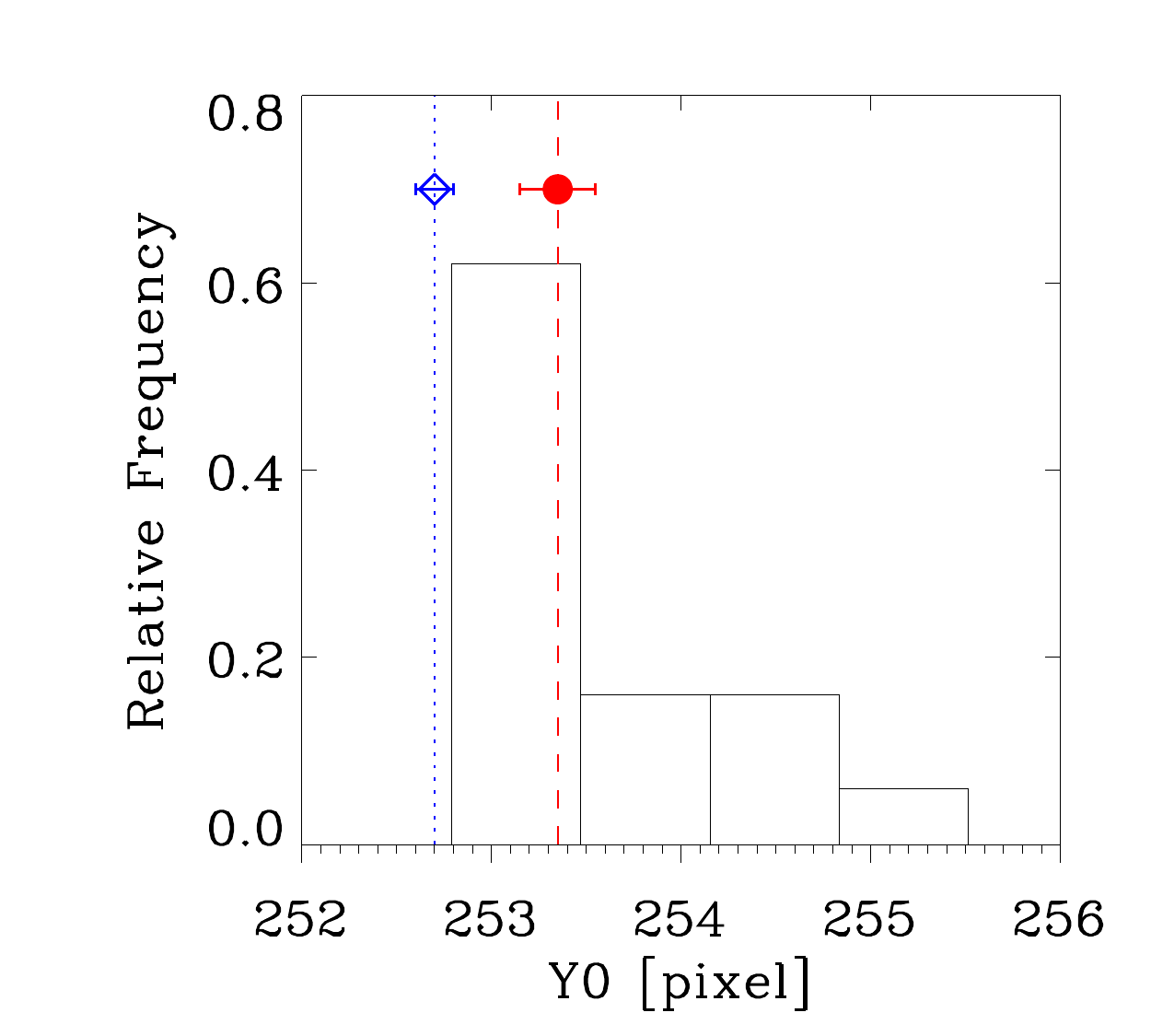}\\
\end{array}$
\end{center}
\caption[IC 4296 (NIC2)]{As in Fig.\ref{fig: NGC4373_W2} for galaxy IC 4296, NICMOS2 - F160W, scale=$0\farcs05$/pxl.}
\label{fig: IC4296_NIC2}
\end{figure*} 

\begin{figure*}[h]
\begin{center}$
\begin{array}{ccc}
\includegraphics[trim=3.75cm 1cm 3cm 0cm, clip=true, scale=0.48]{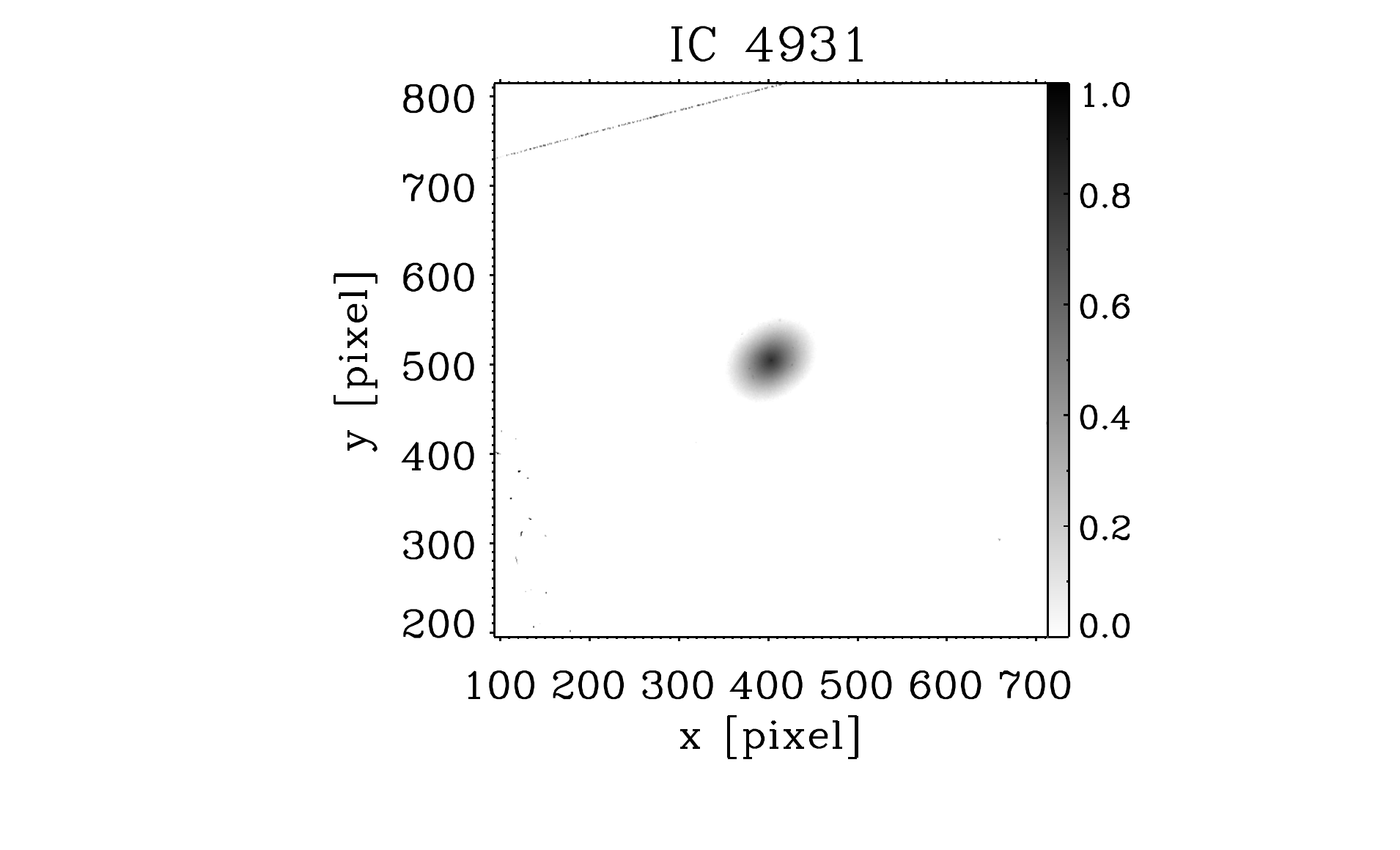} & \includegraphics[trim= 4.cm 1cm 3cm 0cm, clip=true, scale=0.48]{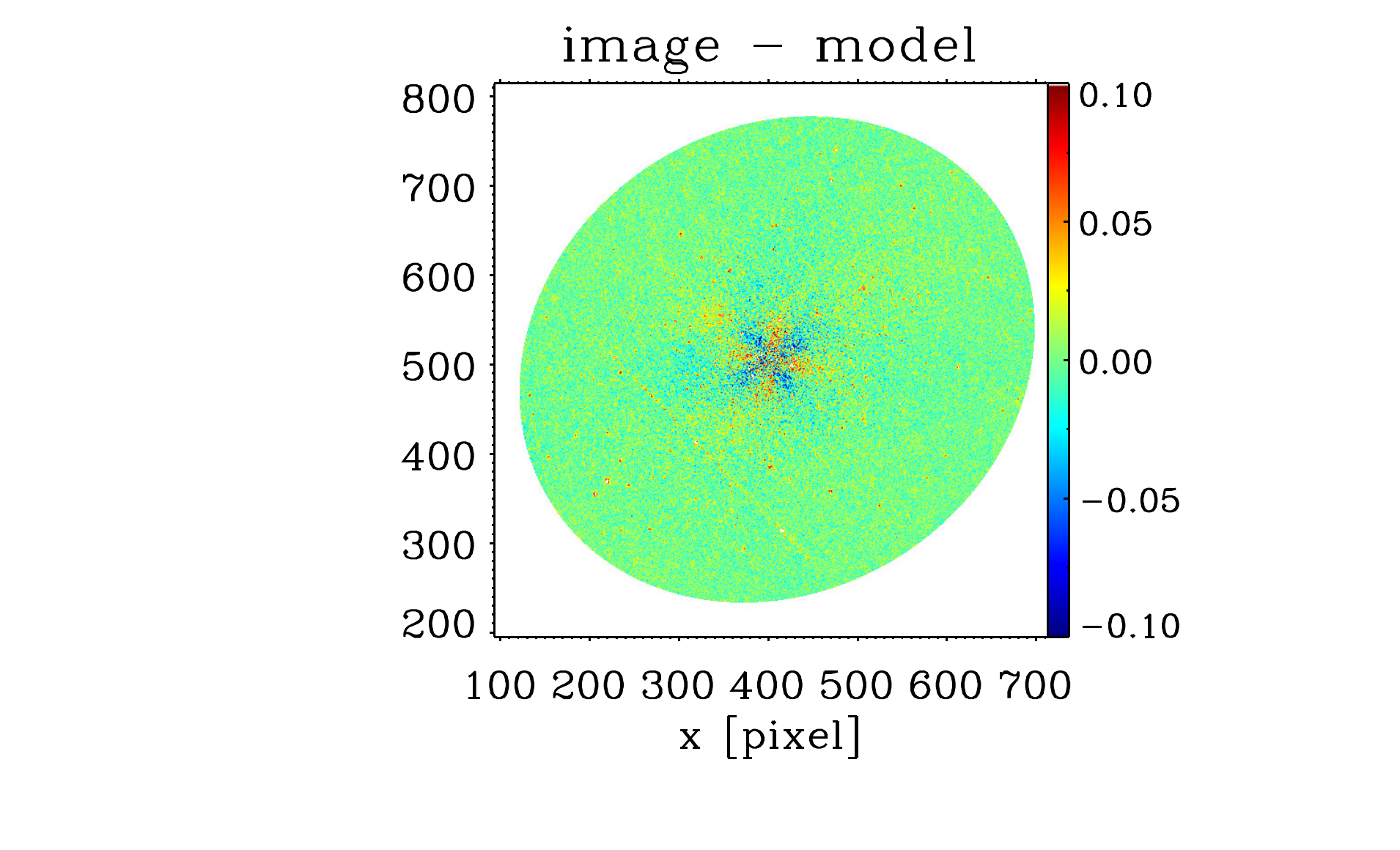}	& \includegraphics[trim= 4.cm 1cm 3cm 0cm, clip=true, scale=0.48]{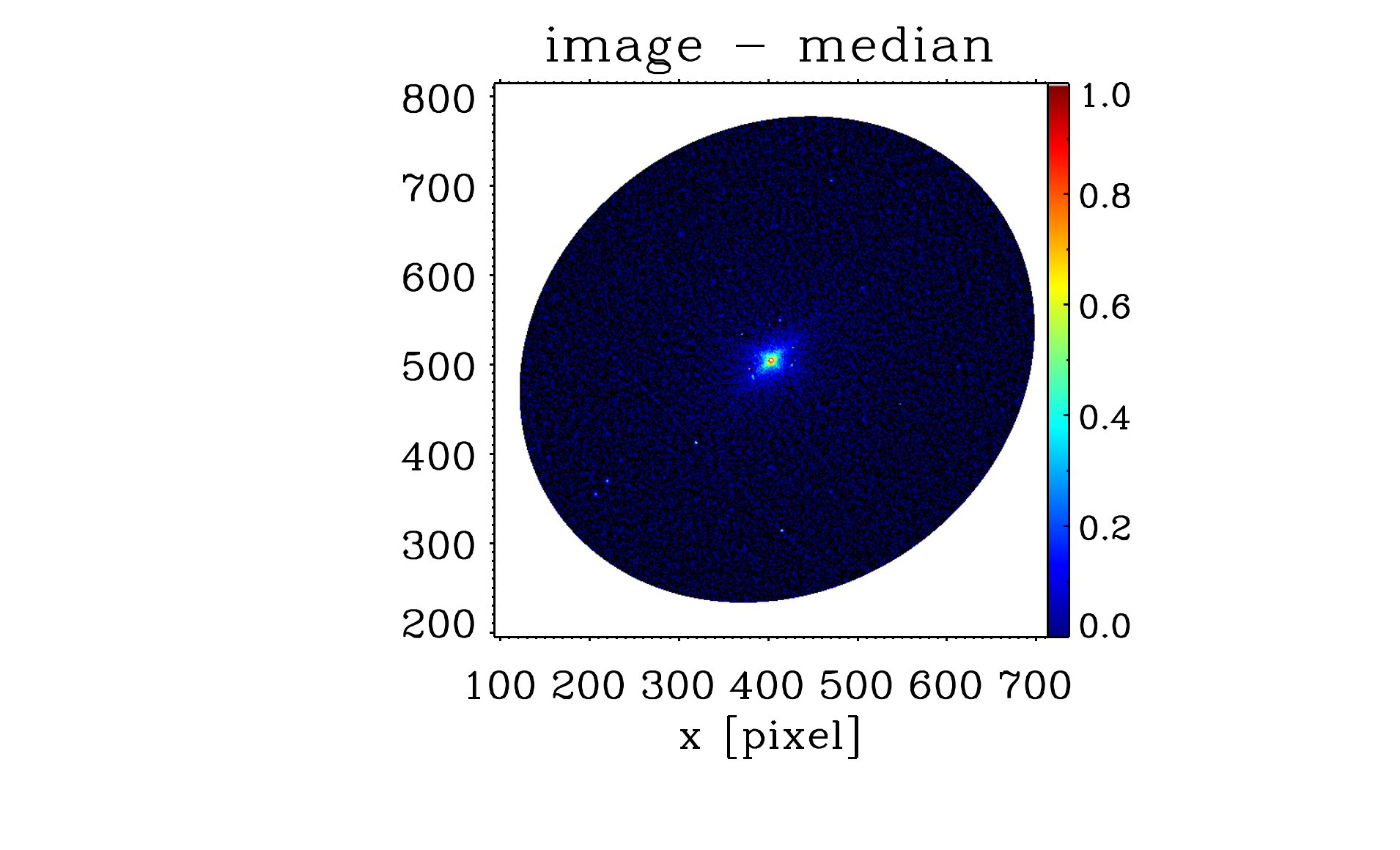} \\
\includegraphics[trim=0.7cm 0cm 0cm 0cm, clip=true, scale=0.46]{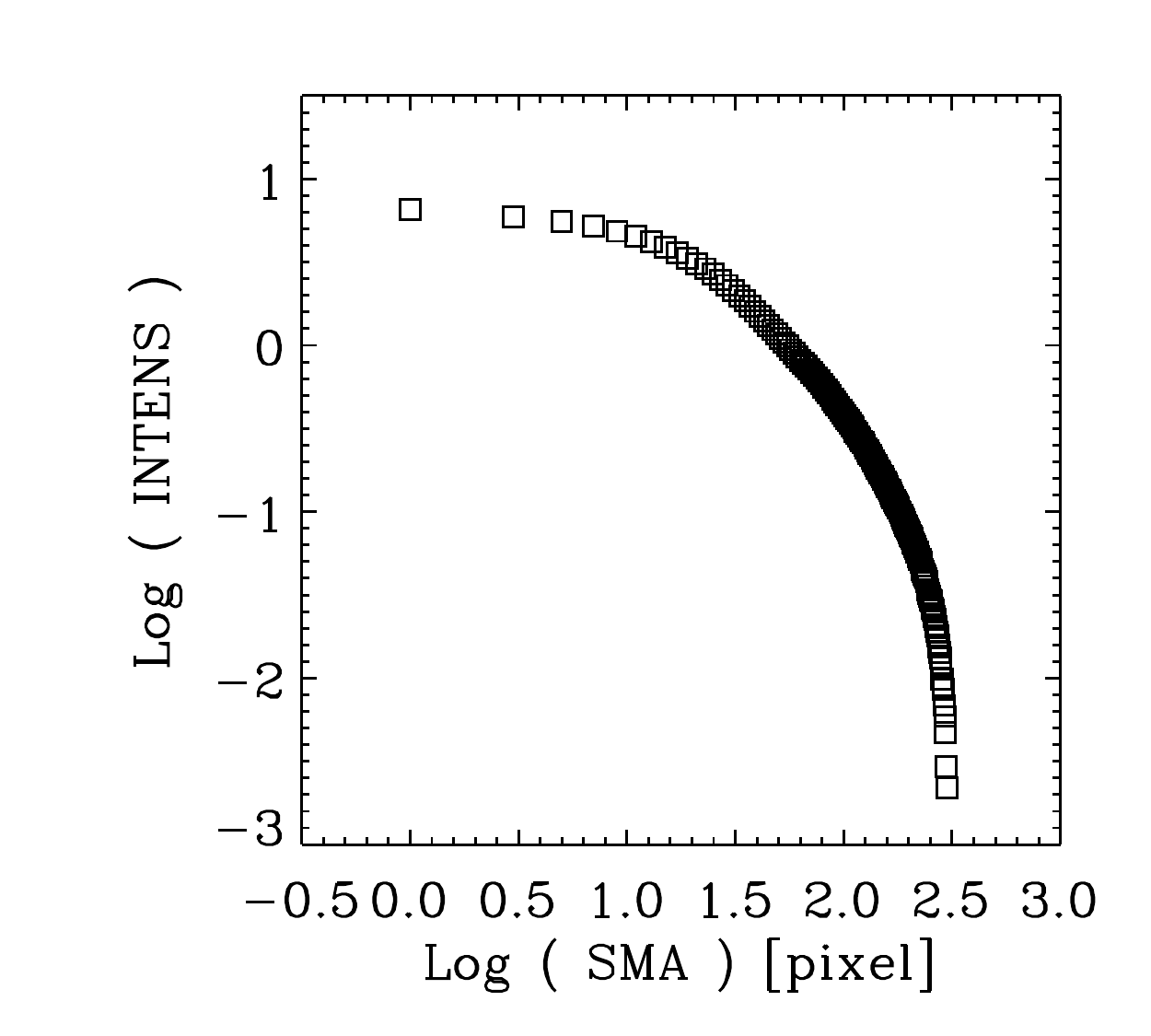}	    &  \includegraphics[trim=0.6cm 0cm 0cm 0cm, clip=true, scale=0.46]{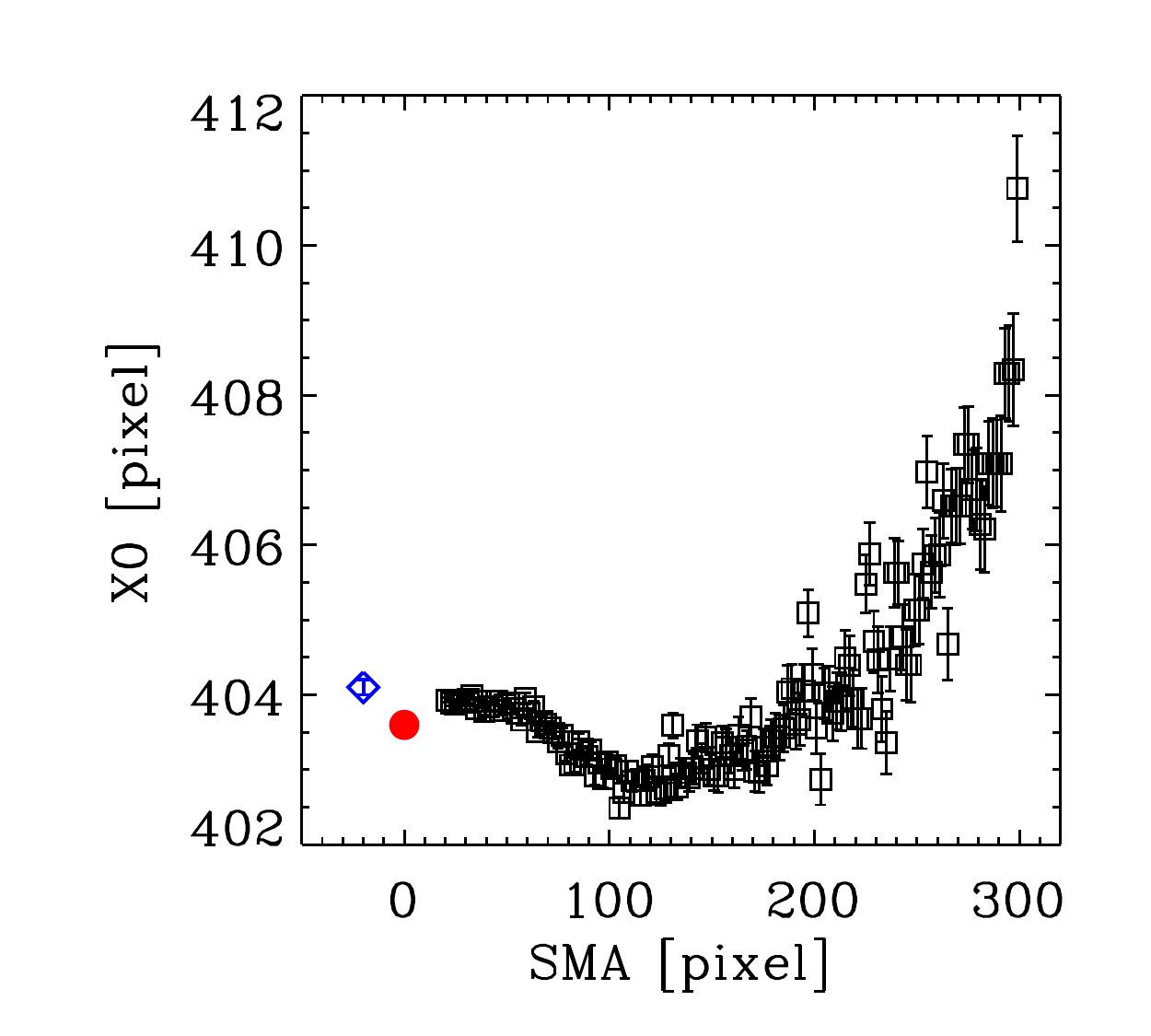}  &  \includegraphics[trim=0.6cm 0cm 0cm 0cm, clip=true, scale=0.46]{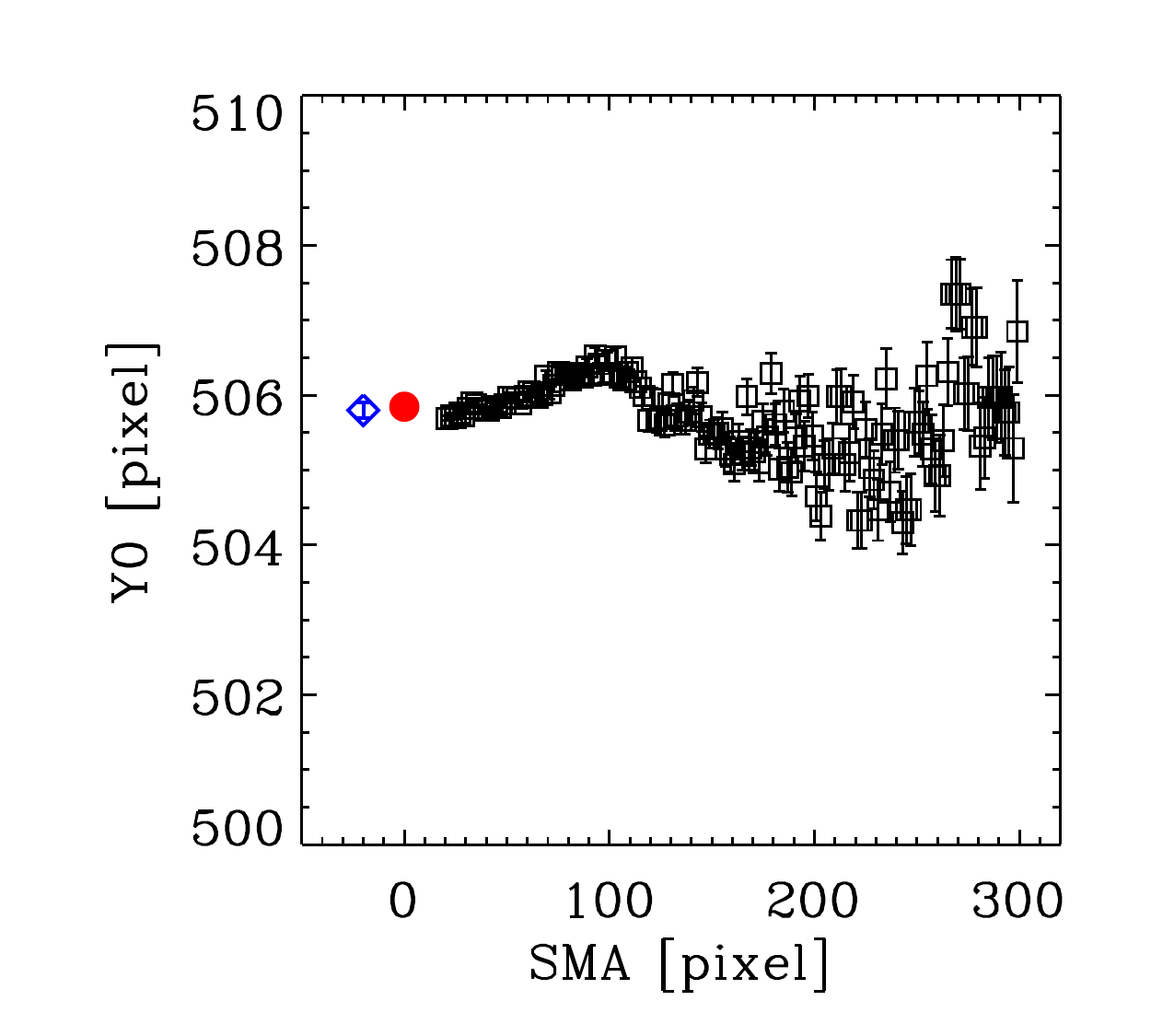} \\	
 \includegraphics[trim=0.65cm 0cm 0cm 0cm, clip=true, scale=0.46]{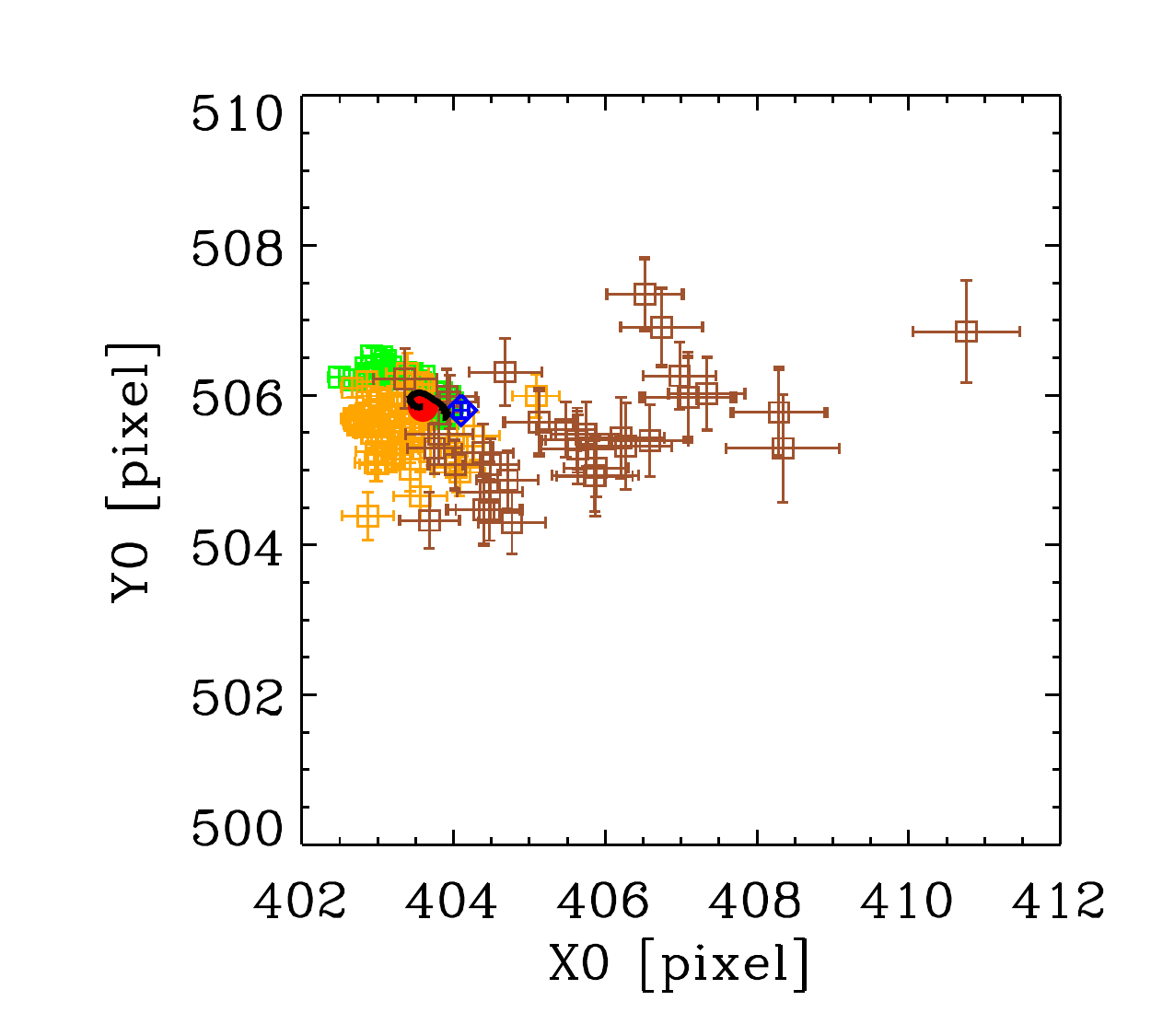}	&  \includegraphics[trim=0.6cm 0cm 0cm 0cm, clip=true, scale=0.46]{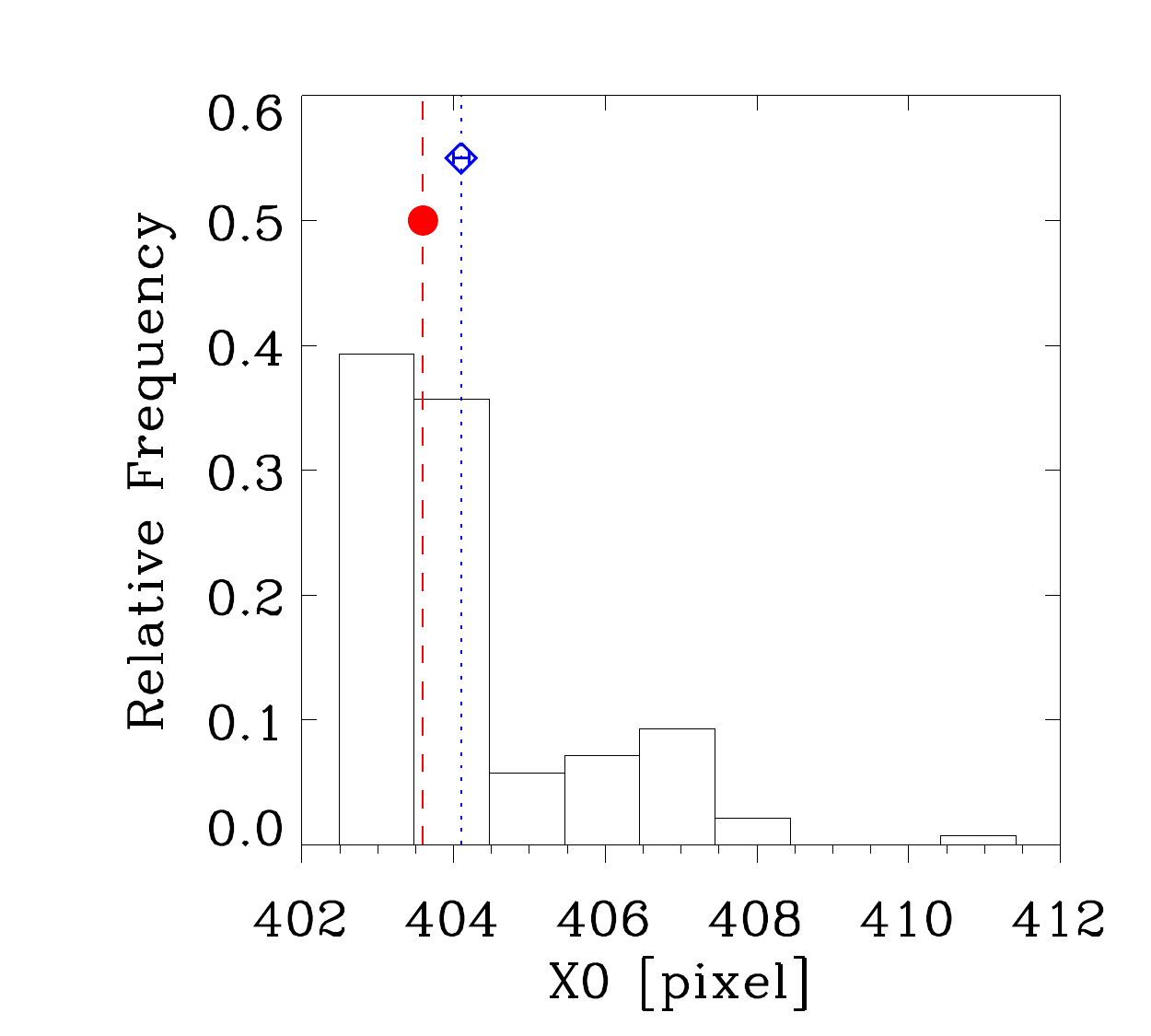}	& \includegraphics[trim=0.6cm 0cm 0cm 0cm, clip=true, scale=0.46]{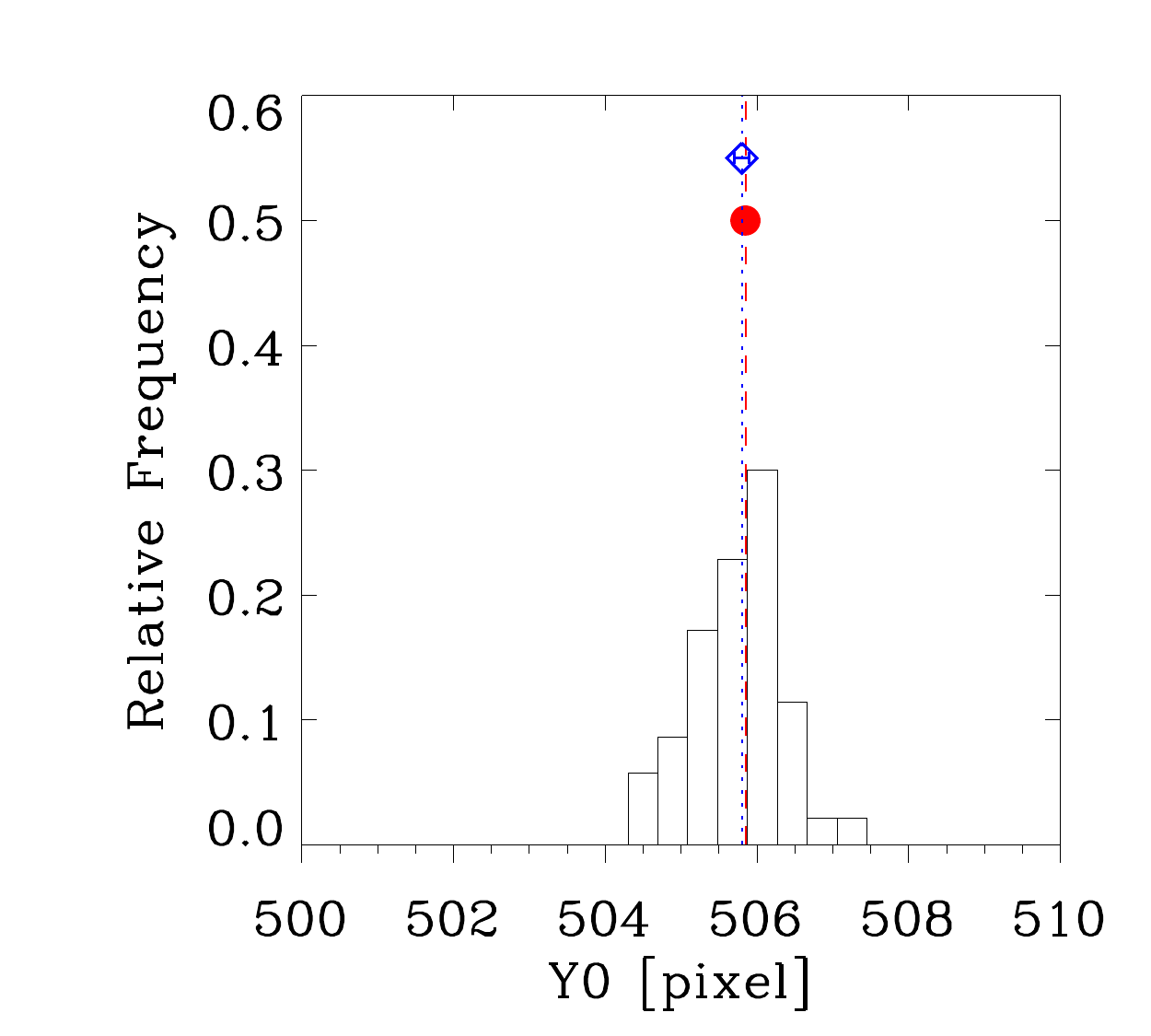}\\
\end{array}$
\end{center}
\caption[IC 4931 (WFPC2_F814W)]{As in Fig.\ref{fig: NGC4373_W2} for galaxy IC 4931, WFPC2/PC - F814W, scale=$0\farcs05$/pxl.}
\label{fig: IC4931_F814W}
\end{figure*} 

\end{appendix}

\end{document}